\documentclass[10pt]{article}
\usepackage[latin1]{inputenc}
\usepackage[english]{babel}
\usepackage{latexsym}
\usepackage{amsmath}
\usepackage{amsfonts}
\usepackage{amssymb}
\usepackage{amsthm}
\usepackage{amsthm}
\usepackage{marvosym} 
\usepackage{fancyhdr}
\usepackage{fancyheadings} 
\usepackage{hyperref}
\usepackage{cleveref}
\ifx\macrosloaded\relax\endinput\else\let\macrosloaded\relax\fi
\newtheorem{theorem}{Theorem}[section]
\newtheorem{prop}[theorem]{Proposition}

\newtheorem{cor}[theorem]{Corollary}

\catcode`\@=11
\newcommand{\eqinsec}{\relax\@addtoreset{equation}{section}}
\renewcommand{\theequation}{\ifx\showlabels\iftrue\the\id\else\thesection.\arabic{equation}\fi}
\catcode`\@=13
\newcounter{supeq}
\newenvironment{subeq}
\stepcounter{equation}
\def\theequation{\ifx\showlabels\iftrue\the\id\else\thesection.\arabic{equation}\fi}
\newtoks\id
\newcommand{\eqlabel}[1]{\label{#1}\global\id={(#1)}} 

\newcommand{\tr}{\mbox{tr}}
\newcommand{\be}{\begin{equation}}
\newcommand{\eeq}{\end{equation}}
\newcommand{\bea}{\begin{eqnarray}}
\newcommand{\eea}{\end{eqnarray}}
\newcommand{\beaa}{\begin{eqnarray*}}
\newcommand{\eeaa}{\end{eqnarray*}}
\newcommand{\bseq}{\begin{subeq}}
\newcommand{\eseq}{\end{subeq}}
\newcommand{\ba}{\begin{array}}
\newcommand{\ea}{\end{array}}
\newcommand{\eql}{\eqlabel}

\def \rectangle#1#2{\hbox{\vrule\vbox to #2
{\hrule\hbox to #1{\hfil}\vfil\hrule}\vrule}}\def\square{\,\,\rectangle{7pt}{7 pt}\,\,}
\newcommand{\edd}{\end{document}}

\renewcommand{\c}{\cdot}
\newcommand{\NI}{\noindent}

\newcommand{\Ubs}{\mbox{${\underline{U}} \mkern-13mu /$\,}}
\newcommand{\Ub}{\underline{U}}

\newcommand{\Lbs}{\mbox{${\underline{L}} \mkern-13mu /$\,}}

\newcommand{\ga}{\gamma}

\newcommand{\Ga}{\Gamma}
\newcommand{\Gab}{\underline{\Gamma}}

\newcommand{\EEb}{\mbox{${\cal E}_1$}}
\newcommand{\EEbb}{\mbox{${\cal E}_2$}}

\newcommand{\gggg}{\mbox{${\bf g}$}}

\newcommand{\dd}{\mbox{${\bf D}$}}

\newcommand{\lapp}{\mbox{$\bigtriangleup  \mkern-13mu / \,$}}
\newcommand{\nab}{\mbox{$\nabla$}}
\newcommand{\nabb}{\mbox{$\nabla \mkern-13mu /$\,}}
\newcommand{\Dbb}{\mbox{$D\mkern-13mu /$\,}}
\newcommand{\partialb}{\mbox{$\partial \mkern-10mu /$\,}}
\newcommand{\Us}{\mbox{$U\mkern-13mu /$\,}}
\newcommand{\Ls}{\mbox{$L\mkern-11mu /$\,}}

\newcommand{\Fs}{\mbox{$F\mkern-13mu /$\,}}

\newcommand{\ddb}{\mbox{$\dd \mkern-13mu /$\,}}

\newcommand{\pr}{\partial}
\newcommand{\hot}{\widehat{\otimes}}

\newtheorem{Le}{Lemma}[section]

\newcommand{\Lie}{\mbox{$\cal L$}}

\newcommand{\lie}{\hat{\Lie}}

\newcommand{\nn}{\nonumber}

\newcommand{\chib}{\underline{\chi}}

\newcommand{\de}{\delta}

\newcommand{\ep}{\epsilon}
\newcommand{\xib}{\underline{\xi}}
\newcommand{\chih}{\hat{\chi}}

\newcommand{\chibh}{\underline{\hat{\chi}}}

\newcommand{\und}[1]{\underline{#1}}

\newcommand{\Nb}{\und{N}}

\newcommand{\Cb}{\und{C}}

\newcommand{\Sch} {Schwarzschild\ }
\newcommand{\ub}{{\und{u}}}
\renewcommand{\c}{\cdot}
\newcommand{\D}{{\cal D}}

\renewcommand{\aa}{\underline{\alpha}}
\newcommand{\bb}{\underline{\beta}}
\renewcommand{\a}{\alpha}
\renewcommand{\b}{\beta}
\newcommand{\dual}{\mbox{}^{\star}\!}

\newcommand{\si}{\sigma}

\newcommand{\ro}{\rho}

\newcommand{\ze}{\zeta}
\newcommand{\divv}{\mbox{div}\mkern-19mu /\,\,\,\,}
\newcommand{\prr}{\mbox{$\pr$}\mkern-11mu /\,}

\newcommand{\curll}{\mbox{curl}\mkern-19mu /\,\,\,\,}

\newcommand{\pih}{\hat{\pi}}

\newcommand{\om}{\omega}
\newcommand{\oom}{\Omega}

\newcommand{\omb}{\underline{\omega}}

\newcommand{\etab}{\underline{\eta}}
\newcommand{\la}{\lambda}

\newcommand{\dddd}{{\bf D} \mkern-13mu /\,}

\newcommand{\ch}{\hat{\chi}}

\def\frac#1#2{{{#1}\over{#2}}}

\newcommand{\QQ}{{\cal Q}}
\newcommand{\QQb}{\underline{\cal Q}}

\renewcommand {\div}{\mbox{div$\mkern 1mu$}}

\newcommand{\acc}{\bar{K}}

\newcommand{\ML}{\!\!\!\!\!\!\!\!\!}
\newcommand{\cbin}[2]{\left(\!\!\!\!\begin{array}{c}{ #1}\\{ #2}\\ \end{array}\!\!\!\!\right)}

\eqinsec
\let\showlabels\iffalse
\begin{document}
\title{ \bf GLOBAL EXISTENCE OF ANALYTIC SOLUTIONS FOR A CLASS OF CHARACTERISTIC EINSTEIN VACUUM EQUATIONS}
\author{G. CACIOTTA\\
 {\small \em Dipartimento di Matematica, Universit\`{a} degli Studi di Roma
``Tor Vergata".}\\
{\small \em Via della Ricerca Scientifica, 00133 Rome, Italy}\\
{\small \em giulioc42\MVAt gmail.com}\\
\\
F. NICOL\` {O}\\
{\small \em Dipartimento di Matematica, Universit\`{a} degli Studi di Roma
``Tor Vergata".}\\
{\small \em Via della Ricerca Scientifica, 00133 Rome, Italy}\\
{\small \em francesco.nicolo\MVAt gmail.com}
\\}\maketitle

\begin {abstract}
\NI {The main goal of this work consists in showing that the analytic solutions for a class of characteristic problems for the Einstein vacuum equations have an existence region larger than the one provided by the Cauchy-Kowalevski theorem, due to the intrinsic hyperbolicity of the Einstein equations. The magnitude of this region depends only on suitable $H_s$ Sobolev norms of the initial data for $s\leq 7$ and if the initial data are sufficiently small  the analytic solution is global. In a previous paper\footnote {See \cite{Ca-Ni:exist2}.}, hereafter ``(I)",  we have described a geometric way of writing the vacuum Einstein equations for the characteristic problems we are considering and a local solution in a suitable ``double null cone gauge" characterised by the use of a double null cone foliation of the spacetime. In this paper, using this ``geometric" gauge, we give a detailed proof of this global existence result. Moreover, as corollary, we prove the global existence of the weak solutions for initial data sufficiently small in  the $H_s$ norms.}
\end{abstract}

\NI {\itshape Keywords:} Vacuum Einstein spacetimes, Characteristic problem; Global solutions.

\bigskip

\NI Mathematics Subject Classification 2010: 83-02

\newpage
\tableofcontents

\newpage
\section{Introduction}\label{S.0}
\NI In this paper we prove a result about the existence region of the analytic solutions of a specific class of characteristic problems, namely those whose ``initial data" are given on a null hypersurface consisting of the union of a truncated outgoing null cone and of a truncated incoming cone intersecting the previous one along a surface  homeomorphic to $S^2$. This class of characteristic problems was studied by different authors, for instance H.Muller Zum Hagen, \cite{Muller},  H.Muller Zum Hagen and H.J.Seifert, \cite{MullerSeifert}, in a series of papers by Dossa, see \cite{Dossa} and references therein, but, in particular,  in the anticipating work by A.Rendall \cite{rendall:charact}, where a thorough examination is done showing how to obtain initial data satisfying the costraint equations and the harmonic conditions and, subsequently, a local existence result is presented. Recently, following, but largely improving the A.Rendall result, one has to recall the paper by Y. Choquet-Bruhat, P.Y.Chrusciel, J.M. Martin-Garcia ``The Cauchy problem on a characteristic cone for the Einstein equations in arbitrary dimensions", see \cite{Choquet Bruhat-Chrusciel-Dossa}. There the authors prove a local existence result for the characteristic problem with initial data on a null cone,  using again the harmonic gauge and proving in a very detailed way how the initial data constraints have to be satisfied and how, relying on Dossa results, \cite{Dossa}, the local existence result can be proved. Moreover the nature of the characteristic problem with initial data on the null cone, adds the extra problem of the ``tip of the cone" which they solve completely.     

\NI In  (I) we showed how to provide an analytic initial data set for a class of characteristic problems and a local solution via the Cauchy-Kowalevski theorem. In the present paper our main goal is to show that the real analytic solutions of the class of problems we are considering have, due to the hyperbolicity of the Einstein equations, a larger existence region than the one proved by the application of the Cauchy-Kowalevski theorem.

\NI More precisely the results of this paper is twofold; first we show how to provide analytic initial data, namely which part of them has to be given in a free way on the whole initial hypersurface and which part has to be given only on the intersection of the incoming and outgoing cones and then obtained on the whole initial hypersurface using the constraint equations. Second, we prove that, given analytic initial data on the null cones, the extension of the analyticity existence region depends only on a finite number of derivatives of the initial data, that is on some appropriate $H_s$ Sobolev norms with a given $s$; hence if we assume the initial data ``small" in these norms, we can prove the global existence for the analytic solutions. Moreover this implies immediately  the global existence of weak solutions with small $H_s$ initial data. At the end of the introduction we will state the two theorems summarising these results. 

\NI Some analogous results have been obtained in the past by  S.Alinhac, G.Metivier,  where they proved the propagation of the analyticity for hyperbolic systems of p.d.e., see \cite{S.Ali} and references therein; \smallskip

\NI In most of the proofs of the global existence a bootstrap argument is exploited: one assumes the spacetime foliated by a family of spacelike hypersurfaces, which amounts basically to choose a gauge, and uses the evolution equations written in this gauge to control the  metric restricted to these hypersurfaces and its derivatives up to a certain order, proving that the solution can be extended beyond the ``last" hypersurface and, by contradiction, the global existence.  Nevertheless when we have to deal with a global existence problem it turns out very problematic, using a foliation made by spacelike hypersurfaces,  to obtain a priori energy-type estimates which can be bounded globally.\footnote{See for instance the Linblad Rodnianski paper, \cite{L-R}, where the ``energy norms" are not bounded, but increase in $t$.} 

\NI Following the seminal ideas of D. Christodoulou and S. Klainerman, \cite{C-K:book},  it turns out that in many instances it appears more appropriate to use a foliation made by a family of null hypersurfaces (generally called cones), outgoing in that case and both outgoing and incoming in \cite{Kl-Ni:book}.  It was in the work of Christodoulou and Klainerman, \cite{C-K:book}, that the crucial use of the foliation in terms of the outgoing cones allowed to prove that  the energy norms, denoted in general with ${\cal Q}$, are bounded everywhere.
\smallskip

\NI The choice of this foliation, we call ``double null canonical gauge",  completely specified later on,  is even more appropriate in the case of characteristic problems where also the ``initial data" are given on null hypersurfaces. Therefore this choice is the first step needed to face and solve our problem. 

\NI These aspects of our result, the construction of the foliation, the explicit expression which the Einstein equations assume, the precise separation between the constraint and the evolution equations and the general strategy to construct the analytic solution and prove its extension to a region larger than the one provided by the Cauchy-Kowalevski theorem are the issues carefully discussed in (I).
The main conclusion proved there is that, assuming the initial data belonging to a function space denoted ${\cal B}_{\a,\ro_0}$\footnote{See \ref{est45abx} for the definition of such space, observe that in (I) ${\cal B}_{\a,\ro}$ is defined with the partial derivatives while here with the covariant ones.} defined in (I) \footnote{Notice that if the initial data belong to ${\cal B}_{\a,\ro_0}$ then they are analytic with radius $\ro_0$ viceversa if they are analytic with radius $\ro_*$ then they belong to ${\cal B}_{\a,\ro_0}$ for suitable $\ro_0<\ro_*$, see lemmas 4.1 and 4.2 of (I) .}, to prove that the Cauchy Kowalevski solution region can be extended, we have to prove that in this region the solution belongs to a function space ${\cal B}_{\a,\ro}$ with $\ro<\ro_0$ depending only on the first derivatives norms, that is to an $H_s$ norms of the solution, with $s\leq J_0<\infty$.\footnote{In the following we choose $J_0=7$,  This is the smallest value we need in order to estimate the first energy norms and apply the inductive mechanism, see also, \cite{Kl-Ni:book} and \cite{Ca-Ni:char}.} 
Moreover if these first derivatives norms are sufficiently small everywhere then the solution is globally analytic. To prove it we need, as said before, that the initial data are small. The proof of this result is the core of this work. 

\newpage

\NI In the remaining part of this introduction we give a survey of all the main technical steps needed to prove this result, which requires the control of all order derivatives of the solution components on the leaves of the ``double null canonical foliation".
\medskip

\NI{\bf 1) The gauge:} This first step was the main content of (I). We just recall, as we said before, that the choice of this foliation made by null outgoing and incoming hypersufaces (cones)  allows to define a natural set of coordinates $\{\nu,\la,\theta,\phi\}$ and look for the Einstein equations solutions as a set of analytic functions in these coordinates. The precise definition of all the quantities defining a solution of the characteristic Einstein problem and the explicit expression of the Einstein equations in this gauge are given in section \ref{S.1}. Here we simply recall that, as discussed at length in (I), the metric of the spacetime has, in this gauge, the following expression, \footnote{Here and in the following we denote in an interchangeable way $u$, $\ub$ and $\la$, $\nu$, the values they take, to denote the null coordinates.}
\bea
{\bf g}=-2{{\oom}}^2(d\la d\nu\!+\!d\nu d\la)\!+\!\ga_{ab}(X^a d\la+d\om^a)(X^bd\la+d\om^b)\ ,\eql{met0int00}
\eea
where $\ga_{ab}$  is the metric induced on the  two-dimensional hypersurface $S(\la,\nu)$, intersection of the outgoing cone $C(\la)$ and the incoming one $\Cb(\nu)$
\[S(\la,\nu)=C(\la)\cap\Cb(\nu)\ ,\]
$\la,\nu$ are the coordinates along the incoming and outgoing cones respectively, $\oom$ has the role of the lapse function in the  gauge associated to the more standard spacelike foliations and $X$ is the analogous of the shift vector and accounts to the non spherical symmetric of the spacetime. The initial data null cones will be indicated as $C_0=C(\la_0)$, for the outgoing one and $\underline{C}_0=\underline{C}(\nu_0)$ for the incoming one. Finally to each point of the spacetime a null orthonormal frame can be associated, $\{e_3,e_4,e_{\theta},e_{\phi}\}$ with $e_3$ and $e_4$ null vectors, ${\bf g}(e_3,e_4)=-2$ and the equivariant vector fields\footnote{See (I) for their main properties.}
\bea
N=\oom e_4=\frac{\pr}{\pr\nu}\ \ ;\ \ \Nb=\oom e_3=\frac{\pr}{\pr\la}+X^a\frac{\pr}{\pr\om^a}
\eea
where $\{\om^a\}=\{{\theta},{\phi}\}$\ .
\smallskip
  
\NI The general solution, denoted by $\Phi$, see (I), is made by various terms,
\[{\Phi}=({\bf g};{\cal O},\underline{\cal O})\equiv({\bf g};{\cal U})=(\ga,\oom,v,\psi, w, X;\ \om,\ze,\chi,\omb,\chib)\ .\]
where
$\ga,\oom,v,\psi, w, X$ are the metric components or their  first angular derivatives, while $\om,\ze,\chi,\omb,\chib$ are the connection coefficients which are, basically, expressions of the remaining first derivatives of the metric components, 
\bea
\chi_{AB}&=&{\bf g}(\dd_{{{e_A}}}{e_4},e_{B})\ \ \ ,\ \chib_{AB}\ =\ {\bf g}(\dd_{{{e_A}}}{e_3},e_{B})\nn\\
\xib_{A}&=&\frac{1}{2}{\bf g}(\dd_{{e_3}}{e_3},e_{A})\ \ \ ,\ 
\xi_{A}=\frac{1}{2}{\bf g}(\dd_{{e_4}}{e_4},e_{A})\nn\\
\omb&=&\frac{1}{4}{\bf g}(\dd_{{e_3}}{e_3},{e_4})\ \ \ ,\ 
\om=\frac{1}{4}{\bf g}(\dd_{{e_4}}{e_4},{e_3})\nn\\
\etab_{A}&=&\frac{1}{2}{\bf g}(\dd_{e_{4}}{e_3},{{e_A}})\ \ \ ,\ 
\eta_{A}=\frac{1}{2}{\bf g}(\dd_{e_{3}}{e_4},{{e_A}})\nn\\
\ze_{A}&=&\frac{1}{2}{\bf g}(\dd_{e_{A}}{e_4},{e_3})\ .\nn
\eea
Moreover we define the initial data set of the metric components and connection coefficients: \[{\Phi^{(0)}}=({\bf g};{\cal O},\underline{\cal O})|_{C_0\cup\underline{C}_0}.\] 
\medskip

\NI {\bf 2) The hierarchy:} As discussed at length in (I) we have to show that our analytic solution defined  in the region ${\cal K}$ is in a ${\cal B}_{\a,\ro}$ function space with a $\ro<\ro_0$ depending only on the first derivatives norms of the solution. More precisely we have to prove the following bounds for all the cone tangential derivatives of the solution components, we denote generically by ${\cal V}$,\footnote{The estimates for $\ga,\oom,v,\psi, w, X$ are slightly different and their precise estimates are given later on.}
\bea
|r^{2+J-\frac{2}{p}}\nab^J{\cal V}|_{p,S}(\la,\nu)\leq C\frac{J!}{J^\a}\frac{e^{(J-2)(\de+\underline{\Ga}(\la))}}{\ro^J}\ .\eql{est45abx}
\eea
The $|\cdot|_{p,S}$ norms are the norms we will use systematically. They are defined as the $L_p$ norms on the surface $S(\la,\nu)$ endowed with the induced metric, with  $p\in[2,4]$, $\nab$ denotes the covariant tangential derivatives\footnote{It is well known that the definition of tangential derivatives along the null cones is in some sense ambiguous.  Here with tangential with respect to the outgoing cone we mean, besides the angular derivatives those along $e_4$ and $e_3$ for the incoming one.}, on the right hand side, $\de>0$ has to be chosen sufficiently large and $\underline{\Ga}(\la)$ is defined in the following section. We can also define the modified $H_s$ spaces associated to the $ L_p$  norms in an obvious way, see  for example (I) section 4.2. \footnote{Notice that it is equivalent to require the smallness in $H_s$ or to require that the constant in front of the right hand side of equation \ref{est45abx} is small for the first $s$ derivatives.}

\NI To prove it we have to  control in a appropriate way the norms of all the tangential derivatives of these components along the outgoing and the incoming cones; therefore along the outgoing cones we have to control the angular derivatives and the derivatives with respect to $e_4$, while for the incoming cones the angular ones and the derivatives with respect to $e_3$.  
 Once these bounds are proved it is immediate to show that the solution belongs to the  ${\cal B}_{\a,\hat{\ro}}$ defined in (I) (with the ordinary partial derivatives)  with $\hat{\ro}<\ro$ and, therefore, to obtain the desired result.

\NI The core of this paper is, therefore, to prove that the estimates \ref{est45abx} hold with $\ro$ depending only on the first $7$ covariant derivatives.
This has to be done in various steps, following a precise order in the estimates of the various $\Phi$ components, which shows a hierarchy in the technical results we need.
\smallskip

\NI We start proving the estimates \ref{est45abx} for ${\cal U}\equiv\{\chi,\ze,\om,\chib, \omb\}$, namely the connection coefficients part of the solution $\Phi$, and more precisely we prove these estimates for their covariant angular derivatives, $\nabb$, see Theorem \ref{L6.1}. This is the first and central result and the largest part of this work; to get it we have to use in a substantial way those structure equations which have the form of transport equations without any ``loss of derivatives"\footnote{see in (I) eqs. (2.39).}; once these estimates are obtained solving these transport equations we can obtain the estimates for all the remaining tangential derivatives in a simpler way using the structure  equations without worrying anymore on the possible ``loss of derivatives" as  we already control all order angular derivatives. Finally once we control the covariant tangential derivatives of the connection coefficients we obtain from them the ordinary partial derivatives of the same quantities completing the estimates we need to prove that the solution belongs to ${\cal B}_{\a,\hat{\ro}}$.
\smallskip

\NI {\bf 3) The extension:} 

\NI As we proved in (I), once the solution belongs to ${\cal B}_{\a,\hat{\ro}}$ in the whole region ${\cal K}$, we assume being the largest analyticity region, we can use these estimates on the upper boundary $\pr{\cal K}_+$ of $\cal K$, to extend, via Cauchy-Kowalewski, the analytic solution in a small strip above it. This implies that ${\cal K}$ is not the largest possible analyticity region and, therefore, the solution is global. It is important to remark that  to claim that the solution is global, we need that the a priori estimates for the energy-type quantities associated to the $H_s$ solutions with $s\leq 7$, hold in the whole spacetime. This requirement is satisfied if the first 7 derivatives initial data are   sufficiently small. If viceversa these initial data are not small, these a priori estimates hold only in a finite region, ${\overline{\cal K}}$,  smaller as the initial data are larger and in this case the analyticity region $\cal K$ cannot be extended beyond ${\overline{\cal K}}$.
\medskip

\NI Due to the complexity of this work we give in the rest of this section some more details of all the main steps quoted before, which will be treated exhaustively in the subsequent sections.
\medskip

\NI{\bf  The angular derivatives:} As we said, to estimate the covariant angular derivatives of the connection coefficients we use those structure equations which have the form of transport equations. To control the not underlined connection coefficients we use the transport equations along the outgoing cones, while for the underlined quantities those along the incoming cones.\footnote{With the exception of $\om$ and $\omb$, see later.} The way to obtain the norm estimates is an inductive one, we assume to control the angular derivatives norms up to order $J\leq N-1$\footnote{To be precise, for technical reasons, we will assume inductive estimates for the $\Lie_O^J$ angular derivatives of the connection coefficients.} and we prove the analogous estimates for $J=N$. This procedure to be satisfied requires, therefore, the use of those transport equations where no ``loss of derivatives" is present. There are three important aspects which have to be emphasised here as they are deeply entangled with the mathematical  structure of the equations we are solving. 
\smallskip

\NI i) A clear example of the first aspect, present in all the situations when  trying to solve the transport equation for the connection coefficients, appears when we look for the estimate of the trace part of $\chi$, $\tr\chi$, and its angular derivatives. Its transport equation, written for $U=\oom^{-1}\tr\chi$,\footnote{The advantage to look at $U$ instead of $\tr\chi$ will be clarified in the next section.}
\bea\label{51}
\frac{\partial{U}}{\partial\nu}+\frac{{\oom}{\tr\chi}}{2}{U}+|\hat{{\chi}}|^2\!=\!0\ ,\eql{109bisaa}
\eea
depends on the traceless part, $\chih$, of $\chi$ and, looking at the structure equations, see for instance (I) subsection 6.1.1, it is immediate to realise that there is not  a transport equation for $\chih$ without a loss of derivatives.

\NI To overcome this problem we remark that the structure equations provide us also with some elliptic Hodge equations defined on the generic two dimensional surface, $S$, intersection of an outgoing cone and an incoming cone of the foliation. In particular for $\chih$ we have the following elliptic equations
\medskip

\bea\label{52}
&&{\divv}\hat{\chi}=\frac{1}{2}\nabb{\tr\chi}-\b-\zeta\!\c\hat{\chi}+\frac{1}{2}\zeta{\tr\chi}\eql{fdef1aa}
\eea
Nevertheless to use this equation to control $\chih$ in \ref{109bisaa} we have to control the null Riemann components, in the present case $\b$. This is where the hyperbolic nature of the Einstein equations plays its important role. In fact to treat \ref{fdef1aa} as an equation and not as the definition of the Riemann component $\b$ we need an independent way to control the null Riemann components. 
\smallskip

\NI ii) The hyperbolic structure is the second important aspect, provided by the hyperbolic nature of the Einstein equations which allows to obtain some a priori estimates for some energy-type quantities, $L^2$ norms of the Bel-Robinson tensor and suitable Lie derivatives, denoted hereafter in general with $\cal Q$ and defined in any detail in Section \ref{S.4.1}. 

\NI These ``a priori energy estimates" allow to control the Riemann components, in the previous case $\b$, in terms of the analogous initial data quantities, therefore, to control via equations \ref{fdef1aa} $\chih$ and its angular derivatives and, finally, through the transport equation \ref{109bisaa}, $\tr\chi$. Clearly these estimates need, to be performed, the boundedness of the derivatives of the connection coefficients and it is here that the bootstrap mechanism enter.
The mechanism just sketched, discussed in greater detail in subsection \ref{SS3.2a}, reproduces itself, with some technical, but not substantial, differences to control all the remaining connection coefficients and their angular derivatives.
It is the basic use of these a priori estimates which implies that the analytic solution can be extended up to the  region where these a priori estimates hold. 
\smallskip

\NI iii) The third aspect to emphasise is that looking at the transport equations, one realises that to control, via the inductive mechanism, all the angular derivatives of the connection coefficients we need to control the asymptotic decay of these quantities along the outgoing cones. This requires, for technical reasons, that we have to integrate the outgoing cones transport equations from above and that, as discussed in any detail in subsections 9.5, 9.6.1 and in subsection 14.11, the foliation of the spacetime has to be chosen in an appropriate way. This is the reason of the introduction of the so called ``double null cone canonical foliation" which, beside its technical need to prove our result, basically satisfies the request that the outgoing cones should coincide asymptotically with those of the \Sch spacetime. \medskip

\NI {\bf  The mixed derivatives and the metric components derivatives:} As said before discussing the natural hierarchy of the steps required to prove the result, once we control the angular derivatives of the connection coefficients, the subsequent step is to control the $\dd_4$, $\dd_3$ derivatives of all order and the mixed ones with the angular derivatives. This will require to introduce the transport equations for the $\Lie_S^p\Lie_O^q$, derivatives of the connection coefficients and to define a class of extended energy norms, in order to repeat all the steps performed for the $\Lie_O^q$ derivatives for the $\Lie_S^p\Lie_O^q$ ones. The last step will be to obtain estimates for the `` not ordered" mixed derivatives.  A different way of obtaining the mixed derivatives estimates can be performed, see the remark after Theorem \ref{L2.1newnew}, emphasising the central role of the angular derivatives.

\NI Once we control all the covariant derivatives of the connection coefficients using equations similar to \ref{51} and \ref{52} it is a long, but easy procedure to obtain the estimates for the metric components; this is the content of Section \ref{S.5nx}.
\medskip

\NI {\bf  The initial data:} Finally, in Section \ref{S.s initial data}, 
we present a careful discussion of how obtaining the class of (global) initial data required to have global analytic solutions. While here we present the explicit way to construct initial data on the whole (unbounded) initial data null hypersurfaces, we refer to (I) for the general discussion of how, in our double null canonical gauge, the constraints appear and have to be satisfied.

\NI We end this introduction stating two theorems summarising our results; they are given here in a compact form and the following sections are devoted to state their precise versions and their proofs.\footnote{see theorem \ref{Thinitialdata}}

\begin{theorem}\label{main1}
Let us assign $\chih$ on $C_0$, $\chibh$ and $X$ on $\Cb_0$ and $\oom$ on $C_0\cup\Cb_0$ analytic in ${\cal B}_{\a;\ro_{0,0,1}}$, moreover assume assigned \footnote{{\bf K} is the curvature of $S_0$} $\{\ga, \zeta,\tr\chi, \tr\chib,\omb,{\bf K}\}$, analytic and satisfying suitable equations on $S_0$. If all these coefficients satisfy suitable smallness bounds in $|\cdot |_{p,S}$ norms together with their first s tangential derivatives, $s\leq 7$, then it is possible to construct analytic initial data $\Phi^0$ on $C_0\cup\Cb_0$ in  ${\cal B}_{\a;\ro_0}$, with $\a>3$, $\ro_0<\ro_{0,0,1}$ such that  the energy type ${\cal Q}^0$  norms\footnote{see definition \ref{qnorms}} on $C_0\cup\Cb_0$ are  finite and small\footnote{the precise version of this theorem is in section \ref{S.s initial data} }.
\end{theorem}
\smallskip

\begin{theorem}\label{main2}
Let  us give $\Phi$, a real analytic solution of the characteristic problem for the Einstein vacuum equations in the region\[{\cal K}(\Lambda_a,\Pi_a)\equiv\{(\la,\nu)\in[\la_0,{\Lambda_a}]\times[\nu_0,{\Pi_a}]\}\ ,\] domain of dependance of $C_0(\la,[\nu_0,\Pi_a])\cup \underline{C}_0([\Lambda_0,\la_0],\nu)$, with initial data set   $\Phi^0$ belonging to the Banach space ${\cal B}_{\a;\ro_0}$, with $\a>3$.
Then there exists a double null foliation, the ``double null canonical foliation",\footnote{Its explicit definition is in \cite{Kl-Ni:book}, Chapter 3, see also \cite{C-K:book} where it was first introduced. Its definition and some specific details for the present case are given in subsections 9.5 and 14.11.} of the region ${\cal K}(\Lambda_a,\Pi_a)$ such that the generic connection coefficient related to $\Phi$ and associated with such foliation, belong to ${\cal B}_{\a,\hat{\ro}}$, with $\hat{\ro}<\ro_0 
$ depending only on the $|\cdot |_{p,S}$ norms of the first $s$ derivatives, $s\leq 7$, of the initial data $\Phi^0$. Moreover if we give $\Phi^0$ such that the energy type ${\cal Q}^0$ norms on $C_0\cup\Cb_0$ are finite and sufficiently small, we can construct an analytic solution in all the domain of dependence of  $C_0\cup \underline{C}_0$ belonging to  ${\cal B}_{\a,\hat{\ro}}$.
\end{theorem}
\NI From these two theorems the following corollary can be derived:

 \begin{cor}
{ Given $\chih$ on $C_0$, $\chibh$ and $X$ on $\Cb_0$ and $\oom$ on $C_0\cup\Cb_0$, satisfying suitable smallness bounds in $|\cdot |_{p,S}$ norms with their first $s$ derivatives, $s\leq 7$, and $\{\ga, \zeta, \tr\chi, \tr\chib, \omb,{\bf K}\}$  on $S_0$, satisfying suitable equations and small in $|\cdot |_{p,S_0}$, with their first $s$ angular derivatives, we can construct a weak global solution, in $|\cdot |_{p,S}$ norms with their first $s$ derivatives, in the whole domain of dependence of $C_0\cup \Cb_0$.}
\end{cor}
\bigskip

\NI{\bf Remark:} 

\NI {\em Notice that in the two theorems and the corollary, the smallness conditions for the initial data are not precisely specified, this is because they are related to the smallness of the ${\cal Q}^0$ energy norms, this smallness, on the other hand, is determined in a precise but not straightforward way in \cite{Kl-Ni:book}, chapter 6.  These to types of smallness are intimately related, hence we we refer to them generally with $O(\ep)$. }

\medskip

\NI{\bf Acknowledgments:} The idea propagate as much as possible the analyticity of the solutions of the Einstein equations was suggested, years ago, to one of the author (F.N.) by S.Klainerman and was implemented together on the ``toy model" of the Burger equation in \cite{Kl-Ni:rew}; moreover both present authors are pleased to thank S.Klainerman one for the long and rich collaboration and (G.C.) for the useful period spent in the Math. $\!\!$Department of the Princeton University and for the long and profitable conversations he had with him.
\newpage

 \section{The Einstein equations in the double null canonical gauge} \label{S.1}
 We collect here some of the results proved in (I) to make this work more complete and to introduce all the notations systematically used in the following. In the chosen gauge, see (I) for details, the evolution part of the Einstein equations is made by  the  following system of first order equations for the various tensors components omitting the indices to simplify the notations, \footnote{In the first line of equations $\om$ denotes the angular coordinates.}
\bea
&&{\ML\ML}\frac{\partial\ga}{\partial\om}-v=0\ \ ,\ \ \frac{\partial\log\oom}{\partial\om}-\psi=0\ \ ,\ \ \frac{\partial\hat{X}}{\partial\om}-w=0\nn\\
&&{\ML\ML}\frac{\partial{\ga}}{\partial\la}-2{\oom}\ \!{\chib}+\Lie_{X}{\ga}=0\nn\\
&&{\ML\ML}\frac{\partial\log{\oom}}{\partial\la}+\psi(X)+{2{\oom}}\ \!{\omb}=0\nn\\
&&{\ML\ML}\frac{\partial v}{\partial\la}+\nabb_Xv+(\partialb X)\c v-S(\partialb\!\otimes\!w)-2\oom\partialb\!\otimes\!\chib-2\oom\ \!\psi\!\otimes\!\chib=0\eql{5.37la}\\
&&{\ML\ML}\frac{\partial\psi}{\partial\la}+\nabb_X\psi+2\oom\omb\psi+\psi(\nabb X)+2\oom\partialb\omb=0\nn\\
&&{\ML\ML}\frac{\partial{\tr\chi}}{\partial\la}+{\oom}{\tr\chib}{\tr\chi}-2{\oom}{\omb}{\tr\chi}+{\nabb}_{\! X}{\tr\chi}
-2{\oom}{\divv}{(\ze+\psi)}-2{\oom}|{\ze+\psi}|^2 +2{\oom}{\bf K}\!=\!0\nn\\
&&{\ML\ML}\frac{\partial\hat{{\chi}}}{\partial\la}+\Lie_{{X}}\hat{{\chi}}-\frac{{\oom}{\tr\chib}}{2}\hat{{\chi}}+\frac{{\oom}{\tr\chi}}{2}\hat{{\chib}}
-{2\oom}\ \!{\omb}\ \!\chih
-{\oom}(\hat{{\chi}}\c\hat{{\chib}})\ga-{\oom}\ \!{\nabb}\hot{(\ze+\psi)} -{\oom}(\ze+\psi)\hot(\ze+\psi)\!=\!0\nn\\
&&{\ML\ML}\frac{\partial{\ze}}{\partial\la}+{\Lie_{X}{\ze}}+{\oom}\ \!{\tr\chib}{\ze}+{\oom}{\divv}{\chibh}-\frac{1}{2}{\oom}{\partialb}{\tr\chib}
\left.+2\oom\omb\psi+2\oom\partialb\omb\right.+{\oom}\psi\!\c\!{\chib}\!=\!0\nn\\
&&{\ML\ML}\frac{\partial{{\om}}}{\partial\la}\!+\!{\partialb}_{\! X}{\om}\!-\!2{\oom}\ \!\!{{\omb}}\
\!{{\om}}\!-\!\frac{3}{2}{{\oom}}|\ze|^2\!+\!\frac{1}{4}{{\oom}}\ze\!\c\!\psi\!+\!\frac{1}{2}{\oom}|\psi|^2
\!+\!\frac{1}{2}{\oom}\!\left({{\bf K}}\!+\!\frac{1}{4}{\tr\chi}{\tr\chib}\!-\!\frac{1}{2}\hat{{\chi}}\!\c\!\hat{{\chib}}\!\right)=0\nn\\
&&\nn\\
&&{\ML\ML}\frac{\partial\hat{X}}{\partial\nu}+4{\oom}^2{\ze}=0\nn\\
&&{\ML\ML}\frac{\partial w}{\partial\nu}+8\oom^2\psi\!\otimes\!\ze+4\oom^2\partialb\!\otimes\!\ze-2\oom\psi\!\otimes\!(\chi\!\c\!X)
-2\oom(\partialb\!\otimes\!\chi)\!\c\!X=0\nn\\
&&{\ML\ML}\frac{\partial{\tr\chib}}{\partial\nu}+{\oom}{\tr\chi}{\tr\chib}
-2{\oom}{\om}{\tr\chib}+2{\oom}{\divv}\ze\!-\!2{\oom}{\divv}\psi-2{\oom}|\ze\!-\!\psi|^2+2{\oom}{\bf K}\!=\!0\eql{2.70gqla}\\
&&{\ML\ML}\frac{\partial\hat{{\chib}}}{\partial\nu}-\frac{{\oom}{\tr\chi}}{2}\hat{{\chib}}
+\frac{{\oom}{\tr\chib}}{2}{\chih}-2\oom\om\chibh-{\oom}(\hat{{\chib}}\c\hat{{\chi}})\ga+{\oom}{\nabb}\hot(\ze\!-\!\psi)
-{\oom}(\ze\!-\!\psi)\hot(\ze\!-\!\psi)\!=\!0\nn\\
&&{\ML\ML}\frac{\partial{{\omb}}}{\partial\nu}-2{\oom}\ \!\!{{\om}}\ \!{{\omb}}-\frac{3}{2}{{\oom}}|\ze|^2
\!-{{\oom}}\ze\!\c\!\psi\!+\!\frac{1}{2}{\oom}|\psi|^2
\!+\!\frac{1}{2}{\oom}\!\left({{\bf K}}\!+\!\frac{1}{4}{\tr\chib}{\tr\chi}
\!-\!\frac{1}{2}\hat{{\chib}}\!\c\!\hat{{\chi}}\!\right)=0\ ,\nn
\eea
where $\ga,\oom,X$ are the components of the metric which has the following expression 
\bea
{{\bf g}}=-2{{\oom}}^2(dud\ub\!+\!d\ub du)\!+\!\ga_{ab}(X^adu+d\om^a)(X^bdu+d\om^b)\ .\eql{met0int0}
\eea
The solution of the characteristic problem specified by the equations \ref{5.37la} and \ref{2.70gqla} becomes a solution of the Einstein equations if the initial data satisfy the following equations, which have to be interpreted as constraint equations,\footnote{See (I), Theorem 2.1, for a detailed discussion of this aspect.}
\bea
&&\ML\ML\mbox{\bf On $C_0$\ :}\ \ \ 
\frac{\partial{\ga}}{\partial\nu}-2{\oom}{\chi}=0\ \ ,\ \ \frac{\partial\!\log{\oom}}{\partial\nu}+{2{\oom}}\om=0\nn\\
&&\ \ \ \ \frac{\partial{{\tr}\chi}}{\partial\nu}+\frac{{\oom}{{{tr}}\chi}}{2}{\tr\chi}+2{\oom}{\om}{\tr\chi}+{\oom}|\hat{{\chi}}|^2\!=\!0\nn\\
&&\ \ \ \ \frac{\partial{\ze}}{\partial\nu}+{\oom}\ \!{\tr\chi}{\ze}-{\oom}{\divv}{\chi}+{\oom}{\partialb}\!{\tr\chi}
+\frac{\partial{\partialb}{\log\oom}}{\partial\nu}-{\oom}{\partialb}\!\log{\oom}\!\c\!{\chi}\!=\!0\nn\\
&&\nn\\
&&\ML\ML\mbox{\bf On $\Cb_0$:}\ \ \  \frac{\partial{\tr\chib}}{\partial\la}+\frac{{\oom}{\tr\chib}}{2}{\tr\chib}
+\partial_X\!{\tr\chib}+2{\oom}{\omb}{\tr\chib}+{\oom}|\hat{{\chib}}|^2\!=\!0\ .
\eea
In the next sections we start proving all the technical lemmas and theorems needed to obtain the final result, following the order required by the hierarchical  structure described in the introduction.

\section{The covariant angular derivatives for the connection coefficients\\}\label{S.2}
 
As said in the introduction, we have to estimate first the angular derivatives of the connection coefficients. The main result of this section is summarised in the following theorem, which was stated in (I) as Lemma (I;4.7). To prove it we assume the a priori estimates for some energy type norms, associated to the Riemann tensor, which are the main content of the hyperbolicity of the Einstein equations and which are proved in Section \ref{S.4.1} adapting to the present problem the results proved in \cite{Kl-Ni:book}, Chapter 6.\footnote{Reminding the analogy discussed in (I) of this problem with the much easier one associated to the Burger equation we observe that here some important differences show up with the way we treat the Burger case. There the a priori estimate provided from  Lemma (I;4.4) can be immediately derived as a straightforward consequence of the hyperbolicity of the Burger equation while in the case of the Einstein equations their hyperbolic nature is somewhat hidden unless we do use some appropriate gauges. Therefore in the present case the order of the proofs is inverted, we prove first Theorem (I;4.6) where we  use the Riemann component estimates; subsequently we prove the Riemann estimates.}

 \begin{theorem}\label{L6.1}
Let ${\Phi}=({\bf g};{\cal O},\underline{\cal O})=({\bf g};{\cal U})=(\ga,\oom,v,\psi, w, X;\ \om,\ze,\chi,\omb,\chib)$ be a real analytic solution in the region,\footnote{See (I) for its precise definition.}
\[{\cal K}(\Lambda_a,\Pi_a)\equiv\{(\la,\nu)\in[\la_0,{\Lambda_a}]\times[\nu_0,{\Pi_a}]\}\ ,\] of the characteristic
problem for the Einstein vacuum equations with analytic initial data, $\{\Phi^{(0)}\}$, on $C_0\cup\Cb_0$  belonging to the Banach space ${\cal B}_{\a;\ro_0}$, with $\a>3$.
Then there exists a double null foliation, the ``double null canonical foliation",\footnote{Its explicit definition is in \cite{Kl-Ni:book}, Chapter 3, see also \cite{C-K:book} where it was first introduced. Its definition and some specific details for the present case are given in subsections \ref{SScanfol} and \ref{SScanfol2}.} of the region ${\cal K}(\Lambda_a,\Pi_a)$ such that the generic connection coefficient we indicate with $\cal V$,  satisfy the following estimates
\footnote{The notations in inequality \ref{est45bx} are a bit symbolic, $J$ is an integer except that in $\nabb^J$ where it has to be considered a multiindex with $|J|=J$, moreover if with $\cal V$ we denote a $\underline{\cal O}$ connection coefficient then the weight factor $r^{2+J-\frac{2}{p}}$ has to be modified, see equations (I;4.52). Finally for some of the connection coefficients the constant $C_1$ has to be substituted by $C_1\log J$, the precise version of this result is in subsection \ref{S.sangest}, Theorem \ref{L2.1new}.}
\bea
|r^{2+J-\frac{2}{p}}\nabb^J{\cal V}|_{p,S}(\la,\nu)\leq C_1\frac{J!}{J^\a}\frac{e^{(J-2)(\de+\underline{\Ga}(\la))}}{\ro^J}\eql{est45bx}
\eea
with $\ro<\ro_0$ depending only on $\ro_0$ and on $\|\Phi^{(0)}\|_J\equiv|\nabla^J\Phi^{(0)}|_{p,S}$ \footnote{To prove this theorem we only need the $|\cdot|_{p,S}$ norms of the angular derivatives of the initial data, the null direction will be important for the estimates of the mixed derivatives. } with $J\leq 7$, with $C_1$  a suitable constant satisfying,
\bea
C_1>||\Phi^{(0)}||_{B_{\a,\ro_0}}\ ,
\eea 

\NI and\footnote{It is appropriate to remark that $\hat{C}$ has the dimension of a length, $[\hat{C}]=L^1$.}

\bea\label{312312}
\ \ \underline{\Ga}(\la)=\hat{C}\frac{(\la-\la_0)}{\la\la_0}\ .\eql{Gadef}
\eea

\NI  
For suitable values of $\de$ and $\hat {C}$.
\end{theorem}

\NI{\bf Remarks:} 

\NI {\em i) As said above and discussed in more detail later on, the constant $C_1$ in \ref{est45bx} has not to be small; the situation is different for $J \leq7$ as in this case, the constant have to be assumed small, $O(\varepsilon)$\footnote{ As said before, the explicit estimate of the constant $\varepsilon$ is not easy to give. It have to be chosen such that the energy norm, ${\cal Q}^0$, defined in section  \ref{S.4.1} and before in \cite{Kl-Ni:book}, chapter 6, is sufficiently small.}, if the solution has to be global. 
 Otherwise if we do not ask this ``smallness" the analytic solution will exist only in the region where the a priori energy estimates are satisfied, see also the remark at the end of subsection \ref{S.sincond}.

\medskip

\NI ii) To complete our proof we have first to obtain estimates  analogous to \ref{est45bx} for the $\Lie_O^N$ derivatives \footnote { With the symbol $\Lie_O$ we mean the sum of the  Lie derivatives with respect to the three $\ ^{(i)}O$ rotation vectors.}
 of the connection coefficients instead of the $\nabb^N$ ones. To do this we have to exploit the structure equations for $\Lie_O^N{\cal O}$ quantities. The reason for this intermediate step lies in the bootstrap argument we have to use to in this approach. In fact, as said before, to estimate those connection coefficients whose structure equations depends on the null Riemann components, we have to use the energy estimates,  (depending on the $\Lie_O$ derivatives of the null Riemann components ) which to be satisfied, require lower order $\Lie_O$ estimates of connection coefficients. At the end we will recover the estimates for the $\nabb$ derivatives lowering the radius of convergence.}

\subsection{The proof of Theorem \ref{L6.1}}\label{S.s2.1}
The estimates we prove refer to the region ${\cal K}(\Lambda_a,\Pi_a)$ where  we have an analytic solution $\Phi$ of the Einstein equations.\footnote{Recall as described carefully in (I) that an analytic solution of the system of equations (I;2.35), (I;2.36) is a solution of the Einstein equations provided the initial data satisfy equations (I;2.48),(I;2.49). It is important to recall, see also Theorem 4.5 of (I), that the solution $\Psi\in B_{\a,\ro''}$ and $\ro''$ is independent, due to the hyperbolicity, from the size of ${\cal K}(\Lambda_a,\Pi_a)$.}  Therefore in this region all the structure equations can be used, both those along the outgoing cones and those along the incoming ones.\footnote{That this region can be foliated by a ``double null cone foliation" follows as  such a foliation is guaranteed even for much less regular solutions, see \cite{Kl-Ni:book} and \cite{Ca-Ni:exist}.} The estimates of these norms follow the same pattern used in \cite{Kl-Ni:book}, Chapter 4, the main difference being that now we have to control all order derivatives. This is done in a recursive way, assuming we control the lower derivatives we show that we can control the higher ones. This implies, again generalising \cite{Kl-Ni:book}, that to obtain this result we need to use those transport equations which do not lose regularity.

\smallskip

\NI This proof is very long and involved, nevertheless once one grasps its general structure  it is easy to realise that the same procedure is repeated over and over to control the angular derivatives of all the various connection coefficients. 

\NI We go by induction. Let us suppose we have already obtained the estimates for $\Lie_O^{N-1}{\cal O}$, therefore we show how to control the $\Lie_O^N$ derivatives of  $\Us=\nabb\oom^{-1}\tr\chi+\oom^{-1}\tr\chi\eta$,\ (the details are in Section 9.)\footnote{$\Us$ is a combination of different connection coefficients. Its introduction is required to avoid some logarithmic divergences which will appear if we consider the transport equation for $\nabb\tr\chi$, see details in \cite{Kl-Ni:book}, Chapter 4. Once we control it with all its angular derivatives we control immediately $\nabb^N\tr\chi$ with $N\geq 1$ and $\nabb^N\chih$.}
  and discuss in the next subsection the general strategy.\footnote{All the definitions not explicitly given here are in (I).} 

\NI Let us consider the transport equation of $\Us$, which can be derived immediately from the transport equation for $U=\oom^{-1}\tr\chi$, \footnote{The trace part of the Raychaudhuri equation. The definition of $U$ allow us to eliminate the dependance on $\om=-2^{-1}\ddb_4\log\oom$ present in the equation for $\tr\chi$.}
\bea
\frac{\partial{U}}{\partial\nu}+\frac{{\oom}{\tr\chi}}{2}{U}+|\hat{{\chi}}|^2\!=\!0\eql{109bisa}
\eea
and the transport equation for $\eta$,
\bea
\frac{\Dbb}{\partial\nu}\eta+\oom\frac{\tr\chi}{2}\eta+\oom\chih\c\eta-\oom\chi\c\etab+\oom\b=0\ ,
\eea
\bea
\frac{\Dbb}{\partial\nu}\Us+\frac{3}{2}\oom\tr\chi\Us=-\nabb|\chih|^2-\eta|\chih|^2-\oom\chih\c{\Us}+\tr\chi(\chih\c\etab)-\tr\chi\b\ ,\eql{Useqp1}
\eea
which we rewrite as
\bea
\frac{\Dbb}{\partial\nu}{\Us}+\frac{3}{2}\oom\tr\chi\Us+\oom\chih\c\Us+2\chih\c(\nabb\chih)=\Fs\ ,
\eea
where
\bea
\Fs=-\tr\chi\b-\eta|\chih|^2+\tr\chi\chih\c\etab\ .\eql{Fsdef}
\eea
From it we derive the structure equations for the angular derivatives\footnote{See next subsection for its precise derivation.} $\Lie_O^{N-1}\Us$ which will allow to control $\Lie_O^N\tr\chi$ and $\Lie_O^{N-1}\nabb\chih$.

\bea\label{Useq1pa}
&&\ML\ML\frac{\Dbb}{\partial\nu}(\Lie_O^{N-1}\Us)+\frac{3}{2}\oom\tr\chi(\Lie_O^{N-1}\Us)=-\oom\chih\c(\Lie_O^{N-1}\Us)-\Lie_O^{N-1}\nabb|\chih|^2+\Lie_O^{N-1}\b+{\cal F}_{l.o.}\nn\\
\eea

\NI where ${\cal F}_{l.o.}$ is a complicated quadratic expression,\footnote{$l.o.$ stays for ``lower order".} see  equation 13.8, of the connection coefficients and of the Riemann tensor components, but with no $\Lie_O$ derivatives of order $N$.\footnote{Of order $N-1$ for the Riemann components.} The strategy is  to use Gronwall inequality to prove the correct bound for the $|\Lie_O^{N-1}\Us|_{p,S}$ norm once we make inductive assumptions for all the lower derivatives of the connection coefficients to control ${\cal F}_{l.o.}$ and once we have estimates for the initial data connection coefficients. The more delicate part, to complete this procedure, is the control of $\Lie_O^{N-1}\nabb\hat{\chi}$ and of 
$\Lie_O^{N-1}\b$. These two contributions have to be estimated separately. 
\smallskip

\NI To control $\Lie_O^{N-1}\nabb\hat{\chi}$, looking at the structure equations, see (I), Appendix to Section 2, we observe that, as we said in the introduction, there are not transport equations for $\chih$ along the outgoing cones with no loss of derivatives, but there are the Hodge equations,\footnote{$\ep$ is the totally antisymmetric tensor.}

\beaa\label{Useq11pa}
&&{\divv}\hat{\chi}=\frac{1}{2}\nabb{\tr\chi}-\b-\zeta\!\c\hat{\chi}+\frac{1}{2}\zeta{\tr\chi}\equiv F^{(1)}\eql{fdef1a}
\eeaa

\NI and from them one derives analogous Hodge systems for the angular derivatives $\Lie_O^{N-1}\nabb\chih$. This will allow to get  for $|\Lie_O^{N-1}\nabb\chih|_{p,S}$ the estimate\footnote{${\cal G}_{l.o.}$ denotes an estimate of terms with lower order derivatives ($<N$).}
\bea
|(\Lie_O^{N-1}\nabb\hat{\chi})|_{p,S}\leq  c\left(|\oom|_{\infty,S}|(\Lie_O^{N-1}\Us)|_{p,S}+|(\Lie_O^{N-1}\b)|_{p,S}\right)\!+{\cal G}_{l.o.} .\ \ \ \ \eql{estaaa}
\eea

\NI{\bf Remarks:}

{ \em \NI i) In order to obtain the equations \ref{Useq1pa}, and \ref{Useq11pa} from the structure equations, we have to estimate the $|\c|_{p,S}$ norms of the commutators $[\Lie_O^K, \Dbb_{\nu}]$,  $[\Lie_O^K, \divv]$ and  $[\Lie_O^K, \nabb]$ applied to $\Us$ and $\nabb\chih$; to do this we need estimates also for the  first derivatives of the connection coefficients, which we include in the  final version of  Theorem \ref{L2.1new}.     

\NI ii) To be useful this estimate requires that we have an independent way of controlling the $|\c|_{p,S}$ norm of $\b$. 
Once this is achieved the expression of \ref{estaaa} of $\Lie_O^{N-1}\nabb\chih$ norm  can be substituted in \ref{Useq1pa}.
 After a careful examination of the structure of the terms in ${\cal F}_{l.o.}$, the right estimate for $|\Lie_O^{N-1}\Us|_{p,S}$ can be obtained and, finally, from the Hodge system the expected estimate for $|\lie_O^{N-1}\nabb\chih|_{p,S}$ and, from the definition of $\Us$, the one of $|\Lie_O^{N-1}\nabb\tr\chi|_{p,S}$. }

\smallskip

\NI Therefore, besides the transport equation and the Hodge system, we have to control $|\Lie_O^{N-1}\b|_{p,S}$. The estimates for $\Lie_O^{N-1}\b$ or, more in general, for the derivatives of the null Riemann components are based on the hyperbolic nature of the Einstein equations implying that we can obtain a priori estimates for ``energy type" norms, as discussed at length,\footnote{Up to the first derivatives of the Riemann tensor.} in \cite{Kl-Ni:book}.  
\NI This procedure reproduces itself for all the remaining connection coefficients, therefore we summarise in the next subsection the general strategy we use consisting, basically, of four different steps.\footnote{We describe it for the non underlined quantities as for the underlined ones the structure is exactly the same with the obvious modifications.}

\subsection{The general strategy for the angular derivatives}\label{SS3.2a}
\NI {\bf I)} We have a transport equation for a quantity  expressed in terms of the first derivatives of the connection coefficients \footnote{or second derivatives...}, let us denote it ${\cal M}$; this implies an analogous equation for its $\Lie_O^{N-1}$ derivatives which has, in general,\footnote{See the details in Section \ref{S.10}.} the following expression:
\bea
\ML\frac{\Dbb{(\Lie_O^{N-1}{\cal M})}}{\partial\nu}+K\frac{\oom\tr\chi}{2}(\Lie_O^{N-1}{\cal M})+{\cal O}\Lie_O^{N-1}\nabb {\cal O}+H({\cal O})\Lie_O^{N-1}\Psi={\cal K}_{l.o.}\nn\\\eql{traspeq}
\eea
where $K$ is some fractional number,  $H({\cal O})\Lie_O^{N-1}\Psi$ denotes a linear expression in the connection coefficients times the  $\Lie^{N-1}_O$ derivatives of Riemann null components and ${\cal K}_{l.o.}$ collects all the terms which depend on lower order derivatives which can be estimated using the inductive assumptions.
\smallskip

\NI {\bf II)} The remaining term  which depends on the highest order derivatives, ${\cal O}\Lie_O^{N-1}\nabb{\cal O}$, has the property that, in general, there is no transport equation for $\nabb{\cal O}$ without loss of derivatives, but ${\cal O}$ satisfies an Hodge equation which, slightly symbolically, can be written as,
\bea
&&\divv{\cal O}={\cal M}+\Psi+{\cal H}_{l.o.}\nn\\
\eea

\NI where with $\Psi$ we denote one or more Riemann null components and ${\cal H}_{l.o.}$ contains lower order terms.
applying $\Lie_O^{N-1}$ times  Hodge system we obtain an equation, similar to \ref{estaaa}, which allows an estimate  $\Lie_O^{N-1}{\nabb\cal O}$ in terms of $\Lie_O^{N-1}{\cal M}$, provided we control the (norms of the) $\Lie_O^{N-1}$ derivatives of the Riemann components $\Psi$. 
\smallskip

\NI {\bf III)} The control of the $\Lie_O$ derivatives of $\Psi$, up to $N\!\!-\!\!1$, is the third step required to get the result which, again, is present in all the connection coefficients norm estimates; it is implemented using the hyperbolic nature of the Einstein equations which shows itself in the a priori estimates for the energy-type $L^2$ norms, we denote them by {\cal Q} norms, whose proof is in Section \ref{S.4.1}; these a priori estimates can be proved in the whole spacetime if the initial data are small, otherwise there will be a bounded region where they hold.\footnote{ Analogous to the time slab 
$R\times[0,T]$ of the Burger equation, see (I), Section 4.}
\smallskip

\NI {\bf IV)} Once we control the $N\!\!-\!\!1$ derivatives of $\Psi$ we integrate \ref{traspeq} along the outgoing cones\footnote{Or along the incoming cones depending on which  transport equation for ${\cal M}$ is without loss of derivatives.} and use the Gronwall's inequality to control $|\Lie_O^{N-1}{\cal M}|_{p,S}$ and finally again the Hodge system to control (the norm of) $\Lie_O^{N-1}\nabb{\cal O}$ and, subsequently, of $\Lie_O^{N}{\cal O}$ .
\smallskip

\NI  {\bf V)}  As final last step we recover the estimates for $\nabb^{N-1}{\cal M}$ and $\nabb^N{\cal O}$  from the previous estimates

\subsection{The last slice and the canonical foliation}\label{S.s3.4}
A technical but important remark is appropriate here and will be discussed in more detail in the following subsection later on. 
To prove Theorem \ref{L6.1}  it is crucial that we assume in the inductive hypothesis the correct asymptotic decay of the (derivatives of the) connection coefficients along the outgoing cones. To obtain it 
it turns out that the integration along the outgoing cones for all the not underlined quantities has to be done  going backward,\footnote{With the exception of $\om$ and $\omb$ whose role is inverted.} starting from the greatest $\ub$ value, $\nu_*=\Pi_a$. 

\NI This implies we have first to know the estimates for all the connection coefficients on the last slice, $\Cb(\nu_*)$. This can be achieved using the incoming transport equations starting from $C_0$ at 
$(\la_0,\nu_*)$ and exploiting the energy estimates starting from the initial data surface $\Cb_0$. The arising problem, proceeding in this way, is that some of these transport equations imply a loss of derivatives which would not allow an inductive procedure to control the higher derivatives. To overcome it  one has to choose, in an appropriate way, on $\Cb(\nu_*)\equiv\Cb_*$ the shift function $\oom$ (see (I) for its definition); this specifies on $\Cb_*$ the function $u=u|_{\Cb_*}$  and a foliation on it. The outgoing cones in ${\cal K}$ are then defined as the level hypersurfaces of a solution of the eikonal equations with $u|_{\Cb_*}$ as  ``initial data". This defines the so called ``double null cone canonical foliation". We discuss carefully how to obtain this foliation in subsections 9.5, 9.6.1 and in subsection 14.11, but we also recall that this procedure is not new and is due to D.Christodoulou and S. Klainerman who introduced it in \cite{C-K:book} and was subsequently readapted  in \cite{Kl-Ni:book} and \cite{Niclast} where more details are given. 
\medskip

\NI{\bf Remark:}

\NI{\em A contradiction seems to appear as we claim that to prove the analytic solution existence, we assign initial data on $C_0\cup \Cb_0$ and, nevertheless, to prove the norm estimates  of Lemma \ref{L6.1} we have to assign ``final data" on the last slice $\Cb(\nu_*)$. Nevertheless this contradiction is only apparent and   the picture of the global strategy to prove the existence of a global analytic solution goes in the following way: 

\NI We assign initial data on $C_0\cup\Cb_0$ satisfying the appropriate norm estimates \ref{123} and \ref{123ab} and applying  Cauchy-Kowalevski we prove, as discussed in $(I)$, the existence of the solution in a small region; then we denote, see Theorem \ref{L6.1},
${\cal K}(\Lambda_a,\Pi_a)\equiv\{(\la,\nu)\in[\nu_0,{\Lambda_a}]\times[\la_0,{\Pi_a}]\}$ as  the larger region where, with the assigned initial data, this analytic solution  does exist. Observe that up to now the double null foliation of the existence region is not specified. 

\NI Introducing on the upper part  $\Cb(\Pi_a)$ of the boundary of ${\cal K}(\Lambda_a,\Pi_a)$, the part we call in the previous subsection  ``the last slice" and denoted $\Cb(\nu_*)$, an appropriate foliation defined through $\oom_*$, we prove, starting from $S(\la_0,\nu_*)\subset C_0$ and exploiting the energy norms norms from $\Cb_0$, appropriate norm estimates for the not underlined connection coefficients on $\Cb(\nu_*)$ and using them and the transport equations on the outgoing cones we prove the norm estimates of Theorem \ref{L6.1}; this is the crucial step to show that the region ${\cal K}(\Lambda_a,\Pi_a)$ can be extended implying, to avoid a contradiction, that this region is in fact unbounded. 

\NI Of course this also implies that the double null foliation of this region required to prove its unboundedness is, see also \cite{Kl-Ni:book}, the ``double null canonical foliation", that is the one determined by the foliation imposed on the last slice to obtain the correct ``final data norms".

\NI Finally the last thing to observe is that, going back from the data on the last slice with the outgoing transport equations to $\Cb_0$ we obtain some norms bounds that the not underlined connection coefficients have to satisfy on $\Cb_0$ and we have to prove that these bounds are compatible with the previously assumed initial data; this is easy to prove for the following reasons: we can now assume all the transport equations along the outgoing cones without worrying anymore of loss of derivatives as we already have the estimates for all order derivatives and due to the fact that the final parameter $\ro$ appearing in these estimates is smaller than the initial one, $\ro_0$, it is immediate to realise that all these estimates are consistent with the initial data estimates.\footnote{In principle the foliation on $\Cb_0$ of the initial data is different from the foliation induced on $\Cb_0$ from the canonical foliation, but it can be proved, proceeding as in \cite{Kl-Ni:book} that, due to the fact that we are considering a ``small initial data problem" the two induced foliations are near and the result follows.}}

\subsection{The initial data conditions}\label{S.sincond}

To prove the inductive estimate for $|r^{N+2-\frac{2}{p}}\nabb^{N-1}\Us|_{p,S}$ and for the $|\c|_{p,S}$ norms of the remaining connection coefficients, Theorem \ref{L6.1}, we need some assumptions on the 
initial data. This is a bit delicate as the initial data have to satisfy some constraint equations.\footnote{See (I) for a detailed discussion of the constraint equations.} As it will be proved in detail later on, the strategy is first to define the initial data in a neighbourhood of $S_{(0)}\equiv C_0\cap \Cb_0$ and then prove, again with a `` Burger type" mechanism, that the analytic initial data can be defined on the whole $C_0\cup\Cb_0$. The proof is a simplified version of the main proof, with the basic difference that in this case we, as we directly assign those connection coefficients which need the Hodge equations to be estimated, we don't need the estimates of the null Riemann components, and hence of the energy type norms. For the same reason we do not need to integrate along $C_0$ ``from above".  Moreover, the exponential factor, $\underline{\Ga}(\la)$, in the initial data assumptions  is different, namely we require,
\bea
\underline{\Ga}_0(\la)={\tilde C}_0\frac{(\la-\la_0)}{\la\la_0}\ .\eql{Gabdef1}
\eea
With suitable $ {\tilde{C}}_0<\hat{C}$.

\NI Therefore we show in Section \ref{S.s initial data} 
that we can construct analytic initial data satisfying the following bounds for any $J$ with $\ep>0$, given suitable conditions. More precisely we prove the following
\smallskip
\begin{theorem}\label{Thinitialdata}
Assuming assigned $\chih$ on $C_0=C(\la_0,\nu)$, $\chibh$ and $X$ on $\Cb_0=\Cb(\la,\nu_0)$ and $\oom$ on $C_0\cup\Cb_0$  analytic, of radius $\ro_{0,0,1}$ satisfying the following estimates\footnote{the estimates for the not derived connection coefficients are more delicate as they involve also the metric $\ga$ with is unknown of the problem and hence have to be done with respect to an auxliary metric, namely $\tilde{\ga}$, the Mikowski metric on $C_0$ with respect to the $\oom$ foliation. see \cite{Ca-Ni:exist}, for all the details.}, 
\bea\label{31}
&&|\nu^{J+2+\ep-\frac{2}{p}}\nabla^J\log\oom|_{p,S}(\la_0,\nu) \leq C^{(0)}_3e^{(J-2)(\underline{\Ga}_0(\la)+\de_0)}\frac{J!}{J^{\a}}\frac {1}{\ro_{0,0,1}^J}\nn\\
&&|\nu^{J+\frac{5}{2}+\ep-\frac{2}{p}}\nabla^J\chih|_{p,S}(\la_0,\nu) \leq C^{(0)}_1e^{(J-2)(\underline{\Ga}_0(\la)+\de_0)}\frac{J!}{J^{\a}}\frac {1}{\ro_{0,0,1}^J}\ ,\eql{chihindata11}\\
&&||\la|^{1+\ep}\nu^{1+J-\frac{2}{p}}\nabla^J\log\oom|_{p,S}(\la,\nu_0) \leq \Cb^{(0)}_3e^{(J-2)(\underline{\Ga}_0(\la)+\de_0)}\frac{J!}{J^{\a}}\frac {1}{\ro_{0,0,1}^J}\nn\\
&&||\la|^{\frac{3}{2}+\ep}\nu^{1+J-\frac{2}{p}}\nabla^J\chibh|_{p,S}(\la,\nu_0)\leq \Cb^{(0)}_1e^{(J-2)(\underline{\Ga}_0(\la)+\de_0)}\frac{J!}{J^{\a}}\frac {1}{\ro_{0,0,1}^J}\nn\\
&&||\la|^{\frac{3}{2}+\ep}\nu^{1+J-\frac{2}{p}}\nabla^J X|_{p,S}(\la,\nu_0)\leq \Cb^{(0)}_1e^{(J-2)(\underline{\Ga}_0(\la)+\de_0)}\frac{J!}{J^{\a}}\frac {1}{\ro_{0,0,1}^J}\nn \ ,
\eea
With \[\underline{\Ga}_0=\tilde{C}_0\frac{\la-\la_0}{\la\la_0},\ \ \tilde{C_0}\ \ \ \mbox{suitable constant,}\]

\medskip

\NI with $C^{(0)}_{1,3}$ and $\Cb^{(0)}_{1,3}$ of order $O(\varepsilon)$ for $J \leq7$ and $O(1)$ otherwise, 
 assuming finally that on $S_0=S(0,\nu_0)$ the connection coefficients \[\{\ga,\tr\chi,\tr\chib, \ze, \om,\omb\}\] 
sufficiently small and such that the following conditions are satisfied\footnote{To completely define the $\nabb$ derivative on $S_0$ we have also to assign $\ga$ on $S_0$ see section \ref{S.s initial data}}\footnote{Clearly, to obtain the smallness of the first $C_{(J)}$ $J\leq 7$,  we have to assume also the angular derivatives $\nabb^s$ of the quantities assigned on $S_0$ of order $\varepsilon$.},
\bea\label{iniz11}
&&\nabb\tr\chi-\ze\tr\chi\leq\varepsilon \nu_0^{-(\frac{7}{2}+\ep)}\nn\\
&&\nabb\tr\chib-\ze\tr\chib\leq\varepsilon \nu_0^{-(\frac{7}{2}+\ep)}\nn\\
&&{\bf K}-\overline{\bf K}+\frac{1}{4}\!\left(\tr\chi\tr\chib-\overline{\tr\chi\ \!\tr\chib}\right)\leq\varepsilon \nu_0^{-\frac{7}{2}}\nn\\
&&\curll\zeta\leq\varepsilon \nu_0^{-\frac{7}{2}}\ .\eql{Socondin1}
\eea
\bea\label{xeq}
X^a(0,\nu_0,\theta,\phi)=\int_{\nu_0}^{\infty}4\oom^2\ga^{ac}\ze_c(\la_1;\nu',\theta,\phi)d\nu'
\eea

\bea\label{omcond}
\ML\omb(0,\nu_0,\theta,\phi)
=-\int_{\nu_0}^{\infty}d\nu'e^{\int_{\nu_0}^{\nu'}(-2\oom\om)d\nu''}\!\left[\ze\c\nabb\log\oom\!+\!\frac{3}{2}|\ze|^2\!-\!\frac{1}{2}|\nabb\log\oom|^2\!-\frac{1}{2}\!\big({\bf K}\!+\!\frac{1}{4}\tr\chi\tr\chib\!-\!\frac{1}{2}\chih\c\chibh\big) \right]\!\!(\nu'),\
\eea
then it is possible to construct analytic initial data on $C_0\cup\Cb_0$ such that suitable energy type ${\cal Q}^0$ norms on $C_0\cup\Cb_0$ are finite and small and the connection coefficients are analytic satisfying the following estimates
\smallskip

\NI{\bf On $C_0$:}
\bea\label{prevres}
&&\big|r^{1+J+\si(J)-\frac{2}{p}}\nabb^J\tr\chi\big|_{p,S}\leq C^{(0)}_{0}e^{(J-2)(\underline{\Ga}_0(\la)+\de_0)}\frac{J!}{J^{\a}}\frac {1}{\ro_{0}^J}\nn\\
&&\big|r^{\frac{5}{2}+J+\ep-\frac{2}{p}}\nabb^J\chih\big|_{p,S}\leq C^{(0)}_{1}e^{(J-2)(\underline{\Ga}_0(\la)+\de_0)}\frac{J!}{J^{\a}}\frac {1}{\ro_{0}^J}\nn\\
&&\big|r^{2+J-\frac{2}{p}}\nabb^J\ze\big|_{p,S}\leq C^{(0)}_{4}e^{(J-2)(\underline{\Ga}_0(\la)+\de_0)}\frac{J!}{J^{\a}}\frac {1}{\ro_{0}^J}\nn\\
&&\big|r^{2+J+\ep-\frac{2}{p}}\nabb^J\om\big|_{p,S}\leq C^{(0)}_{2}e^{(J-2)(\underline{\Ga}_0(\la)+\de_0)}\frac{J!}{J^{\a}}\frac {1}{\ro_{0}^J}\eql{123}\nn\\
&&\big|r^{1+J+\si(J)-\frac{2}{p}}\nabb^J\tr\chib\big|_{p,S}\leq C^{(0)}_{5}e^{(J-2)(\underline{\Ga}_0(\la)+\de_0)}\frac{J!}{J^{\a}}\frac {1}{\ro_{0}^J}\\
&&\big||\la_0|r^{1+J-\frac{2}{p}}\nabb^J\chibh\big|_{p,S}\leq C^{(0)}_{6}e^{(J-2)(\underline{\Ga}_0(\la)+\de_0)}\frac{J!}{J^{\a}}\frac {1}{\ro_{0}^J}\nn\\
&&\big||\la_0|r^{1+J-\frac{2}{p}}\nabb^J\omb\big|_{p,S}\leq C^{(0)}_{7}e^{(J-2)(\underline{\Ga}_0(\la)+\de_0)}\frac{J!}{J^{\a}}\frac {1}{\ro_{0}^J}.\nn
\eea

\NI{\bf On $\Cb_0$:}
\bea\label{prevres1}
&&||\la|^{\si(J)}r^{J+1-\frac{2}{p}}\nabb^J\tr\chib|_{p,S}\leq \Cb^{(0)}_{0}e^{(J-2)(\underline{\Ga}_0(\la)+\de_0)}\frac{J!}{J^{\a}}\frac {1}{\ro_{0}^J}\nn\\
&&||\la|^{\frac{3}{2}+\ep}r^{J+1-\frac{2}{p}}\nabb^J\chibh|_{p,S}\leq \Cb^{(0)}_{1}e^{(J-2)(\underline{\Ga}_0(\la)+\de_0)}\frac{J!}{J^{\a}}\frac {1}{\ro_{0}^J}\nn\\
&& ||\la|^{1+\ep}r^{J+1-\frac{2}{p}}\nabb^J\omb|_{p,S}\leq \Cb^{(0)}_{2}e^{(J-2)(\underline{\Ga}_0(\la)+\de_0)}\frac{J!}{J^{\a}}\frac {1}{\ro_{0}^J}\eql{123ab}\nn\\
&&|r^{1+J+\si(J)-\frac{2}{p}}\nabb^J\tr\chi|_{p,S}\leq \Cb^{(0)}_{5}e^{(J-2)(\underline{\Ga}_0(\la)+\de_0)}\frac{J!}{J^{\a}}\frac {1}{\ro_{0}^J}\nn\\
&&|r^{J+2-\frac{2}{p}}\nabb^J\chih|_{p,S}\leq \Cb^{(0)}_{6}e^{(J-2)(\underline{\Ga}_0(\la)+\de_0)}\frac{J!}{J^{\a}}\frac {1}{\ro_{0}^J}\\
&&|r^{J+2-\frac{2}{p}}\nabb^J\om|_{p,S}\leq \Cb^{(0)}_{7}e^{(J-2)(\underline{\Ga}_0(\la)+\de_0)}\frac{J!}{J^{\a}}\frac {1}{\ro_{0}^J}\nn\\
&&|r^{J+2-\frac{2}{p}}\nabb^J\ze|_{p,S}\leq \Cb^{(0)}_{4}e^{(J-2)(\underline{\Ga}_0(\la)+\de_0)}\frac{J!}{J^{\a}}\frac {1}{\ro_{0}^J}.\nn
\eea
With $\si(J)=0$, $J=0$, $\si(J)=1$, $J>0$, $r(\nu)=\frac{1}{4\pi}|S(0,\nu)|_{\ga}$, $\nabb$ the projection of  the covariant derivative along the angular coordinates, $\underline{\Ga}={C}_0\frac{\la-\la_0}{\la\la_0}$, $\de>\de_0$, $\ro_{0}<\ro_{0,0,1}$, and all the constant $C$ and $\Cb$
of order $O(\varepsilon)$ for $J \leq7$ and $O(1)$ otherwise.

\medskip
\end{theorem}
\NI {\bf Proof:} The proof is substantially the content of  \cite{Ca-Ni:char}, we retrace and improve that result in Section \ref{S.s initial data}.

\NI {\bf Remarks:}  

\NI {\em 
\NI  i) Notice that, as said before, for the initial data, we do not need the energy estimates, as on the initial data we already posses all the tangential derivatives of $\chi$, $\chib$, $\Omega$ and hence for $\om$ and $\omb$ at any order and so we do not have to use  the Hodge estimates and hence the estimates for all the tangential derivatives of the null Riemann components. This is a great advantage, as we do not have to retrace all the machinery we have to use in the internal mixed derivatives estimates. 

\NI ii) Notice moreover that in order to prove Theorem \ref{Thinitialdata} we do not need the full tangential derivatives of the free connection coefficients but only the angular ones in the hypothesis \ref{31}, we require the full ones in order to estimate also the mixed derivatives, see Theorem \ref{T11.1}

\NI iii) We have to require a slightly better decay for some of the connection coefficients, adding an $\ep>0$ on the decay for $r$, in order to assure the boundedness of the ${\cal Q} $ norms.

\NI iv) With respect to the construction of the initial data provided in \cite{Ca-Ni:char}, we  provide the $\zeta$ on $S_0$ instead that on the upper part of the incoming cone $\Cb_0$, this is an improvement.}
\smallskip

\NI From these estimates we obtain the estimates for the $\Lie_O$ derivatives on the initial data, which will be needed to obtain the energy estimates of the null Riemann components. 
Hence we have to prove the following estimates:
\begin{theorem} Under the assumptions of Theorem \ref{Thinitialdata}, the following estimates hold

\NI{\bf On $C_0$:}
\bea\label{iniz1}
&&|r^{2-\frac{2}{p}}\Lie_O^JU|_{p,S}\leq C^{(0)}_{0,0}e^{(J-2)(\de_0+\underline{\Ga}_0(\la))}\frac{J!}{J^\a}\frac{1}{\ro_{0,1}^{J}}\nn\\
&&|r^{\frac{5}{2}+\ep-\frac{2}{p}}\Lie_O^J\chih|_{p,S}\leq C^{(0)}_{0,1}e^{(J-2)(\de_0+\underline{\Ga}_0(\la))}\frac{J!}{J^\a}\frac{1}{\ro_{0,1}^{J}}\nn\\
&&|r^{2+\ep-\frac{2}{p}}\Lie_O^J\om|_{p,S}\leq C^{(0)}_{0,2}e^{(J-2)(\de_0+\underline{\Ga}_0(\la))}\frac{J!}{J^\a}\frac{1}{\ro_ {0,1}^{J}}\nn\\
&&|r^{2-\frac{2}{p}}\Lie_O^J\ze|_{p,S}\leq C^{(0)}_{0,3}e^{(J-2)(\de_0+\underline{\Ga}_0(\la))}\frac{J!}{J^\a}\frac{1}{\ro_ {0,1}^{J}}\  \nn\\
&&\big|r^{2-\frac{2}{p}}\Lie_O^J\tr\chib\big|_{p,S}\leq C^{(0)}_{0,4}e^{(J-2)(\de_0+\underline{\Ga}_0(\la))}\frac{J!}{J^{\a}}\frac{1}{\ro_ {0,1}^J}\\
&&\big||\la_0|r^{1-\frac{2}{p}}\Lie_O^J\chibh\big|_{p,S}\leq C^{(0)}_{0,5}e^{(J-2)(\de_0+\underline{\Ga}_0(\la))}\frac{J!}{J^{\a}}\frac{1}{\ro_ {0,1}^J}\nn\\
&&||\la_0|r^{1-\frac{2}{p}}\Lie_O^J\omb|_{p,S}\leq C^{(0)}_{0,6}e^{(J-2)(\de_0+\underline{\Ga}_0(\la))}\frac{J!}{J^\a}\frac{1}{\ro_ {0,1}^{J}}\nn
\eea

\smallskip

\NI{\bf On $\Cb_0$:}
\bea\label{iniz2}
&&|r^{1-\frac{2}{p}}|\la|\Lie_O^J\underline{U}|_{p,S}\leq \Cb^{(0)}_{0,0}e^{(J-2)(\de_0+\underline{\Ga}_0(\la))}\frac{J!}{J^\a}\frac{1}{\ro_ {0,1}^{J}}\nn\\
&&|r^{1-\frac{2}{p}}|\la|^{\frac{3}{2}+\ep}\Lie_O^J\chibh|_{p,S}\leq \Cb^{(0)}_{0,1}e^{(J-2)(\de_0+\underline{\Ga}_0(\la))}\frac{J!}{J^\a}\frac{1}{\ro_ {0,1}^{J}}\nn\\
&& |r^{1-\frac{2}{p}}|\la|^{1+\ep}\Lie_O^J\omb|_{p,S}\leq \Cb^{(0)}_{0,2}e^{(J-2)(\de_0+\underline{\Ga}_0(\la))}\frac{J!}{J^\a}\frac{1}{\ro_ {0,1}^{J}}\nn\\
&&|r^{2-\frac{2}{p}}\Lie_O^JU|_{p,S}\leq \Cb^{(0)}_{0,3}e^{(J-2)(\de_0+\underline{\Ga}_0(\la))}\frac{J!}{J^\a}\frac{1}{\ro_ {0,1}^{J}}\nn\\
&&|r^{2-\frac{2}{p}}\Lie_O^J\chih|_{p,S}\leq \Cb^{(0)}_{0,4}e^{(J-2)(\de_0+\underline{\Ga}_0(\la))}\frac{J!}{J^\a}\frac{1}{\ro_ {0,1}^{J}}\\
&&|r^{2-\frac{2}{p}}\Lie_O^J\om|_{p,S}\leq \Cb^{(0)}_{0,5}e^{(J-2)(\de_0+\underline{\Ga}_0(\la))}\frac{J!}{J^\a}\frac{1}{\ro_ {0,1}^{J}}\nn\\
&&|r^{2-\frac{2}{p}}\Lie_O^J\ze|_{p,S}\leq \Cb^{(0)}_{0,6}e^{(J-2)(\de_0+\underline{\Ga}_0(\la))}\frac{J!}{J^\a}\frac{1}{\ro_ {0,1} ^{J}}.\ \nn
\eea
\NI With $\ro_ {0,1}<\ro_0$,  and all the constants $C$ and $\Cb$ of order $O(\varepsilon)$ for $J \leq7$ and $O(1)$ otherwise.  
\NI Moreover we have similar estimates for the first derivatives of the connection coefficients:

\smallskip

\NI{\bf On $C_0$:}
\bea\label{iniz3}
&&|r^{3-\frac{2}{p}}\Lie_O^{J-1}\nabb U|_{p,S}\leq C^{(0)}_{1,0}e^{(J-2)(\de_0+\underline{\Ga}_0(\la))}\frac{J!}{J^\a}\frac{1}{\ro_{0,1}^{J}}\nn\\
&&|r^{\frac{7}{2}+\ep-\frac{2}{p}}\Lie_O^{J-1}\nabb\chih|_{p,S}\leq C^{(0)}_{1,1}e^{(J-2)(\de_0+\underline{\Ga}_0(\la))}\frac{J!}{J^\a}\frac{1}{\ro_{0,1}^{J}}\nn\\
&&|r^{3+\ep-\frac{2}{p}}\Lie_O^{J-1}\nabb\om|_{p,S}\leq C^{(0)}_{1,2}e^{(J-2)(\de_0+\underline{\Ga}_0(\la))}\frac{J!}{J^\a}\frac{1}{\ro_ {0,1}^{J}}\nn\\
&&|r^{3-\frac{2}{p}}\Lie_O^{J-1}\nabb\ze|_{p,S}\leq C^{(0)}_{1,3}e^{(J-2)(\de_0+\underline{\Ga}_0(\la))}\frac{J!}{J^\a}\frac{1}{\ro_ {0,1}^{J}}\ \nn\\
&&\big|r^{3-\frac{2}{p}}\Lie_O^{J-1}\nabb\tr\chib\big|_{p,S}\leq C^{(0)}_{1,4}e^{(J-2)(\de_0+\underline{\Ga}_0(\la))}\frac{J!}{J^{\a}}\frac{1}{\ro_ {0,1}^J}\\
&&\big||\la_0|r^{2-\frac{2}{p}}\Lie_O^{J-1}\nabb\chibh\big|_{p,S}\leq C^{(0)}_{1,5}e^{(J-2)(\de_0+\underline{\Ga}_0(\la))}\frac{J!}{J^{\a}}\frac{1}{\ro_ {0,1}^J}\nn\\
&&||\la_0|r^{2-\frac{2}{p}}\Lie_O^{J-1}\nabb\omb|_{p,S}\leq C^{(0)}_{1,6}e^{(J-2)(\de_0+\underline{\Ga}_0(\la))}\frac{J!}{J^\a}\frac{1}{\ro_ {0,1}^{J}}\nn
\eea

\smallskip

\NI{\bf On $\Cb_0$:}
\bea\label{iniz4}
&&|r^{2-\frac{2}{p}}|\la|\Lie_O^{J-1}\nabb\underline{U}|_{p,S}\leq \Cb^{(0)}_{1,0}e^{(J-2)(\de_0+\underline{\Ga}_0(\la))}\frac{J!}{J^\a}\frac{1}{\ro_ {0,1}^{J}}\nn\\
&&|r^{2-\frac{2}{p}}|\la|^{\frac{3}{2}+\ep}\Lie_O^{J-1}\nabb\chibh|_{p,S}\leq \Cb^{(0)}_{1,1}e^{(J-2)(\de_0+\underline{\Ga}_0(\la))}\frac{J!}{J^\a}\frac{1}{\ro_ {0,1}^{J}}\nn\\
&& |r^{2-\frac{2}{p}}|\la|^{1+\ep}\Lie_O^{J-1}\nabb\omb|_{p,S}\leq \Cb^{(0)}_{1,2}e^{(J-2)(\de_0+\underline{\Ga}_0(\la))}\frac{J!}{J^\a}\frac{1}{\ro_ {0,1}^{J}}\nn\\
&&|r^{3-\frac{2}{p}}\Lie_O^{J-1}\nabb U|_{p,S}\leq \Cb^{(0)}_{1,3}e^{(J-2)(\de_0+\underline{\Ga}_0(\la))}\frac{J!}{J^\a}\frac{1}{\ro_ {0,1}^{J}}\nn\\
&&|r^{3-\frac{2}{p}}\Lie_O^{J-1}\nabb\chih|_{p,S}\leq \Cb^{(0)}_{1,4}e^{(J-2)(\de_0+\underline{\Ga}_0(\la))}\frac{J!}{J^\a}\frac{1}{\ro_ {0,1}^{J}}\\
&&|r^{3-\frac{2}{p}}\Lie_O^{J-1}\nabb\om|_{p,S}\leq \Cb^{(0)}_{1,5}e^{(J-2)(\de_0+\underline{\Ga}_0(\la))}\frac{J!}{J^\a}\frac{1}{\ro_ {0,1}^{J}}\nn\\
&&|r^{3-\frac{2}{p}}\Lie_O^{J-1}\nabb\ze|_{p,S}\leq \Cb^{(0)}_{1,6}e^{(J-2)(\de_0+\underline{\Ga}_0(\la))}\frac{J!}{J^\a}\frac{1}{\ro_ {0,1} ^{J}}.\ \nn
\eea

\NI With all the constant $C$ and $\Cb$
of order $O(\varepsilon)$ for $J \leq7$ and $O(1)$ otherwise.
\end{theorem}
\NI{\bf Proof:} The proof is a repetition of lemmas \ref{L4.1} and \ref{L4.2} applied now to the connection coefficients.
\smallskip

\NI {\bf Remark:} 

\NI {\em Notice that the hierarchy among the constants $C$ is obtained from the sequence we adopt to to prove the  inequalities \ref{iniz1} , \ref{iniz2} , \ref{iniz3} and \ref{iniz4}.}
\smallskip

\NI In order to prove Theorem \ref{main2} we need also the estimates for the other tangential derivatives of the initial data null cones, namely $\ddb^J_{\nu}\nabb^p{\cal O}$ for $C_0$ and  $\ddb^J_{\la}{\nabb^p\cal O}$ for $\underline{C}_0$. This is the result of the following theorem:

\begin{theorem}\label{T11.1}
Assuming the hypothesis of Theorem \ref{Thinitialdata}, the following estimates hold for the $\ddb_\nu^p\nabb^q$ derivatives of the connection coefficients on $C_0$,
\bea\label{inizz0}
&&|r^{1+p+q+\si(P)-\frac{2}{p}}\ddb_{\nu}^p\nabb^qU|_{p,S}\leq F_1\!\left(\frac{(p+q)!}{(p+q)^{\a}}\frac{e^{(p+q-2)(\de_0+\underline{\Ga}_0(\la))}}{\ro_0^{p+q}}\right)\nn\\
&&|r^{1+p+q+\si(P)-\frac{2}{p}}\ddb_{\nu}^p\nabb^q\tr\chi|_{p,S}\leq F_2\!\left(\frac{(p+q)!}{(p+q)^{\a}}\frac{e^{(p+q-2)(\de_0+\underline{\Ga}_0(\la))}}{\ro_0^{p+q}}\right)\nn\\
&&|r^{2+p+q-\frac{2}{p}}\ddb_{\nu}^J\nabb^p\chih|_{p,S}\leq F_3\!\left(\frac{(J+P)!}{(p+q)^{\a}}\frac{e^{(p+q-2)(\de_0+\underline{\Ga}_0(\la))}}{\ro_0^{p+q}}\right)\nn\\
&&\big|r^{2+p+q-\frac{2}{p}}\ddb_{\nu}^p\nabb^q\eta\big|_{p,S}\leq F_4\!\left(\frac{(p+q)!}{(p+q)^{\a}}\frac{e^{(p+q-2)(\de_0+\underline{\Ga}_0(\la))}}{\ro_0^{p+q}}\right)\nn\\
&&\big|r^{2+p+q-\frac{2}{p}}\ddb_{\nu}^p\nabb^q\ze\big|_{p,S}\leq F_5\!\left(\frac{(p+q+1)!}{(p+q+1)^{\a}}\frac{e^{(p+q-1)(\de_0+\underline{\Ga}_0(\la))}}{\ro_0^{p+q+1}}\right)\ \ \ \ \ \ \ \ \ \ \eql{miste1}\\ 
&&\big|r^{2+p+q-\frac{2}{p}}\ddb_{\nu}^J\nabb^P\Ub|_{p,S}\leq {\underline F}_1\!\left(\frac{(p+q)!}{(p+q)^{\a}}\frac{e^{(p+q-2)(\de_0+\underline{\Ga}_0(\la))}}{\ro_0^{p+q}}\right)\nn\\
&&\big||\la|r^{1+p+q-\frac{2}{p}}\ddb_{\nu}^J\nabb^P{\chibh}|_{p,S}\leq {\underline F}_3\!\left(\frac{(p+q)!}{(p+q)^{\a}}\frac{e^{(p+q-2)(\de_0+\underline{\Ga}_0(\la))}}{\ro_0^{p+q}}\right)\nn\\
&&\big|r^{2+p+q-\frac{2}{p}}\ddb_{\nu}^J\nabb^P\om\big|_{p,S}\leq F_6\!\left(\frac{(p+q+1)!}{(p+q+1)^{\a}}\frac{e^{(p+q-1)(\de_0+\underline{\Ga}_0(\la))}}{\ro_0^{p+q+1}}\right)\nn\\
&&\big|r^{2+p+q-\frac{2}{p}}\ddb_{\nu}^J\nabb^P\omb\big|_{p,S}\leq {\underline F}_6\!\left(\frac{(p+q)!}{(p+q)^{\a}}\frac{e^{(p+q-2)(\de_0+\underline{\Ga}_0(\la))}}{\ro_0^{p+q}}\right)\ .\nn
\eea
With all the constant $F$ and $\underline{F}$ of order $O(\varepsilon)$ for $J \leq7$ and $O(1)$ otherwise. Clearly analogous estimates with $\ddb_{\la}$ instead of $\ddb_{\nu}$ on $\Cb_0$ hold.
\end{theorem}
\NI {\bf Proof:} The proof is in section \ref{S.s initial data}.

\NI{\bf Remark:}

\NI {\em \NI  As already noticed in Remark i) of  Theorem \ref{Thinitialdata} for the angular derivatives, here we do not need the energy estimates, as on the initial data we already posses all the mixed derivatives of $\chi$ on $C_0$, $\chib$ and $X$ on $\Cb_0$, of  $\Omega$ on $C_0\cup \Cb_0$ and hence for $\om$ and $\omb$ on $C_0\cup\Cb_0$ at any order and so we do not have to use  the Hodge estimates and hence the estimates for all the mixed derivatives of the null Riemann components.
Notice that in order to prove Theorem \ref{T11.1} we need the full tangential derivatives of the free connection coefficients assumed in the hypothesis of Theorem \ref{Thinitialdata}.}

\bigskip

\NI Clearly we can also state the analogous of inequalities \ref{iniz1}, \ref{iniz2}, \ref{iniz3} and \ref{iniz4} for the mixed derivatives, in this case, in order to estimate the right energy norms we have to estimate the $\Lie_S^p\Lie_O^q {\cal O}$, the Lie derivatives of the connection coefficients.\footnote{We can assume the constant $C$ and $\underline {C}$ be the same of the inequalities for the angular derivatives}:
\begin{theorem}\label{midata} Under the hypothesis of Theorem \ref{T11.1}, the following inequalities hold, with $p+q=J$

\NI{\bf On $C_0$:}
\bea\label{inizz1}
&&|r^{2-\frac{2}{p}}\Lie_S^p\Lie_O^qU|_{p,S}\leq F^{(0)}_{0,0}e^{(J-2)(\de_0+\underline{\Ga}_0(\la))}\frac{J!}{J^\a}\frac{1}{\ro_{0,1}^{J}}\nn\\
&&|r^{\frac{5}{2}+\ep-\frac{2}{p}}\Lie_S^p\Lie_O^q\chih|_{p,S}\leq F^{(0)}_{0,1}e^{(J-2)(\de_0+\underline{\Ga}_0(\la))}\frac{J!}{J^\a}\frac{1}{\ro_{0,1}^{J}}\nn\\
&&|r^{2+\ep-\frac{2}{p}}\Lie_S^p\Lie_O^q\om|_{p,S}\leq F^{(0)}_{0,2}e^{(J-2)(\de_0+\underline{\Ga}_0(\la))}\frac{J!}{J^\a}\frac{1}{\ro_ {0,1}^{J}}\nn\\
&&|r^{2-\frac{2}{p}}\Lie_S^p\Lie_O^q\ze|_{p,S}\leq F^{(0)}_{0,3}e^{(J-2)(\de_0+\underline{\Ga}_0(\la))}\frac{J!}{J^\a}\frac{1}{\ro_ {0,1}^{J}}\ \nn\\
&&\big|r^{2-\frac{2}{p}}\Lie_S^p\Lie_O^q\tr\chib\big|_{p,S}\leq F^{(0)}_{0,4}e^{(J-2)(\de_0+\underline{\Ga}_0(\la))}\frac{J!}{J^{\a}}\frac{1}{\ro_ {0,1}^J}\\
&&\big||\la_0|r^{1-\frac{2}{p}}\Lie_S^p\Lie_O^q\chibh\big|_{p,S}\leq F^{(0)}_{0,5}e^{(J-2)(\de_0+\underline{\Ga}_0(\la))}\frac{J!}{J^{\a}}\frac{1}{\ro_ {0,1}^J}\nn\\
&&||\la_0|r^{1-\frac{2}{p}}\Lie_S^p\Lie_O^q\omb|_{p,S}\leq F^{(0)}_{0,6}e^{(J-2)(\de_0+\underline{\Ga}_0(\la))}\frac{J!}{J^\a}\frac{1}{\ro_ {0,1}^{J}}\nn
\eea

\smallskip

\NI{\bf On $\Cb_0$:}
\bea\label{inizz2}
&&|r^{1-\frac{2}{p}}|\la|\Lie_S^p\Lie_O^q\underline{U}|_{p,S}\leq {\underline F}^{(0)}_{0,0}e^{(J-2)(\de_0+\underline{\Ga}_0(\la))}\frac{J!}{J^\a}\frac{1}{\ro_ {0,1}^{J}}\nn\\
&&|r^{1-\frac{2}{p}}|\la|^{\frac{3}{2}+\ep}\Lie_S^p\Lie_O^q\chibh|_{p,S}\leq {\underline F}^{(0)}_{0,1}e^{(J-2)(\de_0+\underline{\Ga}_0(\la))}\frac{J!}{J^\a}\frac{1}{\ro_ {0,1}^{J}}\nn\\
&& |r^{1-\frac{2}{p}}|\la|^{1+\ep}\Lie_S^p\Lie_O^q\omb|_{p,S}\leq {\underline F}^{(0)}_{0,2}e^{(J-2)(\de_0+\underline{\Ga}_0(\la))}\frac{J!}{J^\a}\frac{1}{\ro_ {0,1}^{J}}\nn\\
&&|r^{2-\frac{2}{p}}\Lie_S^p\Lie_O^qU|_{p,S}\leq {\underline F}^{(0)}_{0,3}e^{(J-2)(\de_0+\underline{\Ga}_0(\la))}\frac{J!}{J^\a}\frac{1}{\ro_ {0,1}^{J}}\nn\\
&&|r^{2-\frac{2}{p}}\Lie_S^p\Lie_O^q\chih|_{p,S}\leq {\underline F}^{(0)}_{0,4}e^{(J-2)(\de_0+\underline{\Ga}_0(\la))}\frac{J!}{J^\a}\frac{1}{\ro_ {0,1}^{J}}\\
&&|r^{2-\frac{2}{p}}\Lie_S^p\Lie_O^q\om|_{p,S}\leq {\underline F}^{(0)}_{0,5}e^{(J-2)(\de_0+\underline{\Ga}_0(\la))}\frac{J!}{J^\a}\frac{1}{\ro_ {0,1}^{J}}\nn\\
&&|r^{2-\frac{2}{p}}\Lie_S^p\Lie_O^q\ze|_{p,S}\leq {\underline F}^{(0)}_{0,6}e^{(J-2)(\de_0+\underline{\Ga}_0(\la))}\frac{J!}{J^\a}\frac{1}{\ro_ {0,1} ^{J}}.\ \nn
\eea

\NI With all the constant $F$ and $\underline{F}$ of order $O(\varepsilon)$ for $J \leq7$ and $O(1)$ otherwise.

\NI Moreover we have similar estimates for the first derivatives of the connection coefficients, with $p+q=J-1$:

\smallskip

\NI{\bf On $C_0$:}
\bea\label{inizz3}
&&|r^{3-\frac{2}{p}}\Lie_S^p\Lie_O^q\nabb U|_{p,S}\leq F^{(0)}_{1,0}e^{(J-2)(\de_0+\underline{\Ga}_0(\la))}\frac{J!}{J^\a}\frac{1}{\ro_{0,1}^{J}}\nn\\
&&|r^{\frac{7}{2}+\ep-\frac{2}{p}}\Lie_S^p\Lie_O^q\nabb\chih|_{p,S}\leq F^{(0)}_{1,1}e^{(J-2)(\de_0+\underline{\Ga}_0(\la))}\frac{J!}{J^\a}\frac{1}{\ro_{0,1}^{J}}\nn\\
&&|r^{3+\ep-\frac{2}{p}}\Lie_S^p\Lie_O^q\nabb\om|_{p,S}\leq F^{(0)}_{1,2}e^{(J-2)(\de_0+\underline{\Ga}_0(\la))}\frac{J!}{J^\a}\frac{1}{\ro_ {0,1}^{J}}\nn\\
&&|r^{3-\frac{2}{p}}\Lie_S^p\Lie_O^q\nabb\ze|_{p,S}\leq F^{(0)}_{1,3}e^{(J-2)(\de_0+\underline{\Ga}_0(\la))}\frac{J!}{J^\a}\frac{1}{\ro_ {0,1}^{J}}\ \nn\\
&&\big|r^{3-\frac{2}{p}}\Lie_S^p\Lie_O^q\nabb\tr\chib\big|_{p,S}\leq F^{(0)}_{1,4}e^{(J-2)(\de_0+\underline{\Ga}_0(\la))}\frac{J!}{J^{\a}}\frac{1}{\ro_ {0,1}^J}\\
&&\big||\la_0|r^{2-\frac{2}{p}}\Lie_S^p\Lie_O^q\nabb\chibh\big|_{p,S}\leq F^{(0)}_{1,5}e^{(J-2)(\de_0+\underline{\Ga}_0(\la))}\frac{J!}{J^{\a}}\frac{1}{\ro_ {0,1}^J}\nn\\
&&||\la_0|r^{2-\frac{2}{p}}\Lie_S^p\Lie_O^q\nabb\omb|_{p,S}\leq F^{(0)}_{1,6}e^{(J-2)(\de_0+\underline{\Ga}_0(\la))}\frac{J!}{J^\a}\frac{1}{\ro_ {0,1}^{J}}\nn
\eea

\smallskip

\NI{\bf On $\Cb_0$:}
\bea\label{inizz4}
&&|r^{2-\frac{2}{p}}|\la|\Lie_S^p\Lie_O^q\nabb\underline{U}|_{p,S}\leq {\underline F}^{(0)}_{1,0}e^{(J-2)(\de_0+\underline{\Ga}_0(\la))}\frac{J!}{J^\a}\frac{1}{\ro_ {0,1}^{J}}\nn\\
&&|r^{2-\frac{2}{p}}|\la|^{\frac{3}{2}+\ep}\Lie_S^p\Lie_O^q\nabb\chibh|_{p,S}\leq {\underline F}^{(0)}_{1,1}e^{(J-2)(\de_0+\underline{\Ga}_0(\la))}\frac{J!}{J^\a}\frac{1}{\ro_ {0,1}^{J}}\nn\\
&& |r^{2-\frac{2}{p}}|\la|^{1+\ep}\Lie_S^p\Lie_O^q\nabb\omb|_{p,S}\leq {\underline F}^{(0)}_{1,2}e^{(J-2)(\de_0+\underline{\Ga}_0(\la))}\frac{J!}{J^\a}\frac{1}{\ro_ {0,1}^{J}}\nn\\
&&|r^{3-\frac{2}{p}}\Lie_S^p\Lie_O^q\nabb U|_{p,S}\leq {\underline F}^{(0)}_{1,3}e^{(J-2)(\de_0+\underline{\Ga}_0(\la))}\frac{J!}{J^\a}\frac{1}{\ro_ {0,1}^{J}}\nn\\
&&|r^{3-\frac{2}{p}}\Lie_S^p\Lie_O^q\nabb\chih|_{p,S}\leq {\underline F}^{(0)}_{1,4}e^{(J-2)(\de_0+\underline{\Ga}_0(\la))}\frac{J!}{J^\a}\frac{1}{\ro_ {0,1}^{J}}\\
&&|r^{3-\frac{2}{p}}\Lie_S^p\Lie_O^q\nabb\om|_{p,S}\leq {\underline F}^{(0)}_{1,5}e^{(J-2)(\de_0+\underline{\Ga}_0(\la))}\frac{J!}{J^\a}\frac{1}{\ro_ {0,1}^{J}}\nn\\
&&|r^{3-\frac{2}{p}}\Lie_S^p\Lie_O^q\nabb\ze|_{p,S}\leq {\underline F}^{(0)}_{1,6}e^{(J-2)(\de_0+\underline{\Ga}_0(\la))}\frac{J!}{J^\a}\frac{1}{\ro_ {0,1} ^{J}}.\ \nn
\eea

\NI With all the constant $F$ and $\underline{F}$ of order $O(\varepsilon)$ for $J \leq7$ and $O(1)$ otherwise.

\end{theorem}
\NI {\bf Proof:} The proof  of inequalities \ref{inizz3} and \ref{inizz4} follows as in the proof of lemmas \ref{L4.1} and \ref{L4.2}, with the obvious modifications.
{\subsection{The covariant angular derivatives of the connection coefficients}\label{S.sangest}
Once we have built an analytic initial data set, we write here the precise estimates for the covariant angular derivatives of the connection coefficients on the internal region ${\cal K}$  and provide  the precise version of Theorem \ref{L6.1}, the detailed proof of these results are in Section 9.

\NI First we state a modified version of the main theorem, where the $\nabb$ derivatives are substituted with the $\Lie_O$ derivatives, in order to exploit the energy estimates for the null Riemann components.

\begin{theorem}\label{L2.1new}
Under the hyptoesis  stated in Theorem \ref{Thinitialdata} the following estimates hold on ${\cal K}$,
\bea\label{323}
&&|r^{3-\frac{2}{p}}\Lie_O^{J-1}\nabb U|_{p,S}\leq C_0e^{(J-2)\de}e^{(J-2)\underline{\Ga}(\la)}\frac{J!}{J^\a}\frac{1}{\ro_{0,1}^{J}}\nn\\
&&|r^{3-\frac{2}{p}}\Lie_O^{J-1}\nabb\chih|_{p,S}\leq C_1 e^{(J-2)\de}e^{(J-2)\underline{\Ga}(\la)}\frac{J!}{J^\a}\frac{1}{\ro_{0,1}^{J}}\nn\\
&&|r^{3-\frac{2}{p}}\Lie_O^{J-1}\nabb\om|_{p,S}\leq C_2e^{(J-2)\de}e^{(J-2)\underline{\Ga}(\la)}\frac{J!}{J^\a}\frac{1}{\ro_{0,1}^{J}}\nn\\
&&|r^{3-\frac{2}{p}}\Lie_O^{J-1}\nabb\ze|_{p,S}\leq C_3 e^{(J-2)\de}e^{(J-2)\underline{\Ga}(\la)}\frac{J!}{J^\a}\frac{1}{\ro_{0,1}^{J}}\eql{123a}\\
&&|r^{3-\frac{2}{p}}\Lie_O^{J-1}\nabb\underline{U}|_{p,S}\leq \Cb_0e^{(J-2)\de}e^{(J-2)\underline{\Ga}(\la)}\frac{J!}{J^\a}\frac{1}{\ro_{0,1}^{J}}\nn\\
&&|r^{2-\frac{2}{p}}|\la|\Lie_O^{J-1}\nabb\chibh|_{p,S}\leq \Cb_1e^{(J-2)\de}e^{(J-2)\underline{\Ga}(\la)}\frac{J!}{J^\a}\frac{1}{\ro_{0,1}^{J}}\nn\\
&&|r^{2-\frac{2}{p}}|\la|\Lie_O^{J-1}\nabb\omb|_{p,S}\leq \Cb_2e^{(J-2)\de}e^{(J-2)\underline{\Ga}(\la)}\frac{J!}{J^\a}\frac{1}{\ro_{0,1}^{J}}\ .\nn
\eea

\NI Moreover, the following estimates for  the connection coefficients hold:

\bea\label{324}
&&|r^{2-\frac{2}{p}}\Lie_O^JU|_{p,S}\leq C_0e^{(J-2)\de}e^{(J-2)\underline{\Ga}(\la)}\frac{J!}{J^\a}\frac{1}{\ro_{0,1}^{J}}\nn\\
&&|r^{2-\frac{2}{p}}\Lie_O^J\chih|_{p,S}\leq C_1e^{(J-2)\de}e^{(J-2)\underline{\Ga}(\la)}\frac{J!}{J^\a}\frac{1}{\ro_{0,1}^{J}}\nn\\
&&|r^{2-\frac{2}{p}}\Lie_O^J\om|_{p,S}\leq \Cb_2e^{(J-2)\de}e^{(J-2)\underline{\Ga}(\la)}\frac{J!}{J^\a}\frac{1}{\ro_{0,1}^{J}}\nn\\
&&|r^{2-\frac{2}{p}}\Lie_O^J\underline{U}|_{p,S}\leq \Cb_0e^{(J-2)\de}e^{(J-2)\underline{\Ga}(\la)}\frac{J!}{J^\a}\frac{1}{\ro_{0,1}^{J}}\nn\\
&&|r^{1-\frac{2}{p}}|\la|\Lie_O^J\chibh|_{p,S}\leq \Cb_1e^{(J-2)\de}e^{(J-2)\underline{\Ga}(\la)}\frac{J!}{J^\a}\frac{1}{\ro_{0,1}^{J}}\nn\\
&&|r^{1-\frac{2}{p}}|\la|\Lie_O^J\omb|_{p,S}\leq C_2e^{(J-2)\de}e^{(J-2)\underline{\Ga}(\la)}\frac{J!}{J^\a}\frac{1}{\ro_{0,1}^{J}}\ .\nn
\eea

\NI With a suitable choice of the  constants $C$ and $\Cb$ of order $O(\varepsilon)$ for $J \leq7$ and $O(1)$ otherwise and of the radius of convergence  $\ro_{0,1}<\ro_0<\ro_{0,0,1}$, $\ro_{0,1}$ depending only the first $s$ derivatives of the initial data, with $s\leq 7$.
\end{theorem} 
\NI {\bf Proof:} The proof in Section  9.

\NI Finally, the following estimates hold for the $|\c|_{\infty,S}$ norms of the connection coefficients,
\begin{cor}\label{333}
Under the hypothesis stated in Theorem \ref{Thinitialdata}, the Following estimates hold on ${\cal K}$:
\bea\label{334}
&&|r^{2}\Lie_O^{J-1}U|_{\infty,S}\leq C_0e^{(J-2)\de}e^{(J-2)\underline{\Ga}(\la)}\frac{J!}{J^\a}\frac{1}{\ro_{0,1}^{J}}\nn\\
&&|r^{2}\Lie_O^{J-1}\chih|_{\infty,S}\leq C_1e^{(J-2)\de}e^{(J-2)\underline{\Ga}(\la)}\frac{J!}{J^\a}\frac{1}{\ro_{0,1}^{J}}\nn\\
&&|r^{2}\Lie_O^{J-1}\om|_{\infty,S}\leq \Cb_2e^{(J-2)\de}e^{(J-2)\underline{\Ga}(\la)}\frac{J)!}{J^\a}\frac{1}{\ro_{0,1}^{J}}\nn\\
&&|r^{2}\Lie_O^{J-1}\ze|_{\infty,S}\leq \Cb_3e^{(J-2)\de}e^{(J-2)\underline{\Ga}(\la)}\frac{J!}{J^\a}\frac{1}{\ro_{0,1}^{J}}\\
&&|r^{2}\Lie_O^{J-1}\underline{U}|_{\infty,S}\leq \Cb_0e^{(J-2)\de}e^{(J-2)\underline{\Ga}(\la)}\frac{J!}{J^\a}\frac{1}{\ro_{0,1}^{J}}\nn\\
&&|r|\la|\Lie_O^{J-1}\chibh|_{\infty,S}\leq \Cb_1e^{(J-2)\de}e^{(J-2)\underline{\Ga}(\la)}\frac{J!}{J^\a}\frac{1}{\ro_{0,1}^{J}}\nn\\
&&|r|\la|\Lie_O^{J-1}\omb|_{\infty,S}\leq C_2e^{(J-2)\de}e^{(J-2)\underline{\Ga}(\la)}\frac{J!}{J^\a}\frac{1}{\ro_{0,1}^{J}}\ .\nn
\eea

\bea\label{335}
&&|r^{3}\Lie_O^{J-2}\nabb U|_{\infty,S}\leq C_0e^{(J-2)\de}e^{(J-2)\underline{\Ga}(\la)}\frac{J!}{J^\a}\frac{1}{\ro_{0,1}^{J}}\nn\\
&&|r^{3}\Lie_O^{J-2}\nabb\chih|_{\infty,S}\leq C_1e^{(J-2)\de}e^{(J-2)\underline{\Ga}(\la)}\frac{J!}{J^\a}\frac{1}{\ro_{0,1}^{J}}\nn\\
&&|r^{3}\Lie_O^{J-2}\nabb\om|_{\infty,S}\leq \Cb_2e^{(J-2)\de}e^{(J-2)\underline{\Ga}(\la)}\frac{J)!}{J^\a}\frac{1}{\ro_{0,1}^{J}}\nn\\
&&|r^{3}\Lie_O^{J-2}\nabb\ze|_{\infty,S}\leq \Cb_3e^{(J-2)\de}e^{(J-2)\underline{\Ga}(\la)}\frac{J!}{J^\a}\frac{1}{\ro_{0,1}^{J}}\\
&&|r^{3}\Lie_O^{J-2}\nabb\underline{U}|_{\infty,S}\leq \Cb_0e^{(J-2)\de}e^{(J-2)\underline{\Ga}(\la)}\frac{J!}{J^\a}\frac{1}{\ro_{0,1}^{J}}\nn\\
&&|r^{2}|\la|\Lie_O^{J-2}\nabb\chibh|_{\infty,S}\leq \Cb_1e^{(J-2)\de}e^{(J-2)\underline{\Ga}(\la)}\frac{J!}{J^\a}\frac{1}{\ro_{0,1}^{J}}\nn\\
&&|r^{2}|\la|\Lie_O^{J-2}\nabb\omb|_{\infty,S}\leq C_2e^{(J-2)\de}e^{(J-2)\underline{\Ga}(\la)}\frac{J!}{J^\a}\frac{1}{\ro_{0,1}^{J}}\ .\nn
\eea

\end{cor} }
 
 \NI {\bf Proof:} The proof is in section  9.

\NI {\bf Remarks:} 

\NI {\em i) We will prove the estimates for $\eta=\ze+\nabb\log\oom$. The estimates for $\ze$ follow in a straightforward way.

\NI ii) The estimates \ref{323}, \ref{324} and \ref{335}, have not the same constants $C_i$ and $\Cb_i$, we call them in the same way only for the sake of simplicity. }

\NI Now we can state the final form of Theorem \ref{L6.1}.

\begin{theorem}\label{L2.1neww}
Under the hypothesis stated in Theorem \ref{Thinitialdata}, the following estimates hold  on ${\cal K}$\footnote{ 
All the remaining quantities are defined in Theorem \ref{L6.1}.}
\bea
&&|r^{J+2-\frac{2}{p}}\nabb^JU|_{p,S}\leq C_0e^{(J-2)\de}e^{(J-2)\underline{\Ga}(\la)}\frac{J!}{J^\a}\frac{1}{\ro^{J}}\nn\\
&&|r^{J+2-\frac{2}{p}}\nabb^J\chih|_{p,S}\leq C_1e^{(J-2)\de}e^{(J-2)\underline{\Ga}(\la)}\frac{J!}{J^\a}\frac{1}{\ro^{J}}\nn\\
&&|r^{J+2-\frac{2}{p}}\nabb^J\om|_{p,S}\leq \Cb_2e^{(J-2)\de}e^{(J-2)\underline{\Ga}(\la)}\frac{J)!}{J^\a}\frac{1}{\ro^{J}}\nn\\
&&|r^{J+2-\frac{2}{p}}\nabb^J\ze|_{p,S}\leq \Cb_3e^{(J-2)\de}e^{(J-2)\underline{\Ga}(\la)}\frac{J!}{J^\a}\frac{1}{\ro^{J}}\\
&&|r^{J+2-\frac{2}{p}}\nabb^J\underline{U}|_{p,S}\leq \Cb_0e^{(J-2)\de}e^{(J-2)\underline{\Ga}(\la)}\frac{J!}{J^\a}\frac{1}{\ro^{J}}\nn\\
&&|r^{J+1-\frac{2}{p}}|\la|\nabb^J\chibh|_{p,S}\leq \Cb_1e^{(J-2)\de}e^{(J-2)\underline{\Ga}(\la)}\frac{J!}{J^\a}\frac{1}{\ro^{J}}\nn\\
&&|r^{J+1-\frac{2}{p}}|\la|\nabb^J\omb|_{p,S}\leq C_2e^{(J-2)\de}e^{(J-2)\underline{\Ga}(\la)}\frac{J!}{J^\a}\frac{1}{\ro^{J}}\ .\nn
\eea
\NI with $\ro<\ro_{0,1}$, $\ro$ depending only the first $s$ derivatives of the initial data, with $s\leq 7$
 and with a suitable choice of the  constants $C_i$ , $\Cb_i$ of order $O(\varepsilon)$ for $J \leq7$ and $O(1)$ otherwise
\end{theorem} 
\NI {\bf Proof:} The proof follows from Theorem \ref{L2.1new} and is equivalent, for the connection coefficients, to lemma 10.4 of section 10.1.5 for the null Riemann components .
\subsection{$\ddb_{\nu}$ and $\ddb_{\la}$ derivatives and the mixed ones for the connection coefficients}\label{37}
\NI To complete our proof  we need to estimate the higher derivatives with respect to $e_3$ and $e_4$ of the connection coefficients and consequently the higher derivatives with respect to $e_3$ and $e_4$ of the Riemann components.  One could try to use the Weyl equations to get $\ddb_3^J$ and $\ddb_4^J$ derivatives from the angular ones, for instance, considering the $\b$ component, exploiting the equation
\bea
\dddd_3\b+\tr\chib\b=\nabb\ro+\left[2\omb\b+\dual\nabb\si+2\hat{\chi}\c\bb+3(\eta\ro+\dual\eta\si)\right]
\eea
 We get $\ddb_3\b$ term from the angular derivatives, nevertheless iterating this mechanism for the $\b$ component, in the expression of $\ddb_3^K\b$, with $K>4$, a term 
$\nabb^3\ddb_3^{K-3}\aa$ is present and to control it\footnote{Remember that we do not have a transport equation for $\aa$ along the incoming cones.} we  have to extend the family of the $Q$ norms we are considering.  
\NI More precisely 
we have to extend the $Q$ norms also to terms with ${\tilde W}\!=\!\lie_S^{J}W$ with $J\leq N-1$with $S=\ub e_4+ue_3$, beside those with $\lie_O^{J}W$. More in general to control all the mixed (tangential) derivatives with $\dddd_{4}$, $\dddd_{3}$ $q$ times and the angular ones, $\nabb$, $p$ times with $p+q\leq N-1$, we have to define the $\cal Q$ norms analogous to those defined in $\cite{Kl-Ni:book}$, see also definitions \ref{QQ12} and \ref{QQb12}, for all the Weyl tensors ${\tilde W}_{p,q}=\Lie_S^q\lie_O^{p}W$ with $p+q\leq N\!-\!1$.  In order to estimate these new norms we have to reproduce all the machinery adopted to estimate angular derivatives of both, the connection coefficients and the null Riemann components.

\NI  Finally we will prove the following theorem, analogous to theorem  \ref{L2.1new}.
\begin{theorem}\label{L2.1newnew}
Under the estimates on the initial data of Theorem \ref{T11.1}, the following estimates hold on ${\cal K}$,\footnote{ 
all the remaining quantities are defined in Theorem \ref{L6.1}.}

\NI with $p+q=J$\footnote{We can assume the constant $C$ and $\underline {C}$ be the same of the inequalities for the angular derivatives}:

\bea\label{fin11}
&&|r^{2-\frac{2}{p}}\Lie_S^p\Lie_O^qU|_{p,S}\leq C_{0,0}e^{(J-2)\de}e^{(J-2)\underline{\Ga}(\la)}\frac{J!}{J^\a}\frac{1}{\ro_{0,1}^{J}}\nn\\
&&|r^{\frac{5}{2}+\ep-\frac{2}{p}}\Lie_S^p\Lie_O^q\chih|_{p,S}\leq C_{0,1}e^{(J-2)\de}e^{(J-2)\underline{\Ga}(\la)}\frac{J!}{J^\a}\frac{1}{\ro_{0,1}^{J}}\nn\\
&&|r^{2+\ep-\frac{2}{p}}\Lie_S^p\Lie_O^q\om|_{p,S}\leq C_{0,2}e^{(J-2)\de}e^{(J-2)\underline{\Ga}(\la)}\frac{J!}{J^\a}\frac{1}{\ro_ {0,1}^{J}}\nn\\
&&|r^{2-\frac{2}{p}}\Lie_S^p\Lie_O^q\ze|_{p,S}\leq C_{0,3}e^{(J-2)\de}e^{(J-2)\underline{\Ga}(\la)}\frac{J!}{J^\a}\frac{1}{\ro_ {0,1}^{J}}\ \nn\\
&&\big|r^{2-\frac{2}{p}}\Lie_S^p\Lie_O^q\tr\chib\big|_{p,S}\leq C_{0,4}e^{(J-2)\de}e^{(J-2)\underline{\Ga}(\la)}\frac{J!}{J^{\a}}\frac{1}{\ro_ {0,1}^J}\\
&&\big||\la_0|r^{1-\frac{2}{p}}\Lie_S^p\Lie_O^q\chibh\big|_{p,S}\leq C_{0,5}e^{(J-2)\de}e^{(J-2)\underline{\Ga}(\la)}\frac{J!}{J^{\a}}\frac{1}{\ro_ {0,1}^J}\nn\\
&&||\la_0|r^{1-\frac{2}{p}}\Lie_S^p\Lie_O^q\omb|_{p,S}\leq C_{0,6}e^{(J-2)\de}e^{(J-2)\underline{\Ga}(\la)}\frac{J!}{J^\a}\frac{1}{\ro_ {0,1}^{J}}\nn
\eea

\smallskip

\bea\label{fin12}
&&|r^{1-\frac{2}{p}}|\la|\Lie_S^p\Lie_O^q\underline{U}|_{p,S}\leq \Cb_{0,0}e^{(J-2)\de}e^{(J-2)\underline{\Ga}(\la)}\frac{J!}{J^\a}\frac{1}{\ro_ {0,1}^{J}}\nn\\
&&|r^{1-\frac{2}{p}}|\la|^{\frac{3}{2}}\Lie_S^p\Lie_O^q\chibh|_{p,S}\leq \Cb_{0,1}e^{(J-2)\de}e^{(J-2)\underline{\Ga}(\la)}\frac{J!}{J^\a}\frac{1}{\ro_ {0,1}^{J}}\nn\\
&& |r^{1-\frac{2}{p}}|\la|^{1}\Lie_S^p\Lie_O^q\omb|_{p,S}\leq \Cb_{0,2}e^{(J-2)\de}e^{J\underline{\Ga}(\la)}\frac{J!}{J^\a}\frac{1}{\ro_ {0,1}^{J}}\nn\\
&&|r^{2-\frac{2}{p}}\Lie_S^p\Lie_O^qU|_{p,S}\leq \Cb_{0,3}e^{(J-2)\de}e^{(J-2)\underline{\Ga}(\la)}\frac{J!}{J^\a}\frac{1}{\ro_ {0,1}^{J}}\nn\\
&&|r^{2-\frac{2}{p}}\Lie_S^p\Lie_O^q\chih|_{p,S}\leq \Cb_{0,4}e^{(J-2)\de}e^{(J-2)\underline{\Ga}(\la)}\frac{J!}{J^\a}\frac{1}{\ro_ {0,1}^{J}}\\
&&|r^{2-\frac{2}{p}}\Lie_S^p\Lie_O^q\om|_{p,S}\leq \Cb_{0,5}e^{(J-2)\de}e^{(J-2)\underline{\Ga}(\la)}\frac{J!}{J^\a}\frac{1}{\ro_ {0,1}^{J}}\nn\\
&&|r^{2-\frac{2}{p}}\Lie_S^p\Lie_O^q\ze|_{p,S}\leq \Cb_{0,6}e^{(J-2)\de}e^{(J-2)\underline{\Ga}(\la)}\frac{J!}{J^\a}\frac{1}{\ro_ {0,1} ^{J}}.\ \nn
\eea
\NI With $\ro_{0,1}$, depending only the first $s$ derivatives of the initial data, with $s\leq 7$
 and with a suitable choice of the  constants $C_{0,i}$ of order $O(\varepsilon)$ for $J \leq7$ and $O(1)$ otherwise.
\end{theorem}
\NI {\bf Proof:}  The proof goes in the same way as done for the  $\Lie_O^J {\cal O}$ estimates for the connection coefficients, It suffices to notice that, with $U$ a  generic connection coefficient.
\bea
&&(\Lie_SU)(e_A)=\pr_S(U(e_A))-U([S,e_A])=D_SU(e_A)+U(D_{e_A}S)\nn\\
\eea
Therefore
\bea
\Lie_SU(\c)=D_SU(\c)+{\bf g}(e_B,D_{\c}S)U(e_B)=D_SU(\c)+(U\c{\cal M})(\c)
\eea
where 
\bea
{\cal M}(\c,\c)={\bf g}(\c,D_{\c}S)=u\chib(\c,\c)+{\underline u}\chi(\c,\c)\  \ \ \ .
\eea
and we recall that
\bea
\Lie_OV=D_OV+{\cal H}\c V\ ,
\eea
where 
\bea{\cal H}(\c,\c)={\bf g}(\c,\nabb_{\c}O)\  \ \ \ .\eea
Hence we can pass from $\Lie_O$ to $\nabb_O$ and from $\Lie_S$ to $\ddb_S$ in a formally identical way as discussed in details in Sections 9 and and in its appendix, see Section 14, but now for the $\Lie_S\Lie_O$ derivatives, exploiting the smallness of the initial data \ref{inizz1} and \ref{inizz2} and the smallness of the extended ${\cal Q}$ norms, see section \ref{S.5}, Theorem \ref{324567}.

\NI {\bf Remark:} 

\NI {\em It is important to observe that the mixed derivatives estimates can also be obtained in a different way, without solving again the the transport equations  for $\Lie_S^p\Lie_O^q {\cal O}$. in fact as we already have all the estimates for the $\Lie_O^p {\cal O}$ terms, for any $p$,we can write
\[ \Lie_S^p\Lie_O^p{\cal O}= \Lie_S^{q-1}\Lie_O^p\Lie_S{\cal O}+\Lie_O^{q-1} [\Lie_S^p\Lie_O^q] {\cal O}\] and write explicitly $\Lie_S{\cal O}$ using the structure equations. In this way all the terms on the right side can be estimate, with a long but standard computations, without solving explicitly the transport equations. This is possible as now we can use all the structure equation without worrying of the lost of derivatives, as we already control all the angular derivatives estimates. This procedure is analogous to what has been done for the mixed derivatives  the initial data, see \ref{T11.1},  and points out the fact that the central role played by the estimates of the angular derivatives.  }
\medskip

\NI The next step is to estimates of the $\ddb_{\nu}^p\nabb^q$ mixed derivatives of the connection coefficients.
\begin{Le}\label{le31}
Let us assume the estimates of theorem \ref{L2.1newnew}, then the connection coefficients ${\cal O}$ satisfy the following estimates  for $p+q=J$ on ${\cal K}$,   with $\ro_1<\ro<\ro_{0,1}$ depending only the first $s$ derivatives of the initial data, with $s\leq 7$  and with a suitable choice of the  $D_4$ constant of order $O(\varepsilon)$ for $J \leq7$ and $O(1)$ otherwise.
\bea\label{rof1}
\big| |\la|^{\underline{\phi}(\Psi)}r^{\phi({\cal O})+q-\frac{2}{p}}D_S^p\nabb^{k}{\cal O}\big|_{L^p(S)}\leq D_4e^{(J-2)\de}e^{(J-2)\underline{\Ga}(\la)}\frac{J!}{J^\a}\frac{1}{\ro_1^J}\ ,
\eea
with $\phi({\cal O})$ the suitable weights related to the different connection coefficients, see \ref{inizz1}, \ref{inizz2}. \footnote{For $J=0$ $\ro$ becomes $\ro-{\underline\ro}$, see \cite{Kl-Ni:book}.}
\end{Le}
\NI {\bf Proof:} The proof is analogous to the one of lemma \ref{from|LieTLieOtoDTNabb} but now for the connection coefficients.

\NI We have now to pass from $\ddb_S$ to $\ddb_{\nu}$ and $\ddb_{\la}$ this is the content of the following lemma, analogous to lemma \ref{cor10.3xxx} 
\begin{Le} \label{cor10.3xx}
Given the assumptions of Lemma \ref{le31} the following estimates hold on ${\cal K}$,
\bea
&&\big| |\la|^{\underline{\phi}({\cal O})}r^{\phi({\cal O})+(J-1)-\frac{2}{p}}D_{\nu}^p\nabb^{q}{\cal O}\big|_{L^p(S)}\leq D_5e^{(J-2)\de}e^{(J-2)\underline{\Ga}(\la)}\frac{J!}{J^\a}\frac{1}{\ro_2^p\ro_1^k}\nn\\ 
&&\big| |\la|^{\underline{\phi}(\Psi)+p}r^{\phi({\cal O})+k-\frac{2}{p}}D_{\la}^p\nabb^{q}{\cal O}\big|_{L^p(S)}\leq D_5e^{(J-2)\de}e^{(J-2)\underline{\Ga}(\la)}\frac{J!}{J^\a}\frac{1}{\ro_2^p\ro_1^k}\ .\ \ \ \ \ \ \ 
\eea
With $\ro_2\leq\ro_1$ depending only the first $s$ derivatives of the initial data, with $s\leq 7$ and with a suitable choice of the  constants $D_{i}$of order $O(\varepsilon)$ for $J \leq7$ and $O(1)$ otherwise.

\end{Le}
\NI {\bf Proof:} The proof is a repetition of Lemma \ref{cor10.3xxx}, but now for the connection coefficients .

\NI {\bf Remark:} 

\NI {\em The reason of having  $\ro_2<\ro$ in the previous estimates of lemma \ref{mixedderivatives} and \ref{mixedderivatives2} is due to the fact that in proving these estimates we need the estimates for the mixed derivatives of the Riemann components. These estimates are obtained from the ${\cal Q}$ norm estimates associated to the $\lie_S^k\lie_O^pW$ Weyl fields. The problem arises when from $\lie_S$ or $D_S$ we move to $D_{\nu}$ or to $D_{\la}$ as this  requires lowering $\ro_1$ to $\ro_2$. It is important to remark that we are allowed to do this as we do not have to use the mixed derivatives of the connection coefficients in a bootstrap mechanism. In fact the mixed derivative appear in the error estimates as mixed derivatives of $\pi, {\tilde q}$ or more in general of the connection coefficients, but in this case they appear with $D_S$ or $\lie_S$.}

\subsection{ General mixed derivatives}
\NI Let us discuss the situation when the mixed derivatives are disposed in an arbitrary way, we start considering a specific case as the structure of the proof will be the same for all the remaining connection coefficients.  Therefore we show how to bound the norm of the following term, 

\[\nabb^{p_0}\ddb_{\nu}\nabb^{p_1}\ddb_{\nu}\nabb^{p_2}\c\c\c\ddb_{\nu}\nabb^{p_{J-1}}\ddb_{\nu}\nabb^{p_{J}}U\]
where
\[U=\oom^{-1}\tr\chi\ \ \ ,\ \ \ \sum_{s=0}^Jp_s=P\ \ \ , \ \ \ J+P=N\ .\]
\begin{Le}\label{l3l}
Given the assumptions of Lemma \ref{le31} the following estimates hold on ${\cal K}$,with $\ro_2<\ro$ depending only the first $s$ derivatives of the initial data, with $s\leq 7$ and with a suitable choice of the  constants  ${\hat F}$ of order $O(\varepsilon)$ for $J \leq7$ and $O(1)$ otherwise.
\bea
\ML\ML\ML|r^{1+J+P+\si(P)-\frac{2}{p}}\nabb^{p_0}\ddb_{\nu}\nabb^{p_1}\ddb_{\nu}\nabb^{p_2}\c\c\c\ddb_{\nu}\nabb^{p_{J-1}}\ddb_{\nu}\nabb^{p_{J}}\tr\chi|_{p,S}\leq {\hat F}\!\left(\frac{(J+P)!}{(J+P)^{\a}}\frac{e^{(J+P-2)(\de+\underline{\Ga}(\la))}}{\ro_2^{J+P}}\right)\ .\ \ \ \ 
\eql{Umixedder}
\eea
\NI{\bf Proof:} The proof is in appendix 16.
\end{Le}
\medskip

\NI The proof, can be extended without crucial modification to all the remaining connection coefficients. Notice that to prove the general mixed derivatives estimates for the connection coefficients we need to control also the general mixed derivatives estimates for the Riemann components but at a lower order of derivatives, these estimates can be achieved directly by the knowledge of the mixed derivatives of the connection coefficients already at our disposal by induction, see lemma \ref{mixedderivatives3}. We can state hence the following lemma:\begin{Le}\label{mixedderivatives}
Under the assumptions of Lemma \ref{le31} the following estimates hold on ${\cal K}$, with $\ro_2<\ro$ depending only the first $s$ derivatives of the initial data, with $s\leq 7$  and with a suitable choice of the  constants  $\hat{\underline{ F}}_i$ of order $O(\varepsilon)$ for $J \leq7$ and $O(1)$ otherwise.

\bea
&&\ML\ML\ML|r^{1+J+P+\si(P)-\frac{2}{p}}\nabb^{p_0}\ddb_{\nu}\nabb^{p_1}\ddb_{\nu}\nabb^{p_2}\c\c\c\ddb_{\nu}\nabb^{p_{J-1}}\ddb_{\nu}\nabb^{p_{J}}\tr\chi|_{p,S}\leq {\hat F}_1\!\left(\frac{(J+P)!}{(J+P)^{\a}}\frac{e^{(J+P-2)(\de+\underline{\Ga}(\la))}}{\ro_2^{J+P}}\right)\nn\\
&&\ML\ML\ML|r^{2+J+P-\frac{2}{p}}\nabb^{p_0}\ddb_{\nu}\nabb^{p_1}\ddb_{\nu}\nabb^{p_2}\c\c\c\ddb_{\nu}\nabb^{p_{J-1}}\ddb_{\nu}\nabb^{p_{J}}\chih|_{p,S}\leq {\hat F}_2\!\left(\frac{(J+P)!}{(J+P)^{\a}}\frac{e^{(J+P-2)(\de+\underline{\Ga}(\la))}}{\ro_2^{J+P}}\right)\nn\\
&&\ML\ML\ML|r^{2+J+P-\frac{2}{p}}\nabb^{p_0}\ddb_{\nu}\nabb^{p_1}\ddb_{\nu}\nabb^{p_2}\c\c\c\ddb_{\nu}\nabb^{p_{J-1}}\ddb_{\nu}\nabb^{p_{J}}\ze|_{p,S}\leq {\hat F}_3\!\left(\frac{(J+P+1)!}{(J+P+1)^{\a}}\frac{e^{(J+P-1)(\de+\underline{\Ga}(\la))}}{\ro_2^{J+P+1}}\right)\nn\\
&&\ML\ML\ML|r^{2+J+P-\frac{2}{p}}\nabb^{p_0}\ddb_{\nu}\nabb^{p_1}\ddb_{\nu}\nabb^{p_2}\c\c\c\ddb_{\nu}\nabb^{p_{J-1}}\ddb_{\nu}\nabb^{p_{J}}\om|_{p,S}\leq {\hat F}_4\!\left(\frac{(J+P+1)!}{(J+P+1)^{\a}}\frac{e^{(J+P-1)(\de+\underline{\Ga}(\la))}}{\ro_2^{J+P+1}}\right)\nn\\
&&\ML\ML\ML|r^{1+J+P+\si(P)-\frac{2}{p}}\nabb^{p_0}\ddb_{\nu}\nabb^{p_1}\ddb_{\nu}\nabb^{p_2}\c\c\c\ddb_{\nu}\nabb^{p_{J-1}}\ddb_{\nu}\nabb^{p_{J}}\tr\chib|_{p,S}\leq {\hat F}_5\!\left(\frac{(J+P)!}{(J+P)^{\a}}\frac{e^{(J+P-2)(\de+\underline{\Ga}(\la))}}{\ro_2^{J+P}}\right)\nn\\
&&\ML\ML\ML||\la|r^{1+J+P-\frac{2}{p}}\nabb^{p_0}\ddb_{\nu}\nabb^{p_1}\ddb_{\nu}\nabb^{p_2}\c\c\c\ddb_{\nu}\nabb^{p_{J-1}}\ddb_{\nu}\nabb^{p_{J}}\chibh|_{p,S}\leq {\hat F}_6\!\left(\frac{(J+P)!}{(J+P)^{\a}}\frac{e^{(J+P-2)(\de+\underline{\Ga}(\la))}}{\ro_2^{J+P}}\right)\nn\\
&&\ML\ML\ML||\la|r^{1+J+P-\frac{2}{p}}\nabb^{p_0}\ddb_{\nu}\nabb^{p_1}\ddb_{\nu}\nabb^{p_2}\c\c\c\ddb_{\nu}\nabb^{p_{J-1}}\ddb_{\nu}\nabb^{p_{J}}\omb|_{p,S}\leq {\hat F}_7\!\left(\frac{(J+P)!}{(J+P)^{\a}}\frac{e^{(J+P-2)(\de+\underline{\Ga}(\la))}}{\ro_2^{J+P}}\right)\ \ .\ \ \ \ \ \ \ \ \ \ \ 
\eea
\end{Le}

\NI Proceeding exactly in the same way with the obvious modifications we can prove analogous estimates for the ``tangential" derivatives for the incoming cones. In this case one has to recall that the in the weight factors each $\ddb_{\la}$ derivative brings a $|\la|$ factor instead of a $r$ factor. A specific care has to be done to control the mixed derivatives of $\underline{{\hat\om}}$. 
The final result is summarised in the following lemma:
\begin{Le}\label{mixedderivatives2}
Under the assumptions of Lemma \ref{le31} the following estimates hold on ${\cal K}$, with $\ro_2<\ro$ depending only the first $s$ derivatives of the initial data, with $s\leq 7$  and with a suitable choice of the  constants  $\hat{\underline{ F}}_i$ of order $O(\varepsilon)$ for $J \leq7$ and $O(1)$ otherwise.
\bea
&&\ML\ML\ML\ML||\la|^Jr^{1+P+\si(P)-\frac{2}{p}}\nabb^{p_0}\ddb_{\la}\nabb^{p_1}\ddb_{\la}\nabb^{p_2}\c\c\c\ddb_{\la}\nabb^{p_{J-1}}\ddb_{\la}\nabb^{p_{J}}\tr\chi|_{p,S}\leq {\hat{\underline F}}_1\!\left(\frac{(J+P)!}{(J+P)^{\a}}\frac{e^{(J+P-2)(\de+\underline{\Ga}(\la))}}{\ro_2^{J+P}}\right)\nn\\
&&\ML\ML\ML\ML||\la|^Jr^{2+P-\frac{2}{p}}\nabb^{p_0}\ddb_{\la}\nabb^{p_1}\ddb_{\la}\nabb^{p_2}\c\c\c\ddb_{\la}\nabb^{p_{J-1}}\ddb_{\la}\nabb^{p_{J}}\chih|_{p,S}\leq {\hat{\underline F}}_2\!\left(\frac{(J+P)!}{(J+P)^{\a}}\frac{e^{(J+P-2)(\de+\underline{\Ga}(\la))}}{\ro_2^{J+P}}\right)\nn\\
&&\ML\ML\ML\ML||\la|^Jr^{2+P-\frac{2}{p}}\nabb^{p_0}\ddb_{\la}\nabb^{p_1}\ddb_{\la}\nabb^{p_2}\c\c\c\ddb_{\la}\nabb^{p_{J-1}}\ddb_{\la}\nabb^{p_{J}}\ze|_{p,S}\leq {\hat{\underline F}}_3\!\left(\frac{(J+P+1)!}{(J+P+1)^{\a}}\frac{e^{(J+P-1)(\de+\underline{\Ga}(\la))}}{\ro_2^{J+P+1}}\right)\nn\\
&&\ML\ML\ML\ML||\la|^Jr^{2+P-\frac{2}{p}}\nabb^{p_0}\ddb_{\la}\nabb^{p_1}\ddb_{\la}\nabb^{p_2}\c\c\c\ddb_{\la}\nabb^{p_{J-1}}\ddb_{\la}\nabb^{p_{J}}\om|_{p,S}\leq {\hat{\underline F}}_4\!\left(\frac{(J+P)!}{(J+P)^{\a}}\frac{e^{(J+P-2)(\de+\underline{\Ga}(\la))}}{\ro_2^{J+P}}\right)\nn\\
&&\ML\ML\ML\ML||\la|^Jr^{1+P+\si(P)-\frac{2}{p}}\nabb^{p_0}\ddb_{\la}\nabb^{p_1}\ddb_{\la}\nabb^{p_2}\c\c\c\ddb_{\la}\nabb^{p_{J-1}}\ddb_{\la}\nabb^{p_{J}}\tr\chib|_{p,S}\leq {\hat{\underline F}}_5\!\left(\frac{(J+P)!}{(J+P)^{\a}}\frac{e^{(J+P-2)(\de+\underline{\Ga}(\la))}}{\ro_2^{J+P}}\right)\nn\\
&&\ML\ML\ML\ML||\la|^{J+1}r^{1+P-\frac{2}{p}}\nabb^{p_0}\ddb_{\la}\nabb^{p_1}\ddb_{\la}\nabb^{p_2}\c\c\c\ddb_{\la}\nabb^{p_{J-1}}\ddb_{\la}\nabb^{p_{J}}\chibh|_{p,S}\leq {\hat{\underline F}}_6\!\left(\frac{(J+P)!}{(J+P)^{\a}}\frac{e^{(J+P-2)(\de+\underline{\Ga}(\la))}}{\ro_2^{J+P}}\right)\nn\\
&&\ML\ML\ML\ML||\la|^{J+1}r^{1+P-\frac{2}{p}}\nabb^{p_0}\ddb_{\la}\nabb^{p_1}\ddb_{\la}\nabb^{p_2}\c\c\c\ddb_{\la}\nabb^{p_{J-1}}\ddb_{\la}\nabb^{p_{J}}\omb|_{p,S}\leq {\hat{\underline F}}_7\!\left(\frac{(J+P+1)!}{(J+P+1)^{\a}}\frac{e^{(J+P-1)(\de+\underline{\Ga}(\la))}}{\ro_2^{J+P+1}}\right)\ .\ \ \ \ \ \ \ \ \ \ \ \ 
\eea
\end{Le}
\NI In the next lemma we anticipate the prove of the estimates for the general mixed derivatives of the Riemann null components required in the proof of the previous Lemmas \ref{mixedderivatives} and \ref{mixedderivatives2}. 

\begin{Le}\label{mixedderivatives3}
Under the assumptions of Lemma \ref{cor10.3xxx} the following estimates hold on ${\cal K}$, with $\ro_2<\ro$ depending only the first $s$ derivatives of the initial data, with $s\leq 7$  and with a suitable choice of the  constants  $\hat{\underline{ F}}_i$ of order $O(\varepsilon)$ for $J \leq7$ and $O(1)$ otherwise.

\bea
&&\ML\ML\ML\ML||\la|^Jr^{1+P+\si(P)-\frac{2}{p}}\nabb^{p_0}\ddb_{\la}\nabb^{p_1}\ddb_{\la}\nabb^{p_2}\c\c\c\ddb_{\la}\nabb^{p_{J-1}}\ddb_{\la}\nabb^{p_{J}}\Psi|_{p,S}\leq {\hat{\underline F}}_1\!\left(\frac{(J+P+1)!}{(J+P+1)^{\a}}\frac{e^{(J+P-1)(\de+\underline{\Ga}(\la))}}{\ro_2^{J+P+1}}\right)\nn\\
\eea
\end{Le}
\NI {\bf Proof:} The proof is in appendix 16.

\NI Collecting the results we obtain the precise version of theorem \ref{main2}
 \begin{theorem}\label{main22}
 \NI Under the hypothesis of theorem \ref{Thinitialdata}, where the initial data  $\|\Phi^{(0)}\|_s$, $s\leq 7$ are chosen such that the ${\cal Q}^0$ norms on $C_0\cup\Cb_0$ are sufficiently small, then the solution is analytic and in ${\cal B}_{\a,\ro_2}$ in all the domain of dependance of $C_0\cup\Cb_0$.
\end{theorem} 

\newpage 

\section{The ${\cal Q}$ ``energy norms" and their boundedness\\}\label{S.4.1}
As discussed in the introduction and in subsection \ref{SS3.2a}, step III, to control the angular derivatives of the connection coefficients we need to control the norms of the Riemann components, $\Psi$; this is achieved using the hyperbolicity of the Einstein equations which implies the existence of a priori energy estimates. This is exploited through the introduction of the $\cal Q$ norms defined in the following subsections, a generalisation of the analogous norms defined in \cite{C-K:book} and \cite{Kl-Ni:book}. Repeating the arguments discussed there and also doing some extra work, we prove that these $\cal Q$ norms satisfy some appropriate bounds depending on the initial data's ones. The central point is that now we have to control an infinite family of energy-type norms, we denote $\QQ^{(J-2)}(\la,\nu)$ and $\QQb^{(J-2)}(\la,\nu)$, for all $J$,\footnote{$J-2$ is enough control the $J$ derivatives norms of the connection coefficients as discussed in detail later on. } as we need to control all order derivatives to prove the analyticity of the solution, but, nevertheless, the region where all these a priori estimates are valid depends only on the region where the $\cal Q$ norms with $J = 2$ are bounded.  This, as discussed in remarks below Theorem\ref{main2} and in subsection \ref{S.sincond}, depends on the smallness of the initial data derivatives with $J =2$\footnote{We recall that, in order to bound the ${\cal Q}^0$ norms we need to control the tangential derivatives of the connection coefficients up to order $J\leq 7$ see \cite{Ca-Ni:char}}; if these are of order $O(\varepsilon)$, with suitable $\varepsilon$, we can prove that these a priori estimates, and therefore the a priori estimates for all the $J$, hold everywhere, a crucial step to prove the global existence of our analytic solution. 
\medskip

\NI{\bf Remarks:} 

\NI{\em i) It is exactly to control these $\cal Q$ norms everywhere that the smallness of the initial data is crucial. In fact the proof of Theorem \ref{L6.1} is valid, in the region where the $\cal Q$ norms are bounded, independently from the initial data smallness.

\NI ii) Comparing the strategy discussed in this paper with the analogous result proved in the toy model presented in (I), ``the Burger case", one could ask where the analogous of the $t$ dependance for the Burger inductive assumptions, appears in the present case. Recall that in that case the assumptions we prove are, denoting with $u$ a solution of the Burger equation, 
\bea
||D^Ju(t)||_{L^2}\leq C_0e^{(J-2)\ga t}\frac{J!}{J^{\a}}\frac{1}{\ro^J}\ .\eql{7.101aa}
\eea
The presence of the factor $e^{(J-2)\ga t}$ is an indication of the fact that we are considering a local solution and therefore that these norms increase in time as we do not have an appropriate decay of the solution to make the local solution global.\footnote{Recall that in the Burger case the estimate \ref{7.101aa} does not hold for any $t$ as it can be proved that some norms will blow up in a finite time.}

\NI In the present case the situation is different as in the Einstein equation case we can prove global a priori estimates, provided the initial data are ``small". In fact we use the ``generalized" $Q$ norms where the Bel-Robinson tensor is saturated by the $K$ vector fields which guarantees the appropriate decays; nevertheless in the local case it is enough to use the $Q$ norms ``saturated" with the $T$ vector fields. In this case when we try to obtain the control of the $Q$ norms we have an inequality, formally, of the following type
\beaa
Q(C(\la))=Q(C_0)+{Error}=Q(C_0)+\int_0^{\la}d\la'|F({\cal O})|Q(C(\la'))
\eeaa
where with $F({\cal O})$ we indicate terms, depending on the connection coefficients, which we can estimate through the inductive assumptions. Therefore, assuming we control the $\sup$ norm of $\cal O$, the Gronwall inequality produces an exponential in time which, to be not harmful, requires that a decaying factor in $r\!$ be present in the inductive assumption. An analogous decay factor should be present when we estimate those $Q$ norms made along $\Cb(\nu)$, decay factor which is also required to close the estimates. This remark should explain that, if we do not restrict to small initial data, in our inductive assumption an exponential factor, $e^{(J-2)\ga(|\la|)}$ should be present. Again, if the initial data are sufficiently small and, as it follows, the decay of the connection coefficients is sufficiently good, the previous exponential factor turns out to be substituted by the bounded exponential factor, see \ref{Gadef}, 
 $e^{(J-2)(\underline{\Ga}(\la))}$.
\smallskip

\NI Observe also that we cannot use the Bianchi equations for the Riemann null components to get direct information on their tangential derivatives due to the fact that in these equations there is a loss of tangential derivatives.\footnote{Moreover we do not even have transport equations for $\a$ on the outgoing cones and for $\aa$ on the incoming ones.}
This is an important point as this is exactly the loss of derivatives avoided using the $\cal Q$ norms.
This is exactly where, in the present case, the hyperbolicity plays the role. In fact the boundedness of energy allows to control $\Lie_O^{N-1}\Psi$ on a generic $S(\la,\nu)$ in terms of the analogous quantities on the initial hypersurface.\footnote{The absence of the loss of (tangential) derivatives is due to the fact that in the ``error estimate" for the $\cal Q$ we use $D^{\mu}W_{\mu\nu\ro\si}=0$ which corresponds to the use of $\square\! u=F$ in the wave equation.}
\smallskip

\NI iii) As done before, we proceed looking for the global existence proof; from the previous discussions it should be clear that relaxing the smallness assumptions for the first derivatives initial data, all the subsequent machinery will allow to prove the analyticity in the bounded region ${\cal K}$ where the a priori energy type estimates hold.
}
\medskip

\NI In the following subsection we define the ${\cal Q}$ norms we need  to control the angular derivatives of the null Riemann components and show how to prove their estimates.

\subsection{The ${\cal Q}$ ``energy norms" required to control the angular derivatives}\label{S.s4.1nn}
The ${\cal Q}$ norms we write here and which we will prove bounded are those needed to control all orders of the Lie angular derivatives of the Riemann components, at their turn required to control, as discussed before, the angular derivatives of the connection coefficients. Subsequently, as discussed in section \ref{37} when we estimate the  remaining ``tangential" derivatives of the connection coefficients we need that some other ${\cal Q}$ norms,

\smallskip

\NI We define, for $J\geq 2$,\footnote{For the definitions of the vector fields $O,K,T$ and the modified Lie derivatives $\lie_{\c}$ see \cite{Kl-Ni:book} Chapter 3.}
\bea\label{qnorms}
&&\QQ^{(J-2)}(\la,\nu)=\QQ^{(J-2)}_1(\la,\nu)+\QQ^{(J-2)}_2(\la,\nu)\nn\\
&&\QQb^{(J-2)}(\la,\nu)=\QQb^{(J-2)}_1(\la,\nu)+\QQb^{(J-2)}_2(\la,\nu)
\eea
where,
\bea
\ML\ML\QQ^{(J-2)}_1(\la,\nu)&\equiv&\sum_{i_1,1_2,...,i_{J-2}}^{\{1,3\}}\int_{C(\la)\cap V(\la,\nu)}Q(\lie_{T}{(\lie_{O_{(i_1)}}\lie_{O_{(i_2)}}\c\c\lie_{O_{(i_{J-2})}}W)})(\acc,\acc,\acc,e_4)\nn\\
&&\ML\ML\ +\sum_{i_0,i_1,1_2,...,i_{J-2}}^{\{1,3\}}\int_{C(\la)\cap V(\la,\nu)}Q(\lie_{O_{(i_0)}}(\lie_{O_{(i_1)}}\lie_{O_{(i_2)}}\c\c\lie_{O_{(i_{J-2})}}W))(\acc,\acc,T,e_4)\ .\nn
\eea
{\em we recall that
$Q(\lie_{O}W)=\sum_{i=1}^3Q(\lie_{O_{(i)}}W)$\ . Therefore with our definitions}
\bea\label{163}
\QQ^{(J-2)}_1(\la,\nu)&\equiv&\int_{C(\la)\cap V(\la,\nu)}Q(\lie_{T}{(\lie_{O}^{J-2}W)})(\acc,\acc,\acc,e_4)\nn\\
&&+\int_{C(\la)\cap V(\la,\nu)}Q(\lie_{O}{(\lie_{O}^{J-2}W}))(\acc,\acc,T,e_4)\nn\\
\QQ^{({J-2})}_2(\la,\nu)&\equiv&
\int_{C(\la)\cap V(\la,\nu)}Q(\lie_{O}\lie_{T}{(\lie_{O}^{J-2}W}))(\acc,\acc,\acc,e_4)\nn\\
&&+\int_{C(\la)\cap V(\la,\nu)}Q(\lie^2_{O}{(\lie_{O}^{J-2}W}))(\acc,\acc,T,e_4)\\
&&+\int_{C(\la)\cap V(\la,\nu)}Q(\lie_{S}\lie_{T}{(\lie_{O}^{J-2}W}))(\acc,\acc,\acc,e_4)\nn
\eql{QQ12}
\eea 
\bea\label{164}
\QQb^{(J-2)}_1(\la,\nu)&\equiv&\sup_{V(\la,\nu)\cap C_0}|r^3{\overline\ro(\lie_{O}^{J-2}W)}|^2+
\int_{\Cb(\nu)\cap V(\la,\nu)}Q(\lie_{T}{(\lie_{O}^{J-2}W}))(\acc,\acc,\acc,e_3)\nn\\
&&+\int_{\Cb(\nu)\cap V(\la,\nu)}Q(\lie_{O}{(\lie_{O}^{J-2}W}))(\acc,\acc,T,e_3)\nn\\
\QQb^{(J-2)}_2(\la,\nu)&\equiv&\int_{\Cb(\nu)\cap V(\la,\nu)}Q(\lie_{O}\lie_{T}{(\lie_{O}^{J-2}W}))(\acc,\acc,\acc,e_3)\nn\\
&&+\int_{\Cb(\nu)\cap V(\la,\nu)}Q(\lie^2_{O}{(\lie_{O}^{J-2}W}))(\acc,\acc,T,e_3)\\
&&+\int_{\Cb(\nu)\cap V(\la,\nu)}Q(\lie_{S}\lie_{T}{(\lie_{O}^{J-2}W}))(\acc,\acc,\acc,e_3)\nn
\eql{QQb12}
\eea
where $V(\la,\nu)=J^{(+)}(S(\la_0,\nu_0)\cap J^{(-)}(S(\la,\nu))$.
Analogous expressions hold for the initial data norms defined on $C_0\cup\Cb_0=C(\la_0)\cup\Cb(\nu_0, [\la_0,\la_*])$ which we denote
\bea\label{16555}
&&\QQ_{(0)}^{(J-2)}(\la_0,\nu)=\QQ_{(0),1}^{(J-2)}(\la_0,\nu)+\QQ_{(0),2}^{(J-2)}(\la_0,\nu)\nn\\
&&\QQb_{(0)}^{(J-2)}(\la,\nu_0)=\QQb_{(0),1}^{(J-2)}(\la,\nu_0)+\QQb_{(0),2}^{(J-2)}(\la,\nu_0)\ .\eql{4.4QQ}
\eea
Assuming the initial data sufficiently small, in the sense previously discussed of the ``smallness" of the derivatives up to $J\leq 7$\footnote{Otherwise assuming $|\Pi|$ and $|\Lambda|$ sufficiently small.} we have to prove, mimicking \cite{Kl-Ni:book} that, 

\bea
{{\cal H}^{(N-2)}}(\la,\nu)\equiv\sum_{J=2}^{N}\left(\QQ^{(J-2)}({C(\la;[\nu_0,\nu])})+\QQb^{(J-2)}({\Cb(\nu;[\la_0,\la])})\right)\eql{Hdefx}
\eea
satisfies, everywhere the following bound,
\bea
{{\cal H}^{(N-2)}}(\la, \nu)^{\frac{1}{2}}\leq C^{*(1)}\!\left(\frac{(N)!}{(N)^{\a}}\frac{e^{(N-2)(\de+\underline{\Ga}(\la))}}{\ro_{0,1}^{N}}\right)\ ,
\eea
where defining $C^{(0)}$ a constant which bounds all the initial data constants, see \ref{123} and \ref{123ab},  $C^{(1)}$ does the same for the constants in the inductive assumptions.

\NI The detailed statement of this result is in subsection 10.1.3, Theorem 10.1, and its proof in the Appendix  to Section 10.

\subsection{The estimates of the angular derivatives of the  Riemann components}\label{S.s4.2}

\NI We are now in good position to control the angular derivatives of the Riemann tensor, whose null components are $\Psi\equiv\{\a,\b,\ro,\si,\bb,\aa\}$,\footnote{See (I) and \cite{Kl-Ni:book} for the complete definitions.} in the whole region ${\cal K}$; this is the content of the following theorem, 

\begin{theorem}\label{hypthm2}
Let us assume that 
the initial data connection coefficients satisfy, with all their angular derivatives, the estimates of Theorem \ref{Thinitialdata}, let us assume that the connection coefficients satisfy, in the internal region, ${\cal K}$, with all their angular derivatives up to order $J-1$ the estimates \ref{123a}, then the following estimates hold for the $J-1$ angular derivatives of Riemann tensor null components,
\footnote{Again, if the first derivatives of the initial data are order $O(\varepsilon)$, then the estimates with $J \leq7$ have the same structure of \ref{4.7}, but with the constant $C^{(1)}=O(\varepsilon_0)$ and $\varepsilon_0^2<\varepsilon<\varepsilon_0$.} 
\bea
\big| |\la|^{\underline{\phi}(\Psi)}r^{\phi(\Psi)+(J-1)-\frac{2}{p}}\Lie_O^{J-1}\Psi\big|_{p,S}\leq C^{(1)}e^{(J-2)\de}e^{(J-2)\underline{\Ga}(\la)}\frac{J!}{J^\a}\frac{1}{\ro_{0,1}^J}\ ,\eql{4.7}
\eea
with $C^{(1)}$ of order $O(\varepsilon)$ for $J \leq7$, of order $O(1)$ otherwise, depending on $\nu_0$ and smaller than the constant related to the estimates of the connection coefficients, and
\bea\label{resti}
&&\ML\phi(\a)=\phi(\b)=\frac{7}{2}\ ,\ \phi(\ro)=\phi(\si)=3\ , \phi(\bb)=2\ ,\ \phi(\aa)=1\nn\\
&&\ML\underline{\phi}(\a)=\underline{\phi}(\b)=0\ ,\ \underline{\phi}(\ro)=\underline{\phi}(\si)=\frac{1}{2}\ , \underline{\phi}(\bb)=\frac{3}{2}\ ,\ \underline{\phi}(\aa)=\frac{5}{2}\ .\ \ \eql{phidef}
\eea
\end{theorem}

\NI {\bf Proof:} This long proof is divided in various steps which we list  here and discuss in any detail  in sections 10 and 15
.
\smallskip

\NI {\bf Step 1:} We assume that the initial data connection coefficients satisfy the estimates  of Theorem \ref{Thinitialdata}.

\NI By these assumptions we prove the following estimates for the various null components of the Riemann tensor:

\bea\label{Resti}
&&\ML\ML\ML\ML||\la|^{\underline{\phi}(\Psi)}r^{\phi(\Psi)+\ep+(J-\frac{2}{p})}\nabb^{J-1}\Psi(R)|_{p,S}(\la_0,\nu)\leq \hat{C}_{0,0}\frac{J!}{J^\a}\frac{e^{(J-2)\de_0}e^{(J-2)\underline{\Ga}_0(\la)}}{\ro_0^J}\ ,\eql{step1a}\\
&&\ML\ML\ML\ML||\la|^{\underline{\phi}(\Psi)}r^{\phi(\Psi)+\ep+(J-\frac{2}{p})}\nabb^{J-1}\Psi(R)|_{p,S}(\la,\nu_0)\leq \hat{C}_{0,0}\frac{J!}{J^\a}\frac{e^{(J-2)\de_0}e^{(J-2)\underline{\Ga}_0(\la)}}{\ro_0^J}\ .\nn
\eea 
With $\hat{C}_{0,0}$ of order $O(\varepsilon)$ for $J \leq7$, of order $O(1)$ otherwise.
 From these estimates we derive an analogous estimates for the derivatives with respect to the $O$ rotation generator fields, namely
\bea
&&\ML\ML ||\la|^{\underline{\phi}(\Psi)}r^{\phi(\Psi)+\ep+(J-\frac{2}{p})}\nabb_O^{J-1}\Psi(R)|_{p,S}(\la_0,\nu)\leq |O|_{\infty}^{J-1}\hat{C}_{0,0}\frac{J!}{J^\a}\frac{e^{(J-2)\de_0}e^{(J-2)\underline{\Ga}_0(\la)}}{\ro_{0,1}^J}\ \nn\\
&&\ML\ML ||\la|^{\underline{\phi}(\Psi)}r^{\phi(\Psi)+\ep+(J-\frac{2}{p})}\nabb_O^{J-1}\Psi(R)|_{p,S}(\la,\nu_0)\leq |O|_{\infty}^{J-1}\hat{C}_{0,0}\frac{J!}{J^\a}\frac{e^{(J-2)\de_0}e^{(J-2)\underline{\Ga}_0(\la)}}{\ro_{0,1}^J}\ ,\eql{step1b}\nn\\
\eea 
\NI Namely we prove the following lemma:
\begin{Le}\label{L4.1}
Under the estimates of Theorem \ref{Thinitialdata} for the initial data connection coefficients and  \ref{Resti} for the null Riemann components on $C_0\cup\Cb_0$, the following estimates for the $\nabb_O$ derivatives of the null Riemann components hold,
\bea
||\la|^{\underline{\phi}(\Psi)}r^{\phi(\Psi)+\ep+(J-1)-\frac{2}{p}}\nabb_O^{J-1}\Psi(R)|_{p,S}\bigg|_{C_0\cup\Cb_0}\!\leq |O|_{\infty}^{J-1}\hat{C}_{0,0}\frac{J!}{J^\a}\frac{e^{(J-2)\de_0}e^{(J-2)\underline{\Ga}_0(\la)}}{\ro_{0,1}^J}\ ,\ \ \ \ \ \ \eql{as2}
\eea
where $\ro_{0,1}<\ro_0$ and the $\ep>0$ in the weight factor is required to have the ${\cal Q}$ norms of the initial data finite, see remark below Theorem \ref{main2}.
\end{Le}
\NI {\bf Proof:} See appendix to section 10 \footnote{Remember that, as $O=\sum_{i=1}^3O^{(i)}$ we have to consider also a factor $3^{J-1}$ in the $|O|^{J-1}_{\infty}$ norms. We do not write explicitly it to avoid cumbersome notations.}.

\NI Subsequently we derive an estimate for the $\Lie_O^{J-1}\Psi(R)$ derivatives, an estimate for the $\Psi(\Lie_O^{J-1}R)$ and finally an estimate for $\Psi(\lie_O^{J-1}R)$, in this case we can exploit the initial data estimates for the $\Lie_O$ derivatives of the connection coefficients, see \ref{iniz1}, \ref{iniz2}, \ref{iniz3} and \ref{iniz4}, and avoid the changes of the constant $\hat{C}_{0,0}$ and of the radius of convergence $\ro_{0,1}$, we list here these inequalities,

\bea
&&||\la|^{\underline{\phi}(\Psi)}r^{\phi(\Psi)+\ep-\frac{2}{p})}\Lie_O^{J-1}\Psi(R)|_{p,S}(\la_0,\nu)\leq \hat{C}_{0,0}\frac{J!}{J^\a}\frac{e^{(J-2)\de_0}e^{(J-2)\underline{\Ga}_0(\la)}}{\ro_{0,1}^J}\ \  \eql{step1c}\ \ \ \ \ \ \ \\
&&||\la|^{\underline{\phi}(\Psi)}r^{\phi(\Psi)+\ep-\frac{2}{p})}\Lie_O^{J-1}\Psi(R)|_{p,S}(\la,\nu_0)\leq \hat{C}_{0,0}\frac{J!}{J^\a}\frac{e^{(J-2)\de_0}e^{(J-2)\underline{\Ga}_0(\la)}}{\ro_{0,1}^J}\ ,\nn
\eea
\bea
&&||\la|^{\underline{\phi}(\Psi)}r^{\phi(\Psi)+\ep-\frac{2}{p}}\Psi(\Lie_O^{J-1}R)|_{p,S}(\la_0,\nu)\leq \hat{C}_{0,0}\frac{J!}{J^\a}\frac{e^{(J-2)\de_0}e^{(J-2)\Gab_0(\la)}}{\ro_{0,1}^J}\ \ \ \ \ \ \ \ \ \\
&&||\la|^{\underline{\phi}(\Psi)}r^{\phi(\Psi)+\ep-\frac{2}{p}}\Psi(\Lie_O^{J-1}R)|_{p,S}(\la,\nu_0)\leq \hat{C}_{0,0}\frac{J!}{J^\a}\frac{e^{(J-2)\de_0}e^{(J-2)\underline{\Ga}_0(\la)(\nu)}}{\ro_{0,1}^J}\ ,\nn\eql{step1e}
\eea
\NI and finally
\bea
&&||\la|^{\underline{\phi}(\Psi)}r^{\phi(\Psi)+\ep-\frac{2}{p}}\Psi(\lie_O^{J-1}R)|_{p,S}(\la_0,\nu)\leq \hat{C}_{0,0}\frac{J!}{J^\a}\frac{e^{(J-2)\de_0}e^{(J-2)\underline{\Ga}_0(\la)}}{\ro_{0,1}^J}\ \eql{step1d}\  \ \ \ \ \ \ \   \\
&&||\la|^{\underline{\phi}(\Psi)}r^{\phi(\Psi)+\ep-\frac{2}{p}}\Psi(\lie_O^{J-1}R)|_{p,S}(\la,\nu_0)\leq \hat{C}_{0,0}\frac{J!}{J^\a}\frac{e^{(J-2)\de_0}e^{(J-2)\underline{\Ga}_0(\la)}}{\ro_{0,1}^J}\ ,\nn
\eea
These results are summarised in lemma
\begin{Le}\label{L4.2}
Under the assumptions of lemma \ref{L4.1} on the (initial data) connection coefficients and the previous estimates for $\nabb_O^{J-1}\Psi(R)$, the following bounds hold,
\bea
&&\ML\ML\ML||\la|^{\underline{\phi}(\Psi)}r^{\phi(\Psi)+\ep-\frac{2}{p}}\Lie_O^{J-1}\Psi(R)|_{p,S}\bigg|_{C_0\cup\Cb_0}\leq \hat{C}_{0,0}\frac{J!}{J^\a}\frac{e^{(J-2)\de_0}e^{(J-2)\underline{\Ga}_0(\la)}}{\ro_{0,1}^J}\ \ \eql{as3a}
\eea
\NI moreover
\bea
&&\ML\ML\ML||\la|^{\underline{\phi}(\Psi)}r^{\phi(\Psi)+\ep-\frac{2}{p}}\Psi(\Lie_O^{J-1}R)|_{p,S}\bigg|_{C_0\cup\Cb_0}\leq \hat{C}_{0,0}\frac{J!}{J^\a}\frac{e^{(J-2)\de_0}e^{(J-2)\underline{\Ga}_0(\la)}}{\ro_{0,1}^J}\ \ 
\eea
and finally
\bea
&&\ML\ML\ML||\la|^{\underline{\phi}(\Psi)}r^{\phi(\Psi)+\ep-\frac{2}{p}}\Psi(\lie_O^{J-1}R)|_{p,S}\bigg|_{C_0\cup\Cb_0}\leq \hat{C}_{0,0}\frac{J!}{J^\a}\frac{e^{(J-2)\de_0}e^{(J-2)\underline{\Ga}_0(\la)(\nu)}}{\ro_{0,1}^J}\ ,\eql{as5a}
\eea
\end{Le}
\NI {\bf Step 2:} Once we have the estimates for $\Psi(\lie_O^{J-1}R)$ on $C_0\cup\Cb_0$ we derive easily an estimate for $\QQ_{(0)}^{(J-2)}(\la_0,\nu)$ and for $\QQb_{(0)}^{(J-2)}(\la,\nu_0)$ namely, 
\bea
&&\ML\ML\QQ_{(0)}^{(J-2)}(\la_0,\nu)+\QQb_{(0)}^{(J-2)}(\la,\nu_0)\leq \left(\!\hat{C}_{0,0}\frac{J!}{J^{\a}}\frac{e^{(J-2)\de_{0,1}}e^{(J-2)\underline{\Ga}(\la)}}{\ro_{0,1}^{J}}\right)^{\!2},\ \ \ \eql{step2}\ \ \ \ \ \ \ \ \ \ \
\eea
with $\de_{0,1}>\de_0$.
The proof of estimate \ref{step2} is in Lemma 10.1.
\medskip

\NI {\bf Step 3:} Once we control of the $\QQ_{(0)}^{(J-2)}(\la_0,\nu)$ and the $\QQb_{(0)}^{(J-2)}(\la,\nu_0)$ norms, assuming the initial data sufficiently small  we prove as discussed in the previous subsection and in Appendix 15
 that, in the region where the a priori estimates hold for $J\leq 7$, we have for all $J$,
\bea
\ML\QQ^{\!(J-2)}\!({C(\la;[\nu_0,\nu])})\!+\!\QQb^{\!(J-2)}\!({\Cb(\nu;[\la_0,\la])})\!\leq\! C^{(1)}\!\left(\frac{J!}{J^{\a}}\frac{e^{(J-2)(\de+\underline{\Ga}(\la))}}{{\ro_{0,1}}^J}\right) \eql{4.15}
\eea
where \[C^{(1)}=c\!\left[\hat{C}_{0,0}\!
+\!\left(\varepsilon+CC^{(1)}\frac{e^{-\de}}{\ro_{0,1}}\right)\right]<\hat{C}_{0,0}\ .\] 
\smallskip
\NI This is the content of Theorem 10.1 proved in section 10.

\NI{\bf Remark:}

\NI{\em The estimates of the ${\cal Q}^{(J-2)}$ norms is a substantially a generalization of chapter 6 of \cite{Kl-Ni:book} to the $\Lie_O^{J-2}$ derivatives of the ${\cal Q}$ norms, the proof is very involved, it requires to carefully estimate a great number of therms, It is here that the term $\underline{\Gamma}(\la)$ plays it role to assure the boundedness of some important terms}
\smallskip

\NI {\bf Step 4:} Proceeding again as in \cite{Kl-Ni:book}, Chapter 5, with some more work, we control, using \ref{4.15},
the $\Psi(\lie_O^{J-1}R)$ norms
obtaining 
\bea
&&\ML\ML\ML||u|^{\underline{\phi}(\Psi)}r^{\phi(\Psi)-\frac{2}{p}}\Psi(\lie_O^{J-1}R)|_{p,S}(\la,\nu)\leq C^{*(1)}\!\left(\frac{J!}{J^\a}\frac{e^{(J-2)\de}e^{(J-2)\underline{\Ga}(\la)}}{\ro_{0,1}^J}\right)\eql{step4,1}\ .
\eea
The proof follows from Lemma 10.2.
\smallskip

\NI {\bf Step 5:} First we prove that under the inductive assumptions for the connection coefficients norms up to order $N\!-\!1$ and the previous norm estimates for $\Psi(\lie_O^{J-1}R)$ with $J\leq N$ that the inequality hold, see section 10, lemmas 10.3 and 15.1:
\bea\label{labb}
&&\big| |u|^{\underline{\phi}(\Psi)}r^{\phi(\Psi)-\frac{2}{p}}\Lie_O^{J-1}\Psi\big|_{p,S}(\la,\nu)\leq {C}^{(1)}e^{(J-2)\de}e^{(J-2)\underline{\Ga}(\la)}\frac{J!}{J^\a}\frac{1}{\ro_{0,1}^J}\ ,\nn \ \ \ \\
&&\big| |u|^{\underline{\phi}(\Psi)}r^{\phi(\Psi)-\frac{2}{p}}\nabb_O^{J-1}\Psi\big|_{p,S}(\la,\nu)\leq C^{(1)}e^{(J-2)\de}e^{(J-2)\underline{\Ga}(\la)}\frac{J!}{J^\a}\frac{1}{\ro_{0,1}^J}\ ,\ \ \ \ 
\eea
with $C^{*(1)}< C^{(1)}$. This is obtained substantially inverting lemma \ref{L4.2}. 
\smallskip

\NI{\bf Remark:}

\NI{\em Notice that in order to apply these estimates to the estimates for the connection coefficients the constant $\tilde{C}_4$ has to be smaller of the constants of $C_i$ and $\Cb_i$ of inequalities \ref{324}}
\smallskip

\NI Then we prove following estimates  hold, for $\nabb^{J-1}\Psi(R)$ with  $J\leq N$,
\bea
\big| |u|^{\underline{\phi}(\Psi)}r^{\phi(\Psi)-\frac{2}{p}}\nabb^{J-1}\Psi\big|_{p,S}(\la,\nu)\leq C^{(1)}e^{(J-2)\de}e^{(J-2)\underline{\Ga}(\la)}\frac{J!}{J^\a}\frac{1}{\ro^J}\ ,\ \ \ \ 
\eea
with suitable $\ro<\ro_{0,1}$ and
with $C^{(1)}$ of order $O(\varepsilon)$ for $J \leq7$, of order $O(1)$ otherwise, smaller than the constants related to the estimates of the connection coefficients. This is the content of Lemma 10.4 of section 10, which completes the proof of Theorem \ref{hypthm2} and whose proof is in the Appendix 15.
\smallskip

\subsection{The $\ddb_{\nu}$ and $\ddb_{\la}$ derivatives for the Riemann components}\label{S.5}
\NI Also the mixed derivatives for the Riemann components have to be estimated through the  $\cal Q$ norms.

 \NI These estimates for the connection coefficients are needed as we remarked, see (I) for a more detailed discussion,  that the recursive estimates for the angular derivatives of the connection coefficients are not sufficient to prove the existence of an extended region for the analytic solutions.\footnote{To obtain a Cauchy-Kowalevski solution starting with data on an inner outgoing  or incoming cone we need to prove that they are analytic on them.} 
 In fact on $\Cb(\nu)$ we have to control all the angular $\nabb$ derivatives and all the $\ddb_3$ derivatives mixed together in all the possible ways. Analogously  on the generic outgoing cone $C(\la)$ we have to control all the angular $\nabb$ derivatives and all the $\ddb_4$ derivatives mixed together in all the possible ways.

\NI In section \ref{37}, when we discussed how to control all the ``tangential" derivatives of the connection coefficients, it follows that we have to control all the tangential derivatives of the Riemann components which require  to prove that a larger set of  energy type ${\cal Q}$ norms, always made in terms of the Bell-Robinson tensor, is bounded; this is obtained proceeding in the same way as done for the angular derivatives in Section \ref{S.4.1}. The remaining $\cal Q$ norms we have to introduce have, in fact, the same general structure. The basic difference is that the $\lie_O^{J-2}W$ present there has to be substituted by a sum of analogous  terms with 
$\lie_S^{p}\lie_O^{k}W$  with $p+k\leq N-2$, $p>0$. Observe, as discussed later on, that although we have to control the general mixed derivatives, with arbitrary distribution of the various derivatives for the Riemann  components, nevertheless it is enough to introduce the $\cal Q$ norms relative to the Weyl tensors ${\tilde W}_{k,p}=\lie_S^k\lie_O^{p}W$. The boundedness of these $\cal Q$ norms
have to be proved exactly in the same way obtaining the exact analogous of the estimate 10.8, with $p+k=H$,
\bea
&&\ML\ML\int_{C(\la_0;[\nu_0,\nu])}Q(\lie_S^k\lie_O^{p}W)(\acc,\acc,T,e_4)
\leq \left({\tilde C}^{(1)}\frac{H!}{H^{\a}}\frac{e^{(H-2)\de_{0}}e^{(H-2)\underline{\Ga}(\la)}}{\ro_{0,1}^{H}}\right)^{\!\!2}.\ \ \ \ \ \ \ \ \ \ \ \eql{Qest1abc}
\eea
\smallskip
 
\NI Once we control, via the ``extended $\cal Q$ norms" all the derivatives up to $N-1$ of the null Riemann components we are able to control the $\dddd_{3,4}$ derivatives and the mixed ones for the connection coefficients up to order $N$. 

\NI Repeating the steps of the  section \ref{S.4.1}, discussed in details in section
 10 and appendix 15, but now for the $D_S\nabb_O$ derivatives and exploiting the smallness of mixed derivatives of the initial data, see Theorem \ref{T11.1}, we can state the following  theorem analogous to theorem \ref{L2.1new},

\begin{theorem}\label{324567}
Under the assumptions  \ref{inizz1} and \ref{inizz2} on the mixed derivatives of the initial data connection coefficients on $C_0\cup\Cb_0$, the following estimates hold on ${\cal K}$, with $\tilde{C}^{(1)}$ of order $O(\varepsilon)$ for any $J \leq 7$ and order $O(1)$ otherwise.
\bea
| |\la|^{\underline{\phi}(\Psi)}r^{\phi(\Psi)-\frac{2}{p}}\Psi(\lie_S^p\lie_O^kW)|_{p,S}(\la,\nu)\leq {\tilde C}^{(1)}\frac{J!}{J^{\a}}\frac{e^{(J-2)(\de+\underline{\Ga}(\la))}}{\ro_{0,1}^J}\  .\eql{6.96xx}
\eea
\bea
\big||\la|^{\underline{\phi}(\Psi)}r^{\phi(\Psi)-\frac{2}{p}}\Lie_S^p\Lie_O^{k}\Psi\big|_{p,S}(\la,\nu)\leq {\tilde C}^{(1)}\frac{J!}{J^{\a}}\frac{e^{(J-2)(\de+\underline{\Ga}(\la))}}{\ro_{0,1}^J}\ .
\eea
\smallskip
\end{theorem}

\NI The estimates of the $\ddb_{\nu}^J\nabb^P$ mixed derivatives of the Riemann components proceed in  the same way of Lemma \ref{le31}
\begin{Le}\label{from|LieTLieOtoDTNabb}
Let us assume  the estimates of theorem \ref{324567}, then the null Riemann components  $\Psi$ satisfy the following estimates  for $p+q=J$ on ${\cal K}$,  with $\ro_1<\ro<\ro_{0,1}$ depending only the first $s$ derivatives of the initial data, with $s\leq 7$  and with a suitable choice of the  $ {\tilde C}^{(1)}$ constant of order $O(\varepsilon)$ for $J \leq7$\footnote{If $J\geq 1$, for $J=0$ $\ro$ becomes $\ro-{\underline\ro}$, see \cite{Kl-Ni:book}.} and $O(1)$ otherwise.

\bea\label{rof1}
\big| |\la|^{\underline{\phi}(\Psi)}r^{\phi(\Psi)+k-\frac{2}{p}}D_S^p\nabb^{k}\Psi\big|_{p,S}(\la,\nu)\leq {\tilde C}^{(1)}e^{(J-2)\de}e^{(J-2)\underline{\Ga}(\la)}\frac{J!}{J^\a}\frac{1}{\ro_1^J}\ ,
\eea

\bea
&&\ML\phi(\a)=\phi(\b)=\frac{7}{2}\ ,\ \phi(\ro)=\phi(\si)=3\ , \phi(\bb)=2\ ,\ \phi(\aa)=1\nn\\
&&\ML\underline{\phi}(\a)=\underline{\phi}(\b)=0\ ,\ \underline{\phi}(\ro)=\underline{\phi}(\si)=\frac{1}{2}\ , \underline{\phi}(\bb)=\frac{3}{2}\ ,\ \underline{\phi}(\aa)=\frac{5}{2}\ .\ \ 
\eea
\end{Le}
\NI {\bf Proof:} As in lemma \ref{le31}, also in this case the proof is an adapted version of lemma \ref{L4.1} .
\medskip

\NI From these lemmas the following lemma holds,
\begin{Le} \label{cor10.3xxx}
Under the assumptions of Lemma \ref{from|LieTLieOtoDTNabb} the following estimates hold on ${\cal K}$, with $p+k=J$:
\bea
&&\big| |\la|^{\underline{\phi}(\Psi)}r^{\phi(\Psi)+(J-1)-\frac{2}{p}}D_{\nu}^p\nabb^{k}\Psi\big|_{p,S}(\la,\nu)\leq {\tilde C}^{(1)}e^{(J-2)\de}e^{(J-2)\underline{\Ga}(\la)}\frac{J!}{J^\a}\frac{1}{\ro_2^p\ro_1^k}\nn\\ 
&&\big| |\la|^{\underline{\phi}(\Psi)+p}r^{\phi(\Psi)+k-\frac{2}{p}}D_{\la}^p\nabb^{k}\Psi\big|_{p,S}(\la,\nu)\leq {\tilde C}^{(1)}e^{(J-2)\de}e^{(J-2)\underline{\Ga}(\la)}\frac{J!}{J^\a}\frac{1}{\ro_2^p\ro_1^k}\ .\ \ \ \ \ \ \ 
\eea
\end{Le}
\NI {\bf Proof:}  See subsection 15.7.
\medskip

\NI {\bf Remarks:}

\NI{\em i) Also in this case we can choose $\tilde{C}^{(1)}$ smaller than the constants related to the estimates of the mixed derivatives of the connection coefficients.}

\NI{\em ii) Recall that we have anticipated the estimates of the ``completely' mixed derivatives in section \ref {S.2}, Lemma \ref{mixedderivatives3}}

\section{The estimates for the derivatives of the metric components}\label{S.5nx}
To complete the existence proof of the global analytic solution of the characteristic problems we are considering, we need to control also all the derivatives of the various metric components. In fact  the system of first order equations we have to solve is,\footnote{Remember that eqs. \ref{5.37l}, \ref{2.70gql} are first used to prove the existence of an analytic solution in a, possible small, region, via Cauchy-Kovalevski as discussed in (I) and subsequently used again to extend the assumed maximal analyticity region once we have all the required norm estimates.} omitting the indices to simplify the notations,

\bea
&&{\ML\ML}\frac{\partial\ga}{\partial\om}-v=0\ \ ,\ \ \frac{\partial\log\oom}{\partial\om}-\psi=0\ \ ,\ \ \frac{\partial\hat{X}}{\partial\om}-w=0\nn\\
&&{\ML\ML}\frac{\partial{\ga}}{\partial\la}-2{\oom}\ \!{\chib}+\Lie_{X}{\ga}=0\nn\\
&&{\ML\ML}\frac{\partial\log{\oom}}{\partial\la}+\psi(X)+{2{\oom}}\ \!{\omb}=0\nn\\
&&{\ML\ML}\frac{\partial v}{\partial\la}+\nabb_Xv+(\partialb X)\c v-S(\partialb\!\otimes\!w)-2\oom\partialb\!\otimes\!\chib-2\oom\ \!\psi\!\otimes\!\chib=0\eql{5.37l}\\
&&{\ML\ML}\frac{\partial\psi}{\partial\la}+\nabb_X\psi+2\oom\omb\psi+\psi(\nabb X)+2\oom\partialb\omb=0\nn\\
&&{\ML\ML}\frac{\partial{\tr\chi}}{\partial\la}+{\oom}{\tr\chib}{\tr\chi}-2{\oom}{\omb}{\tr\chi}+{\nabb}_{\! X}{\tr\chi}
-2{\oom}{\divv}{(\ze+\psi)}-2{\oom}|{\ze+\psi}|^2 +2{\oom}{\bf K}\!=\!0\nn\\
&&{\ML\ML}\frac{\partial\hat{{\chi}}}{\partial\la}+\Lie_{{X}}\hat{{\chi}}-\frac{{\oom}{\tr\chib}}{2}\hat{{\chi}}+\frac{{\oom}{\tr\chi}}{2}\hat{{\chib}}
-{2\oom}\ \!{\omb}\ \!\chih
-{\oom}(\hat{{\chi}}\c\hat{{\chib}})\ga-{\oom}\ \!{\nabb}\hot{(\ze+\psi)} -{\oom}(\ze+\psi)\hot(\ze+\psi)\!=\!0\nn\\
&&{\ML\ML}\frac{\partial{\ze}}{\partial\la}+{\Lie_{X}{\ze}}+{\oom}\ \!{\tr\chib}{\ze}+{\oom}{\divv}{\chibh}-\frac{1}{2}{\oom}{\partialb}{\tr\chib}
\left.+2\oom\omb\psi+2\oom\partialb\omb\right.+{\oom}\psi\!\c\!{\chib}\!=\!0\nn\\
&&{\ML\ML}\frac{\partial{{\om}}}{\partial\la}\!+\!{\partialb}_{\! X}{\om}\!-\!2{\oom}\ \!\!{{\omb}}\
\!{{\om}}\!-\!\frac{3}{2}{{\oom}}|\ze|^2\!+\!\frac{1}{4}{{\oom}}\ze\!\c\!\psi\!+\!\frac{1}{2}{\oom}|\psi|^2
\!+\!\frac{1}{2}{\oom}\!\left({{\bf K}}\!+\!\frac{1}{4}{\tr\chi}{\tr\chib}\!-\!\frac{1}{2}\hat{{\chi}}\!\c\!\hat{{\chib}}\!\right)=0\nn\\
&&\nn\\
&&{\ML\ML}\frac{\partial\hat{X}}{\partial\nu}+4{\oom}^2{\ze}=0\nn\\
&&{\ML\ML}\frac{\partial w}{\partial\nu}+8\oom^2\psi\!\otimes\!\ze+4\oom^2\partialb\!\otimes\!\ze-2\oom\psi\!\otimes\!(\chi\!\c\!X)
-2\oom(\partialb\!\otimes\!\chi)\!\c\!X=0\nn\\
&&{\ML\ML}\frac{\partial{\tr\chib}}{\partial\nu}+{\oom}{\tr\chi}{\tr\chib}
-2{\oom}{\om}{\tr\chib}+2{\oom}{\divv}\ze\!-\!2{\oom}{\divv}\psi-2{\oom}|\ze\!-\!\psi|^2+2{\oom}{\bf K}\!=\!0\eql{2.70gql}\\
&&{\ML\ML}\frac{\partial\hat{{\chib}}}{\partial\nu}-\frac{{\oom}{\tr\chi}}{2}\hat{{\chib}}
+\frac{{\oom}{\tr\chib}}{2}{\chih}-2\oom\om\chibh-{\oom}(\hat{{\chib}}\c\hat{{\chi}})\ga+{\oom}{\nabb}\hot(\ze\!-\!\psi)
-{\oom}(\ze\!-\!\psi)\hot(\ze\!-\!\psi)\!=\!0\nn\\
&&{\ML\ML}\frac{\partial{{\omb}}}{\partial\nu}-2{\oom}\ \!\!{{\om}}\ \!{{\omb}}-\frac{3}{2}{{\oom}}|\ze|^2
\!-{{\oom}}\ze\!\c\!\psi\!+\!\frac{1}{2}{\oom}|\psi|^2
\!+\!\frac{1}{2}{\oom}\!\left({{\bf K}}\!+\!\frac{1}{4}{\tr\chib}{\tr\chi}
\!-\!\frac{1}{2}\hat{{\chib}}\!\c\!\hat{{\chi}}\!\right)=0\ .\nn
\eea
Therefore, to complete our proof we need to control all the derivatives of $\ga_{ab}, \oom$ and $X_a$ recalling that we are considering the ``coordinate" components of these quantities. More precisely we need all the angular, $\prr$, and the $\pr_{\nu}$ derivatives for $\ga$ and $\oom$ and the angular and $\pr_{\la}$ derivatives for $X$. The proof of these estimates is in Section 11.

\subsection{The $\oom$ component}
The angular derivatives of this component are easy to control as
\[\prr\log\oom=\nabb\log\oom=2^{-1}(\eta+\etab)\]
and $\eta$ and $\etab$ are already under control. 
The final result is, 
\bea
|\Lie_O^{J-1}\pr\oom|_{p,S}\leq c\!\left(\frac{J!}{J^{\a}}\frac{e^{(J-2)(\de+\underline{\Ga}(\la))}}{\ro^{J}}\right)\ .
\eea
\subsection{The $\ga$ components}\label{gamma}
We recall that $\ga_{ab}$ satisfies the following equation
\bea
\frac{\pr}{\pr\nu}\ga_{ab}-{\oom\tr\chi}\ga_{ab}=2\oom\chih_{ab}\ , \eql{7.13a}
\eea
which we rewrite as
\bea
\frac{\pr}{\pr\nu}\ga_{ab}-\overline{\oom\tr\chi}\ga_{ab}=({\oom\tr\chi}-\overline{\oom\tr\chi})\ga_{ab}+2\oom\chih_{ab}
\eea
and
\bea
&&\ML\ML\frac{\pr}{\pr\nu}(r^{-2}\ga_{ab})=({\oom\tr\chi}-\overline{\oom\tr\chi})(r^{-2}\ga_{ab})+2\oom r^{-2}\chih_{ab}\ .\nn
\eea
The control of the angular non covariant derivatives can be easily obtained applying $\prr^J$ to both sides. The following estimates hold 

\bea
|r^{J-2-\frac{2}{p}}\Lie_O^{J-1}\pr\ga_{ab}|_{p,S}(\la,\nu)\leq c\!\left(\frac{J!}{J^{\a}}\frac{e^{(J-2)(\de+\underline{\Ga}(\la))}}{\ro^{J}}\right)\ .
\eea
{\bf Remark:} 

\NI{\em The estimates for $\prr^J\ga_{ab}$ follow with $J!$ instead of $(J-1)!$ due to the loss of derivatives present in the transport equation \ref{7.13a}. This loss is not present considering the mixed derivatives.}
\subsection{The $X^a$ components}\label{xsec}
We study the equivalent  $X_a=\ga_{ac}X^c$ components; we have the transport equation
\bea
\frac{\pr X_a}{\pr\nu}=-4\oom^2\ze_a\ .
\eea
Considering the norm $|X|=\sqrt{\ga_{ab}X^aX^b}$ after a somewhat cumbersome calculation we obtain the final estimate
for all $J$,\footnote{With some extra work it is possible to prove that,
$$\bigg|\frac{r^{J-\frac{2}{p}}}{\log r}\Lie_O^{J-1}\pr X_a\bigg|_{p,S}\leq c\!\left(\frac{J!}{J^{\a}}\frac{e^{(J-2)(\de+\underline{\Ga}(\la))}}{\ro^J}\right)\ .$$} 

\beaa
|\frac{r^{(J+1-\frac{2}{p})}}{\log r}\Lie_O^{J-1}\pr|X||_{p,S}(\la,\nu)\leq c\frac{J!}{J^{\a}}\frac{e^{(J-2)(\de+\underline{\Ga}(\la))}}{\ro^J}\ , 
\eeaa
\section{The control of the non covariant partial derivatives of the connection coefficients}\label{S.4a}

\NI Assume for simplicity ${\cal O}$ denotes a $S$-tangent vector, as for instance the connection coefficient $\ze$, we are interested to the analyticity of the tensor field $\ze=\ze(e_C)\theta^{C}(\c)$ which means that we have to prove that the various components $\ze(e_C)(\la,\nu,\om^a)$ are analytic functions in the $\la,\nu,\om^a$ variables. To prove it we have to control the (norms of the) mixed derivatives in $\nu,\om^a$ for the quantities defined on an outgoing cone and the mixed derivatives in $\la, \om^a$ for the quantities defined on an incoming cone. The proof of this result is in Section 12
.

\section{The initial data for the global extension}\label{S.s initial data}

\NI In the inductive proofs of the previous sections we implicitely assumed that we can assign analytic initial data on the whole $C_0\cup\Cb_0$ hypersurface satisfying appropriate smallness conditions. This requires a careful discussion  and the goal of the present section is to show how this has to be done. 
\smallskip

\NI The problem we have to confront with is the following one: to prove Theorem \ref{L6.1} we use for the non underlined connection coefficients, ${\cal O}$,\footnote{With the exception of $\om$ and $\omb$ whose roles are inverted.} the transport equations of these quantities along the outgoing cones starting from their values on $\Cb_0$, and the opposite for the underlined quantities $\underline{\cal O}$.\footnote{The problem is even more delicate due to the fact that we have, for the non underlined quantities, to define ``initial data" on the last slice, see the discussion in subsection 9.5.} This requires that we have analytic initial data for $\cal O$ on $\Cb_0$ and analytic initial data for $\underline{\cal O}$ on $C_0$ with their $L_{p,S}$ norms satisfying the estimates \ref{123}, \ref{123ab}. As, on the other side, 
the Cauchy Kowalevski theorem we use for the local solution and in the extension proof, see (I), requires also on $C_0$ analytic initial data for $\cal O$ and the same for $\underline{\cal O}$ on $\Cb_0$, the conclusion is that all the connection coefficients have to be analitically assigned on $C_0$ and on $\Cb_0$ with their tangent derivatives satisfying the norm bounds \ref{123}, \ref{123ab}. 

\NI Moreover, as previously remarked, the initial conditions on $C_0\cup\Cb_0$ for the connection coefficients have to be such that on these initial cones the $\cal Q$ norms are bounded. This requires that the null Riemann components, which are expressed in terms of the connection coefficients, have the appropriate decay and this, at its turn, requires for some of the connection coefficients a decay, on $C_0\cup\Cb_0$,  stronger than the one which will be proved in the internal region. To fulfill it we recall, first of all, the decay conditions required to the null Riemann components to have the ${\cal Q}^{(J-2)}_0$ norms finite and $O(\varepsilon)$ for $J=2$; denoting with $\nab$  both tangential and normal dervatives,\footnote{The factor $|\la_0|$ constant on $C_0$ is, of course, irrelevant here and left just to remind the expected behavior in the interior.} we require, for $J<3$,
\bea
&&\ML\ML\ML\ \ \ \ \ |r^{\frac{7}{2}+J+\ep-\frac{2}{p}}\nab^J\a|_{p=2,S}\leq \varepsilon\ \ ,\ \ |r^{\frac{7}{2}+J+\ep-\frac{2}{p}}\nab^J\b|_{p=2,S}\leq \varepsilon\ \ ,\nn\\
&&\ML\ML\ML\ \ \ \ \  |\la_0|^{\frac{1}{2}}r^{3+J-\frac{2}{p}}\nab^J(\ro-\overline{\ro},\si)|_{p=2,S}\leq \varepsilon\ \ , ||\la_0|^{\frac{3}{2}}r^{2+J+\ep-\frac{2}{p}}\nab^J\bb|_{p=2,S}\leq \varepsilon\eql{rieminidecay}
\eea
and same bounds for $J\geq 3$, with $\varepsilon$ substituted by $C^{(0)}=O(1)$.
The analogous asymptotic conditions required along $\Cb_0$,  to guarantee that also the $\QQb_0$ norms are bounded are, again for $J<3$,

\bea
&&\ML\ML\ML\ \ \ \ \ \ |r^{\frac{7}{2}+J+\ep-\frac{2}{p}}\nab^J\b|_{p=2,S}\leq \varepsilon\ \ , \ ||\la|^{\frac{1}{2}}r^{3+J-\frac{2}{p}}\nab^J(\ro-\overline{\ro},\si)|_{p=2,S}\leq \varepsilon\ , \nn\\
&&\ML\ML\ML\ 
 \ \ \ \ \ ||\la|^{\frac{3}{2}+\ep}r^{2+J-\frac{2}{p}}\nab^J\bb|_{p=2,S}\leq \varepsilon\ , ||\la|^{\frac{5}{2}+\ep}r^{1+J-\frac{2}{p}}\nab^J\aa|_{p=2,S}\leq \varepsilon\ .\eql{rieminidecay2}
\eea
and same bounds for $J\geq 3$, with $\varepsilon$ substituteted by $C^{(0)}=O(1)$.
\smallskip

\NI Differently from what we do in the internal region, on $C_0\cup\Cb_0$ the null Riemann components are directly estimated in terms of the connection coefficients and their first derivatives, this implies that the initial data connection coefficients have to be such that conditions \ref{rieminidecay} and \ref{rieminidecay2} are satisfied.

\NI Due to the constraints for the initial data one has to prove that these initial data can be consistently defined on the whole  $C_0\cup\Cb_0$. This could be a difficulty looking at their transport equations on $C_0$ (the same argument holds on $\Cb_0$ and we do not repeat it here), as, while in the transport equations for the not underlined quantities there is not any ``loss of derivatives", the opposite happens for the underlined ones. The problem is solved observing that the construction of these initial data is done in a well defined order; in other words there is a natural order in the use of the different structure equations which allows to construct the initial data without never facing with the ``loss of derivatives" problem. 
This is the content of the argument we discuss in detail in Section 13, already exploited in \cite{Ca-Ni:char} referring only to the first  few derivatives, implying that we have to choose $J_0=7$. Here we state our final result concerning the initial data, 

\bigskip

\NI {\bf Theorem }\ref{Thinitialdata}
Assuming $\chih$ on $C_0$, $\chibh$ and$X$ on $\Cb_0$ and $\oom$ on $C_0\cup\Cb_0$ satisfying the following estimates, 
\bea
&&|r^{J+2+\ep-\frac{2}{p}}\nabla^J\log\oom|_{p,S}(\la_0,\nu) \leq C^{(0)}_3e^{(J-2)(\de_0+\underline{\Ga}_0(\la))}\frac{J!}{J^\a}\frac{1}{\ro_{0,0,1}^{J}}\nn\\
&&|r^{J+\frac{5}{2}+\ep-\frac{2}{p}}\nabla^J\chih|_{p,S}(\la_0,\nu) \leq C^{(0)}_1e^{(J-2)(\de_0+\underline{\Ga}_0(\la))}\frac{J!}{J^\a}\frac{1}{\ro_{0,0,1}^{J}}\ ,\\
&&||\la|^{1+\ep}r^{1+J-\frac{2}{p}}\nabla^J\log\oom|_{p,S}(\la,\nu_0) \leq \Cb^{(0)}_3e^{(J-2)(\de_0+\underline{\Ga}_0(\la))}\frac{J!}{J^\a}\frac{1}{\ro_{0,0,1}^{J}}\nn\\
&&||\la|^{\frac{3}{2}+\ep}r^{1+J-\frac{2}{p}}\nabla^J\chibh|_{p,S}(\la,\nu_0)\leq \Cb^{(0)}_1e^{(J-2)(\de_0+\underline{\Ga}_0(\la))}\frac{J!}{J^\a}\frac{1}{\ro_{0,0,1}^{J}}\nn \ ,
\eea
with $C^{(0)}_{1,3}$ and $\Cb^{(0)}_{1,3}$ of order $O(\varepsilon)$ for $J \leq7$ and $O(1)$ otherwise, assuming $\omb$ have some definite expressions on $S_0$ depending on the remaining connection coefficients on $S_0$, see \cite{Ca-Ni:char}, assuming, finally, that on $S_0$ the following conditions are satisfied\footnote{To completely define the $\nabb$ derivative on $S_0$ we have also to assign $\ga$ on $S_0$ see section 13.1.1.}\footnote{Clearly, to obtain the smallness of the first $C_{(J)}$ $J\leq 7$,  we have to assume also the angular derivatives $\nabb^s$ of the quantities assigned on $S_0$ of order $\varepsilon$.},
\bea\label{s0}
&&\nabb\tr\chi-\ze\tr\chi\leq\varepsilon r_0^{-(\frac{7}{2}+\ep)}\nn\\
&&\nabb\tr\chib-\ze\tr\chib\leq\varepsilon r_0^{-(\frac{7}{2}+\ep)}\nn\\
&&{\bf K}-\overline{\bf K}+\frac{1}{4}\!\left(\tr\chi\tr\chib-\overline{\tr\chi\ \!\tr\chib}\right)\leq\varepsilon r_0^{-\frac{7}{2}}\nn\\
&&\curll\zeta\leq\varepsilon r_0^{-\frac{7}{2}}\ .
\eea
 then it is possible to construct analytic initial data on $C_0\cup\Cb_0$ such that the energy type ${\cal Q}^0$ norms on $C_0\cup\Cb_0$ are finite and small and the connection coefficients norm satisfy the following estimates, with $\ro_0<\ro_{0,0,1}$ and the constant $C_{(J)}$,  $\Cb_{(J)}$of order $O(\varepsilon)$ for $J \leq7$ and $O(1)$ otherwise, 
\smallskip

\NI{\bf On $C_0$:}
\bea
&&\big|r^{1+J+\si(J)-\frac{2}{p}}\nabb^J\tr\chi\big|_{p,S}\leq C^{(0)}_{0}\!\left(\frac{J!}{J^{\a}}\frac{e^{(J-2)(\de_0+\underline{\Ga}_0(\la))}}{\ro_{0}^J}\right)\nn\\
&&\big|r^{\frac{5}{2}+J+\ep-\frac{2}{p}}\nabb^J\chih\big|_{p,S}\leq C^{(0)}_{1}\!\left(\frac{J!}{J^{\a}}\frac{e^{(J-2)(\de_0+\underline{\Ga}_0(\la))}}{\ro_{0}^J}\right)\\
&&\big|r^{2+J-\frac{2}{p}}\nabb^J\ze\big|_{p,S}\leq C^{(0)}_{4}\!\left(\frac{J!}{J^{\a}}\frac{e^{(J-2)(\de_0+\underline{\Ga}_0(\la))}}{\ro_{0}^J}\right)\nn\\
&&\big|r^{2+J+\ep-\frac{2}{p}}\nabb^J\om\big|_{p,S}\leq C^{(0)}_{2}\!\left(\frac{J!}{J^{\a}}\frac{e^{(J-2)(\de_0+\underline{\Ga}_0(\la))}}{\ro_{0}^J}\right)\nn\\
&&\big|r^{1+J+\si(J)-\frac{2}{p}}\nabb^J\tr\chib\big|_{p,S}\leq C^{(0)}_{5}\!\left(\frac{J!}{J^{\a}}\frac{e^{(J-2)(\de_0+\underline{\Ga}_0(\la))}}{\ro_{0}^J}\right)\nn\\
&&\big||\la_0|r^{1+J-\frac{2}{p}}\nabb^J\chibh\big|_{p,S}\leq C^{(0)}_{6}\!\left(\frac{J!}{J^{\a}}\frac{e^{(J-2)(\de_0+\underline{\Ga}_0(\la))}}{\ro_{0}^J}\right)\nn\\
&&\big||\la_0|r^{1+J-\frac{2}{p}}\nabb^J\omb\big|_{p,S}\leq C^{(0)}_{7}\!\left(\frac{J!}{J^{\a}}\frac{e^{(J-2)(\de_0+\underline{\Ga}_0(\la))}}{\ro_{0}^J}\right)\nn
\eea

\NI{\bf On $\Cb_0$:}
\bea
&&||\la|^{\si(J)}r^{J+1-\frac{2}{p}}\nabb^J\tr\chib|_{p,S}\leq \Cb^{(0)}_{0}\!\left(\frac{J!}{J^{\a}}\frac{e^{(J-2)(\de_0+\underline{\Ga}_0(\la))}}{\ro_{0}^J}\right)\nn\\
&&||\la|^{\frac{3}{2}+\ep}r^{J+1-\frac{2}{p}}\nabb^J\chibh|_{p,S}\leq \Cb^{(0)}_{1}\!\left(\frac{J!}{J^{\a}}\frac{e^{(J-2)(\de_0+\underline{\Ga}_0(\la))}}{\ro_{0}^J}\right)\nn\\
&& ||\la|^{1+\ep}r^{J+1-\frac{2}{p}}\nabb^J\omb|_{p,S}\leq \Cb^{(0)}_{2}\!\left(\frac{J!}{J^{\a}}\frac{e^{(J-2)(\de_0+\underline{\Ga}_0(\la))}}{\ro_{0}^J}\right)\nn\\
&&|r^{1+J+\si(J)-\frac{2}{p}}\nabb^J\tr\chi|_{p,S}\leq \Cb^{(0)}_{5}\!\left(\frac{J!}{J^{\a}}\frac{e^{(J-2)(\de_0+\underline{\Ga}_0(\la))}}{\ro_{0}^J}\right)\nn\\
&&|r^{J+2-\frac{2}{p}}\nabb^J\chih|_{p,S}\leq \Cb^{(0)}_{6}\!\left(\frac{J!}{J^{\a}}\frac{e^{(J-2)(\de_0+\underline{\Ga}_0(\la))}}{\ro_{0}^J}\right)\eql{123b}\\
&&|r^{J+2-\frac{2}{p}}\nabb^J\om|_{p,S}\leq \Cb^{(0)}_{7}\!\left(\frac{J!}{J^{\a}}\frac{e^{(J-2)(\de_0+\underline{\Ga}_0(\la))}}{\ro_{0}^J}\right)\nn\\
&&|r^{J+2-\frac{2}{p}}\nabb^J\ze|_{p,S}\leq \Cb^{(0)}_{4}\!\left(\frac{J!}{J^{\a}}\frac{e^{(J-2)(\de_0+\underline{\Ga}_0(\la))}}{\ro_{0}^J}\right)\ .\nn
\eea

\medskip


\NI The condition for $\underline{\omega}$ on $S_0$ is
\bea
\ML\omb(\nu_0)
=-\int_{\nu_0}^{\infty}d\nu'e^{\int_{\nu_0}^{\nu'}(-2\oom\om)d\nu''}\!\left[\ze\c\nabb\log\oom\!+\!\frac{3}{2}|\ze|^2\!-\!\frac{1}{2}|\nabb\log\oom|^2\!-\frac{1}{2}\!\big({\bf K}\!+\!\frac{1}{4}\tr\chi\tr\chib\!-\!\frac{1}{2}\chih\c\chibh\big)
\right]\!\!(\nu')\ .
\eea

\NI  These are the estimates of Theorem \ref{Thinitialdata}, in order to prove estimates \ref{iniz1}, \ref{iniz2}, \ref{iniz3} and \ref{iniz4}  we only have to reproduce the proofs of lemmas   \ref{L4.1} and  \ref{L4.2}, with the obvious modifications.

\smallskip

\NI Now we have to prove the estimates for the mixed derivatives of the connection coefficients on the initial data.

\NI {\bf Theorem \ref{T11.1}:}
{\em Assuming the hypothesis of Theorem \ref{Thinitialdata}, the estimates \ref{inizz0} for the $\ddb_\nu^J\nabb^P$ derivatives, then the following estimates hold for any $J$ and $P$ on $C_0\cup\underline{C}_0$ where the constants $F_i$ and $\underline{F}_i$  are of order $O(\varepsilon)$ for $J \leq7$ and $O(1)$ otherwise and $\si(P)=0$ for $P=0$ and $\si(P)=1$ for $P>0$. 
\bea
&&|r^{1+J+P+\si(P)-\frac{2}{p}}\ddb_{\nu}^J\nabb^PU|_{p,S}\leq F_1\!\left(\frac{(J+P)!}{(J+P)^{\a}}\frac{e^{(J+P-2)(\de+\underline{\Ga}_0(\la))}}{\ro_0^{J+P}}\right)\nn\\
&&|r^{1+J+P+\si(P)-\frac{2}{p}}\ddb_{\nu}^J\nabb^P\tr\chi|_{p,S}\leq F_2\!\left(\frac{(J+P)!}{(J+P)^{\a}}\frac{e^{(J+P-2)(\de+\underline{\Ga}_0(\la))}}{\ro_0^{J+P}}\right)\nn\\
&&|r^{2+J+P-\frac{2}{p}}\ddb_{\nu}^J\nabb^P\chih|_{p,S}\leq F_3\!\left(\frac{(J+P)!}{(J+P)^{\a}}\frac{e^{(J+P-2)(\de+\underline{\Ga}_0(\la))}}{\ro_0^{J+P}}\right)\nn\\
&&\big|r^{2+J+P-\frac{2}{p}}\ddb_{\nu}^J\nabb^P\eta\big|_{p,S}\leq F_4\!\left(\frac{(J+P)!}{(J+P)^{\a}}\frac{e^{((J+P)-2)(\de+\underline{\Ga}_0(\la))}}{\ro_0^{J+P}}\right)\nn\\
&&\big|r^{2+J+P-\frac{2}{p}}\ddb_{\nu}^J\nabb^P\ze\big|_{p,S}\leq F_5\!\left(\frac{(J+P+1)!}{(J+P+1)^{\a}}\frac{e^{((J+P)-1)(\de+\underline{\Ga}_0(\la))}}{\ro_0^{J+P+1}}\right)\ \ \ \ \ \ \ \ \ \ \eql{miste1}\\ 
&&\big|r^{2+J+P-\frac{2}{p}}\ddb_{\nu}^J\nabb^P\Ub|_{p,S}\leq {\underline F}_1\!\left(\frac{(J+P)!}{(J+P)^{\a}}\frac{e^{((J+P)-2)(\de+\underline{\Ga}_0(\la))}}{\ro_0^{J+P}}\right)\nn\\
&&|r^{1+J+P+\si(P)-\frac{2}{p}}\ddb_{\nu}^J\nabb^P\tr\chib|_{p,S}\leq \underline{F_2}\!\left(\frac{(J+P)!}{(J+P)^{\a}}\frac{e^{(J+P-2)(\de+\underline{\Ga}_0(\la))}}{\ro_0^{J+P}}\right)\nn\\
&&\big||\la|r^{1+J+P-\frac{2}{p}}\ddb_{\nu}^J\nabb^P{\chibh}|_{p,S}\leq {\underline F}_3\!\left(\frac{(J+P)!}{(J+P)^{\a}}\frac{e^{((J+P)-2)(\de+\underline{\Ga}_0(\la))}}{\ro_0^{J+P}}\right)\nn\\
&&\big|r^{2+J+P-\frac{2}{p}}\ddb_{\nu}^J\nabb^P\om\big|_{p,S}\leq F_6\!\left(\frac{(J+P+1)!}{(J+P+1)^{\a}}\frac{e^{((J+P)-1)(\de+\underline{\Ga}_0(\la))}}{\ro_0^{J+P+1}}\right)\nn\\
&&\big|r^{2+J+P-\frac{2}{p}}\ddb_{\nu}^J\nabb^P\omb\big|_{p,S}\leq {\underline F}_6\!\left(\frac{(J+P)!}{(J+P)^{\a}}\frac{e^{((J+P)-2)(\de+\underline{\Ga}_0(\la))}}{\ro_0^{J+P}}\right)\ .\nn
\eea}

\smallskip

\NI {\bf Proof:} First of all notice that the hypothesis of Theorem \ref{Thinitialdata}, already provide the right estimates for $\ddb_{\nu}^J\nabb^P\chi$ on $C_0$, $\ddb_{\nu}^J\nabb^P\chib$ and $\ddb_{\nu}^J\nabb^PX$ on $\Cb_0$, $\ddb_{\nu}^J\nabb^P \Omega$ and hence for $\ddb_{\nu}^J\nabb^P\om$, $\ddb_{\nu}^J\nabb^P\omb$ on $C_0\cup\Cb_0$, for any $\nu$ and $P$. For the other connection coefficients the proof is recursive, we assume these estimates hold for for $(J,P)$ and we prove they hold also for for $(J+1,P-1)$, with $J+P=N$. We sketch the idea of the proof for  $U=\oom^{-1}\tr\chi$ and leave the details  in Appendix.  

\NI Observe that the basis of the induction is exactly the result of Theorem \ref{Thinitialdata}. 
Hence, recalling the transport equation 
\bea
\ddb_{\nu}U+\frac{\oom\tr\chi}{2}{U}+|\chih|^2\!=\!0\ ,\eql{109bisabcd}
\eea
we have the following expression
\bea
&&\ML\ML\ML\ML\ddb_{\nu}^{J}\nabb^PU=\ddb_{\nu}^{J-1}\nabb^P\ddb_{\nu}U+\ddb_{\nu}^{J-1}[\ddb_{\nu},\nabb^P]U
=\ddb_{\nu}^{J-1}\nabb^P\!\left(-\oom\frac{\tr\chi}{2}{U}-|\hat{{\chi}}|^2\right)+\ddb_{\nu}^{J-1}[\ddb_{\nu},\nabb^P]U\nn\\
&&\ML\ML\ML\ML\ML=-\frac{1}{2}\ddb_{\nu}^{J-1}\nabb^P(\oom\tr\chi U)-\ddb_{\nu}^{J-1}\nabb^P(|\chih|^2)+\ddb_{\nu}^{J-1}[\ddb_{\nu},\nabb^P]U\nn\\
&&\ML\ML\ML\ML\ML=-\frac{1}{2}\sum_{q=0}^{J-1}\sum_{h=0}^P\cbin{J-1}{q}\cbin{P}{h}(\ddb_{\nu}^q\nabb^h\oom\tr\chi)(\ddb_{\nu}^{J-1-q}\nabb^{P-h}U)-\sum_{q=0}^{J-1}\sum_{h=0}^P\cbin{J-1}{q}\!\cbin{P}{h}\!(\ddb_{\nu}^q\nabb^h\chih)\c(\ddb_{\nu}^{J-1-q}\nabb^{P-h}\chih)\nn\\
&&\ML\ML\ML\ML\ML\ \ \ \ +\  \ddb_{\nu}^{J-1}([\ddb_{\nu},\nabb^P]U)\ .
\eea
As\ \ 
 $[\nabb^P,\frac{\Dbb}{\partial\nu}]f=\sum_{k=0}^{P-1}\nabb^k[\nabb,\frac{\Dbb}{\partial\nu}]\nabb^{P-k-1}f$ and
from equation 9.1,
\beaa
[\nabb_{\mu},\frac{\Dbb}{\partial\nu}]U_{\nu_1...\nu_k}
=-\sum_{j=1}^kC^{\si_j}_{\mu\nu_j}U_{\nu_1..{\si}_j..\nu_k}+\oom\chi_{\mu}^{\ro}(\nabb_{\ro}U)_{\nu_1..\nu_k}\ ,
\eeaa
it follows 
 \bea
&&\ML[\nabb^P,\frac{\Dbb}{\partial\nu}]f=\sum_{k=0}^{P-1}\nabb^k\left(-(P-k-1)C\nabb^{P-k-1}f+\oom\chi\nabb^{P-k}f\right)\\
&&\ML=-\sum_{k=0}^{P-1}(P-k-1)\sum_{J=0}^k\cbin{k}{J}(\nabb^JC)\nabb^{P-1-J}f
+\sum_{k=0}^{P-1}\sum_{J=0}^k\cbin{k}{J}(\nabb^J\oom\chi) \nabb^{P-J}f\ .\nn
\eea
Therefore we can write
\bea
&&\ML\ML\ddb_{\nu}^{J}\nabb^PU=
-\frac{1}{2}\sum_{q=0}^{J-1}\sum_{h=0}^P\cbin{J-1}{q}\cbin{P}{h}(\ddb_{\nu}^q\nabb^h\oom\tr\chi)(\ddb_{\nu}^{J-1-q}\nabb^{P-h}U)\nn\\
&&\ \ \ \ \ \ -\sum_{q=0}^{J-1}\sum_{h=0}^P\cbin{J-1}{q}\!\cbin{P}{h}\!(\ddb_{\nu}^q\nabb^h\chih)\c(\ddb_{\nu}^{J-1-q}\nabb^{P-h}\chih)\nn\\
&&\ \ \ \ \ \ + \sum_{q=0}^{J-1}\cbin{J-1}{q}\sum_{k=0}^{P-1}(P-k-1)\sum_{j=0}^k\cbin{k}{j}(\ddb_{\nu}^q\nabb^jC)\ddb_{\nu}^{j-1-q}\nabb^{P-1-j}U\nn\\
&&\ \ \ \ \ \ -\sum_{q=0}^{J-1}\sum_{k=0}^{P-1}\sum_{j=0}^k\cbin{J-1}{q}\cbin{k}{j}(\ddb_{\nu}^q\nabb^j\oom\chi)\ddb_{\nu}^{j-1-q}\nabb^{P-j}U\ \eql{5.51a}
\eea
and using the inductive assumptions we can estimate the norms of the right hand side of \ref{5.51a} to prove the expected result. 

\smallskip
\NI Clearly analogous estimates hold for $\ddb_{\la}^J\nabb^P$ on $\Cb_0$.

\smallskip

\newpage
\section {Conclusions}
We can therefore consider totally solved the global characteristic problem associated to two intersecting null cones; by this we mean that we can provide a global initial data set, specifying which quantities can be assigned freely, which  ones have to be given on the intersection of the two cones and finally which constraints they have to satisfy on it. Moreover, we provide, the smallness conditions in appropriate Sobolev norms they have to satisfy to obtain a weak global solution or, alternatively, given an analytic set of initial data, an analytic global solution.
The future program will be to extend this result to the full null cone, extending to the tip the outgoing one and reducing consequently the ingoing cone to a point. This process presents some difficulties, the main ones we can foresee are:
\medskip

\NI i) As the surface intersection of the two null hypersurfaces reduces to a point, how do the constrained quantities have to be assigned on it? 
How to reduce the constraint equations?
\medskip

\NI ii) How  the ``lapse" function $\Omega$ has to be to assigned in order to assure the orthogonal null geodesics starting from the $\nu=constant$ surfaces on the initial data outgoing cone intersect exactly in one point, so to form a null cone ?
\medskip

\NI We are confident that both questions, and other ones that can possibly emerge can find a ``natural solution" in this formalism, allowing to extend the global characteristic problem to the full null cone.
\newpage
\section{ The complete results of Section \ref{S.2}}\label{S.10}
\subsection{ The control of the angular derivatives of all the connection coefficients\label{S.s7.2}, the proof of Theorem \ref{L2.1new}}
\subsubsection{The control of $\Lie_O^{N-1}\Us$}
\NI First of all we state some commutation relations.
\begin{Le}\label{L1.3}
The following relations hold, denoting ${\Dbb}_{\nu}:=\oom\ddb_4$,
\bea
[\nabb_{\mu},\frac{\Dbb}{\partial\nu}]U_{\nu_1...\nu_k}
=-\sum_{j=1}^kC^{\si_j}_{\mu\nu_j}U_{\nu_1..{\si}_j..\nu_k}+\oom\chi_{\mu}^{\ro}(\nabb_{\ro}U)_{\nu_1..\nu_k}\eql{3.11}
\eea
\bea
\mbox{with}\ \ \ \ \ \ \ \ \ \ C^{\si_j}_{\mu\nu_j}={\oom}\!\left[(\chi_{\mu\nu_j}\etab^{\si_j}\!-\!\chi_{\mu}^{\si_j}\etab_{\nu_j})
+\theta^C_{\mu}\theta^D_{\nu_j}R^{{\si}_j}(\c,e_C,e_4,e_D)\right]\ .\ \ \ \ \ \ \ \ \ \ \ \ \ \eql{Cdef}
\eea
in the case of $U$ a 1-form:
\bea\label{dwer}
&&[\nabb_{\mu},\frac{\Dbb}{\partial\nu}]U_{\nu}=\nn \\
&&=
\left[(\chi_{\mu\nu_j}\etab^{\si}\!-\!\chi_{\mu}^{\si}\etab_{\nu})
+\theta^C_{\mu}\theta^D_{\nu}R^{{\si}}(\c,e_C,e_4,e_D)\right]\!U_{\si}
+\oom\chi_{\mu}^{\ro}(\nabb_{\ro}U)_{\nu}\nn\\
&&=-{C}^{\si}_{\mu\nu}U_{\si}+\oom\chi_{\mu}^{\ro}(\nabb_{\ro}U)_{\nu}\  
\eea

\NI We rewrite symbolically the last formula as:
\bea
&&[\nabb_{\mu},\frac{\Dbb}{\partial\nu}]U=-{C}U+D(\nabb U)\ 
\eea

\NI Were with $C$ we mean a combination of null Riemann components and with $D$ we mean a combination of connection coefficients and metric components.

\NI Similarly we have:

\bea
&&[\nabb_{\mu},\frac{\Dbb}{\partial\la}]U_{\nu_1...\nu_k}=\nn \\
&&=-{\oom}\sum_{j=1}^k\!
\left[(\chib_{\mu\nu_j}\eta^{\si_j}\!-\!\chib_{\mu}^{\si_j}\eta_{\nu_j})
+\theta^C_{\mu}\theta^D_{\nu_j}R^{{\si}_j}(\c,e_C,e_3,e_D)\right]\!U_{\nu_1..{\si}_j..\nu_k}
+\oom\chib_{\mu}^{\ro}(\nabb_{\ro}U)_{\nu_1..\nu_k}\nn\\
&&=-\sum_{j=1}^k{\underline{C}}^{\si_j}_{\mu\nu_j}U_{\nu_1..{\si}_j..\nu_k}+\oom\chib_{\mu}^{\ro}(\nabb_{\ro}U)_{\nu_1..\nu_k}\ ,\eql{6.7}
\eea

\NI which in the case of $U$ 1-form becomes:

\bea
&&[\nabb_{\mu},\frac{\Dbb}{\partial\la}]U_{\nu}=\nn \\
&&=
\left[(\chib_{\mu\nu_j}\eta^{\si}\!-\!\chib_{\mu}^{\si}\eta_{\nu})
+\theta^C_{\mu}\theta^D_{\nu}R^{{\si}}(\c,e_C,e_3,e_D)\right]\!U_{\si}
+\oom\chib_{\mu}^{\ro}(\nabb_{\ro}U)_{\nu}\nn\\
&&=-{\underline{C}}^{\si}_{\mu\nu}U_{\si}+\oom\chib_{\mu}^{\ro}(\nabb_{\ro}U)_{\nu}\  \eql{6.6bis}
\eea

\NI Also in this case we rewrite the last formula symbolically as

\bea
&&[\nabb_{\mu},\frac{\Dbb}{\partial\la}]U=-\underline{C}U+\underline{D}(\nabb U)\ 
\eea

\NI We have for the commutator of the tangential derivatives:

\bea
&&\ML[\nabb_{\a},\nabb_{\mu}]U_{\nu_1\nu_2...\nu_q}=\\  \eql{68}
&&\ML=\left[(\theta^3_{\mu}\chib_{\a}^{\b}+\theta^4_{\mu}\chi_{\a}^{\b})-(\theta^3_{\a}\chib_{\mu}^{\b}+\theta^4_{\a}\chi_{\mu}^{\b})\right]
\nabb_\b U_{\nu_1\nu_2...\nu_q}-\theta^C_{\a}\theta^D_{\mu}\sum_{i=1}^qR^{\tilde{\nu}_i}(\c,e_C,\c,e_D)_{\nu_i}U_{\nu_1\nu_2..{\tilde\nu}_i..\nu_q}\nn
\eea

\NI In this case we apply this commutator to a 2-form obtaining:

\bea
&&\ML[\nabb_{\a},\nabb_{\mu}]U_{\nu_1\nu_2}=\nn\\
&&\ML=\left[(\theta^3_{\mu}\chib_{\a}^{\b}+\theta^4_{\mu}\chi_{\a}^{\b})-(\theta^3_{\a}\chib_{\mu}^{\b}+\theta^4_{\a}\chi_{\mu}^{\b})\right]
\nabb_\b U_{\nu_1\nu_2}-\theta^C_{\a}\theta^D_{\mu}R^{\tilde{\nu}}(\c,e_C,\c,e_D)_{\nu_1}U_{\tilde{\nu}\nu_2}\nn\\
&& -\theta^C_{\a}\theta^D_{\mu}R^{\tilde{\nu}}(\c,e_C,\c,e_D)_{\nu_2}U_{\nu_1\tilde{\nu}}\ 
\eea

\NI Which we rewrite symbolically as

\bea
&&[\nabb_{\mu},\frac{\Dbb}{\partial\nu}]U=-\hat{C}U+\hat{D}(\nabb U)\ 
\eea

\NI At last we apply the $[\nabb_{\a},\divv]$ to a 2-form:

\bea
&&\ML[\nabb_{\a},\divv]U_{\nu_1\nu_2}=[\nabb_{\a},\nabb_{\mu}]U^{\mu}_{\nu}=\\
&&\ML=\left[(\theta^3_{\mu}\chib_{\a}^{\b}+\theta^4_{\mu}\chi_{\a}^{\b})-(\theta^3_{\a}\chib_{\mu}^{\b}+\theta^4_{\a}\chi_{\mu}^{\b})\right]
\nabb_\b U^{\mu}_{\nu}-2\theta^C_{\a}\theta^D_{\mu}R^{\tilde{\nu}}(\c,e_C,\c,e_D)_{\nu}U^{\mu}_{\tilde\nu}\ .\nn
\eea

\NI Which we rewrite symbolically as

\bea
&&[\nabb_{\a},{\divv}]U=-\overline{C}U+\overline{D}(\nabb U)\ . 
\eea

\NI Notice that all the commutators have the same structure, namely a combination of products of connection coefficients and metric components times $U$ and a combination of null Riemann components times $\nabb U$.

\NI Moreover the following relations hold for the commutator of the $\Lie_O$ derivatives:

\NI For $U$ 1-form
\bea
&&[\frac{\Dbb}{\partial\nu},\Lie_O]U=\left((O^{\si}C^{\tau}_{\si\mu})+\frac{\Dbb}{\partial\nu}{\cal H}^{\tau}_{\mu}\right)\!U_{\tau}\nn
\eea
which we write in a compact way as
\bea\label{E}
[\frac{\Dbb}{\partial\nu},\Lie_O]U=E^{\tau}_{\mu}U_{\tau}\ ,
\eea
where
\bea
&&E^{\tau}_{\mu}\equiv \left((O^{\si}C^{\tau}_{\si\mu})+\frac{\Dbb}{\partial\nu}{\cal H}^{\tau}_{\mu}\right)\ .\nn
\eea

\NI Where ${\cal H}$ has the following expression 
\bea
{\cal H}_{ab}=\gggg(\nabb_aO,\frac{\pr}{\pr x^b})\ \ 
\eea
With similar calculations we obtain

\bea
[\frac{\Dbb}{\partial\la},\Lie_O]U=\underline{E}^{\tau}_{\mu}U_{\tau}\ ,
\eea

\NI where
\bea\label{unde}
&&\underline{E}^{\tau}_{\mu}\equiv \left((O^{\si}\underline{C}^{\tau}_{\si\mu})+\frac{\Dbb}{\partial\la}{\cal H}^{\tau}_{\mu}\right)\ .\nn
\eea

\medskip

\NI Finally we consider the commutators $[\nabb, \Lie_O]U_{\a\b}$ and  $[\divv,\Lie_O]U_{\a\b}$ with $U_{\a\b}$  a symmetric 2-form.

\bea
&&([\nabb_{\nu},\Lie_O]U)_{ab}\nn\\
&&=O^{\ro}\hat{C}^{\si}_{\ro\nu a }U_{\si b}+2U_{a\ro}\left(\nabb_{\nu}{\cal H}^{\ro}_b\right)+\sum_{A=1,2}(O^{\ro}{\cal H}_{\ro}^{\b}\theta_{\nu}^{ A}\nabb_{\b}U_{ab})
\eea

\NI which we rewrite symbolically as:

\bea
([\nabb_{\nu},\Lie_O]U)_{ab}=\hat{E}^{\si}_{\nu a }U_{\si b} +\sum_{A=12} \hat{\cal H} ^{A\b}_{\nu}\nabb_{\b}U_{ab}
\eea

\NI and consequently

\bea
&&([\divv,\Lie_O]U)_c=[\nabb_{\nu},\Lie_O]U^{\nu}_c\nn\\
&&=O^{\ro}\hat{C}^{\si}_{\ro\nu c }U_{\si}^{ \nu}+2U_{\ro}^{\nu}\left(\nabb_{\nu}{\cal H}^{\ro}_c\right)+\sum_{A=1,2}(O^{\ro}{\cal H}_{\ro}^{\b}\theta^{A\nu}\nabb_{\b}U_{c\nu})
\eea

\NI which we rewrite symbolically as:

\bea
([\divv,\Lie_O]U)_c=\overline{E}^{\si}_{\nu c }U_{\si}^{ \nu} +\sum_{A=1,2} \overline{\cal H} ^{A\b\nu}\nabb_{\b}U_{c\nu}
\eea

\NI Consequenlty, exploiting equation \ref {dwer}, if $N>1$, the following relations holds with $f$ a scalar function:
\bea
[\nabb^N,\frac{\Dbb}{\partial\nu}]f=\sum_{k=0}^{N-1}\left(\!\!\!\begin{array}{c}N\\k\\
\end{array}\!\!\!\right)(\nabb^{N-1-k}\oom\chi)\c\nabb^{k+1}f-\sum_{k=0}^{N-2}\left(\!\!\!\begin{array}{c}N\\k\\
\end{array}\!\!\!\right)(\nabb^{N-2-k}C)\c\nabb^{k+1}f\ ,\ \ \ \ \ \ \ \eql{1.117w}
\eea
the following one if $V$ is an $S$-tangent vector field: 
\bea
[\nabb^{N-1},\frac{\Dbb}{\partial\nu}]V=\sum_{k=0}^{N-2}\left(\!\!\!\begin{array}{c}N-1\\k\\
\end{array}\!\!\!\right)(\nabb^{N-2-k}\oom\chi)\c\nabb^{k+1}V-\sum_{k=0}^{N-2}\left(\!\!\!\begin{array}{c}N\\k\\
\end{array}\!\!\!\right)(\nabb^{N-2-k}C)\c\nabb^{k}V\ .\ \ \ \ \ \ \ \eql{1.117ww}
\eea
\NI in the same way we can calculate $[\Lie_O^{N-1},\frac{\Dbb}{\partial\nu}]V$
\bea
[\Lie_O^{N-1},\frac{\Dbb}{\partial\nu}]V=\sum_{k=0}^{N-2}\left(\!\!\!\begin{array}{c}N-1\\k\\
\end{array}\!\!\!\right)\sum_{k=0}^{N-2}\left(\!\!\!\begin{array}{c}N\\k\\
\end{array}\!\!\!\right)(\nabb^{N-2-k}E)\c\nabb^{k}V\ .\ \ \ \ \ \ \ \eql{1.117ww}
\eea

\end{Le}

\begin{Le}\label{eqforU}
$\Lie_O^{N-1}\Us$ satisfies the following equation along the outgoing cones,

\bea\eql{15.46}
&&\ML\ML\frac{\Dbb}{\partial\nu}(\Lie_O^{N-1}\Us)+\frac{3}{2}\overline{\oom\tr\chi}(\lie_O^{N-1}\Us)=-\oom\chih\c(\Lie_O^{N-1}\Us)- 2\chih\Lie_O^{N-1}(\nabb\chih)-\tr\chi\Lie_O^{N-1}\b+\nn\\
&&\ML\ML+\frac 3 2(\overline{\oom\tr\chi}-\oom\tr\chi)(\Lie_O^{N-1}\Us)+\{{\it (good)_1}\}
\eea

\bea\eql{15.47bis}
&&\ML\ML\{{\it (good)_1}\}=2\sum_{k=1}^{N-1}\cbin{N-1}{k}\Lie_O^{N-1-k}(\nabb\chih)\Lie_O^k\chih\nn\\ 
&&\ML\ML-\frac{3}{2}\sum_{q=1}^{N-1}\cbin{N-1}{q}(\Lie_O^{q}\oom^2 U)(\Lie_O^{N-1-q}\Us)+\sum_{k=0}^{N-1}\cbin{N-1}{k}\Lie_O^{N-1-k}(|\chih|^2)\Lie_O^k\eta\nn\\
&&\ML\ML+\sum_{q=1}^{N-1}\cbin{N-1}{q}(\Lie_O^{q}(\oom\chih)(\Lie_O^{N-1-q}\Us)+\sum_{q=0}^{N-2}\cbin{N-1}{q}(\Lie_O^{N-2-q}E)(\Lie_O^{q}\Us)\nn \\
&&\ML\ML-\sum_{q=0}^{N-1}\cbin{N-1}{q}(\Lie_O^{q}(\oom U)(\Lie_O^{N-1-q}\chih\eta)+\sum_{q=1}^{N-1}\cbin{N-1}{q}(\Lie_O^{N-2-q}\b)(\Lie_O^{q}\tr\chi)\nn \\
\eea

\NI Moreover  the following inequality holds:

\bea\label{925}
-\frac{\partial}{\partial\nu}|r^{(3-\frac{2}{p})}\Lie_O^{N-1}\Us|_{p,S}\leq  |r^{(3-\frac{2}{p})}\Ls|_{p,S}\ \ \ .\eql{diffest1aa}
\eea
where\footnote{In fact we have two inequalities
\bea
-|r^{(3-\frac{2}{p})}\Ls|_{p,S}\leq \frac{\partial}{\partial\nu}|r^{(3-\frac{2}{p})}\nabb^{N-1}\Us|_{p,S}\leq |r^{(3-\frac{2}{p})}\Ls|_{p,S}\ .\eql{xxzz}
\eea
One is used when integrating along the outgoing cones the other ones for the incoming one.}

\bea\eql{15.52}
|\Ls|\!&=&\!\!\left[\big|\oom\chih\c(\Lie_O^{N-1}\Us)\big|+\big|2\chih\c(\Lie_O^{N-1}\nabb\chih)\big|+|\tr\chi|\big|\Lie_O^{N-1}\b\big|
+\big|\big\{(good)_1\big\}\big|\right]\nn\\
&+&\!\!\left[\frac{(3-\frac 2 p)}{2}|{\oom}\tr\chi-\overline{{\oom}\tr\chi}||\Lie_O^{N-1}\Us|\right]\ \ \ \ .\eql{defLa}
\eea

\end{Le}

\NI {\bf Proof:} See Appendix \ref{AS.3}.\footnote{Trying to use the analogous transport equation to estimate $|r^{N+2-\frac{2}{p}}\nabb^NU|_{p,S}$ does not work as the term
$-\frac{U}{2}\sum_{k=1}^{N}\cbin{N}{k}\!(\nabb^k\oom^2)(\nabb^{N-k}U)$
gives rise, when integrated, to logarithmic divergent contributions.}
\smallskip

\NI To obtain from the differential inequality \ref{15.52} an estimate for $\Lie_O^{N-1}\Us$ we need first to control  the term
$(\Lie_O^{N-1}\nabb\hat{\chi})$ present in $\Ls$, which has the maximal order of derivatives.\footnote{Recall that $\Us$ already contain a $\nabb$ derivative.} For it we use the Hodge equation
\bea\label{receqq} 
&&{\divv}\hat{\chi}=\frac{1}{2}\nabb{\tr\chi}-\b-\zeta\!\c\hat{\chi}+\frac{1}{2}\zeta{\tr\chi}\equiv F^{(1)}\eql{fdef1}
\eea
 from the Hodge system \ref{receqq} it is possible to prove the following lemma, 

\begin{Le}\label{estdNchih}

\bea
&&\ML\ML |\Lie_O^{N-1}\nabb\chih|_{p,S}\leq c \left(|\oom|_{\infty,S}|\Lie_O^{N-1}\Us|_{p,S}+|\Lie_O^{N-1}\b|_{p,S}+|{\{\it (good)_2}\}|_{p,S}+|{\{\it (good)_4}\}|_{p,S}\right)\nn\\
\eea

with 
\bea
&&{\{\it (good)_2}\}=[\divv,\Lie_O^{N-1}]\chih\nn\\
&&=\sum_{q=0}^{N-2}\left(\cbin{N-1}{q}(\Lie_O^{N-2-q}\overline{E})(\Lie_O^{q}\Us)+\cbin{N-1}{q}(\Lie_O^{N-2-q}\overline{\cal H})(\Lie_O^{q}\nabb\Us)\right)\nn\\
\eea

\bea
&&\ML\ML|{\{\it (good)_4}\}|_{p,S}=\left(\sum_{k=1}^{N-1}\cbin{N-1}{k}|\Lie_O^k\oom|_{\infty}|\Lie_O^{N-1-k}\chih|_{p,S}+|{\{\it (good)_3}\}|_{p,S}\right)\nn\\
\eea

\bea
&&{\{\it (good)_3}\}=-\Lie_O^{N-1}(\zeta\!\c\hat{\chi})+[\divv,\Lie_O^{N-1}]\chih\nn\\
&&=\Lie_O^{N-1}(\zeta\!\c\hat{\chi})+\sum_{q=0}^{N-2}\left(\cbin{N-1}{q}(\Lie_O^{N-2-q}\overline{E})(\Lie_O^{q}\Us)+\cbin{N-1}{q}(\Lie_O^{N-2-q}\overline{\cal H})(\Lie_O^{q}\nabb\Us)\right)\nn\\
\eea

\end{Le}
\smallskip

\NI The recursive assumption for $\Lie_O^{N-1}\Us$ 
is proved in the following theorem whose detailed proof is in the appendix to Section \ref{S.10}, Appendix \ref{AS.3}, it is a prototype of all the remaining estimates of the connection coefficients angular derivatives.
\medskip

\begin{theorem}\label{T3.1}
Assuming the following norm estimate for the null Riemann component $\b$, with $C^{(1)}=O(\varepsilon)$ for $J\leq 7$, $C^{(1)}=O(1)$ otherwise, $C^{(1)}< C_0$.\footnote {It is important that the constant $C^{(1)}$, is smaller than all the constants estimating the connection coefficients, this is because of the fact that the null Riemann components appears in the structure equations not associated with other terms. }
\bea
|r^{2-\frac{2}{p}}\Lie_O^{N-1}\b|_{p,S}(\la,\nu)\leq C^{(1)}
e^{(N-2)\de}e^{(N-2)\underline{\Ga}(\la)}\frac{N!}{N^\a}\frac{1}{\ro_{0,1}^N}\ ,\eql{betassaa}
\eea
assuming the norm estimates \ref{123} for the initial data connection coefficients with $J\leq N$ and the norm estimates \ref{123a} for the connection coefficients with $J<N$, then the following inequality holds on ${\cal K}$,
\bea
\big|r^{3-\frac{2}{p}}\Lie_O^{N-1}\Us\big|_{p,S}(\la,\nu)\leq C_0e^{(N-2)\de}e^{(N-2)\underline{\Ga}(\la)}\frac{N!}{N^\a}\frac{1}{\ro_{0,1}^{N}}\ ,
\eea

\end{theorem}
{\bf Proof:} See Appendix \ref{AS.3}.
\medskip

\NI {\bf Remark:} 

\NI {\em 

\NI  We are left with proving assumption \ref{betassaa} of Theorem \ref{T3.1}. This is a particular case of the more general Theorem  \ref{hypthm2}  which gives the appropriate estimates for the Riemann null components proved in detail in sections \ref{S.10b} and \ref{AS.4}.}

\subsubsection{The control of $\Lie_O^N{\eta}$ }
The strategy to control the $|\c|_{p,S}$ norms of $\Lie_O^N{\eta}$ is based on the same approach described for the estimates of $\Lie_O^{N-1}\Us$. There are, nevertheless, some differences we are going to point out.
First of all we cannot use in a straightforward way the transport equation for $\eta$,
\bea
\dd_4\eta+\frac{\tr\chi}{2}\eta+\chih\c\eta-\chi\c\etab+\b=0
\eea
as it implies a loss of derivatives (the Riemann coefficient $\b$ depends on second derivatives of the metric while $\eta$ only on first ones). The idea, developed in \cite{C-K:book} and extended in \cite{Kl-Ni:book}, is to define the scalar quantity $\tilde\mu$
\bea
\tilde{\mu}:=-\divv\eta + \frac{1}{2}(\chi\underline{\chi}-\overline{\chi\underline{\chi}})-(\ro-\overline{\ro})\eql{mudef1}
\eea
and observe that, using the structure equations, $\tilde{\mu}$ satisfies the following transport equation,
\bea
\frac{\Dbb}{\partial\nu}\tilde{\mu}+(\Omega\tr\chi)\tilde{\mu}=(\Omega \tilde{F}-\overline{\Omega\tilde{F}})+(\Omega \tilde{H}-\overline{\Omega\tilde{H}})\eql{mutrasp}
\eea
where
\bea
&&\ML\ML\tilde{F}=\hat{\chi}\c(\nabb\hat{\otimes}\eta)+\Omega(\eta-\underline{\eta})\nabb(\Omega^{-1}\tr\chi)+ \frac{1}{2}\tr\chi(|\eta|^2-|\underline{\eta}|^2)-\frac{1}{2}\!\left(\tr\underline{\chi}|\hat{\chi}|^2+\tr\chi(\chih\c\chibh)\right)+2\eta\c\chih\c\etab\nn\\
&&\ML\ML\tilde {H}=\left(-\frac{\tr\chi}{2}\overline{\chib\c\chi}-2\eta\c\b+\tr\chi(\ro+\overline{\ro})\right)\ .
\eea
A long but obvious computation, analogous to the one for $\Lie_O^{N-1}\Us$, proves the following lemma,  
\begin{Le}\label{eqformu}
$\Lie_O^{N-1}{\tilde\mu}$ satisfies the following transport equation along the outgoing cones,  for $N>1$,
\bea
&&\ML\ML\frac{\Dbb{(\Lie_O^{N-1}{\tilde\mu})}}{\partial\nu}+{\oom}\tr\chi\c{(\Lie_O^{N-1}{\tilde\mu})}
-\chih\c\Lie_O^{N-1}\!(\nabb\hat{\otimes}\eta)\nn\\
&&\ \ \ \ \ \ +2\oom\eta\c\Lie_O^{N-1}\b-\oom\tr\chi\Lie_O^{N-1}\ro+\left[\big\{(\widetilde{good})_1\big\}+\big\{(\widetilde{good})_2\big\}\right]=0\ ,\ \ \ \ \ \ \eql{mutransp}
\eea
where 
\bea
&&\ML\ML\ML\big\{(\widetilde{good})_1\big\}=-\sum_{q=0}^{N-2}\cbin{N-1}{q}(\Lie_O^{N-2-q} E)( \Lie_O^{q}\tilde{\mu})\nn\\
&&\ML\ML\ML\big\{(\widetilde{good})_2\big\}=-\big(-\sum_{k=1}^{N-1}\cbin{N-1}{k}\Lie_O^k(\oom\tr\chi)(\Lie_O^{N-1-k}\tilde{\mu})+\Lie_O^{N-1}\!\!\left[(\Omega \tilde{F}-\overline{\Omega\tilde{F}})+(\Omega \tilde{H}-\overline{\Omega\tilde{H}})\right]\! \nn\\
&&\ML\ML\ML-\chih\c\Lie_O^{N-1}\!(\nabb\hat{\otimes}\eta) +2\oom\eta\c\Lie_O^{N-1}\b-\oom\tr\chi\Lie_O^{N-1}\ro\big)\ .\ \ \ \ \ \ \ \ \ \ \eql{tildegood2}
\eea
and
\bea
\big\{(\widetilde{good})_{2+1}\big\}=\big\{(\widetilde{good})_1\big\}+\big\{(\widetilde{good})_2\big\}
\eea
Moreover the following inequality \footnote{Analogous to inequality \ref{925}.} holds
\bea\label{930}
-\frac{\pr|r^{(2-\frac{2}{p})}\Lie_O^{N-1}{\tilde\mu}|_{p,S}}{\pr\nu}\leq |r^{(2-\frac{2}{p})}{\tilde L}|_{p,S}\ ,\eql{diffest1s1}
\eea
where
\bea
|{\tilde L}|\!&\leq&\!\left[\big|\chih\c(\Lie_O^{N-1}{\tilde\mu})\big|+\big|\chih\c(\Lie_O^{N-1}\nabb\hat{\otimes}\eta)\big|+|\eta|\big|\Lie_O^{N-1}\b\big|+|\tr\chi|\big|\Lie_O^{N-1}\ro\big|
+\big|\big\{(\widetilde{good})_{1+2}\big\}\big|\right]\nn\\
&&+\!\left[|{\oom}\tr\chi-\overline{{\oom}\tr\chi}||\Lie_O^{N-1}{\tilde\mu}|\right]\ .\ \ \ \eql{deftildeL1}
\eea
\end{Le}
{\bf Proof:} The proof has the same structure as the proof of Lemma \ref{eqforU} and we do not report it here.
\NI In fact observing that $\eta$ satisfies the following Hodge system
\bea
&&\divv\eta =-\tilde{\mu}+\frac{1}{2}(\chi\underline{\chi}-\overline{\chi\underline{\chi}})-(\ro-\overline{\ro})\equiv F^{(0)}\eql{etahodgea}\\
&&\curll\eta=\sigma-\frac{1}{2}\hat{\underline{\chi}}\wedge\hat{\chi}\equiv G^{(0)}\ .\nn\eql{etahodgebb}
\eea
and examining equations \ref{mutransp}, \ref{etahodgea} it is immediate to recognize to that: 
\smallskip

\NI i)  $\tilde{\mu}$ plays the role of $\Us$ \footnote{There are $N\!-\!1$ derivatives in \ref{mutransp} as, in $\tilde{\mu}$, $\eta$ is already derived once.} and the terms in $\big\{(\widetilde{good})_1\big\}+\big\{(\widetilde{good})_2\big\}$ contain only (angular) derivatives of the connection coefficients lower than $N$ or terms of highest derivative order which have already been controlled in the previous step. In this sense all these terms are ``good terms". 
\smallskip

\NI ii) As in equation \ref{925} for $\Lie_O^{N-1}\Us$  the terms $\Lie_O^{N-1}\nabb\chih$ are controlled deriving the Hodge system \ref{receqq}, here the term $\Lie_O^{N-1}(\nabb\hat{\otimes}\eta)$ is controlled in the same way deriving the Hodge system \ref{etahodgea}. The result of this estimate, 
we do not report here, is in Lemma \ref{nabbNetaest}, the analogous of Lemma \ref{estdNchih}.
\smallskip

\NI iii) Finally, as in the estimate for $\Lie_O^{N-1}\Us$, some Riemann null components with the highest order of derivatives ${N\!-\!1}$ for the Riemann components) appear in the transport equation \ref{mutransp} and in the Hodge system \ref{etahodgea}.  
Again these terms have to be estimated using Theorem  \ref{hypthm2}.

\begin{Le}\label{nabbNetaest}
Assuming that $\eta$ satisfies the Hodge system \ref{etahodgea} then $\Lie_O^{N-1}(\nabb\hat{\otimes}\eta)$ satisfies the following inequality,

\bea
&&\ML\ML |\nabb\Lie_O^{N-1}\eta|_{p,S}\leq c \left(|\Lie_O^{N-1}\tilde{\mu}|_{p,S}+|\Lie_O^{N-1}\ro|_{p,S} +|{\{\it (\widetilde{good})_3}\}|_{p,S})\right)\nn\\
\eea

\NI With $c$ a suitable constant and

\bea
&&{\{\it (\widetilde{good})_3}\}=-\Lie_O^{N-1}\frac{1}{2}(\chi\underline{\chi}-\overline{\chi\underline{\chi}})+\Lie_O^{N-2}\ro+[\divv,\Lie_O^{N-1}]\eta\nn\\
&&=-\Lie_O^{N-1}(\frac{1}{2}(\chi\underline{\chi}-\overline{\chi\underline{\chi}})+\Lie_O^{N-2}\ro+\sum_{q=0}^{N-2}(\cbin{N-1}{q}(\Lie_O^{N-2-q}\overline{E})(\Lie_O^{q}\eta)
\nn\\
&&+\cbin{N-1}{q}(\Lie_O^{N-2-q}\overline{\cal H})(\Lie_O^{q}\nabb\eta))\nn\\
\eea

\NI consequently

\bea\eql{15.60bis}
&&\ML\ML |\Lie_O^{N-1}\nabb\underline{\eta}|_{p,S}\leq c \left(|\Lie_O^{N-1}\tilde{\mu}|_{p,S}+|\Lie_O^{N-1}\ro|_{p,S} +|{\{\it (\widetilde{good})_3}\}|_{p,S}+|{\{\it (\widetilde{good})_4}\}|_{p,S}\right)\nn\\
\eea

\NI With $|{\{\it (\widetilde{good})_4}\}|_{p,S}=[\nabb,\Lie_O^{N-1}]{\eta}$

\smallskip

\NI hence, the same inequality holds for $\Lie_O^{N-1}\nabb\hat{\otimes}\eta$

\bea
|\Lie_O^{N-1}(\nabb\hat{\otimes}\eta)|_{p,S}\leq c \left(|\Lie_O^{N-1}\tilde{\mu}|_{p,S}+|\Lie_O^{N-1}\ro|_{p,S} +|{\{\it (\widetilde{good})_3}\}|_{p,S}+|{\{\it (\widetilde{good})_4}\}|_{p,S}\right)\nn\\
\eea

\end{Le}

\NI Retracing the steps performed for $\Us$ we can hence obtain the estimate for $\Lie_O^{N-1}{\tilde\mu}$, stated in the following theorem.
\begin{theorem}\label{muetaest}
Assuming the norm estimates for the null Riemann components of Theorem \ref{hypthm2},
assuming the norm estimates \ref{123} for the initial data connection coefficients with $J\leq N$ and the norm estimates \ref{123a} for the connection coefficients with $J<N$,
assuming that on the ``last slice" $\Lie_O^{N-1}{\tilde\mu}$ satisfy the following estimate
\bea
\big|r^{(2-\frac{2}{p})}\Lie_O^{N-1}{\tilde\mu}\big|_{p,S}(\la,\nu_*)\leq C_6^{(*)}\left(\frac{N!}{N^\a}e^{(N-2)\de}e^{(N-2)\underline{\Ga}(\la)}\frac{1}{\ro_{0,1}^N}\right) \ ,
\eea
then the following inequality holds with $C_6^{(*)}<C_6$,
\bea\label{999}
&&\big|r^{(2-\frac{2}{p})}\Lie_O^{N-1}{\tilde\mu}\big|_{p,S}(\la,\nu)\leq \tilde{C}_6e^{(N-2)\de}e^{(N-2)\underline{\Ga}(\la)}\frac{N!}{N^\a}\frac{1}{\ro_{0,1}^{N}}\nn\\ .
\eea
And by definition of $\tilde\mu$ we obtain easily the inequality for $\eta$,
\bea
&&\big|r^{(2-\frac{2}{p})}\Lie_O^{N-1}\nabb{\eta}\big|_{p,S}(\la,\nu)\leq C_6e^{(N-2)\de}e^{(N-2)\underline{\Ga}(\la)}\frac{N!}{N^\a}\frac{1}{\ro_{0,1}^{N}}\ 
\eea
With $\tilde{C}_6< C_6$ . 
\end{theorem}
\smallskip 



\subsubsection{The control of $\Lie_O^N\dd_3\log\Omega$}\label{S.S3.5}
To control $\Lie_O^N\omb=-2^{-1}\Lie_O^N{\dd_3\log\Omega}$ we cannot use the transport equation for $\omb$. This is due to the fact that in its transport equation 
\beaa 
\dd_4\omb\!-\!2\om\omb\!-\!\ze\c\nabb\log\oom\!-\!\frac{3}{2}|\ze|^2\!+\!\frac{1}{2}|\nabb\log\oom|^2\!-\frac{1}{2}\ro\!=\!0
\eeaa
there is a loss of derivatives, namely this equation depends on the Riemann component $\ro$. This problem is analogous to the one we faced in the control of $\Lie_O^N\eta$ and the way out\footnote{This is an adaptation of a strategy first used by D.Christodoulou and S.Klainerman in \cite{C-K:book}.} is exactly of the same type as the one used for $\Lie_O^N\eta$. Instead of $\tilde{\mu}$ we introduce the following quantity,
\bea
\tilde{\omb}\equiv\divv{\hat V}\ \ \mbox{where}\ \ {\hat V}=({\tilde V}-\oom\bb)\ , \mbox{and}\ \ \tilde{V}=\nabb\oom \dd_3\log\oom=-2\nabb\oom\omb\eql{omtildedef}\ .
\eea
With standard computations we obtain,
\begin{Le}\label{L3.6}
\NI The structure equation for $\tilde{\omb}$ is
\bea
&&\ML\ML\frac{\partial{\tilde{\omb}}}{\partial\nu}+\oom\tr\chi{\tilde{\omb}}+4\oom\om{\tilde{\omb}}=-2\oom\chih\c\nabb{\hat V}\eql{tildeomb1}\\
&&\ML\ML+\oom\!\left[-2\chih\c\nabb\oom\bb-4\om\divv(\oom\bb)+\oom(\nabb\log\oom)\c\nabb\ro+\frac{1}{2}\tr\chi\oom\divv\bb+\chibh\c\oom\nabb\b+(\nabb\oom)(-\nabb \ro+\dual\nabb\si)\right]\nn\\
&&\ML\ML+{\{\cal{F}\}}\ .\nn
\eea
where ${\{\cal{F}\}}$ collects terms with less derivatives whose explicit expression is in Appendix \ref{AS.3}. Moreover the transport equation for $\Lie_O^{N-2}{\tilde{\omb}}$ is:

\bea\label{3.83}
&&\ML\frac{\partial}{\partial\nu}(\Lie_O^{N-2}{\tilde{\omb}})+(\oom\tr\chi+4\oom\om)(\Lie_O^{N-2}{\tilde{\omb}})+(\oom\chi)\c(\Lie_O^{N-2}{\tilde{\omb}})\nn\\
&&\ML=-2\oom\chih\ \!\Lie_O^{N-2}\nabb{\hat V}+\oom\!\left[-2\oom\chih\c\Lie_O^{N-1}\bb-4\oom\om\Lie_O^{N-2}\divv\bb+\oom(\nabb\log\oom)\c\Lie_O^{N-1}\ro\right.\nn\\
&&\ML\left.\ \ \ +\frac{1}{2}\tr\chi\oom\Lie_O^{N-2}\divv\bb+\chibh\c\oom\Lie_O^{N-1}\b+(\nabb\oom)(-\Lie_O^{N-1}\ro+\Lie_O^{N-2}\dual\nabb\si)\right]\nn\\
&&\ML\ \ \ \ +\{{\cal L}\}+\{{\cal G}\}+{\{\tilde{\cal{ F}}\}}\ ,
\eea

\NI where $\{{\cal L}\}$, $\{{\cal G}\}$, ${\{\tilde{\cal{ F}}\}}$, whose explicit expressions are given in Appendix \ref{AS.3}, collect all the terms with lower order derivatives which can be estimated using the inductive assumptions.
\end{Le}
\NI {\bf Remark:}

\NI{\em Notice that in the terms $\{{\cal L}\}+\{{\cal G}\}+{\{\tilde{\cal{ F}}}\}$ the Lie derivatives of $\ze$ appear, up to order $N-1$, which has been not explicitly estimated, this is not a problem as remember that $\ze=\eta+ \nabb \log \Omega$ and these quantities are already at our disposal by the inductiv assumptions.}


\NI No loss of derivatives is present in equation \ref{tildeomb1}. In fact ${\tilde{\omb}}$ depends on three derivatives of the metric components as all the terms explicitely written on the r.h.s.  while ${\{\tilde{\cal{F}}\}}$ depends on lower derivatives. 

\NI Equation \ref{3.83} plays the same role of the transport equation for $\Lie_O^{N-1}\tilde\mu$ or the one for $\Lie_O^{N-1}\Us$; 
It depends on $\Lie_O^{N-1}$ derivatives of Riemann components 
which  have to be estimated using the hyperbolicity, see Theorem \ref{hypthm2}, and on $\Lie_O^{N-2}\nabb{\hat V}$; this last term is the analogous of the term $\nabb^{N-1}(\nabb\hat{\otimes}\eta)$ present in the equation \ref{mutransp} or to the term $\Lie_O^{N-1}\nabb\chih$ present in equation \ref{15.46} for $\Lie_O^{N-1}\Us$. To control the $\Lie_O^{N-1}{\hat V}$ term in terms of ${\tilde{\omb}}$ and from it to control the norms of $\Lie_O^{N-2}{\tilde{\omb}}$ applying Gronwall's Lemma, we have to use the Hodge system,
\bea
&&\divv{\hat V}={\tilde{\omb}}\ ,\eql{omhodge}
\eea
 which is equivalent, but easier to treat, to the Hodge system \ref{receqq} and to the Hodge system \ref{etahodgea}  used to control the $\eta$ derivatives from the $\tilde\mu$ estimates. Starting from \ref{omhodge} the Hodge system to consider is provided in the following lemma, whose simple proof we do not report here,
\begin{Le}\label{L3.7}
From the Hodge system \ref{omhodge} it follows that $\nabb^{N-1}{\hat V}$
satisfies the following estimates

\bea
|\Lie_O^{N-2}\nabb \hat{V}|_{p,S}\leq |[\div,\Lie_O^{N-2}] \hat{V}|_{p,S} +| \widetilde{\widetilde{\{ good_3\}}}|_{p,S} + |\tilde{\omega}|_{p,S}
\eea
with
\bea
|\widetilde{\widetilde{ \{good_3\}}}|_{p,S}= |[\div,\Lie_O^{N-2}] \hat{V}|_{p,S}
\eea
\end{Le}
\smallskip

\NI Finally once we control the norms of $\Lie_O^{N-2}{\tilde{\omb}}$, as from Theorem \ref{hypthm2} we control $\divv\Lie_O^{N-2}\bb$ it follows that we control $\lapp\Lie_O^{N-2}\om$ 
and from it we control the norm of $\nabb^N\om$ which is the final step required to prove the following theorem, analogous to Theorem  \ref{muetaest},
\begin{theorem}\label{omomtildeest}
Assuming the results of Theorem \ref{hypthm2} for the null Riemann components, the inductive assumptions for the connection coefficients up to $N\!-\!1$ derivatives and the result of previous Theorems \ref{T3.1} and \ref{muetaest}, assuming that on the ``last slice" $\Lie_O^{N-2}\tilde{\omb}$ satisfy the following estimate with ${\underline C}_2> \tilde{\underline C}_2>\tilde{\underline C}_2^{(*)}$,
\bea
&&\ML\ML\ML\big|r^{3-\frac{2}{p}}|\la|\Lie_O^{N-2}\tilde{\omb}\big|_{p,S}(\la,\nu_*)\leq \tilde{\underline C}_2^{(*)}e^{(N-2)\de}e^{(N-2)\underline{\Ga}(\la)}\frac{N!}{N^\a\ro_{0,1}^N} \ ,
\eea
then the following inequality holds
\bea
\ML\big|r^{3-\frac{2}{p}}|\la|\Lie_O^{N-2}{\tilde{\omb}}\big|_{p,S}(\la,\nu)\leq \tilde{\underline C}_2e^{(N-2)\de}e^{(N-2)\underline{\Ga}(\la)}\frac{N!}{N^\a}\frac{1}{\ro_{0,1}^{N}}\ ,
\eea
\bea
 \ML\ \ \ \big|r^{2-\frac{2}{p}}|\la|\Lie_O^N\omb\big|_{p,S}(\la,\nu)\leq {\underline C}_2e^{(N-2)\de}e^{(N-2)\underline{\Ga}(\la)}\frac{N!}{N^\a}\frac{1}{\ro_{0,1}^{N}}\ .
\eea
\end{theorem}

\subsubsection{The control of the $\Lie_O^N$ derivatives of the underlined connection coefficients}
To complete the proof of Theorem \ref{L6.1} we have to obtain the norm estimates for the $\Lie_O^N$ angular derivatives of the underlined connection coefficients, $\chib,\etab, \om$\footnote{To follow previous notation, see \cite{Kl-Ni:book} we denoted with $\om$ a connection coefficient whose transport equation is along the incoming cones.} whose transport equations are along the incoming cones. The procedure is analogous to the one we discussed for $\chi,\eta,\omb$ and we look in detail only to the estimates for $\Lie_O^{N-1}\Ubs$ to point out the main differences, the most important being that in this case the estimates are done ``from below", namely the integration starts form the initial data on $C_0$. As in the $\nabb^N\chi$ case, we start proving the norm estimates for $\nabb^{N-1}\Ubs$ where
$\Ubs=\nabb\oom^{-1}\tr\chib+\oom^{-1}\tr\chib\ \!\etab$\ .

\NI To obtain the transport equation for $\Lie_O^{N-1}\Ubs$ 
we apply $\Lie_O^{N-1}$ to the structure equation\footnote{Observe that equation \ref{Ubseq} has to be considered a tensorial equation. As we are interested to the transport equation for its $L_{p,S}$ norm, we do not need to express ${e_3}$ as $\frac{1}{\oom}\!\left(\frac{\pr}{\pr\la}+X^a\frac{\pr}{\pr\om^a}\right)$, see (I); in fact in this case we can choose on the $S(\la,\nu)$ surfaces an orthonormal frame Fermi transported along the incoming cones. In this way the transport equations along the incoming cones for these norms are exactly the same as those previously obtained for the not underlined coefficients just substituting $\nu$ with $\la$ and interchanging the underlined with the not underlined connection coefficients and Riemann components.}
\bea
\frac{\Dbb}{\partial\la}\Ubs+\frac{3}{2}\oom\tr\chib\Ubs=-\nabb|\chibh|^2-\etab|\chibh|^2-\oom\chibh\c{\Ubs}+\tr\chib(\chibh\c\eta)-\tr\chib\ \!\bb\ .
\eql{Ubseq}
\eea
The result we obtain is, therefore, provided in the following lemma, whose proof is identical to the proof of Lemma \ref{eqforU} with the obvious substitutions,
\begin{Le}\label{eqforUb}
Denoting $\Ubs=\nabb\Ub+\Ub\etab$ where $\Ub=\oom^{-1}\tr\chib$, $\Lie_O^{N-1}\Ubs$ satisfies the following transport equation along the incoming cones,

\bea\eql{15.46bis}
&&\ML\ML\frac{\Dbb}{\partial\la}(\Lie_O^{N-1}\Ubs)+\frac{3}{2}\overline{\oom\tr\chib}(\lie_O^{N-1}\Ubs)=-\oom\chibh\c(\lie_O^{N-1}\Us)- 2\chibh\lie_O^{N-1}(\nabb\chibh)-\tr\chib\Lie_O^{N-1}\bb+\nn\\
&&\ML\ML+\frac 3 2(\overline{\oom\tr\chib}-\oom\tr\chib)(\lie_O^{N-1}\Ubs)+\{{\it( \underline{good})_1}\}
\eea

\NI With

\bea\eql{15.47bis}
&&\ML\ML\{{\it (\underline{good})_1}\}=2\sum_{k=1}^{N-1}\cbin{N-1}{k}\Lie_O^{N-1-k}(\nabb\chibh)\Lie_O^k\chibh\nn\\ 
&&\ML\ML-\frac{3}{2}\sum_{q=1}^{N-1}\cbin{N-1}{q}(\lie_O^{q}\oom^2 \Ub)(\lie_O^{N-1-q}\Ubs)+\sum_{k=0}^{N-1}\cbin{N-1}{k}\Lie_O^{N-1-k}(|\chibh|^2)\Lie_O^k\etab\nn\\
&&\ML\ML+\sum_{q=1}^{N-1}\cbin{N-1}{q}(\Lie_O^{q}(\oom\chibh)(\Lie_O^{N-1-q}\Ubs)+\sum_{q=0}^{N-2}\cbin{N-1}{q}(\Lie_O^{N-2-q}\underline{E})(\Lie_O^{q}\Ubs)\nn \\
&&\ML\ML-\sum_{q=0}^{N-1}\cbin{N-1}{q}(\Lie_O^{q}(\oom \Ub)(\Lie_O^{N-1-q}\chibh\eta)+\sum_{q=1}^{N-1}\cbin{N-1}{q}(\Lie_O^{N-2-q}\bb)(\Lie_O^{q}\tr\chib)\nn 
\eea

\NI and with $\underline{E}$ defined in equation \ref{unde}.

\NI Moreover  the following inequality holds:
\bea
\frac{\partial}{\partial\la}|r^{(3-\frac{2}{p})}\Lie_O^{N-1}\Ubs|_{p,S}\leq  |r^{(3-\frac{2}{p})}\Lbs|_{p,S}\ \ \ ,\eql{diffest2b}
\eea

\bea\eql{15.52bis}
|\Lbs|\!&=&\!\!\left[\big|\oom\chibh\c(\Lie_O^{N-1}\Ubs)\big|+\big|2\chibh\c(\Lie_O^{N-1}\nabb\chibh)\big|+|\tr\chib|\big|\Lie_O^{N-1}\bb\big|
+\big|\big\{(\underline{good})_1\big\}\big|\right]\nn\\
&+&\!\!\left[\frac{(3-\frac 2 p)}{2}|{\oom}\tr\chib-\overline{{\oom}\tr\chib}||\Lie_O^{N-1}\Ubs|\right]\ \ \ \ .\eql{defLabis}
\eea

\end{Le}
\NI Exactly as in the estimate of $\Lie_O^{N-1}\Us$, to obtain from the differential inequality \ref {diffest2b} an estimate for $\Lie_O^{N-1}\Ubs$ we need first to control  the term $\Lie_O\nabb\hat{\chib}$ present in $\Lbs$, which has the same $N$ order of derivatives. As before we use repeatedly the equations
\bea
&&{\divv}{\chibh}=\frac{1}{2}\nabb{\tr\chib}-\bb+\zeta\!\c\hat{\chib}-\frac{1}{2}\zeta{\tr\chib}\equiv {\underline F}^{(1)}\eql{fbdef1}
\eea
and proceeding exactly as in Lemma \ref{estdNchih} we prove the following lemma,
\begin{Le}\label{estdNchibh}
From equation \ref{fbdef1} we obtain the following estimates

\bea
&&\ML\ML |\Lie_O^{N-1}\nabb\chibh|_{p,S}\leq c \left(|\oom|_{\infty,S}|\Lie_O^{N-1}\Ubs|_{p,S}+|\Lie_O^{N-1}\bb|_{p,S}+|{\{\it (\underline{good})_2}\}|_{p,S}+|{\{\it (\underline{good})_4}\}|_{p,S}\right)\nn\\
\eea

with, $c$ a suitable constant and 
\bea
&&{\{\it (\underline{good})_2}\}=[\divv,\Lie_O^{N-1}]\chibh\nn\\
&&=\sum_{q=0}^{N-2}\left(\cbin{N-1}{q}(\Lie_O^{N-2-q}\overline{E})(\Lie_O^{q}\Ubs)+\cbin{N-1}{q}(\Lie_O^{N-2-q}\overline{\cal H})(\Lie_O^{q}\nabb\Ubs)\right)\nn\\
\eea

\bea
&&\ML\ML|{\{\it (\underline{good})_4}\}|_{p,S}=\left(\sum_{k=1}^{N-1}\cbin{N-1}{k}|\Lie_O^k\oom|_{\infty}|\Lie_O^{N-1-k}\chibh|_{p,S}+|{\{\it (\underline{good})_3}\}|_{p,S}\right)\nn\\
\eea

\bea
&&{\{\it (\underline{good})_3}\}=-\Lie_O^{N-1}(\zeta\!\c\hat{\chib})+[\divv,\Lie_O^{N-1}]\chih\nn\\
&&=\Lie_O^{N-1}(\zeta\!\c\hat{\chib})+\sum_{q=0}^{N-2}\left(\cbin{N-1}{q}(\Lie_O^{N-2-q}\overline{E})(\Lie_O^{q}\Ubs)+\cbin{N-1}{q}(\Lie_O^{N-2-q}\overline{\cal H})(\Lie_O^{q}\nabb\Ubs)\right)\nn\\
\eea

\end{Le}

\NI Using the estimate for $\Lie_O^{N-1}\bb$ proved  in Theorem \ref{hypthm2}, 
we prove the analogous of Theorem \ref{T3.1}. 
\begin{theorem}\label{T3.7}
Assuming the following norm estimate for the null Riemann component $\bb$,
\bea
|r^{2-\frac{2}{p}}\Lie_O^{N-1}\bb|_{p,S}(\la,\nu)\leq {C^{(1)}}
e^{(N-2)\de}e^{(N-2)\underline{\Ga}(\la)}\frac{N!}{N^\a}\frac{1}{\ro_{0,1}^N}\ ,\eql{betassa}
\eea
assuming the norm estimates \ref{123} for the initial data connection coefficients with $J\leq N$ and the norm estimates \ref{123a} for the connection coefficients with $J<N$, then the following inequality holds on ${\cal K}$,
\bea
\big|r^{(N+2)-\frac{2}{p}}\Lie_O^{N-1}\Ubs\big|_{p,S}(\la,\nu)\leq \Cb_0e^{(N-2)\de}e^{(N-2)\underline{\Ga}(\la)}\frac{N!}{N^\a}\frac{1}{\ro_{0,1}^{N}}\ .
\eea
\end{theorem}

\NI {\bf Proof:} The proof is similar to the one of theorem \ref{T3.1} and we do not repeat it here.

\NI With this theorem we can consider proved the estimates \ref{323}.
\medskip

\NI In order to prove the estimates \ref{324} we state the following

\begin{cor}\label{cor92}
Assuming the estimates \ref{323} the following estimates hold on ${\cal K}$, with $C$ a suitable constant.
\bea
&&|r^{2-\frac{2}{p}}\Lie_O^N\{{\cal O, \underline{O}}\}|_{p,S}\leq Ce^{(N-2)\de}e^{(N-2)\underline{\Ga}(\la)}\frac{N!}{N^\a}\frac{1}{\ro_{0,1}^{N}}\nn\\
\eea
\end{cor}
{\bf Proof:} See Appendix  \ref{AS.3}.
\medskip


\NI In order to obtain the $\Lie_O^N$ estimates for the connection coefficients, we have to pass from the $\nabb^N$ estimates on the initial data to the $\Lie_O^N$ ones, to do this we have to perform estimates of some auxiliary quantities, with the same structure of the connection coefficients, see equations \ref{equo} and \ref{1687}. Viceversa, once we have the $\Lie_O^N$ estimates for the connection coefficient in all the region, we have to recover the $\nabb^N$, to do this we have to exploit the estimates for some other auxiliary quantities.
We prove these estimates in the following section.

\subsection{The $\gggg(e_A\nabb_{\c} e_B)$ and $\nabb_A O$ coefficients}\label{S.s8.4}
In the proof of our result it follows that many different connection coefficients play some role, even if they do not enter in the Einstein equations. Their estimates are needed when we have to control the angular derivatives of the Riemann components,  first of all, when, starting from the control of the $\nabb$ we pass to the $\nabb_O$ derivatives of the initial data, see equation \ref{equo}, we need the estimates of $\nabb_A O$, and of $\gggg(e_A,\nabb e_C)$  in order to do this we recall that the rotation vectors satisfy the equations, see \cite{Kl-Ni:book}, section 4.6.

\bea
\frac{\Dbb}{\pr\nu}\ ^{(i)}O_b=\oom \chi_{bc}\ ^{(i)}O_c.
\eea

\NI We can observe that these equations can be treated as the structure equations for the connection coefficients, hence with an analogous procedure to the one discussed previously we can obtain the estimates for $\nabb^N \ ^{(i)}O$, on the initial data and estimates for $\Lie_O \ ^{(i)}O$ in the interior region. We do not repeat it here the proofs, finally we obtain the following theorems. 
\footnote{Recall that the sup norm of the $J$ derivative is estimated in term of the $|\c|_{p,S}$ norm of the $J$ and of the $J+1$ derivative. In fact the second line is\newline
$|r^{\phi({\cal O})}\nabb^{J}{\cal O}|_{\infty,S}\leq C\left[\frac{J!}{J^\a}\frac{e^{(J-2)\de_0}e^{(J-2)\underline{\Ga}_0(\la)}}{\ro_0^{J}}+\frac{(J+1)!}{(J+1)^\a}\frac{e^{(J-1)\de_0}e^{(J-1)\underline{\Ga}_0(\la)}}{\ro_0^{(J+1)}}\right]\leq\newline cC\frac{(J+1)!}{(J+1)^\a}\frac{e^{(J-1)\de_0}e^{(J-1)\underline{\Ga}_0(\la)}}{\ro_0^{(J+1)}}$
 and we neglect $c\geq 1+\frac{\ro_0}{(J+1)e^{\de_0+\underline{\Ga}_0(\la)}}$.}
\footnote{The estimate in the second line follows as (assuming $r=1$) although the $O$ generators are at the metric level, nevertheless in the estimates of the angular derivatives of $\ga_{ab}$ and of $X^a$ there is a loss of derivatives, see Section \ref{gamma}, which implies the same estimates as for the connection coefficients.}



\begin{theorem}\label{angulconcoef2}
Assuming the norm estimates for the null Riemann component of Theorem \ref{hypthm2},
assuming the norm estimates \ref{123} for the initial data connection coefficients, the norm estimates \ref{123a} for the connection coefficients with $J<N$ and the following estimate for  $ \ ^{(i)}O_b$, with $J<N$, $c$ a suitable constant
\bea
|r^{(J+1-\frac{2}{p})}\Lie_O^{J} \ ^{(i)}O|_{p,S}\leq c\ \!\!\left(\frac{J!}{J^{\a}}\frac{e^{(J-2)(\de+\underline{\Ga}(\la))}}{\ro_{0,1}^{J}}\right)\nn ,
\eea
then the following inequalities hold on ${\cal K}$,
\bea\label{978}
\big|r^{N+1-\frac{2}{p}}\Lie_O^N \ ^{(i)}O\big|_{p,S}(\la,\nu)\leq c \!\!\left(\frac{N!}{N^\a}\frac{e^{(N-2)\de}e^{(N-2)\underline{\Ga}(\la)}}{\ro_{0,1}^N}\right)\ .
\eea
\bea
\big|r^{N+1-\frac{2}{p}}\Lie_O^{N-1} \ ^{(i)}O\big|_{\infty,S}(\la,\nu)\leq c\!\!\left(\frac{N!}{N^\a}\frac{e^{(N-2)\de}e^{(N-2)\underline{\Ga}(\la)}}{\ro_{0,1}^N}\right)\ .
\eea

\end{theorem}
\begin{cor}
Under the assumptions of Theorem \ref{angulconcoef2} the following estimates hold on ${\cal K}$,
\bea\label{cor910}
\big|r^{N+1-\frac{2}{p}}\nabb_O^N \ ^{(i)}O\big|_{p,S}(\la,\nu)\leq c \!\!\left(\frac{N!}{N^\a}\frac{e^{(N-2)\de}e^{(N-2)\underline{\Ga}(\la)}}{\ro_{0,1}^N}\right)\ .
\eea
\bea
\big|r^{N+1-\frac{2}{p}}\nabb_O^{N-1} \ ^{(i)}O\big|_{\infty,S}(\la,\nu)\leq c\!\!\left(\frac{N!}{N^\a}\frac{e^{(N-2)\de}e^{(N-2)\underline{\Ga}(\la)}}{\ro_{0,1}^N}\right)\ .
\eea
The proof is a repetition of lemma \ref{L4.2}.
\end{cor}

\NI Conversely, once we have the estimates for the $\Lie_O^N$ derivatives of the connection coefficients in all the region, in order to recover the $\nabb^N$ ones, we have to estimates the quantities $\nabb^J e_A$, $J<N$ and hence  $\gggg(e_C,\nabb^J e_A)$, see equation  \ref{nabbea}, from the $\Lie_O$ estimates. Let us proceed step by step: First let us prove the estimates for $\gggg(e_C,\Lie_O^N e_A)$. 
Assuming $e_A$  Fermi transported, $\ddb_4e_A=0$, it follows
\bea
&&\ML\ML\frac{\pr}{\pr\nu}\gggg(e_A,\nabb_Be_C)=\gggg(e_A,\ddb_{\nu}\nabb_Be_C)=\gggg(e_A,[\ddb_{\nu},\nabb_B]e_C)\nn\\
&&\ML\ML=e_B^{\mu}\gggg(e_A,[\ddb_{\nu},\nabb_{\mu}]e_C)
=e_B^{\mu}\gggg(e_A,(-\oom\chi_{\mu\c}\c\nabb e_C)=-\oom\chi(e_B,e_D)\gggg(e_A,\nabb_D e_C)\nn\\
&&\ML\ML=-\frac{\oom\tr\chi}{2}\gggg(e_A,\nabb_B e_C)-\oom\chih_{B\c}\c\gggg(e_A,\nabb e_C)\ .
\eea
Therefore we have the following transport equation
\bea
\frac{\Dbb}{\pr\nu}\gggg(e_A,\nabb e_C)+\frac{\oom\tr\chi}{2}\gggg(e_A,\nabb e_C)
=-\oom\chih\c\gggg(e_A,\nabb e_C)\ \ .\  \ 
\eea
which we rewrite, recalling that $\gggg(e_A,\nabb_{\c} e_C)$ is an $S$-tangent one form, we denote simply by $S_{AC}(\c)\equiv\gggg(e_A,\nabb_{\c} e_C)$ or simpler by $S=S(\c)$,
\bea
\frac{\Dbb}{\pr\nu}S+\frac{\oom\tr\chi}{2}S+\oom\ \!\chih\c S=0\ .
\eea
To obtain the estimate for $\Lie_O^N S$ we proceed as in the cases of the standard connection coefficient, discussed previously, and obtain the following results 
\begin{theorem}\label{angulconcoef}
Assuming the norm estimates for the null Riemann component of Theorem \ref{hypthm2},
assuming the norm estimates \ref{123} for the initial data connection coefficients, the norm estimates \ref{123a} for the connection coefficients with $J<N$ and the following estimate for $\Lie_O^JS$ , with $J<N$, $c$ a suitable constant,
\bea
|r^{(1-\frac{2}{p})}\Lie_O^{J}\gggg(e_A,\nabb e_C)|_{p,S}\leq c\!\!\left(\frac{J!}{J^{\a}}\frac{e^{(J-2)(\de+\underline{\Ga}(\la))}}{\ro_{0,1}^{J}}\right)\ ,
\eea
assuming that on the ``last slice" $\Lie_O^{N}S$ satisfy the following estimate, 
\bea
\big|r^{1-\frac{2}{p}}\Lie_O^{N}S\big|_{p,S}(\la,\nu_*)\leq c\!\!\left(\frac{N!}{N^\a}\frac{e^{(N-2)\de}e^{(N-2)\underline{\Ga}(\la)}}{\ro_{0,1}^N}\right) \ ,
\eea
then the following inequality holds on ${\cal K}$,
\bea\label{985}
\big|r^{1-\frac{2}{p}}\Lie_O^{N}S\big|_{p,S}(\la,\nu)\leq c\!\!\left(\frac{N!}{N^\a}\frac{e^{(N-2)\de}e^{(N-2)\underline{\Ga}(\la)}}{\ro_{0,1}^N}\right)\ .
\eea
\end{theorem}
\begin{cor}\label{cor9.9}
Under the assumptions of Theorem \ref{angulconcoef} the following estimates hold on ${\cal K}$,
\bea\label{lio}
&&|r^{1-\frac{2}{p}}(\Lie_O^{N}e_A)^C|_{p,S}=|r^{(1-\frac{2}{p})}\gggg(e_C,\Lie_O^{N}e_A)|_{p,S}\leq c \!\left(\frac{N!}{N^{\a}}\frac{e^{(N-2)(\de+\underline{\Ga}(\la))}}{\ro_{0,1}^{N}}\right)\ .\nn\\
&&r^{1-\frac{2}{p}}|(\Lie_O^{N-1}e_A)^C|_{\infty,S}=|r^{(1-\frac{2}{p})}\gggg(e_C,\Lie_O^{N}e_A)|_{p,S}\leq c \!\left(\frac{N!}{N^{\a}}\frac{e^{(N-2)(\de+\underline{\Ga}(\la))}}{\ro_{0,1}^{N}}\right)\nn\\ .
\eea
The proof is in subsection \ref{sec158}
\end{cor}
\NI As second step we estimates, in the same way as done in lemma \ref{L4.2}, inverting the estimates \ref{1685}  the quantities $r^{1-\frac{2}{p}}(\nabb_O^{N-1}e_A)^C$, namely we have, with $c$ a suitable constant.
\begin{cor}\label{corlio2}
\bea\label{lio2}
&&|r^{1-\frac{2}{p}}(\nabb_O^{N}e_A)^C|_{p,S}=|r^{(1-\frac{2}{p})}\gggg(e_C,\Lie_O^{N}e_A)|_{p,S}\leq c \!\left(\frac{N!}{N^{\a}}\frac{e^{(N-2)(\de+\underline{\Ga}(\la))}}{\ro_{0,1}^{N}}\right)\ \nn\\
&&|r^{1-\frac{2}{p}}(\nabb_O^{N-1}e_A)^C|_{\infty,S}=|r^{(1-\frac{2}{p})}\gggg(e_C,\Lie_O^{N}e_A)|_{p,S}\leq c \!\left(\frac{N!}{N^{\a}}\frac{e^{(N-2)(\de+\underline{\Ga}(\la))}}{\ro_{0,1}^{N}}\right).\nn\\
\eea
\end{cor}
\NI As third and last step we recover the estimates for $\gggg(e_E,\nabb^{N}e_A)$ from estimates for $\gggg(e_C,\nabb_O^Ne_A)$. To do this we have substantially to invert lemma \ref{L4.1}, hence we introduce the quantities $h_{(i),A}$ and  $h_{(i),B}$, such that
\bea
\sum_{1=1}^{3}h_{(i),A}\ ^{{i}}O=e_{A}\ \ ,\ \ \ \ \ \sum_{1=1}^{3}h_{(i),B}\ ^{{i}}O=e_{B}
\eea
Exploiting the equations for $S$ and the Leibnitz rule we have the equation
\bea
\nabb_{(\c)} h_{(i)},A=\gggg{\nabb_{(\c)}, e_A) -\gggg(\nabb_{(\c)}\ ^{(i)}O)=S_{(\c)},A}-\gggg(\nabb_{(\c)}\ ^{{i}}O)
\eea
end hence we have the transport equation for $  h_{(i)}$
\bea
\frac{\pr}{\pr\nu} h_{(i)}=\frac{\oom\tr\chi}{2}S+\oom\ \!\chih\c S-\oom \chi\ ^{(i)}O
\eea
Hence, exploiting the previous estimates for $\ ^{(i)}O$ and $S$ , see \ref{985} and \ref{978}, we can estimate $h$ on $C_0\cup \Cb_0$ and consequently on all the region. We have the 

\begin{theorem}\label{angh}
Assuming the norm estimates for the null Riemann component of Theorem \ref{hypthm2},
assuming the norm estimates \ref{123} for the initial data connection coefficients, the norm estimates \ref{123a} for the connection coefficients with $J<N$ and the following estimate for  $h$, with $J<N$, $c$ a suitable constant,
\bea
|r^{(1-\frac{2}{p})}\Lie_O^{J} h|_{p,S}\leq c\ \!\!\left(\frac{J!}{J^{\a}}\frac{e^{(J-2)(\de+\underline{\Ga}(\la))}}{\ro_{0,1}^{J}}\right)\nn ,
\eea
then the following inequality holds on ${\cal K}$,
\bea
\big|r^{1-\frac{2}{p}}\Lie_O^N h\big|_{p,S}(\la,\nu)\leq c \!\!\left(\frac{N!}{N^\a}\frac{e^{(N-2)\de}e^{(N-2)\underline{\Ga}(\la)}}{\ro_{0,1}^N}\right)\ .
\eea
\bea
\big|r^{1-\frac{2}{p}}\Lie_O^{N-1} h\big|_{\infty,S}(\la,\nu)\leq c\!\!\left(\frac{N!}{N^\a}\frac{e^{(N-2)\de}e^{(N-2)\underline{\Ga}(\la)}}{\ro_{0,1}^N}\right)\ .
\eea

\end{theorem}

\begin{cor}\label{cor9.9bis}
Under the assumptions of Theorem \ref{angulconcoef2} the following estimates hold on ${\cal K}$, with $c$ a suitable constant,
\bea
\big|r^{1-\frac{2}{p}}\nabb_O^N h\big|_{p,S}(\la,\nu)\leq c \!\!\left(\frac{N!}{N^\a}\frac{e^{(N-2)\de}e^{(N-2)\underline{\Ga}(\la)}}{\ro_{0,1}^N}\right)\ .
\eea
\bea\label{9910}
\big|r^{1-\frac{2}{p}}\nabb_O^{N-1} h\big|_{\infty,S}(\la,\nu)\leq c\!\!\left(\frac{N!}{N^\a}\frac{e^{(N-2)\de}e^{(N-2)\underline{\Ga}(\la)}}{\ro_{0,1}^N}\right)\ .
\eea

\NI The proof is also in this case a  repetition of lemma \ref{L4.2} and we do not report it here.
\end{cor}

\NI Now in order to estimate $|\nabb_B^N e_A|_{\infty,S}$  we notice that $\nabb_B e_A=\nabb_{h_{(i),B}\ ^{(i)}O}e_A$, hence, we can exploit the estimates of corollary \ref{cor9.9bis} and lemma \ref{L4.1} with ${\nabb_O}$ instead of $\nabb_B$ and $h_{(i),B}$ instead of $\ ^{(i)}O$. to prove the following 
 
\begin{theorem}\label{angh2}
Assuming the norm estimates for the null Riemann component of Theorem \ref{hypthm2},
assuming the norm estimates \ref{123} for the initial data connection coefficients, the norm estimates \ref{123a} for the connection coefficients with $J<N$ and the  estimate of corollary \ref{cor9.9bis}  for $h$, the following estimates hold on ${\cal K}$,

\bea\label{9911}
\big|r^{1-\frac{2}{p}}\nabb_{(.)}^{N-1} e_A\big|_{\infty,S}(\la,\nu)\leq c\!\!\left(\frac{N!}{N^\a}\frac{e^{(N-2)\de}e^{(N-2)\underline{\Ga}(\la)}}{\ro_{0,3}^N}\right)\ .
\eea
With $(.)=e_{(A,B)}$ and $\ro_{0,3}<\ro_{0,1}$, $c$ a suitable constant.
\end{theorem}
\NI In the same way we can estimate the $\nabb_A$ derivatives of the null Riemann components, and of the connection coefficients:
\begin{cor}\label{angcorr}
\NI Under the assumptions of theorem \ref{angh2}, the following estimates hold on ${\cal K}$,\footnote{Recall that the sup norm of the $J$ derivative is estimated in term of the $|\c|_{p,S}$ norm of the $J$ and of the $J+1$ derivative. In fact the second line is
$$|r^{\phi({\cal O})}\nabb^{J}{\cal O}|_{\infty,S}\leq C\left[\frac{J!}{J^\a}\frac{e^{(J-2)\de_0}e^{(J-2)\underline{\Ga}_0(\la)}}{\ro_{0,3}^{J}}+
\frac{(J+1)!}{(J+1)^\a}\frac{e^{(J-1)\de_0}e^{(J-1)\underline{\Ga}_0(\la)}}{\ro_{0,3}^{(J+1)}}\right]\leq C\frac{(J+1)!}{(J+1)^\a}\frac{e^{(J-1)\de_0}e^{(J-1)\underline{\Ga}_0(\la)}}{\ro_{0,3}^{(J+1)}}$$ and we neglect $c\geq 1+\frac{\ro_{0,3}}{(J+1)e^{\de_0+\underline{\Ga}_0(\la)}}$.}

\bea\label{99111}
&&\big|r^{{\phi(\cal O)}}\nabb_{(.)}^{N-1} {\cal O}\big|_{\infty,S}(\la,\nu)\leq c\!\!\left(\frac{N!}{N^\a}\frac{e^{(N-2)\de}e^{(N-2)\underline{\Ga}(\la)}}{\ro_{0,3}^N}\right) \nn\\
&&\big|r^{{\phi(R)-\frac{2}{p}}}\nabb_{(.)}^{N-1} R\big|_{p,S}(\la,\nu)\leq c\!\!\left(\frac{N!}{N^\a}\frac{e^{(N-2)\de}e^{(N-2)\underline{\Ga}(\la)}}{\ro_{0,3}^N}\right)
\eea
With $(.)=e_{(A,B)}$ and $\ro_{0,3}<\ro_{0,1}$ and $c$ a suitable constant.
\end{cor}

\subsection{Proof of theorem \ref{L2.1new}}

\NI In order to prove theorem \ref{L2.1new} we have to retrace all the steps of section \ref {165}, in order to do this we have to obtain the results of lemma \ref{new1}
but now for the connection coefficients. This can be obtained applying lemma \ref{L4.2} to the estimates \ref{324}, see corollary \ref{cor92}. Hence we can state the 

\begin{cor}
\NI Assuming the estimates \ref{324}, and the estimates of corollary  \ref{cor92}.
\bea
\big|r^{\phi({\cal O})-\frac{2}{p}}\Lie_O^{J} \{{\cal O, \underline{O}}\}\big|_{p,S}(\la,\nu)\leq c\!\!\left(\frac{J!}{J^\a}\frac{e^{(J-2)\de}e^{(J-2)\underline{\Ga}(\la)}}{\ro_{0,1}^J}\right)\ .
\eea
\NI for $J<N$, the following estimates hold on ${\cal K}$, with $c$ a suitable constant,
\bea\label{9913}
\big|r^{\phi({\cal O})-\frac{2}{p}}\nabb_O^{N}  \{{\cal O, \underline{O}}\}\big|_{p,S}(\la,\nu)\leq c\!\!\left(\frac{N!}{N^\a}\frac{e^{(N-2)\de}e^{(N-2)\underline{\Ga}(\la)}}{\ro_{0,1}^N}\right)\ .
\eea
\NI The proof follows, as in corollary \ref{cor92}, from Theorem \ref{L4.2}.
\end{cor}
\NI Once we have stated the estimates \ref{9913} we can perform the steps of lemma \ref{fromLieOtoNabb} and, exploiting inequalities \ref{9911}, obtain the estimates:
\bea\label{9914}
\big|r^{N+\phi({\cal O})-\frac{2}{p}}\nabb^{N}  \{{\cal O, \underline{O}}\}\big|_{p,S}(\la,\nu)\leq c\!\!\left(\frac{N!}{N^\a}\frac{e^{(N-2)\de}e^{(N-2)\underline{\Ga}(\la)}}{\ro^N}\right)\ .
\eea
\NI Hence we can consider proved the theorem \ref{L2.1new}.
\subsection {Proof of corollary \ref{333}} 

{\NI In order to estimate $|r^{(\c)}\Lie_O^{h}{\cal O}|_\infty(\la,\nu)$ with $h \leq N-1$ and    $|r^{(\c)+1}\Lie_O^{h}\nabb{\cal O}|_\infty(\la,\nu)$ with $h \leq N-2$ we  recall propositions 3.3.3 and 3.3.4 of \cite{C-K:book} for a a 1-form and symmetric 2-tensor, let us recall them. 

\NI {\bf Proposition}

\NI Let $U_{\a}$ be 1-form, then  it does exists a constant $C$ such that

\bea\label{nabb4}
|\nabb U|^2\leq C \sum_i|\Lie_{\ ^{(i)}O}U|^2
\eea

\NI Let $U_{\a\b}$ be a symmetric two tensor, then  it does exists a constant $C$ such that

\bea\label{nabb5}
|\nabb U|^2\leq C \sum_i|\Lie_{\ ^{(i)}O}U|^2
\eea





\NI We will proceed by induction.
\NI Moreover from the Sobolev lemma
\bea\eql{2345}
&&\ML|r^{(\c)}\Lie_O^{h}{\cal O}|_\infty(\la,\nu) \leq C\left[|r^{(\c)-\frac{2}{4}}\Lie_O^{h}{\cal O}|_{4,S}(\la,\nu)
+|r^{((\c)+1-\frac{2}{4})}\nabb\Lie_O^{h}{\cal O}|_{4,S}(\la,\nu)\right]\nn\\
\eea


\NI Hence, to estimate $|r^{(\c)}\Lie_O^{h}{\cal O}|_\infty(\la,\nu)$, we have to estimate $|r^{(\c)-\frac{2}{4}}\Lie_O^{h}{\cal O}|_{4,S}$ and $|r^{((\c)+1-\frac{2}{4})}\nabb\Lie_O^{h}{\cal O}|_{4,S}$
The first term can be estimated inductively while for the second, it holds

\bea\label{15277}
&&\ML\ML|\nabb\Lie_O^h{\cal O}|_{4,S}\leq|\Lie^h_O\nabb\chih-[\Lie^h_O,\nabb]{\cal O}|_{4,S}\nn\\
&&\ML\ML\leq|\Lie^h_O\nabb{\cal O}|_{4,S}+|[\Lie_O^h,\nabb]{\cal O}|_{4,S}
\eea

\NI and both terms can be estimated inductively. Hence we obtain 

\bea
|r^{(\c)+1}\Lie_O^{h-1}{\cal O}|_\infty(\la,\nu)\leq C_0e^{(h-2)\de}e^{(h-2)\underline{\Ga}(\la)}\frac{h!}{h^\a}\frac{1}{\ro_{0,1}^{h}}.\nn\\
\eea

\smallskip

\NI Let us now estimate $|r^{(\c)}\Lie_O^{h}\nabb{\cal O}|_\infty$:

\bea
&&|r^{(\c)+1}\Lie_O^{h}\nabb{\cal O}|_\infty\leq|r^{(\c)+1}\nabb\Lie_O^{h}{\cal O}|_{\infty,S}+|r^{(\c)+1}[\nabb,\Lie_O^h]{\cal O}|_{\infty,S}\nn\\
&&\leq C|r^{(\c)+1}\Lie_O^{h+1}{\cal O}|_{\infty,S}+|r^{(\c)}[\nabb,\Lie_O^h]{\cal O}|_{\infty,S}
\eea

\NI Where in the second line we used the inequalities \ref{nabb4} or \ref{nabb5}.  Now it is sufficient to observe that the first term can be estimated in the same way of $|\Lie_O^k{\cal O}|_{\infty, S}$ while the second term can be estimated inductively as  the $[\nabb, \Lie_O^h]$ commutator involve only $\Lie_O$ derivatives up to order $h-1$. }

\NI Hence, finally, we obtain the estimate
\bea
|r^{(\c)}\Lie_O^{k-2}\nabb{\cal O}|_\infty(\la,\nu)\leq C e^{(k-2)\de}e^{(k-2)\underline{\Ga}(\la)}\frac{k!}{k^\a}\frac{1}{\ro_{0,1}^{k}}.\nn
\eea

\subsection{The definition of the canonical foliation}\label{SScanfol}
We shortly motivated in subsection \ref{S.s3.4} 
the introduction  of the ``double null cone canonical foliation" to prove our result; we add here a more detailed discussion to show why it is needed and the way it is built.

\NI From equation \ref{mutransp} in Lemma \ref{eqformu} it is immediate to see that the estimate for ${\tilde\mu}$ and subsequently the one for $\eta$ have a worst decay than expected, namely it follows that, integrating from $\Cb_0$ we could only prove,
\bea
\big|r^{1-\frac{2}{p}}\Lie_O^{N-1}{\tilde \mu}\big|_{p,S}(\la,\nu)\leq \tilde{C}_3e^{(N-2)\de}e^{(N-2)\underline{\Ga}(\la)}\frac{N!}{N^\a}\frac{1}{\ro_{0,1}^{N}}\ ,
\eea
\bea
\big|r^{1-\frac{2}{p}}\Lie_O^N\eta\big|_{p,S}(\la,\nu)\leq C_3e^{(N-2)\de}e^{(N-2)\underline{\Ga}(\la)}\frac{N!}{N^\a}\frac{1}{\ro_{0,1}^{N}}\ .
\eea
provided that in all the previous proofs no problem is encountered assuming for $J<N$,
\bea
&&\big|r^{1-\frac{2}{p}}\Lie_O^J\zeta\big|_{p,S}(\la,\nu)\leq C_3e^{(J-2)\de}e^{(J-2)\underline{\Ga}(\la)}\frac{J!}{J^\a}\frac{1}{\ro_{0,1}^{J}}\ \ .\ \ \ \ \ \ \ \ 
\eea
This result is not sufficient as it is not difficult to realize in the course of the proof of Theorem \ref{L2.1new} that, to obtain that the right estimates for the order $N$ angular derivatives of all the connection coefficients are all consistent, we are forced to assume the appropriate (and optimal) decays for all of them together with their derivatives up to order $N-1$. To obtain this decay we are, therefore, forced to integrate the transport equations, for the not underlined connection coefficients,  from above which requires to know in advance their estimates on $\Cb_*$. This can be done provided we choose an appropriate foliation on this last slice proceeding as in the following steps,  
\smallskip

\NI{\bf Step 1:} We assume the previous inductive assumptions for $J<N$ hold in the region ${\cal K}(\Lambda_a,\Pi_a)$, see (I), where we have the analytic solution. 
\smallskip

\NI{\bf Step 2:} We assume that there are initial data for $\oom$ on $\Cb_0$ such that on the last slice $\Cb_*\equiv\Cb(\nu_*=\Pi_a)$ the following equation is satisfied
\bea
{\tilde\mu}|_{\Cb_*}=\frac{1}{4}(\tr\chi\tr\chib-\overline{\tr\chi\tr\chib})\ .\eql{mustardef}
\eea
Observe that, recalling the definition of ${\tilde \mu}$, see \ref{mudef1}, this equation is equivalent to the following one on $\Cb_*$,
\bea
\lapp\log\oom=-\divv\ze+\frac{1}{2}(\chih{\chibh}-\overline{\chih{\chibh}})-(\ro-\overline{\ro})\ .\eql{lasteq1}
\eea
Let us denote $\oom_*\equiv\oom|_{C_*}$ the solution of \ref{lasteq1} .
\smallskip

\NI{\bf Step 3:} Provided we know the estimates for the $|\c|_{p,S}$ norms of ${\tilde\mu}|_{\Cb_*}=\frac{1}{4}(\tr\chi\tr\chib-\overline{\tr\chi\tr\chib})$ and for its angular derivatives $\Lie_O^J$ with $J\leq N-1$, we can integrate the equation for $\tilde\mu$ from the last slice going down which implies that we can gain an extra $r$ decay factor. This allows to conclude that the following estimates hold
\bea
&&\ML\big|r^{2-\frac{2}{p}}\Lie_O^{N-1}{\tilde \mu}\big|_{p,S}(\la,\nu)\leq \tilde{C}_3e^{(N-2)\de}e^{(N-2)\underline{\Ga}(\la)}\frac{N!}{N^\a}\frac{1}{\ro_{0,1}^{N}}\ ,\ \ \ \ \ \\
&&\ML\big|r^{2-\frac{2}{p}}\Lie_O^N\eta\big|_{p,S}(\la,\nu)\leq C_3\log Ne^{(N-2)\de}e^{(N-2)\underline{\Ga}(\la)}\frac{N!}{N^\a}\frac{1}{\ro_{0,1}^{N}}\ .\nn
\eea
\smallskip

\NI{\bf Step 4:} Integrating from above the equation for $\tilde{\omb}$, again knowing the estimates for the $|\c|_{p,S}$ norms of $\tilde{\omb}|_{\Cb_*}$ and its angular derivatives $\nabb^J$ with $J\leq N-2$, it follows that all the remaining estimates for the angular derivatives of the connection coefficients can be obtained, those for $\chi$ integrating 
from $\Cb_*$
 and those from the underlined quantities integrating from $C_0$. This result requires that the solution $\oom_*$ does exist. Once this has been proved the existence of the appropriate initial data on $\Cb_0$ follows observing that
\bea
\oom(\la,\nu_0)=\oom(\la,\nu_*)+2\int_{\nu_0}^{\nu_*}\!\!d\nu\ \!\oom^2\om\ 
\eea
and the integral is finite due to the asymptotic behaviour of $\om$.
\smallskip

\NI Summarizing, to prove Lemma \ref{L6.1} we need to achieve the following preliminary results:

\NI a) The proof that on the whole $\Cb_*$ the solution $\oom_*$ exists.

\NI b) The proof that we control on the whole $\Cb_*$ the $|\c|_{p,S}$ norms of $\tilde \mu$ and of $\tilde{\omb}$ and their angular derivatives with $J\leq N-1$ and $J\leq N-2$ respectively. Let us discuss how to obtain these results. 
\smallskip

\NI{\bf Step 5:} To prove that the solution $\oom_*$ exists one proceeds, basically, as in \cite{Kl-Ni:book} and \cite{Niclast}. We prove the existence of the solution $\oom_*$ on $\Cb_*$ in two steps: first we prove that a solution, $\oom$, exists in a small interval of $\Cb_*$, $\Cb_*([\la_0,\la_0+\de])$ and that in this interval $\Lie_O^J\eta$, $J\leq N$ satisfies the following bounds, with a given ${\tilde C}_0$, 
\bea
\big|r^{(J+2)-\frac{2}{p}}\Lie_O^J\eta\big|_{p,S}(\la,\nu)\leq {\tilde C}_0\log Je^{(J-2)\de}e^{(J-2)\underline{\Ga}(\la)}\frac{J!}{J^\a}\frac{1}{\ro_{0,1}^{J}}\ ;\eql{bootstcond1}
\eea
we consider then the solution $\oom$ in the largest possible interval of $\Cb_*$, $\Cb_*([\la_0,\la_1))$ where \ref{bootstcond1} is still valid. We prove that, in fact, in this interval the previous bounds hold with ${\tilde C}_0$ substitued by $C_3\leq\frac{{\tilde C}_0}{2}$. This implies that the estimate \ref{bootstcond1} and the solution holds on the whole $\Cb_*$. For it we have to prove, see next step, that we control $\chih$ and $\chibh$ so that the estimates \ref{bootstcond1} can be improved and moreover we have to prove that the estimate for $\ro-\overline{\ro}$ gives a small contribution which is possible to do as in this region\footnote{Bounded by $C(\la_0)$, $C(\la_0+\de)$, $C(\nu_0;[\la_0,\la_0+\de])$ and $C(\nu_*;[\la_0,\la_0+\de])$.  } the canonical foliation does exist and, therefore, we can express the Riemann components in terms of the ``small" initial data Riemann components.

\NI
The local existence has been proved in \cite{Niclast}, it is easy to extend it to a solution for all its $\nabb^J$ derivatives. To find the ``global solution again one has to proceed as in \cite{Kl-Ni:book} extending the result for all the derivatives $\nabb^J\eta$. Observe that equation \ref{lasteq1} can be rewritten as
\bea
\divv\eta=\frac{1}{2}(\chih{\chibh}-\overline{\chih{\chibh}})-(\ro-\overline{\ro})=\frac{1}{4}(\tr\chi\tr\chib-\overline{\tr\chi\tr\chib})+2(K-\overline{K})\ ,\eql{lasteq2}
\eea
therefore we have to control $\chi,\chib$ 
and their derivatives up to $N-1$ on $\Cb_*([\la_0,\la_1))$. This can be done looking at the transport equations for 
these quantities along the incoming cones and observing that, as in this interval $\oom_*$ exists, the transport equation for $\tr\chi$ which usually have a ``loss of derivatives" does not have it due to the choice $\oom=\oom_*$. This equation is in fact,
\bea
\ML\dddd_3\tr\chi\!+\!\tr\chib\tr\chi\!-\!2\omb\tr\chi\!-\!2\divv\eta\!-\!2|\eta|^2\!+\!2{K}=0\ 
\eea
and it becomes on $\Cb_*$, due to \ref{lasteq2},
\bea
&&\ML\ML\dddd_3\oom\tr\chi\!+\!\frac{1}{2}\oom\tr\chib\tr\chi\!-\!2\oom|\eta|^2\!+\oom\overline{\chih\chibh}-2\oom\overline{\ro}=0\ ;\eql{traspc*}
\eea
it is immediate to recognize that there is no loss of derivatives for the angular derivatives of $\tr\chi$, for $\nabb^J\tr\chi$, as $\nabb\overline{\ro}=0$. The situation is different for $\chih$ whose transport equation
\bea
\ML\dddd_3\chih\!+\!\frac{1}{2}\tr\chib\chih\!+\!\frac{1}{2}\tr\chi\chibh\!-\!2\omb\chih
\!-\!\nabb\hot\eta\!-\!\eta\hot\eta=0\ ,\nn
\eea
transforms into
\bea
\ML\dddd_3\oom\chih\!+\!\frac{1}{2}\oom\tr\chib\chih\!+\!\frac{1}{2}\oom\tr\chi\chibh\!-\!\oom\nabb\hot\eta\!-\!\oom\eta\hot\eta=0\ .\nn
\eea
This equation has still a loss of derivatives, but in fact is not required to control the last slice.\footnote{In the sense that we need only lower derivatives, assumed by induction, see the following remark.} In fact to control $\eta$ we need to control
${\tilde\mu}|_{\Cb_*}$,  \ref{mustardef}, which requires to control on $\Cb_*$ only $\tr\chi$ (and of course $\tr\chib$). 

\NI To control the angular derivatives of $\omb$ we have to control on $\Cb_*$, $\nabb^{N-2}\divv{\hat V}$ where ${\hat V}=({\tilde V}-\oom\bb)$ and $\tilde{V}=2\nabb\oom\omb$; to do it we proceed as follows: we start from equation \ref{lasteq1} which we rewrite as
\bea
\lapp\log\oom=\frac{1}{2}\left[\divv\etab+\frac{1}{2}(\chih{\chibh}-\overline{\chih{\chibh}})-(\ro-\overline{\ro})\right]\ .
\eea
Proceeding as in \cite{Kl-Ni:book} we differentiate this equation with respect to $\dd_3$ and obtain an elliptic equation, see \cite{Kl-Ni:book} eq. (7.4.21),
\bea
\lapp(\oom\omb)=\divv F_1+G_1-\overline{G}_1
\eea
and applying to it $\Lie_O^{N-2}\divv$ we obtain the elliptic equation for $\Lie_O^{N-2}\tilde{\omb}$ whose solution gives the required $\Lie_O^{N-2}\tilde{\omb}$ on $\Cb_*$ which is what we need to complete the proof of Theorem \ref {omomtildeest}.
\smallskip

\NI{\bf Remarks:} 

\NI{\em Observe that to solve equation \ref{lasteq1} on the last slice and the analogous elliptic equations for the tangential derivatives up to $N$, we need the estimates for the derivatives of $\chih$ only up to order $N-1$ which are already provided by the inductive assumptions. Moreover in any case the estimates of $\Lie_O^N\chih$ on $\Cb_*$ are provided by the Hodge equations obtained deriving
\[{\divv}\hat{\chi}=\frac{1}{2}\nabb{\tr\chi}-\b-\zeta\!\c\hat{\chi}+\frac{1}{2}\zeta{\tr\chi}=\frac{\oom}{2}\Us-\ze\chih-\b\]
and therefore do not  use the transport equation along $\Cb_*$, but, viceversa, require the previous knowledge of the angular derivatives of $\tr\chi$.
\smallskip

\NI 
What we are left to do is to show that $\nabb^{N-2}\tilde{\omb}\big|_{\Cb_*}$ and $\nabb^{N-1}\tilde{\mu}\big|_{\Cb_*}$ do exist and satisfy the expected estimates. The estimates we need for it are the estimates along $\Cb_*$ of $\nabb^N\tr\chi$ which are provided by the transport equation for $\tr\chi$ along $\Cb_*$ in the appropriate canonical foliation.
\smallskip
 
\NI A last remark has to be done about the local solution of equation \ref{lasteq1} in $\Cb_*([\la_0,\la_0+\de))$. There we need a background foliation at our disposal to solve it. Recall that this equation is not linear and that the fixed point solution requires some smallness of the various quantities the connection coefficients and the Riemann components. In particular we need to control the Riemann component $\ro$ in the canonical foliation in terms of those relative to the background foliation. This has been discussed at length in \cite{Kl-Ni:book} and \cite{Niclast}.}
\smallskip

\subsection{The estimates of the ``initial data" on the last slice}\label{lasts}
To complete the proof of 
Theorem \ref{muetaest} and Theorem \ref{omomtildeest} and, therefore, of Theorem \ref{L2.1new}
we need to control, as required in steps 3 and 4, the estimates on $\Cb_*$ of $\Lie_O^{N-1}\tilde{\mu}$ and $\Lie_O^{N-2}\tilde{\omb}$, more specifically we have to prove the following,
\bea
&&\ML\ML\ML\big|r^{2-\frac{2}{p}}\Lie_O^{N-1}{\tilde\mu}\big|_{p,S}(\la,\nu_*)\leq C_6^{(*)}\!\left(e^{(N-2)\de}e^{(N-2)\underline{\Ga}(\la)}\frac{N!}{N^\a\ro_{0,1}^N}\right)\eql{mustarest}\\
&&\ML\ML\ML\big|r^{1-\frac{2}{p}}\tau_{-}\Lie_O^{N-2}\tilde{\omb}\big|_{p,S}(\la,\nu_*)\leq \tilde{\underline C}_2^{(*)}\!\left(e^{(N-2)\de}e^{(N-2)\underline{\Ga}(\la)}\frac{N!}{N^\a\ro_{0,1}^N}\right) .\eql{omtildestarest}
\eea
{\bf Sketch of the proof:} 
The steps required to prove these estimates and all the remaining ones on the last slice are the following: we do not give all the details as they are a repetition of the previous discussion:
\smallskip

0) We assume inductively all the connection coefficients angular derivatives up to $J\leq N-1$.
\medskip

1) To control $\Lie_O^{N-1}{\tilde\mu}$ on $\Cb_*$ we have to control $\Lie_O^{N-1}\tr\chi$ and $\Lie_O^{N-1}\tr\chib$,  already at disposal by inductive assumptions.
\medskip

2) From the definition of ${\tilde\mu}|_{\Cb_*}$ it follows that to control $\Lie_O^N\eta$ we need $\Lie_O^{N-1}\chih$, 
$\Lie_O^{N-1}\chibh$, $\Lie_O^{N-1}(\ro,\si)$, the first two by inductive assumptions, the last two via the control of the ${\cal Q}$ norms.
\medskip

3) From the equation $\oom_*$ satisfies on $\Cb_*$ it follows that $\Lie_O^N\log\oom_*$ is under control once we control $\Lie_O^{N-1}\ze$, $\Lie_O^{N-1}\chih$,  $\Lie_O^{N-1}\chibh$, $\Lie_O^{N-1}\ro$, therefore once we control $\Lie_O^{N}\log\oom_*$ we control $\Lie_O^N\ze$ and $\Lie_O^N\etab$.
\medskip

4) To control $\omb$ on $\Cb_*$ we apply $\ddb_3$ to the elliptic equation satisfied by $\log\oom$ and we control 
$\Lie_O^N\omb$ as we already control $\Lie_O^N\etab$ and the remaining quantities are known by inductive assumptions or through the $\cal Q$ norms.
\medskip

5) We are left to control $\Lie_O^N\tr\ch$ and $\Lie_O^N\chih$, the second one is controlled by the Hodge equations for 
$\Lie_O^{N-1}\chih$  which at their turn require the control of $\Lie_O^N\tr\chi$; therefore this is the norm estimate we are left to control and this is done through the transport equation \ref{traspc*} where, as we are on $\Cb_*$, there is no loss of derivatives.

\subsubsection{The initial data and the ``last slice problem"}\label{S.sInandLast}
A  contradiction seems to appear as we claim that to prove the analytic solution existence, we assign initial data on $C_0\cup \Cb_0$ and, nevertheless, to prove the norm estimates  of Lemma \ref{L6.1} we have to assign ``final data" on the last slice $\Cb(\nu_*)$. Nevertheless this contradiction is only apparent and   the picture of the global strategy to prove the existence of a global analytic solution goes in the following way: 

\NI We assign initial data on $C_0\cup\Cb_0$ satisfying the appropriate norm estimates \ref{123} and \ref{123ab} and applying  Cauchy-Kowalevski we prove, as discussed in $(I)$, the existence of the solution in a small region; then we denote, see Theorem \ref{L6.1},
${\cal K}(\Lambda_a,\Pi_a)\equiv\{(\la,\nu)\in[\nu_0,{\Lambda_a}]\times[\la_0,{\Pi_a}]\}$ as  the larger region where, with the assigned initial data, this analytic solution  does exist. Observe that up to now the double null foliation of the existence region is not specified. 

\NI Introducing on the upper part  $\Cb(\Pi_a)$ of the boundary of ${\cal K}(\Lambda_a,\Pi_a)$, the part we call in the previous subsection  ``the last slice" and denoted $\Cb(\nu_*)$, an appropriate foliation defined through $\oom_*$, we prove, starting from $S(\la_0,\nu_*)\subset C_0$ and exploiting the energy norms norms from $\Cb_0$, appropriate norm estimates for the not underlined connection coefficients and using them and the transport equations on the outgoing cones we prove the norm estimates of Theorem \ref{L6.1}; this is the crucial step to show that the region ${\cal K}(\Lambda_a,\Pi_a)$ can be extended implying, to avoid a contradiction, that this region is in fact unbounded. 

\NI Of course this also implies that the double null foliation of this region required to prove its unboundedness is, see also \cite{Kl-Ni:book}, the ``double null canonical foliation", that is the one determined by the foliation imposed on the last slice to obtain the correct ``final data norms".


\NI Finally the last thing to observe is that, going back from the data on the last slice with the outgoing transport equations to $\Cb_0$ we obtain some norms bounds that the not underlined connection coefficients have to satisfy on $\Cb_0$ and we have to prove that these bounds are compatible with the previously assumed initial data; this is easy to prove for the following reasons: we can now assume all the transport equations along the outgoing cones without worrying anymore of loss of derivatives as we already have the estimates for all order derivatives and due to the fact that the final parameter $\ro$ appearing in these estimates is smaller than the initial one, $\ro_0$, it is immediate to realize that all these estimates are consistent with the initial data estimates.\footnote{In principle the foliation on $\Cb_0$ of the initial data is different from the foliation induced on $\Cb_0$ from the canonical foliation, but it can be proved, proceeding as in \cite{Kl-Ni:book} that, due to the fact that we are considering a ``small initial data problem" the two induced foliations are near and the result follows.}

\section{The complete results of Section \ref{S.4.1}}\label{S.10b}
\subsection{The detailed steps in the proof of Theorem \ref{hypthm2}.}\label{S.s3.9}

The proof of this theorem has been discussed in subsection \ref{S.s4.2} where the various steps of the proof are listed, here we give more details for each of them. 

\subsubsection{Step 1: Initial data estimates for $\nabb_O^{J-1}\Psi$}
\NI We start with the initial data, see Theorem\ref{Thinitialdata}, from them and the expression of the Riemann components in terms of the connection coefficients, see for example \cite{Kl-Ni:book} chapter 3, we obtain immediately the following estimates,  where with $\Psi(R)$ we denote a generic null Riemann component and with $\hat{C}_{0,0}$ a suitable  common constant,
\bea
||u|^{\underline{\phi}(\Psi)}r^{\phi(\Psi)+\ep+(J-1)-\frac{2}{p}}\nabb^{J-1}\Psi(R)|_{p,S}\leq \hat{C}_{0,0}\frac{J!}{J^\a}\frac{e^{(J-2)\de_0}e^{(J-2)\underline{\Ga}_0(\la)}}{\ro_0^J}\ ,\eql{as1}
\eea 
where $\ro_0$ is the ``convergence radius" for the initial data and  $\phi(\Psi)$ has been defined in \ref{phidef}.
\smallskip

\NI{\bf Remark:}\

\NI{\em 
We use here and in the following the convention that if the constant of the connection coefficient or the Riemann null component estimates are multiplied by a constant $c$ which does not depend on these estimates nor on the combinatorial sums we omit it.}
\smallskip

\NI  As sketched in subsection \ref{S.s4.2} 
we move from $\nabb^{J-1}\Psi(R)$ to $\nabb_O^{J-1}\Psi(R)$, proving the following lemma:

\NI {\bf Lemma \ref{L4.1}}
{ \em Under the previous estimates for the connection coefficients and for the null Riemann components on $C_0\cup\Cb_0$, the following estimates for the $\nabb_O$ derivatives of the null Riemann components hold:
\bea
||\la|^{\underline{\phi}(\Psi)}r^{\phi(\Psi)+\ep+(J-1)-\frac{2}{p}}\nabb_O^{J-1}\Psi(R)|_{p,S}\bigg|_{C_0\cup\Cb_0}\!\leq |O|_{\infty}^{J-1}\hat{C}_{0,0}\frac{J!}{J^\a}\frac{e^{(J-2)\de_0}e^{(J-2)\underline{\Ga}_0(\la)}}{\ro_{0,1}^J}\ ,\ \ \ \ \ \ \eql{as2}
\eea
where $\ro_{0,1}<\ro_0$ and $|O|_{\infty}$ is the $\sup$ norm over the whole initial cones\ .} \footnote{Remember that, as $O=\sum_{i=1}^3O^{(i)}$ we have to consider also a factor $3^{J-1}$ in the $|O|^{J-1}_{\infty}$ norms. We do not write explicitly it to avoid cumbersome notations.}

\NI {\bf Proof:} See appendix to section \ref{S.10b}.

\NI Proceeding exactly in the same way we also prove the following corollary,
Moving from the estimate for $\nabb_O^{J-1}\Psi(R)$ to the estimate for $\Psi(\lie_O^{J-1}R)$ requires some intermediate steps: first we go from $\nabb_O^{J-1}\Psi(R)$ to $\Lie_O^{J-1}\Psi(R)$, then from $\Lie_O^{J-1}\Psi(R)$ to $\Psi(\Lie_O^{J-1}R)$ and finally from $\Psi(\Lie_O^{J-1}R)$ to $\Psi(\lie_O^{J-1}R)$\ .
The following lemma is proved in the appendix to section \ref{S.10b}.

\NI {\bf Lemma \ref{L4.2}}
\NI{ \em Under the previous assumptions on the (initial data) connection coefficients and the previous estimates for $\nabb_O^{J-1}\Psi(R)$, the following bounds hold,
\bea
&&\ML\ML\ML||\la|^{\underline{\phi}(\Psi)}r^{\phi(\Psi)+\ep-\frac{2}{p}}\Lie_O^{J-1}\Psi(R)|_{p,S}\bigg|_{C_0\cup\Cb_0}\leq \hat{C}_{0,0}\frac{J!}{J^\a}\frac{e^{(J-2)\de_0}e^{(J-2)\underline{\Ga}_0(\la)}}{\ro_{0,1}^J}\ \ 
\eea
\NI moreover
\bea
&&\ML\ML\ML||\la|^{\underline{\phi}(\Psi)}r^{\phi(\Psi)+\ep-\frac{2}{p}}\Psi(\Lie_O^{J-1}R)|_{p,S}\bigg|_{C_0\cup\Cb_0}\leq \hat{C}_{0,0}\frac{J!}{J^\a}\frac{e^{(J-2)\de_0}e^{(J-2)\underline{\Ga}_0(\la)}}{\ro_{0,1}^J}\ \ 
\eea
and finally}
\bea
&&\ML\ML\ML||\la|^{\underline{\phi}(\Psi)}r^{\phi(\Psi)+\ep-\frac{2}{p}}\Psi(\lie_O^{J-1}R)|_{p,S}\bigg|_{C_0\cup\Cb_0}\leq \hat{C}_{0,0}\frac{J!}{J^\a}\frac{e^{(J-2)\de_0}e^{(J-2)\underline{\Ga}_0(\la)(\nu)}}{\ro_{0,1}^J}\ 
\eea
where $\ro_{0,1}<\ro_{0}$


\subsubsection{Step 2: The initial data energy type norms}
\begin{Le}\label{L4.3}
From the previous estimates for the null Riemann components, we derive the following estimates for the $\QQ_{(0),2}^{(J-2)}(\la_0,\nu)$ and $\QQb_{(0),2}^{(J-2)}(\la,\nu_0)$, part of the ${\cal Q}^{(J-2)}_0$ norms,\footnote{We omit the  $\QQ_{(0),1}^{(J-2)}$ norms }\footnote{$(\varepsilon c_{0,4})$ instead of $C_{(0,4)}$ if $J<J_0$.}

\bea
&&\ML\ML\QQ_{(0),2}^{(J-2)}(\la_0,\nu)+\QQb_{(0),2}^{(J-2)}(\la,\nu_0)\leq 
\left(C^{(1)}\frac{J!}{J^{\a}}\frac{e^{(J-2)\de_{0}}e^{(J-2)\underline{\Ga}(\la)}}{\ro_{0,1}^{J}}\right)^{\!\!2}.\ \ \ \eql{step2a}\ \ \ \ \ \ 
\eea
\end{Le}

\NI{\bf Proof:} 
Recalling, see definition \ref{16555}, the explicit expressions, 
\beaa
\QQ^{(J-2)}_{(0),2}(\la_0,\nu)&\equiv&\int_{C(\la_0;[\nu_0,\nu])}Q(\lie_{O}\lie_{T}{(\lie_{O}^{J-2}W}))(\acc,\acc,\acc,e_4)\nn\\
&&+\int_{C(\la_0;[\nu_0,\nu])}Q(\lie^2_{O}{(\lie_{O}^{J-2}W}))(\acc,\acc,T,e_4)\\
&&+\int_{C(\la_0;[\nu_0,\nu])}Q(\lie_{S}\lie_{T}{(\lie_{O}^{J-2}W}))(\acc,\acc,\acc,e_4)\nn
\eeaa
\beaa
\QQb^{(J-2)}_{(0),2}(\la,\nu_0)&\equiv&\int_{\Cb(\nu_0;[\la_0,\la])}Q(\lie_{O}\lie_{T}{(\lie_{O}^{J-2}W}))(\acc,\acc,\acc,e_3)\nn\\
&&+\int_{\Cb(\nu_0;[\la_0,\la])}Q(\lie^2_{O}{(\lie_{O}^{J-2}W}))(\acc,\acc,T,e_3)\\
&&+\int_{\Cb(\nu_0;[\la_0,\la])}Q(\lie_{S}\lie_{T}{(\lie_{O}^{J-2}W}))(\acc,\acc,\acc,e_3)\nn
\eeaa
we look at the second term of $\QQ^{(J-2)}_{(0),2}(\la_0,\nu)$, all the remaining terms are treated, basically, in the same way. Recalling the explicit expression of $Q(W))(\acc,\acc,T,e_4)$, see \cite{Kl-Ni:book} Chapter 3, eq. (3.5.1),
\beaa
&&\ML Q(W))(\acc,\acc,T,e_4)=\\
&&\ML\frac{1}{4}\nu^4|\a(W)|^2
+\frac{1}{2}(\nu^4+2\nu^2 \la^2)|\b(W)|^2+\frac{1}{2}(\la^4+2\nu^2\la^2)(\ro(W)^2+\si(W)^2)
+\frac{1}{2}\la^4|\bb(W)|^2
\eeaa
we have the following inequality, 
\bea
&&\ML\ML\ML\ML\int_{C(\la_0;[\nu_0,\nu])}Q(\lie_{O}^{H}W)(\acc,\acc,T,e_4)
=\frac{1}{2}\int_{\nu_0}^{\nu}d\nu'\int_{S(\la_0,\nu')}\left[\frac{1}{2}\nu'^4|\a(\lie_{O}^{H}W)|^2+(\nu'^4+2\nu'^2 \la_0^2)|\b(\lie_{O}^{H}W)|^2+\right.\nn\\
&&\left.\ \ \ \ \ \ \ \ \ \ \ \ \ \ \ \ \ \ \ \ \ \ \ \ \ \ \ \ \ \ \ \ \ \ \ \ \ \ \ \  +(\la_0^4+2\nu'^2\la_0^2)(\ro(\lie_{O}^{H}W)^2+\si(\lie_{O}^{H}W)^2)+\la_0^4|\bb(\lie_{O}^{H}W)|^2\right]\nn\\
&&\ML\ML\ML\ML\leq c\int_{\nu_0}^{\nu}d\nu'\left[\big|\nu'^2\a(\lie_{O}^{H}W)\big|_{(2,S(\la_0,\nu'))}^2+\big|\nu'^2\b(\lie_{O}^{H}W)\big|_{(2,S(\la_0,\nu'))}^2
+\big|\nu'|\la_0|\ro(\lie_{O}^{H}W)\big|_{(2,S(\la_0,\nu'))}^2\right.\nn\\
&&\left.\ \ \ \ \ \ +\big|\nu'|\la_0|\si(\lie_{O}^{H}W)\big|_{(2,S(\la_0,\nu'))}^2+\big|\la_0^2\bb(\lie_{O}^{H}W)\big|_{(2,S(\la_0,\nu'))}^2\right]\nn\\
&&\ML\ML\ML\ML\leq c\int_{\nu_0}^{\nu}d\nu'\left[\frac{1}{\nu'^{1+\ep}}\big|r^{\frac{7}{2}+\ep-\frac{2}{2}}\a(\lie_{O}^{H}W)\big|_{(2,S(\la_0,\nu'))}^2
+\frac{1}{\nu'^{1+\ep}}\big|r^{\frac{7}{2}+\ep-\frac{2}{2}}\b(\lie_{O}^H W)\big|_{(2,S(\la_0,\nu'))}^2\right.\nn\\
&&\ML\left.+\frac{|\la_0|}{\nu'^{2+\ep}}\big|\nu'^{3+\ep-\frac{2}{2}}|\la_0|^{\frac{1}{2}}\ro(\lie_{O}^H W)\big|_{(2,S(\la_0,\nu'))}^2+\frac{|\la_0|}{\nu'^{2+\ep}}\big|\nu'^{3+\ep-\frac{2}{2}}|\la_0|^{\frac{1}{2}}\si(\lie_{O}^{H}W)\big|_{(2,S(\la_0,\nu'))}^2\right.\nn\\
&&\ML\left.+\frac{|\la_0|}{\nu'^{1+\ep}}\big|\nu'^{2+\ep-\frac{2}{2}}|\la_0|^{\frac{3}{2}}\bb(\lie_{O}^{H}W)\big|_{(2,S(\la_0,\nu'))}^2\right]\ \ .\eql{ineq1a}
\eea
\NI{\bf Remarks:} { \em

\NI i) The weight for the $\ro$ null component is correct provided $H>0$, this is the case from the explicit expression of the $\cal Q$ norms. The non derived $\ro$ has a decay $O(r^{-3})$ which cannot be improved.

\NI ii) As it appears from the previous estimates we could assume a weaker decay for $\ro$ and $\si$ on the initial data (without the $\ep$ factor) as also in this case the $\cal Q$ norms are bounded on $C_0\cup\Cb_0$. See the discussion in section \ref{S.s initial data}.}
\smallskip

\NI Inserting in \ref{ineq1a} the estimates \ref{as5a} proved in Lemma \ref{L4.2} we obtain
\bea\label{910}
&&\ML\ML\int_{C(\la_0;[\nu_0,\nu])}Q(\lie_{O}^{H}W)(\acc,\acc,T,e_4)
\leq c\!\left(\hat{C}_{0,0}\frac{(H+1)!}{(H+1)^{\a}}\frac{e^{(H-1)\de_0}}{\ro_{0,1}^{H+1}}\right)^{\!\!2}\!\!\int_{\nu_0}^{\nu}\frac{d\nu'}{\nu'^{1+\ep}}e^{2(H-1)\underline{\Ga}_0(\la)}\nn\\
&&\ML\ML\leq \!\left(\hat{C}_{0,0}\frac{(H+1)!}{(H+1)^{\a}}\frac{e^{(H-1)\de_0}e^{(H-1)\underline{\Ga}_0(\la)}}{\ro_{0,1}^{H+1}}\right)^{\!\!2}\!\!
\int_{\nu_0}^{\nu}\frac{d\nu'}{\nu'^{1+\ep}}\nn\\
&&\ML\ML\leq \!\left(\hat{C}_{0,0}\frac{(H+1)!}{(H+1)^{\a}}\frac{e^{(H-1)\de_0}e^{(H-1)\underline{\Ga}_0(\la)(\nu)}}{\ro_{0,1}^{H+1}}\right)^{\!\!2}\nn\\
&&\ML\ML\leq \!\left(\hat{C}_{0,0}\frac{(H+1)!}{(H+1)^{\a}}\frac{e^{(H-1)\de_{0}}e^{(H-1)\underline{\Ga}_0(\la)}}{\ro_{0,1}^{H+1}}\right)^{\!\!2}\eea

Finally we can rewrite the previous estimate as
\bea
&&\int_{C(\la_0;[\nu_0,\nu])}Q(\lie^2_{O}{(\lie_{O}^{J-2}W}))(\acc,\acc,T,e_4)=\int_{C(\la_0;[\nu_0,\nu])}Q(\lie_{O}^{J}W)(\acc,\acc,T,e_4)\nn\\
&&\leq \left(\hat{C}_{0,0}\frac{(J+1)!}{(J+1)^{\a}}\frac{e^{(J-1)\de_{0}}e^{(J-1)\underline{\Ga}_0(\la)}}{\ro_{0,1}^{J+1}}\right)^{\!\!2}.\ \ \ \ \ \ \ \ \ \ \ \eql{Qest1ab}\nn\\
&&\leq \left(\hat{C}_{0,0}\frac{J!}{J^{\a}}\frac{e^{(J-2)\de}e^{(J-2)\underline{\Ga}_0(\la)}}{\ro_{0,1}^{J}}\right)^{\!\!2},\ \ \ \ \ \ \ \ \ \ \ \nn\\
\eea 
with suitable $C^{(2)}$  and  $\de>\de_0$ .

\NI Let us calculate also $\int_{\Cb(\nu_);[\la_0,\la])}Q(\lie_{O}{(\lie_{O}^{J-1}W}))(\acc,\acc,T,e_3)$, exploiting the estimates of \cite{Kl-Ni:book} Chapter 3, we obtain the, as before, estimates
\bea
&&\ML\ML\int_{\Cb(\nu_);[\la_0,\la])}Q(\lie^2_{O}{(\lie_{O}^{J-2}W}))(\acc,\acc,T,e_3)
\leq \!\left(\hat{C}_{0,0}\frac{(J+1)!}{(J+1)^{\a}}\frac{e^{(J-1)\de_0}}{\ro_{0,1}^{J+1}}\right)^{\!\!2}\!\!\int_{\la_0}^{\la}\frac{d\la'}{\la'^{1+\ep}}e^{2(J-1)\underline{\Ga}_0(\la')}\nn\\
&&\ML\ML\leq \!\left(\hat{C}_{0,0}\frac{(J+1)!}{(J+1)^{\a}}\frac{e^{(J-1)\de_0}e^{(J-1)\underline{\Ga}_0(\la)}}{\ro_{0,1}^{J+1}}\right)^{\!\!2}\!\!
\int_{\la_0}^{\la}\frac{d\la'}{\la'^{1+\ep}}e^{2(J-1)(\underline{\Ga}_0(\la')-\underline{\Ga}_0(\la))}\nn\\
&&\ML\ML\leq '\!\left(\hat{C}_{0,0}\frac{(J+1)!}{(J+1)^{\a}}\frac{e^{(J-1)\de_0}e^{(J-1)\underline{\Ga}_0(\la)}}{\ro_{0,1}^{J+1}}\right)^{\!\!2}\!\!
\int_{\la_0}^{\la}\frac{d\la'}{\la'^{1+\ep}}e^{2(J-1)(\frac{\la'-\la}{\la'\la})}\nn\\
&&\ML\ML\leq \!\left(\hat{C}_{0,0}\frac{(J+1)!}{(J+1)^{\a}}\frac{e^{(J-1)\de_{0}}e^{(J-1)\underline{\Ga}_0(\la)}}{\ro_{0,1}^{J+1}}\right)^{\!\!2}\!\!\nn\\
&&\ML\ML\leq \!\left(\hat{C}_{0,0}\frac{J!}{J^{\a}}\frac{e^{(J-2)\de}e^{(J-2)\underline{\Ga}_0(\la)}}{\ro_{0,1}^{J}}\right)^{\!\!2}\!\!
\eea


\NI{\bf Remark:} 

\NI {\em It is important to recall, see appendix later, that to control $|\lie_O^{J-1}\Psi(W)|_{p,S}$ we need to control ${\cal Q}^{(J-2)}$ and not ${\cal Q}^{(J-1)}$.}


\subsubsection{Step 3: The conservation of the $\cal Q$ norms}\label{S.S4.5}
 ``Step 3" is a generalization of the boundedness of the $\cal Q$ norms proved in \cite{Kl-Ni:book}, Chapter 6, see also \cite{C-K:book}. The main difference is that we have now a larger family of $\cal Q$ norms to control all the derivatives (tangential and null) of the  Riemann components; this makes the control of these norms technically more complicated, but the final result is that, provided the ${\cal Q}^{(0)}$ norms are bounded in a region ${\cal K}$, which can also be unbounded, in terms of the initial data  ${\cal Q}^{(0)}$ norms, then we control all the $\cal Q$ norms in the same region.
  
\NI As we said at the beginning of subsection \ref{S.s4.1nn} we consider for the moment only the norms needed to control all the angular derivatives, see \ref{QQ12} and \ref{QQb12} and prove the following crucial result:

\begin{theorem}\label{Qconserv1}
Let us define
\bea
{{\cal H}^{(J-2)}}(\la,\nu)\equiv\sum_{J=2}^{N}\left(\QQ^{(J-2)}({C(\la;[\nu_0,\nu])})+\QQb^{(J-2)}({\Cb(\nu;[\la_0,\la])})\right)\ ,\eql{Hdefx1}
\eea
let assume that ${{\cal H}^{(0)}}(\la,\nu)$ satisfies the following estimates:
\bea
{{\cal H}^{(0)}}(\la,\nu)< C_{0}^{(1)}\left(\QQ_{(0,2)}^{(0)}(\la_0,\nu)+\QQb^{(0)}_{(0,2)}(\la,\nu_0)\right)\ ,
\eea
 in ${\cal K}$, with $C_{0}^{(1)}$ a suitable constant, let assume the initial data satisfy the inequalities \ref{123} and \ref{123ab} for  all $J\leq N-1$, let assume the inductive assumptions \ref{123a} for the connection coefficients hold with $J\leq N-1$, then 
the following estimate holds, 
\bea
{{\cal H}^{(N-2)}}(\la, \nu)^{\frac{1}{2}}\leq 
C^{*(1)}\!\left(\frac{N!}{N^{\a}}\frac{e^{(N-2)(\de+\underline{\Ga}(\la))}}{\ro_{0,1}^{N}}\right)\ , \eql{10.15www}
\eea
with $C_{0}^{(1)}<C^{*(1)}$ a suitable constant.
\end{theorem}
{\bf Proof:} The proof goes by induction, see Appendix to Section \ref{S.10b}.
\smallskip


 
\subsubsection{Step 4: The internal estimates of $\Psi(\lie_O^{J-1}W)$ , $\Psi(\Lie_O^{J-1}W)$ and  $\Lie_O^{J-1}\Psi(W)$}

\NI 
Due to the estimates we have proved for the various $\cal Q$ norms, see Theorem \ref{Qconserv1}, we have immediatly
\bea\label{rem}
\sqrt{\QQ^{(J-2)}(\la,\nu)+\QQb^{(J-2)}(\la,\nu)}\leq C^{*(1)}\frac{J!}{J^{\a}}\frac{e^{(J-2)(\de+\underline{\Ga}(\la))}}{\ro_{0,1}^J}.
\eea
The following lemma holds, with $C^{*(1)}< C^{(1)}$,\footnote{The standard estimate is with $p=4$, for $p\in [2,4)$ we need an interpolation.}\footnote{Remember we need $C^{(1)}$ to be smaller than the constants estimating the connection coefficients. }
\begin{Le}\label{Psi(LieO)boundedfromQ0}
\NI Under the estimates \ref{rem} and the inductive assumptions, up to $J-1$ for the connection coefficients and up to $J-2$ for the null Riemann components, the following estimates hold:
\bea
| |\la|^{\underline{\phi}(\Psi)}r^{\phi(\Psi)-\frac{2}{p}}\Psi(\lie_O^{J-1}W)|_{p,S}(\la,\nu)\leq C^{(1)}\frac{J!}{J^{\a}}\frac{e^{(J-2)(\de+\underline{\Ga}(\la))}}{\ro_{0,1}^J}\  .\eql{6.96}
\eea
\end{Le}

\NI{\bf Proof:} The proof of this lemma is analogous to the one in  \cite{Kl-Ni:book}. It is discussed in detail in Section \ref{AS.4}.
we recall its main lines as it is a crucial step in proving our global result. We consider for simplicity just one of the null Riemann component, namely $\b({\lie_O^{J-1}W})$\ .

 
\NI We define ${\tilde W}= \lie^{J-1}_{\cal O}W$ and consider the $\cal Q$ norms introduced in \cite{Kl-Ni:book}, but now relative to the Weyl tensor  ${\tilde W}$ instead that $W$. Namely        
with these $Q({\tilde W})$ norms, proceeding as in Chapter 5  of \cite{Kl-Ni:book} we control
$\|r^3\nabb\b({\tilde W})\|_{L^2(C)}$ and $\|r^3\ddb_4\b({\tilde W})\|_{L^2(C)}$, more precisely, for $J > 7$, recalling Lemma 4.1.2 of \cite{Kl-Ni:book}, we have,\footnote{This inequality is slightly stronger than the one proved in \cite{Kl-Ni:book} due to the fact that we are considering angular derivatives greater than $3$ which implies that to prove the boundedness of the $\QQ_{(1)}^{(J-1)}({C(\la;[\nu_0,\nu])})+\QQb_{(1)}^{(J-1)}({\Cb(\nu;[\la_0,\la])})$ norms we can use the inductive assumptions for the connection coefficients up to third order.}
\bea
&&\ML\ML\ \ \ |r^{\frac{7}{2}-\frac{2}{4}}\b({\tilde W})|_{p=4,s}(\la,\nu))\nn\\
&&\ML\ML\leq |r^{\frac{7}{2}-\frac{2}{4}}\b({\tilde W})|_{p=4,s}(\la,\nu_0)+\left[\|r^3\nabb\b({\tilde W})\|_{L^2(C(\la;[\nu_0,\nu]))}+\|r^3\ddb_{e_4}\b({\tilde W})\|_{L^2(C(\la;[\nu_0,\nu]))}\right]\nn\\
&&\ML\leq |r^{\frac{7}{2}-\frac{2}{4}}\b({\tilde W})|_{p=4,s}(\la,\nu_0)+\sqrt{\QQ^{(J-2)}(\la,\nu)+\QQb^{(J-2)}(\la,\nu)}\eql{ineq1}\\
&&\ML\leq \hat{C}_{0,0}\frac{J!}{J^{\a}}\frac{e^{(J-2)(\de_0+\underline{\Ga}_0(\la))}}{\ro_0^J}+C^{(1)}\frac{J!}{J^{\a}}\frac{e^{(J-2)(\de+\underline{\Ga}(\la))}}{\ro_{0,1}^J}
\leq C^{(1)}\frac{J!}{J^{\a}}\frac{e^{(J-2)(\de+\underline{\Ga}(\la))}}{\ro_{0,1}^J}\ ,\nn
\eea
where the term $|r^{\frac{7}{2}-\frac{2}{4}}\b({\tilde W})|_{p=4,s}(\la,\nu_0)$ has been estimated using the initial data assumptions for the connection coefficients and its derivatives.

\subsubsection{Step 5: Internal estimate for $\nabb^{J-1}\Psi(W)$ }\label{sec5}
 The last  step is moving from $\Psi(\lie_O^{J-1}W)$ to $\nabb^{J-1}\Psi(W)$. This is the content of the following lemmas
\begin{Le}\label{new1}
Assuming the estimates for the null Riemann components of lemma \ref{Psi(LieO)boundedfromQ0},
\bea
\big|\Psi(\lie_O^{J-1}W)\big|_{p,S}\leq C^{(1)}\frac{J!}{J^{\a}}\frac{e^{(J-2)(\de+\underline{\Ga}(\la)}}{\ro_{0,1}^J}\ ,
\eea
the following estimate holds for the norms of $\Lie_O^{J-1}\Psi(W)$,
\bea\label{zizi}
\big|\Lie_O^{J-1}\Psi\big|_{p,S}\leq C^{(1)}\frac{J!}{J^{\a}}\frac{e^{(J-2)(\de+\underline{\Ga}(\la)}}{\ro_{0,1}^J}\
\eea
 
 \end{Le} 
 which completes the proof of Theorem \ref{hypthm2}.
\begin{Le}\label{fromLieOtoNabb} 
Let us assume that for any $J$ the estimates \ref{zizi} hold, let us assume that the connection coefficients satisfy, in the internal region, with all their angular derivatives up to order $J-1$ the estimates \ref{123a}, 
then the null components $\Psi$ of the Riemann tensor satisfy the following estimates  for $6< N\leq J-1$  with $\ro<\ro_{0,1}$, 
\bea
\big| |\la|^{\underline{\phi}(\Psi)}r^{\phi(\Psi)+(J-1)-\frac{2}{p}}\nabb^{J-1}\Psi\big|_{L^p(S)}\leq C^{(1)}e^{(J-2)\de}e^{(J-2)\underline{\Ga}(\la)}\frac{J!}{J^\a}\frac{1}{\ro^J}\ ,
\eea
with $C^{(1)}$ of order $O(1)$ for $J\geq 7$, and $O(\varepsilon)$
\ for $J<7$,\footnote{If $J\geq 1$, for $J=0$ $\ro$ becomes $\ro-{\underline\ro}$} see \cite{Kl-Ni:book}, chapter 6.
\bea
&&\ML\phi(\a)=\phi(\b)=\frac{7}{2}\ ,\ \phi(\ro)=\phi(\si)=3\ , \phi(\bb)=2\ ,\ \phi(\aa)=1\nn\\
&&\ML\underline{\phi}(\a)=\underline{\phi}(\b)=0\ ,\ \underline{\phi}(\ro)=\underline{\phi}(\si)=\frac{1}{2}\ , \underline{\phi}(\bb)=\frac{3}{2}\ ,\ \underline{\phi}(\aa)=\frac{5}{2}\ .\ \ 
\eea
\end{Le}
\NI{\bf Proof:} See Appendix to section \ref{S.10b}.

\section{The complete results of Section \ref{S.5nx}}\label{S.12}

\subsubsection{The $\oom$ component}

The angular derivatives of this component are easy to control as
\[\prr\log\oom=\nabb\log\oom=2^{-1}(\eta+\etab)\]
and $\eta$ and $\etab$ are already under control. Therefore 
\bea
\nabb^N\log\oom=\frac{1}{2}(\nabb^{N-1}\eta+\nabb^{N-1}\etab)
\eea
and
\bea
&&\ML\ML\nabb^N\oom=\nabb^Ne^{\log\oom}=\sum_{k=0}\frac{1}{k!}\nabb^N(\log\oom)^k
=\sum_{k=1}\frac{1}{k!}\nabb^N(\log\oom)^k\nn\\
&&\ML\ML=N!\sum_{k=1}^{\infty}\frac{1}{k!}\sum_{\ga_1,\c,\c,\c,\ga_k}^{\sum_{s=1}^k\ga_s=N; \ga_s\in[0,N]}\frac{1}{\ga_1!\ga_2!\c\c\ga_k!}(\nabb^{\ga_1}\log\oom)(\nabb^{\ga_2}\log\oom)\c\c\c(\nabb^{\ga_k}\log\oom)\nn
\eea
Therefore 
\bea
&&\ML\ML|\nabb^N\oom|_{p,S}\leq 
N!\sum_{k=1}^{\infty}\frac{1}{k!}\left[\sum_{\ga_1=0}^{\frac{N}{2}}\sum_{\ga_2,\c,\c,\c,\ga_k}^{\sum_{s=2}^k\ga_s=N-\ga_1; \ga_s\in[0,N]}\frac{1}{\ga_1!\ga_2!\c\c\ga_k!}\big|(\nabb^{\ga_1}\log\oom)(\nabb^{\ga_2}\log\oom)\c\c\c(\nabb^{\ga_k}\log\oom)\big|_{p,S}\right.\nn\\
&&\ML\ML\ \ \ \ \ \ \ \ \ \ \ \ \ \ \ \ \ \ \ \ \ \ \ \ \left.\sum_{\ga_1=\frac{N}{2}+1}^{N}\sum_{\ga_2,\c,\c,\c,\ga_k}^{\sum_{s=2}^k\ga_s=N-\ga_1; \ga_s\in[0,N]}\frac{1}{\ga_1!\ga_2!\c\c\ga_k!}\big|(\nabb^{\ga_1}\log\oom)(\nabb^{\ga_2}\log\oom)\c\c\c(\nabb^{\ga_k}\log\oom)\big|_{p,S}\right]\ .\nn
\eea
From it we obtain, see Section \ref{S.17}, the following estimate,
\bea
|\nabb^N\oom|_{p,S}\leq \left({\tilde C}_4\frac{(N-1)!}{(N-1)^{\a}}\frac{e^{(N-3)(\de+\underline{\Ga}(\la))}}{\ro^{N-1}}\right)\ .\eql{oomestxx}
\eea
Adapting the same techniques used before to prove the estimates for the angular derivatives of the connection coefficients we prove the following estimate,
\bea
|\pr^N_{\nu}\oom|_{p,S}\leq c\!\left(\frac{N!}{N^{\a}}\frac{e^{(N-2)(\de+\underline{\Ga}(\la))}}{\ro^{N}}\right)\ .\eql{oomest1}
\eea
The detailed proof of \ref{oomest1} is in Section \ref{S.17}.

\subsubsection{The $\ga$ components}
We recall that $\ga_{ab}$ satisfies the following equation
\bea
\frac{\pr}{\pr\nu}\ga_{ab}-{\oom\tr\chi}\ga_{ab}=2\oom\chih_{ab}\ , \eql{7.13}
\eea
which we rewrite as
\bea
\frac{\pr}{\pr\nu}\ga_{ab}-\overline{\oom\tr\chi}\ga_{ab}=({\oom\tr\chi}-\overline{\oom\tr\chi})\ga_{ab}+2\oom\chih_{ab}
\eea
and
\bea
&&\ML\ML\frac{\pr}{\pr\nu}(r^{-2}\ga_{ab})=
({\oom\tr\chi}-\overline{\oom\tr\chi})(r^{-2}\ga_{ab})+2\oom r^{-2}\chih_{ab}\ .\nn
\eea
The control of the angular non covariant derivatives can be easily obtained applying $\prr^J$ to both sides. The following estimates hold
\bea
|r^{J-2-\frac{2}{p}}\prr^J\ga_{ab}|_{p,S}\leq c\left(\frac{J!}{J^{\a}}\frac{e^{(J-2)(\de+\underline{\Ga}(\la))}}{\ro^{J}}\right)\ .
\eea
{\bf Remark:} 

\NI{\em The estimates for $\prr^J\ga_{ab}$ follow with $J$ instead of $J-1$ due to the loss of derivatives present in the transport equation \ref{7.13}. This loss is not present considering the mixed derivatives.}
\subsubsection{The $X^a$ components}
We study the $X_a=\ga_{ac}X^c$ components which is equivalent; we have the transport equation
\bea
\frac{\pr X_a}{\pr\nu}=-4\oom^2\ze_a\ .
\eea
Let us look at the transport equation for the norm $|X|=\sqrt{\ga_{ab}X^aX^b}$, 
\bea
&&\ML\ML\frac{\pr|X|^2}{\pr\nu}=\frac{\pr\ga_{ab}}{\pr\nu}X^aX^b+2\ga_{ab}X^a\frac{\pr X^b}{\pr\nu} 
=\oom\tr\chi|X|^2+2\oom X\c\chih\c X+2\ga_{ab}X^a\frac{\pr \ga^{bc}X_c}{\pr\nu}\nn\\
&&\ML\ML=\oom\tr\chi|X|^2+2\oom X\c\chih\c X+2\ga_{ab}X^a\frac{\pr \ga^{bc}}{\pr\nu}X_c+2\ga_{ab}\ga^{bc}X^a\frac{\pr X_c}{\pr\nu}\nn\\
&&\ML\ML=\oom\tr\chi|X|^2+2\oom X\c\chih\c X-2\frac{\pr\ga_{ab}}{\pr\nu}X^aX^b+2X^c\ze_c\nn\\
&&\ML\ML=\oom\tr\chi|X|^2+2\oom X\c\chih\c X-2\oom\tr\chi|X|^2-4\oom X\c\chih\c X+2X\c{\tilde\ze}\nn\\
&&\ML\ML=-\oom\tr\chi|X|^2-2\oom X\c\chih\c X+2X\c{\tilde\ze}\ , 
\eea
where ${\tilde\ze}=-4\oom^2\ze$. Therefore
\bea
\frac{\pr |X|^2}{\pr\nu}+\oom\tr\chi|X|^2=-2\oom X\c\chih\c X+2X\c{\tilde\ze}
\eea
To obtain the transport equation for $\nabb^J|X|^2$ we proceed in the following way,
\bea
\nabb^J\frac{\Dbb}{\pr\nu}|X|^2=\frac{\Dbb}{\pr\nu}(\nabb^J|X|^2)+[\nabb^J,\frac{\Dbb}{d\nu}]|X|^2
\eea
and
\bea
&&\ML\ML\frac{\Dbb}{\pr\nu}(\nabb^J|X|^2)=\nabb^J\frac{\Dbb}{\pr\nu}|X|^2-[\nabb^J,\frac{\Dbb}{d\nu}]|X|^2\nn\\
&&\ML\ML=\nabb^J\frac{\Dbb}{\pr\nu}|X|^2-\left[\sum_{k=0}^{J-1}\cbin{J}{k}(\nabb^{J-1-k}\oom\chi)\c\nabb^{k+1}|X|^2-\sum_{k=0}^{J-1}\cbin{J+1}{k}(\nabb^{J-1-k}C)\nabb^k|X|^2\right]\nn\\
&&\ML\ML=\ -J\ \!\frac{\oom\tr\chi}{2}\c\nabb^J|X|^2-J(\oom\chih)\c\nabb^J|X|^2+\nabb^J\frac{\Dbb}{\pr\nu}|X|^2\nn\\
&&\ML-\left[\sum_{k=0}^{J-2}\cbin{J}{k}(\nabb^{J-1-k}\oom\chi)\c\nabb^{k+1}|X|^2-\sum_{k=0}^{J-1}\cbin{J+1}{k}(\nabb^{J-1-k}C)\nabb^k|X|^2\right]\nn\\
\eea
As
\bea
&&\ML\ML\nabb^J\frac{\Dbb}{\pr\nu}|X|^2=-\nabb^J\!\left[\oom\tr\chi|X|^2+2\oom X\c\chih\c X-2X\c\ze
\right]\nn\\
&&\ML\ML=-\sum_{k=0}^J\cbin{J}{k}(\nabb^k\oom\tr\chi)\nabb^{J-k}|X|^2
-\sum_{k=0}^J\sum_{l=0}^{J-k}\frac{J!}{k!l!(J-k-l)!}(\nabb^kX)(\nabb^l\oom\chih)(\nabb^{J-k-l}X)\nn\\
&&\ML\ML\ \ +2\sum_{k=0}^J\sum_{l=0}^{J-k}\frac{J!}{k!(J-k)!}(\nabb^kX)\nabb^{J-k}\ze\nn\\
&&\ML\ML=-{\oom\tr\chi}(\nabb^J|X|^2)-2(\nabb^JX)\c\chih\c X+2(\nabb^JX)\c{\tilde\ze}\nn\\
&&\ML\ML-\left[\sum_{k=1}^J\cbin{J}{k}(\nabb^k\oom\tr\chi)\nabb^{J-k}|X|^2+\sum_{\ga_1\ga_2\ga_3}^{\sum_{s=1}^4\!\ga_s=J; \ga_1,\ga_3\neq J}\frac{J!}{\ga_1!\ga_2!\ga_3!}(\nabb^{\ga_1}X)\c(\nabb^{\ga_2}\oom\chih)\c(\nabb^{\ga_3}X)\right.\nn\\
&&\ML\ML\ \ \left.+2\sum_{k=0}^{J-1}\sum_{l=0}^{J-k}\frac{J!}{k!(J-k)!}(\nabb^kX)\nabb^{J-k}{\tilde\ze}\right]\nn\\
\eea
substituting in the previous expression we have
\bea
&&\ML\ML\frac{\Dbb}{\pr\nu}(\nabb^J|X|^2)+(J+2)\ \!\frac{\oom\tr\chi}{2}\c\nabb^J|X|^2+J(\oom\chih)\c\nabb^J|X|^2
-2(\nabb^JX)\c\chih\c X+2(\nabb^JX)\c{\tilde\ze}\nn\\
&&\ML\ML=-\left\{\left[\sum_{k=0}^{J-2}\cbin{J}{k}(\nabb^{J-1-k}\oom\chi)\c\nabb^{k+1}|X|^2-\sum_{k=0}^{J-1}\cbin{J+1}{k}(\nabb^{J-1-k}C)\nabb^k|X|^2\right]\right.\nn\\
&&\ML\ML\left.-\left[\sum_{k=1}^J\cbin{J}{k}(\nabb^k\oom\tr\chi)\nabb^{J-k}|X|^2+\sum_{\ga_1\ga_2\ga_3}^{\sum_{s=1}^4\!\ga_s=J; \ga_1,\ga_3\neq J}\frac{J!}{\ga_1!\ga_2!\ga_3!}(\nabb^{\ga_1}X)\c(\nabb^{\ga_2}\oom\chih)\c(\nabb^{\ga_3}X)\right.\right.\nn\\
&&\ML\ML\ \ \left.\left.+2\sum_{k=0}^{J-1}\sum_{l=0}^{J-k}\frac{J!}{k!(J-k)!}(\nabb^kX)\nabb^{J-k}{\tilde\ze}\right]\right\}
\eea
which we rewrite in a more compact way as
\bea
\ML\ML\ML\frac{\Dbb}{\pr\nu}(\nabb^J|X|^2)+(J+2)\ \!\frac{\oom\tr\chi}{2}\c\nabb^J|X|^2+J(\oom\chih)\c\nabb^J|X|^2
-2(\nabb^JX)\c\oom\chih\c X+2(\nabb^JX)\c{\tilde\ze}=\bigg\{good\bigg\}_{\!X}\ \ 
\eea
where
\bea
&&\ML\ML\bigg\{good\bigg\}_{\!X}
=-\left\{\left[\sum_{k=0}^{J-2}\cbin{J}{k}(\nabb^{J-1-k}\oom\chi)\c\nabb^{k+1}|X|^2-\sum_{k=0}^{J-1}\cbin{J+1}{k}(\nabb^{J-1-k}C)\nabb^k|X|^2\right]\right.\nn\\
&&\ML\ML\left.-\left[\sum_{k=1}^J\cbin{J}{k}(\nabb^k\oom\tr\chi)\nabb^{J-k}|X|^2+\sum_{\ga_1\ga_2\ga_3}^{\sum_{s=1}^4\!\ga_s=J; \ga_1,\ga_3\neq J}\frac{J!}{\ga_1!\ga_2!\ga_3!}(\nabb^{\ga_1}X)\c(\nabb^{\ga_2}\oom\chih)\c(\nabb^{\ga_3}X)\right.\right.\nn\\
&&\ML\ML\ \ \left.\left.+2\sum_{k=0}^{J-1}\sum_{l=0}^{J-k}\frac{J!}{k!(J-k)!}(\nabb^kX)\nabb^{J-k}{\tilde\ze}\right]\right\}.\nn
\eea
From this equation one gets  the transport equation for the $|\c|_{p,S}$ norms with the appropriate weights and proceeding as done many times before, we prove the following estimate for all $J$,
\bea
|\frac{r^{(J+1-\frac{2}{p})}}{\log r}\nabb^{J}|X||_{p,S}(\la,\nu)\leq c\varepsilon_0\left(\frac{J!}{J^{\a}}\frac{e^{(J-2)(\Ga_1+\de)}}{\ro^J}\right)\ .
\eql{12.14az} 
\eea

\section{The complete results of Section \ref{S.4a}}\label{S.6}
\subsection{The control of the (non covariant) partial derivatives}
\NI Assume for simplicity ${\cal O}$ denotes a $S$-tangent vector, for instance the connection coefficient $\ze$; we are interested to the analyticity of the tensor field $\ze=\ze(e_C)\theta^{C}(\c)$ which means that we have to prove that the various components $\ze(e_C)(\la,\nu,\om^a)$ are analytic functions in the $\la,\nu,\om^a$ variables. To prove it we have to control the norm of the derivatives of these components. As discussed in (I) this implies that we have to control the (norms of the) mixed derivatives in $\nu,\om^a$ for the quantities defined on an outgoing cone and the mixed derivatives in $\la, \om^a$ for the quantities defined on an incoming cone. We start looking at the first situation observing that the second case can be treated exactly in the same way.

\subsubsection{The non covariant mixed derivatives in $\{\nu,\om^a\}$\ .}
\NI Preliminary we recall some trivial facts about the covariant derivatives, let $V$ be a covariant vector field,
\bea
&& |\nabb V|^2=\nabb_{\mu}V_{\nu}\nabb_{\ro}V_{\si}g^{\mu\ro}g^{\nu\si}
=\sum_{A,B}\nabb_{\mu}V_{\nu}\nabb_{\ro}V_{\si}e_A^{\mu}e_A^{\ro}e_B^{\nu}e_B^{\si}\nn\\
&&=\sum_{A,B}(\nabb_{A}V)_{\nu}e_B^{\nu}(\nabb_{A}V)_{\si}e_B^{\si}
=\sum_{A,B}\left((\pr_AV)_{\nu}e_B^{\nu}-\Ga_{AB}^{\nu}V_{\nu}\right)\left((\pr_AV)_{\si}e_B^{\si}-\Ga_{AB}^{\si}V_{\si}\right)\nn\\
&&=\sum_{A,B}\left((\pr_AV)_{\nu}e_B^{\nu}-\Ga_{AB}^{\nu}V_{\nu}\right)^2\ .
\eea 
Observe that 
\bea
&&\left((\pr_AV)_{\nu}e_B^{\nu}-\Ga_{AB}^{\nu}V_{\nu}\right)=\pr_A(V(e_B))-(\pr_Ae_B)^{\nu}V_{\nu}
-\Ga_{AB}^{\nu}V_{\nu}\nn
\eea
and 
\[(\pr_Ae_B)^{\nu}\neq \Ga_{AB}^{\nu}\ .\]
\bigskip

We have the following relations,
\bea
\pr_{e_A}\ze(e_B)\!&=&\!(\dd_{e_A}\ze)(e_B)+\ze(e_C)\gggg(e_C,\dd_{e_A}e_B)\nn\\
\!&=&\!(\nabb_{e_A}\ze)(e_B)-\ze(e_C)\gggg(\nabb_{e_A}e_C,e_B)\nn\\
&&\nn\\
\pr_{\nu}\ze(e_B)\!&=&\!(\oom\dd_{e_4}\ze)(e_B)+\oom\ze(e_C)\gggg(e_C,\dd_{e_4}e_B)\nn\\
\!&=&\!(\ddb_{\nu}\ze)(e_B)-\ze(e_C)\oom\gggg(\dd_{e_4}e_C,e_B)\nn\\
\!&=&\!(\ddb_{\nu}\ze)(e_B)-\ze(e_C)\oom\gggg(\dd_{e_C}e_4,e_B)-\ze(e_C)(\c)\oom\gggg([e_C,e_4],e_B)\nn\\
\!&=&\!(\ddb_{\nu}\ze)(e_B)-\oom\ze(e_C)\chi(e_C,e_B)-\oom\ze(e_C)\gggg([e_C,e_4],e_B)\nn\\
\!&=&\!(\ddb_{\nu}\ze)(e_B)-2\oom\ze(e_C)\chi(e_C,e_B)\nn
\eea
where we used the relation
\bea
[e_C,e_4]=-\ddb_4e_C+\chi(e_C,e_D)e_D-\nabb_A(\log\oom)e_4
\eea
and assumed we have chosen a Fermi transported frame, implying $\ddb_4e_C=0$. Therefore we can write 
\bea
\pr_{\a}\ze=\nab_{\a}\ze-{\tilde c}_{\a}\c\ze=(\nab_{\a}-{\tilde c}_{\a}\c)\ze
\eea
where
\bea
&&\nab_A=\nabb_{e_A}\ \ ,\ \ \nab_{\nu}=\oom\ddb_4\nn\\
&&{\tilde c}_A=-\gggg(\nabb_{e_A}e_C,e_B)(\c)\ \ ,\ \ {\tilde c}_{\nu}=2\oom\chi\ .
\eea
Therefore we write in a symbolic way,
\bea
\pr{\cal O}=(\nab-c){\cal O}
\eea
and, symbolically,
\bea
\pr^{J-1}{\cal O}=(\nab-c)^{J-1}{\cal O}\ ``="\sum_{k=0}^{J-1}\cbin{J-1}{k}\nab^kc^{J-1-k}{\cal O}
\eea
where in the right hand side we have to be careful at the position of the $\nab$'s as they can operate on the $c$'s or on the ${\cal O}$. Therefore fixed $k$ one has to interpret the right hand side as made by $J-k$ slots, each one at the left of one of the $c$'s we think as numbered and the last one immediately at the left of the ${\cal O}$. In these $J-k$ slots one has to distribute $k$ $\nab$'s, each one operating exclusively on the tensor $c$ at its immediate right, or on
${\cal O}$ if we are considering the last slot. Computing all the possible ways of distributing these $\nab$'s one is considering more terms of the real existing ones, but this over estimate should not be harmful. Therefore we write, omitting the minus signs as at the end we are doing a norm estimate, the following expression\footnote{Recall that the value of $\ro$ has been lowered to have also for the mixed derivatives an estimate with $\frac{(J+P)!}{(J+P)^{\a}}$ instead of  $\frac{(J+P+1)!}{(J+P+1)^{\a}}$.}
 \medskip
 
\bea
\ML\ML\pr^{J-1}{\cal O} ``=" \sum_{k=0}^{J-1}\cbin{J-1}{k}\sum_{\ga_1,\ga_2,...,\ga_{J-k}}^{\sum_{s=1}^{J-k}\ga_s=k}\frac{k!}{\ga_1!,\ga_2!,...,\ga_{J-k}!}
(\nab^{\ga_1} c)(\nab^{\ga_2}c)\c\c\c(\nab^{\ga_{J-1-k}}c)(\nab^{\ga_{J-k}}{\cal O})\nn
\eea
and its norm estimate is, omitting for notation simplicity the $r$ weights and using the estimates proved in Theorem \ref{angulconcoef}
\bea
\ML\ML\ML\big|\pr^{J-1}{\cal O}\big|_{p,S} \leq \sum_{k=0}^{J-1}\cbin{J-1}{k}\sum_{\ga_1,\ga_2,...,\ga_{J-k}}^{\sum_{s=1}^{J-k}\ga_s=k}\frac{k!}{\ga_1!,\ga_2!,...,\ga_{J-k}!}
|\nab^{\ga_1} c|_{\infty,S} |(\nab^{\ga_2}c|_{\infty,S}\c\c\c|\nab^{\ga_{J-1-k}}c|_{\infty,S} |\nab^{\ga_{J-k}}{\cal O}|_{p,S}\nn
\eea
which we rewrite, denoting $q=\ga_{J-k}$,
\bea
&&\ML\ML\ML\ML\ML\ML\big|\pr^{J-1}{\cal O}\big|_{p,S} \leq \sum_{k=0}^{J-1}\sum_{q=0}^{k}\cbin{J-1}{k}\cbin{k}{q}|\nab^{q}{\cal O}|_{p,S}\sum_{\ga_1,\ga_2,...,\ga_{J-1-k}}^{\sum_{s=1}^{J-1-k}\ga_s=k-q}\frac{(k-q)!}{\ga_1!,\ga_2!,...,\ga_{J-1-k}!}
|\nab^{\ga_1} c|_{\infty,S} |\nab^{\ga_2}c|_{\infty,S}\c\c\c|\nab^{\ga_{J-1-k}}c|_{\infty,S}\nn\\
&&\ML\ML\ML\ML\ML
 \leq {(J-1)!}\sum_{k=0}^{J-1}\frac{1}{(J-1-k)!}\sum_{q=0}^{k}\frac{|\nab^{q}{\cal O}|_{p,S}}{q!}\sum_{\ga_1,\ga_2,...,\ga_{J-1-k}}^{\sum_{s=1}^{J-1-k}\ga_s=k-q}\frac{1}{\ga_1!,\ga_2!,...,\ga_{J-1-k}!}
|\nab^{\ga_1} c|_{\infty,S} |\nab^{\ga_2}c|_{\infty,S}\c\c\c|\nab^{\ga_{J-1-k}}c|_{\infty,S}\nn\\
&&\ML\ML\ML \leq\left(c\frac{(J-1)!}{(J-1)^\a}\right)
\left\{\sum_{k=0}^{J-1}\frac{1}{(J-1-k)!}\sum_{q=0}^{k}\frac{(J-1)^{\a}}{(\ep(q)+q)^{\a}}\frac{e^{(q-2)(\de_0+\Ga)}}{\ro^q}\right.\nn\\
&&\ML\ML\ML\ML\ML\ML \left.
 \c\!\left[\sum_{\ga_1,\ga_2,...,\ga_{J-1-k}}^{\sum_{s=1}^{J-1-k}\ga_s=k-q}\frac{1}{\ga_1!,\ga_2!,...,\ga_{J-1-k}!}
\frac{(\ga_1+1)!}{(\ga_1+1)^{\a}}\frac{e^{(\ga_1-1)(\de+\underline{\Ga}(\la))}}{\ro^{\ga_1+1}}\frac{(\ga_2+1)!}{(\ga_2+1)^{\a}}\frac{e^{(\ga_2-1)(\de+\underline{\Ga}(\la))}}{\ro^{\ga_2+1}}\c\c\c\frac{(\ga_{J-1-k}+1)!}{(\ga_{J-1-k}+1)^{\a}}\frac{e^{(\ga_{J-1-k}-1)(\de+\underline{\Ga}(\la))}}{\ro^{\ga_{J-1-k+1}}}\right]\right\}\nn\\
\eea
where $\ep(0)=1, \ep(q)=0, q>0$. We write
\bea
&&\ML\ML\ML\ML\ML\ML\big|\pr^{J-1}{\cal O}\big|_{p,S}
\leq c\!\left(\frac{(J-1)!}{(J-1)^{\a}}\right)\c\!\left(\frac{e^{[(q-2)+(k-q)-(J-1-k)](\de+\underline{\Ga}(\la))}}{\ro^{q+(k-q)+J-1-k}}\right)
\left\{\sum_{k=0}^{J-1}\frac{1}{(J-1-k)!}\sum_{q=0}^{k}\frac{(J-1)^{\a}}{(\ep(q)+q)^{\a}}\ \ \c\right.\nn\\
&&\ML\ML \left.
 \c \left[\sum_{\ga_1,\ga_2,...,\ga_{J-1-k}}^{\sum_{s=1}^{J-1-k}\ga_s=k-q}\frac{1}{\ga_1!,\ga_2!,...,\ga_{J-1-k}!}
\frac{(\ga_1+1)!}{(\ga_1+1)^{\a}}\frac{(\ga_2+1)!}{(\ga_2+1)^{\a}}\c\c\c\frac{(\ga_{J-1-k}+1)!}{(\ga_{J-1-k}+1)^{\a}}\right]\right\}\nn\\
&&\ML\ML\leq c\!\left(\frac{(J-1)!}{(J-1)^{\a}}\frac{e^{((J-1)-2)(\de+\underline{\Ga}(\la))}}{{\ro}^{J-1}}\right)
\left\{\sum_{k=0}^{J-1}\frac{1}{(J-1-k)!}\sum_{q=0}^{k}\frac{(J-1)^{\a}}{(\ep(q)+q)^{\a}}
 \c \left[\sum_{\ga_1,\ga_2,...,\ga_{J-1-k}}^{\sum_{s=1}^{J-1-k}\ga_s=k-q}
\frac{1}{\ga_1^{\a-1}\ga_2^{\a-1}\c\c\c\ga_{J-1-k}^{\a-1}}\right]\right\}\nn\\
&&\ML\ML\leq c\!\left(\frac{(J-1)!}{(J-1)^{\a}}\frac{e^{((J-1)-2)(\de+\underline{\Ga}(\la))}}{{\ro}^{J-1}}\right)
\left\{\sum_{k=0}^{J-1}\frac{c_1^{J-1-k}}{(J-1-k)!}\sum_{q=0}^{k}\frac{(J-1)^{\a}}{(\ep(q)+q)^{\a}}\right\}\ .\nn
\eea
Observe that 
\bea
&&\left\{\sum_{k=0}^{J-1}\frac{c_1^{J-1-k}}{(J-1-k)!}\sum_{q=0}^{k}\frac{(J-1)^{\a}}{(\ep(q)+q)^{\a}}\right\}\\
&&\ML=\left\{\sum_{k=0}^{\left[\frac{J-1}{2}\right]}\frac{c_1^{J-1-k}}{(J-1-k)!}\sum_{q=0}^{k}\frac{(J-1)^{\a}}{(\ep(q)+q)^{\a}}\right\}+
\left\{\sum_{k=\left[\frac{J-1}{2}\right]+1}^{J-1}\frac{c_1^{J-1-k}}{(J-1-k)!}\sum_{q=0}^{k}\frac{(J-1)^{\a}}{(\ep(q)+q)^{\a}}\right\}\nn\\
&&\ML\leq c\left\{\sum_{k=0}^{\left[\frac{J-1}{2}\right]}\frac{c_1^{J-1-k}}{(J-1-\a-k)!}\sum_{q=0}^{k}\frac{1}{(\ep(q)+q)^{\a}}\right\}
+\left\{\sum_{k=\left[\frac{J-1}{2}\right]+1}^{J-1}\frac{c_1^{J-1-k}}{(J-1-k)!}\sum_{q=0}^{k}\frac{(J-1)^{\a}}{(\ep(q)+q)^{\a}}\right\}\nn
\eea
and, choosing ${{\hat\ro}}/{\ro}$ sufficiently small, it follows that
\bea
\big|\pr^{J-1}{\cal O}\big|_{p,S}\leq c\!\left(\frac{(J-1)!}{(J-1)^{\a}}\frac{e^{((J-1)-2)(\de+\underline{\Ga}(\la))}}{{\hat\ro}^{J-1}}\right)\ .
\eea

\subsection{The $O^{(i)}$ components}
 In the estimates for the null Riemann components, see the appendix to Section \ref{S.10b}, we use repeatedly the estimates of the tangential derivatives of the $O^{(i)}$ components. To prove these estimates we proceed in the following way; let us considered first the simplified case where the $O^{(i)}$ have the same expression as in the Minkowski spacetime, therefore
 \bea
 &&O^{(1)}=-r\sin\phi\ \!e_{\theta}-r\cos\phi{\cos\theta}\ \!e_{\phi}\nn\\
 &&O^{(2)}=r\cos\phi\ \! e_{\theta}-r\sin\phi{\cos\theta}\ \!e_{\phi}\nn\\
&&O^{(3)}= r\sin\theta\ \! e_{\phi}\ 
 \eea
 and summarizing
 \bea
^{(i)\!}O^c=\sum_Cf^{(i)}_C(r,\theta,\phi) e^c_{C}
 \eea
 where $f^{(i)}_C(r,\theta,\phi)$ are rational functions in $r,\theta,\phi$. let us consider the norm of $\nabb\ \!\! ^{(i)\!}O$,
 \bea
 &&\ML\ML|\nabb\ \!\! ^{(i)\!}O|^2=\nabb_a\ \!\! ^{(i)\!}O^b\nabb_{\tilde a}\ \!\! ^{(i)\!}O^{\tilde b}\ga^{a{\tilde a}}\ga_{b{\tilde b}}
 =\nabb_a\ \!\! ^{(i)\!}O^b\nabb_{\tilde a}\ \!\! ^{(i)\!}O^{\tilde b}\sum_Ce^a_Ce^{\tilde a}_C\sum_D\theta^D_b\theta^D_{\tilde b}\nn\\
 &&\ML\ML=\sum_C\sum_D\nabb_C\ \!\! ^{(i)\!}O^b\nabb_C\ \!\! ^{(i)\!}O^{\tilde b}\theta^D_b\theta^D_{\tilde b}
 =\sum_C\sum_D\gggg(e_D,\nabb_C\ \!\! ^{(i)\!}O)\gggg(e_D,\nabb_C\ \!\! ^{(i)\!}O)\nn\\
 &&\ML\ML=\sum_D\gggg(e_D,\nabb\ \!\! ^{(i)\!}O)\gggg(e_D,\nabb\ \!\! ^{(i)\!}O)=\sum_D|\gggg(e_D,\nabb\ \!\! ^{(i)\!}O)|^2\nn\\
 &&\ML\ML=\sum_D|\sum_C\gggg(e_D,\nabb\ \!\!f^{(i)}_Ce_{C})|^2
 =\sum_D|\sum_C\nabb\ \!\!f^{(i)}_C\gggg(e_D,e_{C})+\sum_C f^{(i)}_C\gggg(e_D,\nabb\ \!e_{C})|^2\nn\\
 &&\ML\ML\leq 2\sum_D|\nabb f^{(i)}_D|^2+2\sum_D\sum_C|f^{(i)}_C|^2|\gggg(e_D,\nabb\ \!e_{C})|^2\leq 2^2|\nabb f^{(i)}|^2
 +2^3|f^{(i)}|^2|\gggg(e_{\c},\nabb\ \!e_{\c})|^2\nn\\
 \eea
 where
 \bea
 &&|\gggg(e_{\c},\nabb\ \!e_{\c})|^2=\sup_{C,D}|\gggg(e_D,\nabb\ \!e_{C})|^2\nn\\
 &&|\nabb f^{(i)}|^2=\sup_C|\nabb f^{(i)}_C|^2\ \ ,\ \ |f^{(i)}|^2=\sup_C|f^{(i)}_C|^2\ .
 \eea
 The estimate for $|\nabb f^{(i)}|^2$ and for $ |f^{(i)}|^2$ are better than those inductively assumed,\footnote{Apart from corrections at the level of metric components.} therefore we are left only with the estimate of  $|\gggg(e_D,\nabb\ \!e_{C})|^2$ obtained in subsection \ref{S.s8.4}. More in general we have
 \bea
 &&\ML\ML
\nabb^J{^{(i)\!}O}=\sum_C\nabb^Jf^{(i)}_C(r,\theta,\phi) e_{C}=\sum_C\sum_{k=0}^J\cbin{J}{k}(\nabb^kf^{(i)}_C(r,\theta,\phi))\nabb^{J-k}e_{C}\ \ \ \ 
 \eea
and, for  a generic norm,
\bea
 &&\ML\ML|\nabb^J{^{(i)\!}O}|\leq \sum_C\sum_{k=0}^J\cbin{J}{k}|\nabb^kf^{(i)}_C| |\nabb^{J-k}e_{C}|
 \leq \sum_C\sum_{k=0}^J\cbin{J}{k}|\nabb^kf^{(i)}_C|\sqrt{\sum_D\gggg(e_D,\nabb^{J-k}e_C)^2}\nn\\
 &&\ \ \leq \sum_C\sum_D\sum_{k=0}^J\cbin{J}{k}|\nabb^kf^{(i)}_C||\gggg(e_D,\nabb^{J-k}e_C)| \ .
\eea
 Therefore
 \bea
 &&\ML\ML|\nabb^J{^{(i)\!}O}|_{p,S} \leq \sum_C\sum_D\sum_{k=0}^J\cbin{J}{k}|\nabb^kf^{(i)}_C|_{\infty,S}|\gggg(e_D,\nabb^{J-k}e_C)|_{p,S}\nn\\
 &&\ \ \ \ \ \  \leq2^2\sum_{k=0}^J\cbin{J}{k}|\nabb^kf^{(i)}|_{\infty,S}\left(\sup_{C,D}|\gggg(e_D,\nabb^{J-k}e_C)|_{p,S}\right)\ .\ \ 
\eea
As we already have the control of the $|\gggg(e_D,\nabb^{J-k}e_C)|_{p,S}$ norms, recalling that these norms are different from zero even in the Minkowski case, the final result is 
 \bea
 &&\ML\ML|\nabb^J{^{(i)\!}O}|_{p,S} \leq c\!\left(\frac{J!}{J^{\a}}\frac{e^{(J-2)(\de+\underline{\Ga}(\la))}}{\ro^{J}}\right)\ .
 \eea
 
 \section{The complete results of Section \ref{S.s initial data}}\label{S.13}
 \subsection{The initial data on the outgoing cone}
\NI Let us write the transport equations which the initial data have to satisfy along $C_0$, see equations (I;2.28), (I;2.30),
\beaa
&&
\frac{\pr}{\pr\nu}\ga_{ab}=2\oom\chi_{ab}\ \ ,\ \ 
\frac{\pr}{\pr\nu}\log\oom=-2\oom\om\nn
\eeaa
\beaa
&&\ML\ML\dd_4\tr\chi+\frac{1}{2}(\tr\chi)^2+2\om\tr\chi+|\chih|^2=0\nn\\
&&\ML\ML\dddd_4\zeta+\zeta\chi+\tr\chi\zeta-\divv\chi+\nabb\tr\chi+\ddb_4\nabb\log\oom=0\nn\\
&&\ML\ML\dd_4\tr\chib+\tr\chi \tr\chib-2\om \tr\chib+2\divv(\ze\!-\!\nabb\log\oom)\!-\!2|\ze\!-\!\nabb\log\oom|^2=
-2{\bf K}\nn\\
&&\ML\ML\dddd_4\chibh\!+\!\frac{1}{2}\tr\chi\chibh\!+\!\frac{1}{2}\tr\chib\chih\!-\!2\om\chibh\!+\!\nabb\hot(\ze\!-\!\nabb\!\log\oom)
\!-\!(\ze\!-\!\nabb\!\log\oom)\hot(\ze\!-\!\nabb\!\log\oom)\!=\!0\nn\\
&&\ML\ML\dd_4\omb\!-\!2\om\omb\!-\!\ze\c\nabb\log\oom\!-\!\frac{3}{2}|\ze|^2\!+\!\frac{1}{2}|\nabb\log\oom|^2\!+\frac{1}{2}\!\big({\bf
K}\!+\!\frac{1}{4}\tr\chi\tr\chib\!-\!\frac{1}{2}\chih\c\chibh\big)\!=\!0\ .\nn
\eeaa
On $C_0$ we assign freely $\oom$ and $\chih$. We require them to be analytic functions on the whole $C_0$ satisfying, for any $J>0$, the following bounds,  see also inequalities \ref{123}, 
\bea\label{freest}
&&|r^{J+2+\ep-\frac{2}{p}}\nabla^J\log\oom|_{p,S} \leq C^{(0)}_3e^{(J-2)(\de_0+\underline{\Ga}_0(\la))}\frac{J!}{J^\a}\frac{1}{\ro_{0,0,1}^{J}}\nn\\
&&|r^{J+\frac{5}{2}+\ep-\frac{2}{p}}\nabla^J\chih|_{p,S}\leq C^{(0)}_1e^{(J-2)(\de_0+\underline{\Ga}_0(\la))}\frac{J!}{J^\a}\frac{1}{\ro_{0,0,1}^{J}}\ .\eql{chihindata}
\eea
The strategy to prove the existence of the analytic initial data on $C_0$ is the following one:
\smallskip

\NI i) We assume that on $S_0$ is assigned as a real analytic function satisfying a p.d.e. equation in the angular variables we describe later on and such that, for any $J>0$, satisfies the following norm bound, where $U=\oom^{-1}\tr\chi$, with $\ro_{0,0}<\ro_{0,0,1}$,
\bea
|r^{J+2-\frac{2}{p}}\nabb^JU|_{p,S_0}\leq C^{(0,0)}_0e^{(J-2)(\de_0+\underline{\Ga}_0(\la))}\frac{J!}{J^\a}\frac{1}{\ro_{0,0}^{J}}\ .\eql{Soboundtrchi}
\eea
\smallskip

\NI ii) We prove a simplified version of Theorem \ref{T3.1}, namely we assume inequality \ref{Soboundtrchi} 
and on $C_0$ we make the following inductive assumptions, with $J<N$, for $U$ and for $\ze$  and prove 
the following theorem,
\begin{theorem}\label{T3.1initialdata1}
Assume that on $S_0$ is assigned as a real analytic function satisfying for any $J$ the bound, 
\bea
|r^{J+2-\frac{2}{p}}\nabb^JU|_{p,S_0}\leq C^{(0,0)}_0e^{(J-2)(\de_0+\underline{\Ga}_0(\la))}\frac{J!}{J^\a}\frac{1}{\ro_{0,0}^{J}}\ .\eql{Soboundtrchi1}
\eea
Assume that on $C_0$, $U$ and $\ze$ satisfy the following initial conditions with $J<N$
\bea
&&|r^{J+2-\frac{2}{p}}\nabb^JU|_{p,S}(\la_0,\nu)\leq C^{(0)}_0{(J-2)(\de_0+\underline{\Ga}_0(\la))}\frac{J!}{J^\a}\frac{1}{\ro_{0,0,1}^{J}}\\
&&|r^{J+2-\frac{2}{p}}\nabb^J\ze|_{p,S}(\la_0,\nu)\leq C^{(0)}_4e^{(J-2)(\de_0+\underline{\Ga}_0(\la))}\frac{(J)!}{(J)^\a}\frac{1}{\ro_{0,0,1}^{J}}\nn
\eea
 then $\nabb^NU(\la_0,\nu)$ satisfies the following estimate
\bea
|r^{N+2-\frac{2}{p}}\nabb^NU|_{p,S}(\la_0,\nu)\leq C^{(0)}_0e^{(J-2)(\de_0+\underline{\Ga}_0(\la))}\frac{N!}{N^\a}\frac{1}{\ro_{0,0,1}^{N}}
\eea
for any $\nu$,  with $\ro_{0,0,1}<\ro_{0,0}$, provided that the initial data are small. 
\end{theorem}
\NI{\bf Proof:} The proof of this theorem mimicks the proof of Theorem \ref{T3.1}, but is significantly simpler for the following reasons:
\smallskip

\NI a) In Theorem \ref{T3.1} to control $\nabb^NU$ we have to look at the transport equation for $\nabb^N\Us$ to avoid when integrated, logarithmic divergent contributions. here due to the better decay assumed in \ref{chihindata} for the initial data, we can directly use the transport equation for  $\nabb^NU$ as we show explicitly in the following remark.
Moreover, looking at that theorem, we needed to express the $\nabb^N\chih$ norms in terms of the $\nabb^NU$ norms. This required the use of the Hodge equation for $\chih$ and  a control of the $\nabb^{N-1}\b$ norm. 
Here this is avoided as $\chih$ is assigned freely on the whole $C_0$ and $\nabb^{N-1}\b$ does not appear in the transport equation for $\nabb^NU$. Finally we do not have, in the present case, to integrate from the last slice.
\medskip

\NI b) In the control of the terms denoted collectively by $\{(good)_1\}$, see equation \ref{frog}, contribution due to Riemann components appear, with a lower derivative order; due to a signature argument the only contribution is due to $\nabb^J\b$ with $J\leq N-2$; again we do not need any hyperbolicity argument as we can use the relation,
 \[\nabb^J\b=\nabb^J\big[\nabb\tr\chi-\divv\chi-\ze\c\chi+\ze\tr\chi\big]\]
 and control this term using the previous assumptions, \ref{123},  for $U$ and $\ze$ and the knowledge of $\chih$ and its angular derivatives. Moreover in the estimates of lower order terms even $\nabb^J\oom$ factors are present, which, again, can be controlled easily as $\oom$ is assigned freely on $C_0\cup\Cb_0$.
 \smallskip
 
 \NI c) All this guarantees that the proof of Theorem \ref{T3.1initialdata1} follows in a simpler way from the proof of Theorem \ref{T3.1}. Still one step has to be done. Observe, in fact,  that this inductive result requires a starting point namely that for $J=1$ we have separately proved the estimate, this issue is delicate and has been proved in \cite{Ca-Ni:char}, see also subsection \ref{gase}\footnote{The analogous estimate for $J=0$ which we also need is an estimate for $\oom^{-1}\!\left(\tr\chi-\frac{\overline{\oom\tr\chi}}{\oom}\right)$. }
\bea
|r^{1+2-\frac{2}{p}}\nabb U|_{p,S}(\la_0,\nu)\leq C^{(0)}_0e^{(1-2)(\de_0+\underline{\Ga}_0(\la))}\frac{1}{\ro_{0,0,1}}\ .
\eea
\smallskip

\NI {\bf Remark:}\ 

\NI {\em As said in a) we prove Theorem \ref{T3.1initialdata1} in a different way from the proof of Theorem  \ref{T3.1} which presents some advantages when we prove the norms of $\nabb^J\ze$ on $C_0$. This is based on the fact that the $r$ decay required \footnote{To have the ${\cal Q}_0$ norms bounded.} in the initial conditions \ref{chihindata} is, due to $\ep>0$, stronger than the one proved in the internal region. This allows to control directly the $\nabb$ derivatives of $U=\oom^{-1}\tr\chi$ instead the to obtain it through $\Us=\nabb\oom^{-1}\tr\chi+\oom^{-1}\tr\chi\eta$. In fact the transport equation for $\nabb^JU$ is,
\bea
&&\ML\frac{\Dbb{(\nabb^NU)}}{\partial\nu}+{\oom}\big((N+2)\frac{\tr\chi}{2}+N\chih\big)\c{(\nabb^NU)}+2\chih\c(\nabb^N\hat{\chi})
=-\bigg\{(good)_1\bigg\}\ \ \ \ \ \ \ \ \ \ \ \ \ \eql{110L}
\eea
where
\bea\label{frog}
&&\bigg\{(good)_1\!\bigg\}=\left\{\sum_{k=1}^{N-1}\left(\!\!\!\!\begin{array}{c}N\\k\\
\end{array}\!\!\!\!\right)(\nabb^{k}\chih)\!\c\!(\nabb^{N-k}\chih)-\frac{U}{2}\sum_{k=1}^{N}
\left(\!\!\!\begin{array}{c}N\\k\\\end{array}\!\!\!\right)(\nabb^k\oom^2)(\nabb^{N-k}U)
\right.\nn\\
&&\ML\ML\ML\ML\ML\ML\left.+\frac{1}{2}\sum_{k=1}^{N-1}\left(\!\!\!\!\begin{array}{c}N\\k\\
\end{array}\!\!\!\!\right)(\nabb^k\oom\tr\chi)\!\c\!(\nabb^{N-k}U)
+\sum_{k=1}^{N-1}\left(\!\!\!\begin{array}{c}N\\k+1\\\end{array}\!\!\!\right)(\nabb^{k}\oom\chi)\!\c\!(\nabb^{N-k}U)
-\sum_{k=1}^{N-1}\left(\!\!\!\begin{array}{c}N\\k+1\\\end{array}\!\!\!\right)(\nabb^{k-1}C)\c(\nabb^{N-k}U)
\bigg\}\ .\ \ \ \ \ \ \ \ \ \right.\eql{good1L}
\eea
Moreover  the following inequality holds:
\bea
\frac{\partial}{\partial\nu}|r^{(N+2-\frac{2}{p})}\nabb^NU|_{p,S}\leq  |r^{(N+2-\frac{2}{p})}L|_{p,S}\ \ \ \eql{diffest1a}
\eea
where
\bea
\ML|L|=\left[N|\chih\c(\nabb^NU)|+2|\chih\c(\nabb^N\hat{\chi})|+\big|\!\big\{(good)_1\big\}\!\big|\right]\!+\!\left[\frac{(N+2)}{2}|{\oom}\tr\chi-\overline{{\oom}\tr\chi}||\nabb^NU|\right].\ \ \ \ \ \ \ \eql{defL}
\eea
\bea
&&\ML\ML\ML\big|r^{(N+2)-\frac{2}{p}}\nabb^NU\big|_{p,S}(\la,\nu)\leq \big|r^{(N+2)-\frac{2}{p}}\nabb^NU\big|_{p,S}(\la,\nu_0)\nn\\
&&\ML\ML\ML+N\!\!\int_{\nu_0}^{\nu}d\nu'\big|\chih\c r^{(N+2)-\frac{2}{p}}(\nabb^NU)|_{p,S}(\la,\nu')
+\frac{(N+2)}{2}\!\!\int_{\nu_0}^{\nu}d\nu'\big|({\oom}\tr\chi-\overline{{\oom}\tr\chi}) r^{(N+2)-\frac{2}{p}}(\nabb^NU)|_{p,S}(\la,\nu')\nn\\
&&\ML\ML\ML+2\!\int_{\nu_0}^{\nu}d\nu'\big|r^{(N+2)-\frac{2}{p}}\chih\c(\nabb^N\hat{\chi})\big|_{p,S}(\la,\nu')
+\int_{\nu_0}^{\nu}d\nu'\big|r^{(N+2)-\frac{2}{p}}\big\{(good)_1\big\}\big|_{p,S}(\la,\nu')\ .\eql{122yL}
\eea

\smallskip

\NI As already said in the footnote in the proof of Lemma \ref{L1.3} it is the term
\[\frac{U}{2}\sum_{k=1}^{N}\left(\!\!\!\begin{array}{c}N\\k\\\end{array}\!\!\!\right)(\nabb^k\oom^2)(\nabb^{N-k}U)\]
which in the transport equation in the interior produces a $\log$ divergence forcing us to use $\Us$. Here $\nabb^k\oom^2$ decaying as $r^{-(k+1)-\ep}$ avoids the problem and allows to prove in a direct way the estimate\ \footnote{Nevertheless the initial data estimate for $\nabb^J\tr\chi$ cannot have a decay better than $O\left(r^{-(3+J)}\right)$ and no extra $r^{-\ep}$ factor is possible, due to the fact that we integrate starting from $S_0$.} 
\bea
|r^{N+2-\frac{2}{p}}\nabb^NU|_{p,S}(\la_0,\nu)\leq C^{(0)}_0e^{(N-2)(\de_0+\underline{\Ga}_0(\la))}\frac{N!}{N^\a}\frac{1}{\ro_{0,0,1}^{N}}\ .
\eea
More important is the fact that in the recursive proof of $\nabb^N\tr\chi$ the Riemann components appear only at order $\nabb^{N-2}\Psi$ which, at its turn, implies that when $\Psi=\b$, $\nabb^{N-2}\b$ depends only on $\nabb^{N-1}\tr\chi$ and on $\nabb^{N-2}\ze$ which we know by inductive assumptions.}

\subsubsection{The complete control of $\nabb^J\tr\chi$ on $C_0$}\label{gase}

The previous discussion about how we control on $C_0$ $\nabb^J\tr\chi$ suffers a bit of oversimplification, in fact the situation is slightly more complicated as discussed in \cite{Ca-Ni:char}, let us recall its main aspect. The transport equation for $U\!=\!\oom^{-1}\tr\chi$,
\[\frac{\pr}{\pr\nu}U+\frac{1}{2}(\oom\tr\chi)U+\chih^{ac}\chih^{bd}\ga_{ab}\ga_{cd}=0\]
depends on the $S^2$ metric $\ga$ on $C_0$ which, at its turn, has to satisfy the transport equation
\[\frac{\pr}{\pr\nu}\ga_{ab}=(\oom\tr\chi)\ga_{ab}+2\oom\chih_{ab}.\]
Therefore we have to deal with the coupled system of equations,
\bea
&&\frac{\pr}{\pr\nu}\ga_{ab}-(\oom\tr\chi)\ga_{ab}-2\oom\chih_{ab}=0\eql{syst1}\\
&&\frac{\pr}{\pr\nu}U+\frac{1}{2}(\oom\tr\chi)U+\chih^{ac}\chih^{bd}\ga_{ab}\ga_{cd}=0\ .\nn
\eea
The strategy mimics the one in \cite{Ca-Ni:char}, we assign $\chih$ on $C_0$ with the due regularity then we solve on $C_0$ the system \ref{syst1} obtaining $\ga_{ab}$ and $U$ (provided we assign them on $S_0$). Once we control $\ga$ the previous discussion shows how to control $\nabb^JU$ for any $J$ on $C_0$ and, using again the second equation of \ref{syst1}, we control all the mixed derivatives of $U$ with respect to $\nabb$ and to $\ddb_{\nu}$.

\NI Therefore to complete the initial data construction, also the quantities at the ``metric level" have to be assigned. To control  the analyticity of the metric components, for instance of $\ga_{ab}$; the procedure is a little different as, being $\nabb\ga=0$, we have to use non covariant derivatives from the beginning. To control the non covariant partial derivatives of $\ga_{ab}$ on $C_0$ we derive the first evolution equation of \ref{syst1},
\bea
\frac{\pr}{\pr\nu}\pr^J\ga_{ab}-\sum_{k=0}^J\cbin{J}{k}(\pr^k\oom\tr\chi)\pr^{J-k}\ga_{ab}-2\sum_{k=0}^J\cbin{J}{k}(\pr^k\oom)\pr^{J-k}\chih_{ab}=0\ ,\ \ \ \ \ 
\eea
and proceed in a standard way, recalling that we are able to control of the non covariant partial derivatives for the connection coefficients once we control the covariant ones. \footnote{Of course in the inductive procedure we do not need to consider the equations for $\nabb^J U$ and $\pr^J \ga$ as coupled.}
\smallskip

\NI Next step is to control on $C_0$ the norm of $\ze$ and its tangential derivatives; we proceed  using the fact that the control of $\nabb^NU$ requires only the inductive assumptions for $\nabb^J\ze$ with $J\leq N-2$. The evolution equation of $\ze$ along the outgoing cones is \footnote{this equation cannot be used for the internal estimates.}
\bea
\dddd_{\nu}\zeta+\frac{3}{2}\oom\tr\chi\zeta+\oom(\zeta\chih-\divv\chih+\frac{1}{2}\nabb\tr\chi)+\ddb_{\nu}\nabb\log\oom=0\ .
\eea
From it we easily derive the transport equation along $C_0$ for $\nabb^N\ze$,
\bea
&&\ML\ML\ML\dddd_{\nu}\nabb^N\zeta-[\dddd_{\nu},\nabb^N]\ze=\nabb^N\dddd_{\nu}\ze
=-\nabb^N\left(\frac{3}{2}\oom\tr\chi\zeta+\oom(\zeta\chih-\divv\chih+\frac{1}{2}\nabb\tr\chi)+\ddb_{\nu}\nabb\log\oom\right)\ ,\ \ \ \ \ \ \ \ \ \ 
\eea
therefore
\bea
&&\ML\ML\ML\dddd_{\nu}\nabb^N\zeta=-\nabb^N\left(\frac{3}{2}\oom\tr\chi\zeta+\oom(\zeta\chih-\divv\chih+\frac{1}{2}\nabb\tr\chi)+\ddb_{\nu}\nabb\log\oom\right)+[\dddd_{\nu},\nabb^N]\ze\\
&&\ML\ML\ML\ML=-3\frac{\oom\tr\chi}{2}\nabb^N\ze-\frac{3}{2}\sum_{k=1}^N\cbin{N}{k}(\nabb^k\oom\tr\chi)\nabb^{N-k}\ze
-\nabb^N\left(\oom(\zeta\chih-\divv\chih+\frac{1}{2}\nabb\tr\chi+\ddb_4\nabb\log\oom)\right)+[\dddd_{\nu},\nabb^N]\ze\ .\nn
\eea
As, see Lemma \ref{L1.3},
\bea
&&\ML\ML[\dddd_{\nu},\nabb^N]\ze=\sum_{k=0}^{N-1}\left(\!\!\!\begin{array}{c}N\\k\\
\end{array}\!\!\!\right)(\nabb^{N-1-k}\oom\chi)\c\nabb^{k+1}\ze-\sum_{k=0}^{N-1}\left(\!\!\!\begin{array}{c}{N+1}\\k\\
\end{array}\!\!\!\right)(\nabb^{N-1-k}C)\c\nabb^{k}\ze\nn\\
&&\ML\ML=N\frac{\oom\tr\chi}{2}\nabb^N\ze+\frac{1}{2}\sum_{k=0}^{N-2}\cbin{N}{k}(\nabb^{N-1-k}\oom\tr\chi)\c\nabb^{k+1}\ze+
\sum_{k=0}^{N-1}\left(\!\!\!\begin{array}{c}N\\k\\
\end{array}\!\!\!\right)(\nabb^{N-1-k}\oom\chih)\c\nabb^{k+1}\ze\nn\\
&&\ML\ML-\sum_{k=0}^{N-1}\left(\!\!\!\begin{array}{c}{N+1}\\k\\ \end{array}\!\!\!\right)(\nabb^{N-1-k}C)\c\nabb^{k}\ze\ ,
\eea
the final expression for the transport equation is
\bea
&&\ML\ML\ML\ML\dddd_{\nu}(\nabb^N\zeta)+\frac{(N+3)}{2}\oom\tr\chi(\nabb^N\zeta)+\oom\chih(\nabb^N\zeta)=
-\nabb^N\!\!\left(\oom(-\divv\chih+\frac{1}{2}\nabb\tr\chi+\ddb_4\nabb\log\oom)\right)\!+\!\{[l.o.t]\}\ \ \ \ \ \ \ \ \ \ \ \ \eql{6.24}
\eea
where
\bea
&&\ML\ML\bigg\{[l.o.t]\bigg\}=\!\frac{1}{2}\!\sum_{k=0}^{N-2}\cbin{N}{k}(\nabb^{N-1-k}\oom\tr\chi)\c\nabb^{k+1}\ze+\left\{-\frac{3}{2}\sum_{k=1}^N\cbin{N}{k}(\nabb^k\oom\tr\chi)\nabb^{N-k}\ze\right.\nn\\
&&\ML\ML\left.-\sum_{k=1}^N\cbin{N}{k}(\nabb^k\oom\chih)\nabb^{N-k}\ze+\sum_{k=0}^{N-1}\left(\!\!\!\begin{array}{c}N\\k\\
\end{array}\!\!\!\right)(\nabb^{N-1-k}\oom\chih)\c\nabb^{k+1}\ze-\sum_{k=0}^{N-1}\left(\!\!\!\begin{array}{c}{N+1}\\k\\ \end{array}\!\!\!\right)(\nabb^{N-1-k}C)\c\nabb^{k}\ze\right\} .\nn
\eea
From this equation we obtain the associated equation for the $\big|\c\big|_{p,S}$ norm and integrating it from below\footnote{The initial conditions for $\ze$ and the remaining connection coefficients on $S_0$ are fully discussed in Section\ref{S.18}.} we obtain, see also \cite{Ca-Ni:char}, an inequality of the following kind, with $C_{(J)}=O(1)$ for $J>J_0$, and $O(\varepsilon)$ otherwise,
\bea
\big|r^{2+N-\frac{2}{p}}\nabb^N\ze\big|_{p,S}(\la_0,\nu)\leq C_{(J)}\!\left(\frac{(N+1)!}{(N+1)^{\a}}\frac{e^{(N-1)(\de_0+\underline{\Ga}_0(\la))}}{\ro_{0,0,1}^{N+1}}\right)
\eea
And, increasing the constant $C_{(J)}$  and exploiting the factor $\frac 1 {(N+1)^{\a}}$ we obtain:
\bea
\big|r^{2+N-\frac{2}{p}}\nabb^N\ze\big|_{p,S}(\la_0,\nu)\leq C_{(J)}\!\left(\frac{N!}{N^{\a}}\frac{e^{(N-2)(\de_0+\underline{\Ga}_0(\la))}}{\ro_{0,0,1}^{N}}\right)
\eea

assuming as initial data estimates, for $J<N$,
\bea
\big|r^{2+J-\frac{2}{p}}\nabb^J\ze\big|_{p,S}(\la_0,\nu)\leq C_{(J)}\!\left(\frac{J!}{J^{\a}}\frac{e^{(J-2)(\de_0+\underline{\Ga}_0(\la))}}{\ro_{0,0}^{J}}\right)\ .
\eea


\NI{\bf Remark:} 

\NI {\em Observe that $\ze$ on $C_0$ cannot gain an extra $r^{-\ep}$ decay due to the term $\nabb^{N+1}\tr\chi$ present in its transport equation. Nevertheless from the required behavior of $\si$ on $C_0\cup\Cb_0$, see later, we require that on $S_0$ $\ze$ satisfies, 
for any $J$,
\bea
\big|r^{\frac{5}{2}+J+\ep-\frac{2}{p}}\nabb^{J-1}\curll\ze\big|_{p,S}(\la_0,\nu_0)\leq c\!\left(\frac{J!}{J^{\a}}\frac{e^{(J-2)(\de_{0}+\Ga_{0})}}{\ro_{0,0}^J}\right)\ .
\eea}

\NI To control $\nabb^N\chib$ on $C_0$ we use the transport equation, along $C_0$, for $\tr\chib$ and for $\chibh$; in this case the loss of derivatives implied by these equations is not harmful as we have already the control of all the $\nabb$ derivatives of $\ze$ (and the $\pr$ derivatives of $\ga$) which requires only the previous control of $\chi$. 

\NI The control of $\omb$ and its angular derivatives is, viceversa, a bit more delicate and require that $\omb$ satisfies a condition on $S_0$ as discussed in \cite{Ca-Ni:char}. 
\smallskip

\NI{More precisely if we try to obtain the initial data norm estimates on $C_0$ for $\omb$ simply requiring that the transport equation,
\[\dd_4\omb\!-\!2\om\omb\!-\!\ze\c\nabb\log\oom\!-\!\frac{3}{2}|\ze|^2\!+\!\frac{1}{2}|\nabb\log\oom|^2\!+\frac{1}{2}\!\big({\bf
K}\!+\!\frac{1}{4}\tr\chi\tr\chib\!-\!\frac{1}{2}\chih\c\chibh\big)\!=\!0\]
be satisfied we obtain the estimate than the one we assumed in \ref{123}, 
\bea
\big||\la_0|^2r^{J-\frac{2}{p}}\nabb^J\omb\big|_{p,S}(\la_0,\nu)\leq C_{(J)}\!\left(\frac{J!}{J^{\a}}\frac{e^{(J-2)(\de_0+\underline{\Ga}_0(\la))}}{\ro_{0}^J}\right)\eql{weakest}
\eea
with $\ro_{0}<\ro_{0.0}$ and which a decay apparently in disagreement with the internal estimates in \ref{123a}. This apparent contradiction has the following solution: in the original assignment of the initial data and in the Cauchy-Kowalevski solution in a finite region the double null foliation is not specified and the data are given on $C_0\cup\Cb_0$ in the way we have presented now; in particular $\omb$  satisfies the weaker estimate on $C_0$, see subsection \ref{S.sInandLast}. As discussed in subsection \ref{SScanfol}, to extend the region and prove the global existence we integrate the transport equations along the outgoing cones from above and in particular we obtain $\omb$ on the last slice through the choice of a specific $\oom$ on the last slice and the connection coefficients previously estimated independently from $\omb$. In this way the decay estimates for $\omb$ in the internal region and, of course also on $C_0$ are stronger and in particular on $C_0$ we obtain 
\[\big||\la_0|r^{1+J-\frac{2}{p}}\nabb^J\omb\big|_{p,S}(\la_0,\nu)\leq C_{(J)}\!\left(\frac{J!}{J^{\a}}\frac{e^{(J-2)(\de_0+\underline{\Ga}_0(\la))}}{\ro_{0}^J}\right)\ .\]
On the other side, as we have a transport equation for $\omb$ along the outgoing cones and therefore, also on $C_0$, this implies that going, along $C_0$ from $S(\la_0,\nu_*)$ to $S(\la_0,\nu_0)\equiv S_0$ the value of $\omb$ on $S_0$ is assigned, see also  \cite{Ca-Ni:char}.  }
\smallskip

\NI Collecting all these results and the ones for the missing connection coefficients, whose proof we do not report here,  being  simple extensions of the analogous results in \cite{Ca-Ni:char}, Lemma 2.3, we obtain,\footnote{If, as discussed above, we assume an arbitrary foliation on $\Cb_0$, not the specific one induced by the canonical foliation, the estimate on $C_0$ for $\omb$ would be the weaker one, \ref{weakest}.}
\bea
&&\big|r^{1+J+\si(J)-\frac{2}{p}}\nabb^J\tr\chib\big|_{p,S}(\la_0,\nu)\leq C_{(J)}\!\left(\frac{J!}{J^{\a}}\frac{e^{(J-2)(\de_0+\underline{\Ga}_0(\la))}}{\ro_{0}^J}\right)\nn\\
&&\big||\la_0|r^{1+J-\frac{2}{p}}\nabb^J\chibh\big|_{p,S}(\la_0,\nu)\leq C_{(J)}\!\left(\frac{J!}{J^{\a}}\frac{e^{(J-2)(\de_0+\underline{\Ga}_0(\la))}}{\ro_{0}^J}\right)\\
&&\big||\la_0|r^{1+J-\frac{2}{p}}\nabb^J\omb\big|_{p,S}(\la_0,\nu)\leq C_{(J)}\!\left(\frac{J!}{J^{\a}}\frac{e^{(J-2)(\de_0+\underline{\Ga}_0(\la))}}{\ro_{0}^J}\right)\ , \nn\\
&&\big|r^{2+J-\frac{2}{p}}\nabb^J\ze\big|_{p,S}(\la_0,\nu)\leq C_{(J)}\!\left(\frac{J!}{J^{\a}}\frac{e^{(J-2)(\de_0+\underline{\Ga}_0(\la))}}{\ro_{0}^J}\right)\ ,\nn
\eea
with $\ro_{0}<\ro_{0,0,1}<\ro_{0,0}$.
Observe that we can also conclude that on $C_0$ all the remaining connection coefficients satisfy the following estimates, for any $J$,\footnote{Recalling, anyway, that  the first connection coefficients can have a better estimate, with $\ro_{0,0,1}>\ro_{0}$.}
\bea
&&\big|r^{1+J+\si(J)-\frac{2}{p}}\nabb^J\tr\chi\big|_{p,S}(\la_0,\nu)\leq C_{(J)}\!\left(\frac{J!}{J^{\a}}\frac{e^{(J-2)(\de_0+\underline{\Ga}_0(\la))}}{\ro_{0}^J}\right)\nn\\
&&\big|r^{\frac{5}{2}+J+\ep-\frac{2}{p}}\nabb^J\chih\big|_{p,S}(\la_0,\nu)\leq C_{(J)}\!\left(\frac{J!}{J^{\a}}\frac{e^{(J-2)(\de_0+\underline{\Ga}_0(\la))}}{\ro_{0}^J}\right)\nn\\
&&\big|r^{2+J+\ep-\frac{2}{p}}\nabb^J\om\big|_{p,S}(\la_0,\nu_0)\leq C^{(0)}_{2}\!\left(\frac{J!}{J^{\a}}\frac{e^{(J-2)(\de_0+\underline{\Ga}_0(\la))}}{\ro_{0}^J}\right)
\eea
proving the first part of Theorem \ref{main1}.

 \medskip
 
 \NI{\bf Remark:} 
 
 \NI{\em Observe that $\chibh$ on $C_0$ cannot gain an extra $r^{-\ep}$ decay due to the factor $\frac{1}{2}\tr\chi\chibh$ present in its transport equation. Nevertheless as required from the estimates of $\chibh$ on $\Cb_0$, see later, we ask that on $S_0$ $\chibh$ satisfies, for any $J$,
\bea
\big|r_0^{\frac{5}{2}+J+\ep-\frac{2}{p}}\nabb^J\chibh\big|_{p,S}(\la_0,\nu_0)\leq c_{(J)}\!\left(\frac{J!}{J^{\a}}\frac{e^{(J-2)(\de_{0}+\Ga_{0,0})}}{\ro_{0,0}^J}\right)\ .
\eea}

\subsection{The initial data on the incoming cone}
\subsubsection{The connection coefficient $\zeta$ on the initial incoming cone}
Observe first of all that on $\Cb_0$ $\chibh$, $X$ and $\oom$, are assigned in a free way therefore also $\omb$ is assigned freely on $\Cb_0$, the requirement on the norms for these quantities are similar to those for $\chih$ and $\oom$ on $C_0$ namely,
\bea\label{freest2}
&&||\la|^{1+\ep}r^{1+J-\frac{2}{p}}\nabla^J\log\oom|_{p,S} \leq \Cb^{(0)}_3e^{(J-2)(\de_0+\underline{\Ga}_0(\la))}\frac{J!}{J^\a}\frac{1}{\ro_{0,0,1}^{J}}\nn\\
&&||\la|^{\frac{3}{2}+\ep}r^{1+J-\frac{2}{p}}\nabla^J\chibh|_{p,S}\leq \Cb^{(0)}_1e^{(J-2)(\de_0+\underline{\Ga}_0(\la))}\frac{J!}{J^\a}\frac{1}{\ro_{0,0,1}^{J}}\nn\\ .\eql{chihindataint}
&&||\la|^{\frac{3}{2}+\ep}r^{1+J-\frac{2}{p}}\nabla^J X|_{p,S}\leq \Cb^{(0)}_1e^{(J-2)(\de_0+\underline{\Ga}_0(\la))}\frac{J!}{J^\a}\frac{1}{\ro_{0,0,1}^{J}}\ .
\eea
Proceeding as for $\tr\chi$ we control  $\nabb^J\tr\chib$ on $\Cb_0$. The next step is to control $\ze$ whose transport equation is on $\Cb_0$
\bea
\dddd_3\zeta+\frac{3}{2}\tr\chib\zeta+\zeta\c\chibh+\divv\chib-\nabb\tr\chib-\ddb_3\nabb\log\oom=0\ .
\eea
The factor $3/2$ is a problem as it implies that we have a transport equation for $\big|r^{N+3-\frac{2}{p}}\nabb^N\ze\big|_{p,S}$, while we expect that the possible bound is for  $\big|r^{N+2-\frac{2}{p}}\nabb^N\ze\big|_{p,S}$ and to lower the weight factor is problematic as, differently from the $C_0$ case, moving along the incoming cone starting from $S_0$ the radius of $S$ decreases. We proceed in a different way writing the transport equation for $\ze$ along $\Cb_0$,
\bea
\dddd_3\zeta+\tr\chib\zeta+2\ze\c\chibh-(\nabb\tr\chib-\divv\chib-\zeta\tr\chib+\ze\c\chib)-\ddb_3\nabb\log\oom=0\ .
\eea
as
\bea
\dddd_3\zeta+\tr\chib\zeta+2\ze\c\chibh-\ddb_3\nabb\log\oom+\bb=0
\eea
using the relation,
\bea
\bb=-(\nabb\tr\chib-\divv\chib-\zeta\tr\chib+\ze\c\chib)\ .
\eea
The crucial remark is that on $\Cb_0$ all the factors composing $\bb$ are already under control with the exception of $\ze$, therefore we solve simultaneously the two transport equations for $\ze$ and $\bb$,
\bea
&&\ddb_3\zeta+\tr\chib\zeta+2\ze\c\chibh+\bb-(\ddb_3\nabb\log\oom)=0\nn\\
&&\ddb_3\bb+2\tr\chib\ \!\bb+2\omb\bb-\ze\c\aa-(\divv\aa+\nabb\log\oom\c\aa)=0\ ,
\eea
where it is crucial to observe that 
\bea
\aa=-\big(\ddb_3\chibh+\tr\chib+2\omb\chibh\big)
\eea
can be considered as a given data as, on $\Cb_0$ , $\chib$ and $\omb$ are already assigned. These two equations can be rewritten in a slightly different way as
\bea
&&\ddb_{\la}\zeta+\oom\tr\chib\zeta+2\oom\chibh\c\ze+\oom\bb-(\ddb_{\la}\nabb\log\oom)=0\nn\\
&&\ddb_{\la}\bb+2\oom\tr\chib\ \!\bb+2\oom\omb\bb-\oom\aa\c\ze-[\oom(\divv\aa+\nabb\log\oom\c\aa)]=0\ .\ \ \ \ \ 
\eea
We expect that the solution to this system produces the correct estimates for the $\ze$ and $\bb$ norms. In fact we prove the following lemma
\begin{Le}\label{Lemma14.1}
Assuming that on $\Cb_0$ the following estimates hold with ${\tilde\de}>\de\geq 0$, 
\bea
\ML\ML\ML \ddb_{\la}\nabb\oom=O\left(r^{-(2+\ep)}|\la|^{-(1+{\tilde\de})}\right)\ \ ,\ \ 
\aa=O\left(r^{-(1+\ep)}|\la|^{-(\frac{5}{2}+{\tilde\de})}\right) .\ \ \ \ \ \ \eql{Cb0cond1}
\eea
and that on $S_0$ we have
\bea
|r^{2-\frac{2}{p}}\ze|_{p,S}(\la_0,\nu_0)\leq \varepsilon\ \ ,\ \ ||\la|^{\frac{3}{2}}r^{2+\ep-\frac{2}{p}}\bb|_{p,S}(\la_0,\nu_0)\leq \varepsilon\ .\eql{Socond1}
\eea
then the following estimates hold on $\Cb_0$
\bea
|r^{2-\frac{2}{p}}\ze|_{p,S}(\la,\nu_0)\leq \varepsilon\ \ ,\ \ ||\la|^{\frac{3}{2}}r^{2+\ep-\frac{2}{p}}\bb|_{p,S}(\la,\nu_0)\leq \varepsilon\ .\eql{Cbocond1}
\eea
\end{Le}
\NI{\bf Proof:} See Appendix \ref{S.18}\ .
\smallskip

\NI As before we do not report the proofs for the decays of the other connection coefficients, being them only a repetition of the ones given in \cite{Ca-Ni:char}
\medskip
\subsubsection{The control of the required decay for the Riemann components on $C_0$}
To complete the discussion about the initial data on $C_0$ we have to show that the proved results allow to fulfill the asymptotic behaviour \ref{rieminidecay}. Let us start to consider $\a$ whose expression is\footnote{$\a$ (and $\aa$) plays a different role between the null Riemann components. In fact here its behavior can be assigned simply assigning $\oom$ and $\chih$ on $C_0$. The same does not happen for $\b$ as, in its case, $\tr\chi$ and $\ze$ cannot be assigned freely as they have to satisfy some constraint equations, see the discussion in (I).}
\[\a=-[\dddd_4\chih+\tr\chi\chih-(\dd_4\log\oom)\chih]\ .\]
The expected asymptotic behavior for $\a$ is satisfied provided that the estimates \ref{chihindata} for $\chih$ and $\oom$ on $C_0$ are satisfied. To prov the expected asymptotic behavior for $\b$ requires more work and an extra condition; in fact its explicit expression,
\beaa
\b=\nabb\tr\chi-\divv\chi-\zeta\c\chi+\zeta\tr\chi\ ,\eql{betaexpr}
\eeaa
seems not to provide the right decay, but if we integrate its evolution equation along the outgoing cone, $C_0$,
\beaa
\ddb_{\nu}\b+2\oom\tr\chi\b=2\oom\om\b-\oom\left[\divv\a-(\ze+\nabb\log\oom)\a\right]\ ,\eql{1.70cz}
\eeaa
we obtain
\bea
|r^{4-\frac{2}{p}}\b|_{p,S}(\la_0,\nu)\leq c\left(|r^{4-\frac{2}{p}}\b|_{p,S}(\la_0,\nu_0)
+\int_{\nu_0}^{\nu}d\nu'\left(\big|r^{4-\frac{2}{p}}\oom\ \!\ze\c\a\big|_{p,S}+\big|r^{4-\frac{2}{p}}L_2\big|_{p,S}\right)\!(\la_0,\nu')\right)\nn
\eea
where
\[L_2=\oom\left(\divv\a-\nabb\log\oom\c\a\right)\ .\]
Dividing both sides by ${r^{(\frac{1}{2}-\ep)}(\la_0,\nu)}$ and recalling that, for $\nu'\leq \nu$,
\[r(\la_0,\nu)^{-1}<r(\la_0,\nu')^{-1}<r(\la_0,\nu_0)^{-1}\ ,\]
we obtain
\bea
\ML\ML|r^{\frac{7}{2}+\ep-\frac{2}{p}}\b|_{p,S}(\la_0,\nu)\leq c\!\left(|r^{\frac{7}{2}+\ep-\frac{2}{p}}\b|_{p,S}(\la_0,\nu_0)
+\frac{1}{r^{(\frac{1}{2}-\ep)}(\la_0,\nu)}\!\int_{\nu_0}^{\nu}d\nu'\left(\big|r^{4-\frac{2}{p}}\oom\ \!\ze\c\a\big|_{p,S}+\big|r^{4-\frac{2}{p}}L_2\big|_{p,S}\right)\!\right)\nn
\eea
and the integral part is bounded due to the previous estimates. The same procedure can be applied to estimate all the angular derivatives of $\b$.\footnote{In fact to all the tangential derivatives on $C_0$, therefore also for the $\ddb_{\nu}$ derivatives, this last problem has already been treated in the general case and we do not have to repeat here.} It remains, nevertheless, to prove that on $S_0$ we have
\[|r^{\frac{7}{2}+\ep-\frac{2}{p}}\b|_{p,S}(\la_0,\nu_0)\leq c\varepsilon\ .\]
Looking again at the expression for $\b$ in terms of the connection coefficients, \ref{betaexpr}, on $S_0$ the following condition must hold, where $r_0=r(\la_0,\nu_0)=|\la_0|=\nu_0$,
\bea
\nabb\tr\chi-\ze\tr\chi=O(\varepsilon r_0^{-(\frac{7}{2}+\ep)})\ .
\eea

\NI Next step is devoted to the the decay of the null Riemann components $\ro$ and $\si$; from their evolution equations and their explicit expressions, see also \cite{Ca-Ni:char}, it follows that we obtain the required decay for $(\ro-\overline{\ro})$ provided that on $S_0$
\bea
{\bf K}-\overline{\bf K}+\frac{1}{4}\!\left(\tr\chi\tr\chib-\overline{\tr\chi\ \!\tr\chib}\right)=O(\varepsilon r_0^{-\frac{7}{2}})\ .
\eea
and repeating the same procedure for $\si$ we obtain its correct decay provided that on $S_0$ the following condition holds,\footnote{We already know from the previous requirements that on $S_0$, $(\chibh\wedge\chih)$ behaves much better, as $r_0^{-(\frac{11}{2}+\ep)}$.}
\bea
\curll\zeta=O(\varepsilon r_0^{-\frac{7}{2}})\ .
\eea 
Finally the same procedure used for estimating $\b$ holds for $\bb$, 
 \footnote{Analogous condition is not required for $\aa$ on $C_0$ as $\aa$ is not present in the part of ${\cal Q}_0$ made by an integration along $C_0$.} obtaining the appropriate decay requiring that, on $S_0$, the following condition is satisfied
 \bea
 \nabb\tr\chib+\ze\tr\chib=O(\varepsilon r_0^{-(\frac{7}{2}+\ep)})\ .
 \eea
 Summarizing we need that on $S_0$ the initial data satisfy the following conditions,
\bea
&&\nabb\tr\chi-\ze\tr\chi=O(\varepsilon r_0^{-(\frac{7}{2}+\ep)})\nn\\
&&{\bf K}-\overline{\bf K}+\frac{1}{4}\!\left(\tr\chi\tr\chib-\overline{\tr\chi\ \!\tr\chib}\right)=O(\varepsilon r_0^{-\frac{7}{2}})\nn\\
&&\curll\zeta=O(\varepsilon r_0^{-\frac{7}{2}})\ .\eql{Socondin}
\eea
\smallskip
\subsection{The mixed derivatives for the initial data. proof of theorem \ref{T11.1}:}\label{mixinix}

\NI \subsubsection{Some useful computations:}

\NI We have to compute $[\ddb_{\nu},\ddb_{\la}]$, let $U=U_{\si_1\nu_2...\si_k}d\om^{\si_1}\otimes d\om^{\si_2}\otimes\c\c\c\otimes d\om^{\si_k}$ a $(0,k)$ $S$-tangent tensor, we have
\bea
&&\ML\ML\ML\ML \ddb_{\nu}\ddb_{\la}U=\oom\ddb_4\Pi^{\si_1..\si_k}_{\nu_1..\nu_k}\oom \dd_3U_{\si_1..\si_k}
=\oom\Pi_{\tau_1..\tau_k}^{\nu_1..\nu_k}\dd_4\Pi^{\si_1..\si_k}_{\nu_1..\nu_k}\oom \dd_3U_{\si_1..\si_k}
\eea
\bea
&&\ML\ML\ML\ML \ddb_{\la}\ddb_{\nu}U=\oom\ddb_3\Pi^{\si_1..\si_k}_{\nu_1..\nu_k}\oom \dd_4U_{\si_1..\si_k}
=\oom\Pi_{\tau_1..\tau_k}^{\nu_1..\nu_k}\dd_3\Pi^{\si_1..\si_k}_{\nu_1..\nu_k}\oom \dd_4U_{\si_1..\si_k}\ .
\eea
We have 
\beaa
&&\ML\Pi^{\nu_1..\nu_k}_{\tau_1..\tau_k}\dd_4\Pi^{\si_1..\si_k}_{\nu_1..\nu_k}(\oom\dd_3U)_{\si_1..\si_j..\si_k}\nn\\
&&\ML\ML=\sum_{j=1}^k\Pi^{\si_1..\hat{\si}_j..\si_k}_{\tau_1..\hat{\tau}_j..\tau_k}(\Pi^{\nu_j}_{\tau_j}\dd_4\Pi^{\si_j}_{\nu_j})(\oom\dd_3U)_{\si_1..\si_j..\si_k}+\Pi^{\si_1..\si_k}_{\tau_1..\tau_k}\dd_4(\oom\dd_3U)_{\si_1..\si_k}\nn\\
&&\ML\ML=-\sum_{j=1}^k\Pi^{\si_1..\hat{\si}_j..\si_k}_{\tau_1..\hat{\tau}_j..\tau_k}\left(\Pi^{\nu_j}_{\tau_j}(\dd_4\theta^4)_{\nu_j}e_4^{\si_j}
+\Pi^{\nu_j}_{\tau_j}(\dd_4\theta^3)_{\nu_j}e_3^{\si_j}\right)(\oom\dd_3U)_{\si_1..\si_j..\si_k}\nn\\
&&\ML\ML\ \ \ +\Pi^{\si_1..\si_k}_{\tau_1..\tau_k}\dd_4(\oom\dd_3U)_{\si_1..\si_k}\nn\\
&&\ML\ML\ML\ML=-\sum_{j=1}^k\Pi^{\si_1..\hat{\si}_j..\si_k}_{\tau_1..\hat{\tau}_j..\tau_k}\left(-\frac{1}{2}\gggg(\dd_4e_3,e_C)\theta^C_{\tau_j}\right)e_4^{\si_j}(\oom\dd_3U)_{\si_1..\si_j..\si_k}+\Pi^{\si_1..\si_k}_{\tau_1..\tau_k}\dd_4(\oom\dd_3U)_{\si_1..\si_k}\nn\\
&&\ML\ML\ML\ML=-\frac{1}{2}\sum_{j=1}^k\Pi^{\si_1..\hat{\si}_j..\si_k}_{\tau_1..\hat{\tau}_j..\tau_k}\left(\gggg(\dd_4e_3,e_C)\theta^C_{\tau_j}\right)(\oom\dd_3e_4)^{\si_j}U_{\si_1..\si_j..\si_k}+\Pi^{\si_1..\si_k}_{\tau_1..\tau_k}\dd_4(\oom\dd_3U)_{\si_1..\si_k}\nn\\
&&\ML\ML\ML\ML=-\frac{1}{2}\sum_{j=1}^k\Pi^{\si_1..\hat{\si}_j..\si_k}_{\tau_1..\hat{\tau}_j..\tau_k}\etab_{\tau_j}(\oom\dd_3e_4)^{\si_j}U_{\si_1..\si_j..\si_k}+\Pi^{\si_1..\si_k}_{\tau_1..\tau_k}\dd_4(\oom\dd_3U)_{\si_1..\si_k}\nn\\
\eeaa
Therefore
\bea
&&\ML\ML\ML\ML \ddb_{\nu}\ddb_{\la}U=\oom\ddb_4\Pi^{\si_1..\si_k}_{\tau_1..\tau_k}\oom \dd_3U_{\si_1..\si_k}
=\oom\Pi_{\tau_1..\tau_k}^{\nu_1..\nu_k}\dd_4\Pi^{\si_1..\si_k}_{\nu_1..\nu_k}\oom \dd_3U_{\si_1..\si_k}\\
&&\ML\ML\ML\ML=-\frac{\oom}{2}\sum_{j=1}^k\Pi^{\si_1..\hat{\si}_j..\si_k}_{\tau_1..\hat{\tau}_j..\tau_k}\etab_{\tau_j}(\oom\dd_3e_4)^{\si_j}U_{\si_1..\si_j..\si_k}+\Pi^{\si_1..\si_k}_{\tau_1..\tau_k}\oom\dd_4(\oom\dd_3U)_{\si_1..\si_k}\nn\\
&&\ML\ML\ML\ML=-\oom^2\sum_{j=1}^k\Pi^{\si_1..\hat{\si}_j..\si_k}_{\tau_1..\hat{\tau}_j..\tau_k}\etab_{\tau_j}(\eta\c U)_{\si_1..{\hat\si}_j..\si_k}+\Pi^{\si_1..\si_k}_{\tau_1..\tau_k}\oom\dd_4(\oom\dd_3U)_{\si_1..\si_k}\nn
\eea
Analogously
\bea
&&\ML\ML\ML\ML \ddb_{\la}\ddb_{\nu}U=\oom\ddb_3\Pi^{\si_1..\si_k}_{\tau_1..\tau_k}\oom \dd_4U_{\si_1..\si_k}
=\oom\Pi_{\tau_1..\tau_k}^{\nu_1..\nu_k}\dd_4\Pi^{\si_1..\si_k}_{\nu_1..\nu_k}\oom \dd_3U_{\si_1..\si_k}\\
&&\ML\ML\ML\ML=-\frac{\oom}{2}\sum_{j=1}^k\Pi^{\si_1..\hat{\si}_j..\si_k}_{\tau_1..\hat{\tau}_j..\tau_k}\eta_{\tau_j}(\oom\dd_4e_3)^{\si_j}U_{\si_1..\si_j..\si_k}+\Pi^{\si_1..\si_k}_{\tau_1..\tau_k}\oom\dd_3(\oom\dd_4U)_{\si_1..\si_k}\nn\\
&&\ML\ML\ML\ML=-\oom^2\sum_{j=1}^k\Pi^{\si_1..\hat{\si}_j..\si_k}_{\tau_1..\hat{\tau}_j..\tau_k}\eta_{\tau_j}(\etab\c U)_{\si_1..{\hat\si}_j..\si_k}+\Pi^{\si_1..\si_k}_{\tau_1..\tau_k}\oom\dd_3(\oom\dd_4U)_{\si_1..\si_k}\ .\nn
\eea
Therefore
\bea
&&\ML\ML[\ddb_{\nu},\ddb_{\la}]U_{\tau_1..\tau_k}
=-\oom^2\sum_{j=1}^k\Pi^{\si_1..\hat{\si}_j..\si_k}_{\tau_1..\hat{\tau}_j..\tau_k}\left(\etab_{\tau_j}(\eta\c U)-\eta_{\tau_j}(\etab\c U)\right)_{\si_1..{\hat\si}_j..\si_k}+\Pi^{\si_1..\si_k}_{\tau_1..\tau_k}[\dd_{\nu},\dd_{\la}]U_{\si_1..\si_k}\nn\\
&&\ML\ML\ \ \ \ \ \ \ \ \ \ \ \ \ \ =-\oom^2\sum_{j=1}^k\left(\etab_{\tau_j}(\eta\c U)-\eta_{\tau_j}(\etab\c U)\right)_{\tau_1..{\hat\tau}_j..\tau_k}\\
&&\ \ \ \ \ +\ \!\oom^2\sum_{i=1}^{k}R^{ \tilde{\tau}_i}(e_4,e_3)_{\tau_i}U_{\tau_1..\tilde{\tau}_i..\tau_k}+\Pi^{\si_1..\si_k}_{\tau_1..\tau_k}[\oom e_4,\oom e_3]^{\mu}D_{\mu}U_{\si_1..\si_k}\nn\\
&&\ML\ML\ \ \ \ \ \ \ \ \ \ \ \ \ \ =-\oom^2\sum_{j=1}^k\left(\etab_{\tau_j}(\eta\c U)-\eta_{\tau_j}(\etab\c U)\right)_{\tau_1..{\hat\tau}_j..\tau_k}\\
&&\ \ \ \ \ +\ \!\oom^2\sum_{i=1}^{k}R^{ \tilde{\tau}_i}(e_4,e_3)_{\tau_i}U_{\tau_1..\tilde{\tau}_i..\tau_k}+4\oom^2(\ze\c\nabb U)_{\tau_1..\tau_k}\ .\nn
\eea
Therefore
\bea
&&\ML\ML\ML\ML\sum_{k=0}^{J-1}\ddb_{\nu}^{J-1-k}[\ddb_{\nu},\ddb_{\la}]\ddb^{k}_{\nu}\nabb^P{\hat\om}_{\tau_1..\tau_P}\\
&&\ML\ML\ML\ML=\sum_{k=0}^{J-1}\ddb_{\nu}^{J-1-k}\left[-\oom^2\sum_{j=1}^P\left(\etab_{\tau_j}(\eta\c\ddb^{k}_{\nu}\nabb^P{\hat\om})_{\tau_1..\tilde{\tau}_j..\tau_P}-\eta_{\tau_j}(\etab\c\ddb^{k}_{\nu}\nabb^P{\hat\om})_{\tau_1..\tilde{\tau}_j..\tau_P}\right)\right.\nn\\
&&\ML\ML\ML\left.\ \ \ \ +\oom^2\sum_{j=1}^{P}R^{ \tilde{\tau}_j}(e_4,e_3)_{\tau_j}(\ddb^{k}_{\nu}\nabb^P{\hat\om})_{\tau_1..\tilde{\tau}_j..\tau_P}
+4\oom^2(\ze\c\nabb(\ddb^{k}_{\nu}\nabb^P{\hat\om})_{\tau_1..\tau_P}\right]\ ,\nn
\eea
which we rewrite as
\bea
&&\ML\ML\sum_{k=0}^{J-1}\ddb_{\nu}^{J-1-k}[\ddb_{\nu},\ddb_{\la}]\ddb^{k}_{\nu}\nabb^P{\hat\om}
=\sum_{k=0}^{J-1}\ddb_{\nu}^{J-1-k}\left[-\oom^2 P\left(\etab\otimes(\eta\c\ddb^{k}_{\nu}\nabb^P{\hat\om})-\eta\otimes(\etab\c\ddb^{k}_{\nu}\nabb^P{\hat\om})\right)\right.\nn\\
&&\left.\ \ \ \ \ \ \ \ \ \ \ \ \ \ \ \ \ \ \ \ \ \ \ \ \ \ \ \ \ \ \ +\ \!\oom^2PR(e_4,e_3)\c(\ddb^{k}_{\nu}\nabb^P{\hat\om})+4\oom^2\ze\c\nabb(\ddb^{k}_{\nu}\nabb^P{\hat\om})\right]\ .\ \ \ \ \ \ \ \ \ \ \ \ \ \ \ \ \ \ 
\eea
Let us write explicitly the term 
\[\sum_{h=0}^{P-1}\ddb_{\nu}^J\nabb^{P-1-h}[\nabb,\ddb_{\la}]\nabb^h{\hat\om}\ ,\]
we recall that, as done in the appendix, see \ref{6.6},
\bea
&&[\nabb_{\mu},{\ddb}_{\la}]U_{\nu_1...\nu_k}=-{\oom}\bigg[\sum_{j=1}^k\!
\left(\chib_{\mu\nu_j}\eta^{\si_j}-\chib_{\mu}^{\si_j}\eta_{\nu_j}\right)\!U_{\nu_1..{\si}_j..\nu_k}
-\chib_{\mu}^{\ro}(\nabb_{\ro}U)_{\nu_1..\nu_k}\bigg]\nn\\
&&\ \ \ \ \ \ \ \ \ \ \ \ \ \ \ \ \ \ \ \ \ \ \ \ \ \ 
-\oom\theta^C_{\mu}\sum_{j=1}^k\theta^D_{\nu_j}R^{{\si}_j}(\c,e_C,e_3,e_D)U_{\nu_1..{\si}_j..\nu_k}\nn\\
&&=-{\oom}\sum_{j=1}^k\!
\left[(\chib_{\mu\nu_j}\eta^{\si_j}\!-\!\chib_{\mu}^{\si_j}\eta_{\nu_j})
+\theta^C_{\mu}\theta^D_{\nu_j}R^{{\si}_j}(\c,e_C,e_3,e_D)\right]\!U_{\nu_1..{\si}_j..\nu_k}
+\oom\chib_{\mu}^{\ro}(\nabb_{\ro}U)_{\nu_1..\nu_k}\nn\\
&&=-\sum_{j=1}^k\oom{\underline{C}}^{\si_j}_{\mu\nu_j}U_{\nu_1..{\si}_j..\nu_k}+\oom\chib_{\mu}^{\ro}(\nabb_{\ro}U)_{\nu_1..\nu_k}
\eea
where
\bea
{\underline{C}}^{\si_j}_{\mu\nu_j}=(\chib_{\mu\nu_j}\eta^{\si_j}\!-\!\chib_{\mu}^{\si_j}\eta_{\nu_j})
+\theta^C_{\mu}\theta^D_{\nu_j}R^{{\si}_j}(\c,e_C,e_3,e_D)\ .
\eea
Therefore
\bea
[\nabb_{\mu},{\ddb}_{\la}](\nabb^h{\hat\om})_{\nu_1...\nu_h}=-\sum_{j=1}^h{\underline{C}}^{\si_j}_{\mu\nu_j}(\nabb^h{\hat\om})_{\nu_1..{\si}_j..\nu_h}+\oom\chib_{\mu}^{\ro}(\nabb_{\ro}\nabb^h{\hat\om})_{\nu_1..\nu_h}\ \ \ \ \ \ 
\eea
and
\bea
&&\ML\ML\sum_{h=0}^{P-1}\ddb_{\nu}^J\nabb^{P-1-h}[\nabb,\ddb_{\la}]\nabb^h{\hat\om}_{\nu_1..\nu_h}
=\sum_{h=0}^{P-1}\ddb_{\nu}^J\nabb^{P-1-h}\!\left[-\sum_{j=1}^h{\underline{C}}^{\si_j}_{\mu\nu_j}(\nabb^h{\hat\om})_{\nu_1..{\si}_j..\nu_h}+\oom\chib_{\mu}^{\ro}(\nabb_{\ro}\nabb^h{\hat\om})_{\nu_1..\nu_h}\right]\nn
\eea
which we rewrite in a more compact way as
\bea
&&\ML\ML\sum_{h=0}^{P-1}\ddb_{\nu}^J\nabb^{P-1-h}[\nabb,\ddb_{\la}]\nabb^h{\hat\om}
=\sum_{h=0}^{P-1}\ddb_{\nu}^J\nabb^{P-1-h}\!\left[-\oom h\left(\chib\otimes(\eta\c\nabb^h{\hat\om})-\eta\otimes(\chib\c\nabb^h{\hat\om})\right)\right.\nn\\
&&\left.\ \ \ \ \ \ \ \ \ \ \ \ \ \ \ \ \ \ \ \ \ \ \ \ \ \ \ \ \ \ \ -\oom h R(\c,\c,e_3,\c)\c(\nabb^h{\hat\om})+\oom\chib\c\nabb(\nabb^h{\hat\om})\right]\ .
\eea
In conclusion we have
\bea
&&\ML\ML\sum_{k=0}^{J-1}\ddb_{\nu}^{J-1-k}[\ddb_{\nu},\ddb_{\la}]\ddb^{k}_{\nu}\nabb^P{\hat\om}
=\sum_{k=0}^{J-1}\ddb_{\nu}^{J-1-k}\left[\oom^2 P\left(\etab\otimes(\eta\c\ddb^{k}_{\nu}\nabb^P{\hat\om})-\eta\otimes(\etab\c\ddb^{k}_{\nu}\nabb^P{\hat\om})\right)\right.\nn\\
&&\left.\ \ \ \ \ \ \ \ \ \ \ \ \ \ \ \ \ \ \ \ \ \ \ \ \ \ \ \ \ \ \ +\ \!\oom^2PR(e_4,e_3)\c(\ddb^{k}_{\nu}\nabb^P{\hat\om})+4\oom^2\ze\c\nabb(\ddb^{k}_{\nu}\nabb^P{\hat\om})\right]\nn\\
&&\ML\ML\sum_{h=0}^{P-1}\ddb_{\nu}^J\nabb^{P-1-h}[\nabb,\ddb_{\la}]\nabb^h{\hat\om}
=\sum_{h=0}^{P-1}\ddb_{\nu}^J\nabb^{P-1-h}\!\left[-\oom h\left(\chib\otimes(\eta\c\nabb^h{\hat\om})-\eta\otimes(\chib\c\nabb^h{\hat\om})\right)\right.\nn\\
&&\left.\ \ \ \ \ \ \ \ \ \ \ \ \ \ \ \ \ \ \ \ \ \ \ \ \ \ \ \ \ \ \ -\oom h R(\c,\c,e_3,\c)\c(\nabb^h{\hat\om})+\oom\chib\c\nabb(\nabb^h{\hat\om})\right]\ .
\eea

\subsection{proof of Theorem \ref{T11.1}}\label{mixinix1}

{\bf Theorem \ref{T11.1}}

\NI {\em Assuming the estimates of Theorem \ref{Thinitialdata} for any $N$, and the estimates \ref{freest}, \ref{freest2} for $\{\Omega,\chi,\chib\}$ on $C_0\cup\underline{C}_0$, the estimates \ref{inizz0} for the $\ddb_\nu^J\nabb^P$for $(J-1, P+1)$ with $J+P\leq N$, then the estimates \ref{inizz0} hold for $J+P\leq N$ on $C_0\cup\underline{C}_0$ with the various $F$ constants $O(1)$.}

 \medskip

\NI{\bf Proof:}\ First of all, by the Bianchi equations, see \cite{Kl-Ni:book} equations 2.2.13,  the inductive assumptions and exploiting those of the structure equation which depends on the null Riemann components, see \cite{Kl-Ni:book} equations 3.1.46, we can assume the following estimate on the null Riemann components. 

\bea
&&\big| |\la|^{\underline{\phi}(\Psi)}r^{\phi(\Psi)+N-1-\frac{2}{p}}\ddb_{\nu}^{J-1}\nabb^{P}\Psi\big|_{L^p(S)}\leq C^{(1)}e^{(J-3)\de_0}e^{(J-3)\underline{\Ga}_0(\la)}\frac{N-1!}{(N-1)^\a}\frac{1}{\ro_0^{N-1}}\ ,\ \ \nn\\ 
&&\big| |\la|^{\underline{\phi}(\Psi)+J-1}r^{\phi(\Psi)+P-\frac{2}{p}}\ddb_{\la}^{J-1}\nabb^{P}\Psi\big|_{L^p(S)}\leq C^{(1)}e^{(J-3)\de_0}e^{(J-3)\underline{\Ga}_0(\la)}\frac{N-1!}{(N-1)^\a}\frac{1}{\ro_0^{N-1}}\nn\\
\eea

\NI We go by induction on the number of $\ddb_{\nu}$ derivatives. The base of induction is nothing more that estimates of Theorem{Thinitialdata}. 

\NI First of alll notice that we already have the right estimates for $\ddb_{\nu}^J\nabb^P\chih$, $\ddb_{\nu}^J\nabb^P\underline{\omega}$ and  $\ddb_{\nu}^J\nabb^P\omega$, see remark a) of Theorem \ref{T3.1initialdata1}. 

\NI Let us start with

\medskip 

\NI{\bf  \{$\ddb_{\nu}^J\nabb^P\tr\chi$:\}} 

\smallskip

\NI We start trying to control $\ddb_{\nu}^J\nabb^P\tr\chi$ for $(J,P)$ such that $(J+P)=N$.
Therefore we make the following
assumptions for $({\tilde J},{\tilde P})=(J-k,P+k)$ with $k\geq 1$ and all couples $({\tilde J},{\tilde P})$ such that $({\tilde J}+{\tilde P})<N$.  \footnote{We use for notational simplicity an overall constant $F>F_i$.}
\bea
&&\ML\ML\ML|r^{1+{\tilde J}+{\tilde P}+\si({\tilde P})-\frac{2}{p}}\ddb_{\nu}^{\tilde J}\nabb^{\tilde P}\tr\chi|_{p,S}\leq F\!\left(\frac{({\tilde J}+{\tilde P})!}{({\tilde J}+{\tilde P})^{\a}}\frac{e^{({\tilde J}+{\tilde P}-2)(\de_0+\underline{\Ga}_0(\la))}}{\ro_0^{{\tilde J}+{\tilde P}}}\right)\eql{assmiste1}\\
&&\ML\ML\ML|r^{1+{\tilde J}+{\tilde P}+\si({\tilde P})-\frac{2}{p}}\ddb_{\nu}^{\tilde J}\nabb^{\tilde P}U|_{p,S}\leq F\!\left(\frac{({\tilde J}+{\tilde P})!}{({\tilde J}+{\tilde P})^{\a}}\frac{e^{({\tilde J}+{\tilde P}-2)(\de_0+\underline{\Ga}_0(\la))}}{\ro_0^{{\tilde J}+{\tilde P}}}\right)\nn\\
&&\ML\ML\ML|r^{2+{\tilde J}+{\tilde P}-\frac{2}{p}}\ddb_{\nu}^{\tilde J}\nabb^{\tilde P}\chih|_{p,S}\leq F\!\left(\frac{({\tilde J}+{\tilde P})!}{({\tilde J}+{\tilde P})^{\a}}\frac{e^{({\tilde J}+{\tilde P}-2)(\de_0+\underline{\Ga}_0(\la))}}{\ro_0^{{\tilde J}+{\tilde P}}}\right)\ .\nn
\eea
Starting from the transport equation for  $U=\oom^{-1}\tr\chi$
\bea
\ddb_{\nu}U+\frac{\oom\tr\chi}{2}{U}+|\chih|^2\!=\!0\ ,\eql{109bisabc}
\eea
\bea
&&\ML\ML\ML\ML\ML\ddb_{\nu}^{J}\nabb^PU=\ddb_{\nu}^{J-1}\nabb^P\ddb_{\nu}U+\ddb_{\nu}^{J-1}[\ddb_{\nu},\nabb^P]U
=\ddb_{\nu}^{J-1}\nabb^P\!\left(-\oom\frac{\tr\chi}{2}{U}-|\hat{{\chi}}|^2\right)+\ddb_{\nu}^{J-1}[\ddb_{\nu},\nabb^P]U\nn\\
&&\ML\ML\ML\ML\ML=-\frac{1}{2}\ddb_{\nu}^{J-1}\nabb^P(\oom\tr\chi U)-\ddb_{\nu}^{J-1}\nabb^P|\chih|^2+\ddb_{\nu}^{J-1}[\ddb_{\nu},\nabb^P]U\nn\\
&&\ML\ML\ML\ML\ML=-\frac{1}{2}\sum_{q=0}^{J-1}\sum_{h=0}^P\cbin{J-1}{q}\cbin{P}{h}(\ddb_{\nu}^q\nabb^h\oom\tr\chi)(\ddb_{\nu}^{J-1-q}\nabb^{P-h}U)-\sum_{q=0}^{J-1}\sum_{h=0}^P\cbin{J-1}{q}\!\cbin{P}{h}\!(\ddb_{\nu}^q\nabb^h\chih)\c(\ddb_{\nu}^{J-1-q}\nabb^{P-h}\chih)\nn\\
&&\ML\ML\ML\ML\ML\ \ \ \ +\  \ddb_{\nu}^{J-1}([\ddb_{\nu},\nabb^P]U)\ .
\eea
Recall that 
the following relation holds
\bea
[\nabb^P,\frac{\Dbb}{\partial\nu}]f=\sum_{k=0}^{P-1}\nabb^k[\nabb,\frac{\Dbb}{\partial\nu}]\nabb^{P-k-1}f\ 
\eea
and from equation \ref{3.11}
\beaa
[\nabb_{\mu},\frac{\Dbb}{\partial\nu}]U_{\nu_1...\nu_k}
=-\sum_{j=1}^kC^{\si_j}_{\mu\nu_j}U_{\nu_1..{\si}_j..\nu_k}+\oom\chi_{\mu}^{\ro}(\nabb_{\ro}U)_{\nu_1..\nu_k}\ ,
\eeaa
it follows 
 \bea
&&\ML[\nabb^P,\frac{\Dbb}{\partial\nu}]f=\sum_{k=0}^{P-1}\nabb^k\left(-(P-k-1)C\nabb^{P-k-1}f+\oom\chi\nabb^{P-k}f\right)\\
&&\ML=-\sum_{k=0}^{P-1}(P-k-1)\sum_{J=0}^k\cbin{k}{J}(\nabb^JC)\nabb^{P-1-J}f
+\sum_{k=0}^{P-1}\sum_{J=0}^k\cbin{k}{J}(\nabb^J\oom\chi) \nabb^{P-J}f\ .\nn
\eea
As the following relation holds
\bea
\sum_{k=0}^{P-1}(P-k-1)\sum_{j=0}^k\cbin{k}{j}=\sum_{j=0}^{P-1}\sum_{k=j}^{P-1}(P-k-1)\cbin{k}{j}=\sum_{j=0}^{P-2}\cbin{P}{P-2-j}=\sum_{l=0}^{P-2}\cbin{P}{l}\ \ \ \ \eql{bin1a}
\eea
we can write
 \bea
&&\ML[\frac{\Dbb}{\partial\nu},\nabb^P]f=\sum_{k=0}^{P-1}\nabb^k\left((P-k-1)C\nabb^{P-k-1}f-\oom\chi\nabb^{P-k}f\right)\\
&&\ML=\sum_{j=0}^{P-2}\cbin{P}{P-2-j}(\nabb^jC)\nabb^{P-1-j}f
-\sum_{k=0}^{P-1}\sum_{j=0}^k\cbin{k}{j}(\nabb^j\oom\chi) \nabb^{P-j}f\ .\nn
\eea
Therefore we can write
\bea
&&\ML\ML\ML\ML\ddb_{\nu}^{J}\nabb^PU=\nn\\
&&\ML\ML\ML\ML\ML\ \ \ -\frac{1}{2}\sum_{q=0}^{J-1}\sum_{h=0}^P\cbin{J-1}{q}\cbin{P}{h}(\ddb_{\nu}^q\nabb^h\oom\tr\chi)(\ddb_{\nu}^{J-1-q}\nabb^{P-h}U)\nn\\
&&\ML\ML\ML\ML\ML\ \ \ -\sum_{q=0}^{J-1}\sum_{h=0}^P\cbin{J-1}{q}\!\cbin{P}{h}\!(\ddb_{\nu}^q\nabb^h\chih)\c(\ddb_{\nu}^{J-1-q}\nabb^{P-h}\chih)\nn\\
&&\ML\ML\ML\ML\ML\ \ \ + \sum_{q=0}^{J-1}\cbin{J-1}{q}\sum_{k=0}^{P-1}(P-k-1)\sum_{j=0}^k\cbin{k}{j}(\ddb_{\nu}^q\nabb^jC)\ddb_{\nu}^{J-1-q}\nabb^{P-1-j}U\nn\\
&&\ML\ML\ML\ML\ML\ \ \ -\sum_{q=0}^{J-1}\sum_{k=0}^{P-1}\sum_{j=0}^k\cbin{J-1}{q}\cbin{k}{j}(\ddb_{\nu}^q\nabb^j\oom\chi)\ddb_{\nu}^{J-1-q}\nabb^{P-j}U\ .\eql{5.51}
\eea
We start estimating the last sum which turns out to be the more delicate and which we rewrite as
\bea
&&\ML\ML\ML\ML\ML\ \ \ \sum_{q=0}^{J-1}\sum_{k=0}^{P-1}\sum_{j=0}^k\cbin{J-1}{q}\cbin{k}{j}(\ddb_{\nu}^q\nabb^j\oom\chi)\ddb_{\nu}^{J-1-q}\nabb^{P-j}U\nn\\
&&\ML\ML\ML\ML\ML=\sum_{q=0}^{J-1}\sum_{h=0}^{P-1}\cbin{J-1}{q}\cbin{P}{h+1}(\ddb_{\nu}^q\nabb^h\oom\chi)\ddb_{\nu}^{J-1-q}\nabb^{P-h}U\ .
\eea
In fact the more delicate part of this last sum is the one with $\tr\chi$, part of $\chi$, omitting $\oom$. Therefore we look at
\bea
\sum_{q=0}^{J-1}\sum_{h=0}^{P-1}\cbin{J-1}{q}\cbin{P}{h+1}(\ddb_{\nu}^q\nabb^h\tr\chi)(\ddb_{\nu}^{J-1-q}\nabb^{P-h}U)
\eea
We observe, preliminary, that, denoting $N=J+P$,
\bea
&&\ML\ML\nab^NAB=\sum_{k=0}^N\cbin{N}{k}\nab^kA\nab^{N-k}B=\nab^J\nab^PAB=\sum_{q=0}^{J}\sum_{h=0}^P\cbin{J}{q}\cbin{P}{h}\nab^{q+h}A\nab^{N-(q+h)}B\nn\\
&&\ML\ML=\sum_{k=0}^N\left(\sum_{q=0}^{J}\sum_{h=0}^P\de(q+h=k)\cbin{J}{q}\cbin{P}{h}\right)\nab^kA\nab^{N-k}B
\eea
therefore for any couple $(J,P)$ such that $J+P=N$ the following relation holds which we use later on,
\bea
&&\cbin{N}{k}=\left(\sum_{q=0}^{J}\sum_{h=0}^P\de(q+h=k)\cbin{J}{q}\cbin{P}{h}\right)\nn\\
&&\sum_{k=0}^N\cbin{N}{k}=\sum_{k=0}^N\left(\sum_{q=0}^{J}\sum_{h=0}^P\de(q+h=k)\cbin{J}{q}\cbin{P}{h}\right)\nn\\
&&\sum_{q=0}^{J}\sum_{h=0}^P\cbin{J}{q}\cbin{P}{h}\left(\sum_{k=0}^N\de(q+h=k)\right)=\sum_{q=0}^{J}\sum_{h=0}^P\cbin{J}{q}\cbin{P}{h}\ . \nn
\eea
Therefore in conclusion
\bea
&&\sum_{k=0}^N\cbin{N}{k}=\sum_{q=0}^{J}\sum_{h=0}^P\cbin{J}{q}\cbin{P}{h}\ . \eql{esttouse}
\eea
We have the following  estimate, recalling that $\si(0)=0\ , \si(P)=1, \mbox{for} \  P\geq 1$, 
 \bea
 &&\ML\ML\ML\ML\sum_{q=0}^{J-1}\sum_{h=0}^{P-1}\cbin{J-1}{q}\cbin{P}{h+1}\big|r^{1+J+P+\si(P)-\frac{2}{p}}(\ddb_{\nu}^q\nabb^h\tr\chi)(\ddb_{\nu}^{J-1-q}\nabb^{P-h}U)\big|_{p,S}\eql{5.100}\\
 &&\ML\ML\ML\ML\ML\leq\sum_{q=0}^{J-1}\sum_{h=0}^{P-1}\chi\left(J+P=N\right)\cbin{J-1}{q}\cbin{P}{h+1}\left|r^{1+J+P+\si(P)-\frac{2}{p}}(\ddb_{\nu}^q\nabb^h\tr\chi)(\ddb_{\nu}^{J-1-q}\nabb^{P-h}U)\right|_{p,S}\nn\\
&&\ML\ML\ML\ML\ML\leq\sum_{q=0}^{J-1}\sum_{h=0}^{P-1}\chi\!\left(J+P=N\ ;\ q+h\leq\left[\frac{N-1}{2}\right]\right)\cbin{J-1}{q}\cbin{P}{h+1}\left|r^{1+J+P+\si(P)-\frac{2}{p}}(\ddb_{\nu}^q\nabb^h\tr\chi)(\ddb_{\nu}^{J-1-q}\nabb^{P-h}U)\right|_{p,S}\nn\\
&&\ML\ML\ML\ML\ML+\sum_{q=0}^{J-1}\sum_{h=0}^{P-1}\chi\!\left(J+P=N\ ;\ q+h\geq\left[\frac{N-1}{2}\right]+1\right)\cbin{J-1}{q}\cbin{P}{h+1}\left|r^{1+J+P+\si(P)-\frac{2}{p}}(\ddb_{\nu}^q\nabb^h\tr\chi)(\ddb_{\nu}^{J-1-q}\nabb^{P-h}U)\right|_{p,S}\ .\nn
 \eea
Let us consider the first sum
\medskip

\bea
&&\ML\ML\ML\ML\ML\ML\ML\sum_{q=0}^{J-1}\sum_{h=0}^{P-1}\chi\!\left(q+h\leq\left[\frac{N-1}{2}\right]\right)\cbin{J-1}{q}\cbin{P}{h+1}\left|r^{1+J+P+\si(P)-\frac{2}{p}}(\ddb_{\nu}^q\nabb^h\tr\chi)(\ddb_{\nu}^{J-1-q}\nabb^{P-h}U)\right|_{p,S}\nn\\
&&\ML\ML\ML\ML\ML\ML\ML\ML\ \leq \sum_{q=0}^{J-1}\sum_{h=0}^{P-1}\chi\!\left(q+h\leq\left[\frac{N-1}{2}\right]\right)\cbin{J-1}{q}\cbin{P}{h+1}
\big|r^{1+q+h+\si(h)}(\ddb_{\nu}^q\nabb^h\tr\chi)\big|_{\infty,S}\big|r^{J+P-q-h+\si(P-h)-\frac{2}{p}}(\ddb_{\nu}^{J-1-q}\nabb^{P-h}U)\big|_{p,S}\nn\\
&&\ML\ML\ML\ML\ML\ML\ML\ML\leq\left(F\frac{N!}{N^{\a}}\frac{e^{(N-2)(\de+\underline{\Ga}_0(\la))}}{\ro_0^{N}}\right)\!
\left(Fe^{-2(\de_0+\underline{\Ga}_0(\la))}\right)\c\nn\\
&&\ML\ML\ML\ML\ML\ML\ML\sum_{q=0}^{J-1}\sum_{h=0}^{P-1}\chi\!\left(q+h\leq\left[\frac{N-1}{2}\right]\right)\cbin{J-1}{q}\cbin{P}{h+1}
\left(\frac{(q+h+1)!}{((q+h+1)^{\a}}\right)\left(\frac{(J-1+P-(h+q))!}{(J-1+P-(h+q))^{\a}}\right)\!\frac{N^{\a}}{N!}\ .
\eea
Denoting ${\tilde h}=h+1$
\bea
&&\ML\ML\ML\ML\ML\leq\left(F\frac{N!}{N^{\a}}\frac{e^{(N-2)(\de+\underline{\Ga}_0(\la))}}{\ro_0^{N}}\right)\!\left(Fe^{-2(\de_0+\underline{\Ga}_0(\la))}\right)\c\nn\\
&&\ML\ML\ML\ML\ML\sum_{q=0}^{J-1}\sum_{{\tilde h}=1}^{P}\chi\!\left(q+{\tilde h}\leq\left[\frac{N-1}{2}\right]+1\right)\cbin{J-1}{q}\cbin{P}{{\tilde h}}
\left(\frac{(q+{\tilde h})!}{(q+{\tilde h})^{\a}}\right)\left(\frac{(J+P-(q+{\tilde h}))!}{(J+P-(q+{\tilde h}))^{\a}}\right)\!\frac{N^{\a}}{N!}\ \ \ \ \ \ \ \ \ \ \ \ \ \ \ \ \ \ \\
&&\ML\ML\ML\ML\ML\leq\left(F\frac{N!}{N^{\a}}\frac{e^{(N-2)(\de+\underline{\Ga}_0(\la))}}{\ro_0^{N}}\right)\!\left(fe^{-2(\de+\underline{\Ga}_0(\la))}\right)\c\nn\\
&&\ML\ML\ML\ML\ML\sum_{q=0}^{J-1}\sum_{{\tilde h}=0}^{P}\chi\!\left(q+{\tilde h}\leq\left[\frac{N-1}{2}\right]+1\right)\cbin{J-1}{q}\cbin{P}{{\tilde h}}
\left(\frac{(q+{\tilde h})!}{((q+{\tilde h})^{\a}}\right)\left(\frac{(J+P-(q+{\tilde h}))!}{(J+P-(q+{\tilde h}))^{\a}}\right)\!\frac{N^{\a}}{N!}\ ,\nn
\eea
using the previous relation \ref{esttouse}, with $J+P=N$, denoting $k=q+{\tilde h}$
\[\sum_{k=0}^N\cbin{N}{k}=\sum_{q=0}^{J}\sum_{h=0}^P\cbin{J}{q}\cbin{P}{h}\ , \]
\bea
&&\ML\ML\ML\ML\ML\ML\leq\c\left(F\frac{N!}{N^{\a}}\frac{e^{(N-2)(\de_0+\underline{\Ga}_0(\la))}}{\ro_0^{N}}\right)\!\left(Fe^{-2(\de_0+\underline{\Ga}_0(\la))}\right)\c\nn\\
&&\ML\ML\ML\ML\ML\sum_{q=0}^{J-1}\sum_{{\tilde h}=0}^{P}\chi\!\left(k=q+{\tilde h}\leq\left[\frac{N-1}{2}\right]+1\right)\cbin{J-1}{q}\cbin{P}{{\tilde h}}
\left(\frac{k!}{k^{\a}}\right)\left(\frac{(N-k)!}{(N-k)^{\a}}\right)\!\frac{N^{\a}}{N!}\nn\\
&&\ML\ML\ML\ML\ML\ML\leq\left(F\frac{N}{N^{\a}}\frac{e^{(N-2)(\de_0+\underline{\Ga}_0(\la))}}{\ro_0^{N}}\right)\!\left(Fe^{-2(\de_0+\underline{\Ga}_0(\la))}\right)\c\sum_{k=0}^{\left[\frac{N-1}{2}\right]+1}\cbin{N-1}{k}\left(\frac{k!}{k^{\a}}\right)\left(\frac{(N-k)!}{(N-k)^{\a}}\right)\!\frac{N^{\a}}{N!}\nn\\
&&\ML\ML\ML\ML\ML\ML\leq\left(F\frac{N!}{N^{\a}}\frac{e^{(N-2)(\de_0+\underline{\Ga}_0(\la))}}{\ro_0^{N}}\right)\!\left(Fe^{-2(\de_0+\underline{\Ga}_0(\la))}\right)\c
\sum_{k=0}^{\left[\frac{N-1}{2}\right]+1}\frac{(N-1)!}{(N-1-k)!}\left(\frac{1}{k^{\a}}\right)\left(\frac{(N-k)!}{(N-k)^{\a}}\right)\!\frac{N^{\a}}{N!}\nn\\
&&\ML\ML\ML\ML\ML\ML\leq\left(F\frac{N!}{N^{\a}}\frac{e^{(N-2)(\de_0+\underline{\Ga}_0(\la))}}{\ro_0^{N}}\right)\!\left(Fe^{-2(\de_0+\underline{\Ga}_0(\la))}\right)\c
\sum_{k=0}^{\left[\frac{N-1}{2}\right]+1}\left(\frac{1}{k^{\a}}\right)\left(\frac{(N-k)}{(N-k)^{\a}}\right)\!\frac{N^{\a}}{N}\nn\\
&&\ML\ML\ML\ML\ML\ML\leq\left(F\frac{N!}{N^{\a}}\frac{e^{(N-2)(\de_0+\underline{\Ga}_0(\la))}}{\ro_0^{N}}\right)\!\left(Fe^{-2(\de_0+\underline{\Ga}_0(\la))}\right)\c
\sum_{k=0}^{\left[\frac{N-1}{2}\right]+1}\left(\frac{1}{k^{\a}}\right)\left(\frac{N^{\a}}{(N-k)^{\a}}\right)\!\nn\\
&&\ML\ML\ML\ML\ML\ML\leq\left(F\frac{N!}{N^{\a}}\frac{e^{(N-2)(\de_0+\underline{\Ga}_0(\la))}}{\ro_0^{N}}\right)\ ,
\eea
provided that $\a>1$ and $\de$ sufficiently large such that,
\bea
\left(Fe^{-2(\de_0+\underline{\Ga}_0(\la))}\right)\c
\sum_{k=0}^{\left[\frac{N-1}{2}\right]+1}\left(\frac{1}{k^{\a}}\right)\left(\frac{N^{\a}}{(N-k)^{\a}}\right)<<1\ .
\eea

Let us consider now the second sum
\bea
&&\ML\ML\ML\ML\ML\ML\sum_{q=0}^{J-1}\sum_{h=0}^{P-1}\chi\!\left(q+h\geq\left[\frac{N-1}{2}\right]+1\right)\cbin{J-1}{q}\cbin{P}{h+1}\left|r^{1+J+P+\si(P)-\frac{2}{p}}(\ddb_4^q\nabb^h\tr\chi)(\ddb_4^{J-1-q}\nabb^{P-h}U)\right|_{p,S}\nn\\
&&\ML\ML\ML\ML\ML\ML\ML\leq \sum_{q=0}^{J-1}\sum_{h=0}^{P-1}\chi\!\left(q+h\geq\left[\frac{N-1}{2}\right]+1\right)\cbin{J-1}{q}\cbin{P}{h+1}
\big|r^{1+q+h+\si(h)-\frac{2}{p}}(\ddb_4^q\nabb^h\tr\chi)\big|_{p,S}\big|r^{J+P-q-h+\si(P-h)}(\ddb_4^{J-1-q}\nabb^{P-h}U)\big|_{\infty,S}\nn\\
&&\ML\ML\ML\ML\ML\ML\leq\left(F\frac{N!}{N^{\a}}\frac{e^{(N-2)(\de_0+\underline{\Ga}_0(\la)}}{\ro_2^{N}}\right)\!\left(Fe^{-2(\de_0+\underline{\Ga}_0(\la))}\right)\c\nn\\
&&\ML\ML\ML\ML\ML\ML\sum_{q=0}^{J-1}\sum_{h=0}^{P-1}\chi\!\left(q+h\geq\left[\frac{N-1}{2}\right]+1\right)\cbin{J-1}{q}\cbin{P}{h+1}
\left(\frac{(q+h)!}{((q+h)^{\a}}\right)\left(\frac{(J+P+1-(h+q))!}{(J+P+1-(h+q))^{\a}}\right)\!\frac{N^{\a}}{N!}\ .
\eea
\bea
&&\ML\ML\ML\ML\ML\leq\left(F\frac{N!}{N^{\a}}\frac{e^{(N-2)(\de_0+\underline{\Ga}_0(\la))}}{\ro_0^{N}}\right)\!\left(Fe^{-2(\de_0+\underline{\Ga}_0(\la))}\right)\c\nn\\
&&\ML\ML\ML\ML\ML\sum_{q=0}^{J-1}\sum_{h=0}^{P-1}\chi\!\left(q+h\geq\left[\frac{N-1}{2}\right]+1\right)\cbin{J-1}{q}\cbin{P}{h}
\left(\frac{(q+h)!}{((q+h)^{\a}}\right)\left(\frac{(J+P+1-(q+h))!}{(J+P+1-(q+h))^{\a}}\right)\!\frac{N^{\a}}{N!}\ .\ \ \ \ \ \ \  \ \ \ \ \ \ \ 
\eea
Using the previous relation \ref{esttouse}, with $J+P=N$,
\[\sum_{k=0}^N\cbin{N}{k}=\sum_{q=0}^{J}\sum_{h=0}^P\cbin{J}{q}\cbin{P}{h}\ , \]
denoting $k=q+h$
\bea
&&\ML\ML\ML\ML\ML\ML\ML\leq\left(F\frac{N!}{N^{\a}}\frac{e^{(N-2)(\de_0+\underline{\Ga}_0(\la))}}{\ro_0^{N}}\right)\!\left(Fe^{-2(\de_0+\underline{\Ga}_0(\la))}\right)\c
\sum_{k=\left[\frac{N-1}{2}\right]+1}^{N-1}\cbin{N-1}{k}
\left(\frac{k!}{k^{\a}}\right)\left(\frac{(N+1-k)!}{(N+1-k)^{\a}}\right)\!\frac{N^{\a}}{N!}\nn\\
&&\ML\ML\ML\ML\ML\ML\ML\leq\left(F\frac{N!}{N^{\a}}\frac{e^{(N-2)(\de_0+\underline{\Ga}_0(\la))}}{\ro_0^{N}}\right)\!\left(Fe^{-2(\de_0+\underline{\Ga}_0(\la))}\right)\c
\sum_{k=\left[\frac{N-1}{2}\right]+1}^{N-1}\frac{(N-1)!}{k!(N-1-k)!}
\left(\frac{k!}{k^{\a}}\right)\left(\frac{(N+1-k)!}{(N+1-k)^{\a}}\right)\!\frac{N^{\a}}{N!}\nn\\
&&\ML\ML\ML\ML\ML\ML\ML\leq\left(F\frac{N!}{N^{\a}}\frac{e^{(N-2)(\de_0+\underline{\Ga}_0(\la))}}{\ro_0^{N}}\right)\!\left(Fe^{-2(\de_0+\underline{\Ga}_0(\la))}\right)\c
\sum_{k=\left[\frac{N-1}{2}\right]+1}^{N-1}
\left(\frac{1}{k^{\a}}\right)\left(\frac{(N+1-k)(N-k)}{(N+1-k)^{\a}}\right)\!\frac{N^{\a}}{N}\nn\\
&&\ML\ML\ML\ML\ML\ML\leq\left(F\frac{N!}{N^{\a}}\frac{e^{(N-2)(\de_0+\underline{\Ga}_0(\la))}}{\ro_0^{N}}\right)\!\left(Fe^{-2(\de_0+\underline{\Ga}_0(\la))}\right)\c
\sum_{k=\left[\frac{N-1}{2}\right]+1}^{N-1}
\left(\frac{1}{k^{\a}}\right)\left(\frac{(N+1-k)(N-k)}{(N+1-k)^{\a}}\right)\!\frac{N^{\a}}{N}\nn\\
&&\ML\ML\ML\ML\ML\ML\leq\left(F\frac{N!}{N^{\a}}\frac{e^{(N-2)(\de_0+\underline{\Ga}(\la))}}{\ro_0^{N}}\right)\!\left(Fe^{-2(\de_0+\underline{\Ga}_0(\la))}\right)\c
2^{\a}\sum_{k=\left[\frac{N-1}{2}\right]+2}^{N-1}
\frac{1}{(N+1-k)^{\a-1}}\!\nn\\
&&\ML\ML\ML\ML\ML\ML\leq\left(F\frac{N!}{N^{\a}}\frac{e^{(N-2)(\de_0+\underline{\Ga}(\la))}}{\ro_0^{N}}\right)\!\left(c2^{\a}Fe^{-2(\de_0+\underline{\Ga}_0(\la))}\right)
\leq\left(F\frac{N!}{N^{\a}}\frac{e^{(N-2)(\de_0+\underline{\Ga}_0(\la)}}{\ro_0^{N}}\right)\ ,
\eea
choosing $\de$ sufficiently large.

\NI To complete these estimates we look at the less dangerous part of the commutator part, let us recall first the complete expression of the commutator, see \ref{5.51}
\bea
&&\ML\ML\ddb_{\nu}^{J-1}[\ddb_{\nu},\nabb^P]U= \sum_{q=0}^{J-1}\cbin{J-1}{q}\sum_{h=0}^{P-1}(P-h-1)\sum_{j=0}^h\cbin{h}{j}(\ddb_{\nu}^q\nabb^jC)\ddb_{\nu}^{J-1-q}\nabb^{P-1-j}U\nn\\
&&\ \ \ \ \ \ \ \ \ \ \ \ \ \ -\sum_{q=0}^{J-1}\sum_{h=0}^{P-1}\sum_{j=0}^h\cbin{J-1}{q}\cbin{h}{j}(\ddb_{\nu}^q\nabb^j\chi)\ddb_{\nu}^{J-1-q}\nabb^{P-j}U\ .\eql{5.92z}
\eea
As the following relation holds
\bea
\sum_{h=0}^{P-1}(P-h-1)\sum_{j=0}^h\cbin{h}{j}=\sum_{j=0}^{P-1}\sum_{h=j}^{P-1}(P-h-1)\cbin{h}{j}=\sum_{j=0}^{P-2}\cbin{P}{P-2-j}=\sum_{l=0}^{P-2}\cbin{P}{l}\ \ \ \ \ \ \ \eql{bin1}
\eea
we can write
\bea
\ML\ML\ddb_{\nu}^{J-1}[\ddb_{\nu},\nabb^P]U\!&=&\! \sum_{q=0}^{J-1}\sum_{j=0}^{P-2}\cbin{J-1}{q}\cbin{P}{P-2-j}(\ddb_{\nu}^q\nabb^jC)\ddb_{\nu}^{J-1-q}\nabb^{P-1-j}U\nn\\
\ML\ML\ML\ML\ \ \ \!&-&\!\sum_{q=0}^{J-1}\sum_{h=0}^{P-1}\sum_{j=0}^h\cbin{J-1}{q}\cbin{h}{j}(\ddb_{\nu}^q\nabb^j\chi)\ddb_{\nu}^{J-1-q}\nabb^{P-j}U\nn\\
\ML\ML\ML\ML\!&=&\! \sum_{q=0}^{J-1}\sum_{h=0}^{P-2}\cbin{J-1}{q}\cbin{P}{h}(\ddb_{\nu}^q\nabb^{h}C)\ddb_{\nu}^{J-1-q}\nabb^{P-h-1}U\nn\\
\ML\ML\ML\ML\ \ \ \!&-&\!\sum_{q=0}^{J-1}\sum_{h=0}^{P-1}\cbin{J-1}{q}\cbin{P}{h}(\ddb_{\nu}^q\nabb^h\chi)(\ddb_{\nu}^{J-1-q}\nabb^{P-h-1}U)
\eql{5.92z1}
\eea
The terms in the second line have already been estimated, the term in the first line has exactly the same structure and therefore produces the same estimate.
\smallskip

\NI The first two lines of $\ddb_{\nu}^J\nabb^PU$, \ref{5.51}, can be estimated in the same way and they are even easier to bound,\footnote{Due to the fact that, comparing the first and the fourth line of \ref{5.51} a factor $\cbin{P}{h}$ is present, in the first one, instead of $\cbin{P}{h+1}$.} therefore we do not report their estimates.
\medskip

\NI{\bf Remarks:} {\em 

\NI i) Observe that in the estimates of the $|\c|_{\infty,S}$ norms in terms of the $|\c|_{p=4,S}$ norms with an extra $\nabb$ we have neglected the position of this extra $\nabb$; it is easy to see that taking it into account does not change the final estimate (at most we have to choose a large $\a$).

\NI ii) As already said, notice that in the case of the initial data we do not need the energy estimates due to the fact that all the mixed derivatives of $\chih$ are already at our disposal. This is crucial as in the case of the initial data we do not have any possibility to exploit the energy estimates to bound the null Riemann components.}
\smallskip

\NI  
Let us consider the structure  equation,
\[\dddd_4\zeta+\zeta\chi+\tr\chi\zeta-\divv\chi+\nabb\tr\chi+\ddb_4\nabb\log\oom=0\ .\]
All the terms can be treated in the same way as before and we do not repeat here, therefore we can prove the estimates, with $J+P=N$,
\medskip

\NI Moreover, by the identity \[\eta=\zeta+ \nabb\log\Omega,\] see equations 3.1.33 \cite{Kl-Ni:book}, and the assumptions on $\Omega$,
\medskip
 
\NI  we obtain
\bea
&&\big|r^{2+J+P-\frac{2}{p}}\ddb_{\nu}^J\nabb^P\eta\big|_{p,S}\leq F\!\left(\frac{(J+P)!}{(J+P)^{\a}}\frac{e^{((J+P)-2)(\de_0+\underline{\Ga}_0(\la))}}{\ro_0^{J+P}}\right)\nn\\
&&\big|r^{2+J+P-\frac{2}{p}}\ddb_{\nu}^J\nabb^P\ze\big|_{p,S}\leq F\!\left(\frac{(J+P+1)!}{(J+P+1)^{\a}}\frac{e^{((J+P)-1)(\de_0+\underline{\Ga}_0(\la))}}{\ro_0^{J+P+1}}\right)\ .\ \ \ \ \ \ \ \ \ \eql{zeest1}
\eea

\subsubsection{Estimate of the underlined quantities}
As already discussed we can use also the transport equations implying a loss of derivatives as the angular derivatives estimates are already under control.
\medskip

\NI {\bf \{$\ddb_{\nu}^J\nabb^P\tr\chib$\}}\ :
\medskip

\NI In this case we use the transport equation for $\tr\chib$ along the outgoing cones
which we write, denoting $\Ub=\oom\tr\chib$,
\bea
&&\ML\ML\dddd_{\nu}\Ub+\frac{1}{2}\oom\tr\chi\Ub +\oom^2\chih\c\chibh-2\oom^2\divv\etab\!-\!2\oom^2|\etab|^2-2\oom^2\ro=0\ .\ \ \ \ \ \ \ \ \ \ \ \ \ \eql{subsetstrout1}
\eea
Proceeding as before, we write
\bea
&&\ML\ML\ML\ddb_{\nu}^J\nabb^P\Ub=\ddb_{\nu}^{J-1}\nabb^P\big(\ddb_{\nu}\Ub+\ddb_{\nu}^{J-1}[\ddb_{\nu},\nabb^P]\Ub\nn\\
&&\ML\ML\ML=-\ddb_{\nu}^{J-1}\nabb^P\!\big( \frac{1}{2}\oom\tr\chi\Ub +\oom^2\chih\c\chibh-2\oom^2\divv\etab\!-\!2\oom^2|\etab|^2-2\oom^2\ro\big)+\ddb_{\nu}^{J-1}[\ddb_{\nu},\nabb^P]\Ub\nn\\
&&\nn\\
&&\ML\ML\ML=-\ddb_{\nu}^{J-1}\nabb^P\!\bigg( \frac{1}{2}\oom\tr\chi\Ub +\oom^2\chih\c\chibh-2\oom^2\divv\etab\!-\!2\oom^2|\etab|^2\bigg)+2\ddb_{\nu}^{J-1}\nabb^P\!(\oom^2\ro)\nn\\
&&\ML\ML\ML\ \ \ +\sum_{q=0}^{J-1}\sum_{j=0}^{P-2}\cbin{J-1}{q}\cbin{P}{h+1}(\ddb_{\nu}^q\nabb^{(h-1)}C)\ddb_{\nu}^{J-1-q}\nabb^{P-h}{\hat U}\nn\\
&&\ML\ML\ML\ \ \ -\sum_{q=0}^{J-1}\sum_{h=0}^{P-1}\cbin{J-1}{q}\cbin{P}{h+1}(\ddb_{\nu}^q\nabb^h\chi)(\ddb_{\nu}^{J-1-q}\nabb^{P-h}{\hat U})\nn\eql{5.92z3}
\eea
It is immediate to realize that all the estimates can be done as the previous ones and, therefore, the final result is, 
\bea
\big|r^{2+J+P-\frac{2}{p}}\ddb_{\nu}^J\nabb^P\Ub|_{p,S}\leq F\left(\frac{(J+P)!}{(J+P)^{\a}}\frac{e^{((J+P)-2)(\de_0+\underline{\Ga}_0(\la))}}{\ro_0^{J+P}}\right) \ .
\eea
\medskip

\NI {\bf \{$\ddb_{\nu}^J\nabb^P\chibh$\}}\ :
\medskip

\NI In this case we use the transport equation
\bea
&&\ML\ML\ddb_{\nu}\chibh\!+\!\frac{\oom}{2}\tr\chi\chibh\!+\!\frac{\oom}{2}\tr\chib\chih\!-\!2\oom\om\chibh\!-\!\nabb\hot\etab\!-\!\etab\hot\etab\!=\!0\nn
\eea
which we rewrite, denoting ${\hat{\Ub}}=\oom\chibh$,
\bea
&&\ML\ML\ddb_{\nu}{\hat{\Ub}}+\!\frac{\oom}{2}\tr\chi{\hat{\Ub}}\!+\!\frac{\oom}{2}\tr\chib(\oom\chih)\!-\!\oom\nabb\hot\etab\!-\!\oom\etab\hot\etab\!=\!0\ .\nn
\eea
Once again
\bea
&&\ML\ML\ML\ddb_{\nu}^J\nabb^P{\hat{\Ub}}=\ddb_{\nu}^{J-1}\nabb^P\big(\ddb_{\nu}{\hat{\Ub}}+\ddb_{\nu}^{J-1}[\ddb_{\nu},\nabb^P]{\hat{\Ub}}\nn\\
&&\ML\ML\ML=-\ddb_{\nu}^{J-1}\nabb^P\!\big( \!\frac{\oom}{2}\tr\chi{\hat{\Ub}}\!+\!\frac{\oom}{2}\tr\chib(\oom\chih)\!-\!\oom\nabb\hot\etab\!-\!\oom\etab\hot\etab\big)+\ddb_{\nu}^{J-1}[\ddb_{\nu},\nabb^P]\Ub\nn\\
&&\nn\\
&&\ML\ML\ML=-\ddb_{\nu}^{J-1}\nabb^P\!\big( \!\frac{\oom}{2}\tr\chi{\hat{\Ub}}\!+\!\frac{\oom}{2}\tr\chib(\oom\chih)\!-\!\oom\nabb\hot\etab\!-\!\oom\etab\hot\etab\big)+\ddb_{\nu}^{J-1}[\ddb_{\nu},\nabb^P]{\hat{\Ub}}\nn\\
&&\ML\ML\ML\ \ \ +\sum_{q=0}^{J-1}\sum_{j=0}^{P-2}\cbin{J-1}{q}\cbin{P}{h}(\ddb_{\nu}^q\nabb^{h}C)\ddb_{\nu}^{J-1-q}\nabb^{P-h}{\hat{\Ub}}\nn\\
&&\ML\ML\ML\ \ \ -\sum_{q=0}^{J-1}\sum_{h=0}^{P-1}\cbin{J-1}{q}\cbin{P}{h}(\ddb_{\nu}^q\nabb^h\chi)(\ddb_{\nu}^{J-1-q}\nabb^{P-h}{\hat{\Ub}})\nn\eql{5.92z4}
\eea
and the inductive estimates are proved exactly in the same way obtaining as final result, 
\bea
\big||\la|r^{1+J+P-\frac{2}{p}}\ddb_{\nu}^J\nabb^P{\hat{\Ub}}|_{p,S}\leq F\!\left(\frac{(J+P)!}{(J+P)^{\a}}\frac{e^{((J+P)-2)(\de_0+\underline{\Ga}_0(\la)}}{\ro_0^{J+P}}\right)\ .\ \ \eql{5.163}
\eea
Proceeding exactly in the same way we obtain the analogous estimates for the mixed derivatives with $\ddb_{\la}$ instead of $\ddb_{\nu}$.

\section{Appendix to Section \ref{S.10} }\label{AS.3}
{ {\bf Some commutation relations}
\medskip

\NI The following relations hold:
\bea
&&\ML\ML[\nabb_{\mu},\frac{\Dbb}{\partial\nu}]U_{\nu_1...\nu_k}=\nn \\
&&\ML\ML=-{\oom}\sum_{j=1}^k\!
\left[(\chi_{\mu\nu_j}\etab^{\si_j}\!-\!\chi_{\mu}^{\si_j}\etab_{\nu_j})
+\theta^C_{\mu}\theta^D_{\nu_j}R^{{\si}_j}(\c,e_C,e_4,e_D)\right]\!U_{\nu_1..{\si}_j..\nu_k}
+\oom\chi_{\mu}^{\ro}(\nabb_{\ro}U)_{\nu_1..\nu_k}\nn\\
&&\ML\ML=-\sum_{j=1}^k{C}^{\si_j}_{\mu\nu_j}U_{\nu_1..{\si}_j..\nu_k}+\oom\chi_{\mu}^{\ro}(\nabb_{\ro}U)_{\nu_1..\nu_k}\ ,\eql{6.6}
\eea
\bea
\mbox{with}\ \ \ \ \ \ \ \ \ \ {C}^{\si}_{\mu\nu}=\oom\!\left[(\chi_{\mu\nu}\etab^{\si}\!-\!\chi_{\mu}^{\si}\etab_{\nu})
+\theta^C_{\mu}\theta^D_{\nu}R^{{\si}}(\c,e_C,e_4,e_D)\right]\ .\ \ \ \ \ \ \ \ \ \ \ \ \ \ \ \ 
\eea
If $U$ is a 1-form we have,
\bea
&&\ML\ML[\nabb_{\mu},\frac{\Dbb}{\partial\nu}]U_{\nu}=\left[(\chi_{\mu\nu_j}\etab^{\si}\!-\!\chi_{\mu}^{\si}\etab_{\nu})
+\theta^C_{\mu}\theta^D_{\nu}R^{{\si}}(\c,e_C,e_4,e_D)\right]\!U_{\si}
+\oom\chi_{\mu}^{\ro}(\nabb_{\ro}U)_{\nu}\nn\\
&&\ \ \ \ \ \ \ =-{C}^{\si}_{\mu\nu}U_{\si}+\oom\chi_{\mu}^{\ro}(\nabb_{\ro}U)_{\nu}\  
\eea

\NI We rewrite symbolically the last formula as:
\bea
&&[\nabb_{\mu},\frac{\Dbb}{\partial\nu}]U=-{C}U+D(\nabb U)\ 
\eea

\NI Were with $C$ we mean a combination of null Riemann components and with $F$ we mean a combination of connection coefficients and metric components.
Similarly we have:

\bea
&&\ML\ML[\nabb_{\mu},\frac{\Dbb}{\partial\la}]U_{\nu_1...\nu_k}=\nn \\
&&\ML\ML=-{\oom}\sum_{j=1}^k\!
\left[(\chib_{\mu\nu_j}\eta^{\si_j}\!-\!\chib_{\mu}^{\si_j}\eta_{\nu_j})
+\theta^C_{\mu}\theta^D_{\nu_j}R^{{\si}_j}(\c,e_C,e_3,e_D)\right]\!U_{\nu_1..{\si}_j..\nu_k}
+\oom\chib_{\mu}^{\ro}(\nabb_{\ro}U)_{\nu_1..\nu_k}\nn\\
&&\ML\ML=-\sum_{j=1}^k{\underline{C}}^{\si_j}_{\mu\nu_j}U_{\nu_1..{\si}_j..\nu_k}+\oom\chib_{\mu}^{\ro}(\nabb_{\ro}U)_{\nu_1..\nu_k}\ ,
\eea

\NI which, in the case of $U$ 1-form, becomes:

\bea
&&\ML\ML[\nabb_{\mu},\frac{\Dbb}{\partial\la}]U_{\nu}=
\left[(\chib_{\mu\nu_j}\eta^{\si}\!-\!\chib_{\mu}^{\si}\eta_{\nu})
+\theta^C_{\mu}\theta^D_{\nu}R^{{\si}}(\c,e_C,e_3,e_D)\right]\!U_{\si}
+\oom\chib_{\mu}^{\ro}(\nabb_{\ro}U)_{\nu}\nn\\
&&\ \ \ \ \ \ \ =-{\underline{C}}^{\si}_{\mu\nu}U_{\si}+\oom\chib_{\mu}^{\ro}(\nabb_{\ro}U)_{\nu}\  
\eea

\NI Again we rewrite the last formula symbolically as

\bea
&&[\nabb_{\mu},\frac{\Dbb}{\partial\la}]U=-\underline{C}U+\underline{D}(\nabb U)\ .
\eea

\NI The commutator of the tangential derivatives is,

\bea
&&\ML\ML\ML[\nabb_{\a},\nabb_{\mu}]U_{\nu_1\nu_2...\nu_q}=\\  
&&\ML\ML\ML=\left[(\theta^3_{\mu}\chib_{\a}^{\b}+\theta^4_{\mu}\chi_{\a}^{\b})-(\theta^3_{\a}\chib_{\mu}^{\b}+\theta^4_{\a}\chi_{\mu}^{\b})\right]
\nabb_\b U_{\nu_1\nu_2...\nu_q}-\theta^C_{\a}\theta^D_{\mu}\sum_{i=1}^qR^{\tilde{\nu}_i}(\c,e_C,\c,e_D)_{\nu_i}U_{\nu_1\nu_2..{\tilde\nu}_i..\nu_q}\nn
\eea

\NI and when  the commutator operates on a 2-form,

\bea
&&\ML\ML\ML\ML[\nabb_{\a},\nabb_{\mu}]U_{\nu_1\nu_2}=\left[(\theta^3_{\mu}\chib_{\a}^{\b}+\theta^4_{\mu}\chi_{\a}^{\b})-(\theta^3_{\a}\chib_{\mu}^{\b}+\theta^4_{\a}\chi_{\mu}^{\b})\right]
\nabb_\b U_{\nu_1\nu_2}\\
&&-\theta^C_{\a}\theta^D_{\mu}R^{\tilde{\nu}}(\c,e_C,\c,e_D)_{\nu_1}U_{\tilde{\nu}\nu_2}\ -\theta^C_{\a}\theta^D_{\mu}R^{\tilde{\nu}}(\c,e_C,\c,e_D)_{\nu_2}U_{\nu_1\tilde{\nu}}\ \nn
\eea

\NI which we rewrite symbolically as

\bea
&&[\nabb_{\mu},\frac{\Dbb}{\partial\nu}]U=-\hat{C}U+\hat{F}(\nabb U)\ 
\eea

\NI Finally applying  $[\nabb_{\a},\divv]$ to a 2-form we have,

\bea
&&\ML[\nabb_{\a},\divv]U_{\nu_1\nu_2}=[\nabb_{\a},\nabb_{\mu}]U^{\mu}_{\nu}=\\
&&\ML=\left[(\theta^3_{\mu}\chib_{\a}^{\b}+\theta^4_{\mu}\chi_{\a}^{\b})-(\theta^3_{\a}\chib_{\mu}^{\b}+\theta^4_{\a}\chi_{\mu}^{\b})\right]
\nabb_\b U^{\mu}_{\nu}-2\theta^C_{\a}\theta^D_{\mu}R^{\tilde{\nu}}(\c,e_C,\c,e_D)_{\nu}U^{\mu}_{\tilde\nu}\ ,\nn
\eea
which we rewrite symbolically as
\bea
&&[\nabb_{\a},{\divv}]U=-\overline{C}U+\overline{F}(\nabb U)\ 
\eea

\NI Notice that all the commutators have the same structure, namely a combination of products of connection coefficients and metric components time $\nabb U$ and a combination of null Riemann components times $U$.} 
\medskip

\NI {\bf Detailed derivation of all the previous relations:}

Consider first  the commutator
$[\frac{\Dbb}{\partial\nu},\nabb_{\mu}]$ where with $\frac{\Dbb}{\partial\nu}$ we denote the covariant derivative $\oom\dddd_{e_4}$. Let
$U_{\nu_1...\nu_k}$ be a covariant tensor tangent to $S$
\bea
[\frac{\Dbb}{\partial\nu},\nabb_{\mu}]U_{\nu_1...\nu_k}
=\frac{\Dbb}{\partial\nu}\nabb_{\mu}U_{\nu_1...\nu_k}-\nabb_{\mu}\frac{\Dbb}{\partial\nu}U_{\nu_1...\nu_k}\nn
\eea
and
\bea
&&\frac{\Dbb}{\partial\nu}\nabb_{\mu}U_{\nu_1...\nu_k}=\oom\ddb_4\Pi^{\ro}_{\mu}\Pi^{\si_1..\si_k}_{\nu_1..\nu_k}D_{\ro}U_{\si_1..\si_k}\nn\\
&&\nabb_{\mu}\frac{\Dbb}{\partial\nu}U_{\nu_1...\nu_k}=
\Pi^{\ro}_{\mu}\Pi^{\la_1..\la_k}_{\nu_1..\nu_k}D_{\ro}\oom\Pi^{\si_1..\si_k}_{\la_1..\la_k}e_4^{\tau}D_{\tau}U_{\si_1...\si_k}
\eea
where we defined
\beaa
&&\Pi^{\si_1..\si_k}_{\nu_1..\nu_k}\equiv\Pi^{\si_1}_{\nu_1}\c\c\Pi^{\si_k}_{\nu_k}\\
&&\Pi^{\si_i}_{\nu_i}=\de^{\si_i}_{\nu_i}-(e_3\theta_3+e_4\theta_4)^{\si_i}_{\nu_i}
=\de^{\si_i}_{\nu_i}+\frac{1}{2}g_{\nu_j\ro}\!\left(e_3^{\si_i}e_4^{\ro}+e_4^{\si_i}e_3^{\ro}\right)\ .
\eeaa
Therefore
\bea
\frac{\Dbb}{\partial\nu}\nabb_{\mu}U_{\nu_1...\nu_k}
\!&=&\!\oom(\Pi^{\mu'}_{\mu}\dd_4\Pi^{\ro}_{\mu'})\Pi^{\si_1..\si_k}_{\nu_1..\nu_k}D_{\ro}U_{\si_1..\si_k}+
\oom\Pi^{\ro}_{\mu}(\Pi^{\nu'_1..\nu'_k}_{\nu_1..\nu_k}\dd_4\Pi^{\si_1..\si_k}_{\nu'_1..\nu'_k})D_{\ro}U_{\si_1..\si_k}\nn\\
&+&\!\oom\Pi^{\ro}_{\mu}\Pi^{\si_1..\si_k}_{\nu_1..\nu_k}e_4^{\tau}D_{\tau}D_{\ro}U_{\si_1..\si_k}\ ,
\eea
\bea
\nabb_{\mu}\frac{\Dbb}{\partial\nu}U_{\nu_1...\nu_k}
\!&=&\!\Pi^{\ro}_{\mu}(D_{\ro}\oom)\Pi^{\si_1..\si_k}_{\nu_1..\nu_k}e_4^{\tau}D_{\tau}U_{\si_1...\si_k}
+\oom\Pi^{\ro}_{\mu}(\Pi^{\la_1..\la_k}_{\nu_1..\nu_k}D_{\ro}\Pi^{\si_1..\si_k}_{\la_1..\la_k})e_4^{\tau}D_{\tau}U_{\si_1...\si_k}\nn\\
&+&\!\oom\Pi^{\ro}_{\mu}\Pi^{\si_1..\si_k}_{\nu_1..\nu_k}(D_{\ro}e_4^{\tau})D_{\tau}U_{\si_1...\si_k}
+\oom\Pi^{\ro}_{\mu}\Pi^{\si_1..\si_k}_{\nu_1..\nu_k}e_4^{\tau}D_{\ro}D_{\tau}U_{\si_1...\si_k}\ .\ \ 
\eea
From it
\bea
[\nabb_{\mu},\frac{\Dbb}{\partial\nu}]U_{\nu_1...\nu_k}\!&=&\!\Pi^{\ro}_{\mu}(D_{\ro}\oom)\Pi^{\si_1..\si_k}_{\nu_1..\nu_k}e_4^{\tau}D_{\tau}U_{\si_1...\si_k}
+\oom\Pi^{\ro}_{\mu}(\Pi^{\la_1..\la_k}_{\nu_1..\nu_k}D_{\ro}\Pi^{\si_1..\si_k}_{\la_1..\la_k})e_4^{\tau}D_{\tau}U_{\si_1...\si_k}\nn\\
&+&\!\oom\Pi^{\ro}_{\mu}\Pi^{\si_1..\si_k}_{\nu_1..\nu_k}(D_{\ro}e_4^{\tau})D_{\tau}U_{\si_1...\si_k}
+\oom\Pi^{\ro}_{\mu}\Pi^{\si_1..\si_k}_{\nu_1..\nu_k}e_4^{\tau}D_{\ro}D_{\tau}U_{\si_1...\si_k}\nn\\
&-&\oom(\Pi^{\mu'}_{\mu}\dd_4\Pi^{\ro}_{\mu'})\Pi^{\si_1..\si_k}_{\nu_1..\nu_k}D_{\ro}U_{\si_1..\si_k}
-\oom\Pi^{\ro}_{\mu}(\Pi^{\nu'_1..\nu'_k}_{\nu_1..\nu_k}\dd_4\Pi^{\si_1..\si_k}_{\nu'_1..\nu'_k})D_{\ro}U_{\si_1..\si_k}\nn\\
&-&\!\oom\Pi^{\ro}_{\mu}\Pi^{\si_1..\si_k}_{\nu_1..\nu_k}e_4^{\tau}D_{\tau}D_{\ro}U_{\si_1..\si_k}\nn\\
&&\nn\\
\!&=&\!\Pi^{\ro}_{\mu}(D_{\ro}\log\oom)\Pi^{\si_1..\si_k}_{\nu_1..\nu_k}\frac{\Dbb}{\partial\nu}U_{\si_1...\si_k}
+\Pi^{\ro}_{\mu}(\Pi^{\la_1..\la_k}_{\nu_1..\nu_k}D_{\ro}\Pi^{\si_1..\si_k}_{\la_1..\la_k})\frac{\Dbb}{\partial\nu}U_{\si_1...\si_k}\nn\\
&-&\!\oom(\Pi^{\mu'}_{\mu}\dd_4\Pi^{\ro}_{\mu'})\Pi^{\si_1..\si_k}_{\nu_1..\nu_k}D_{\ro}U_{\si_1..\si_k}
-\oom\Pi^{\ro}_{\mu}(\Pi^{\nu'_1..\nu'_k}_{\nu_1..\nu_k}\dd_4\Pi^{\si_1..\si_k}_{\nu'_1..\nu'_k})D_{\ro}U_{\si_1..\si_k}\nn\\
&+&\!\oom\Pi^{\ro}_{\mu}\Pi^{\si_1..\si_k}_{\nu_1..\nu_k}(D_{\ro}e_4^{\tau})D_{\tau}U_{\si_1...\si_k}
+\oom\Pi^{\ro}_{\mu}\Pi^{\si_1..\si_k}_{\nu_1..\nu_k}e_4^{\tau}[D_{\ro},D_{\tau}]U_{\si_1...\si_k}\nn\\
&&\nn\\
&=&\!\left[\Pi^{\ro}_{\mu}(D_{\ro}\log\oom)\Pi^{\si_1..\si_k}_{\nu_1..\nu_k}\frac{\Dbb}{\partial\nu}U_{\si_1...\si_k}
-\oom(\Pi^{\mu'}_{\mu}\dd_4\Pi^{\ro}_{\mu'})\Pi^{\si_1..\si_k}_{\nu_1..\nu_k}D_{\ro}U_{\si_1..\si_k}\right.\nn\\
&&\left.
+\oom\Pi^{\ro}_{\mu}\Pi^{\si_1..\si_k}_{\nu_1..\nu_k}(D_{\ro}e_4^{\tau})D_{\tau}U_{\si_1...\si_k}\right]\nn\\
&+&\!\left[\Pi^{\ro}_{\mu}(\Pi^{\la_1..\la_k}_{\nu_1..\nu_k}D_{\ro}\Pi^{\si_1..\si_k}_{\la_1..\la_k})\frac{\Dbb}{\partial\nu}U_{\si_1...\si_k}
-\oom\Pi^{\ro}_{\mu}(\Pi^{\nu'_1..\nu'_k}_{\nu_1..\nu_k}\dd_4\Pi^{\si_1..\si_k}_{\nu'_1..\nu'_k})D_{\ro}U_{\si_1..\si_k}\right]\nn\\
&+&\!\oom\Pi^{\ro}_{\mu}\Pi^{\si_1..\si_k}_{\nu_1..\nu_k}e_4^{\tau}[D_{\ro},D_{\tau}]U_{\si_1...\si_k}\nn
\eea
Observe that the first square bracket simplifies. In fact recalling that
\[\Pi^{\mu'}_{\mu}\dd_4\Pi^{\ro}_{\mu'}=\etab_{\mu}e_4^{\ro}\ \ ,\ \ \oom\Pi^{\ro}_{\mu}D_{\ro}e_4^{\tau}
=\oom\chi_{\mu}^{\tau}-\oom\ze_{\mu}e_4^{\tau}\]
\bea
\bigg[(1)\bigg]\!&=&\!(\nabb_{\mu}\log\oom)\Pi^{\si_1..\si_k}_{\nu_1..\nu_k}\frac{\Dbb}{\partial\nu}U_{\si_1...\si_k}
-\etab_{\mu}\Pi^{\si_1..\si_k}_{\nu_1..\nu_k}\frac{\Dbb}{\partial\nu}U_{\si_1...\si_k}\nn\\
&+&\!\oom\chi_{\mu}^{\tau}\Pi^{\si_1..\si_k}_{\nu_1..\nu_k}D_{\tau}U_{\si_1...\si_k}
-\ze_{\mu}\Pi^{\si_1..\si_k}_{\nu_1..\nu_k}\frac{\Dbb}{\partial\nu}U_{\si_1...\si_k}\nn\\
&=&\!\oom\Pi^{\si_1..\si_k}_{\nu_1..\nu_k}\chi_{\mu}^{\ro}D_{\ro}U_{\si_1...\si_k}\ ,
\eea
therefore
\bea
[\nabb_{\mu},\frac{\Dbb}{\partial\nu}]U_{\nu_1...\nu_k}\!&=&
\left[\Pi^{\ro}_{\mu}(\Pi^{\la_1..\la_k}_{\nu_1..\nu_k}D_{\ro}\Pi^{\si_1..\si_k}_{\la_1..\la_k})\frac{\Dbb}{\partial\nu}U_{\si_1...\si_k}
-\oom\Pi^{\ro}_{\mu}(\Pi^{\nu'_1..\nu'_k}_{\nu_1..\nu_k}\dd_4\Pi^{\si_1..\si_k}_{\nu'_1..\nu'_k})D_{\ro}U_{\si_1..\si_k}\right.\nn\\
&&\!\left.\ +\oom\Pi^{\si_1..\si_k}_{\nu_1..\nu_k}\chi_{\mu}^{\ro}D_{\ro}U_{\si_1...\si_k}\right.\bigg]
+\oom\Pi^{\ro}_{\mu}\Pi^{\si_1..\si_k}_{\nu_1..\nu_k}e_4^{\tau}[D_{\ro},D_{\tau}]U_{\si_1...\si_k}\ .\nn
\eea
Observe now that 
\beaa
\Pi^{\ro}_{\mu}(\Pi^{\la_1..\la_k}_{\nu_1..\nu_k}D_{\ro}\Pi^{\si_1..\si_k}_{\la_1..\la_k})
=\sum_{j=1}^k\Pi^{\si_1..\hat{\la}_j..\si_k}_{\nu_1..\hat{\nu}_j..\nu_k}(\Pi^{\la_j}_{\nu_j}\nabb_{\mu}\Pi^{\si_j}_{\la_j})
=\frac{1}{2}\sum_{j=1}^k\Pi^{\si_1..\hat{\la}_j..\si_k}_{\nu_1..\hat{\nu}_j..\nu_k}(\chib_{\mu\nu_j}e_4^{\si_j}+\chi_{\mu\nu_j}e_3^{\si_j})
\eeaa
Therefore
\beaa
&&\Pi^{\ro}_{\mu}(\Pi^{\la_1..\la_k}_{\nu_1..\nu_k}D_{\ro}\Pi^{\si_1..\si_k}_{\la_1..\la_k})\frac{\Dbb}{\partial\nu}U_{\si_1...\si_k}=
\frac{1}{2}\sum_{j=1}^k\Pi^{\si_1..\hat{\si}_j..\si_k}_{\nu_1..\hat{\nu}_j..\nu_k}(\chib_{\mu\nu_j}e_4^{\si_j}+\chi_{\mu\nu_j}e_3^{\si_j})
\frac{\Dbb}{\partial\nu}U_{\si_1..\si_j..\si_k}\nn\\
&&=\frac{\oom}{2}\sum_{j=1}^k\Pi^{\si_1..\hat{\si}_j..\si_k}_{\nu_1..\hat{\nu}_j..\nu_k}
\left(\chib_{\mu\nu_j}(D_4U)_{\si_1..{\si}_j..\si_k}e_4^{\si_j}+\chi_{\mu\nu_j}(D_4U)_{\si_1..{\si}_j..\si_k}e_3^{\si_j}\right)\nn\\
&&=-\frac{\oom}{2}\sum_{j=1}^k\Pi^{\si_1..\hat{\si}_j..\si_k}_{\nu_1..\hat{\nu}_j..\nu_k}
\left(\chib_{\mu\nu_j}U_{\si_1..{\si}_j..\si_k}(D_4e_4)^{\si_j}+\chi_{\mu\nu_j}U_{\si_1..{\si}_j..\si_k}(D_4e_3)^{\si_j}\right)\nn\\
&&=-\frac{\oom}{2}\sum_{j=1}^k\Pi^{\si_1..\hat{\si}_j..\si_k}_{\nu_1..\hat{\nu}_j..\nu_k}
\chi_{\mu\nu_j}U_{\si_1..{\si}_j..\si_k}(D_4e_3)^{\si_j}=
-{\oom}\sum_{j=1}^k\Pi^{\si_1..\hat{\si}_j..\si_k}_{\nu_1..\hat{\nu}_j..\nu_k}
\chi_{\mu\nu_j}\etab^{\si_j}U_{\si_1..{\si}_j..\si_k}\ .\nn\\
\eeaa
Moreover, as 
\beaa
&&\Pi^{\nu'_1..\nu'_k}_{\nu_1..\nu_k}\dd_4\Pi^{\si_1..\si_k}_{\nu'_1..\nu'_k}  
=\sum_{j=1}^k\Pi^{\si_1..\hat{\si}_j..\si_k}_{\nu_1..\hat{\nu}_j..\nu_k}(\Pi^{\nu'_j}_{\nu_j}\dd_4\Pi^{\si_j}_{\nu'_j})\nn\\
&&=-\sum_{j=1}^k\Pi^{\si_1..\hat{\si}_j..\si_k}_{\nu_1..\hat{\nu}_j..\nu_k}\left(\Pi^{\nu'_j}_{\nu_j}(\ddb_4\theta^4)_{\nu'_j}e_4^{\si_j}
+\Pi^{\nu'_j}_{\nu_j}(\ddb_4\theta^3)_{\nu'_j}e_3^{\si_j}\right)\ \ ,
\eeaa
\beaa
&&\Pi^{\ro}_{\mu}(\Pi^{\nu'_1..\nu'_k}_{\nu_1..\nu_k}\dd_4\Pi^{\si_1..\si_k}_{\nu'_1..\nu'_k})D_{\ro}U_{\si_1..\si_k}
=(\Pi^{\nu'_1..\nu'_k}_{\nu_1..\nu_k}\dd_4\Pi^{\si_1..\si_k}_{\nu'_1..\nu'_k})(\nabb_{\mu}U_{\si_1..\si_k})\nn\\
&&
=-\sum_{j=1}^k\Pi^{\si_1..\hat{\si}_j..\si_k}_{\nu_1..\hat{\nu}_j..\nu_k}\left(\Pi^{\nu'_j}_{\nu_j}(\dd_4\theta^4)_{\nu'_j}e_4^{\si_j}
+\Pi^{\nu'_j}_{\nu_j}(\dd_4\theta^3)_{\nu'_j}e_3^{\si_j}\right)(\nabb_{\mu}U_{\si_1..\si_k})\nn\\
&&=-\sum_{j=1}^k\Pi^{\si_1..\hat{\si}_j..\si_k}_{\nu_1..\hat{\nu}_j..\nu_k}\left(\Pi^{\nu'_j}_{\nu_j}(\dd_4\theta^4)_{\nu'_j}
\right)(\nabb_{\mu}U_{\si_1..\si_j..\si_k})e_4^{\si_j}\nn\\
&&=\sum_{j=1}^k\Pi^{\si_1..\hat{\si}_j..\si_k}_{\nu_1..\hat{\nu}_j..\nu_k}\left(\Pi^{\nu'_j}_{\nu_j}(\dd_4\theta^4)_{\nu'_j}
\right)U_{\si_1..\si_j..\si_k}(\nabb_{\mu}e_4)^{\si_j}\nn\\
&&=-\frac{1}{2}\sum_{j=1}^k\Pi^{\si_1..\hat{\si}_j..\si_k}_{\nu_1..\hat{\nu}_j..\nu_k}\left(\gggg(\dd_4e_3,e_C)\theta^C_{\nu_j}
\right)U_{\si_1..\si_j..\si_k}(\nabb_{\mu}e_4)^{\si_j}\nn\\
&&=-\sum_{j=1}^k\Pi^{\si_1..\hat{\si}_j..\si_k}_{\nu_1..\hat{\nu}_j..\nu_k}\etab_{\nu_j}U_{\si_1..\si_j..\si_k}(\nabb_{\mu}e_4)^{\si_j}
=-\sum_{j=1}^k\Pi^{\si_1..\hat{\si}_j..\si_k}_{\nu_1..\hat{\nu}_j..\nu_k}\chi_{\mu}^{\si_j}\etab_{\nu_j}U_{\si_1..\si_j..\si_k}.\nn
\eeaa
Therefore
\bea
&&\oom\Pi^{\ro}_{\mu}(\Pi^{\nu'_1..\nu'_k}_{\nu_1..\nu_k}\dd_4\Pi^{\si_1..\si_k}_{\nu'_1..\nu'_k})D_{\ro}U_{\si_1..\si_k}
=-\oom\sum_{j=1}^k\Pi^{\si_1..\hat{\si}_j..\si_k}_{\nu_1..\hat{\nu}_j..\nu_k}\chi_{\mu}^{\si_j}\etab_{\nu_j}U_{\si_1..\si_j..\si_k}\nn
\eea
and the final result is:\footnote{In our notations $(\nabb_{\mu}U_{\si_1..\si_k})$ is not in general $S$-tangent, while $(\nabb_{\mu}U)_{\si_1..\si_k}$ is.}
\bea
[\nabb_{\mu},\frac{\Dbb}{\partial\nu}]U_{\nu_1...\nu_k}\!&=&\!
\left[\Pi^{\ro}_{\mu}(\Pi^{\la_1..\la_k}_{\nu_1..\nu_k}D_{\ro}\Pi^{\si_1..\si_k}_{\la_1..\la_k})\frac{\Dbb}{\partial\nu}U_{\si_1...\si_k}
-\oom\Pi^{\ro}_{\mu}(\Pi^{\nu'_1..\nu'_k}_{\nu_1..\nu_k}\dd_4\Pi^{\si_1..\si_k}_{\nu'_1..\nu'_k})D_{\ro}U_{\si_1..\si_k}\right.\nn\\
&&\!\left.\ +\oom\Pi^{\si_1..\si_k}_{\nu_1..\nu_k}\chi_{\mu}^{\ro}D_{\ro}U_{\si_1...\si_k}\right.\bigg]
+\oom\Pi^{\ro}_{\mu}\Pi^{\si_1..\si_k}_{\nu_1..\nu_k}e_4^{\tau}[D_{\ro},D_{\tau}]U_{\si_1...\si_k}\nn\\
&&\nn\\
\!&=&\!-{\oom}\bigg[\sum_{j=1}^k\Pi^{\si_1..\hat{\si}_j..\si_k}_{\nu_1..\hat{\nu}_j..\nu_k}\!
\left(\chi_{\mu\nu_j}\etab^{\si_j}-\chi_{\mu}^{\si_j}\etab_{\nu_j}\right)\!U_{\si_1..{\si}_j..\si_k}
-\Pi^{\si_1..\si_k}_{\nu_1..\nu_k}\chi_{\mu}^{\ro}(\nabb_{\ro}U)_{\si_1...\si_k}\bigg]\nn\\
&&-\oom\Pi^{\si_1..\si_k}_{\nu_1..\nu_k}\Pi^{\ro}_{\mu}e_4^{\tau}[D_{\tau},D_{\ro}]U_{\si_1...\si_k}\ \ .
\eea
The last term can be rewritten in the following way
\bea
&&-\oom\Pi^{\si_1..\si_k}_{\nu_1..\nu_k}\Pi^{\ro}_{\mu}e_4^{\tau}[D_{\tau},D_{\ro}]U_{\si_1...\si_k}=
-\oom\Pi^{\si_1..\si_k}_{\nu_1..\nu_k}\theta^C_{\mu}e_C^{\ro}e_4^{\tau}[D_{\tau},D_{\ro}]U_{\si_1...\si_k}\nn\\
&&=-\oom\Pi^{\si_1..\si_k}_{\nu_1..\nu_k}\theta^C_{\mu}\sum_{i=1}^kR^{\tilde{\si}_i}(e_C,e_4)_{\si_i}U_{\si_1..\tilde{\si}_i..\si_k}
=-\oom\sum_{i=1}^k\Pi^{\si_1..\hat{\si}_i..\si_k}_{\nu_1..\hat{\nu}_i..\nu_k}
\theta^C_{\mu}\Pi^{\si_i}_{\nu_i}R^{\tilde{\si}_i}(e_C,e_4)_{\si_i}U_{\si_1..\tilde{\si}_i..\si_k}\nn\\
&&=-\oom\theta^C_{\mu}\sum_{i=1}^k\Pi^{\si_1..\hat{\si}_i..\si_k}_{\nu_1..\hat{\nu}_i..\nu_k}
\Pi^{\si_i}_{\nu_i}R^{\tilde{\si}_i}(e_C,e_4)_{\si_i}U_{\si_1..\tilde{\si}_i..\si_k}
=-\oom\theta^C_{\mu}\sum_{i=1}^k\theta^D_{\nu_i}R^{{\si}_i}(\c,e_C,e_4,e_D)U_{\nu_1..{\si}_i..\nu_k}\nn
\eea
and
\bea
&&[\nabb_{\mu},\frac{\Dbb}{\partial\nu}]U_{\nu_1...\nu_k}=-{\oom}\bigg[\sum_{j=1}^k\!
\left(\chi_{\mu\nu_j}\etab^{\si_j}-\chi_{\mu}^{\si_j}\etab_{\nu_j}\right)\!U_{\nu_1..{\si}_j..\nu_k}
-\chi_{\mu}^{\ro}(\nabb_{\ro}U)_{\nu_1..\nu_k}\bigg]\nn\\
&&\ \ \ \ \ \ \ \ \ \ \ \ \ \ \ \ \ \ \ \ \ \ \ \ \ \ 
-\oom\theta^C_{\mu}\sum_{j=1}^k\theta^D_{\nu_j}R^{{\si}_j}(\c,e_C,e_4,e_D)U_{\nu_1..{\si}_j..\nu_k}\nn\\
&&=-{\oom}\sum_{j=1}^k\!
\left[(\chi_{\mu\nu_j}\etab^{\si_j}\!-\!\chi_{\mu}^{\si_j}\etab_{\nu_j})
+\theta^C_{\mu}\theta^D_{\nu_j}R^{{\si}_j}(\c,e_C,e_4,e_D)\right]\!U_{\nu_1..{\si}_j..\nu_k}
+\oom\chi_{\mu}^{\ro}(\nabb_{\ro}U)_{\nu_1..\nu_k}\nn\\
&&=-\sum_{j=1}^k{C}^{\si_j}_{\mu\nu_j}U_{\nu_1..{\si}_j..\nu_k}+\oom\chi_{\mu}^{\ro}(\nabb_{\ro}U)_{\nu_1..\nu_k}\ ,
\eea
\bea
\mbox{with}\ \ \ \ \ \ \ \ \ \ {C}^{\si}_{\mu\nu}=\oom\!\left[(\chi_{\mu\nu}\etab^{\si}\!-\!\chi_{\mu}^{\si}\etab_{\nu})
+\theta^C_{\mu}\theta^D_{\nu}R^{{\si}}(\c,e_C,e_4,e_D)\right]\ \ \ \ \ \ \ \ \ \ \ \ \ \ \ \ \ 
\eea
and the proof of \ref{6.6} is achieved.

\medskip

\NI The proof  of equation \ref{6.7} is similar and we do not report it here.

\medskip

\NI Next we consider the commutator $[\nabb_{\a},\nabb_{\mu}]$ applied to the generic tensor $U_{\nu_1\nu_2...\nu_q}$

\bea
&&\ML\nabb_{\a}\nabb_{\mu}U_{\nu_1\nu_2...\nu_q}=\nabb_{\a}\Pi^{\si}_{\mu}\Pi^{\si_1..\si_q}_{\nu_1..\nu_q}D_{\si}U_{\si_1\si_2...\si_q}
=\Pi^{\ro}_{\a}\Pi^{\la_1..\la_q}_{\nu_1..\nu_q}D_{\ro}\Pi^{\si}_{\mu}\Pi^{\si_1..\si_q}_{\la_1..\la_q}D_{\si}U_{\si_1\si_2...\si_q}\nn\\
&&\ML\ \ \ \ \ \ =(\Pi^{\ro}_{\a}\Pi^{\la_1..\la_q}_{\nu_1..\nu_q}D_{\ro}\Pi^{\si}_{\mu}\Pi^{\si_1..\si_q}_{\la_1..\la_q})D_{\si}U_{\si_1\si_2...\si_q}
+\Pi^{\ro}_{\a}\Pi^{\la_1..\la_q}_{\nu_1..\nu_q}\Pi^{\si}_{\mu}\Pi^{\si_1..\si_q}_{\la_1..\la_q}D_{\ro}D_{\si}U_{\si_1\si_2...\si_q}\nn\\
&&\ML\nabb_{\mu}\nabb_{\a}U_{\nu_1\nu_2...\nu_q}=\nabb_{\mu}\Pi^{\si}_{\a}\Pi^{\si_1..\si_q}_{\nu_1..\nu_q}D_{\si}U_{\si_1\si_2...\si_q}
=\Pi^{\si}_{\mu}\Pi^{\la_1..\la_q}_{\nu_1..\nu_q}D_{\si}\Pi^{\ro}_{\a}\Pi^{\si_1..\si_q}_{\la_1..\la_q}D_{\ro}U_{\si_1\si_2...\si_q}\nn\\
&&\ML\ \ \ \ \ \ =(\Pi^{\ro}_{\mu}\Pi^{\la_1..\la_q}_{\nu_1..\nu_q}D_{\ro}\Pi^{\si}_{\a}\Pi^{\si_1..\si_q}_{\la_1..\la_q})D_{\si}U_{\si_1\si_2...\si_q}
+\Pi^{\si}_{\mu}\Pi^{\la_1..\la_q}_{\nu_1..\nu_q}\Pi^{\ro}_{\a}\Pi^{\si_1..\si_q}_{\la_1..\la_q}D_{\si}D_{\ro}U_{\si_1\si_2...\si_q}\ .\nn
\eea
Therefore
\bea
&&[\nabb_{\a},\nabb_{\mu}]U_{\nu_1\nu_2...\nu_q}=\nn\\
&&\left[(\Pi^{\ro}_{\a}\Pi^{\la_1..\la_q}_{\nu_1..\nu_q}D_{\ro}\Pi^{\si}_{\mu}\Pi^{\si_1..\si_q}_{\la_1..\la_q})
-(\Pi^{\ro}_{\mu}\Pi^{\la_1..\la_q}_{\nu_1..\nu_q}D_{\ro}\Pi^{\si}_{\a}\Pi^{\si_1..\si_q}_{\la_1..\la_q})\right]D_{\si}U_{\si_1\si_2...\si_q}\nn\\
&&\ \ \ +\Pi^{\ro}_{\a}\Pi^{\si}_{\mu}\Pi^{\si_1..\si_q}_{\nu_1..\nu_q}[D_{\ro},D_{\si}]U_{\si_1\si_2...\si_q}\nn\\
&&=\left[(\Pi^{\ro}_{\a}D_{\ro}\Pi^{\si}_{\mu})-(\Pi^{\ro}_{\mu}D_{\ro}\Pi^{\si}_{\a})\right]\Pi^{\si_1..\si_q}_{\nu_1..\nu_q}D_{\si}U_{\si_1\si_2...\si_q}
+\Pi^{\ro}_{\a}\Pi^{\si}_{\mu}\Pi^{\si_1..\si_q}_{\nu_1..\nu_q}[D_{\ro},D_{\si}]U_{\si_1\si_2...\si_q}\nn\\
&&=\left[(\Pi^{\ro}_{\a}D_{\ro}\Pi^{\si}_{\mu})-(\Pi^{\ro}_{\mu}D_{\ro}\Pi^{\si}_{\a})\right]\Pi^{\si_1..\si_q}_{\nu_1..\nu_q}D_{\si}U_{\si_1\si_2...\si_q}
+\Pi^{\ro}_{\a}\Pi^{\si}_{\mu}[D_{\ro},D_{\si}]\Pi^{\si_1..\si_q}_{\nu_1..\nu_q}U_{\si_1\si_2...\si_q}\nn\\
&&=\left[(\Pi^{\ro}_{\a}D_{\ro}\Pi^{\si}_{\mu})-(\Pi^{\ro}_{\mu}D_{\ro}\Pi^{\si}_{\a})\right]\Pi^{\si_1..\si_q}_{\nu_1..\nu_q}D_{\si}U_{\si_1\si_2...\si_q}
+\Pi^{\ro}_{\a}\Pi^{\si}_{\mu}[D_{\ro},D_{\si}]U_{\nu_1\nu_2...\nu_q}\nn
\eea
Computing the first terms we obtain after an easy, but long computation
\bea
\left[(\Pi^{\ro}_{\a}D_{\ro}\Pi^{\si}_{\mu})-(\Pi^{\ro}_{\mu}D_{\ro}\Pi^{\si}_{\a})\right]
=\left[(\theta^3_{\mu}\chib_{\a}^{\b}+\theta^4_{\mu}\chi_{\a}^{\b})-(\theta^3_{\a}\chib_{\mu}^{\b}+\theta^4_{\a}\chi_{\mu}^{\b})\right]\Pi^{\si}_{\b}\ ,
\eea
therefore
\bea
&&\left[(\Pi^{\ro}_{\a}D_{\ro}\Pi^{\si}_{\mu})-(\Pi^{\ro}_{\mu}D_{\ro}\Pi^{\si}_{\a})\right]\Pi^{\si_1..\si_q}_{\nu_1..\nu_q}D_{\si}U_{\si_1\si_2...\si_q}\nn\\
&&=\left[(\theta^3_{\mu}\chib_{\a}^{\b}+\theta^4_{\mu}\chi_{\a}^{\b})-(\theta^3_{\a}\chib_{\mu}^{\b}+\theta^4_{\a}\chi_{\mu}^{\b})\right]\Pi^{\si}_{\b}\Pi^{\si_1..\si_q}_{\nu_1..\nu_q}D_{\si}U_{\si_1\si_2...\si_q}\nn\\
&&=\left[(\theta^3_{\mu}\chib_{\a}^{\b}+\theta^4_{\mu}\chi_{\a}^{\b})-(\theta^3_{\a}\chib_{\mu}^{\b}+\theta^4_{\a}\chi_{\mu}^{\b})\right]
\nabb_\b U_{\nu_1\nu_2...\nu_q}
\eea
and the last term of 
\bea
&&\ML\Pi^{\ro}_{\a}\Pi^{\si}_{\mu}[D_{\ro},D_{\si}]U_{\nu_1\nu_2...\nu_q}
=\theta^C_{\a}\theta^D_{\mu}e_C^{\ro}e_D^{\si}[D_{\ro},D_{\si}]U_{\nu_1\nu_2...\nu_q}\\
&&\ML=\theta^C_{\a}\theta^D_{\mu}\sum_{i=1}^qR^{\tilde{\nu}_i}(\c,e_C,e_D,\c)_{\nu_i}U_{\nu_1\nu_2..{\tilde\nu}_i..\nu_q}=
-\theta^C_{\a}\theta^D_{\mu}\sum_{i=1}^qR^{\tilde{\nu}_i}(\c,e_C,\c,e_D)_{\nu_i}U_{\nu_1\nu_2..{\tilde\nu}_i..\nu_q}\nn
\eea
so that finally
\bea
&&\ML[\nabb_{\a},\nabb_{\mu}]U_{\nu_1\nu_2...\nu_q}\\
&&\ML=\left[(\theta^3_{\mu}\chib_{\a}^{\b}+\theta^4_{\mu}\chi_{\a}^{\b})-(\theta^3_{\a}\chib_{\mu}^{\b}+\theta^4_{\a}\chi_{\mu}^{\b})\right]
\nabb_\b U_{\nu_1\nu_2...\nu_q}-\theta^C_{\a}\theta^D_{\mu}\sum_{i=1}^qR^{\tilde{\nu}_i}(\c,e_C,\c,e_D)_{\nu_i}U_{\nu_1\nu_2..{\tilde\nu}_i..\nu_q}\ .\nn
\eea

\NI Which is formula \ref{68}.

\medskip

{ \NI {\bf Some more commutation relations.}{

\NI  We start considering \[[\frac{\Dbb}{\partial\nu},\Lie_O]U\]
where $U$ is a one form, $U=U_{\mu}dx^{\mu}$, $\Lie_O$ a Lie derivative with respect to a $O^{(i)}$, $\Lie_O\equiv\Lie_{O^{(i)}}$, and $\frac{\Dbb}{\partial\nu}=\oom D_4$.
We have
\bea
&&\ML\ML(\Lie_OU)(e_A)=\pr_O(U(e_A))-U([O,e_A])=D_OU(e_A)+U(D_{e_A}O)\nn\\
&&\ML\ML=(\nabb_OU)(e_A)+\gggg(e_B,\nabb_{e_A}O)U(e_B)\ .
\eea
Therefore
\bea
\Lie_OU(\c)=\nabb_OU(\c)+\gggg(e_B,\nabb_{\c}O)U(e_B)=\nabb_OU(\c)+(U\c{\cal H})(\c)
\eea
\NI Observe that ${\cal H}^{ab}$ has the structure of a connection coefficient, in fact
\bea
{\cal H}_{ab}=\gggg(\nabb_aO,\frac{\pr}{\pr x^b})\ \ 
\eea
\bea
&&\frac{\Dbb}{\partial\nu}\Lie_OU=\frac{\Dbb}{\partial\nu}\nabb_OU+\frac{\Dbb}{\partial\nu}U\c{\cal H}+U\c\frac{\Dbb}{\partial\nu}{\cal H}\nn\\
&&\Lie_O\frac{\Dbb}{\partial\nu}U=\nabb_O\frac{\Dbb}{\partial\nu}U+(\frac{\Dbb}{\partial\nu}U\c{\cal H})\ .
\eea
Therefore
\bea
[\frac{\Dbb}{\partial\nu},\Lie_O]U=[\frac{\Dbb}{\partial\nu},\nabb_O]U+U\c\frac{\Dbb}{\partial\nu}{\cal H}
=(\frac{\Dbb}{\partial\nu}O)\c\nabb U+(O\c[\frac{\Dbb}{\partial\nu},\nabb])U+U\c\frac{\Dbb}{\partial\nu}{\cal H}\nn
\eea
and expressing the components
\bea
[\frac{\Dbb}{\partial\nu},\Lie_O]U_{\mu}=\left(\frac{\Dbb}{\partial\nu}O\right)^{\!\!\la}\nabb_{\la} U_{\mu}+O^{\si}[\frac{\Dbb}{\partial\nu},\nabb]_{\si}U_{\mu}+U_{\ro}\left(\frac{\Dbb}{\partial\nu}{\cal H}\right)^{\!\ro}_{\!\mu}\nn
\eea

Moreover
\bea
O^{\si}[\frac{\Dbb}{\partial\nu},\nabb]_{\si}U_{\mu}=(O^{\si}C^{\tau}_{\si\mu})U_{\tau}-(O^{\si}\oom\chi^{\ro}_{\si})\nabb_{\ro}U_{\mu}\ ,
\eea
\bea
&&\ML\ML\left(\frac{\Dbb}{\partial\nu}O\right)^{\la}=\oom (D_4O)_{\mu}\de^{\mu\la}=-\frac{1}{2}\oom (D_4O)_{\mu}(e_3^{\mu}e_4^{\la}+e_4^{\mu}e_3^{\la})+\oom (D_4O)_{\mu}e_A^{\mu}e_A^{\la}\nn\\
&&\ML\ML=-\frac{1}{2}\oom\gggg(D_4O,e_3)e_4^{\la}+\oom\gggg(D_4O,e_A)e_A=\oom\gggg(\etab,O)e_4^{\la}+\oom\chi_{\mu\la}O^{\mu}e_A^{\la}
\eea
and
\bea
\left(\frac{\Dbb}{\partial\nu}O\right)^{\!\!\la}\nabb_{\la} U_{\mu}=(O^{\nu}\oom\chi_{\nu}^{\la})\nabb_{\la}U_{\mu}
+(g(\etab,O)e_4^{\la})\nabb_{\la}U_{\mu}=(O^{\nu}\oom\chi_{\nu}^{\la})\nabb_{\la}U_{\mu}\ .\ \ \ 
\eea
Therefore
\bea
&&[\frac{\Dbb}{\partial\nu},\Lie_O]U=\left((O^{\si}C^{\tau}_{\si\mu})+\frac{\Dbb}{\partial\nu}{\cal H}^{\tau}_{\mu}\right)\!U_{\tau}+(O^{\nu}\oom\chi_{\nu}^{\la})\nabb_{\la}U_{\mu}
-(O^{\si}\oom\chi^{\ro}_{\si})\nabb_{\ro}U_{\mu}\nn\\
&&\ \ \ \ \ \ \ \ \ \ \ \ \ \ =\left((O^{\si}C^{\tau}_{\si\mu})+\frac{\Dbb}{\partial\nu}{\cal H}^{\tau}_{\mu}\right)\!U_{\tau}\nn
\eea
which we write in a compact way as
\bea
[\frac{\Dbb}{\partial\nu},\Lie_O]U=E^{\tau}_{\mu}U_{\tau}\ ,
\eea
where
\bea
&&E^{\tau}_{\mu}\equiv \left((O^{\si}C^{\tau}_{\si\mu})+\frac{\Dbb}{\partial\nu}{\cal H}^{\tau}_{\mu}\right)\ .\nn
\eea

\NI With similar calculations we obtain

\bea
[\frac{\Dbb}{\partial\la},\Lie_O]U=\underline{E}^{\tau}_{\mu}U_{\tau}\ ,
\eea

\NI where
\bea
&&\underline{E}^{\tau}_{\mu}\equiv \left((O^{\si}\underline{C}^{\tau}_{\si\mu})+\frac{\Dbb}{\partial\nu}{\cal H}^{\tau}_{\mu}\right)\ .\nn
\eea

\medskip

\NI Finally we consider $[\nabb, \Lie_O]U_{\a\b}$ and  $[\divv,\Lie_O]U_{\a\b}$ with $U_{\a\b}$  a symmetric 2-form.

\bea
&&\ML\ML([\nabb_{\nu},\Lie_O]U)_{ab}=\left(\nabb_{\nu} O\right)^{\!\!\la}\nabb_{\la} U_{ab}+O^{\si}[\nabb_{\nu},\nabb_{\si}]U_{ab}+2U_{a\ro}\left(\nabb_{\nu}{\cal H}^{\ro}_b\right)=\nn \\
&&\ML\ML=O^{\ro}\hat{C}^{\si}_{\ro\nu a }U_{\si b}+2U_{a\ro}\left(\nabb_{\nu}{\cal H}^{\ro}_b\right)+\sum_{A=1,2}(O^{\ro}{\cal H}_{\ro}^{\b}\theta_{\nu A}\nabb_{\b}U_{ab})
\eea

\NI which we rewrite symbolically as:

\bea
([\nabb_{\nu},\Lie_O]U)_{ab}=\hat{E}^{\si}_{\nu a }U_{\si b} +\sum_{A=12}\ O^{\ro}\hat{\cal H} ^{\b}_{\ro}\theta_{\nu A}\nabb_{\b}U_{ab}
\eea

\NI and from it,

\bea
&&([\divv,\Lie_O]U)_c=[\nabb_{\nu},\Lie_O]U^{\nu}_c=O^{\ro}\hat{C}^{\si}_{\ro\nu}\ ^{\nu}U_{\si c}+2U_{c\ro}\left(\nabb_{\nu}{\cal H}^{\ro\nu}\right)+\sum_{A=1,2}(O^{\ro}{\cal H}_{\ro}^{\b}\theta_{\nu A}\nabb_{\b}U^{\nu}_{c})\nn\\
\eea


\NI which we rewrite symbolically as:

\bea
([\divv,\Lie_O]U)_c=\overline{E}^{\si}_{\nu}\ ^{\nu}U_{\si c} +\sum_{A=12} \ ^A\overline{\cal H} ^{\b\nu}\nabb_{\b}U_{\nu c}
\eea

\NI Performing analogous steps we obtain for $U_{\a}$ 1-form  

\bea\label{1535}
&&\ML\ML([\nabb_{\nu},\Lie_O]U)_{c}=\left(\nabb_{\nu} O\right)^{\!\!\la}\nabb_{\la} U_{c}+O^{\si}[\nabb_{\nu},\nabb_{\si}]U_{c}+U_{\ro}\left(\nabb_{\nu}{\cal H}^{\ro}_c\right)=\nn \\
&&\ML\ML=O^{\ro}\hat{C}^{\si}_{\ro\nu c }U_{\si }+U_{\ro}\left(\nabb_{\nu}{\cal H}^{\ro}_c\right)+\sum_{A=1,2}(O^{\ro}{\cal H}_{\ro}^{\b}\theta_{\nu A}\nabb_{\b}U_{c})
\eea

\NI which we rewrite symbolically as:

\bea
[\nabb_{\nu},\Lie_O]U_{c}=\widehat{E}^{\si}_{\nu c }U_{\si } +\sum_{A=12}\ O^{\ro}\widehat{\cal H} ^{\b}_{\ro}\theta_{\nu A}\nabb_{\b}U_{c}
\eea

\NI and from it,

\bea
&&([\divv,\Lie_O]U)=[\nabb_{\nu},\Lie_O]U^{\nu}=O^{\ro}\hat{C}^{\si}_{\ro\nu}\ ^{\nu}U_{\si }+U_{\ro}\left(\nabb_{\nu}{\cal H}^{\ro\nu}\right)+\sum_{A=1,2}(O^{\ro}{\cal H}_{\ro}^{\b}\theta_{\nu A}\nabb_{\b}U^{\nu})\nn\\
\eea

\NI which we rewrite symbolically as:

\bea
([\divv,\Lie_O]U)=\widetilde{E}^{\si}_{\nu}\ ^{\nu}U_{\si} +\sum_{A=12} \ ^A\widetilde{\cal H} ^{\b\nu}\nabb_{\b}U_{\nu}
\eea

\NI Moreover we need an expression for $[\curll,\Lie_O]U$, with $U$ 1-form, to do this we exploit the expression \ref{1535} 

\bea
&&\ML\ML([\curll,\Lie_O]U)=\ep^{\nu c}\left(\nabb_{\nu} O\right)^{\!\!\la}\nabb_{\la} U_{c}+\ep^{\nu c}O^{\si}[\nabb_{\nu},\nabb_{\si}]U_{c}+\ep^{\nu c}U_{\ro}\left(\nabb_{\nu}{\cal H}^{\ro}_c\right)=\nn \\
&&\ML\ML=\ep^{\nu c}O^{\ro}\hat{C}^{\si}_{\ro\nu c }U_{\si }+\ep^{\nu c}U_{\ro}\left(\nabb_{\nu}{\cal H}^{\ro}_c\right)+\ep^{\nu c}\sum_{A=1,2}(O^{\ro}{\cal H}_{\ro}^{\b}\theta_{\nu A}\nabb_{\b}U_{c})
\eea

\NI As before we write the last expression in a symbolical way

\bea
([\curll,\Lie_O]U)=\hat{\hat{{E}}}^{\si}_{\nu}\ ^{\nu}U_{\si} +\sum_{A=12} \ ^A\hat{\hat{{\cal H}}} ^{\b c}\nabb_{\b}U_{c}
\eea

\NI{\bf Remarks:} {\em 

\NI i) It is important to remark that  the difference of the commutator $[\Lie_O^{N-1},\frac{\Dbb}{\partial\nu}]$ with respect to $[\nabb^{N-1},\frac{\Dbb}{\partial\nu}]$ is the reason of the missing of the factor $(N\!-\!1)$ in front of $\tr\chi$ in the transport equation of $\Lie_O^{N-1}\Us$, as expected and connected to the absence of extra $r$-decays for $\lie_O^{N-1}\Us$.
\medskip

\NI ii) We expect, as we show next, that we could perform all the estimates for the $\Lie_O^N$ of the connection coefficients in the same way obtaining for their $|\c|_{p,S}$ norms the estimates assumed before for the $\nabb^N$ derivatives, with a different constant in front.
In fact, also in this case, the commutators have the same structure of the commutators with the $\nabb$ derivatives, namely a combination of products of connection coefficients and metric components time $\nabb U$ and a combination of null Riemann components times $ U$.}} 

\subsection{Proof of Lemma \ref{eqforU}:} 
\smallskip

Let us look at the transport equation of $\Lie_O^{N-1}\nabb\tr\chi$ or more carefully to the one of $\Lie_O^{N-1}\Us$, the equation for $\Us$ is
\bea
\frac{\Dbb}{\partial\nu}\Us+\frac{3}{2}\oom\tr\chi\Us=-\nabb|\chih|^2-\eta|\chih|^2-\oom\chih\c{\Us}+\tr\chi(\chih\c\etab)-\tr\chi\b\ .\eql{Useqa}
\eea
Applying $\Lie_O^{N-1}$ to equation \ref{Useqa} we obtain:
\bea
\ML\Lie_O^{N-1}\frac{\Dbb\Us}{\partial\nu}=\frac{\Dbb}{\partial\nu}(\Lie_O^{N-1}\Us)+[\Lie_O^{N-1},\frac{\Dbb}{\partial\nu}]\Us=\frac{\Dbb}{\partial\nu}(\Lie_O^{N-1}\Us)+\sum_{k=0}^{N-2}\Lie_O^k[\Lie_O,\frac{\Dbb}{\partial\nu}](\Lie_O^{N-2-k}\Us)\ \ \ .\nn
\eea
Therefore
\bea
&&\ML\ML\ML\ML\frac{\Dbb}{\partial\nu}(\Lie_O^{N-1}\Us)=\Lie_O^{N-1}\frac{\Dbb\Us}{\partial\nu}-\sum_{k=0}^{N-2}\Lie_O^k[\frac{\Dbb}{\partial\nu},\Lie_O](\Lie_O^{N-2-k}\Us)\nn\\
&&\ML\ \ =\Lie_O^{N-1}\frac{\Dbb\Us}{\partial\nu}-\sum_{k=0}^{N-2}\Lie_O^kE(\Lie_O^{N-2-k}\Us)
\ .\ \ \ \ \ \ \ \ \ \ \ 
\eea
Observe that
\bea
&&\ML\ML\sum_{k=0}^{N-2}\sum_{s=0}^{k}\cbin{k}{s}=\sum_{s=0}^{N-2}\sum_{k=s}^{N-2}\cbin{k}{s}=\sum_{s=0}^{N-2}\cbin{N-1}{N-2-s}\ ,
\eea
therefore \eql{15.49}
\bea
&&\ML\ML\sum_{k=0}^{N-2}\Lie_O^{k}E(\Lie_O^{N-2-k}\Us)=\sum_{k=0}^{N-2}\sum_{s=0}^{k}\cbin{k}{s}(\Lie_O^{s}E)(\Lie_O^{N-2-s}\Us)\\
&&\ML\ML=\sum_{s=0}^{N-2}\sum_{k=s}^{N-2}\cbin{k}{s}(\Lie_O^{s}E)(\Lie_O^{N-2-s}\Us)
=\sum_{s=0}^{N-2}\cbin{N-1}{N-2-s}(\Lie_O^{s}E)(\Lie_O^{N-2-s}\Us)\nn
\eea
and
\bea \eql{15.43}
&&\ML\ML\frac{\Dbb}{\partial\nu}(\Lie_O^{N-1}\Us)=\Lie_O^{N-1}\frac{\Dbb\Us}{\partial\nu}-\sum_{k=0}^{N-2}\Lie_O^k[\frac{\Dbb}{\partial\nu},\Lie_O](\Lie_O^{N-2-k}\Us)\nn\\
&&\ML\ML=\Lie_O^{N-1}\frac{\Dbb\Us}{\partial\nu}-\sum_{k=0}^{N-2}\Lie_O^kE(\Lie_O^{N-2-k}\Us)=
\Lie_O^{N-1}\frac{\Dbb\Us}{\partial\nu}-\sum_{s=0}^{N-2}\cbin{N-1}{N-2-s}(\Lie_O^{s}E)(\Lie_O^{N-2-s}\Us)\nn\\
&&\ML\ML=\Lie_O^{N-1}\frac{\Dbb\Us}{\partial\nu}-\sum_{q=0}^{N-2}\cbin{N-1}{q}(\Lie_O^{N-2-q}E)(\Lie_O^{q}\Us)
\ .\ \ \ \ \ \ \ \ \ \ \ 
\eea
Therefore
\bea\label{[]}
&&\ML\ML\frac{\Dbb}{\partial\nu}(\Lie_O^{N-1}\Us)=\Lie_O^{N-1}\!\left[-\frac{3}{2}\oom\tr\chi\Us-\nabb|\chih|^2-\eta|\chih|^2-\oom\chih\c{\Us}+\tr\chi(\chih\c\etab)-\tr\chi\b\right]\nn\\
&&\ML\ML\ \ \ \ \ \ \ \ \ \ \ \ \ \ \ \ \ \ -\sum_{q=0}^{N-2}\cbin{N-1}{q}(\Lie_O^{N-2-q}E)(\Lie_O^{q}\Us)\ .
\eea
which we rewrite as

\bea
&&\ML\ML\frac{\Dbb}{\partial\nu}(\Lie_O^{N-1}\Us)+\frac{3}{2}\oom\tr\chi(\Lie_O^{N-1}\Us)=-\oom\chih\c(\Lie_O^{N-1}\Us)-\Lie_O^{N-1}\nabb|\chih|^2\nn\\
&&\ML\ML-\frac{3}{2}\sum_{q=1}^{N-1}\cbin{N-1}{q}(\Lie_O^{q}\oom\tr\chi)(\Lie_O^{N-1-q}\Us)-\Lie_O^{N-1}\!\left[-\eta|\chih|^2-\oom\chih\c{\Us}+\tr\chi(\chih\c\etab)-\tr\chi\b\right]\nn\\
&&\ML\ML-\sum_{q=0}^{N-2}\cbin{N-1}{q}(\Lie_O^{N-2-q}E)(\Lie_O^{q}\Us)=\nn\\
&&\ML\ML=-\oom\chih\c(\Lie_O^{N-1}\Us)- 2\chih\Lie_O^{N-1}(\nabb\chih)-\tr\chi\Lie_O^{N-1}\b+2\sum_{k=1}^{N-1}\cbin{N-1}{k}\Lie_O^{N-1-k}(\nabb\chih)\Lie_O^k\chih\nn\\
&&\ML\ML-\frac{3}{2}\sum_{q=1}^{N-1}\cbin{N-1}{q}(\Lie_O^{q}\oom\tr\chi)(\Lie_O^{N-1-q}\Us)+\sum_{k=0}^{N-1}\cbin{N-1}{k}\Lie_O^{N-1-k}(|\chih|^2)\Lie_O^k\eta\nn\\
&&\ML\ML+\sum_{q=1}^{N-1}\cbin{N-1}{q}(\Lie_O^{q}(\oom\chih)(\Lie_O^{N-1-q}\Us)-\sum_{q=0}^{N-2}\cbin{N-1}{q}(\Lie_O^{N-2-q}E)(\Lie_O^{q}\Us)\nn \\
&&\ML\ML-\sum_{q=0}^{N-1}\cbin{N-1}{q}(\Lie_O^{q}(\tr\chi)(\Lie_O^{N-1-q}\chih\eta)+\sum_{q=1}^{N-1}\cbin{N-1}{q}(\Lie_O^{N-2-q}\b)(\Lie_O^{q}\tr\chi)
\eea

\NI We rewrite the last formula in a more convenient way:

\bea
&&\ML\ML\ML\ML\ML\frac{\Dbb}{\partial\nu}(\Lie_O^{N-1}\Us)+\frac{3}{2}\overline{\oom\tr\chi}(\Lie_O^{N-1}\Us)=-\oom\chih\c(\Lie_O^{N-1}\Us)- 2\chih\Lie_O^{N-1}(\nabb\chih)-\tr\chi\Lie_O^{N-1}\b\nn\\
&&\ \ \ \ \ \ \ \ \ \ \ \ \ \ \ \ \ \ \ +\frac 3 2(\overline{\oom\tr\chi}-\oom\tr\chi)(\Lie_O^{N-1}\Us)+\{{\it (good)_1}\}\eql{15.46x}
\eea

\bea
&&\ML\ML\{{\it (good)_1}\}=2\sum_{k=1}^{N-1}\cbin{N-1}{k}\Lie_O^{N-1-k}(\nabb\chih)\Lie_O^k\chih\nn\\ 
&&\ML\ML-\frac{3}{2}\sum_{q=1}^{N-1}\cbin{N-1}{q}(\Lie_O^{q}\oom^2 U)(\Lie_O^{N-1-q}\Us)+\sum_{k=0}^{N-1}\cbin{N-1}{k}\Lie_O^{N-1-k}(|\chih|^2)\Lie_O^k\eta\nn\\
&&\ML\ML+\sum_{q=1}^{N-1}\cbin{N-1}{q}(\Lie_O^{q}(\oom\chih)(\Lie_O^{N-1-q}\Us)+\sum_{q=0}^{N-2}\cbin{N-1}{q}(\Lie_O^{N-2-q}E)(\Lie_O^{q}\Us)\nn \\
&&\ML\ML-\sum_{q=0}^{N-1}\cbin{N-1}{q}(\Lie_O^{q}(\oom U)(\Lie_O^{N-1-q}\chih\eta)+\sum_{q=1}^{N-1}\cbin{N-1}{q}(\Lie_O^{N-2-q}\b)(\Lie_O^{q}\tr\chi)\nn \\
\eea

\NI and equation \ref{15.46} of lemma \ref{eqforU} is proved. In order to prove inequality \ref{diffest1aa} we have to consider the $|\ \ |_{p.S}$ norms, equation \ref{15.46} implies the following one:

\bea
&&\ML\ML\frac{\partial|(\Lie_O^{N-1}\Us)|^p}{\partial\nu}+p\frac{3}{2}\overline{\oom\tr\chi}|(\Lie_O^{N-1}\Us)|^p\geq -p|\Lie_O^{N-1}\Us|^{p}\frac 3 2|(\oom\tr\chi)-(\overline{\oom\tr\chi})|\nn\\
&&\ML\ML-p|\Lie_O^{N-1}\Us|^{p-1}[\oom|\chih\c(\Lie_O^{N-1}\Us)| +2|\chih\Lie_O^{N-1}(\nabb\chih)|+|\tr\chi||\Lie_O^{N-1}\b|+|\{{\it (good)_1}\}|] \ \ \ \ \ \ \ \ \ \ \ \
\eea

\NI which we summarize as 

\bea\eql{15.47}
&&\ML\ML\frac{\partial|(\Lie_O^{N-1}\Us)|^p}{\partial\nu}+p\frac{3}{2}\overline{\oom\tr\chi}|(\Lie_O^{N-1}\Us)|^p\geq -p\frac 3 2|(\oom\tr\chi)-(\overline{\oom\tr\chi})|
|\Lie_O^{N-1}\Us|^{p}-p|\Lie_O^{N-1}\Us|^{p-1}|F|,\nn\\
\eea

\NI with

\bea
&&\ML\ML |F|=[\oom|\chih\c(\Lie_O^{N-1}\Us)| +2|\chih\Lie_O^{N-1}(\nabb\chih)|+|\tr\chi||\Lie_O^{N-1}\b|+|\{{\it (good)_1}\}|]\ .
\ \ \ \ \ \ \ \ \ \ \ \ \eea

\NI From \ref{15.47} and by lemma 4.1.5 \cite{Kl-Ni:book}

\bea
&&\ML\ML\ML\frac{\partial|r^{(3-\frac{2}{p})}\Lie_O^{N-1}\Us|^p_{p,S}}{\partial\nu}=\int_S\left[r^{(3-\frac{2}{p})p}\!\!\left(\frac{\pr|\Lie_O^{N-1}{\Us}|^p}{\pr\nu}
+p\frac{3\overline{\oom\tr\chi}}{2}|\Lie_O^{N-1}{\Us}|^p\right)\right]d\mu_{\ga}\nn\\
&&\ML\ML\ML\geq -p\int_S\left[r^{(3-\frac{2}{p})p}|\Lie_O^{N-1}{\Us}|^{p-1}|F|+\left(\frac{3-\frac{2}{p}}{2}|\oom\tr\chi-{\overline{{\oom}\tr\chi}}|\right)|r^{(3-\frac{2}{p})}\Lie_O^{N-1}{\Us}|^p\right]\ ,\nn
\eea
which we rewrite
\bea
-\frac{\partial|r^{(3-\frac{2}{p})}\Lie_O^{N-1}\Us|^p_{p,S}}{\partial\nu}\leq p\!\int_S r^{(3-\frac{2}{p})p} |\Lie_O^{N-1}\Us|^{p-1}|\Ls|d\mu_{\ga}\ ,
\eea
where
\bea
|\Ls|\!&=&\!\!\left[\big|\oom\chih\c(\Lie_O^{N-1}\Us)\big|+\big|2\chih\c(\Lie_O^{N-1}\nabb\chih)\big|+|\tr\chi|\big|\Lie_O^{N-1}\b\big|
+\big|\big\{(good)_1\big\}\big|\right]\nn\\
&+&\!\!\left[\frac{(3-\frac 2 p)}{2}|{\oom}\tr\chi-\overline{{\oom}\tr\chi}||\Lie_O^{N-1}\Us|\right]\ \ \ \ .
\eea
Applying the Holder inequality to the right hand side we obtain
\bea
\int_S r^{(3-\frac{2}{p})p} |\Lie_O^{N-1}\Us|^{p-1}|\Ls|d\mu_{\ga}\leq |r^{(3-\frac{2}{p})}\Ls|_{p,S}|r^{(3-\frac{2}{p})}\Lie_O^{N-1}\Us|^{{p-1}}_{p,S}\ \ \ \ 
\eea
and from it inequality \ref{xxzz},
\bea
-\frac{\partial}{\partial\nu}|r^{(3-\frac{2}{p})}\Lie_O^{N-1}\Us|_{p,S}\leq  |r^{(3-\frac{2}{p})}\Ls|_{p,S}\ \ \ .\eql{diffest1aax}
\eea

\NI and Lemma \ref{eqforU} is proved\ .}

\subsection{Proof of Lemma \ref{estdNchih}:}\label{SS3.2}

{ \NI We are left with the problem of estimating the term $|\Lie_O^{N-1}\nabb|\chih|^2|_{p,S}$. We have
\bea
&&\ML\ML \Lie_O^{N-1}\chih\c\nabb\chih=\sum_{q=0}^{N-1}(\Lie_O^{N-1-q}\chih)\c(\Lie_O^q\nabb\chih)
=\chih\c(\Lie_O^{N-1}\nabb\chih)+\sum_{q=0}^{N-2}(\Lie_O^{N-1-q}\chih)\c(\lie_O^q\nabb\chih)\ .\nn
\eea
which implies that we need estimates for the norms of $\Lie_O^q\nabb\chih$. We  have them inductively when $q<N-1$, while for $q=N-1$ we have to proceed differently using the Hodge system. Observe that we can write
\bea\eql{15.55}
\Lie_O^{N-1}\nabb\chih=\nabb(\Lie_O^{N-1}\chih)+[\Lie_O^{N-1},\nabb]\chih
\eea
\NI The first term of the r.h.s. can be estimated using the  Hodge system for $(\Lie_O^{N-1}\chih)$
derived from
\bea\eql{15.56}
\divv\chih=\frac{1}{2}\nabb{\tr\chi}-\b-\zeta\!\c\hat{\chi}+\frac{1}{2}\zeta{\tr\chi}=\frac{\oom}{2}\Us-\b-\zeta\!\c\hat{\chi}\ ,\eql{fdef2}
\eea
 
\NI while the second term  written in the following way,
\bea
[\Lie_O^{N-1},\nabb]\chih=\sum_{h=0}^{N-2}\Lie_O^h[\Lie_O,\nabb](\lie_O^{N-2-h}\chih)\ .
\eea
 can be treated as done for  the term $[\frac{\ddb}{d\nu}, \Lie_O^{N-1}]\Us$, see \ref{15.43}, and it involves lower order terms which can be estimated inductively. 
Let us regroup all the terms involving the commutator  $[\nabb, \Lie_O^{N-1}]\chih_{ab}$ and denoting it ${\{\it (good)_2}\}$. 

\NI We are left with estimating $\nabb(\Lie_O^{N-1}\chih)$ and to do this, we have to estimate $\divv\Lie_O^{N-1}\chih$. From equation \ref{15.56} it follows,
\bea 
&&\ML\ML \divv\Lie_O^{N-1}\chih=\Lie_O^{N-1}\divv\chih+[\divv,\Lie_O^{N-1}]\chih=\Lie_O^{N-1}\{\frac{\oom}{2}\Us-\b-\zeta\!\c\hat{\chi}\ \}+[\divv,\Lie_O^{N-1}]\chih\eql{fdef2bis}\nn\\
&&\ \ \ \ \ \ =\Lie_O^{N-1}\frac{\oom}{2}\Us-\Lie_O^{N-1}\b + {\{\it (good)_3}\}\ ,
\eea 
with 
\bea
&&\ML\ML {\{\it (good)_3}\}=-\Lie_O^{N-1}(\zeta\!\c\hat{\chi})+[\divv,\Lie_O^{N-1}]\chih\\
&&\ML\ML =\Lie_O^{N-1}(\zeta\!\c\hat{\chi})+\sum_{q=0}^{N-2}\left(\cbin{N-1}{q}(\Lie_O^{N-2-q}\overline{E})(\Lie_O^{q}\Us)+\cbin{N-1}{q}(\Lie_O^{N-2-q}\overline{\cal H})(\Lie_O^{q}\nabb\Us)\right)\nn
\eea
By proposition 4.1.3 of \cite{Kl-Ni:book} we have that
\bea\eql{15.60}
&&\ML\ML |\nabb\Lie_O^{N-1}\chih|_{p,S}\leq c \left(|\oom|_{\infty,S}|\Lie_O^{N-1}\Us|_{p,S}+|\Lie_O^{N-1}\b|_{p,S}+\sum_{k=1}^{N-1}\cbin{N-1}{k}|\Lie_O^k\oom|_{\infty}|\Lie_O^{N-1-k}\Us|_{p,S}+|{\{\it (good)_3}\}|_{p,S}\right)\nn\\
&&= c \left(|\oom|_{\infty,S}|\Lie_O^{N-1}\Us|_{p,S}+|\Lie_O^{N-1}\b|_{p,S}+|{\{\it (good)_4}\}|_{p,S}\right)
\eea
with $c$ a generic constant and 
\bea
&&\ML\ML|{\{\it (good)_4}\}|_{p,S}=\left(\sum_{k=1}^{N-1}\cbin{N-1}{k}|\Lie_O^k\oom|_{\infty}|\Lie_O^{N-1-k}\chih|_{p,S}+|{\{\it (good)_3}\}|_{p,S}\right)\ \ \ \ \ \ \ \ \ \ \ \ \ \ \ \ \ \  
\eea
Collecting equations \ref{15.55} and \ref{15.60} we obtain
\bea
&&\ML\ML |\Lie_O^{N-1}\nabb\chih|_{p,S}\leq c \left(|\oom|_{\infty,S}|\Lie_O^{N-1}\Us|_{p,S}+|\Lie_O^{N-1}\b|_{p,S}+|{\{\it (good)_2}\}|_{p,S}+|{\{\it (good)_4}\}|_{p,S}\right)\ \ \ \ \ \ \ \ \ \ \ \ \ \ \ \ \ \ 
\eea
and Lemma \ref{estdNchih} is proved.}

{ \subsection{Proof of Theorem \ref{T3.1}}\label{15.4xx}

\NI In order to prove theorem \ref{T3.1} we have to estimate inductively $|\Ls|_{p,s}$, applying the result of lemma \ref{estdNchih} to the expression of $|\Ls|$, see \ref{15.52}

\beaa
&&\ML\ML |\Ls|_{p,s}\leq \!\!\left(\big|\oom\chih\c(\Lie_O^{N-1}\Us)\big|_{p,S}+\big|2\chih|_{\infty,S}|(\Lie_O^{N-1}\nabb\hat{\chi})\big|_{p,S}+|\tr\chi|_{\infty,S}\big|\Lie_O^{N-1}\b\big|_{p,S}
+\big|\big\{(good)_1\big\}\big|_{p,S}\right)\nn\\
&+&\!\!\left(\frac{(3-\frac 2 p)}{2}|{\oom}\tr\chi-\overline{{\oom}\tr\chi}|_{\infty,S}|\Lie_O^{N-1}\Us|_{p,S}\right)\ \ \ \ .\nn\\
&&\ML\ML \leq \!\!\left(\big|\oom\chih\c(\Lie_O^{N-1}\Us)\big|_{p,S}+|\tr\chi|_{\infty,S}\big|\Lie_O^{N-1}\b\big|_{p,S}
+\big|\big\{(good)_1\big\}\big|_{p,S}\right)\nn\\
&&\ML\ML+\big|2\chih|_{\infty,S} \left(|\oom|_{\infty,S}|\Lie_O^{N-1}\Us|_{p,S}+|\Lie_O^{N-1}\b|_{p,S}+|{\{\it (good)_4}\}|_{p,S}\right)+\!\!\left(\frac{(3-\frac 2 p)}{2}|{\oom}\tr\chi-\overline{{\oom}\tr\chi}|_{\infty,S}|\Lie_O^{N-1}\Us|_{p,S}\right)=\ \ \ \ .\nn\\
&&\ML\ML\leq c \Big(\left(\big|\chih|_{\infty,S} |\oom|_{\infty,S}+ \frac{(3-\frac 2 p)}{2}|{\oom}\tr\chi-\overline{{\oom}\tr\chi}|_{\infty,S}\right) \big|(\Lie_O^{N-1}\Us)\big|_{p,S} +\left(2\big|\chih|_{\infty,S}+|\tr\chi|_{\infty,S}\right)|\Lie_O^{N-1}\b|_{p,S}\nn\\
&&\ML\ML+\big|\big\{(good)_1\big\}\big|_{p,S}+\big|\big\{(good)_4\big\}\big|_{p,S}\Big)
\eeaa

\NI which we rewrite, defining,
\bea
&&\ML|{\cal H}|_{\infty, S}\equiv |\oom|_{\infty,S}||\chih|_{\infty,S}+\frac{(3-\frac 2 p)}2|\oom\tr\chi-\overline{\oom\tr\chi}|_{\infty,S}\ \eql{Hdef}
\eea
\bea
&&\ML|r^{3-\frac{2}{p}}\{(good)_{1+4}\}|_{p,S}\equiv |r^{3-\frac{2}{p}}\big\{(good)_{1}\big\}|_{p,S}\!+|r^{3-\frac{2}{p}}\big\{(good)_{4}\big\}|_{p,S}\ ,\nn\eql{good123def}
\eea

\NI We have

\bea
&&\ML|r^{(3-\frac{2}{p})}\Ls|_{p,S}\leq c|{\cal H}|_{\infty,S}(\la,\nu)\big|r^{3-\frac{2}{p}}(\Lie_O^{N-1}\Us)|_{p,S}\nn\\
&&\ML+c(2\big|\chih|_{\infty,S}+\big|\tr\chi\big|_{\infty,S})|r^{3-\frac{2}{p}}(\Lie_O^{N-1}\b)|_{p,S}
+c|r^{3-\frac{2}{p}}\{(good)_{1+4}\}|_{p,S}\ .\nn
\eea

\NI Hence inequality \ref{diffest1aax} can be written as,

\bea
&&\ML\ML -\frac{\partial}{\partial\nu}|r^{(3-\frac{2}{p})}\Lie_O^{N-1}\Us|_{p,S}-c|{\cal H}|_{\infty,S}|r^{(3-\frac{2}{p}}(\Lie_O^{N-1}\Us)|_{p,S}\eql{diffest1b}\\
&&\ML\ML\leq  \!\left[(2\big|\chih\big|_{\infty,S}+\big|\tr\chi\big|_{\infty,S})|r^{3-\frac{2}{p}}(\Lie_O^{N-1}\b)|_{p,S}
+c_0|r^{3-\frac{2}{p}}\big\{(good)_{1+4}\big\}|_{p,S}\right]\ ,\nn
\eea

\NI Equation \ref{diffest1b} is the one we use to prove the estimate for $\Lie_O^{N-1}\Us$.

\smallskip

\NI We are left with the problem of estimating $|(\Lie_O^{N-1}\b)|_{p,S}$. We postpone its estimate and prove the result for $|r^{(N+2-\frac{2}{p})}\Lie_O^{N-1}\Us|_{p,S}(\la,\nu)$ 
using as inductive assumptions for the connection coefficients with $J<N$, the results of Theorem \ref{L2.1new}.\footnote{These inductive assumptions imply the inductive assumption for $\Us$,
$$
|r^{3-\frac{2}{p}}\Lie_O^{J-1}\Us|_{p,S}\leq C_0e^{(J-2)\de}e^{(J-2)\underline{\Ga}(\la)}\frac{J!}{J^\a}\frac{1}{\ro_{0,1}^{J}}\ .
$$}

\NI Integrating inequality \ref{diffest1b} from $\nu$ to $\nu_*$ we obtain, applying the Gronwall lemma: 
\bea
&&\ML\ML\ML|r^{(3-\frac{2}{p})}\Lie_O^{N-1}\Us|_{p,S}(\la,\nu)
\leq \left(\exp{c\!\int_{\nu}^{\nu_*}\big|{\cal H}\big|_{\infty,S}(\nu')d\nu'}\right)\!|r^{(3-\frac{2}{p})}\Lie_O^{N-1}\Us|_{p,S}(\la,\nu_*)\eql{130y}\\
&&\ML\ML\ML+C\bigg(e^{\!c\!\int_{\nu}^{\nu_*}\big|{\cal H}\big|_{\infty,S}(\nu')d\nu'}\bigg)\!\c\!
\left[\int_{\nu}^{\nu_*}d\nu'\bigg(e^{\!-c\!\int_{\nu'}^{\nu_*}\big|{\cal H}\big|_{\infty,S}(\nu'')d\nu''}\bigg)\!\c
 \!\big(2\big|\chih\big|_{\infty,S}+\big|\tr\chi\big|_{\infty,S})\big)|r^{3-\frac{2}{p}}(\Lie_O^{N-1}\b)|_{p,S}\right.\nn\\
&&\ \ \ \ \ \ \ \ \ \ \ \ \ \ \ \ \ \left.\!+\int_{\nu}^{\nu_*}d\nu'\bigg(e^{\!-c\!\int_{\nu'}^{\nu_*}\big|{\cal H}\big|_{\infty,S}(\nu'')d\nu''}\bigg)\!\right.
\left.|r^{3-\frac{2}{p}}\big\{(good)_{1+4}\big\}|_{p,S}\right.\bigg]\nn\\
&&\ML\ML\leq \left(\exp{c\!\int_{\nu}^{\nu_*}\big|{\cal H}\big|_{\infty,S}(\nu')d\nu'}\right)\!|r^{(3-\frac{2}{p})}\Lie_O^{N-1}\Us|_{p,S}(\la,\nu_*)\ \ \ \ \ \ \ \ \ \ \ \eql{130y2}\\
&&\ML\ML\ML+C\!\left(\int_{\nu}^{\nu_*}d\nu' e^{c\left(\int_{\nu}^{\nu_*}\big|{\cal H}\big|_{\infty,S}(\nu')d\nu'-\int_{\nu'}^{\nu_*}\big|{\cal H}\big|_{\infty,S}(\nu'')d\nu''\right)}(2\big|\chih\big|_{\infty,S}+\big|\tr\chi\big|_{\infty,S})
|r^{3-\frac{2}{p}}(\Lie_O^{N-1}\b)|_{p,S}\right.\nn\\
&&\ML\ML\ML\left.+\!\int_{\nu}^{\nu_*}\!d\nu'\! e^{c\left(\int_{\nu}^{\nu_*}\big|{\cal H}\big|_{\infty,S}(\nu')d\nu'-\int_{\nu'}^{\nu_*}\big|{\cal H}\big|_{\infty,S}(\nu'')d\nu''\right)}\left[|r^{3-\frac{2}{p}}\big\{(good)_{1+4}\big\}|_{p,S}\!\right]\!(\la,\nu')\right.\bigg)\c\nn
\eea

\NI We start proving that the recursive assumption is satisfied considering only the $\big\{(good)_{1}\big\}$ term. Let us consider, therefore, the following simplified inequality,
\bea
&&\ML\ML|r^{(3-\frac{2}{p})}\Lie_O^{N-1}\Us|_{p,S}(\la,\nu)
\leq \left(\exp{c\!\int_{\nu}^{\nu_*}\big|{\cal H}\big|_{\infty,S}(\nu')d\nu'}\right)\!|r^{(3-\frac{2}{p})}\Lie_O^{N-1}\Us|_{p,S}(\la,\nu_*)\ \ \ \ \ \nn\\
&&\ML\ML+C\!\left(\int_{\nu}^{\nu_*}d\nu' e^{c\left(\int_{\nu}^{\nu_*}\big|{\cal H}\big|_{\infty,S}(\nu')d\nu'-\int_{\nu'}^{\nu_*}\big|{\cal H}\big|_{\infty,S}(\nu'')d\nu''\right)}|r^{(3-\frac{2}{p})}\big\{(good)_{1}\big\}|_{p,S}\right)\eql{130y2aa}\\
&&\ML\ML \leq e^{\Ga_*(\nu)}|r^{(3-\frac{2}{p})}\Lie_O^{N-1}\Us|_{p,S}(\la,\nu_*)
+C\!\left(\int_{\nu}^{\nu_*}d\nu' e^{(\Ga_*(\nu)-\Ga_*(\nu'))}|r^{3-\frac{2}{p}}\big\{(good)_{1}\big\}|_{p,S}\right)\nn
\eea
\NI where, with ${\tilde C}>0$,
\bea
\Ga_*(\nu)\equiv c\!\int_{\nu}^{\nu_*}|{\cal H}|_{\infty,S}(\nu')d\nu'={\tilde C}\frac{\nu_*-\nu}{\nu_*\nu}\ .
\eea
\NI Let us estimate now  $|r^{3-\frac{2}{p}}\big\{(good)_{1}\big\}|_{p,S}$, subsequently we will use this estimate as paradigmatic  of all the estimates involving a product of two or more derived connection coefficients, with order of derivatives less than $N$ in order to apply the inductive estimates,
 \bea
&&\ML|r^{3-\frac{2}{p}}\big\{(good)_1\!\big\}|_{p,S}\leq2\sum_{k=1}^{N-1}\cbin{N-1}{k}|r^{3-\frac{2}{p}}(\Lie_O^{N-1-k})(\nabb\chih)(\Lie_O^k\chih)|\eql{15.46xx}\\ 
&&\ML\ML-\frac{3}{2}\sum_{q=1}^{N-1}\cbin{N-1}{q}|r^{3-\frac{2}{p}}(\lie_O^{q}\oom^2 U)(\lie_O^{N-1-q}\Us)|+|r^{3-\frac{2}{p}}\sum_{k=0}^{N-1}\cbin{N-1}{k}(\Lie_O^{N-1-k}|\chih|^2)(\Lie_O^k\eta)|\nn\\
&&\ML\ML+\sum_{q=1}^{N-1}\cbin{N-1}{q}|r^{3-\frac{2}{p}}(\Lie_O^{q}(\oom\chih)(\Lie_O^{N-1-q}\Us)|+\sum_{q=0}^{N-2}\cbin{N-1}{q}|r^{3-\frac{2}{p}}(\Lie_O^{N-2-q}E)(\Lie_O^{q}\Us)|\nn \\
&&\ML\ML-\sum_{q=0}^{N-1}\cbin{N-1}{q}|r^{3-\frac{2}{p}}(\Lie_O^{q}(\oom U)(\Lie_O^{N-1-q}\chih\eta)|+\sum_{q=1}^{N-1}\cbin{N-1}{q}|r^{3-\frac{2}{p}}(\Lie_O^{N-2-q}\b)(\Lie_O^{q}\tr\chi)|\nn
\eea

\NI To apply the inductive assumptions to this expression we need a recursive assumption also for $\oom$. We assume, see the estimate \ref{oomestxx},\footnote{For $J> 0$ the decay of $\Lie_O^J\oom$ is better.}
\smallskip

\bea
|\Lie_O^J\oom|_{p,S}\leq C_4e^{((J-1)-2)\de}e^{(J-1)\Ga}\frac{(J-1)!}{(J-1)^\a}\frac{1}{\ro_{0,1}^{(J-1)}}\ .
\eea
Let us start looking at the first sum of \ref{15.46xx}, if $N-1$ is even,
\bea\label{1572}
&&\ML\sum_{k=1}^{N-1}\cbin{N-1}{k}|r^{3-\frac{2}{p}}(\Lie_O^{N-1-k})(\nabb\chih)(\Lie_O^k\chih)|\leq\\
&&\ML\frac{ 1}{ r^2}\!\!\left[\sum_{k=1}^{\frac{N-1}{2}}\left(\!\!\!\!\begin{array}{c}N\\k\\
\end{array}\!\!\!\!\right)|r^{3}\Lie_O^{k}\nabb\chih|_\infty|r^{2-\frac{2}{p}}\Lie_O^{N-1-k}\chih|_{p,S}
+\!\!\sum_{k>\frac{N-1}{2}}^{N-1}\left(\!\!\!\!\begin{array}{c}N\\k\\
\end{array}\!\!\!\!\right)|r^{3-\frac{2}{p}}\Lie_O^{k}\nabb\chih|_{p,S}|r^{2}\Lie_O^{N-1-k}\chih|_\infty\right]\nn
\eea
In order to estimate inductively the right hand sides of these equations  we recall the inductive assumptions on the right hand side terms
\bea
&&\ML|r^{3}\Lie_O^{k}\nabb\chih|_{\infty,S}\leq C e^{k\de}e^{k\underline{\Ga}(\la)}\frac{(k+2)!}{(k+2)^\a}\frac{1}{\ro_{0,1}^{k+2}}\nn\\
&&\ML|r^{2-\frac{2}{p}}\Lie_O^{N-1-k}\chih|_{p,S}\leq C e^{((N-3-k)\de}e^{(N-3-k)\underline{\Ga}(\la)}\frac{(N-1-k)!}{(N-1-k)^\a}\frac{1}{\ro_{0,1}^{(N-1-k)}}\nn\\
&&\ML|r^{3-\frac{2}{p}}\Lie_O^{k}\nabb\chih|_{p,S}\leq C e^{(k-1)\de}e^{(k-1)\underline{\Ga}(\la)}\frac{(k+1)!}{(k+1)^\a}\frac{1}{\ro_{0,1}^{k+1}}\\
&&\ML| r^{2}\Lie_O^{N-1-k}\chih|_{\infty,S}\leq  C e^{(N-1-k)\de}e^{(N-1-k)\underline{\Ga}(\la)}\frac{(N+1-k)!}{(N+1-k)^\a}\frac{1}{\ro_{0,1}^{N+1-k}}\ .\nn
 \eea
\NI Therefore,
\bea
&&\ML\ML\ML\frac{1}{r^2}\sum_{k=1}^{\left[\frac{N}{2}\right]}\left(\!\!\!\!\begin{array}{c}N\\k\\
\end{array}\!\!\!\!\right)|r^{3}\Lie^k_O\nabb\chih|_\infty|r^{2-\frac{2}{p}}\Lie_O^{N-1-k}\chih|_{p,S}\nn\\
&&\ML\ML\ML\leq \frac{C^2}{r^2}\sum_{k=1}^{\left[\frac{N}{2}\right]}\left(\!\!\!\!\begin{array}{c}N\\k\\\end{array}\!\!\!\!\right)
e^{k\de}e^{k\underline{\Ga}(\la)}\frac{(k+2)!}{(k+2)^\a}\frac{1}{\ro_{0,1}^{k+2}}e^{(N-1-k)\de}e^{(N-1-k)\underline{\Ga}(\la)}\frac{(N+1-k)!}{(N+1-k)^\a}\frac{1}{\ro_{0,1}^{N+1-k}}\nn\\
&&\ML\ML\ML\leq  \left(C\frac{N!}{N^\a}\frac{1}{\ro_{0,1}^{N}}e^{(N-2)\de}e^{(N-2)\underline{\Ga}(\la)}\right)\left[\sum_{k=1}^{\left[\frac{N}{2}\right]}C N^\a \frac{(k+2)!}{k!}\frac{e^{-\de}e^{-\underline{\Ga}(\la)}}{(k+2)^\a(N+1-k)^\a}\frac{1}{\ro^2_{0,1}}\right]\nn\\
&&\ML\ML\ML\leq \frac {1}{ r^2} \left(C\frac{N!}{N^\a}\frac{1}{\ro_{0,1}^{N}}e^{(N-2)\de}e^{(N-2)\underline{\Ga}(\la)}\right)\left[\frac{N^{\a}}{\left(\frac{N}{2}\right)^{\a}}\sum_{k=1}^{\left[\frac{N}{2}\right]}C  \frac{(k+2)!}{k!}\frac{e^{-\de}e^{-\underline{\Ga}(\la)}}{(k+2)^\a}\frac{1}{\ro^2_{0,1}}\right]\nn\\
&&\ML\ML\ML\leq \frac {1}{ r^2}  \left(C\frac{N!}{N^\a}\frac{1}{\ro_{0,1}^{N}}e^{(N-2)\de}e^{(N-2)\underline{\Ga}(\la)}\right)\left[\frac{C 2^{\a}e^{-\de}e^{-\underline{\Ga}(\la)}}{\ro^2_{0,1}}\sum_{k=1}^{\left[\frac{N}{2}\right]}\frac{1}{(k+2)^{\a-2}}\right]\nn\\
&&\ML\ML\ML \leq \frac {1}{ r^2} \left(CCe^{-\de}\frac{N!}{N^\a}\frac{1}{\ro_{0,1}^{N}}e^{(N-2)\de}e^{(N-2)\underline{\Ga}(\la)}\right)
\leq\frac {1}{ r^2} \left(C e^{-\de}\frac{N!}{N^\a}\frac{1}{\ro_{0,1}^{N}}e^{(N-2)\de}e^{(N-2)\underline{\Ga}(\la)}\right)\ \ \ \ \ \ \ \ \eea
with
\bea
\left[\frac{C 2^{\a}e^{-\de}e^{-\underline{\Ga}(\la)}}{\ro^2_{0,1}}\sum_{k=1}^{\left[\frac{N}{2}\right]}\frac{1}{(k+2)^{\a-2}}\right]\leq Ce^{-\de}\nn
\eea
and $\a>3$. Clearly the term 
\bea
\sum_{k>\frac{N-1}{2}}^{N-1}\left(\!\!\!\!\begin{array}{c}N\\k\\
\end{array}\!\!\!\!\right)|r^{3-\frac{2}{p}}\Lie_O^{k}\nabb\chih|_{p,S}\big|r^{2}\Lie_O^{N-1-k}\chih\big|_{\infty,S}\nn
\eea
\NI Can be estimated in the same way. Therefore, considering only this contribution from  $\big\{(good)_{1}\big\}$, it follows, recalling the last slice estimates of the connection coefficients, see subsection \ref{lasts},
\bea\label{1578}
&&\ML\big|r^{3-\frac{2}{p}}\Lie_O^{N-1}\Us\big|_{p,S}(\la,\nu)\leq e^{\Ga_*(\nu)}|r^{(3-\frac{2}{p})}\Lie_O^{N-1}\Us|_{p,S}(\la,\nu_*)\nn\\
&&\ML\ \ \ \ \ \ \ \ \ \ \ \ \ \ \ \ \ \ \ \ \ \ \ \ \ \ +C\!\left(\int_{\nu}^{\nu_*}d\nu' e^{(\Ga_*(\nu)-\Ga_*(\nu'))}|r^{3-\frac{2}{p}}\big\{(good)_{1}\big\}|_{p,S}\right)\nn\\
&&\ML\ML\ML\leq C_{0}^{(0)} e^{\Ga_*(\nu)}\!\left(e^{(N-2)\de}e^{(N-2){\underline{\Ga}}(\la)}\frac{N!}{N^\a}\frac{1}{\ro_{0,1}^{N}}\right)\eql{130y2a}\\
&&\ML\ML\ML+\left(C\frac{N!}{N^\a}\frac{1}{\ro_{0,1}^{N}}e^{(N-2)\de}e^{(N-2)\underline{\Ga}(\la)}\right)\!\left( Ce^{-\de} \int_{\nu}^{\nu_*}d\nu' \frac{e^{(\Ga_*(\nu)-\Ga_*(\nu')}}{r^2}\right)\nn\\
&&\ML\ML\ML\leq C_{0}^{(0)} e^{\Ga_*(\nu)}\!\left(e^{(N-2)\de}e^{(N-2){\underline{\Ga}}(\la)}\frac{N!}{N^\a}\frac{1}{\ro_{0,1}^{N}}\right)\\
&&\ML\ML\ML+\left(C\frac{N!}{N^\a}\frac{1}{\ro_{0,1}^{N}}e^{(N-2)\de}e^{(N-2)\underline{\Ga}(\la)}\right)\!\left( C e^{-\de}\int_{\nu}^{\nu_*}\frac{d\nu'}{{\nu'}^2} e^{(\Ga_*(\nu)-\Ga_*(\nu'))}\right)\nn\\
&&\ML\ML\ML\leq\left(C\frac{N!}{N^\a}\frac{1}{\ro_{0,1}^{N}}e^{(N-2)\de}e^{(N-2)\underline{\Ga}(\la)}\right)\!\left(\frac{C_{0}^{(0)}}{C} e^{(\Ga_*(\nu))}\right.\nn\\
&&\ML\ML\ML\left.\ \ \ \ \ \ \ \ \ \ \ \ \ \ \ \ \ \ \ \ \ \ \ \ \ \ \ \ \ \ \ \ \ \ \ \ \ \ \ \ \ \ \ \ \ \ \ \ \ + Ce^{-\de}\int_{\nu}^{\nu_*}\frac{d\nu'}{{\nu'}^2} e^{(\Ga_*(\nu)-\Ga_*(\nu'))}\right)\nn\\
&&\ML\ML\ML\leq\left(C\frac{N!}{N^\a}\frac{1}{\ro_{0,1}^{N}}e^{(N-2)\de}e^{(N-2)\underline{\Ga}(\la)}\right)\eql{130z2a},
\eea
 
 \NI choosing suitable $C>C_0$ and $\de$ sufficiently large.
 
 \NI The other terms in $\{good\}_1$ , and  $ \{good\}_4$ can be estimated in the same way, for the sake of completeness we only consider another term, namely

\bea\label{1579}
\sum_{k=1}^{N-2}\left(\!\!\!\begin{array}{c}N-1\\k\\\end{array}\!\!\!\right)\big|r^{3-\frac{2}{p}}(\Lie_O^{k-1}E)\c(\Lie_O^{N-1-k}\Us)\big|_{p,S}
\eea
recalling \ref{E} we write, a bit symbolically,
\bea
\Lie_O^{k-1}E\ ``=" \  \Lie_O^{k-1}\chih\etab+\Lie_O^{k-1}\tr\chi\etab+ \Lie_O^{k-1}\chibh\eta+\Lie_O^{k-1}\tr\chib\eta+\nabb^{k-1}\Psi+\Lie_O^{k-1}{\cal H}\nn\\
\eea
where with $\Psi$ we denote a generic null Riemann component. We consider the contribution of the first term, the ones associated to the other term will be of the same kind, but easier and the one associated to the last term will be examinated later on. We have
\bea
&&\ML\ML\sum_{k=1}^{N-2}\cbin{N-1}{k}\big|r^{3-\frac{2}{p}}(\Lie_O^{k-1}\chih\etab)\c(\Lie_O^{N-1-k}\Us)\big|_{p,S}\eql{2.76a}\\
&&\ML\ML=\frac{1}{r^4}\sum_{k=1}^{\left[\frac{N}{2}\right]}\cbin{N-1}{k}\big|r^2(\Lie_O^{k-1}\chih\etab)\big|_{\infty}\big|r^{3-\frac{2}{p}}(\Lie_O^{N-1-k}\Us)\big|_{p,S}\nn\\
&&\ML\ML+\frac{1}{r^4}\!\!\!\sum_{k=\left[\frac{N}{2}\right]+1}^{N-2}\cbin{N-1}{k}
\big|r^{2}(\Lie_O^{N-1-k}\Us)\big|_{\infty}\big|r^{3-\frac{2}{p}}(\Lie_O^{k-1}\chih\etab)\big|_{p,S}\nn\\
&&\ML\ML\ML\leq\frac{1}{r^4}\sum_{k=1}^{\left[\frac{N}{2}\right]}\sum_{q=0}^{k-1}\cbin{N-1}{k}\cbin{k-1}{q}\big|r^{3}\Lie_O^{q}\chih\big|_{\infty}\big|r^{2}\Lie_O^{k-1-q}\etab\big|_{\infty}\big|r^{3-\frac{2}{p}}\Lie_O^{N-1-k}\Us\big|_{p,S}\nn\\
&&\ML\ML\ML+\frac{1}{r^4}\sum_{k\left[\frac{N}{2}\right]+1}^{N-2}\sum_{q=0}^{k-1}\cbin{N-1}{k}\cbin{k-1}{q}\big|r^{2}(\Lie_O^{N-1-k}\Us)\big|_{\infty}\big|r^2\nabb^{q}\chih\big|_{\infty}\big|r^{3-\frac{2}{p}}\Lie_O^{k-1-q}\etab\big|_{p,S}\nn
\eea
Let us estimate the first term of the last sum in \ref{2.76a},
\bea
&&\ML\ML\ML\frac{1}{r^4}\sum_{k=1}^{\left[\frac{N}{2}\right]}\sum_{q=0}^{k-1}\cbin{N-1}{k}\cbin{k-1}{q}\big|r^{2}\Lie_O^{q}\chih\big|_{\infty}\big|r^{2}\Lie_O^{k-1-q}\etab\big|_{\infty}\big|r^{(3-\frac{2}{p}}\Lie_O^{N-1-k}\Us\big|_{p,S}\nn\\
\eea

\NI Using the inductive assumptions:

\bea
&&\ML\ML\ML\leq\frac{C^3N!}{r^4}\sum_{k=1}^{\left[\frac{N}{2}\right]}\frac{(k-1)!}{(N-k-1)!(k+1)!}\sum_{q=0}^{k-1}\frac{1}{(k-1-q)!q!}
\left[\frac{(q+1)!}{(q+1)^\a}\frac{e^{(q-1)\de}e^{(q-1)\underline{\Ga}(\la)}}{\ro_{0,1}^{q+1}}\right]\c\nn\\
&&\ML\ML\ML\c\left[\frac{(k-q)!}{(k-q)^\a}\frac{e^{(k-q-2)\de}e^{(k-q-2)\underline{\Ga}(\la)}}{\ro_{0,1}^{(k-q)}}\right]
\frac{(N-k)!}{(N-k)^\a}\frac{e^{(N-k-2)\de}e^{(N-k-2)\underline{\Ga}(\la)}}{\ro_{0,1}^{N-k}}\nn\\
&&\ML\ML\ML\leq\frac{C^3}{r^4}\frac{N!}{N^\a}\frac{e^{(N-5)\de}e^{(N-5)\underline{\Ga}(\la)}}{\ro_{0,1}^{N}}\sum_{k=1}^{\left[\frac{N}{2}\right]}\frac{(N-k)}{Nk(k+1)}
\sum_{q=0}^{k-1}
\left[\frac{1}{\ro_{0,1}}\frac{(q+1)}{(q+1)^\a}\right]\c\nn\\
&&\ML\ML\ML\left[\frac{(k-q)}{(k-q)^\a}\right]\frac{N^\a}{\left[\frac{N}{2}\right]^\a}\nn\\
&&\ML\ML\ML\leq\frac{1}{r^4}\left(\frac{N!}{N^\a}\frac{e^{(N-2)\de}e^{(N-2)\underline{\Ga}(\la)}}{\ro_{0,1}^{N}}\right) e^{-3\de}e^{-3\underline{\Ga}(\la)}2^{\a}\sum_{k=1}^{\left[\frac{N}{2}\right]}\frac{1}{k(k+1)}
\sum_{q=0}^{k-1}
\left[\frac{1}{\ro_{0,1}}\frac{(q+1)}{(q+1)^\a}\right]\c\nn\\
&&\ML\ML\ML\c\left[\frac{(k-q)}{(k-q)^\a}\right]\nn\\
&&\ML\ML\ML\leq\frac{C^3}{r^4}\left(\frac{N!}{N^\a}\frac{e^{(N-2)\de}e^{(N-2)\underline{\Ga}(\la)}}{\ro_{0,1}^{N}}\right)\left( e^{-3\de}e^{-3\underline{\Ga}(\la)}2^{\a}C\right)\eql{2.79}
\eea
where
\bea
\sum_{k=1}^{\left[\frac{N}{2}\right]}\frac{1}{k(k+1)}
\sum_{q=0}^{k-1}
\left[\frac{1}{\ro_{0,1}}\frac{(q+1)}{(q+1)^\a}\right]\c\left[\frac{(k-q)}{(k-q)^\a}\right]\leq C\nn\\
\eea
\NI Choosing $\de$ sufficiently large we obtain

\bea\label{1584}
\sum_{k=1}^{N-2}\left(\!\!\!\begin{array}{c}N-1\\k\\\end{array}\!\!\!\right)\big|r^{3-\frac{2}{p}}(\Lie_O^{k-1}E)\c(\Lie_O^{N-1-k}\Us)\big|_{p,S}
\leq\frac{1}{r^4}\left(\frac{N!}{N^\a}\frac{e^{(N-2)\de}e^{(N-2)\underline{\Ga}(\la)}}{\ro_{0,1}^{N}}\right)\nn\\
\eea
From the estimate \ref{1584} it follows that this contribution to the integral of \ref{1579} is similar to the previous terms.
\medskip

We are left with the problem of estimating 
\NI At last let us control the following term of the integral in the inequality \ref{130y2}, 
\bea\label{15888}
\int_{\nu}^{\nu_*}d\nu'e^{(\Ga_*(\nu)-\Ga_*(\nu'))}(\big|2\chih\big|_{\infty,S}+\big|\tr\chi\big|_{\infty,S})|r^{(3-\frac{2}{p})}\Lie_O^{N-1}\b|_{p,S}\ .\ \ \ \ \ \ \eql{estax}
\eea
To estimate it we need to control $|r^{3-\frac{2}{p}}\Lie_O^{N-1}\b|_{p,S}$, this is the content of Theorem \ref{hypthm2}.
From it the following estimate holds for the $\Lie_O^{N-1}$ angular derivatives of the Riemann tensor null component $\b$: 
\bea
\big| r^{\frac{7}{2}+(N-1)-\frac{2}{p}}\Lie_O^{N-1}\b\big|_{p,S}\leq C^{(1)}e^{(N-2)\de}e^{(N-2)\underline{\Ga}(\la)}\frac{N!}{N^\a}\frac{1}{\ro_{0,1}^N}\  \eql{betaestx} 
\eea
and the integral \ref{estax} satisfies:
\bea
&&\ML\ML\int_{\nu}^{\nu_*}d\nu'e^{(\Ga_*(\nu)-\Ga_*(\nu'))}
(\big|2\chih\big|_{\infty,S}+\big|\tr\chi\big|_{\infty,S})|r^{(3-\frac{2}{p})}\Lie_O^{N-1}\b|_{p,S}\nn\\
&&\ML\ML\leq \ep\left(\!\int_{\nu}^{\nu_*}\frac{d\nu'}{\nu' r^{\frac 1 2}}e^{(\Ga_*(\nu)-\Ga_*(\nu'))}
|r^{(\frac{7}{2}-\frac{2}{p})}\Lie_O^{N-1}\b|_{p,S}(\nu')\right.\nn\\
&&\ML\ML\left.\ \ \ \ +\!\int_{\nu}^{\nu_*}\frac{d\nu'}{\nu'r^{\frac 3 2}}e^{(\Ga_*(\nu)-\Ga_*(\nu'))}
|r^{\frac{7}{2}-\frac{2}{p}}\Lie_O^{N-1}\b|_{p,S}(\nu')\right)\nn\\
&&\ML\ML\leq c C^{(1)}e^{(N-2)\underline{\Ga}(\la)}e^{(N-2)\de}\frac{N!}{N^\a}\frac{1}{\ro_{0,1}^N}\left(\int_{\nu}^{\nu_*}\frac{d\nu'}{\nu'r^\frac{1}{2}}e^{(\Ga_*(\nu)-\Ga_*(\nu'))}\right.\nn\\
&&\ML\ML\left.\ \ \ \ \ \ \ \ \ \ \ \ \ \ \ \ \ \ \ \ \ \ \ \ \ \ \ \ \ \ \ \ \ \ \ \ +\!\int_{\nu}^{\nu_*}\frac{d\nu'}{r^\frac{1}{2}\nu'^2}e^{(\Ga_*(\nu)-\Ga_*(\nu'))}\right)\nn\\
&&\ML\ML
\leq C_0e^{(N-2)\de}e^{(N-2)\underline{\Ga}(\la)}\frac{N!}{N^\a}\frac{1}{\ro_{0,1}^N}\ ,\nn
\eea

\medskip

\bigskip
 
{\bf Remark:}

\NI {\em Observe that these estimates are consistent due to the fact that the null Riemann components can be bounded by a constant $C^{(1)}$ smaller than the constants bounding the connection coefficients.} 
\smallskip

\NI Looking at the remaining contributions to the integral in the right hand side of \ref{130y2} it is easy to realise that they all have the same structure and can be estimated in the same way. Therefore, collecting all these estimates together it follows that Theorem  \ref{T3.1} is proved choosing $\de$ sufficiently large.}

\medskip

 \subsection{Proof of Lemma \ref{eqformu}}
Applying $\Lie_O^{N-1}$ to the transport equation \ref{mutrasp},\footnote{We apply $\nabb^{N-1}$ instead of $\nabb^N$ as $\tilde \mu$ depends on $\divv\eta$ and, therefore, the control of $\Lie_O^{N-1}{\tilde\mu}$ allows to control $\Lie_O^N\eta$. } and using the result of Lemma \ref{L1.3}, we obtain an equation with the same structure as the transport equation \ref{eqforU} for $\Lie_O^{N-1}\Us$, 
\bea
&&\frac{\Dbb{(\Lie_O^{N-1}{\tilde\mu})}}{\partial\nu}+\oom\tr\chi\c(\Lie_O^{N-1}\tilde{\mu})+\Lie_O^{N-1}\!\!\left[(\Omega \tilde{F}-\overline{\Omega\tilde{F}})+(\Omega \tilde{H}-\overline{\Omega\tilde{H}})\right]
=\big\{(\widetilde{good})_1\big\}\ \ \ \ \ \ \ \ \ \ \ \ \ \eql{110La}
\eea

\NI With, see equation \ref{[]}

\bea
\big\{(\widetilde{good})_1\big\}=[ \Dbb_{\nu},\Lie_O^{N-1}]\tilde{\mu}=\sum_{q=0}^{N-2}\cbin{N-1}{q}(\Lie_O^{N-2-q} E)( \Lie_O^{q}\tilde{\mu})
\eea
Rewriting $\Lie_O^{N-1}\!\!\left[(\Omega \tilde{F}-\overline{\Omega\tilde{F}})+(\Omega \tilde{H}-\overline{\Omega\tilde{H}})\right]$, extracting the higher derivative terms,
\bea\label{1588}
\ML\Lie_O^{N-1}\!\!\left[(\Omega \tilde{F}-\overline{\Omega\tilde{F}})+(\Omega \tilde{H}-\overline{\Omega\tilde{H}})\right]\!=\oom\chih\c\Lie_O^{N-1}(\nabb\hat{\otimes}\eta)-2\oom\eta\c\Lie_O^{N-1}\!\b+\oom\tr\chi\Lie_O^{N-1}\!\ro+\big\{(\widetilde{good})_2\big\} ,\ \ \ \ \ \ \eql{FHder}
\eea
we obtain the transport equation \ref{mutransp}.
\subsection{Proof of Lemma \ref{nabbNetaest}}
\NI We start from the following Hodge system
\bea
&&\divv\eta =-\tilde{\mu}+\frac{1}{2}(\chi\underline{\chi}-\overline{\chi\underline{\chi}})-(\ro-\overline{\ro})\equiv F^{(0)}\nn\\
&&\curll\eta=\sigma-\frac{1}{2}\hat{\underline{\chi}}\wedge\hat{\chi}\equiv G^{(0)}\ .\nn
\eea
 
 \NI Following the steps of lemma \ref {estdNchih} we can write,
 
\bea
&&\divv\Lie_O^{N-1}\eta =[\divv,\Lie_O^{N-1}]\eta+\Lie_O^{N-1}\big(-\tilde{\mu}+\frac{1}{2}(\chi\underline{\chi}-\overline{\chi\underline{\chi}})-(\ro-\overline{\ro})\big)\nn\\
&&=-\Lie_O^{N-1}\tilde{\mu}-\Lie_O^{N-1}\ro +{\{\it (\widetilde{good})_{3one}}\}
\eea

\NI With 

\bea
&&{\{\it (\widetilde{good})_{3one}}\}=-\Lie_O^{N-1}\frac{1}{2}(\chi\underline{\chi}-\overline{\chi\underline{\chi}})+\Lie_O^{N-2}\ro+[\divv,\Lie_O^{N-1}]\eta\nn\\
&&=-\Lie_O^{N-1}(\frac{1}{2}(\chi\underline{\chi}-\overline{\chi\underline{\chi}})+\Lie_O^{N-2}\overline{\ro}+\sum_{q=0}^{N-2}(\cbin{N-1}{q}(\Lie_O^{N-2-q}\overline{E})(\Lie_O^{q}\eta)
\nn\\
&&+\cbin{N-1}{q}(\Lie_O^{N-2-q}\overline{\cal H})(\Lie_O^{q}\nabb\eta))\nn
\eea
In the same way we can write
\bea
&&\curll\Lie_O^{N-1}\eta =[\curll,\Lie_O^{N-1}]\eta+\Lie_O^{N-1}\big(\sigma-\frac{1}{2}\hat{\underline{\chi}}\wedge\hat{\chi}\big)\eql{etahodgeaa}\nn\\
&&=-\Lie_O^{N-1}\si +{\{\it (\widetilde{good})_{3bis}}\}
\eea

\NI With 

\bea
&&{\{\it (\widetilde{good})_{3bis}}\}=\Lie_O^{N-1}\big(\sigma-\frac{1}{2}\hat{\underline{\chi}}\wedge\hat{\chi}\big)+[\curll,\Lie_O^{N-1}]\eta\nn\\
\eea

\NI Now by proposition 4.1.3 of \cite{Kl-Ni:book} we have that, defining 
\bea
{\{\it (\widetilde{good})_{3}}\}={\{\it (\widetilde{good})_{3one}}\}+{\{\it (\widetilde{good})_{3bis}}\}
\eea
\bea\eql{15.60x}
&&\ML\ML |\nabb\Lie_O^{N-1}{\eta}|_{p,S}\leq c \left(|\Lie_O^{N-1}\tilde{\mu}|_{p,S}+|\Lie_O^{N-1}\ro|_{p,S} +|\Lie_O^{N-1}\si|_{p,S}+|{\{\it (\widetilde{good})_3}\}|_{p,S}\right)\nn\\
\eea

\NI With $c$ a suitable constant. Consequently, following the same strategy of lemma \ref{estdNchih}  the same inequality holds for $|\Lie_O^{N-1}\nabb\eta|_{p,S}$ and consequently for $|\Lie_O^{N-1}\nabb\hat{\otimes}\eta|_{p,S}$, hence we obtain 

\bea
|\Lie_O\tilde{\mu}|_{p,S}\leq c \left(|\Lie_O^{N-1}\tilde{\mu}|_{p,S}+|\Lie_O^{N-1}\ro|_{p,S} +|\Lie_O^{N-1}\si|_{p,S}+|{\{\it (\widetilde{good})_3}\}|_{p,S}\right)\nn\\
\eea

\NI Substituting this last inequality in equation \ref{15888} 

\bea
&&|\frac{\partial{(r^{2-\frac{2}{p}}\Lie_O^{N-1}\tilde{\mu})}}{\partial\nu}|_{p,S}\leq C\left(|\big\{(\widetilde{good})_1\big\}|_{p,S}+|\big\{(\widetilde{good})_2\big\}|_{p,S}\right. \nn\\
&&\left.+|\oom\chih\big\{(\widetilde{good})_3\big\}|_{p,S}+|\oom\eta\Lie_O^{N-1}\beta|_{p,S}+|\Lie_O^{N-1}\ro|_{p,S}\right)
\eea

\medskip

\NI Now performing the same steps of theorem \ref{T3.1} we obtain inequality  \ref{nabbNetaest} and proof of
 lemma \ref{nabbNetaest} is achieved.
 

\subsection{Proof of Lemma \ref{L3.6}}
\NI We prove first the following
\begin{Le}\label{VtildetraspLe}
The structure equations, see $(I)$, imply that ${\tilde V}$ satisfies the following transport equation, 
\bea
\dd_4\tilde{V}+\frac{\tr\chi}{2}{\tilde V}+\chih\c{\tilde V}+4\om{\tilde V}\!&=&\!\oom(\nabb\ro)-\oom(\nabb\log\oom)\ro+\nabb[\oom(\etab\cdot\eta-2\zeta^2-2\zeta\cdot\nabb\log\oom)]\nn\\
\!&+&\!\big[4\oom(\eta+\etab)\om\omb\big] \ .\eql{Vtildetrasp}
\eea
\end{Le}
\NI{\bf Proof of Lemma \ref{VtildetraspLe}:}
We start computing $\dd_4\tilde{V}$:
\bea
&&\ML\dd_4\tilde{V}=\dd_4\nabb(\oom\dd_3\log\oom)=\dd_4\left[ (\nabb\Omega)\dd_3\log\Omega)+\Omega(\nabb\dd_3\log\Omega)\right]\nn\\
&&\ML=(\dd_4\nabb\Omega)(\dd_3\log\Omega)+(\nabb\Omega)(\dd_4\dd_3\log\Omega)+(\dd_4\Omega)(\nabb\dd_3\log\Omega)+\Omega\dd_4(\nabb\dd_3\log\Omega)\nn\\
&&\ML=(\dd_4\nabb\Omega)(\dd_3\log\Omega)+(\nabb\Omega)(\dd_4\dd_3\log\Omega)+(\dd_4\Omega)(\nabb\dd_3\log\Omega)
+\Omega\nabb\dd_4\dd_3\log\Omega\nn\\
&&\ML\ \ \ +\Omega[\dd_4,\nabb]\dd_3\log\Omega\\
&&\ML=(\dd_4\nabb\Omega)(\dd_3\log\Omega)+(\nabb\Omega)(\dd_4\dd_3\log\Omega)+(\dd_4\Omega)(\nabb\dd_3\log\Omega)
+\Omega\nabb\dd_4\dd_3\log\Omega\nn\\
&&\ML\ \ \ +\oom(\nabb\log\oom)\dd_4\dd_3\log\Omega-\oom\chi\c\nabb\dd_3\log\oom\nn\\
&&\ML=(\dd_4\nabb\Omega)(\dd_3\log\Omega)+2(\nabb\Omega)(\dd_4\dd_3\log\Omega)+(\dd_4\Omega)(\nabb\dd_3\log\Omega)
+\Omega\nabb\dd_4\dd_3\log\Omega\nn\\
&&\ML\ \ \ -\oom\chi\c\nabb\dd_3\log\oom\nn\\
&&\ML=(\dd_4\nabb\Omega)(\dd_3\log\Omega)+2(\nabb\Omega)(\dd_4\dd_3\log\Omega)+(\dd_4\Omega)(\nabb\dd_3\log\Omega)
+\Omega\nabb\dd_4\dd_3\log\Omega\nn\\
&&\ML\ \ \ -\chi\c\nabb(\oom\dd_3\log\oom)+\chi\c(\nabb\oom)\dd_3\log\oom\nn\\
&&\ML=(\dd_4\nabb\Omega)(\dd_3\log\Omega)+2(\nabb\Omega)(\dd_4\dd_3\log\Omega)+(\dd_4\Omega)(\nabb\dd_3\log\Omega)
+\Omega\nabb\dd_4\dd_3\log\Omega\nn\\
&&\ML\ \ \ -\chi\c{\tilde V}+\oom\chi\c(\nabb\log\oom)\dd_3\log\oom\ .\nn
\eea
Summarizing
\bea
&&\ML\dd_4\tilde{V}=(\dd_4\nabb\Omega)(\dd_3\log\Omega)+2(\nabb\Omega)(\dd_4\dd_3\log\Omega)+(\dd_4\Omega)(\nabb\dd_3\log\Omega)
+\Omega\nabb\dd_4\dd_3\log\Omega\nn\\
&&\ML\ \ \ -\chi\c{\tilde V}+\oom\chi\c(\nabb\log\oom)\dd_3\log\oom\nn\\
&&\ML=\left(\nabb\dd_4\Omega+[\dd_4,\nabb]\Omega\right)(\dd_3\log\Omega)+2(\nabb\Omega)(\dd_4\dd_3\log\Omega)+(\dd_4\Omega)(\nabb\dd_3\log\Omega)\nn\\
&&\ML+\oom\nabb(\dd_4\dd_3\log\oom)-\chi\c\left({\tilde V}-\oom(\nabb\log\oom)\dd_3\log\oom\right)\nn\\
&&\ML\ML= -\chi\c{\tilde V}+\oom\nabb(\dd_4\dd_3\log\oom)+(\nabb\dd_4\Omega)(\dd_3\log\Omega)
+\oom(\dd_4\log\Omega)(\nabb\dd_3\log\Omega)+2(\nabb\oom)(\dd_4\dd_3\log\oom)\nn\\
&&\ML\ML\ \ \ +\bigg[\left((\nabb\log\oom)\dd_4\oom-\chi\c\nabb\oom\right)(\dd_3\log\Omega)+\oom\chi\c(\nabb\log\oom)(\dd_3\log\oom)\bigg]\nn\\
&&\ML\ML=-\chi\c{\tilde V}+\oom\nabb(\dd_4\dd_3\log\oom)+(\dd_3\log\oom)(\nabb\oom\dd_4\log\oom)+(\dd_4\log\Omega)(\nabb\oom\dd_3\log\Omega)
\nn\\
&&\ML\ML\ \ \ +2\oom(\nabb\log\oom)(\dd_4\dd_3\log\oom) +\bigg[-\oom(\dd_4\log\oom)(\nabb\log\oom)(\dd_3\log\oom)+\left(\oom(\nabb\log\oom)(\dd_4\log\oom)\right.\nn\\
&&\ML\ML\ \ \ \ \ \ -\oom\chi\c(\nabb\log\oom)(\dd_3\log\Omega)+\oom\chi\c(\nabb\log\oom)(\dd_3\log\oom)\bigg]\eql{D4Vtilde}
\eea
We reexpress now the term $\oom\nabb(\dd_4\dd_3\log\oom)$ in the following way: recall that these equations hold
\bea
&&\dd_3\dd_4\log\oom+\dd_4\dd_3\log\oom=-2(\dd_3\log\oom)(\dd_4\log\oom)+2(\etab\cdot\eta-2\zeta^2-\ro)\nn\\
&&\dd_4\dd_3\log\oom-\dd_3\dd_4\log\oom=-4\zeta\cdot\nabb\log\oom
\eea
which imply
\bea
\dd_4\dd_3\log\oom=-(\dd_3\log\oom)(\dd_4\log\oom)+(\etab\cdot\eta-2\zeta^2-2\zeta\cdot\nabb\log\oom)-\ro\ .\ \ \ \ 
\eea
Therefore
\bea
&&\ML\ML\oom\nabb(\dd_4\dd_3\log\oom)=-\oom\nabb\!\left(\dd_3\log\oom\dd_4\log\oom\right)+\oom\nabb(\etab\cdot\eta-2\zeta^2-2\zeta\cdot\nabb\log\oom)
-\oom\nabb\ro\nn\\
&&\ML\ML=-\oom(\nabb\dd_3\log\oom)(\dd_4\log\oom)-(\dd_3\log\oom)\oom(\nabb\dd_4\log\oom)-\oom\nabb\ro+\oom\nabb(\etab\cdot\eta-2\zeta^2-2\zeta\cdot\nabb\log\oom)\nn\\
&&\ML\ML=-(\nabb\oom\dd_3\log\oom)(\dd_4\log\oom)-(\dd_3\log\oom)(\nabb\oom\dd_4\log\oom)+2\oom(\nabb\log\oom)(\dd_3\log\oom)(\dd_4\log\oom)\nn\\
&&\ML\ML\ \ \ -\oom\nabb\ro+\oom\nabb(\etab\cdot\eta-2\zeta^2-2\zeta\cdot\nabb\log\oom)\nn\\
&&\ML\ML=(\dd_4\log\oom){\tilde V}-(\dd_3\log\oom)(\nabb\oom\dd_4\log\oom)-\oom\nabb\ro+\oom\nabb(\etab\cdot\eta-2\zeta^2-2\zeta\cdot\nabb\log\oom)\nn\\
&&\ML\ML\ \ \ +2\oom(\nabb\log\oom)(\dd_3\log\oom)(\dd_4\log\oom)\ .
\eea
Substituting this expression in \ref{D4Vtilde} we obtain 
\bea
&&\ML\ML\dd_4\tilde{V}=-\chi\c{\tilde V}+2(\dd_4\log\oom){\tilde V}-\oom\nabb\ro+\oom\nabb(\etab\cdot\eta-2\zeta^2-2\zeta\cdot\nabb\log\oom)
+2\oom(\nabb\log\oom)(\dd_4\dd_3\log\oom)\nn\\
&&\ML\ML\ \ \  +\bigg[2\oom(\nabb\log\oom)(\dd_3\log\oom)(\dd_4\log\oom)-\oom(\dd_4\log\oom)(\nabb\log\oom)(\dd_3\log\oom)\nn\\
&&\ML\ML\ \ \ +\left(\oom(\nabb\log\oom)(\dd_4\log\oom)-\oom\chi\c(\nabb\log\oom)\right)(\dd_3\log\Omega)+\oom\chi\c(\nabb\log\oom)(\dd_3\log\oom)\bigg]\nn\\
&&\ML\ML=-\chi\c{\tilde V}+2(\dd_4\log\oom){\tilde V}-\nabb(\oom\ro)+\nabb[\oom(\etab\cdot\eta-2\zeta^2-2\zeta\cdot\nabb\log\oom)]\nn\\
&&\ML\ML\ \ \ \ +\bigg[2\oom(\nabb\log\oom)(\dd_3\log\oom)(\dd_4\log\oom)\bigg]\ ,\eql{D4Vtildea}
\eea
which, recalling the definition of $\om$ and $\omb$ becomes
\bea
&&\ML\dd_4\tilde{V}=-\chi\c{\tilde V}-4\om{\tilde V}-\nabb(\oom\ro)+\nabb[\oom(\etab\cdot\eta-2\zeta^2-2\zeta\cdot\nabb\log\oom)]
+4\oom(\eta+\etab)\om\omb\ \ \ \ \ \ \ \ \ \ \ 
\eea
which we rewrite as
\bea
&&\ML\dd_4\tilde{V}+\frac{\tr\chi}{2}{\tilde V}+\chih\c{\tilde V}+4\om{\tilde V}=\\
&&\ML=-\oom(\nabb\ro)-\oom(\nabb\log\oom)\ro+\nabb[\oom(\etab\cdot\eta-2\zeta^2-2\zeta\cdot\nabb\log\oom)]
+\big[4\oom(\eta+\etab)\om\omb\big] \ .\ \ \nn 
\eea
This equation has the positive characteristic that it does not depend on $\underline{\tilde V}$ (exactly as the equation for ${\tilde\mu}$ does not depend on for $\underline{\tilde\mu}$. Still, due to the presence of $\oom\nabb\ro$, there is a loss of derivative which we have to cure. To face this problem we we look at the transport equation of  ${\tilde{\omb}}=\divv({\tilde V}-\oom\bb)$. 
\smallskip

\NI We start looking at the transport equation of $\dd_4\nabb\tilde{V}$,
\bea
&&\ML\ML\dd_4\nabb_a\tilde{V}_b=\nabb_a\dd_4\tilde{V}_b+[\dd_4,\nabb_a]\tilde{V}_b\nn\\
&&\ML\ML=\nabb_a\dd_4\tilde{V}_b+\left[-\chi_{ac}\nabb_c\tilde{V}_b-\etab_b\chi_{ac}\tilde{V}_c+\chi_{ab}(\etab\c \tilde{V})
+(\nabb_a\log\oom)\dddd_4\tilde{V}_b+\left([D_{\tau},D_{\ro}]\tilde{V}_{\si}\right)e_4^{\tau}e_a^{\ro}e_b^{\si}\right]\nn\\
&&\ML\ML=-\nabb_a(\chi_{bc}\c{\tilde V}_c)-4\nabb_a(\om{\tilde V}_b)+\nabb_a\!\left[-\oom(\nabb_b\ro)-\oom(\nabb_b\log\oom)\ro
+\nabb_b[\oom(\etab\cdot\eta-2\zeta^2-2\zeta\cdot\nabb\log\oom)]\right]\nn\\
&&\ML\ML\ \ \ +\nabb_a\big[4\oom(\eta+\etab)\om\omb\big]+\left[-\chi_{ac}\nabb_c\tilde{V}_b-\etab_b\chi_{ac}\tilde{V}_c+\chi_{ab}(\etab\c \tilde{V})
+(\nabb_a\log\oom)\dddd_4\tilde{V}_b+\left([D_{\tau},D_{\ro}]\tilde{V}_{\si}\right)e_4^{\tau}e_a^{\ro}e_b^{\si}\right]\nn\\
&&\ML\ML=-\tr\chi\nabb_a{\tilde V}_b-(\chih_{bc}\nabb_a{\tilde V}_c+\chih_{ac}\nabb_c\tilde{V}_b)-4\om\nabb_a{\tilde V}_b-\oom\nabb_a\nabb_b\ro
+\oom(\nabb_a\log\oom)\nabb_b\ro
-\oom(\nabb_b\log\oom)\nabb_a\ro\nn\\
&&\ML\ML\ \ \ +(\nabb_a\log\oom)\dddd_4\tilde{V}_b+\left[-(\nabb_a\chi_{bc})\c{\tilde V}_c-4(\nabb_a\om){\tilde V}_b-\oom(\nabb_a\nabb_b\log\oom)\ro+\nabb_a\nabb_b[\oom(\etab\cdot\eta-2\zeta^2-2\zeta\cdot\nabb\log\oom)]\right]\nn\\
&&\ML\ML\ \ \ +\oom(\nabb_a\log\oom)(\nabb_b\log\oom)\ro+\nabb_a\big[4\oom(\eta+\etab)\om\omb\big]+\left[-\etab_b\chi_{ac}\tilde{V}_c+\chi_{ab}(\etab\c \tilde{V})
+\left([D_{\tau},D_{\ro}]\tilde{V}_{\si}\right)e_4^{\tau}e_a^{\ro}e_b^{\si}\right]\nn\\
&&\ML\ML=-\tr\chi\nabb_a{\tilde V}_b-(\chih_{bc}\nabb_a{\tilde V}_c+\chih_{ac}\nabb_c\tilde{V}_b)-4\om\nabb_a{\tilde V}_b-\oom\nabb_a\nabb_b\ro
+2\oom(\nabb_a\log\oom)\nabb_b\ro-\oom(\nabb_b\log\oom)\nabb_a\ro\nn\\
&&\ML\ML\ \ \ +(\nabb_a\log\oom)\!\left[-\chi_{bc}{\tilde V}_c-4\om{\tilde V}_b-\oom(\nabb_b\log\oom)\ro+\nabb_b[\oom(\etab\cdot\eta-2\zeta^2-2\zeta\cdot\nabb\log\oom)]+4\oom(\eta+\etab)_b\om\omb\right]\nn\\
&&\ML\ML\ \ \ +\left[-(\nabb_a\chi_{bc})\c{\tilde V}_c-4(\nabb_a\om){\tilde V}_b-\oom(\nabb_a\nabb_b\log\oom)\ro+\nabb_a\nabb_b[\oom(\etab\cdot\eta-2\zeta^2-2\zeta\cdot\nabb\log\oom)]\right]\nn\\
&&\ML\ML\ \ \ +\oom(\nabb_a\log\oom)(\nabb_b\log\oom)\ro+\nabb_a\big[4\oom(\eta+\etab)\om\omb\big]+\left[-\etab_b\chi_{ac}\tilde{V}_c+\chi_{ab}(\etab\c \tilde{V})
+\left([D_{\tau},D_{\ro}]\tilde{V}_{\si}\right)e_4^{\tau}e_a^{\ro}e_b^{\si}\right]\nn
\eea
which we rewrite ordering the r.h.s terms following the number of derivatives (of the metric components)
\bea
&&\ML\ML\dd_4\nabb_a\tilde{V}_b+\tr\chi\nabb_a{\tilde V}_b+(\chih_{bc}\nabb_a{\tilde V}_c+\chih_{ac}\nabb_c\tilde{V}_b)+4\om\nabb_a{\tilde V}_b
=-\oom\nabb_a\nabb_b\ro\nn\\
&&\ML\ML +2\oom(\nabb_a\log\oom)\nabb_b\ro-\oom(\nabb_b\log\oom)\nabb_a\ro+{\cal Q}_{ab}
\eea
where ${\cal Q}_{ab}$ collects all the terms which depend at most on two derivatives or on three derivatives of terms whose estimates have been already obtained as, for instance, the second derivatives of $\ze$,
\bea
&&\ML\ML{\cal Q}_{ab}=(\nabb_a\log\oom)\!\!\left[-\chi_{bc}{\tilde V}_c-4\om{\tilde V}_b-\oom(\nabb_b\log\oom)\ro+\nabb_b[\oom(\etab\cdot\eta-2\zeta^2-2\zeta\cdot\nabb\log\oom)]+4\oom(\eta+\etab)_b\om\omb\right]\nn\\
&&\ML\ML\ \ \ \ \ +\left[-(\nabb_a\chi_{bc})\c{\tilde V}_c-4(\nabb_a\om){\tilde V}_b-\oom(\nabb_a\nabb_b\log\oom)\ro+\nabb_a\nabb_b[\oom(\etab\cdot\eta-2\zeta^2-2\zeta\cdot\nabb\log\oom)]\right]\nn\\
&&\ML\ML\ \ \ \ \ +\oom(\nabb_a\log\oom)(\nabb_b\log\oom)\ro+\nabb_a\big[4\oom(\eta+\etab)\om\omb\big]+\left[-\etab_b\chi_{ac}\tilde{V}_c+\chi_{ab}(\etab\c \tilde{V})+\left([D_{\tau},D_{\ro}]\tilde{V}_{\si}\right)e_4^{\tau}e_a^{\ro}e_b^{\si}\right]\nn
\eea
Taking the trace we obtain
\bea
\dd_4\divv\tilde{V}+\tr\chi\divv{\tilde V}+2\chih\c\nabb{\tilde V}+4\om\divv{\tilde V}
=-\oom\lapp\ro+\oom(\nabb\log\oom)\c\nabb\ro+\tr{\cal Q}\ .\ \ \ \ \eql{trdivvVtilde}
\eea
As expected the term $-\oom\lapp\ro$ implies a loss of derivatives. To cure it we look at the transport equation for $\divv\bb$. From eq. (4.4.5) of \cite{Kl-Ni:book} we have,
\bea
\ML\dd_4(\oom\nabb_a\bb_b)+\tr\chi\oom\nabb_a\bb_b=-\oom\nabb_a\nabb_b\ro+\oom\nabb_a\dual\nabb_b\si-\frac{1}{2}\tr\chi\oom\nabb_a\bb_b
-\chibh_{ac}\oom\nabb_c\b_b+{\cal H}\ ,\ \ \ \ \ \ \eql{445}
\eea
\bea
\dd_4(\oom\divv\bb)+\tr\chi(\oom\divv\bb)=-\oom\lapp\ro-\frac{1}{2}\tr\chi\oom\divv\bb
-\chibh\c\oom\nabb\bb+\tr{\cal H}\ .\nn
\eea
Moreover
\bea
\ML\dd_4\divv(\oom\bb)\!&=&\!\dd_4(\oom\divv\bb)+\dd_4(\nabb\oom\c\bb)
\eea
and
\bea
&&\ML\ML\dd_4\nabb\oom=[\dd_4,\nabb]\oom+\nabb\dd_4\oom=[\dd_4,\nabb]\oom+\nabb\oom\dd_4\log\oom\nn\\
&&\ML\ML=[\dd_4,\nabb]\oom-2\nabb\oom\om=[\dd_4,\nabb]\oom-2\oom(\nabb\log\oom)\om-2\oom\nabb\om\\
&&\ML\ML=[\dd_4,\nabb]\oom-\oom(\eta+\etab)\om-2\oom\nabb\om\nn\\
&&\ML\ML=(\ze+\etab)\ddb_4\oom-\chi\c\nabb\oom-\oom(\eta+\etab)\om-2\oom\nabb\om\nn\\
&&\ML\ML=-2\oom(\ze+\etab)\om-\oom\chi\c(\zeta+\etab)-\oom(\eta+\etab)\om-2\oom\nabb\om\nn\\
&&\ML\ML=-4\oom(\nabb\log\oom)\om-\oom\chi\c(\nabb\log\oom)-2\oom\nabb\om\nn
\eea
therefore
\bea
(\dd_4\nabb\oom)\c\bb=-\bigg[(4\oom\om(\nabb\log\oom)+\oom(\nabb\log\oom)\c\chi)+2\oom\nabb\om\bigg]\c\bb
\eea
moreover
\bea
\nabb\oom\c\dd_4\bb+\nabb\oom\tr\chi\bb=\nabb\oom\c\big[-\nabb\ro+[\c\c\c\c]\big]
\eea
it follows
\bea
&&\ML\ML\dd_4\divv(\oom\bb)+\tr\chi\divv(\oom\bb)=\dd_4(\oom\divv\bb)+(\dd_4\nabb\oom)\c\bb+\nabb\oom\c\dd_4\bb
+\tr\chi(\oom\divv\bb)+(\nabb\oom)\tr\chi\bb\nn\\
&&\ML\ML=\left(\dd_4(\oom\divv\bb)+\tr\chi(\oom\divv\bb)\right)+(\nabb\oom)\c\!\left(\dd_4\bb+\tr\chi\bb\right)+(\dd_4\nabb\oom)\c\bb\nn\\
&&\ML\ML\ML\ML=\left[-\oom\lapp\ro-\frac{1}{2}\tr\chi\oom\divv\bb-\chibh\c\oom\nabb\b+\tr{\cal H}\right]
+(\nabb\oom)\left[-\nabb\ro+[\c\c\c\c]\right]-\bigg[(4\oom\om(\nabb\log\oom)+\oom(\nabb\log\oom)\c\chi)+2\oom\nabb\om\bigg]\c\bb\nn\\
&&\ML\ML\ML\ML=\left[-\oom\lapp\ro-\frac{1}{2}\tr\chi\oom\divv\bb-\chibh\c\oom\nabb\b+\tr{\cal H}\right]
+(\nabb\oom)\left[-\nabb \ro+\left(2\om\bb+2\hat{\chib}\c\b+\dual\nabb\si
-3(\etab\ro-\dual\etab\si)\right)\right]\nn\\
&&\ML\ML\ML\ML\ -\bigg[(4\oom\om(\nabb\log\oom)+\oom(\nabb\log\oom)\c\chi)+2\oom\nabb\om\bigg]\c\bb\nn
\eea
which implies
\bea
&&\ML\ML\ML\dd_4[\divv(\tilde{V}-\oom\bb)]+\tr\chi[\divv(\tilde{V}-\oom\bb)]\\
&&\ML\ML\ML=[-2\chih\c\nabb{\tilde V}-4\om\divv{\tilde V}+\oom(\nabb\log\oom)\c\nabb\ro+\tr{\cal Q}]-\left[-\frac{1}{2}\tr\chi\oom\divv\bb-\chibh\c\oom\nabb\b+\tr{\cal H}\right]\nn\\
&&\ML\ML\ML+(\nabb\oom)\left[-\nabb \ro+\left(2\om\bb+2\hat{\chib}\c\b+\dual\nabb\si
-3(\etab\ro-\dual\etab\si)\right)\right]+\bigg[(4\oom\om(\nabb\log\oom)+\oom(\nabb\log\oom)\c\chi)+2\oom\nabb\om\bigg]\c\bb\nn\\
&&\ML\ML\ML=-4\om[\divv({\tilde V}-\oom\bb)]-2\chih\c\nabb{\tilde V}\\
&&\ML\ML\ML\ \ \ +\!\left[-4\om\divv(\oom\bb)+\oom(\nabb\log\oom)\c\nabb\ro+\frac{1}{2}\tr\chi\oom\divv\bb+\chibh\c\oom\nabb\b+(\nabb\oom)(-\nabb \ro+\dual\nabb\si)\right]\nn\\
&&\ML\ML\ML\ \ +\left[\tr{\cal Q}+\tr{\cal H}+(\nabb\oom)(2\om\bb+2\hat{\chib}\c\b-3(\etab\ro-\dual\etab)+(4\oom\om(\nabb\log\oom)+\oom(\nabb\log\oom)\c\chi)\c\bb+2\oom\nabb\om\c\bb\right]\nn
\eea 
which we rewrite in a more compact way
\bea
&&\ML\ML\dd_4\divv{\hat V}+\tr\chi\divv{\hat V}=-4\om\divv{\hat V}-2\chih\c\nabb{\hat V}\nn\\
&&\ML\ML\ \ \ +\!\left[-2\chih\c\nabb\oom\bb-4\om\divv(\oom\bb)+\oom(\nabb\log\oom)\c\nabb\ro+\frac{1}{2}\tr\chi\oom\divv\bb+\chibh\c\oom\nabb\b+(\nabb\oom)(-\nabb \ro+\dual\nabb\si)\right]\nn\\
&&\ML\ML\ \ \ +\oom^{-1}{\{\cal{F}\}}
\eea
where $\oom^{-1}\{\cal{F}\}$ collect all lower order terms which can be estimated using the inductive assumptions, more precisely
\bea
\ML\ML\ML\oom^{-1}\{{\cal{F}}\}=\tr({\cal Q}+{\cal H})+\!\left[(\nabb\oom)(2\om\bb+2\hat{\chib}\c\b-3(\etab\ro-\dual\etab)+(4\oom\om(\nabb\log\oom)+\oom(\nabb\log\oom)\c\chi)\c\bb+2\oom\nabb\om\c\bb\right].\ \ \ \ 
\eea
Therefore
\bea
&&\ML\ML\frac{d{\tilde{\omb}}}{d\nu}+\oom\tr\chi{\tilde{\omb}}+4\oom\om{\tilde{\omb}}=-\oom\left[2\chih\c\nabb{\hat V}
-2\chih\c\nabb\oom\bb-4\om\divv(\oom\bb)+\oom(\nabb\log\oom)\c\nabb\ro\right.\ \ \ \ \ \ \ \ \ \ \ \eql{tildeomb}\\
&&\ \ \ \ \ \ \ \ \ \ \ \ \ \ \ \ \ \ \ \ \ \ \left.+\frac{1}{2}\tr\chi\oom\divv\bb+\chibh\c\oom\nabb\b+(\nabb\oom)(-\nabb \ro+\dual\nabb\si)\right]
+{\{\cal{F}\}}\ .\nn
\eea
{To prove the transport equation \ref{3.83} for $\Lie_O^{N-2}{\tilde{\omb}}$ we proceed exactly in the same way as we did for $\Lie_O^{N-1}\Us$.  Starting from
\bea
&&\Lie_O^{N-2}\frac{\Dbb{\tilde{\omb}}}{\partial\nu}=\frac{\Dbb}{\partial\nu}(\Lie_O^{N-2}{\tilde{\omb}})+[\Lie_O^{N-2},\frac{\Dbb}{\partial\nu}]{\tilde{\omb}}
\eea
and using \ref{1.117w} it follows 
\bea
\ML\ML\frac{\Dbb}{\partial\nu}(\Lie_O^{N-2}{\tilde{\omb}})\!&=&\!
\Lie_O^{N-2}\frac{\Dbb{\tilde{\omb}}}{\partial\nu}-\sum_{q=0}^{N-2}\cbin{N-2}{q}(\Lie_O^{N-3-q}E)(\Lie_O^{q}{\tilde{\omb}})
\eea
\NI From equation \ref{tildeomb} it follows: 
\bea
&&\ML\ML\Lie_O^{N-2}\frac{\Dbb{\tilde{\omb}}}{\partial\nu}=-\Lie_O^{N-2}\!\left((\oom\tr\chi+4\oom\om){\tilde{\omb}}\right)
-2\Lie_O^{N-2}\oom\chih\c\nabb{\hat V}\nn\\
&&\ML\ML+\Lie_O^{N-2}\!\left\{\oom\!\left[-2\chih\c\nabb\oom\bb-4\om\divv(\oom\bb)+\oom(\nabb\log\oom)\c\nabb\ro+\frac{1}{2}\tr\chi\oom\divv\bb+\chibh\c\oom\nabb\b+(\nabb\oom)(-\nabb \ro+\dual\nabb\si)\right]\right\}\nn\\
&&\ML\ML+\Lie_O^{N-2}{\{\cal{F}\}}\nn\\
&&\ML\ML=-(\oom\tr\chi+4\oom)\Lie_O^{N-2}{\tilde{\omb}}-2\oom\chih\ \!\Lie_O^{N-2}\nabb{\hat V}
+\oom\!\left[-2\oom\chih\c\Lie_O^{N-2}\nabb\bb-4\oom\om\Lie_O^{N-2}\divv\bb+\oom(\nabb\log\oom)\c\Lie_O^{N-2}\nabb\ro\right.\nn\\
&&\ML\ML\left.\ \ \ +\frac{1}{2}\tr\chi\oom\Lie_O^{N-2}\divv\bb+\chibh\c\oom\Lie_O^{N-2}\nabb\b+(\nabb\oom)(-\Lie_O^{N-2}\nabb\ro+\Lie_O^{N-2}\dual\nabb\si)\right]
+\{{\cal G}\}+\Lie_O^{N-2}{\{\cal{F}\}}\ ,\eql{nabbNomtildetrasp}
\eea
where
\bea
&&\ML\ML\ML\{{\cal G}\}=\Lie_O^{N-2}\!\left\{\oom\!\left[-2\chih\c\nabb\oom\bb-4\om\divv(\oom\bb)+\oom(\nabb\log\oom)\c\nabb\ro+\frac{1}{2}\tr\chi\oom\divv\bb+\chibh\c\oom\nabb\b+(\nabb\oom)(-\nabb \ro+\dual\nabb\si)\right]\right\}\nn\\
&&\ML\ML\ML\ \ \ -\oom\!\left[-2\oom\chih\c\Lie_O^{N-2}\nabb\bb-4\oom\om\Lie_O^{N-2}\divv\bb+\oom(\nabb\log\oom)\c\Lie_O^{N-2}\nabb\ro\right.\nn\\
&&\ML\ML\left.\ \ \ +\frac{1}{2}\tr\chi\oom\Lie_O^{N-2}\divv\bb+\chibh\c\oom\Lie_O^{N-2}\nabb\b+(\nabb\oom)(-\Lie_O^{N-2}\nabb\ro+\Lie_O^{N-2}\dual\nabb\si)\right]\nn\\
&&\ML\ML\ML\ \ \ -\left[2\Lie_O^{N-2}\oom\chih\c\nabb{\hat V}-2\oom\chih\ \!\Lie_O^{N-1}{\hat V}-(\oom\tr\chi+4\oom)\Lie_O^{N-2}{\tilde{\omb}}-2\oom\chih\ \!\Lie_O^{N-2}\nabb{\hat V}\right]\ .
\eea
It is clear that the terms $\{{\cal G}\}$ and ${\{\tilde{\cal{F}}\}}\equiv\Lie_O^{N-2}{\{\cal{F}\}}$ do not contain the highest derivatives and can be estimated inductively in the standard way, the higher order derivatives terms in \ref{nabbNomtildetrasp} are the more important ones. Collecting all the terms we have for $\Lie_O^{N-2}{\tilde{\omb}}$ the following transport equation,
\bea
&&\ML\frac{\Dbb}{\partial\nu}(\Lie_O^{N-2}{\tilde{\omb}})+(\oom\tr\chi+4\oom\om)(\Lie_O^{N-2}{\tilde{\omb}})+(\oom\chi)\c(\Lie_O^{N-2}{\tilde{\omb}})\nn\\
&&\ML=-2\oom\chih\ \!\Lie_O^{N-2}\nabb{\hat V}+\oom\!\left[-2\oom\chih\c\Lie_O^{N-1}\bb-4\oom\om\Lie_O^{N-2}\divv\bb+\oom(\nabb\log\oom)\c\Lie_O^{N-1}\ro\right.\nn\\
&&\ML\left.\ \ \ +\frac{1}{2}\tr\chi\oom\Lie_O^{N-2}\divv\bb+\chibh\c\oom\Lie_O^{N-1}\b+(\nabb\oom)(-\Lie_O^{N-1}\ro+\Lie_O^{N-2}\dual\nabb\si)\right]\nn\\
&&\ML\ \ \ \ +\{{\cal L}\}+\{{\cal G}\}+{\{\tilde{\cal{ F}}\}}\ ,
\eea
where 
\bea
&&\ML{\{\tilde{\cal{ F}}\}}=\Lie_O^{N-2}{\{{\cal{ F}}\}}\\
&&\ML\{{\cal L}\}=-\sum_{q=0}^{N-2}\cbin{N-2}{q}(\Lie_O^{N-3-q}E)(\Lie_O^{q}{\tilde{\omb}})
\eea}
\subsection{ Proof of Lemma \ref{L3.7}}
\NI As done previously for lemma \ref{nabbNetaest}, we start from the following Hodge system
\bea
&&\divv{\hat V}={\tilde{\omb}}\equiv F^{(0)}\eql{Vhodgeaa}\nn\\
&&\curll{\hat V}=\oom(\nabb\log\oom\wedge\bb)+\oom\curll\bb\equiv G^{(0)}\ \eql{Vhodgebb}
\eea
and we proceed exactly as in the proof of Lemma \ref{nabbNetaest} with the substitution of $\eta$ with ${\hat V}=2\nabb\oom\omb-\oom\bb$. As ${\hat V}$ is at the level of a Riemann component (a derivative of a connection coefficient) we have to control its $\Lie_O^{N-1}$ angular derivative. We repeat the first steps of the previous lemma with the quoted substitutions and obtain.

\bea
|\nabb\Lie_O^{N-2}\hat{V}|_{p,S}\leq| \widetilde{\widetilde{ \{good_3\}}} |_{p,S}+ |\tilde{\omega}|_{p,S}
\eea

\NI with 

\bea
|\widetilde{\widetilde{ \{good_3\}}}|_{p,S}= |[\divv,\Lie_O^{N-2}] \hat{V}|_{p,S} +|[\curll,\Lie_O^{N-2}] \hat{V}|_{p,S}+\Lie_O^{N-1}(\oom(\nabb\log\oom\wedge\bb)+\oom\curll\bb)\nn\\
\eea

\NI Are terms with can be easily estimated by the inductive assumptions.

\NI Hence

\bea
|\Lie_O^{N-2}\nabb \hat{V}|_{p,S}\leq | \widetilde{\widetilde{\{ good_3\}}}|_{p,S} + |\tilde{\omega}|_{p,S}
\eea
\subsection{The estimates for the underlined connection coefficients}
\NI The estimates for the underlined connection coefficients can be substantially obtained in the same way as done for the non underlined ones, the only difference occours for the integral estimates of the terms classified as $\{{\it (\underline{good})_1}\}$, let us calculate in details the corresponding estimates in this case, defining
\bea
\tilde{\Ga}(\la)\equiv c\!\int_{\la_0}^{\la}|{\cal H}|_{\infty,S}(\la')d\la'={\tilde C}\frac{\la-\la_0}{\la\la'}\ .
\eea
\

\bea
&&\ML\big|r^{3-\frac{2}{p}}\Lie_O^{N-1}\Ubs\big|_{p,S}(\la,\nu)\leq e^{\tilde{\Ga}(\la)}|r^{(3-\frac{2}{p})}\Lie_O^{N-1}\Ubs|_{p,S}(\la_0,\nu)\nn\\
&&\ML\ \ \ \ \ \ \ \ \ \ \ \ \ \ \ \ \ \ \ \ \ \ \ \ \ \ +C\!\left(\int_{\la_0}^{\la}d\la' e^{(\tilde{\Ga}(\la_0)-\tilde{\Ga}(\la'))}|r^{3-\frac{2}{p}}\big\{{\it (\underline{good})_1}\big\}|_{p,S}\right)\nn\\
&&\ML\ML\ML\leq C_{0}^{(0)} e^{\tilde{\Ga}(\la)}\!\left(e^{(N-2)\de}e^{(N-2){\underline{\Ga}}(\la)}\frac{N!}{N^\a}\frac{1}{\ro_{0,1}^{N}}\right)\\
&&\ML\ML\ML+\left(C\frac{N!}{N^\a}\frac{1}{\ro_{0,1}^{N}}e^{(N-2)\de}e^{(N-2)\underline{\Ga}(\la)}\right)\!\left( Ce^{-\de} \int_{\la_0}^{\la}d\la' \frac{e^{(-(N-2)\underline{\Ga}(\la)+\tilde{\Ga}(\la)-\tilde{\Ga}(\la'))}}{r^2}\right)\nn\\
&&\ML\ML\ML\leq C_{0}^{(0)} e^{\tilde{\Ga}(\la)}\!\left(e^{(N-2)\de}e^{(N-2){\underline{\Ga}}(\la)}\frac{N!}{N^\a}\frac{1}{\ro_{0,1}^{N}}\right)\\
&&\ML\ML\ML+\left(C\frac{N!}{N^\a}\frac{1}{\ro_{0,1}^{N}}e^{(N-2)\de}e^{(N-2)\underline{\Ga}(\la)}\right)\!\left( C e^{-\de} \int_{\la_0}^{\la}d\la' \frac{e^{(-(N-2)\underline{\Ga}(\la')+\tilde{\Ga}(\la)-\tilde{\Ga}(\la'))}}{r^2}\right)\nn\\
&&\ML\ML\ML\leq\left(C\frac{N!}{N^\a}\frac{1}{\ro_{0,1}^{N}}e^{(N-2)\de}e^{(N-2)\underline{\Ga}(\la)}\right)\!\left(\frac{C_{0}^{(0)}}{C} e^{(\tilde{\Ga}(\la))}\right.\nn\\
&&\ML\ML\ML\left.\ \ \ \ \ \ \ \ \ \ \ \ \ \ \ \ \ \ \ \ \ \ \ \ \ \ \ \ \ \ \ \ \ \ \ \ \ \ \ \ \ \ \ \ \ \ \ \ \ + C'e^{-\de}\int_{\la_0}^{\la}\frac{d\la'}{{\la'}^2} e^{-(N-2)C\frac{\la'-\la_0}{\la'\la_0}}\right)\nn\\
&&\ML\ML\ML\leq\left(C''\frac{N!}{N^\a}\frac{1}{\ro_{0,1}^{N}}e^{(N-2)\de }e^{(N-2)\underline{\Ga}(\la)}\right)
\eea
\NI Choosing suitably $C_0^{(0)}<C$ and $\de$  and were the last integral has been estimated as in equation \ref{130z2a}.
\NI At las let us estimate the corresponding of integral \ref{1588}
\bea
\int_{\la_0}^{\la}d\la' e^{(\tilde{\Ga}(\la)-\tilde{\Ga}(\la'))}(\big|2\chibh\big|_{\infty,S}+\big|\tr\chib\big|_{\infty,S})|r^{(3-\frac{2}{p})}\Lie_O^{N-1}\bb|_{p,S}\ .\ \ \ \ \ \ \
\eea
In order to estimate this term we can perform the same steps of the previous term, with  the difference thet the decay of the $\tr\chib$ is worst. Actually we obtain
\bea
&&\int_{\la_0}^{\la}d\la' e^{(\tilde{\Ga}(\la)-\tilde{\Ga}(\la'))}(\big|2\chibh\big|_{\infty,S}+\big|\tr\chib\big|_{\infty,S})|r^{(3-\frac{2}{p})}\Lie_O^{N-1}\bb|_{p,S}\ .\ \ \ \ \ \ \eql{estaxb}\\
&&\ML\ML\ML\leq \left.+ C'e^{-\de}\la^{\frac{1}{2}}\int_{\la_0}^{\la}\frac{d\la'}{{\la'}^2} e^{-(N-2)C\frac{\la-\la_0}{\la'\la_0}}\right)\leq\left(C''\frac{N!}{N^\a}\frac{1}{\ro_{0,1}^{N}}e^{(N-2)\de }e^{(N-2)\underline{\Ga}(\la)}\right)\nn\\
\eea
\subsection{ Proof of corollary \ref{cor92}}

\NI Once we have the estimate for $\Lie_O^{N-1}\Us"="\Lie_O^{N-1}\nabb U$, we have to estimate $\Lie_O^{N}U$. To do this we have to exploit a modified version of proposition 3.3.3 and 3.3.4 of \cite{Kl-Ni:book}, let us recall it.   

\begin{prop}\label{nabb}
\NI Let $U_{\a}$ be 1-form, then  it does exists a constant $C$ such that

\bea
|\nabb U|^2\leq C \sum_i|\Lie_{\ ^{(i)}O}U|^2
\eea

\NI Let $U_{\a\b}$ be a symmetric two tensor, then  it does exists a constant $C$ such that

\bea
|\nabb U|^2\leq C \sum_i|\Lie_{\ ^{(i)}O}U|^2
\eea

\end{prop}

\NI We add to this proposition the 

\begin{prop}\label{nabb6}
\NI Let $U_{\a}$ be 1-form, then  it does exists a constant $C$ such that

\bea\label{nabb7}
|\nabb U|^4\leq C \sum_i|\Lie_{\ ^{(i)}O}U|^4
\eea

\NI Let $U_{\a\b}$ be a symmetric two tensor, then  it does exists a constant $C$ such that

\bea
|\nabb U|^4\leq C \sum_i|\Lie_{\ ^{(i)}O}U|^4
\eea
The proof is a straightforward modification of the propositions 3.3.3 and 3.3.4 of \cite{C-K:book}. 

\end{prop}

\NI Then, exploiting the previous propositions, we obtain

\bea
&&|\Lie_O^{N-1}\Us| "="|\Lie_O^{N-1}\nabb U|_{p,S}=|\nabb\Lie_O U|+|[\Lie_O^{N-1},\nabb]U|_{p,S}\leq C |\Lie_O^N U|_{p,S} + \widetilde{good_{6}}\nn\\
\eea

\NI Hence we have the right estimates for $|\Lie_O^{N} U|_{p,S}$ with $p=2, 4$. The other values of $p$ are obtained by interpolation. Clearly the same estimate can be done for $\Lie_O^{N-1}\nabb\chib$.

\subsection{Proof of corollary \ref{cor9.9}}\label{sec158}

\NI In the following we prove corollary \ref{cor9.9}, to do it we proceed inductively in the following way, we observe that
\bea
&&\ML\ML\Lie_O^JS_{AC}=\Lie_O^{J}\gggg(e_A,\nabb e_C)=\sum_{k=0}^J\cbin{J}{k}\gggg(\Lie_O^ke_A,\Lie_O^{J-k}\nabb e_C)
=\sum_{k=0}^J\cbin{J}{k}\ga_{ac}\Lie_O^ke_A^a\Lie_O^{J-k}\nabb e_C^c\nn\\
&&\ML\ML=\sum_{k=0}^J\cbin{J}{k}\sum_E\theta^E_a\ga_{ac}\Lie_O^ke_A^a\theta^E_c\Lie_O^{J-k}\nabb e_C^c
=\sum_E\sum_{k=0}^J\cbin{J}{k}\gggg(e_E\Lie_O^ke_A)\gggg(e_E\Lie_O^{J-k}\nabb e_C)\ .\nn
\eea
Therefore
\bea
\gggg(e_A,\Lie_O^J\nabb e_C)=\Lie_O^JS-\sum_E\sum_{k=1}^J\cbin{J}{k}\gggg(e_E\Lie_O^ke_A)\gggg(e_E\Lie_O^{J-k}\nabb e_C)
\eea
and
\bea
&&\ML\ML|r^{(J+1-\frac{2}{p})}\gggg(e_A,\Lie_O^J\nabb e_C)|_{p,S}\\
&&\ML\ML\leq |r^{(J+1-\frac{2}{p})}\Lie_O^JS_{AC}|_{p,S}
+\sum_E\sum_{k=1}^J\cbin{J}{k}|r^{(J+1-\frac{2}{p})}\gggg(e_E,\Lie_O^ke_A)\gggg(e_E,\Lie_O^{J-k}\nabb e_C)|_{p,S}\nn\\
&&\ML\ML\leq |r^{(J+1-\frac{2}{p})}\Lie_O^JS_{AC}|_{p,S}
+\sum_E\sum_{k=1}^{\left[\frac{J}{2}\right]}\cbin{J}{k}|r^{k}\gggg(e_E,\Lie_O^ke_A)|_{\infty,S}|r^{(J+1-k-\frac{2}{p})}\gggg(e_E,\Lie_O^{J-k}\nabb e_C)|_{p,S}\nn\\
&&\ \ \ \ \ \ \ \ \ \ \ \ \ \ \ +\sum_E\!\!\sum_{k=\left[\frac{J}{2}\right]+1}^J\cbin{J}{k}|r^{(k-\frac{2}{p})}\gggg(e_E,\Lie_O^ke_A)|_{p,S}|r^{(J+1-k)}\gggg(e_E,\Lie_O^{J-k}\nabb e_C)|_{\infty,S}\ .\nn
\eea
We make the inductive assumptions for $|r^{(k-\frac{2}{p})}\gggg(e_E,\Lie_O^ke_A)|_{p,S}$, with $k\leq J$
\bea
|r^{(k-\frac{2}{p})}\gggg(e_E,\Lie_O^ke_A)|_{p,S}\leq c\!\left(\frac{(k-1)!}{(k-1)^{\a}}\frac{e^{(k-3)(\de+\underline{\Ga}(\la))}}{\ro_{0,1}^{k-1}}\right)
\eea
therefore
\bea
&&\ML\ML|r^{(J+1-\frac{2}{p})}\gggg(e_A,\Lie_O^J\nabb e_C)|_{p,S}\leq c\!\left(\frac{J!}{J^{\a}}\frac{e^{(J-2)(\de+\underline{\Ga}(\la))}}{\ro_{0,1}^{J}}\right)\\
&&\ML\ML+\sum_E\sum_{k=1}^{\left[\frac{J}{2}\right]}\cbin{J}{k}c^2\!\left(\frac{k!}{k^{\a}}\frac{e^{(k-2)(\de+\underline{\Ga}(\la))}}{\ro_{0,1}^{k}}\right)\!\left(\frac{(J-k)!}{(J-k)^{\a}}\frac{e^{((J-k)-2)(\de+\underline{\Ga}(\la))}}{\ro_{0,1}^{J-k}}\right)\nn\\
&&\ML\ML+\sum_E\!\!\sum_{k=\left[\frac{J}{2}\right]+1}^J\cbin{J}{k}c^2\!\left(\frac{(k-1)!}{(k-1)^{\a}}\frac{e^{(k-3)(\de+\underline{\Ga}(\la))}}{\ro_{0,1}^{k-1}}\right)\!\left(\frac{(J+1-k)!}{(J+1-k)^{\a}}\frac{e^{((J+1-k)-2)(\de+\underline{\Ga}(\la))}}{\ro_{0,1}^{J+1-k}}\right)\nn\\
&&\ML\ML\leq c\!\left(\frac{J!}{J^{\a}}\frac{e^{(J-2)(\de+\underline{\Ga}(\la))}}{\ro_{0,1}^{J}}\right)\nn\\
&&\ML\ML+2\sum_E\sum_{k=0}^{\left[\frac{J}{2}\right]}\cbin{J}{k}c^2\!\left(\frac{k!}{k^{\a}}\frac{e^{(k-2)(\de+\underline{\Ga}(\la))}}{\ro_{0,1}^{k}}\right)\!\left(\frac{(J-k)!}{(J-k)^{\a}}\frac{e^{((J-k)-2)(\de+\underline{\Ga}(\la))}}{\ro_{0,1}^{J-k}}\right)\nn\\
&&\ML\ML\leq c\!\left(\frac{J!}{J^{\a}}\frac{e^{(J-2)(\de+\underline{\Ga}(\la))}}{\ro_{0,1}^{J}}\right)\nn\\
&&\ML\ML+\!\left(\frac{J!}{J^{\a}}\frac{e^{(J-2)(\de+\underline{\Ga}(\la))}}{\ro_{0,1}^{J}}\right)\left[\left(\frac{J!}{J^{\a}}\frac{e^{(J-2)(\de+\underline{\Ga}(\la))}}{\ro_{0,1}^{J}}\right)^{-1}\right.\nn\\
&&\ML\ML\left.
\frac{e^{(J-2)(\de+\underline{\Ga}(\la))}}{\ro_{0,1}^{J}}
\left(2^2c^2{e^{-2(\de+\underline{\Ga}(\la))}}\right)\sum_{k=0}^{\left[\frac{J}{2}\right]}
\cbin{J}{k}\!\left(\frac{k!}{k^{\a}}\right)\!\left(\frac{(J-k)!}{(J-k)^{\a}}\right)\right]\nn\\
&&\ML\ML\leq c\!\left(\frac{J!}{J^{\a}}\frac{e^{(J-2)(\de+\underline{\Ga}(\la))}}{\ro_{0,1}^{J}}\right)\nn\\
&&\ML\ML+\!\left(\frac{J!}{J^{\a}}\frac{e^{(J-2)(\de+\underline{\Ga}(\la))}}{\ro_{0,1}^{J}}\right)\left[
\left(2^2c^2{e^{-2(\de+\underline{\Ga}(\la))}}\right)\sum_{k=0}^{\left[\frac{J}{2}\right]}\!\left(\frac{1}{k^{\a}}\right)\!\left(\frac{{J^{\a}}}{(J-k)^{\a}}\right)\right]\nn\\
&&\ML\ML\ML\ML\leq c\!\left(\frac{J!}{J^{\a}}\frac{e^{(J-2)(\de+\underline{\Ga}(\la))}}{\ro_{0,1}^{J}}\right)+\!\left(\frac{J!}{J^{\a}}\frac{e^{(J-2)(\de+\underline{\Ga}(\la))}}{\ro_{0,1}^{J}}\right)\!\!\left[
\left(2^{2+\a}c^2{e^{-2(\de+\underline{\Ga}(\la))}}\right)\sum_{k=0}^{\left[\frac{J}{2}\right]}\!\frac{1}{k^{\a}}\right]\nn\\
&&\ML\ML\leq (c+\frac{c}{2})\!\left(\frac{J!}{J^{\a}}\frac{e^{(J-2)(\de+\underline{\Ga}(\la))}}{\ro_{0,1}^{J}}\right),
\eea
provided that $\de$ is sufficiently large so that,
\bea
\left[\left(2^{2+\a}c{e^{-2(\de+\underline{\Ga}(\la))}}\right)\sum_{k=0}^{\left[\frac{J}{2}\right]}\!\frac{1}{k^{\a}}\right]
\leq \frac{1}{2}\ .
\eea  
Choosing suitably the constants $c$, we have proved the inequality
\bea\label{15138}
|r^{((J+1)-\frac{2}{p})}\gggg(e_E,\Lie_O^{J}\nabb e_A)|_{p,S}\leq c\!\left(\frac{J!}{J^{\a}}\frac{e^{(J-2)(\de+\underline{\Ga}(\la))}}{\ro_{0,1}^{J}}\right)\ ,\eql{8.59}
\eea

\NI Once we have the estimates \ref{15138} we obtain the estimates for $|r^{((J+1)-\frac{2}{p})}\gggg(e_E,\Lie_O^{J+1}e_A)|_{p,S}$ applying corollary \ref{cor92}.

\subsection{ Some extra remarks on the canonical foliation}\label{SScanfol2}
To see in a more clear way why the canonical foliation has to be used we have first to discuss why technically this problem arises and subsequently look for a more ``deep" justification. Let us start with the technical aspects. As discussed before to get the correct decays we need to control some connection coefficient norms integrating from above. This requires, specifically for $\tilde\mu$ and $\tilde{\omb}$, the control of these quantities on $\Cb_*$. To do it we have integrate them along $\Cb_*$, therefore along an incoming direction which is not the natural one for these quantities associated to non underlined connection coefficients. In fact if one tries to do it along $\Cb_*$ looking at its transport equation in a generic foliation it follows immediately that a loss of derivatives appears. In fact as
${\tilde\mu}=-\divv\eta + \frac{1}{2}(\chi\underline{\chi}-\overline{\chi\underline{\chi}})-(\ro-\overline{\ro})$ it seems evident that computing $\ddb_3{\tilde\mu}$ on the right hand side we obtain terms of the order of angular derivatives of a Riemann component, this happens both $\ddb_3$ deriving $-\divv\eta$ and $\ro$; on the other side $\tilde\mu$ is of order of a Riemann component (one derivative of connection coefficients) and therefore a loss of derivative is present. The choice of the canonical foliation solves this problem and, moreover, guarantees the better decay for the connection components along the null outgoing directions. Moreover no loss of derivatives is present imposing the initial data on $\Cb_*$ for $\tilde{\omb}$. Once we realize that to estimate $\tilde\mu$ we have to integrate from above and, therefore, for getting the right estimates for ${\tilde\mu}|_{\Cb_*}$ we need to choose an appropriate (canonical) foliation, it turns out that we have to use this foliation for the transport of all the not underlined quantities which, therefore, all of them, have to be integrated from above. This implies that the chosen foliation, that is $\oom$, has to be such that even $\tr\chi$ and its derivatives (or $\Us$ which is basically the same thing) when transported along $\Cb_*$ must avoid any loss of derivatives. This is possible to obtain with the right choice of $\oom|_{\Cb_*}$ as we discuss in more detail in the following.
Let us start writing the transport equation for $\tilde\mu$ in a generic foliation. In that case 
\[{\tilde\mu}=-\divv\eta + \frac{1}{2}(\chi\underline{\chi}-\overline{\chi\underline{\chi}})-(\ro-\overline{\ro})\ .\]
Therefore
\bea
&&\ML\ddb_3\divv\eta=\divv\ddb_3(-\etab+2\nabb\log\oom)+[\ddb_3,\divv]\eta\nn\\
&&=\divv\left[\chib\c\etab-\chib\c\eta-\bb\right]+2\divv\ddb_3\nabb\log\oom+[\ddb_3,\divv]\eta\nn\\
&&=\divv\left[\chib\c\etab-\chib\c\eta-\bb\right]+2\lapp\ddb_3\log\oom+2\divv[\ddb_3,\nabb]\log\oom+[\ddb_3,\divv]\eta\nn\\
&&=-\divv\bb+\divv\left[\chib\c\etab-\chib\c\eta\right]+2\lapp\ddb_3\log\oom+2\divv[\ddb_3,\nabb]\log\oom+[\ddb_3,\divv]\eta\nn\\
&&=-\divv\bb-4\lapp\omb+\left\{\divv\left[\chib\c\etab-\chib\c\eta\right]+2\divv[\ddb_3,\nabb]\log\oom+[\ddb_3,\divv]\eta\right\}\nn
\eea
which we can rewrite as 
\bea
\ddb_3\divv\eta=-\divv\bb-4\lapp\omb+[{\tilde\mu}.o.t]\ .
\eea
Recall that $\tilde\mu$ is at the level of the first derivative of the connection coefficients, that is at the ``Riemann level", moreover
\bea
\frac{1}{2}\ddb_3(\chi\underline{\chi}-\overline{\chi\underline{\chi}})=[{\tilde\mu}.o.t]
\eea
where with $[{\tilde\mu}.o.t]$ we indicate terms at the order of ${\tilde\mu}$ which, therefore do not imply any loss of derivatives in the transport equation. Finally
\bea
\ddb_3(\ro-\overline{\ro})=\ddb_3\ro+[{\tilde\mu}.o.t]=-\divv\bb+[{\tilde\mu}.o.t]
\eea
and finally 
\bea
&&\ddb_3{\tilde\mu}=-\ddb_3\divv\eta+\frac{1}{2}\ddb_3(\chi\underline{\chi}-\overline{\chi\underline{\chi}})-\ddb_3(\ro-\overline{\ro})\nn\\
&&=\divv\bb+4\lapp\omb+\divv\bb+[{\tilde\mu}.o.t]=4\lapp\omb+2\divv\bb+[{\tilde\mu}.o.t]\nn\\
&&=-2\lapp\ddb_3\log\oom+2\divv\bb+[{\tilde\mu}.o.t]=-2\ddb_3\lapp\log\oom+2\divv\bb+[{\tilde\mu}.o.t]\ .\nn
\eea
Therefore to avoid the loss of derivatives in the transport equation for ${\tilde\mu}|_{\Cb_*}$ we have to choose $\oom|_{\Cb_*}$ in such a way that $-2\ddb_3\lapp\log\oom+2\divv\bb$ is of the same order as ${\tilde\mu}$. This still leaves many possibilities for the choice of $\oom$, but, as we said, once we choose $\oom|_{\Cb_*}$ and we define the outgoing cones of the foliation we need to transport all the non underlined quantities starting from above, therefore we do it also for $\tr\chi$ and its angular derivatives or for $\Us$ even if in principle they have the right decay even if integrating them from below. This imposes a choice of $\oom|_{\Cb_*}$ such that the transport equation of $\tr\chi$ and its derivatives along $\Cb_*$ does not have any loss of derivatives and this defines $\oom|_{\Cb_*}$ as done in \cite{Kl-Ni:book}, a choice which simultaneously control also the evolution equation for ${\tilde\mu}|_{\Cb_*}$. Let us see it in a constructive way: the equation along the incoming cones for $\tr\chi$ is
\bea
\ML\dddd_3\tr\chi\!+\!\tr\chib\tr\chi\!-\!2\omb\tr\chi\!-\!2\divv\eta\!-\!2|\eta|^2\!+\!2{\bf K}=0\ .
\eea
which we rewrite using the relation
\bea
{\bf K}+\frac{1}{4}\tr\chi\tr\chib=\frac{1}{2}\chih\c\chibh-\ro\ ,
\eea
as
\bea
\ML\dddd_3\tr\chi\!+\!\frac{1}{2}\tr\chib\tr\chi\!-\!2\omb\tr\chi\!-\!2|\eta|^2\!-\!2\divv\eta\!+\chih\c\chibh-2\ro=0\ .
\eea
Recalling that
\bea
2\tilde{\mu}:=-2\divv\eta +(\chih{\chibh}-\overline{\chih{\chibh}})-2(\ro-\overline{\ro})
+\frac{1}{2}(\tr\chi\tr\chib-\overline{\tr\chi\tr\chib})
\eea
we rewrite the transport equation for $\tr\chi$ as
\bea
&&\ML\ML\dddd_3\tr\chi\!+\!\frac{1}{2}\tr\chib\tr\chi\!-\!2\omb\tr\chi\!-\!2|\eta|^2\!+\!\chih\chibh-2\ro+2\tilde{\mu}-(\chih{\chibh}-\overline{\chih{\chibh}})
+2(\ro-\overline{\ro})-\frac{1}{2}(\tr\chi\tr\chib-\overline{\tr\chi\tr\chib})\nn\\
&&\ML\ML=\dddd_3\tr\chi+\frac{1}{2}\overline{\tr\chi\tr\chib}-2\omb\tr\chi-2|\eta|^2+\overline{\chih{\chibh}}-2\overline{\ro}+2\tilde{\mu}=0\ .
\eea
Therefore the equation which $\tr\chi$ satisfies, in a general foliation, is
\bea
\dddd_3\tr\chi+\frac{1}{2}\overline{\tr\chi\tr\chib}-2\omb\tr\chi-2|\eta|^2+\overline{\chih{\chibh}}-2\overline{\ro}+2\tilde{\mu}=0\ ,
\eea
and if we want that there is no loss of derivatives we need that the order of $\tilde{\mu}$ be that of a connection coefficient. Therefore looking at the expression for $\tilde\mu$,
\beaa
{\tilde\mu}=-\divv\eta + \frac{1}{2}(\chi\underline{\chi}-\overline{\chi\underline{\chi}})-(\ro-\overline{\ro})\ ,
\eeaa
we are forced to impose that
\bea
-\divv\eta-\ro=0+O(\overline{\ro}+[conn.coe\!f.order])\eql{cancond1}
\eea
implying for $\tr\chi$ the transport equation
\bea
&&\ML\dddd_3\tr\chi+\frac{1}{2}\overline{\tr\chi\tr\chib}-2\omb\tr\chi-2|\eta|^2+\overline{\chih{\chibh}}-2\overline{\ro}+(\chi\underline{\chi}-\overline{\chi\underline{\chi}})+O(\overline{\ro}+[conn.coe\!f.order])\nn\\
&&\ML=\dddd_3\tr\chi+\frac{1}{2}{\tr\chib\tr\chi}-2\omb\tr\chi-2|\eta|^2+{\chih\chibh}+O(\overline{\ro}+[conn.coe\!f.order])=0\ .
\eea
If we assume in \ref{cancond1}, $O(\overline{\ro}+[conn.coe\!f.order])=0$ then the equation satisfied on $\Cb_*$ by $\tr\chi$ is
\bea
\dddd_3\tr\chi+\frac{1}{2}{\tr\chib\tr\chi}-2\omb\tr\chi-2|\eta|^2+{\chih\chibh}=0\eql{ee1}
\eea
and $\oom$ has to satisfy
\bea
\lapp\oom=-\divv\ze-\ro \ .\eql{cancond2}
\eea
If we assume in \ref{cancond1}, $O(\overline{\ro}+[conn.coe\!f.order])=-\overline{\ro}$ then the equation satisfied on $\Cb_*$ by $\tr\chi$ is
\bea
\dddd_3\tr\chi+\frac{1}{2}{\tr\chib\tr\chi}-2\omb\tr\chi-2|\eta|^2+{\chih\chibh}-2\overline{\ro}=0 \eql{ee2}
\eea
and $\oom$ has to satisfy
\bea
\lapp\oom=-\divv\ze-(\ro-\overline{\ro}) \ .\eql{cancond3}
\eea
{\bf Remarks:} {\em 

\NI a) Observe that here what we are really interested are the estimates of  inductive estimates of the angular derivatives, the estimates for the first 3 derivaties are already done in \cite{Kl-Ni:book}. For the higher derivative estimates of $\tr\chi$ we use equations \ref{ee1} or \ref{ee2} and for the angular derivatives of $\eta$ and of $\chih$ and $\chibh$ in \ref{ee2} we use the Hodge systems for $\chih$ and $\chibh$ and for $\eta$ the Hodge system valid on $\Cb_*$ \ref{etahodgea} 
\beaa
&&\divv\eta =-\tilde{\mu}|_{\Cb_*}+\frac{1}{2}(\chi\underline{\chi}-\overline{\chi\underline{\chi}})-(\ro-\overline{\ro})\\
&&\curll\eta=\sigma-\frac{1}{2}\hat{\underline{\chi}}\wedge\hat{\chi}\ .
\eeaa

\NI b) From the previous considerations it seems that there is still a certain freedom in specifying the $\oom|_{\Cb_*}$ which defines the canonical foliation. In fact in principle it seems that up to now requiring that $\oom|_{\Cb_*}$ satisfies the elliptic equations \ref{cancond1} or \ref{cancond2} or even \ref{cancond3} is at this level completely equivalent. A more stringent condition can be required once we look at the asymptotic behaviours of the Riemann tensor components. }


\newpage
\section{Appendix to Section \ref{S.10b}}\label{AS.4}

\subsection{The definition and the control of the ${\cal Q}$ norms}
We discuss first of all which are the norms we need to control for $J>3$; let us recall, first of all, their definitions given in Section \ref{S.4.1}:

\NI We define, for $J\geq 1$,\footnote{For the definitions of the vector fields $O,K,T$ and the modified Lie derivatives $\lie_{\c}$ see \cite{Kl-Ni:book} Chapter 3.}
\bea
&&\QQ^{(J-1)}(\la,\nu)=\QQ^{(J-1)}_1(\la,\nu)+\QQ^{(J-1)}_2(\la,\nu)\nn\\
&&\QQb^{(J-1)}(\la,\nu)=\QQb^{(J-1)}_1(\la,\nu)+\QQb^{(J-1)}_2(\la,\nu)
\eea
{\bf Remark:}\

\NI {\em Recall that $Q(\lie_{O}W)=\sum_{i=1}^3Q(\lie_{O_{(i)}}W)$ .} 

\NI Therefore with our definitions
\bea
\ML\ML\QQ^{(J-1)}_1(\la,\nu)&\equiv&\sum_{i_1,1_2,...,i_{J-1}}^{\{1,3\}}\int_{C(\la)\cap V(\la,\nu)}Q(\lie_{T}{(\lie_{O_{(i_1)}}\lie_{O_{(i_2)}}\c\c\lie_{O_{(i_{J-1})}}W)})(\acc,\acc,\acc,e_4)\nn\\
&&\ML\ML\ +\sum_{i_0,i_1,1_2,...,i_{J-1}}^{\{1,3\}}\int_{C(\la)\cap V(\la,\nu)}Q(\lie_{O_{(i_0)}}(\lie_{O_{(i_1)}}\lie_{O_{(i_2)}}\c\c\lie_{O_{(i_{J-1})}}W))(\acc,\acc,T,e_4)\ .\nn
\eea
We write in a compact way, but recalling the previous remark,
\bea
\QQ^{(J-1)}_1(\la,\nu)&\equiv&\int_{C(\la)\cap V(\la,\nu)}Q(\lie_{T}{(\lie_{O}^{J-1}W)})(\acc,\acc,\acc,e_4)\nn\\
&&+\int_{C(\la)\cap V(\la,\nu)}Q(\lie_{O}{(\lie_{O}^{J-1}W}))(\acc,\acc,T,e_4)\nn\\
\QQ^{({J-1})}_2(\la,\nu)&\equiv&
\int_{C(\la)\cap V(\la,\nu)}Q(\lie_{O}\lie_{T}{(\lie_{O}^{J-1}W}))(\acc,\acc,\acc,e_4)\nn\\
&&+\int_{C(\la)\cap V(\la,\nu)}Q(\lie^2_{O}{(\lie_{O}^{J-1}W}))(\acc,\acc,T,e_4)\\
&&+\int_{C(\la)\cap V(\la,\nu)}Q(\lie_{S}\lie_{T}{(\lie_{O}^{J-1}W}))(\acc,\acc,\acc,e_4)\nn
\eea 
\bea
\QQb^{(J-1)}_1(\la,\nu)&\equiv&\sup_{V(\la,\nu)\cap C_0}|r^3{\overline\ro(\lie_{O}^{J-1}W)}|^2+
\int_{\Cb(\nu)\cap V(\la,\nu)}Q(\lie_{T}{(\lie_{O}^{J-1}W}))(\acc,\acc,\acc,e_3)\nn\\
&&+\int_{\Cb(\nu)\cap V(\la,\nu)}Q(\lie_{O}{(\lie_{O}^{J-1}W}))(\acc,\acc,T,e_3)\nn\\
\QQb^{(J-1)}_2(\la,\nu)&\equiv&\int_{\Cb(\nu)\cap V(\la,\nu)}Q(\lie_{O}\lie_{T}{(\lie_{O}^{J-1}W}))(\acc,\acc,\acc,e_3)\nn\\
&&+\int_{\Cb(\nu)\cap V(\la,\nu)}Q(\lie^2_{O}{(\lie_{O}^{J-1}W}))(\acc,\acc,T,e_3)\\
&&+\int_{\Cb(\nu)\cap V(\la,\nu)}Q(\lie_{S}\lie_{T}{(\lie_{O}^{J-1}W}))(\acc,\acc,\acc,e_3)\nn
\eql{QQb12}
\eea
where $V(\la,\nu)=J^{(+)}(S(\la_0,\nu_0)\cap J^{(-)}(S(\la,\nu))$.
\bigskip

\bigskip

\NI{\bf Proof of lemma \ref{Psi(LieO)boundedfromQ0}:}
Let us consider a generic value $H\leq J-1$, with ${\tilde W}^{(H)}=\lie_O^HW$ we have, 
\bea
&&\ML\ML{Q}({\tilde W}^{(H)})(\acc,\acc,T,e_4)
=\frac{1}{4}\tau_+^4\bigg(|\a({\tilde W}^{(H)})|^2\bigg)+\c\c\c\c
\eea
and the analogous expressions for ${Q}({\tilde W}^{(H)})(\acc,\acc,\acc,e_4)$ , ${Q}({\tilde W}^{(H)})(\acc,\acc,T,e_3)$, 
 ${Q}({\tilde W}^{(H)})(\acc,\acc,T,e_4)\ , ....$\ . 
 Recall that 
 \bea
&&\ML |rF|_{4,S(\la,\nu)}\leq |rF|_{4,S(\la,\nu_0)}+c\left(+\int_{C(\la)\cap V(\la,\nu)}\left[|F|^2+|r\nabb F|^2+|r\ddb_4F|^2\right]\right)^{\frac{1}{2}}\\
&&\ML |r^2\nabb F|_{4,S(\la,\nu)}\leq|r^2\nabb F|_{4,S(\la,\nu_0)}+c\left(+\int_{C(\la)\cap V(\la,\nu)}\left[|r\nabb F|^2+|r^2\nabb^2 F|^2+|r^2\nabb\ddb_4F|^2\right]\right)^{\frac{1}{2}}\ .\nn
 \eea
Therefore
\bea
&&\ML |r^{\frac{9}{2}-\frac{2}{4}}\nabb \a|_{4,S(\la,\nu)}\leq |r^{\frac{9}{2}-\frac{2}{4}}\nabb \a|_{4,S(\la,\nu_0)}+c\left(+\int_{C(\la)\cap V(\la,\nu)}\left[|r^3\nabb\a|^2+|r^4\nabb^2\a|^2+|r^4\nabb\ddb_4\a|^2\right]\right)^{\frac{1}{2}}\nn
\eea
 and, omitting the constants in front of the integrals,  using Lemma 5.1.1 of \cite{Kl-Ni:book} and omitting correction terms,\footnote{Arising from passing from $\Lie_O$ to $\lie_O$ and from the commutation terms arising interchanging $\Lie_O$ and $r\nabb$.}
 \bea
 &&\ML\ML\int_{C(\la)\cap V(\la,\nu)}|r^3\nabb\a|^2=\int_{\nu_0}^{\nu}d\nu'\int_{S(\la,\nu')}r^6|\nabb\a|^2=
 \int_{\nu_0}^{\nu}d\nu'r^4\int_{S(\la,\nu')}|r\nabb\a|^2\nn\\
 &&\ML\ML \leq\int_{\nu_0}^{\nu}d\nu' r^4\int_{S(\la,\nu')}|\lie_O\a|^2\ ,\nn\\
 &&\ML\ML\int_{C(\la)\cap V(\la,\nu)}|r^4\nabb^2\a|^2=\int_{\nu_0}^{\nu}d\nu'\int_{S(\la,\nu')}r^8|\nabb^2\a|^2
 =\int_{\nu_0}^{\nu}d\nu'r^4\int_{S(\la,\nu')}|r^2\nabb^2\a|^2\nn\\
 &&\ML\ML \leq\int_{\nu_0}^{\nu}d\nu' r^4\int_{S(\la,\nu')}|\lie^2_O\a|^2\ .\nn
 \eea
 In an analogous way  we can show that
  \bea
 &&\ML\ML\int_{C(\la)\cap V(\la,\nu)}|r^4\nabb\ddb_4\a|^2\leq 
 \int_{\nu_0}^{\nu}d\nu' r^6\int_{S(\la,\nu')}|\lie_O\lie_T\a|^2\ .\nn
 \eea
 Therefore
 \bea
 &&\ML |r^{\frac{9}{2}-\frac{2}{p}}\nabb \a|_{p=4,S}(\la,\nu)\leq |r^{\frac{9}{2}-\frac{2}{p}}\nabb \a|_{p=4,S}(\la,\nu_0)\eql{16.3aas}\\
 &&\ML\ML+\int_{\nu_0}^{\nu}d\nu' r^4\int_{S(\la,\nu')}|\lie_O\a|^2+\int_{\nu_0}^{\nu}d\nu'r^4\int_{S(\la,\nu')}|\lie_O^2\a|^2+ \int_{\nu_0}^{\nu}d\nu' r^6\int_{S(\la,\nu')}|\lie_O\lie_T\a|^2\ .\nn
 \eea
It follows, 
 \bea
 &&\ML |r^{\frac{9}{2}-\frac{2}{p}}\nabb \a({\tilde W}^{(H)})|_{p=4,S(\la,\nu)}
 =\left(\int_{S(\la,\nu)}d\mu \bigg(\left|r^4\nabb\a({\tilde W}^{(H)})\right|^2\bigg)^2\right)^{\frac{1}{4}}\nn\\
&&\ML = \left(\int_{S(\la,\nu)}d\mu |r^4\nabb\a({\tilde W}^{(H)})|^4\right)^{\frac{1}{4}}
 \eea
The relation becomes, considering ${\tilde W}^{(H)}=\lie_O^HW$,
\bea
 &&\ML\ML\ML |r^{\frac{9}{2}-\frac{2}{p}}\nabb \a({\tilde W}^{(H)})|_{p=4,S}(\la,\nu)\leq |r^{\frac{9}{2}-\frac{2}{p}}\nabb \a({\tilde W}^{(H)})|_{p=4,S}(\la,\nu_0)+\int_{\nu_0}^{\nu}d\nu' r^4\int_{S(\la,\nu')}|\lie_O\a({\tilde W}^{(H)})|^2\nn\\
 &&\ML\ML\ML+\int_{\nu_0}^{\nu}d\nu'r^4\int_{S(\la,\nu')}|\lie_O^2\a({\tilde W}^{(H)})|^2+ \int_{\nu_0}^{\nu}d\nu' r^6\int_{S(\la,\nu')}|\lie_O\lie_T\a({\tilde W}^{(H)})|^2\eql{16.3abs}
 \eea
and, more explicitly,
\bea
&&\ML\ML\int_{\nu_0}^{\nu}d\nu' r^4\int_{S(\la,\nu')}|\lie_O\a({\tilde W}^{(H)})|^2
\leq \int_{\nu_0}^{\nu}d\nu' r^4|\a(\lie_O{\tilde W}^{(H)})|^2
\nn\\
&&\ML\ML
\leq \int_{\nu_0}^{\nu}d\nu' Q(\lie_O{\tilde W}^{(H)})(\acc,\acc,T,e_4)\ .\nn
\eea

Therefore in conclusion we have
\bea
 &&\ML\ML\ML |r^{\frac{9}{2}-\frac{2}{p}}\nabb \a({\tilde W}^{(H)})|_{p=4,S}(\la,\nu)\leq |r^{\frac{9}{2}-\frac{2}{p}}\nabb \a({\tilde W}^{(H)})|_{p=4,S}(\la,\nu_0)\nn\\
 &&\ML\ML\ML+\int_{\nu_0}^{\nu}d\nu' Q(\lie_O{\tilde W}^{(H)})(\acc,\acc,T,e_4)+\int_{\nu_0}^{\nu}d\nu' Q(\lie_O^2{\tilde W}^{(H)})(\acc,\acc,T,e_4)\nn\\
 &&\ML\ML\ML+\int_{\nu_0}^{\nu}d\nu' Q(\lie_O\lie_T{\tilde W}^{(H)})(\acc,\acc,\acc,e_4)\ .
\eea
 Therefore if we control ${\cal Q}_{1,2}^{(J-2)}$ we control the norm $|r^{\frac{9}{2}-\frac{2}{p}}\nabb \a({\tilde W}^{(J-2)})|_{p,S}(\la,\nu)$ and with a slight modification 
 it follows that we control $\big|r^{\frac{7}{2}-\frac{2}{p}+(J-1)}\lie_O^{J-1}\a(W)\big|_{p,S}$.\footnote{
 On the other side if we consider only ${\cal Q}_{1}^{(J-1)}$ we can also control directly $|r^{\frac{9}{2}-\frac{2}{p}}\a({\tilde W}^{(J-1)})|_{p,S}(\la,\nu)$.}

\subsection{Proof of Lemma \ref{L4.1}}

\NI To estimate $\nabb_O^{J-1}\Psi$ we proceed in the following way, we write symbolically,
\[\nabb_O^{J-1}\Psi=O\nabb O\nabb O\c\c\c\c O\nabb\Psi\ .\]
We consider the $O$'s as endowed with a given name, numbered, each one, with the exception of the first, defining a ``slot" namely the position at its left, therefore there are $J-2$ slots associated to the $(J\!-\!2)$ $O$'s we are considering; we add an extra slot which corresponds to the position immediately at the left of $\Psi$. Therefore we have $(J\!-\!1)$ slots and we imagine to distribute in all the possible ways the $(J\!-\!1)$ $\nabb$ in these slots. Clearly in this way we are counting more than the possible terms we obtain writing explicitely $\nabb_O^{J-1}\Psi$ as each $\nabb$  cannot operate on the $O$'s at its left. Assuming the described distribution of the $\nabb$'s we have (the equality is formal and it has to be interpreted as an upper bounds when we consider the norms),
\bea
\ML\ML\nabb_O^{J-1}\Psi``=" O\left\{\sum_{\ga_1,\ga_2,...,\ga_{J-2}\ga_{J-1}}^{\sum_{s=1}^{J-1}\!\!\ga_s=(J-1);\ga_{J-1}\geq1}
\frac{(J\!-\!1)!}{\ga_1!\ga_2!\c\c\c\ga_{J-2}!\ga_{J-1}!}\nabb^{\ga_1}O\nabb^{\ga_2}O\c\c\c\nabb^{\ga_{J-2}}O\nabb^{\ga_{J-1}}\Psi\right\}\ .\ \ \ 
\eea
We denote $\ga_{J-1}=q$ and rewrite the previous expression as
\bea
\ML\ML\nabb_O^{J-1}\Psi``=" O\left\{\sum_{q=1}^{J-1}\cbin{J-1}{q}\nabb^{q}\Psi\!\!\!\!\!\!\!\!\sum_{\ga_1,\ga_2,...,\ga_{J-2}}^{\ \ \ \ \ \sum_{s=1}^{J-2}\!\!\ga_s=(J-1-q)}
\!\!\!\!\!\!\!\!\frac{(J\!-\!1\!-\!q)!}{\ga_1!\ga_2!\c\c\c\ga_{J-2}!}\nabb^{\ga_1}O\nabb^{\ga_2}O\c\c\c\nabb^{\ga_{J-2}}O\right\}\ .\ \ \ 
\eea
Denoting $k=J-1-q$ we rewrite the previous expression as
\bea
&&\ML\ML\nabb_O^{J-1}\Psi``=" O\left\{\sum_{k=0}^{J-2}\cbin{J-1}{k}\nabb^{J-1-k}\Psi\!\!\!\!\!\!\!\!\sum_{\ga_1,\ga_2,...,\ga_{J-2}}^{\ \ \ \ \ \sum_{s=1}^{J-2}\!\!\ga_s=k}
\!\!\!\!\frac{k!}{\ga_1!\ga_2!\c\c\c\ga_{J-2}!}\nabb^{\ga_1}O\nabb^{\ga_2}O\c\c\c\nabb^{\ga_{J-2}}O\right\}\nn\\
&&\ML\ML=O^{J-1}\nabb^{J-1}\Psi+O\left\{\sum_{k=1}^{J-2}\cbin{J-1}{k}\nabb^{J-1-k}\Psi\!\!\!\!\!\!\!\!\sum_{\ga_1,\ga_2,...,\ga_{J-2}}^{\ \ \ \ \ \sum_{s=1}^{J-2}\!\!\ga_s=k}
\!\!\!\!\frac{k!}{\ga_1!\ga_2!\c\c\c\ga_{J-2}!}\nabb^{\ga_1}O\nabb^{\ga_2}O\c\c\c\nabb^{\ga_{J-2}}O\right\}\ .\nn
\eea
Going to the norms the estimate becomes
\bea
&&\ML\ML|\nabb_O^{J-1}\Psi(R)|_{p,S}\leq |O|^{J-1}_{\infty}|\nabb^{J-1}\Psi(R)|_{p,S}\\
&&\ML\ML+|O|_{\infty}\!\left\{\sum_{k=1}^{J-2}\cbin{J-1}{k}|\nabb^{J-1-k}\Psi(R)|_{p,S}\!\!\!\!\!\!\!\!\sum_{\ga_1,\ga_2,...,\ga_{J-2}}^{\ \ \ \ \ \sum_{s=1}^{J-2}\!\!\ga_s=k}
\!\!\!\!\frac{k!}{\ga_1!\ga_2!\c\c\c\ga_{J-2}!}\bigg[|\nabb^{\ga_1}O|_{\infty,S}|\nabb^{\ga_2}O|_{\infty,S}\c\c\c|\nabb^{\ga_{J-2}}O|_{\infty}\bigg]\right\}\ .\nn
\eea
Recalling  theorem \ref{angulconcoef2} we have the estimates
\bea\label{equo}
&&\ML|\nabb^{\ga}O|_{\infty,S}\leq C_{0,0}\frac{(\ga+1)!}{(\ga+1)^\a}\frac{e^{(\ga-1)\de_0}e^{(\ga-1)\underline{\Ga}_0(\la)}}{\ro_0^{\ga+1}}\ , \ga>0\ ;\ |O|_{\infty,S}\leq c\ \ \ \ .\nn
\eea 


\NI From these estimates it follows that 
\bea
&&\de(\sum_{s=1}^{J-2}\!\!\ga_s=k)\bigg[|\nabb^{\ga_1}O|_{\infty,S}|\nabb^{\ga_2}O|_{\infty,S}\c\c\c|\nabb^{\ga_{J-2}}O|_{\infty}\bigg]\\
&&\leq
|O|^{\left(J-1-k\right)}_{\infty,S}\c\frac{e^{k(\de_0+\underline{\Ga}_0(\la))}}{\ro_0^k}\frac{(\ga_1+1)!(\ga_2+2)!\c\c\c(\ga_{J-2}+1)!e^{-(\de_0+\underline{\Ga}_0(\la))(\ep{(\ga_1)}+\ep{(\ga_2)}+\c\c\c+\ep{(\ga_{J-2})})}}
{((1-\ep{(\ga_1)})+\ga_1^\a)((1-\ep{(\ga_2)})+\ga_2^\a)\c\c\c((1-\ep{(\ga_{J-2})})+\ga_{J-2}^\a)}\nn
\eea
and we can rewrite the previous estimate, recalling that  $ |O|_{\infty,S}\geq 1$, as
\bea
&&\ML\ML\ML|\nabb_O^{J-1}\Psi(R)|_{p,S}\leq |O|^{J-1}_{\infty,S}|\nabb^{J-1}\Psi(R)|_{p,S}
+|O|^{J-2}_{\infty,S}\!\left\{\sum_{k=1}^{J-2}\cbin{J-1}{k}|\nabb^{J-1-k}\Psi(R)|_{p,S}\frac{e^{k(\de_0+\underline{\Ga}_0(\la))}}{\ro_0^k}k!\ \c\right.\nn\\
&&\ \ \ \ \left.\c\left[\sum_{\ga_1,\ga_2,...,\ga_{J-2}}^{\ \ \ \ \ \sum_{s=1}^{J-2}\!\!\ga_s=k}
\!\!\!\!\frac{C_{0,0}^{(\ep{(\ga_1)}+\ep{(\ga_2)}+\c\c\c+\ep{(\ga_{J-2}))}}e^{-(\de_0+\underline{\Ga}_0(\la))(\ep{(\ga_1)}+\ep{(\ga_2)}+\c\c\c+\ep{(\ga_{J-2})})}}{((1-\ep{(\ga_1)})+\ga_1^{\a-1})((1-\ep{(\ga_2)})+\ga_2^{\a-1})\c\c\c((1-\ep{(\ga_{J-2})})+\ga_{J-2}^{\a-1})}\right]\right\}\nn\ .
\eea
Applying the previous estimates for the various terms we have
\bea
&&\ML\ML\ML\ML|\nabb_O^{J-1}\Psi(R)|_{p,S}\leq  |O|^{J-1}_{\infty,S}\left\{C_{0,0}\frac{J!}{J^\a}\frac{e^{(J-2)\de_0}e^{(J-2)\underline{\Ga}_0(\la)}}{\ro_0^J}\right.\eql{4.18}\\
&&\ML\ML\ML\ML\left.+\sum_{k=1}^{J-2}\cbin{J-1}{k}C_{0,0}\frac{(J-k)!}{(J-k)^\a}\frac{e^{((J-k)-2)\de_0}e^{((J-k)-2)\underline{\Ga}_0(\la)}}{\ro_0^{J-k}}\frac{e^{k(\de_0+\underline{\Ga}_0(\la))}}{\ro_0^k}k!\big|\left[{\cal F}\right]\big|\right\}\nn
\eea
where
\bea
\ML\ML\left[{\cal F}\right]\equiv\left[\sum_{\ga_1,\ga_2,...,\ga_{J-2}}^{\ \ \ \ \ \sum_{s=1}^{J-2}\!\!\ga_s=k}
\!\!\!\!\frac{C_{0,0}^{(\ep{(\ga_1)}+\ep{(\ga_2)}+\c\c\c+\ep{(\ga_{J-2}))}}e^{-(\de_0+\underline{\Ga}_0(\la))(\ep{(\ga_1)}+\ep{(\ga_2)}+\c\c\c+\ep{(\ga_{J-2})})}}{((1-\ep{(\ga_1)})+\ga_1^{\a-1})((1-\ep{(\ga_2)})+\ga_2^{\a-1})\c\c\c((1-\ep{(\ga_{J-2})})+\ga_{J-2}^{\a-1})}\right]\ .\ \ \ \ 
\eea
Therefore we rewrite \ref{4.18} as
\bea
&&\ML\ML\ML\ML|\nabb_O^{J-1}\Psi(R)|_{p,S}\leq  |O|^{J-1}_{\infty,S}\left(C_{0,0}\frac{J!}{J^\a}\frac{e^{(J-2)\de_0}e^{(J-2)\underline{\Ga}_0(\la)}}{\ro_0^J}\right)\!\!\left\{1+\sum_{k=1}^{J-2}\cbin{J-1}{k}\frac{(J-k)!k!}{J!}\frac{J^{\a}}{(J-k)^{\a}}
\big|\left[{\cal F}\right]\big|\right\}\nn\eql{4.18a}\\
&&\ML\ML\leq  |O|^{J-1}_{\infty,S}\left(C_{0,0}\frac{J!}{J^\a}\frac{e^{(J-2)\de_0}e^{(J-2)\underline{\Ga}_0(\la)}}{\ro_0^J}\right)\!
\!\left\{1+\sum_{k=1}^{J-2}\frac{(J-k)}{J}\frac{J^{\a}}{(J-k)^{\a}}\big|\left[{\cal F}\right]\big|\right\}\nn\\
&&\ML\ML\leq  |O|^{J-1}_{\infty,S}\left(C_{0,0}\frac{J!}{J^\a}\frac{e^{(J-2)\de_0}e^{(J-2)\underline{\Ga}_0(\la)}}{\ro_0^J}\right)\!
\!\left\{1+\sum_{k=1}^{J-2}\frac{J^{\a-1}}{(J-k)^{\a-1}}\big|\left[{\cal F}\right]\big|\right\}\ .\eql{4.16ss}
\eea
We have to estimate $\big|\left[{\cal F}\right]\big|$, the simplest and rougher estimate is:
\bea
\big|\left[{\cal F}\right]\big|\leq C_{0,0}e^{-(\de_0+\underline{\Ga}_0(\la))}C_1^{J-2}\ ,
\eea
a better estimate is obtained writing
\bea
&&\ML\ML\big|\left[{\cal F}\right]\big|=C_{0,0}e^{-(\de_0+\underline{\Ga}_0(\la))}\!\left(\sum_{l=1}^k\cbin{J-2}{l}C_{0,0}^{l-1}e^{-(l-1)(\de_0+\underline{\Ga}_0(\la))}\!\!\left(\!\!\!\!\!\!\!\!\!\sum_{\ga_1,\ga_2,...,\ga_l}^{\ \ \ \ \ \sum_{s=1}^{l}\!\!\ga_s=k, \ga_s\geq 1}\frac{1}{\ga_1^{\a-1}\ga_2^{\a-1}\c\c\c\ga_{l}^{\a-1}}\right)\right]\nn\\
&&\ML\ML\leq C_{0,0}e^{-(\de_0+\underline{\Ga}_0(\la))}\left(1+c_1C_{0,0}e^{-(\de_0+\underline{\Ga}_0(\la))}\right)^{J-2}\leq \varepsilon c_{0,0}e^{-(\de_0+\underline{\Ga}_0(\la))}
\left(1+c_1\varepsilon c_{0,0}e^{-(\de_0+\underline{\Ga}_0(\la))}\right)^{\!J-2}.\ \ \ \ \ \ \ \ \ \ 
\eea
Finally
\bea
\big|\left[{\cal F}\right]\big|\leq \varepsilon c_{0,0}e^{-(\de_0+\underline{\Ga}_0(\la))}\left(1+c_2\varepsilon\right)^{\!J-2}
\eea
where
\bea
\!\!\!\!\!\!\!\!\!\sum_{\ga_1,\ga_2,...,\ga_l}^{\ \ \ \ \ \sum_{s=1}^{l}\!\!\ga_s=k, \ga_s\geq 1}\frac{1}{\ga_1^{\a-1}\ga_2^{\a-1}\c\c\c\ga_{l}^{\a-1}}\leq c_1^l\ \ \ ,
\ \ \ c_2=c_1c_{0,0}e^{-(\de_0+\underline{\Ga}_0(\la))}\ .
\eea
Therefore
\bea
\bigg|\!\left\{1+\sum_{k=1}^{J-2}\frac{J^{\a-1}}{(J-k)^{\a-1}}\big|\left[{\cal F}\right]\big|\right\}\bigg|
\leq \left(1+J^{\a-1}\varepsilon c_{0,0}e^{-(\de_0+\underline{\Ga}_0(\la))}\left(1+c_2\varepsilon\right)^{\!J-2}\right)\ \ \ \ \ 
\eea
and finally
\bea
&&\ML\ML\ML\ML|\nabb_O^{J-1}\Psi(R)|_{p,S}\bigg|_{C_0\cup\Cb_0}\leq  \left(C_{0,0}\frac{J!}{J^\a}\frac{e^{(J-2)\de_0}e^{(J-2)\underline{\Ga}_0(\la)}}{\ro_0^J}\right)\!\left[\!|O|^{J-1}_{\infty,S}\!\left(1+J^{\a-1}\varepsilon c_{0,0}e^{-(\de_0+\underline{\Ga}_0(\la))}\left(1+c_2\varepsilon\right)^{\!J-2}\right)\right]\nn\\
&&\ \ \ \leq |O|^{J-1}_{\infty}\!\left(C_{0,0}\frac{J!}{J^\a}\frac{e^{(J-2)\de_0}e^{(J-2)\underline{\Ga}_0(\la)}}{\ro_{0,1}^J}\right)
\eea
where $\ro_{0,1}$ is chosen in such a way that:
\bea
\left(\frac{\ro_{0,1}}{\ro_0}\right)^{\!\!J}\!\!\left(\!1+J^{\a-1}\varepsilon c_{0,0}e^{-(\de_0+\underline{\Ga}_0(\la))}
\left(1+c_2\varepsilon\right)^{\!J-2}\right)\leq 1\ \ .\eql{4.23}
\eea

\subsection{Proof of Lemma \ref{L4.2}:}\label{subxx} 

\NI Assume for simplicity $\Psi(R)$ denotes a $S$-tangent vector, like for instance the null component $\b$,  with $O$ a rotation generator vector,
\bea
\Lie_O\Psi(\c)=\nabb_O\Psi(\c)-\Psi_b\nabb_{\c}O^b\equiv \nabb_O\Psi(\c)-\Psi_bc^b(\c)
\eea
where
\[c^b(\c)=\ \nabb_{\c}O^b.\]
Recalling theorem \ref{angulconcoef2}, the following estimates for the initial data hold
\bea
|r^{(J+1-\frac{2}{p})}\nabb^{J-1} \ ^{(i)}O|_{\infty,S}\leq c\ \!\!\left(\frac{J!}{J^{\a}}\frac{e^{(J-2)(\de_0+\underline{\Ga}_0(\la))}}{\ro_{0}^{J}}\right)\ ,
\eea

\bea\label{ccrto}
\Lie_{O}\Psi(\c)=\nabb_O\Psi(\c)-(\Psi\c c)(\c)\ ,
\eea
we write symbolically as
\bea\label{1685}
\Lie_{O}\Psi=(\nabb_O-c)\Psi
\eea
and 
\bea
\Lie^{J-1}_{O}\Psi=(\nabb_O-c)^{J-1}\Psi=\sum_{k=0}^{J-1}\cbin{J-1}{k}\nabb_O^kc^{J-1-k}\Psi
\eea
where in the right hand side we have to be careful at the position of the $\nabb_O$'s as they can operate on the $c$'s or on the $\Psi$. Therefore fixed $k$ one has to interpret the right hand side as made by $J-k$ slots, each one on the left of one $c$ and the last slot on the left of the $\Psi$. In these $J-k$ slots one has to distribute $k$ $\nabb_O$'s, each one operating exclusively on the tensor $c$ at its immediate right, or on $\Psi$ for the last slot. Computing all the possible ways of distributing these $\nabb_O$'s one is considering more terms of the real ones, but this over estimate should not be harmful. Therefore, omitting the minus signs as at the end we are doing an upper bound norm estimate, we write the following expression\footnote{To avoid too cumbersome notations we avoided the weight factors in the norm estimates; following the previous considerations ti will be easy to reinsert them.}

\bea
\ML\ML \Psi(\Lie_O^{J-1}(\c)) ``=" \sum_{k=0}^{J-1}\cbin{J-1}{k}\sum_{\ga_1,\ga_2,...,\ga_{J-k}}^{\sum_{s=1}^{J-k}\ga_s=k}\frac{k!}{\ga_1!,\ga_2!,...,\ga_{J-k}!}
(\nabb_O^{\ga_1} \hat{c})(\nabb_O^{\ga_2}\hat{c})\c\c\c(\nabb_O^{\ga_{J-1-k}}\hat{c})(\nabb_O^{\ga_{J-k}}\Psi)\nn
\eea
and the norm estimate is
\bea
&&\ML\ML\ML\ML\leq \frac{J!}{J^{\a}}\left[\sum_{k=0}^{{J-1}}\frac{J^{\a-1}}{(J-1-k)!}\sum_{\ga_1,\ga_2,...,\ga_{J-k}}^{\sum_{s=1}^{J-k}\ga_s=k}\frac{1}{\ga_1!,\ga_2!,...,\ga_{J-k}!}
\bigg|(\nabb_O^{\ga_1} \hat{c})(\nabb_O^{\ga_2}\hat{c})\c\c\c(\nabb_O^{\ga_{J-1-k}}\hat{c})(\nabb_O^{\ga_{J-k}}{\Psi})\bigg|_{p,S}
\right]\nn
\eea
\NI In order to estimate $\nabb_O c$ we apply lemma  \ref{L4.1} to $c$ and recalling theorem  \ref{angulconcoef2}, it follows that with a suitable $c$
\bea\label{1687}
|\nabb_O^{\ga}c|_{\infty,S}\leq  \left(\frac{(\ga+1)!}{(\ga+1)^{\a}}\frac{e^{((\ga+1)-2)(\de_0+\underline{\Ga}_0(\la))}}{\ro_{0,1}^{\ga+1}}\right)
\eea
Now notice that at least one among the $\ga_i $ is greater than $\frac{k}{J-k}$ hence, without lost of generality, we can suppose  $\ga_{J-k}\geq\frac{k}{J-k}$
\bea
&&\ML\ML\ML\ML\ML\ \ \  \frac{J!}{J^{\a}}\left[\sum_{k=0}^{J-1}\frac{J^{\a-1}}{(J-1-k)!}\sum_{\ga_1,\ga_2,...,\ga_{J-k}}^{\sum_{s=1}^{J-k}\ga_s=k}\frac{1}{\ga_1!,\ga_2!,...,\ga_{J-k}!}
\bigg|(\nabb_O^{\ga_1} \hat{c})(\nabb_O^{\ga_2}\hat{c})\c\c\c(\nabb_O^{\ga_{J-1-k}}\hat{c})(\nabb_O^{\ga_{J-k}}{\Psi})\bigg|_{p,S}
\right]\nn\\
&&\ML\ML\ML\ML\ML\ML \leq \frac{J!}{J^{\a}}\left[\sum_{k=0}^{J-1}\frac{J^{\a-1}}{(J-1-k)!}\sum_{\ga_1,\ga_2,...,\ga_{J-k}}^{\sum_{s=1}^{J-k}\ga_s=k}\frac{1}{\ga_1!,\ga_2!,...,\ga_{J-k}!}
\big|(\nabb_O^{\ga_1}\hat{c})(\nabb_O^{\ga_2}\hat{c})\c\c\c(\nabb_O^{\ga_{J-1-k}}\hat{c})\big|_{\infty,S}\big|(\nabb_O^{\ga_{J-k}}{\Psi})\big|_{p,S}
\right]\nn\\
&&\ML\ML\ML\ML\ML\ML\leq \frac{J!}{J^{\a}}
\left[\sum_{k=0}^{J-1}\frac{J^{\a-1}}{(J-1-k)!}\sum_{\ga_1,\ga_2,...,\ga_{J-k}}^{\sum_{s=1}^{J-k}\ga_s=k}c^{J-k}\frac{({\ga_1}+1)}{({\ga_1}+1)^{\a}}\frac{({\ga_2}+1)}{({\ga_2}+1)^{\a}}\c\c\c\frac{({\ga_{J-1-k}}+1)}{({\ga_{J-1-k}}+1)^{\a}}\frac{(\ga_{J-k}+1)}{(\ga_{J-k}+1)^{\a}}\frac{e^{(k-(J-k))(\de+\Ga)}}{\ro_{0,1}^{J}}\right]\nn\\
&&\ML\ML\ML\ML\ML\ML\leq 
 \frac{J!}{J^{\a}}\!\left(\frac{e^{(J-2)(\de+\Ga)}}{\ro_{0,1}^J}\right)\!
\left[\sum_{k=0}^{J-1}c^{J-k}\frac{J^{\a-1}}{(J-1-k)!}\cdot\right.\nn\\
&&\ML\ML\ML\ML\ML\left.\sum_{\ga_1,\ga_2,...,\ga_{J-k}}^{\sum_{s=1}^{J-k}\ga_s=k}\frac{1}{({\ga_1}+1)^{\a-1}}\frac{1}{({\ga_2}+1)^{\a-1}}\c\c\c\frac{1}{({\ga_{J-1-k}}+1)^{\a-1}}\frac{1}{(\ga_{J-k}+1)^{\a-1}}{e^{-2J)(\de+\Ga)}}\right]\nn\\
&&\ML\ML\ML\ML\ML\leq 
 \frac{J!}{J^{\a}}\!\left(\frac{e^{(J-2)(\de+\Ga)}}{\ro_{0,1}^J}\right)\!
\left[\sum_{k=0}^{J-1}\frac{J^{\a-1}(J-k)^{\a-1}}{k^{\a-1}}\frac{c^{J-k}{e^{-2(J-1)(\de+\Ga)}}}{(J-1-k)!\ro_{0,1}^{J-k-1}}\right.\nn\\
&&\left.\left(\sum_{\ga_1,\ga_2,...,\ga_{J-k}}^{\sum_{s=1}^{J-k}\ga_s=k}\frac{1}{({\ga_1}+1)^{\a-1}}\frac{1}{({\ga_2}+1)^{\a-1}}\c\c\c\frac{1}{({\ga_{J-1-k}}+1)^{\a-1}}\right)\right]\nn\\
&&\ML\ML\ML\ML\ML\leq 
 c\frac{J!}{J^{\a}}\!\left(\frac{e^{(J-2)(\de+\Ga)}}{\ro_{0,1}^J}\right)\!
\left[\sum_{k=0}^{J-1}\frac{J^{\a-1}(J-k)^{\a-1}}{k^{\a-1}}\frac{2^{J-k}c^{J-k}{e^{-2(J-1-k)(\de+\Ga)}}}{(J-1)!}\right]\nn\\
\eea
provided $\a>3$\footnote{We need  $\a>3$ due to the possible estimate of $\big|(\Lie_O^{\ga_{J-k}}{\Psi})\big|_{\infty}$ which give a term $\frac{1}{(\ga_{J-k})^{\a-2}}$} . We rearrange the terms in the following way
\bea
&&\ML\ML\ML\ML\ML\leq 
 c\frac{J!}{J^{\a}}\!\left(\frac{e^{(J-2)(\de+\Ga)}}{{\ro}_{0,1}^J}\right)\!
\left[\sum_{k=0}^{J-1}\frac{J^{\a-1}}{k^{\a-1}(J-1-k)!}\frac{(J-k)^{\a-1}c^{J-k}{e^{-2(J-1-k)(\de+\Ga)}}}{\ro_{0,1}^{J-k-1}}\right]\nn\\
&&\ML\ML\ML\ML\ML\leq 
 c'\frac{J!}{J^{\a}}\!\left(\frac{e^{(J-2)(\de+\Ga)}}{\ro_{0,1}^J}\right)\!
\left[\sum_{k=0}^{J-1}\frac{J^{\a-1}}{k^{\a-1}(J-1-k)!}\right]\nn\\
&&\ML\ML\ML\ML\ML\leq 
 C\frac{J!}{J^{\a}}\!\left(\frac{e^{(J-2)(\de+\Ga)}}{\ro_{0,1}^J}\right),\!
\eea
provided $\de$ sufficiently large.
Clearly the same lemma can be used to pass from  $\lie_O(\Psi(R))$ to $\Lie_O(\Psi(R))$ and from  $\Lie_O(\Psi(R))$ to $\Psi(\Lie_OR)$  

\NI Let us write down symbolically the formula to pass from  $\Lie_O\Psi(R)$ to $\Psi(\Lie_OR)$, see \cite{Kl-Ni:book} proposition 5.1.1 equation 5.1.4 as example,

\bea\label{1690}
\psi(\Lie_OR)=(\Lie_{O}-\hat{c})\Psi(R),
\eea

\NI where $\hat{c}$ is a combination of connection coefficients.

\NI At its turn we can write

\bea\label{1691}
\Psi(\lie_OR)=(\Lie_O-\tilde{c})\Psi(R),
\eea

\NI where $\tilde{c}$ is a combination of connection coefficients, see \cite{Kl-Ni:book}, proposition 5.1.1. equation 5.1.6 as example.

\subsection{ Theorem \ref{Qconserv1}, he Proof of the boundedness of the ${\cal Q}$ norms. The estimate of the Error }\label{Qxxx}
\smallskip

\NI The Proof of the boundedness is more complicated than the one discussed in \cite{Kl-Ni:book}, due to the presence to terms associated to all values of $J$. 
\medskip

\NI The way to prove our result mimics what done in \cite{Kl-Ni:book} and more in general each time we have a global existence proof, namely we assume the existence of the solution satisfying some specific properties in a region we consider the largest possible and, subsequently, we prove that this region can be extended. This implies, to avoid a contradiction, that the assumed ``largest region" is in fact the whole space where the problem is well defined. This argument requires, not to be empty, that one could prove, usually via a locally existence result, that a solution with the ``specific properties" does exist, at least in a small region. 

\NI Therefore we have to look carefully at the following two points:
\smallskip

i) Which are the specific properties we have to assume.

ii) How we extend the assumed ``largest possible region".
\smallskip

\NI The second point is the heart of the problem, but the choice of the first is crucial to allow the second one.
\medskip

\NI {\bf The specific properties: } 
\smallskip

\NI We assume that in the ``largest possible region" ${\cal K}$:
\smallskip

\NI a) The norms of the connection coefficients with all their derivatives satisfy some bounds which are those expressed in the initial sections and we are going to recall later on, with a well definite constant in front  which we assume small, $\varepsilon$, for all  derivatives up to the $J_0$ order.
\smallskip

\NI b) The norms of the Riemann components with all their derivatives satisfy some bounds which are those expressed in the initial sections and we are going to recall later on, with a well definite constant in front which we assume small, $O(\varepsilon)$, for all  derivatives up to the $J_0$ order.

\NI c) The ``largest possible region" has a diamond shape and the last slice is $C_*\equiv C(\nu_*;[\la_0,\la_1])$. The energy-type norms $\cal Q$ defined on the  ``largest possible region" have to satisfy the following bounds, for any $J$ and any $\la\in[\la_0,\overline{\la}]$, and $\nu\in[\nu_0,\overline{\nu}]$, with $c$ sufficiently small
\bea
&&\left[\sum_{H=2}^J(\QQ^{(H-2)}(\la,[\nu_0,\nu])+\QQb^{(H-2)}([\la_0,\la],\nu)\right] \leq c\left[\sum_{H=2}^J(\QQ_{(0),2}^{(H-2)}(\la_0,[\nu_0,\nu])+\QQb_{(0),2}^{(H-2)}([\la_0,\la],\nu_0)\right]\nn .\\ \eql{asbound}
\eea

\NI {\bf Remark:} {\em
\smallskip


\NI We expect that  requirement b) is redundant as all these norm estimates, provided requirement a) is satisfied, can be derived from the boundedness of the energy-type norm estimates, requirement c).}
\medskip

\NI{\bf How to extend the region:}

\NI As said in subsection \ref{S.s4.1nn} we have to prove the following inequality in ${\cal K}$ ( here we write only the one in $\la$, the one in $\nu$ is similar ):
\beaa
Q(C(\la))=Q(C_0)+{Error}=Q(C_0)+\int_0^{\la}d\la'|F({\cal O})|Q(C(\la'))
\eeaa
If we can prove the $Error$ is sufficiently small to lower the constant $C^{(1)}$, of the inductive estimates in ${\cal K}$ for the null Riemann components, estimate \ref{labb}, we can extend the region ${\cal K}$ and obtain the global existence.
This result can be obtained estimating carefully the $Error$ of the ${\cal Q}^{(J-2)}$ norms for every $J$.

\NI The $Error$ term, we will divide it in $\EEb$ and $\EEbb$, we have to estimate have the following structure, see \cite{Kl-Ni:book},
\bea
\EEb(u,\ub)&=&\int_{V_{(u,\ub)}}DivQ(\lie_{T}{\tilde W})_{\b\ga\de}(\acc^{\b}\acc^{\ga}\acc^{\de})\nn\\
&&+\int_{V_{(u,\ub)}}DivQ(\lie_{O}{\tilde W})_{\b\ga\de}(\acc^{\b}\acc^{\ga}T^{\de})\nn\\
&&+\frac{3}{2}\int_{V_{(u,\ub)}}Q(\lie_{T}{\tilde W})_{\a\b\ga\de}
(^{(\acc)}\pi^{\a\b}\acc^{\ga}\acc^{\de})\eql{6.0.6ffg}\\
&&+\int_{V_{(u,\ub)}}Q(\lie_{O}{\tilde W})_{\a\b\ga\de}(^{(\acc)}\pi^{\a\b}\acc^{\ga}T^{\de})\nn\\
&&+\frac{1}{2}\int_{V_{(u,\ub)}}Q(\lie_{O}{\tilde W})_{\a\b\ga\de}
(^{(T)}\pi^{\a\b}\acc^{\ga}\acc^{\de})\nn
\eea
\bea
\EEbb(u,\ub)
&=&\int_{V_{(u,\ub)}}DivQ(\lie_{O}^2 {\tilde W})_{\b\ga\de}(\acc^{\b}\acc^{\ga}T^{\de})\nn\\
&&+\int_{V_{(u,\ub)}}DivQ(\lie_{O}\lie_{T}{\tilde W})_{\b\ga\de}(\acc^{\b}\acc^{\ga}\acc^{\de})\nn\\
&&+\int_{V_{(u,\ub)}}DivQ(\lie_{S}\lie_{T}{\tilde W})_{\b\ga\de}
(\acc^{\b}\acc^{\ga}\acc^{\de})\nn\\
&&+\int_{V_{(u,\ub)}}Q(\lie_{O}^2 {\tilde W})_{\a\b\ga\de}
(^{(\acc)}\pi^{\a\b}\acc^{\ga}T^{\de})\nn\\
&&+\frac{1}{2}\int_{V_{(u,\ub)}}Q(\lie_{O}^2 {\tilde W})_{\a\b\ga\de}
(^{(T)}\pi^{\a\b}\acc^{\ga}\acc^{\de})\nn\\
&&+\frac{3}{2}\int_{V_{(u,\ub)}}Q(\lie_{O}\lie_{T} {\tilde W})_{\a\b\ga\de}
(^{(\acc)}\pi^{\a\b}\acc^{\ga}\acc^{\de})\eql{6.0.7ffg}\\
&&+\frac{3}{2}\int_{V_{(u,\ub)}}Q(\lie_{S}\lie_{T} {\tilde W})_{\a\b\ga\de}
(^{(\acc)}\pi^{\a\b}\acc^{\ga}\acc^{\de})\ ,\nn
\eea
where ${\tilde W}=\lie_O^{J-2}W$ and $W$ (but not $\tilde W$) satisfies the Bianchi identities.

\NI Looking at these expressions it follows that the terms without the divergence can be treated exactly as in \cite{Kl-Ni:book} as the presence of ${\tilde W}$ instead of $W$ does not play any role, in fact the only property which will be used is that ${\tilde W}$ is a Weyl tensor exactly as $W$. The situation is, viceversa, different for the derived terms as in the present case $D^{\mu}{\tilde W_{\mu\nu\ro\si}}$ is different from zero while $D^{\mu}W_{\mu\nu\ro\si}=0$. This implies that some more terms are present in the error. 
\smallskip

\NI Let us look at it in greater detail considering separately the parts relative to 
$\EEb(u,\ub)$,
\bea
&&\int_{V_{(u,\ub)}}DivQ(\lie_{O}{\tilde W})_{\b\ga\de}(\acc^{\b}\acc^{\ga}T^{\de})\nn\\
&&\int_{V_{(u,\ub)}}DivQ(\lie_{T}{\tilde W})_{\b\ga\de}(\acc^{\b}\acc^{\ga}\acc^{\de})\ 
\eea
and those relative to $\EEbb(u,\ub)$,
\bea
&&\int_{V_{(u,\ub)}}DivQ(\lie_{O}^2 {\tilde W})_{\b\ga\de}(\acc^{\b}\acc^{\ga}T^{\de})\nn\\
&&\int_{V_{(u,\ub)}}DivQ(\lie_{O}\lie_{T}{\tilde W})_{\b\ga\de}(\acc^{\b}\acc^{\ga}\acc^{\de})\nn\\
&&\int_{V_{(u,\ub)}}DivQ(\lie_{S}\lie_{T}{\tilde W})_{\b\ga\de}(\acc^{\b}\acc^{\ga}\acc^{\de})\ .
\eea
For simplicity we restrict to the first terms of $\EEb(u,\ub)$ and of $\EEbb(u,\ub)$ as we expect that the remaining terms can be treated in the same way,

\beaa
&&\int_{V_{(u,\ub)}}DivQ(\lie_{O}{\tilde W}^{(J-2)})_{\b\ga\de}(\acc^{\b}\acc^{\ga}T^{\de})\nn\\
&&\nn\\
&&\int_{V_{(u,\ub)}}DivQ(\lie^2_{O}{\tilde W}^{(J-2)})_{\b\ga\de}(\acc^{\b}\acc^{\ga}T^{\de})\ .
\eeaa

\NI From Proposition 7.1.1 of \cite{C-K:book} we have the following expressions
\bea
&&\ML\ML{Div}Q(\lie_{O}{\tilde W}^{(J-2)})_{\b\ga\de}=\eql{div1}\\
&&\ML\ML=(\lie_O{\tilde W}^{(J-2)})_{\b\ \de\ }^{\ \mu\ \nu}\dd^{\a}(\lie_O{\tilde W}^{(J-2)})_{\a\mu\ga\nu}+(\lie_O{\tilde W}^{(J-2)})_{\b\ \ga\ }^{\ \mu\ \nu}\dd^{\a}(\lie_O{\tilde W}^{(J-2)})_{\a\mu\de\nu}\nn\\
&&\ML\ML\ \ +{^*}(\lie_O{\tilde W}^{(J-2)})_{\b\ \de\ }^{\ \mu\ \nu}\dd^{\a}(\lie_O{^*}{\tilde W}^{(J-2)})_{\a\mu\ga\nu}+{^*}(\lie_O{\tilde W}^{(J-2)})_{\b\ \ga\ }^{\ \mu\ \nu}\dd^{\a}(\lie_O{^*}{\tilde W}^{(J-2)})_{\a\mu\de\nu}\nn
\eea
\bea
&&\ML\ML{Div}Q(\lie^2_{O}{\tilde W}^{(J-2)})_{\b\ga\de}=\eql{div2}\\
&&\ML\ML=(\lie^2_O{\tilde W}^{(J-2)})_{\b\ \de\ }^{\ \mu\ \nu}\dd^{\a}(\lie^2_O{\tilde W}^{(J-2)})_{\a\mu\ga\nu}+(\lie^2_O{\tilde W}^{(J-2)})_{\b\ \ga\ }^{\ \mu\ \nu}\dd^{\a}(\lie^2_O{\tilde W}^{(J-2)})_{\a\mu\de\nu}\nn\\
&&\ML\ML\ {^*}(\lie^2_O{\tilde W}^{(J-2)})_{\b\ \de\ }^{\ \mu\ \nu}\dd^{\a}(\lie^2_O{^*}{\tilde W}^{(J-2)})_{\a\mu\ga\nu}+{^*}(\lie^2_O{\tilde W}^{(J-2)})_{\b\ \ga\ }^{\ \mu\ \nu}\dd^{\a}(\lie^2_O{^*}{\tilde W}^{(J-2)})_{\a\mu\de\nu}\nn
\eea
We consider the term associated to the $\EEbb$ error as,\footnote{In fact it turns out that the estimate for $\EEb$ can also be done in a easier way.} once we prove the right estimate for this term, the estimate for the term associated to $\EEb$ follows. Therefore we consider the first term of \ref{div2},
\bea
&&\ML\ML{Div}Q(\lie^2_{O}{\tilde W}^{(J-2)})_{\b\ga\de}
=(\lie^2_O{\tilde W}^{(J-2)})_{\b\ \de\ }^{\ \mu\ \nu}\dd^{\a}(\lie^2_O{\tilde W}^{(J-2)})_{\a\mu\ga\nu}
\eea
and the part of the error term $\EEbb$ we consider is,
\bea
&&\ML\ML\int_{V_{(\la,\nu)}}(\dd^{\mu}\lie^2_O{\tilde W}^{(J-2)})_{\mu\ro\ga\si}(\lie^2_O{\tilde W}^{(J-2)})_{\b\ \de\ }^{\ \ro\ \si}
(\acc^{\b}\acc^{\ga}T^{\de})\ .\eql{ee2a}
\eea
Considering only this part as ``error term" we can write,
\bea
{\cal Q}_2^{(J-2)}(\la,\nu)\!={\cal Q}_{(0),2}^{(J-2)}\!\!\!\!\!&\!+\!&\!\!\!\!\!\int_{V_{(\la,\nu)}}\!\!(\dd^{\mu}\lie^2_O{\tilde W}^{(J-2)})_{\mu\ro\ga\si}(\lie^2_O{\tilde W}^{(J-2)})_{\b\ \de\ }^{\ \ro\ \si}(\acc^{\b}\acc^{\ga}T^{\de})\ \ \ \ \ \ \ \ \ \ \ \ \ \\
\!\!&\!+\!&\!\!\!\!\!\int_{V_{(\la,\nu)}}\!\!(\dd^{\mu}\lie^2_O{\tilde W}^{(J-2)})_{\mu\ro\de\si}(\lie^2_O{\tilde W}^{(J-2)})_{\b\ \ga\ }^{\ \ro\ \si}(\acc^{\b}\acc^{\ga}T^{\de})\nn
\eea
and the problem is reduced to showing that the second term and  the third one can be bounded by $\varepsilon_1({\cal Q}_1^{(J-2)}+{\cal Q}_2^{(J-2)})(\la,\nu)$ where with $\varepsilon_1$ we mean a small quantity (not a priori related to $\varepsilon$). The two parts of the error we are considering are basically equal and, therefore, we estimate only the first one. 
\smallskip

\NI We start computing $(\dd^{\mu}\lie^2_O{\tilde W}^{(J-2)})_{\mu\b\ga\de}$: From Proposition 7.1.2 in \cite{C-K:book} it follows, 
\bea
&&\ML\ML\ML(\dd^{\mu}\lie^2_O{\tilde W}^{(J-2)})_{\mu\b\ga\de}=\dd^{\mu}(\lie_O{\tilde W}^{(J-1)})_{\mu\b\ga\de}\nn\\
&&\ML\ML\ML=\Lie_O\dd^{\mu}{\tilde W}^{(J-1)}_{\mu\b\ga\de}+\left[-\frac{1}{2}\left({\!\ ^{(O)}\pih}_{\ \b}^{\ro}\dd^{\mu}({\tilde W}^{(J-1)})_{\mu\ro\ga\de}+{\!\ ^{(O)}\pih}_{\ \ga}^{\ro}\dd^{\mu}({\tilde W}^{(J-1)})_{\mu\b\ro\de}+{\!\ ^{(O)}\pih}_{\ \de}^{\ro}
\dd^{\mu}({\tilde W}^{(J-1)})_{\mu\b\ga\ro}\right)\right.\nn\\
&&\left.\ \ \ \ \ \ \ \ \ \ \ \ \ \ \ \ +\frac{1}{8}(\tr{^{(O)}}\pi)\dd^{\mu}({\tilde W}^{(J-1)})_{\mu\b\ga\de}\right]\nn\\
&&\ML\ML+\frac{1}{2}{\!\ ^{(O)}\pih}^{\mu\nu}\dd_{\nu}{\tilde W}^{(J-1)}_{\mu\b\ga\de}
+\frac{1}{2}\dd^{\a}({\!\ ^{(O)}\pih}_{\a\la}{{\tilde W}^{(J-1)\la}}_{\ \ \ \ \b\ga\de})+\frac{1}{2}\left\{(\dd_{\b}{\!\ ^{(O)}\pih}_{\mu\la}-
\dd_{\la}{\!\ ^{(O)}\pih}_{\mu\b}){{\tilde W}^{(J-1)\mu\la}}_{\ \ \ \ \ \ \ga\de}+\c\c\right\}\nn\\
&&\nn\\
&&\ML\ML\ML=\Lie_O\dd^{\mu}{\tilde W}^{(J-1)}_{\mu\b\ga\de}+\left[-\frac{1}{2}\left({\!\ ^{(O)}\pih}_{\ \b}^{\ro}\dd^{\mu}({\tilde W}^{(J-1)})_{\mu\ro\ga\de}+{\!\ ^{(O)}\pih}_{\ \ga}^{\ro}\dd^{\mu}({\tilde W}^{(J-1)})_{\mu\b\ro\de}+{\!\ ^{(O)}\pih}_{\ \de}^{\ro}
\dd^{\mu}({\tilde W}^{(J-1)})_{\mu\b\ga\ro}\right)
\right.\nn\\
&&\ML\ML\left.\ \ \ \ \ \ \ \ \ \ \ \ \ \ \ \ \ \ \ \ +\frac{1}{8}(\tr{^{(O)}}\pi)\dd^{\mu}({\tilde W}^{(J-1)})_{\mu\b\ga\de}+{\!\ ^{(O)}\pih}^{\mu\nu}\dd_{\nu}{\tilde W}^{(J-1)}_{\mu\b\ga\de}\right]\nn\\
&&\ML\ML\ML+\bigg\{{\!\ ^{(O)}p}_{\la}{{\tilde W}^{(J-1)\la}}_{\ \ \ \ \b\ga\de}
+\left({\!\ ^{(O)}q}_{\a\b\la}{{\tilde W}^{(J-1)\a\la}}_{\ \ \ \ \ \ \ga\de}+{\!\ ^{(O)}q}_{\a\ga\la}{{\tilde W}^{(J-1)\a\ \la}}_{\ \ \ \ \ \b \ \de}+{\!\ ^{(O)}q}_{\a\de\la}{{\tilde W}^{(J-1)\a\ \ \ \la}}_{\ \ \ \ \ \ \b\ga}\right)\bigg\}\eql{4.17}
\eea
where ${\!\ ^{(O)}p}_{\la}$ and the ${\!\ ^{(O)}q}_{\a\de\la}$ are defined in \cite{Kl-Ni:book}.
\medskip

\NI Observe that the first term has $(J-1)+2$ derivations, the terms in square brackets have $(J-1)+1$ derivations and the terms in the curly brackets have $(J-1)$ derivations. 
It is the first term we cannot estimate with ${\cal Q}_2^{(J-2)}$ while the second and third term can be easily estimated in terms of ${\cal Q}_2^{(J-2)}$. We have, therefore, to iterate the procedure computing again as before
\[\dd^{\mu}{\tilde W}^{(J-1)}_{\mu\b\ga\de}=(\dd^{\mu}\lie_O{\tilde W}^{(J-2)})_{\mu\b\ga\de}\ \]
and then repeating the procedure up to the moment we have 
\bea
&&\Lie_O^{J-1}\dd^{\mu}{\tilde W}^{(1)}_{\mu\b\ga\de}=\Lie_O^{J-1}(\dd^{\mu}\lie_O{\tilde W}^{(0)})_{\mu\b\ga\de}=\Lie_O^{J-1}(\dd^{\mu}\lie_O W)_{\mu\b\ga\de}\\
&&=\Lie_O^{J-1}([\dd^{\mu},\lie_O] W)_{\mu\b\ga\de}+\Lie_O^{J-1}(\lie_O \dd^{\mu}W)_{\mu\b\ga\de}=\Lie_O^{J-1}([\dd^{\mu},\lie_O] W)_{\mu\b\ga\de}\ ,\nn
\eea
 as $W={\tilde W}^{(0)}$ has divergence zero.
 Therefore we have to repeat the procedure $J$ times which implies that at every step the terms in square brackets and curly brackets reproduce themselves, obtaining
 \bea
&&\ML\ML\ML(\dd^{\mu}\lie^2_O{\tilde W}^{(J-2)})_{\mu\b\ga\de}=(\dd^{\mu}\lie_O{\tilde W}^{(J-1)})_{\mu\b\ga\de}=\nn\\
&&\ML\ML\ML=\sum_{k=0}^{J-1}\Lie_O^k\left[-\frac{1}{2}\left({\!\ ^{(O)}\pih}_{\ \b}^{\ro}\dd^{\mu}({\tilde W}^{(J-1-k)})_{\mu\ro\ga\de}+{\!\ ^{(O)}\pih}_{\ \ga}^{\ro}\dd^{\mu}({\tilde W}^{(J-1-k)})_{\mu\b\ro\de}+{\!\ ^{(O)}\pih}_{\ \de}^{\ro}
\dd^{\mu}({\tilde W}^{(J-1-k)})_{\mu\b\ga\ro}\right)
\right.\nn\\
&&\ML\ML\left.\ \ \ \ \ \ \ \ \ \ \ \ \ \ \ \ \ \ \ \ +\frac{1}{8}(\tr{^{(O)}}\pi)\dd^{\mu}({\tilde W}^{(J-1-k)})_{\mu\b\ga\de}+{\!\ ^{(O)}\pih}^{\mu\nu}\dd_{\nu}{\tilde W}^{(J-1-k)}_{\mu\b\ga\de}\right]\nn\\
&&\ML\ML\ML+\sum_{k=0}^{J-1}\Lie_O^k\bigg\{{\!\ ^{(O)}p}_{\la}{{\tilde W}^{(J-1-k)\la}}_{\ \ \ \ \b\ga\de}
+\left({\!\ ^{(O)}q}_{\a\b\la}{{\tilde W}^{(J-1-k)\a\la}}_{\ \ \ \ \ \ \ga\de}+{\!\ ^{(O)}q}_{\a\ga\la}{{\tilde W}^{(J-1-k)\a\ \la}}_{\ \ \ \ \ \b \ \de}+{\!\ ^{(O)}q}_{\a\de\la}{{\tilde W}^{(J-1-k)\a\ \ \ \la}}_{\ \ \ \ \ \ \b\ga}\right)\bigg\}\nn\\
\eea
which we write symbolically, omitting the constants, as
\bea
&&\ML\ML \sum_{k=0}^{J-2}\Lie_O^k\left({\!\ ^{(O)}\pi}_{\ \b}^{\ro}\dd^{\mu}({\tilde W}^{(J-1-k)})_{\mu\ro\ga\de}\right)+\sum_{k=0}^{J-1}\Lie_O^k\left({\!\ ^{(O)}\pih}^{\mu\nu}\dd_{\nu}{\tilde W}^{(J-1-k)}_{\mu\b\ga\de}\right)\nn\\
&&\ML\ML+\sum_{k=0}^{J-1}\Lie_O^k\left({\!\ ^{(O)}{\tilde q}}_{\a\b\la}{{\tilde W}^{(J-1-k)\a\la}}_{\ \ \ \ \ \ \ga\de}\right)\ ,\eql{testbb}
\eea
where 
\[{\tilde q}_{\a\b\la}= \de_{\a\b}{^{(O)}p}_{\la} \ \ \ \ \mbox{or}\ \ \ \ {\!\ ^{(O)}{ q}}_{\a\b\la}\ .\]
Let us estimate the first sum, the second one should be estimated in the same way, the third one will be considered later on.
Before doing this estimate let us look at what we would like to get.
The error, considering only the first sum, has the form
\bea
&&\ML\ML\int_{V_{(\la,\nu)}}(\dd^{\mu}\lie^2_O{\tilde W}^{(J-2)})_{\mu\ro\ga\si}(\lie^2_O{\tilde W}^{(J-2)})_{\b\ \de\ }^{\ \ro\ \si}
(\acc^{\b}\acc^{\ga}T^{\de})\eql{ee2aa}\\
&&\ML\ML=\sum_{k=0}^{J-2}\int_{V_{(\la,\nu)}} \left(\Lie_O^k{\!\ ^{(O)}\pi}_{\ \ro}^{\tau}\dd^{\mu}({\tilde W}^{(J-1-k)})_{\mu\tau\ga\si}\right)(\lie^2_O{\tilde W}^{(J-2)})_{\b\ \de\ }^{\ \ro\ \si}(\acc^{\b}\acc^{\ga}T^{\de})\ .\nn
\eea
\NI Therefore neglecting all the other parts of the error which we will consider later on we have to deal with the following inequality,
\bea
&&\ML\ML{\cal Q}_2^{(J-2)}(\la,\nu)+\QQb_2^{(J-2)}(\la,\nu)\!={\cal Q}_{(0),2}^{(J-2)}(\la,\nu)\nn\\
&&\ML\ML+\sum_{k=0}^{J-2}\int_{V_{(\la,\nu)}} \left(\Lie_O^k{\!\ ^{(O)}\pi}_{\ \ro}^{\tau}\dd^{\mu}({\tilde W}^{(J-1-k)})_{\mu\tau\ga\si}\right)(\lie^2_O{\tilde W}^{(J-2)})_{\b\ \de\ }^{\ \ro\ \si}(\acc^{\b}\acc^{\ga}T^{\de})\nn
\eea
where
\bea
{\cal Q}_{(0),2}^{(J-2)}(\la,\nu)= \QQ_{(0),2}^{(J-2)}(\la_0,\nu)+\QQb_{(0),2}^{(J-2)}(\la,\nu_0)\ .\eql{4.4QQa}
\eea
To estimate this part of the error we proceed as done in \cite{Kl-Ni:book}, we consider one term of the sum of terms in the integrand
\bea
\ML\ML \sum_{k=0}^{J-2}\int_{V_{(\la,\nu)}} \!\!\tau_+^4\left(\Lie_O^k{\!\ ^{(O)}\pi}_{\ \ro}^{\tau}\dd^{\mu}({\tilde W}^{(J-1-k)})_{\mu\tau4\si}\right)(\lie^2_O{\tilde W}^{(J-2)})_{4\ 4\ }^{\ \ro\ \si}\eql{testxx}
\eea
 which at its turn can be decomposed in two terms 
 \bea
&&\ML\ML\ML\ML \sum_{k=0}^{J-2}\int_{V_{(\la,\nu)}} \!\!\tau_+^4\!\left(\Lie_O^k{\!\ ^{(O)}\pi}_{\ \ro}^{\tau}\dd^{\mu}({\tilde W}^{(J-1-k)})_{\mu\tau4\si}\right)(\lie^2_O{\tilde W}^{(J-2)})_{4\ 4\ }^{\ \ro\ \si}\nn\\
&&\ML\ML\ML\ML=\sum_{k=0}^{J-2}\int_{V_{(\la,\nu)}} \!\!\tau_+^4\!\left(\Lie_O^k{\!\ ^{(O)}\pi}_{\ \ro}^{\tau}\dd({\tilde W}^{(J-1-k)})\right)\c\a(\lie^2_O{\tilde W}^{(J-2)})\nn\\
&&\ML\ML\ML\!\!\!\!+\sum_{k=0}^{J-2}\int_{V_{(\la,\nu)}} \!\!\tau_+^4\!\left(\Lie_O^k{\!\ ^{(O)}\pi}_{\ \ro}^{\tau}\dd({\tilde W}^{(J-1-k)})\right)\c\b(\lie^2_O{\tilde W}^{(J-2)})
\eea
The two terms have the same structure therefore we estimate only the first one. Let us consider a generic term of the first sum
\bea
&&\ML\ML\ML\ML \int_{V_{(\la,\nu)}} \!\!\tau_+^4\!\left(\Lie_O^k{\!\ ^{(O)}\pi}_{\ \ro}^{\tau}\dd({\tilde W}^{(J-1-k)})\right)\c\a(\lie^2_O{\tilde W}^{(J-2)})\nn\\
&&\ML\ML\ML\ML=\sum_{l=0}^k\cbin{k}{l} \int_{V_{(\la,\nu)}} \!\!\tau_+^4\!\left((\Lie_O^l{\!\ ^{(O)}\pi}_{\ \ro}^{\tau})\lie_O^{k-l}\dd({\tilde W}^{(J-1-k)})\right)\c\a(\lie^2_O{\tilde W}^{(J-2)})\nn\\
&&\ML\ML\ML\ML=\sum_{l=0}^k\cbin{k}{l} \int_{\la_0}^{\la}d\la'\int_{C(\la';[\nu_0,\nu])}\!\!\tau_+^4\!\left((\Lie_O^l{\!\ ^{(O)}\pi}_{\ \ro}^{\tau})\lie_O^{k-l}\dd({\tilde W}^{(J-1-k)})\right)\c\a(\lie^2_O{\tilde W}^{(J-2)})\nn\\
&&\ML\ML\ML\ML= \sum_{l=0}^k\cbin{k}{l}\int_{\la_0}^{\la}d\la'\left(\int_{C(\la';[\nu_0,\nu])}\!\!\tau_+^4|(\Lie_O^l{\!\ ^{(O)}\pi}_{\ \ro}^{\tau})\lie_O^{k-l}\dd({\tilde W}^{(J-1-k)})|^2\right)^{\!\!\frac{1}{2}}\!\left(\int_{C(\la';[\nu_0,\nu])}\!\!\tau_+^4|\a(\lie^2_O{\tilde W}^{(J-2)})|^2\right)^{\!\!\frac{1}{2}}\nn\\
&&\ML\ML\ML\ML\leq \left(\sup_{\la'\in[\la_0,\la]}{\cal Q}_2^{(J-2)}(\la',\nu)\right)^{\!\!\frac{1}{2}}\left[\sum_{l=0}^k\cbin{k}{l}
 \int_{\la_0}^{\la}d\la'\left(\int_{C(\la';[\nu_0,\nu])}\!\!\tau_+^4|(\Lie_O^l{\!\ ^{(O)}\pi}_{\ \ro}^{\tau})\lie_O^{k-l}\dd({\tilde W}^{(J-1-k)})|^2\right)^{\!\!\frac{1}{2}}\right]\ .\ \ \ \ \ \ \ \ \ \ \ \ \ \ \ \eql{17.31}
\eea
Therefore assuming for the moment this as the only error contribution we have
\bea
&&\ML\ML\ML{\cal Q}_2^{(J-2)}(\la,\nu)+\QQb_2^{(J-2)}(\la,\nu)\!\leq {\cal Q}_{(0),2}^{(J-2)}(\la,\nu)\nn\\
&&\nn\\
&&\ML\ML\ML+\left(\sup_{\la'\in[\la_0,\la]}{\cal Q}_2^{(J-2)}(\la',\nu)\right)^{\!\!\frac{1}{2}}\sum_{k=0}^{J-2}\left[\sum_{l=0}^k\cbin{k}{l}
 \int_{\la_0}^{\la}d\la'\left(\int_{C(\la';[\nu_0,\nu])}\!\!\tau_+^4|(\Lie_O^l{\!\ ^{(O)}\pi}_{\ \ro}^{\tau})\lie_O^{k-l}\dd({\tilde W}^{(J-1-k)})|^2\right)^{\!\!\frac{1}{2}}\right]\ .\nn
\eea
Let us state now the estimate assumptions for ${\cal Q}_2^{(H)}(\la,\nu)$ and for $\QQb_2^{(H)}(\la,\nu)$ for $H< J-2$ , which will be used later on, see eqs. \ref{10.15www},\footnote { The consistency follows recalling that $\QQ_2^{(H)}$ allows to estimate $|\Lie_OW^{(H)}|_{p,S}$.}\footnote{Notice that, as the constant $C^{(1)}$ bounds the sum of the first $H$ norms we should choose a different, smaller constant. We call it still $C^{(1)}$ for the sake of simplicity. }
\smallskip


\NI \footnote{ We have to remember that we are proving the estimate for the ${\cal Q}^{(H)}$ in an inductive way which implies that these estimates (we expect correct) have been already proved for $H<J-2$ and we prove now for $H=J-2$.}
 \bea\label{indqnorms}
&&(\QQ_2^{(H)}(\la,\nu))^{\frac{1}{2}}\leq C^{*(1)}\frac{(H+2)!}{(H+2)^{\a}}\frac{e^{H(\de+\underline{\Ga}(\la))}}{\ro_{0,1}^{(H+2)}}\nn\\
&&(\QQb_2^{(H)}(\la,\nu))^{\frac{1}{2}}\leq {C}^{*(1)}\frac{(H+2)!}{(H+2)^{\a}}\frac{e^{H(\de+\underline{\Ga}(\la))}}{\ro_{0,1}^{(H+2)}}\ .\eql{estaax}
\eea

\NI We write
\bea\label{35498}
&&\ML\ML{\cal Q}_2^{(J-2)}(\la,\nu)+\QQb_2^{(J-2)}(\la,\nu)\!\leq {\cal Q}_{(0),2}^{(J-2)}(\la,\nu)\nn\\
&&\nn\\
&&\ML\ML+\left(\sup_{\la'\in[\la_0,\la]}{\cal Q}_2^{(J-2)}(\la',\nu)\right)^{\!\!\frac{1}{2}}\left\{\sum_{k=0}^{J-2}\left[
 \int_{\la_0}^{\la}d\la'\left(\int_{C(\la';[\nu_0,\nu])}\!\!\tau_+^4|({\!\ ^{(O)}\pi}_{\ \ro}^{\tau})\lie_O^{k}\dd({\tilde W}^{(J-1-k)})|^2\right)^{\!\!\frac{1}{2}}\right]\right.\nn\\
 &&\ML\ML\left.+\sum_{k=0}^{J-2}\left[\sum_{l=1}^k\cbin{k}{l}
 \int_{\la_0}^{\la}d\la'\left(\int_{C(\la';[\nu_0,\nu])}\!\!\tau_+^4|(\Lie_O^l{\!\ ^{(O)}\pi}_{\ \ro}^{\tau})\lie_O^{k-l}\dd({\tilde W}^{(J-1-k)})|^2\right)^{\!\!\frac{1}{2}}\right]\right\}
\eea
Let us consider the first sum in the curled brackets, first of all we write
\bea
&&\ML\ML\left(\int_{C(\la';[\nu_0,\nu])}\!\!\tau_+^4|({\!\ ^{(O)}\pi}_{\ \ro}^{\tau})\lie_O^{k}\dd({\tilde W}^{(J-1-k)})|^2\right)^{\!\!\frac{1}{2}}\nn\\
&&\ML\ML\leq \left(\sup_{V(\la,\nu)}|r^{\si}{\!\ ^{(O)}\pi}|^2\right)^{\!\!\frac{1}{2}}\!\left(\int_{C(\la';[\nu_0,\nu])}\!\!\tau_+^4r^{-2\si}|\lie_O^{k}\dd\lie_O^{(J-1-k)}W|^2\right)^{\!\!\frac{1}{2}} .\ \ \ \ \ \ \ \ \ \ \ \ 
\eea
 We assume (easy to prove\footnote{ Notice, in fact, that an extra  factor $\leq O(J)$ is produced for each term when we commute $\dd$ with the $\Lie_O$, implying we have also a term to estimate multiplied by a factor $\leq O(J^2)$, but there is one $\lie_O$ less which allows to control this extra $O(J)$ factor. Therefore this terms are controlled as the more delicate one with the difference the  we can use the inductive assumption on the ${\cal Q}^{(J-2)}$.}) that we can move $\dd$ on the left paying a price of harmless lower order terms. Moreover the weight $r^{\si}$ is not specified as it depends on which component of ${\!\ ^{(O)}\pi}$ one is considering, but it will turn out that it is systematically the appropriate one.  Observe that, see for instance \cite{C-K:book} eq. (7.5.12a), we can expect the following estimate we have to investigate carefully later on,
 \bea
 &&\ML\ML\int_{C(\la';[\nu_0,\nu])}\!\!\tau_+^4r^{-2\si}|\lie_O^{k}\dd{\tilde W}^{(J-1-k)}|^2
 \leq \int_{\nu_0}^{\nu}\tau_+^4r^{-2\si}\int_{S(\la',\nu)}\frac{1}{r^2(\la',\nu')}|\lie^2_O{\tilde W}^{(J-2)}|^2\nn\\
 &&\ML\ML\leq \frac{1}{r^{2+2\si}(\la',\nu_0)}\!\int_{\nu_0}^{\nu}\tau_+^4\!\int_{S(\la',\nu)}|\lie^2_O\Psi({\tilde W}^{(J-2)})|^2
 \leq \frac{1}{r^{2+2\si}(\la',\nu_0)}\!{\cal Q}_2^{(J-2)}(\la',\nu)\nn\\
 &&\ML\ML\leq \frac{c}{|\la'|^{2+2\si}}{\cal Q}_2^{(J-2)}(\la',\nu)\ . \eql{17.33x}
 \eea
and from it, observing that
\[\left(\sup_{\la'\in[\la_0,\la]}|r^{\si}{\!\ ^{(O)}\pi}|^2\right)^{\!\!\frac{1}{2}}=O(\varepsilon)\ ,\]
\bea
&&\ML\ML\sum_{k=0}^{J-2}\left[
 \int_{\la_0}^{\la}d\la'\left(\int_{C(\la';[\nu_0,\nu])}\!\!\tau_+^4|({\!\ ^{(O)}\pi}_{\ \ro}^{\tau})\lie_O^{k}\dd({\tilde W}^{(J-1-k)})|^2\right)^{\!\!\frac{1}{2}}\right]\nn\\
 &&\ML\ML\leq  c\left(\sup_{V(\la,\nu)}|r^{\si}{\!\ ^{(O)}\pi}|^2\right)^{\!\!\frac{1}{2}}\!(J-1)\!\int_{\la_0}^{\la}d\la'
  \frac{1}{|\la'|^{1+1\si}}\!\left({\cal Q}_2^{(J-2)}(\la',\nu)\right)^{\frac{1}{2}}\\
  &&\ML\ML\leq  c\ep_0 (J-1)\!\int_{\la_0}^{\la}d\la'
  \frac{1}{|\la'|^{1+1\si}}\!\left({\cal Q}_2^{(J-2)}(\la',\nu)\right)^{\frac{1}{2}}\nn,
  \eea
  so that, we have,
  \bea\
&&\ML\ML{\cal Q}_2^{(J-2)}(\la,\nu)+\QQb_2^{(J-2)}(\la,\nu)\!\leq {\cal Q}_{(0),2}^{(J-2)}(\la,\nu)\eql{17.39xx}\\
&&\nn\\
&&\ML\ML+\left(\sup_{\la'\in[\la_0,\la]}{\cal Q}_2^{(J-2)}(\la',\nu)\right)^{\!\!\frac{1}{2}}\left\{c\ep (J-1)\!\int_{\la_0}^{\la}d\la'
  \frac{1}{|\la'|^{1+\si}}\!\left({\cal Q}_2^{(J-2)}(\la',\nu)\right)^{\frac{1}{2}}\right.\nn\\
 &&\ML\ML\left.+\sum_{k=0}^{J-2}\left[\sum_{l=1}^k\cbin{k}{l}
 \int_{\la_0}^{\la}d\la'\left(\int_{C(\la';[\nu_0,\nu])}\!\!\tau_+^4|(\Lie_O^l{\!\ ^{(O)}\pi}_{\ \ro}^{\tau})\lie_O^{k-l}\dd({\tilde W}^{(J-1-k)})|^2\right)^{\!\!\frac{1}{2}}\right]\right\}\nn\\.\eql{166555}
\eea
{\bf Remark:} {\em 

\NI  Let us forget for a moment the second sum in the curly brackets and observe that many terms of the error can be controlled via the inductive estimates, but this cannot be done for the first part as, due to the factor $\ep(J-1)$, this term cannot be written as $\QQ_{1,2}^{(J-2)}$ multiplied by a small factor. Therefore we need an internal bootstrap mechanism: we prove that the estimate \ref{estaax} for $\QQ_{1,2}^{(H=J-2)}$ is satisfied in a small region starting from the the initial data surface $C_0\cup\Cb_0$  going up in $\nu$ from $\nu_0$ to $\nu_0+\de$ and in $\la$ from $\la_0$ to $\la_0+\de$ then we assume that the largest region where this estimate holds with the constant $C^{(1)}$ be the one with $\nu\in [\nu_0,{\overline\nu}]$ and $\la \in[\la_0,{\overline\la}]$ and prove that this region can be extended going up in $\nu$ and up in $\la$ . 
We do it now considering only the contribution of the first term in the curly brackets, but the procedure can be done in the same way considering all the error terms.}
\smallskip




\NI Let us go back to our proof of the boundedness of the ${\cal Q}$ norms via the error estimates.
As we said the estimates we want to prove are,
\bea
({ \QQ_2}^{(J-2)}(\la,\nu))^{\frac{1}{2}}\leq C^{*(1)}\frac{J!}{J^{\a}}\frac{e^{(J-2)(\de+\underline{\Ga}(\la))}}{\ro_{0,1}^J}\ ,
\eea
\smallskip

 \NI Recalling that we are estimating the second term of inequality \ref{166555} and that $\Ga(\la)$  is increasing in $\la$, and using the bootstrap assumption
 \bea
 &&\ML{ \QQ_2}^{(J-2)}(\la,\nu))+\QQb_2^{(J-2)}(\la,\nu))\!\leq {\QQ}_{(0),2}^{(J-2)}({\la,\nu}))\\
&&\nn\\
&&\ML+\left(C^{*(1)}\frac{J!}{J^{\a}}\frac{e^{(J-2)(\de+\underline{\Ga}_0(\la))}}{\ro_{0,1}^J}\right)\left\{c\ep (J-1)\!\int_{\la_0}^{\la}d\la'
  \frac{1}{|\la'|^{1+\si}}\!\left({ \QQ_2}^{(J-2)}(\la',\nu)\right)^{\!\!\frac{1}{2}}\right\}\nn\\
  &&\nn\\
&&\ML\leq \left(C_{0}^{(1)}\frac{J!}{J^{\a}}\frac{e^{(J-2)(\de+\underline{\Ga}_0(\la))}}{\ro_{0,1}^J}\right)^{\!\!2}\nn\\
&&\ML+\left(C^{*(1)}\frac{J!}{J^{\a}}\frac{e^{(J-2)(\de+\underline{\Ga}(\la))}}{\ro_{0,1}^J}\right)\!\left\{c\ep (J-1)\!\int_{\la_0}^{\la}d\la'
  \frac{1}{|\la'|^{1+\si}}\!\left({ \QQ_2}^{(J-2)}(\la',\nu)\right)^{\!\!\frac{1}{2}}\right\}\nn\\
  &&\nn\\
&&\ML\leq \left(C_{0}^{(1)}\frac{J!}{J^{\a}}\frac{e^{(J-2)(\de+\underline{\Ga}_0(\la))}}{\ro_{0,1}^J}\right)^{\!\!2}\nn\\
&&\ML+\left[(c\varepsilon)\!\left(C^{*(1)}\frac{J!}{J^{\a}}\frac{e^{(J-2)(\de+\underline{\Ga}(\la))}}{\ro^J}\right)^{\!\!2}\!(J-1)\!\!\int_{\la_0}^{\la}d\la'\frac{1}{|\la'|^{1+\si}}e^{(J-2)(-\underline{\Ga}(\la)+\underline{\Ga}(\la'))}\right]\nn\\
&&\ML\ML\leq  \left(C_{0}^{(1)}\frac{J!}{J^{\a}}\frac{e^{(J-2)(\de+\underline{\Ga}_0(\la))}}{\ro_{0,1}^J}\right)^{\!\!2}
+\left[(c\varepsilon)\!\left(C^{*(1)}\frac{J!}{J^{\a}}\frac{e^{(J-2)(\de+\underline{\Ga}(\la))}}{\ro_{0,1}^J}\right)^{\!\!2}\!(J-1)\!\!\int_{\la_0}^{\la}d\la'\frac{1}{|\la'|^2}e^{(J-2)\frac{(\la'-\la)}{\la'\la}}\right]\ .\nn
\eea
Therefore
\bea\label{1562}
 &&\ML\ML\left({ \QQ_2}^{(J-2)}(\la,\nu))+\QQb_2^{(J-2)}(\la,\nu))\right)^{\frac{1}{2}}\\
 &&\ML\ML\leq \left(C_{0}^{(1)}\frac{J!}{J^{\a}}\frac{e^{(J-2)(\de+\underline{\Ga}_0(\la))}}{\ro_{0,1}^J}\right)
+(c\varepsilon)\!\left(C^{*(1)}\frac{J!}{J^{\a}}\frac{e^{(J-2)(\de+\underline{\Ga}(\la))}}{\ro_{0,1}^J}\right)\!\left[(J-1)\!\!\int_{\la_0}^{\la}d\la'\frac{1}{|\la'|^2}e^{(J-2)\frac{(\la'-\la)}{\la'\la}}\right]^{\!\frac{1}{2}}\nn
 \eea
Observe that denoting, 
\[x=\frac{(\la'-\la)}{\la'\la}\ \ ;\ \ \frac{dx}{d\la'}=\frac{1}{\la'^2}\ ,\]
therefore
\bea
&&\ML\int_{\la_0}^{\la}d\la'\frac{1}{\la'^2}e^{(J-2)\frac{(\la'-{\la})}{\la'{\la}}}
=-\int_0^{x_0} dxe^{(J-2)x}=\frac{c}{J-2}
\eea
as $x_0=\frac{(\la_0-{\la})}{\la_0{\la}}<0$\ and, finally,

\bea
&&\left({ \QQ_2}^{(J-2)}(\la,\nu))+\QQb_2^{(J-2)}(\la,\nu))\right)^{\frac{1}{2}}\leq \left[C_{0}^{(1)}+(c'\varepsilon)C^{*(1)}\right]\!\left(\frac{J!}{J^{\a}}\frac{e^{(J-2)(\de+\underline{\Ga}({\la}))}}{\ro_{0,1}^J}\right)\nn\\
&&<\left(C^{*(1)}\frac{J!}{J^{\a}}\frac{e^{(J-2)(\de+\underline{\Ga}(\la))}}{\ro_{0,1}^J}\right)\ ,
 \eea
choosing $\varepsilon$ sufficiently small, which implies that it can be extended to any $\nu$ up to $\nu_*$ and to the largest value of $\la=\la_1$. Therefore the result holds everywhere and the control of the more delicate term of the error is done.
 \smallskip
 
 
\NI  There is, see later on, another term which requires the same internal bootstrap which therefore has to be done considering all the terms, but now we investigate the second sum in \ref{35498} which can be estimated via the inductive assumptions, namely
 \[\sum_{k=0}^{J-2}\left[\sum_{l=1}^k\cbin{k}{l}
 \int_{\la_0}^{\la}d\la'\left(\int_{C(\la';[\nu_0,\nu])}\!\!\tau_+^4|(\Lie_O^l{\!\ ^{(O)}\pi}_{\ \ro}^{\tau})\lie_O^{k-l}\dd({\tilde W}^{(J-1-k)})|^2\right)^{\!\!\frac{1}{2}}\right]\ .\]

 \NI To control this sum, assuming 
  the estimates \ref{estaax} 
 \beaa
&&({\QQ}_2^{(H)}(\la,\nu))^{\frac{1}{2}}\leq C^{*(1)}\frac{(H+2)!}{(H+2)^{\a}}\frac{e^{H(\de+\underline{\Ga}(\la))}}{\ro_{0,1}^{(H+2)}}\nn\\
&&({{\QQb}}_2^{(H)}(\la,\nu))^{\frac{1}{2}}\leq { C}^{*(1)}\frac{(H+2)!}{(H+2)^{\a}}\frac{e^{H(\de+\underline{\Ga}(\la))}}{\ro_{0,1}^{(H+2)}}
\eeaa 
and the following estimate,
  \bea
 |r^{\si-\frac{2}{p}}\lie_O^l{\!\ ^{(O)}\pi}|_{p,S(\la',\nu')}\leq c_1\frac{l!}{l^{\a}}\frac{e^{(l-2)(\de+\underline{\Ga}(\la'))}}{\ro_{0,1}^l}\ ,
 \eql{17.34xx}
 \eea
we have, assuming for the moment $k\leq J/2$,
 \bea
 &&\ML\ML\ML\ML\left[\sum_{l=1}^k\cbin{k}{l}
 \int_{\la_0}^{{\la}}d\la'\left(\int_{C(\la';[\nu_0,{\nu}])}\!\!\tau_+^4|(\Lie_O^l{\!\ ^{(O)}\pi}_{\ \ro_{0,1}}^{\tau})\lie_O^{k-l}\dd({\tilde W}^{(J-1-k)})|^2\right)^{\!\!\frac{1}{2}}\right]\nn\\
 &&\ML\ML\ML\ML\leq c\sum_{l=1}^k\cbin{k}{l} \left(\sup_{ V({\overline\la},{\overline\nu})}|r^{\si}\lie_O^l{\!\ ^{(O)}\pi}|^2\right)^{\!\!\frac{1}{2}}\int_{\la_0}^{{\overline\la}}d\la'\frac{1}{|\la'|^{2+2\si}}\left({\QQ}_2^{(J-2-l)}(\la',{\nu})\right)^{\!\frac{1}{2}}\nn\\
 &&\ML\ML\ML\ML\leq cc_1\sum_{l=1}^k\cbin{k}{l}\frac{(l+2)!}{(l+2)^{\a}}\frac{e^{(l-1)(\de+\underline{\Ga}(\la))}}{\ro_{0,1}^{l+1}}\int_{\la_0}^{{\la}}d\la'\frac{1}{|\la'|^{2+2\si}}C^{(1)}\frac{(J-l)!}{(J-l)^{\a}}\frac{e^{(J-l-2)(\de+\underline{\Ga}(\la'))}}{\ro_{0,1}^{(J-l)}}\nn\\
  &&\ML\ML\ML\ML\leq cc_1C^{(1)}\sum_{l=1}^k\cbin{k}{l}\frac{(l+2)!}{(l+2)^{\a}}\frac{e^{(l-1)(\de+\underline{\Ga}(\la))}}{\ro_{0,1}^{l+1}}\int_{\la_0}^{{\la}}d\la'\frac{1}{|\la'|^{2+2\si}}\frac{(J-l)!}{(J-l)^{\a}}\frac{e^{(J-l-2)(\de+\underline{\Ga}(\la'))}}{\ro_{0,1}^{(J-l)}}\nn\\
 &&\ML\ML\ML\ML\leq cc_1C^{(1)}\sum_{l=1}^k\cbin{k}{l}\frac{(l+2)!}{(l+2)^{\a}}\frac{(J-l)!}{(J-l)^{\a}}\frac{e^{(l-1)(\de+\underline{\Ga}(\la))}}{\ro_{0,1}^{J+1}}\int_{\la_0}^{{\la}}d\la'\frac{1}{|\la'|^{2+2\si}}{e^{(J-l-2)(\de+\underline{\Ga}(\la'))}}\nn\\
 &&\ML\ML\ML\ML\leq C^{(1)}\left(\frac{e^{(J-2)(\de+\underline{\Ga}(\la))}}{\ro_{0,1}^{J}}\right)\!\left(cc_1\frac{e^{-\de}}{\ro_{0,1}}\right)\sum_{l=1}^k\cbin{k}{l}\frac{(l+2)!}{(l+2)^{\a}}\frac{(J-l)!}{(J-l)^{\a}}\int_{\la_0}^{\la}d\la'\frac{1}{|\la'|^{2+2\si}}{e^{-(J-l-2)(\underline{\Ga}({\la})-\underline{\Ga}(\la'))}}\ .\nn
 \eea
Denoting, 
\[x=\frac{(\la'-{\la})}{\la'{\la}}\ \ ;\ \ \frac{dx}{d\la'}=\frac{1}{\la'^2}\ ,\]
we have
\bea
&&\ML\ML \int_{\la_0}^{{\la}}d\la'\frac{1}{|\la'|^{2+2\si}}{e^{-(J-l-2)(\underline{\Ga}({\la})-\underline{\Ga}(\la'))}}
\leq\int_{\la_0}^{{\la}}d\la'\frac{1}{|\la'|^{2}}e^{(J-2-l)\frac{(\la'-{\la})}{\la'{\la}}}\nn\\
&&\ML\ML=-\int_0^{x_0} dxe^{(J-2-l)x}=\frac{c'}{(J-2-l)}\ .
\eea
Therefore
\bea
&&\ML\ML\ML\sum_{k=0}^{\left[\frac{J}{2}\right]}\left[\sum_{l=1}^k\cbin{k}{l}
 \int_{\la_0}^{{\la}}d\la'\left(\int_{C(\la';[{\nu_0},\nu])}\!\!\tau_+^4|(\Lie_O^l{\!\ ^{(O)}\pi}_{\ \ro}^{\tau})\lie_O^{k-l}\dd({\tilde W}^{(J-1-k)})|^2\right)^{\!\!\frac{1}{2}}\right]\nn\\
 &&\ML\ML\ML\leq C^{(1)}\left(\frac{J!}{J^{\a}}\frac{e^{(J-2)(\de+\underline{\Ga}(\la))}}{\ro_{0,1}^{J}}\right)\!\left(cc'c_1\frac{e^{-\de}}{\ro_{0,1}}\right)
 \left\{\sum_{k=0}^{\left[\frac{J}{2}\right]}\sum_{l=1}^k\cbin{k}{l}\frac{(l+2)!}{(l+2)^{\a}}\frac{(J-l)!}{(J-l)^{\a}}\frac{J^{\a}}{J!}\frac{1}{(J-2-l)}\right\}\ .\nn
\eea
Observe that
\bea
&&\ML\ML\left\{\c\c\c\c\right\}=\sum_{k=0}^{\left[\frac{J}{2}\right]}\sum_{l=1}^k\frac{k!}{(k-l)!l!}\frac{(l+2)!}{(l+2)^{\a}}\frac{(J-l)!}{(J-l)^{\a}}\frac{J^{\a}}{J!}\frac{1}{(J-2-l)}\\
&&\ML\ML=\sum_{k=0}^{\left[\frac{J}{2}\right]}\sum_{l=1}^k\frac{1}{(l+2)^{\a}}\frac{(l+2)!}{l!}\left[\frac{k!}{(k-l)!}\frac{(J-l)!}{J!}\frac{J^{\a}}{(J-l)^{\a}}\frac{1}{(J-2-l)}\right]\nn\\
&&\ML\ML\leq\frac{1}{J}\sum_{k=0}^{\left[\frac{J}{2}\right]}\sum_{l=1}^k\frac{1}{(l+2)^{\a}}\frac{(l+2)!}{l!}\left[\frac{k!}{(k-l)!}\frac{(J-l)!}{J!}\frac{J^{\a}}{(J-l)^{\a}}\frac{J}{(J-2-l)}\right]\nn\\
&&\ML\ML\leq\frac{1}{J}\sum_{k=0}^{\left[\frac{J}{2}\right]}\sum_{l=1}^k\frac{1}{(l+2)^{\a}}\frac{(l+2)!}{l!}\left[2\frac{k!}{(k-l)!}\frac{(J-l)!}{J!}2^{\a}\right]\nn\\
&&\ML\ML\leq c{\left(\frac{2^{\a+1}}{J}\right)}\sum_{k=0}^{\left[\frac{J}{2}\right]}\left[\sum_{l=1}^k\frac{1}{(l+2)^{\a-2}}\right]\leq c_2\nn
\eea
choosing $\a>3$. Therefore finally
\bea
&&\ML\ML\ML\sum_{k=0}^{\left[\frac{J}{2}\right]}\left[\sum_{l=1}^k\cbin{k}{l}
 \int_{\la_0}^{{\la}}d\la'\left(\int_{C(\la';[{\nu_0},\nu])}\!\!\tau_+^4|(\Lie_O^l{\!\ ^{(O)}\pi}_{\ \ro}^{\tau})\lie_O^{k-l}\dd({\tilde W}^{(J-1-k)})|^2\right)^{\!\!\frac{1}{2}}\right]\nn\\
 &&\ML\ML\ML\leq C^{(1)}\left(\frac{J!}{J^{\a}}\frac{e^{(J-2)(\de_0+\underline{\Ga}_0(\la))}}{\ro_{0,1}^{J}}\right)\!\left(cc'c_1c_2\frac{e^{-\de}}{\ro_{0,1}}\right)\ .
 \eea
Let us investigate the same sum with $k\geq J/2$. We have to divide it in two sums, the first with $l\leq J/2$ and the second with $l>J/2$. The first sum is treated exactly as the previous one so we are left with estimating,
\[\sum_{k=\left[\frac{J}{2}\right]+1}^{J-2}\sum_{l=\left[\frac{J}{2}\right]+1}^k\cbin{k}{l}
 \int_{\la_0}^{{\la}}d\la'\left(\int_{C(\la';[{\nu_0},\nu])}\!\!\tau_+^4|(\Lie_O^l{\!\ ^{(O)}\pi}_{\ \ro}^{\tau})\lie_O^{k-l}\dd({\tilde W}^{(J-1-k)})|^2\right)^{\!\!\frac{1}{2}}\ .\]
\medskip

\NI 
 We write
 \bea
&&\ML\ML\int_{C(\la';[{\nu_0},\nu])}\!\!\tau_+^4|(\Lie_O^l{\!\ ^{(O)}\pi}_{\ \ro}^{\tau})\lie_O^{k-l}\dd({\tilde W}^{(J-1-k)})|^2\nn\\
&&\ML\ML=\int_{{\nu_0}}^{\nu}d\nu'\tau_+^4\int_{\!S(\la',\nu')}\!\!\!\!d\mu\ \!|\Lie_O^l{\!\ ^{(O)}\pi}_{\ \ro}^{\tau}|^2|\lie_O^{k-l}\dd({\tilde W}^{(J-1-k)})|^2\nn\\
&&\ML\ML\leq\int_{{\nu_0}}^{\nu}d\nu'\tau_+^4\left(\int_{\!S(\la',\nu')}\!\!\!\!d\mu\ \!|\Lie_O^l{\!\ ^{(O)}\pi}_{\ \ro}^{\tau}|^4\right)^{\frac{1}{2}}\left(\int_{\!S(\la',\nu')}\!\!\!\!d\mu\ \!|\lie_O^{k-l}\dd({\tilde W}^{(J-1-k)})|^4\right)^{\frac{1}{2}}\nn\\
&&\ML\ML\leq\int_{{\nu_0}}^{\nu}d\nu'\tau_+^4|\Lie_O^l{\!\ ^{(O)}\pi}_{\ \ro}^{\tau}|^2_{4,S(\la',\nu')}|\lie_O^{k-l}\dd({\tilde W}^{(J-1-k)})|^2_{4,S(\la',\nu')}\ .
\eea
Therefore
\bea
&&\ML\ML\sum_{k=\left[\frac{J}{2}\right]+1}^{J-2}\sum_{l=\left[\frac{J}{2}\right]+1}^k\cbin{k}{l}
 \int_{\la_0}^{{\la}}d\la'\left(\int_{C(\la';[{\nu_0},\nu])}\!\!\tau_+^4|(\Lie_O^l{\!\ ^{(O)}\pi}_{\ \ro}^{\tau})\lie_O^{k-l}\dd({\tilde W}^{(J-1-k)})|^2\right)^{\!\!\frac{1}{2}}\eql{17.46xxa}\\
&&\ML\ML\leq \sum_{k=\left[\frac{J}{2}\right]+1}^{J-2}\sum_{l=\left[\frac{J}{2}\right]+1}^k\cbin{k}{l}\int_{\la_0}^{{\la}}d\la'\left(\int_{{\nu_0}}^{\nu}d\nu'\tau_+^4|\Lie_O^l{\!\ ^{(O)}\pi}_{\ \ro}^{\tau}|^2_{4,S(\la',\nu')}|\lie_O^{k-l}\dd({\tilde W}^{(J-1-k)})|^2_{4,S(\la',\nu')}\right)^{\!\!\frac{1}{2}}\ .\nn
\eea
To estimate \ref{17.46xxa} we have to make an assumption about the estimate of  $|r^{\si}\lie_O^l{\!\ ^{(O)}\pi}|^2$. 
As ${\!\ ^{(O)}\pi}$ has to be treated as a connection coefficient we can assume the following estimates coherent with the inductive assumptions for the Lie derivatives of the connection coeffcients, see later on.

 \bea
 |r^{\si-\frac{2}{p}}\Lie_O^l{\!\ ^{(O)}\pi}|_{p,S(\la',\nu')}\leq c_1\frac{l!}{l^{\a}}\frac{e^{(l-2)(\de+\underline{\Ga}(\la'))}}{\ro_{0,1}^l}
 \eql{17.34xxaa}
 \eea
 with $\si\geq 2$ if $l>0$, $=1$ if $l=0$ and $c_1=O(\ep_0)$ if $l\leq 7$. Moreover
\bea
&&|\lie_O^{k-l}\dd({\tilde W}^{(J-1-k)})|_{4,S(\la',\nu')}\leq |r^{-1}\Psi({\tilde W}^{(J-l)})|_{4,S(\la',\nu')}\nn\\
&&\leq r^{-1-(1-\frac{2}{4})}|\la'|^{-\frac{5}{2}}\left(C^{(1)}\frac{(J-l)!}{(J-l)^{\a}}\frac{e^{(J-l-1)(\de+\underline{\Ga}(\la'))}}{\ro_{0,1}^{J-l+1}}\right)
\eea
Therefore
\bea
&&\ML\ML\ML\ML\sum_{k=\left[\frac{J}{2}\right]+1}^{J-2}\sum_{l=\left[\frac{J}{2}\right]+1}^k\cbin{k}{l}
 \int_{\la_0}^{{\la}}d\la'\left(\int_{C(\la';[{\nu_0},\nu])}\!\!\tau_+^4|(\Lie_O^l{\!\ ^{(O)}\pi}_{\ \ro}^{\tau})\lie_O^{k-l}\dd({\tilde W}^{(J-1-k)})|^2\right)^{\!\!\frac{1}{2}}\eql{17.46xx}\\
&&\ML\ML\ML\ML\leq \sum_{k=\left[\frac{J}{2}\right]+1}^{J-2}\sum_{l=\left[\frac{J}{2}\right]+1}^k\cbin{k}{l}\int_{\la_0}^{{\la}}d\la'\left(\int_{{\nu_0}}^{\nu}d\nu'\tau_+^4|\Lie_O^l{\!\ ^{(O)}\pi}_{\ \ro}^{\tau}|^2_{4,S(\la',\nu')}|\lie_O^{k-l}\dd({\tilde W}^{(J-1-k)})|^2_{4,S(\la',\nu')}\right)^{\!\!\frac{1}{2}}\nn\\
&&\ML\ML\ML\ML\leq \sum_{k=\left[\frac{J}{2}\right]+1}^{J-2}\sum_{l=\left[\frac{J}{2}\right]+1}^k\cbin{k}{l}\left(c_1\frac{l!}{l^{\a}}\frac{1}{\ro_{0,1}^l}\right)\left(C^{(1)}\frac{(J-l)!}{(J-l)^{\a}}\frac{1}{\ro_{0,1}^{J-l+1}}\right)\c\nn\\
&&\ML\ML\ML\ML\c\int_{\la_0}^{\la}d\la'\left(\int_{\nu_0}^{\nu}d\nu'\tau_+^4\left(\frac{1}{r^3(\la',\nu')}e^{2(l-2)\underline{\Ga}(\la')}\right)
\left(\frac{1}{|\la'|^5r^3}e^{2(J-l-1)\underline{\Ga}(\la')}\right)\right)^{\!\!\frac{1}{2}}\nn\\
&&\ML\ML\ML\ML\ML\leq \sum_{k=\left[\frac{J}{2}\right]+1}^{J-2}\sum_{l=\left[\frac{J}{2}\right]+1}^k\cbin{k}{l}\left(c_1C^{(1)}\frac{l!}{l^{\a}}\frac{(J-l)!}{(J-l)^{\a}}\frac{e^{(J-3)\de}}{\ro_{0,1}^{J+1}}\right)
\int_{\la_0}^{\la}d\la'\frac{1}{|\la'|^{\frac{5}{2}}}e^{(J-3)\Ga(\la')}\left(\int_{{\nu_0}}^{\nu}d\nu'\frac{1}{\nu'^2}
\right)^{\!\!\frac{1}{2}}\nn\\
&&\ML\ML\ML\ML\leq\left(C^{(1)}\frac{J!}{J^{\a}}\frac{e^{(J-2)(\de+\underline{\Ga}({\la},{\nu}))}}{\ro_{0,1}^{J}}\right)\!\left(c_1\frac{e^{-\de}}{\ro_{0,1}}\right)\!\left[\sum_{k=\left[\frac{J}{2}\right]+1}^{J-2}\sum_{l=\left[\frac{J}{2}\right]+1}^k\cbin{k}{l}\frac{J^{\a}}{J!}
\frac{l!}{l^{\a}}\frac{(J-l)!}{(J-l)^{\a}}\right]\c\nn\\
&&\ML\ML\ML\c\left(\int_{\la_0}^{{\la}}\frac{d\la'}{|\la'|^{\frac{5}{2}}}e^{-(J-2)(\Ga({\la})-\Ga(\la'))}e^{-\Ga(\la')}\right)\nn\\
&&\ML\ML\ML\ML\leq\left(C^{(1)}\frac{J!}{J^{\a}}\frac{e^{(J-2)(\de+\Ga({\la},{\nu}))}}{\ro_{0,1}^{J}}\right)\!\left(c_1\frac{e^{-\de}e^{\frac{1}{\nu_0}}}{\ro_{0,1}}\right)\!\left[\sum_{k=\left[\frac{J}{2}\right]+1}^{J-2}\sum_{l=\left[\frac{J}{2}\right]+1}^k\cbin{k}{l}\frac{J^{\a}}{J!}
\frac{l!}{l^{\a}}\frac{(J-l)!}{(J-l)^{\a}}\right]\c\nn\\
&&\ML\ML\ML\c\left(\int_{\la_0}^{{\la}}\frac{d\la'}{|\la'|^{\frac{5}{2}}}e^{-(J-2)(\Ga({\la})-\Ga(\la'))}\right) ,\nn
\eea
the last inequality as $e^{-\Ga(\la')}<1$.
Observe that, for suitable $c"$
\bea
&&\ML\ML\int_{\la_0}^{{\la}}\frac{d\la'}{|\la'|^{\frac{5}{2}}}e^{(J-2)(-\Ga({\la})-\Ga(\la'))}
=-\int_0^{x_0}dxx^{\frac{1}{2}}e^{(J-2)x}\leq\frac{c'}{(J-2)^{\frac{3}{2}}}\ \ \ 
\eea
and substituting in the previous estimate we obtain

\bea
&&\ML\ML\ML\sum_{k=\left[\frac{J}{2}\right]+1}^{J-2}\sum_{l=\left[\frac{J}{2}\right]+1}^k\cbin{k}{l}
 \int_{\la_0}^{{\la}}d\la'\left(\int_{C(\la';[{\nu_0},\nu])}\!\!\tau_+^4|(\Lie_O^l{\!\ ^{(O)}\pi}_{\ \ro}^{\tau})\lie_O^{k-l}\dd({\tilde W}^{(J-1-k)})|^2\right)^{\!\!\frac{1}{2}}\eql{17.46xxzz}\\
 &&\ML\ML\ML\ML\leq\left(C^{(1)}\frac{J!}{J^{\a}}\frac{e^{(J-2)(\de+\Ga({\la},{\nu}))}}{\ro_{0,1}^{J}}\right)\!\left(c_1c'\frac{e^{-\de}e^{\frac{1}{\nu_0}}}{\ro_{0,1}}\right)\!\left[\sum_{k=\left[\frac{J}{2}\right]+1}^{J-2}\sum_{l=\left[\frac{J}{2}\right]+1}^k\cbin{k}{l}\frac{J^{\a}}{J!}
\frac{l!}{l^{\a}}\frac{(J-l)!}{(J-l)^{\a}}\frac{1}{(J-2)^{\frac{3}{2}}}\right]\ .\nn
\eea
We have to prove that the factor $\big[\c\c\big]$ is bounded by a constant $c$. Choosing $\de$ sufficiently large this term is under control. In fact
\bea
&&\ML\ML\ML\ML\left[\sum_{k=\left[\frac{J}{2}\right]+1}^{J-2}\sum_{l=\left[\frac{J}{2}\right]+1}^k\cbin{k}{l}\frac{J^{\a}}{J!}
\frac{l!}{l^{\a}}\frac{(J-l)!}{(J-l)^{\a}}\frac{1}{(J-2)^\frac{3}{2}}\right]\nn\\
&&\ML\ML\ML\ML=\frac{J^{\a}}{J!}\frac{1}{(J-2)^\frac{3}{2}}\sum_{k=\left[\frac{J}{2}\right]+1}^{J-2}\sum_{l=\left[\frac{J}{2}\right]+1}^k\frac{k!}{(k-l)!}\frac{1}{l^{\a}}\frac{(J-l)!}{(J-l)^{\a}}\nn\\
&&\ML\ML\ML\ML=\frac{1}{J!}\frac{1}{(J-2)^{\frac{3}{2}}}\sum_{k=\left[\frac{J}{2}\right]+1}^{J-2}\sum_{l=\left[\frac{J}{2}\right]+1}^k\frac{k!}{(k-l)!}\frac{J^{\a}}{l^{\a}}\frac{(J-l)!}{(J-l)^{\a}}\nn\\
&&\ML\ML\ML\ML\leq \frac{1}{(J-2)^{\frac{3}{2}}}\sum_{k=\left[\frac{J}{2}\right]+1}^{J-2}\sum_{l=\left[\frac{J}{2}\right]+1}^k\left[\frac{k!(J-l)!}{J!(k-l)!}\right]\!\left[\frac{J^{\a}}{l^{\a}}\right]\frac{1}{(J-l)^{\a}}\nn\\
&&\ML\ML\ML\ML\leq\left(\sup\left[\frac{k!(J-l)!}{J!(k-l)!}\right]\!\left[\frac{1}{(J-2)^{\frac{3}{2}}}\frac{J^{\a}}{l^{\a}}\right]\right)\sum_{k=\left[\frac{J}{2}\right]+1}^{J-2}\left(\sum_{l=\left[\frac{J}{2}\right]+1}^k\frac{1}{(J-l)^{\a}}\right)\nn\\
&&\ML\ML\ML\ML\leq c\frac{1}{(J-2)^{\frac{3}{2}}}\sum_{k=\left[\frac{J}{2}\right]+1}^{J-2}c'\leq \frac{c''}{(J-2)^{\frac{1}{2}}}\ .\nn
\eea

\smallskip


\subsubsection{Going back to the error terms:}
Recalling the previous expressions, see eq. \ref{testbb} 
written symbolically, omitting the constants, as
\bea
&&\ML\ML(\dd^{\mu}\lie^2_O{\tilde W}^{(J-2)})_{\mu\b\ga\de}= \sum_{k=0}^{J-2}\Lie_O^k\left({\!\ ^{(O)}\pi}_{\ \b}^{\ro}\dd^{\mu}({\tilde W}^{(J-1-k)})_{\mu\ro\ga\de}\right)+\sum_{k=0}^{J-1}\Lie_O^k\left({\!\ ^{(O)}\pih}^{\mu\nu}\dd_{\nu}{\tilde W}^{(J-1-k)}_{\mu\b\ga\de}\right)\nn\\
&&\ML\ML+\sum_{k=0}^{J-1}\Lie_O^k\left({\!\ ^{(O)}{\tilde q}}_{\a\b\la}{{\tilde W}^{(J-1-k)\a\la}}_{\ \ \ \ \ \ \ga\de}\right)\ ,\nn
\eea
where 
\[{\tilde q}_{\a\b\la}= \de_{\a\b}{^{(O)}p}_{\la} \ \ \ \ \mbox{or}\ \ \ \ {\!\ ^{(O)}{\tilde q}}_{\a\b\la}\ .\]
We have  estimated the contribution of the first sum, the second one should be estimated in the same way, we look here at the contribution of the third one. Therefore we  proceed as we had
\bea
(\dd^{\mu}\lie^2_O{\tilde W}^{(J-2)})_{\mu\b\ga\de}= \sum_{k=0}^{J-1}\Lie_O^k\left({\!\ ^{(O)}{\tilde q}}_{\a\b\la}{{\tilde W}^{(J-1-k)\a\la}}_{\ \ \ \ \ \ \ga\de}\right)+[terms\ under\ control]\ .\nn
\eea
 The error, considering only the first sum, has the form
\bea
&&\ML\ML\int_{V_{(\la,\nu)}}(\dd^{\mu}\lie^2_O{\tilde W}^{(J-2)})_{\mu\ro\ga\si}(\lie^2_O{\tilde W}^{(J-2)})_{\b\ \de\ }^{\ \ro\ \si}
(\acc^{\b}\acc^{\ga}T^{\de})\\
&&\ML\ML=\sum_{k=0}^{J-1}\int_{V_{(\la,\nu)}} \Lie_O^k\left({\!\ ^{(O)}{\tilde q}}_{\a\ro\la}{{\tilde W}^{(J-1-k)\a\la}}_{\ \ \ \ \ \ \ga\si}\right)(\lie^2_O{\tilde W}^{(J-2)})_{\b\ \de\ }^{\ \ro\ \si}(\acc^{\b}\acc^{\ga}T^{\de})\ .\nn
\eea
\NI Therefore neglecting all the other parts of the error  we have to deal with the following inequality,
\bea
&&\ML\ML{{\cal Q}}_2^{(J-2)}(\la,\nu)+{{\QQb}}_2^{(J-2)}(\la,\nu)\!={{{\cal Q}}}_{(0),2}^{(J-2)}(\la,\nu)\nn\\
&&\ML\ML=\sum_{k=0}^{J-1}\int_{V_{(\la,\nu)}} \Lie_O^k\left({\!\ ^{(O)}{\tilde q}}_{\a\ro\la}{{\tilde W}^{(J-1-k)\a\la}}_{\ \ \ \ \ \ \ga\si}\right)(\lie^2_O{\tilde W}^{(J-2)})_{\b\ \de\ }^{\ \ro\ \si}(\acc^{\b}\acc^{\ga}T^{\de})\nn\\
&&\nn\\
&&\ML\ML=\int_{V_{(\la,\nu)}} \Lie_O^{J-1}\left({\!\ ^{(O)}{\tilde q}}_{\a\ro\la}{{\tilde W}^{(0)\a\la}}_{\ \ \ \ \ \ \ga\si}\right)(\lie^2_O{\tilde W}^{(J-2)})_{\b\ \de\ }^{\ \ro\ \si}(\acc^{\b}\acc^{\ga}T^{\de})\nn\\
&&\ML\ML+\sum_{k=0}^{J-2}\int_{V_{(\la,\nu)}} \Lie_O^k\left({\!\ ^{(O)}{\tilde q}}_{\a\ro\la}{{\tilde W}^{(J-1-k)\a\la}}_{\ \ \ \ \ \ \ga\si}\right)(\lie^2_O{\tilde W}^{(J-2)})_{\b\ \de\ }^{\ \ro\ \si}(\acc^{\b}\acc^{\ga}T^{\de})\nn
\eea
where
Proceeding as before, we start looking at the first term which is the one we have to add to the ``internal bootstrap mechanism", at its turn it can be decomposed in two terms
\bea
 &&\ML\ML\ML\int_{V_{(\la,\nu)}} \left(\Lie_O^{J-1}{\!\ ^{(O)}{\tilde q}}_{\a\ro\la}{{\tilde W}^{(0)\a\la}}_{\ \ \ \ \ \ \ga\si}\right)(\lie^2_O{\tilde W}^{(J-2)})_{\b\ \de\ }^{\ \ro\ \si}(\acc^{\b}\acc^{\ga}T^{\de})\nn\\
 &&\ML\ML\ML=\int_{V_{(\la,\nu)}} \!\!\tau_+^4 \left(\Lie_O^{J-1}{\!\ ^{(O)}{\tilde q}}_{\a\ro\la}{{\tilde W}^{(0)\a\la}}_{\ \ \ \ \ \ 4\si}\right)(\lie^2_O{\tilde W}^{(J-2)})_{4\ 4\ }^{\ \ro\ \si}(\acc^{\b}\acc^{\ga}T^{\de})\nn\\
&&\ML\ML\ML=\int_{V_{(\la,\nu)}} \!\!\tau_+^4\ \!\left(\Lie_O^{J-1}{\!\ ^{(O)}{\tilde q}}_{\a\c\la}{{\tilde W}^{(0)\a\la}}_{\ \ \ \ \ \ 4\c}\right)\c\a(\lie^2_O{\tilde W}^{(J-2)})\nn\\
&&\ML\ML\ML+\int_{V_{(\la,\nu)}} \!\!\tau_+^4\ \!\left(\Lie_O^{J-1}{\!\ ^{(O)}{\tilde q}}_{\a\c\la}{{\tilde W}^{(0)\a\la}}_{\ \ \ \ \ \ 4\c}\right)\c\b(\lie^2_O{\tilde W}^{(J-2)})\ .
\eea
These two terms have the same structure and it is enough to estimate the first one as the same estimate holds for the second,
\bea
&&\ML\ML\ML\ML\int_{V_{(\la,\nu)}} \!\!\tau_+^4\ \!\left(\Lie_O^{J-1}{\!\ ^{(O)}{\tilde q}}_{\a\c\la}{{\tilde W}^{(0)\a\la}}_{\ \ \ \ \ \ 4\c}\right)\c\a(\lie^2_O{\tilde W}^{(J-2)})\nn\\
&&\ML\ML\ML\ML=\sum_{l=0}^{J-1}\cbin{J-1}{l} \int_{V_{(\la,\nu)}} \!\!\tau_+^4\!\left((\Lie_O^{l}{\!\ ^{(O)}{\tilde q}}_{\a\c\la})\lie_O^{J-1-l}{{\tilde W}^{(0)\a\la}}_{\ \ \ \ \ \ 4\c}\right)\c\a(\lie^2_O{\tilde W}^{(J-2)})\nn\\
&&\ML\ML\ML\ML=\sum_{l=0}^{J-1}\cbin{J-1}{l}\int_{\la_0}^{\la}d\la'\left(\int_{C(\la';[\nu,\nu_*])}\!\!\tau_+^4\big|(\Lie_O^{l}{\!\ ^{(O)}{\tilde q}})\c(\lie_O^{J-1-l}{{\tilde W}^{(0)}})\big|^2\right)^{\!\!\frac{1}{2}}\!\left(\int_{C(\la';[\nu,\nu_*])}\!\!\tau_+^4|\a(\lie^2_O{\tilde W}^{(J-2)})|^2\right)^{\!\!\frac{1}{2}}\nn\\
&&\ML\ML\ML\ML\leq \left(\sup_{\la'\in[\la_0,\la]}{\cal Q}_2^{(J-2)}(\la',\nu)\right)^{\!\!\frac{1}{2}}\left[\sum_{l=0}^{J-1}\cbin{J-1}{l} 
 \int_{\la_0}^{\la}d\la'\left(\int_{C(\la';[\nu,\nu_*])}\!\!\tau_+^4\big|(\Lie_O^{l}{\!\ ^{(O)}{\tilde q}})\c(\lie_O^{J-1-l}{{\tilde W}^{(0)}})\big|^2\right)^{\!\!\frac{1}{2}}\right]\nn\\
 &&\nn\\
 &&\ML\ML\ML\ML\leq \left(\sup_{\la'\in[\la_0,\la]}{\cal Q}_2^{(J-2)}(\la',\nu)\right)^{\!\!\frac{1}{2}}\left[
 \int_{\la_0}^{\la}d\la'\left(\int_{C(\la';[\nu,\nu_*])}\!\!\tau_+^4\big|(\Lie_O^{J-1}{\!\ ^{(O)}{\tilde q}})\c({{\tilde W}^{(0)}})\big|^2\right)^{\!\!\frac{1}{2}}\right]\nn\\
 &&\ML\ML\ML\ML+\left(\sup_{\la'\in[\la_0,\la]}{\cal Q}_2^{(J-2)}(\la',\nu)\right)^{\!\!\frac{1}{2}}\left[\sum_{l=0}^{J-2}\cbin{J-1}{l} 
 \int_{\la_0}^{\la}d\la'\left(\int_{C(\la';[\nu,\nu_*])}\!\!\tau_+^4\big|(\Lie_O^{l}{\!\ ^{(O)}{\tilde q}})\c(\lie_O^{J-1-l}{{\tilde W}^{(0)}})\big|^2\right)^{\!\!\frac{1}{2}}\right]\ .\ \ \ \ \ \ \ \ \ \ \ \ \eql{use1}
 \eea
Only the first term enters the internal bootstrap mechanism, let us estimate it. 
Mimicking the previous computations we have
\bea
&&\ML{ \QQ_2}^{(J-2)}({\la},{\nu}))+{\QQb_2}^{(J-2)}({\la},{\nu}))\!\leq {{\QQ}}_{(0),2}^{(J-2)}({\la},{\nu}))\\
&&\nn\\
&&\ML+\left(C^{*(1)}\frac{J!}{J^{\a}}\frac{e^{(J-2)(\de+\underline{\Ga}({\la}))}}{\ro_{0,1}^J}\right)
\left\{ \int_{\la_0}^{{\la}}d\la'\left(\int_{C(\la';[{\nu_0},\nu])}\!\!\tau_+^4\big|(\Lie_O^{J-1}{\!\ ^{(O)}{\tilde q}})\c{{\tilde W}^{(0)}}\big|^2\right)^{\!\!\frac{1}{2}}\right\}\nn
\eea
 and, see \ref{1562}
 \bea
 &&\ML\big({\QQ_2}^{(J-2)}({\la},{\nu}))+\QQb_2^{(J-2)}({\la},{\nu}))\big)^{\frac{1}{2}}\\
 &&\ML\leq \left(C_{0}^{(1)}\frac{J!}{J^{\a}}\frac{e^{(J-2)(\de+\underline{\Ga}_0({\la}))}}{\ro_{0,1}^J}\right)+\int_{\la_0}^{{\la}}d\la'\left(\int_{C(\la';[{\nu_0},\nu])}\!\!\tau_+^4\big|(\Lie_O^{J-1}{\!\ ^{(O)}{\tilde q}})\c{{\tilde W}^{(0)}}\big|^2\right)^{\!\!\frac{1}{2}}\ .\nn
 \eea
 Moreover, proceeding as before,
 \bea
 &&\ML\ML\left(\int_{C(\la';[\nu_0,\nu])}\!\!\tau_+^4\big|(\Lie_O^{J-1}{\!\ ^{(O)}{\tilde q}})\c{{\tilde W}^{(0)}}\big|^2\right)^{\frac{1}{2}}
 =\left(\int_{{\nu_0}}^{\nu}d\nu'\tau_+^4\int_{S(\la',\nu')}d\nu\big|(\Lie_O^{J-1}{\!\ ^{(O)}{\tilde q}})\c{{\tilde W}^{(0)}}\big|^2\right)^{\frac{1}{2}}\nn\\
 &&\ML\ML\leq \left(\int_{{\nu_0}}^{\nu}d\nu'\tau_+^4|\Lie_O^{J-1}{\!\ ^{(O)}{\tilde q}}|^2_{4,S(\la',\nu')}|{\tilde W}^{(0)}|^2_{4,S(\la',\nu')}\right)^{\frac{1}{2}}
  \eea
Using the estimates,\footnote{Again recalling the previous discussion about the estimate for the $\Lie_O^J$ of the connection coefficients.}
 \bea
|r^{2-\frac{1}{2}}\Lie_O^{J-1}{\!\ ^{(O)}{\tilde q}}|_{4,S(\la',\nu')} \leq c_1\frac{J!}{J^{\a}}\frac{e^{(J-2)(\de+\underline{\Ga}(\la'))}}{\ro_{0,1}^J}
 \eql{17.34xxac}
 \eea
and,\footnote{The following estimate is correct due to the index $4$ in the beginning of \ref{use1} which forbids to $W^{(0)}$ of being $\aa(W)$.}
\bea
|({\tilde W}^{(0)})|_{4,S(\la',\nu')}\leq |\Psi(W)|_{4,S(\la',\nu')}
\leq r^{-(2-\frac{2}{4})}|\la'|^{-\frac{3}{2}}\left(c\varepsilon\right)
\eea 
we have
\bea
 &&\ML\ML\left(\int_{C(\la';[\nu_0,\nu])}\!\!\tau_+^4\big|(\Lie_O^{J-1}{\!\ ^{(O)}{\tilde q}})\c{{\tilde W}^{(0)}}\big|^2\right)^{\frac{1}{2}}
 \leq \left(\int_{{\nu_0}}^{\nu}d\nu'\tau_+^4|\Lie_O^{J-1}{\!\ ^{(O)}{\tilde q}}|^2_{4,S(\la'\nu')}|{\tilde W}^{(0)}|^2_{4,S(\la',\nu')}\right)^{\frac{1}{2}}\nn\\
 &&\ML\ML\leq\frac{1}{|\la'|^{\frac{3}{2}}}\left(c_1\frac{J!}{J^{\a}}\frac{e^{(J-2)(\de+\underline{\Ga}(\la'))}}{\ro_{0,1}^J}
\right)\!\left(c\varepsilon\right)\left(\int_{{\nu_0}}^{\nu}d\nu'\frac{\tau_+^4}{r^{3+3}} \right)^{\frac{1}{2}}\nn\\
&&\ML\ML\leq\frac{1}{|\la'|^{\frac{3}{2}}}\left(c_1\frac{J!}{J^{\a}}\frac{e^{(J-2)(\de+\underline{\Ga}({\la'}))}}{\ro_{0,1}^J}
\right)\!\left(c\varepsilon\right)\left(\int_{{\nu_0}}^{\nu}\frac{d\nu'}{\nu'^2} \right)^{\frac{1}{2}}\nn\\
&&\ML\ML\leq\frac{c\varepsilon}{|\la'|^{\frac{3}{2}}}\left(c_1\frac{J!}{J^{\a}}\frac{e^{(J-2)(\de+\underline{\Ga}({\la'}))}}{\ro_{0,1}^J}
\right)\!\!\nn
 \eea
and
\bea
&&\ML\ML\int_{\la_0}^{{\la}}d\la'\left(\int_{C(\la';[{\nu_0},\nu])}\!\!\tau_+^4\big|(\Lie_O^{J-1}{\!\ ^{(O)}{\tilde q}})\c{{\tilde W}^{(0)}}\big|^2\right)^{\!\!\frac{1}{2}}\\
&&\ML\ML\leq c\varepsilon\!\left(c_1\frac{J!}{J^{\a}}\frac{e^{(J-2)(\de+\Ga({\la},{\nu}))}}{\ro_{0,1}^J}
\right)\!\left(\int_{\la_0}^{{\la}}\frac{1}{|\la'|^{\frac{3}{2}}}\right)\leq c'\varepsilon\!\left(c_1\frac{J!}{J^{\a}}\frac{e^{(J-2)(\de+\underline{\Ga}({\la}))}}{\ro_{0,1}^J}\right)\ .\nn
\eea
Therefore the terms which have a role in the internal bootstrap are under control and we are left only with terms which can be estimated by the inductive assumptions namely,
\bea
&&\ML\ML\sum_{l=0}^{J-2}\cbin{J-1}{l} \int_{V_{(\la,\nu)}} \!\!\tau_+^4\!\left((\Lie_O^{l}{\!\ ^{(O)}{\tilde q}}_{\a\c\la})\lie_O^{J-1-l}{{\tilde W}^{(0)\a\la}}_{\ \ \ \ \ \ 4\c}\right)\c\a(\lie^2_O{\tilde W}^{(J-2)})\nn\\
&&\ML\ML+\sum_{k=0}^{J-2}\int_{V_{(\la,\nu)}} \!\!\tau_+^4\ \!\Lie_O^k\!\left({\!\ ^{(O)}{\tilde q}}_{\a\c\la}{{\tilde W}^{(J-1-k)\a\la}}_{\ \ \ \ \ \ 4\c}\right)\c\a(\lie^2_O{\tilde W}^{(J-2)})\nn
\eea
and the analogous term with $\b$. Let us look at the first sum, proceeding as before we have,
\bea
&&\ML\ML\sum_{l=0}^{J-2}\cbin{J-1}{l} \int_{V_{(\la,\nu)}} \!\!\tau_+^4\!\left((\Lie_O^{l}{\!\ ^{(O)}{\tilde q}}_{\a\c\la})\lie_O^{J-1-l}{{\tilde W}^{(0)\a\la}}_{\ \ \ \ \ \ 4\c}\right)\c\a(\lie^2_O{\tilde W}^{(J-2)})\nn\\
&&\ML\ML\leq \left(\sup_{\la'\in[\la_0,\la]}{\cal Q}_2^{(J-2)}(\la',\nu)\right)^{\!\!\frac{1}{2}}\left[\sum_{l=0}^{J-2}\cbin{J-1}{l} 
 \int_{\la_0}^{\la}d\la'\left(\int_{C(\la';[\nu_0,\nu])}\!\!\tau_+^4\big|(\Lie_O^{l}{\!\ ^{(O)}{\tilde q}})\c(\lie_O^{J-1-l}{{\tilde W}^{(0)}})\big|^2\right)^{\!\!\frac{1}{2}}\right]\nn\\
\eea
the second sum can be treated analogously, finally we have 
  \bea
 &&\ML\ML\ML\ML\big({\QQ_2}^{(J-2)}({\la},{\nu}))+\QQb_2^{(J-2)}({\la},{\nu}))\big)^{\frac{1}{2}}\leq\nn\\
  &&\ML\ML\ML\ML\leq \left[\left(C_{0}^{(1)}\frac{J!}{J^{\a}}\frac{e^{(J-2)(\de+\underline{\Ga}({\la}))}}{\ro_{0,1}^J}\right)+c'\varepsilon\!\left(c_1\frac{J!}{J^{\a}}\frac{e^{(J-2)(\de+\underline{\Ga}({\la}))}}{\ro_{0,1}^J}\right)\right]\nn\\
  &&\ML\ML\ML+\left[\sum_{l=0}^{J-2}\cbin{J-1}{l} 
 \int_{\la_0}^{{\la}}d\la'\left(\int_{C(\la';[\nu_0,\nu])}\!\!\tau_+^4\big|(\Lie_O^{l}{\!\ ^{(O)}{\tilde q}})\c({{\tilde W}^{(J-1-l)}})\big|^2\right)^{\!\!\frac{1}{2}}\right] \left(\sup_{\la'\in[\la_0,\la]}{\cal Q}_2^{(J-2)}(\la',\nu)\right)^{\!\!\frac{1}{2}}\nn\\
 &&\ML\ML\ML+\left[\sum_{k=0}^{J-2}\sum_{l=0}^k\cbin{k}{l}\int_{\la_0}^{{\la}}d\la'\left(\int_{C(\la';[\nu_0,\nu])}\!\!\tau_+^4\big|(\Lie_O^l{\!\ ^{(O)}{\tilde q}})\c({{\tilde W}^{(J-1-l)}})\big|^2\right)^{\!\!\frac{1}{2}}\right] \left(\sup_{\la'\in[\la_0,\la]}{\cal Q}_2^{(J-2)}(\la',\nu)\right)^{\!\!\frac{1}{2}}\ .\nn\\ \ 
  \eea
 The last two sums can be written in a more compact way as
 \bea
 \left[\sum_{k=0}^{J-1}\sum_{l=0}^{k\wedge(J-2)}\cbin{k}{l}\int_{\la_0}^{{\la}}d\la'\left(\int_{C(\la';[\nu_0,\nu])}\!\!\tau_+^4\big|(\Lie_O^l{\!\ ^{(O)}{\tilde q}})\c({{\tilde W}^{(J-1-l)}})\big|^2\right)^{\!\!\frac{1}{2}}\right]\ ,\nn
 \eea
 where $k\wedge(J-2)=min[k,(J-2)]$.

Therefore assuming for the moment this as the only error contribution we have
\bea
&&\ML\ML\ML{\cal Q}_2^{(J-2)}(\la,\nu)+\QQb_2^{(J-2)}(\la,\nu)\!\leq \left[\left(C_{0}^{(1)}\frac{J!}{J^{\a}}\frac{e^{(J-2)(\de+\Ga({\la},{\nu}))}}{\ro_{0,1}^J}\right)+c'\varepsilon\!\left(c_1\frac{J!}{J^{\a}}\frac{e^{(J-2)(\de+\Ga({\la},{\nu}))}}{\ro_{0,1}^J}\right)\right]\nn\\
&&\nn\\
&&\ML\ML\ML+\left(\sup_{\la'\in[\la_0,\la]}{\cal Q}_2^{(J-2)}(\la',\nu)\right)^{\!\!\frac{1}{2}}\sum_{k=0}^{k\wedge(J-2)}\left[\sum_{l=0}^k\cbin{k}{l}
 \int_{\la_0}^{\la}d\la'\left(\int_{C(\la';[\nu_0,\nu])}\!\!\tau_+^4| (\Lie_O^l{\!\ ^{(O)}{\tilde q}})({\tilde W}^{(J-1-l)})|^2\right)^{\!\!\frac{1}{2}}\right]\ .\nn
\eea

\NI We write
\bea
&&\ML\ML{\cal Q}_2^{(J-2)}(\la,\nu)+\QQb_2^{(J-2)}(\la,\nu)\!\leq \left[\left(C_{0}^{(1)}\frac{J!}{J^{\a}}\frac{e^{(J-2)(\de+\Ga({\la},{\nu}))}}{\ro_{0,1}^J}\right)+c'\varepsilon\!\left(c_1\frac{J!}{J^{\a}}\frac{e^{(J-2)(\de+\Ga({\la},{\nu}))}}{\ro_{0,1}^J}\right)\right]\nn\\
&&\nn\\
&&\ML\ML+\left(\sup_{\la'\in[\la_0,\la]}{\cal Q}_2^{(J-2)}(\la',\nu)\right)^{\!\!\frac{1}{2}}\left\{\sum_{k=0}^{J-2}\left[
 \int_{\la_0}^{\la}d\la'\left(\int_{C(\la';[\nu_0,\nu])}\!\!\tau_+^4| ({\!\ ^{(O)}{\tilde q}})({\tilde W}^{(J-1)})|^2\right)^{\!\!\frac{1}{2}}\right]\right.\nn\\
 &&\ML\ML\left.+\sum_{k=0}^{k\wedge(J-2)}\left[\sum_{l=1}^k\cbin{k}{l}
 \int_{\la_0}^{\la}d\la'\left(\int_{C(\la';[\nu_0,\nu])}\!\!\tau_+^4| (\Lie_O^l{\!\ ^{(O)}{\tilde q}}))({\tilde W}^{(J-1-l)})|^2\right)^{\!\!\frac{1}{2}}\right]\right\}\nn\\
\eea
Let us consider the first sum in the curly brackets, first of all we write
\bea
&&\ML\ML\left(\int_{C(\la';[\nu_0,\nu])}\!\!\tau_+^4|({\!\ ^{(O)}{\tilde q}})({\tilde W}^{(J-1)})|^2\right)^{\!\!\frac{1}{2}}\nn\\
&&\ML\ML\leq \left(\sup_{V(\la,\nu)}|r^{\si}{\!\ ^{(O)}{\tilde q}}|^2\right)^{\!\!\frac{1}{2}}\!\left(\int_{C(\la';[\nu_0,\nu])}\!\!\tau_+^4r^{-2\si}|\lie_O^{(J-1)}W|^2\right)^{\!\!\frac{1}{2}} .\ \ \ \ \ \ \ \ \ \ \ \ 
\eea
 As before, the weight $r^{\si}$ is not specified as it depends on which component of ${\!\ ^{(O)}\pi}$ one is considering, but it will turn out that it is systematically the appropriate one, hence we obtain the following estimate,
 \bea
 &&\ML\ML\int_{C(\la';[\nu_0,\nu])}\!\!\tau_+^4r^{-2\si}|{\tilde W}^{(J-1)}|^2
 \leq \int_{\nu_0}^{\nu}\tau_+^4r^{-2\si}\int_{S(\la',\nu)}\frac{1}{r^2(\la',\nu')}|{\tilde W}^{(J-1)}|^2\nn\\
 &&\ML\ML\leq \frac{1}{r^{2+2\si}(\la',\nu_0)}\!\int_{\nu_0}^{\nu}\tau_+^4\!\int_{S(\la',\nu)}|\Psi({\tilde W}^{(J-1)})|^2
 \leq \frac{1}{r^{2+2\si}(\la',\nu_0)}\!{\cal Q}_2^{(J-3)}(\la',\nu)\nn\\
 &&\ML\ML\leq \frac{c}{|\la'|^{2+2\si}}{\cal Q}_2^{(J-2)}(\la',\nu)\ . 
 \eea
and from it, observing that
\[\left(\sup_{\la'\in[\la_0,\la]}|r^{\si}{\!\ ^{(O)}{\tilde q}}|^2\right)^{\!\!\frac{1}{2}}=O(\varepsilon)\ ,\]
\bea
&&\ML\ML\sum_{k=0}^{J-2}\left[
 \int_{\la_0}^{\la}d\la'\left(\int_{C(\la';[\nu_0,\nu])}\!\!\tau_+^4|({\!\ ^{(O)}{\tilde q}})({\tilde W}^{(J-1)})|^2\right)^{\!\!\frac{1}{2}}\right]\nn\\
 &&\ML\ML\leq  c\left(\sup_{V(\la,\nu)}|r^{\si}{\!\ ^{(O)}{\tilde q}}|^2\right)^{\!\!\frac{1}{2}}\!(J-1)\!\int_{\la_0}^{\la}d\la'
  \frac{1}{|\la'|^{1+1\si}}\!\left({\cal Q}_2^{(J-3)}(\la',\nu)\right)^{\frac{1}{2}}\\
  &&\ML\ML\leq  c\ep_0 (J-1)\!\int_{\la_0}^{\la}d\la'
  \frac{1}{|\la'|^{1+1\si}}\!\left({\cal Q}_2^{(J-3)}(\la',\nu)\right)^{\frac{1}{2}}\nn,
  \eea
  so that, we have,
  \bea\
&&\ML\ML{\cal Q}_2^{(J-2)}(\la,\nu)+\QQb_2^{(J-2)}(\la,\nu)\!\leq  \left[\left(C_{0}^{(1)}\frac{J!}{J^{\a}}\frac{e^{(J-2)(\de+\Ga({\la},{\nu}))}}{\ro_{0,1}^J}\right)+c'\varepsilon\!\left(c_1\frac{J!}{J^{\a}}\frac{e^{(J-2)(\de+\Ga({\la},{\nu}))}}{\ro_{0,1}^J}\right)\right]\nn\\
\nn\\
&&\ML\ML+\left(\sup_{\la'\in[\la_0,\la]}{\cal Q}_2^{(J-2)}(\la',\nu)\right)^{\!\!\frac{1}{2}}\left\{c\ep_0 (J-1)\!\int_{\la_0}^{\la}d\la'
  \frac{1}{|\la'|^{1+\si}}\!\left({\cal Q}_2^{(J-3)}(\la',\nu)\right)^{\frac{1}{2}}\right.\nn\\
 &&\ML\ML\left.+\sum_{k=0}^{k\wedge(J-2)}\left[\sum_{l=1}^k\cbin{k}{l}
 \int_{\la_0}^{\la}d\la'\left(\int_{C(\la';[\nu_0,\nu])}\!\!\tau_+^4|(\Lie_O^l{\!\ ^{(O)}{\tilde q}})\lie_O^{l}({\tilde W}^{(J-1-l)})|^2\right)^{\!\!\frac{1}{2}}\right]\right\}\nn\\ \eql{1665}
\eea
{

1) Let us estimate the first part,
\smallskip

 \bea
 &&\left(\sup_{\la'\in[\la_0,\la]}{\cal Q}_2^{(J-2)}(\la',\nu)\right)^{\!\!\frac{1}{2}}\left\{c\ep_0 (J-1)\!\int_{\la_0}^{\la}d\la'
  \frac{1}{|\la'|^{1+\si}}\!\left({\cal Q}_2^{(J-3)}(\la',\nu)\right)^{\frac{1}{2}}\right.\nn\\
&&\leq\left(C^{*(1)}\frac{J!}{J^{\a}}\frac{e^{(J-3)(\de+\underline{\Ga}(\la))}}{\ro_{0,1}^J}\right)\!\left\{c\ep_0 (J-1)\!\int_{\la_0}^{\la}d\la'
  \frac{1}{|\la'|^{1+\si}}\!\left({ \QQ_2}^{(J-3)}(\la',\nu)\right)^{\!\!\frac{1}{2}}\right\}\nn\\
  &&\nn\\
&&\ML\leq \left[(c\varepsilon)\!\left(C^{*(1)}\frac{\ro}{e^{(\de+\underline{\Ga}_0(\la))}}\frac{J!}{J^{\a}}\frac{e^{(J-2)(\de+\underline{\Ga}(\la))}}{\ro^J}\right)^{\!\!2}\!\frac{(J-1)}{J}\!\!\int_{\la_0}^{\la}d\la'\frac{1}{|\la'|^{1+\si}}e^{(J-2)(-\underline{\Ga}(\la)+\underline{\Ga}(\la'))}\right]\nn\\
&&\ML\ML\leq \left[(c\varepsilon)\!\left(C^{*(1)}\frac{\ro}{e^{(\de+\underline{\Ga}_0(\la))}}\frac{J!}{J^{\a}}\frac{e^{(J-2)(\de+\underline{\Ga}(\la))}}{\ro_{0,1}^J}\right)^{\!\!2}\!\frac{(J-1)}{J}\!\!\int_{\la_0}^{\la}d\la'\frac{1}{|\la'|^2}e^{(J-2)\frac{(\la'-\la)}{\la'\la}}\right]\ .\nn
\eea
Therefore
\bea
 &&\ML\ML\left({ \QQ_2}^{(J-2)}(\la,\nu))+\QQb_2^{(J-2)}(\la,\nu))\right)^{\frac{1}{2}}\\
 &&\ML\ML\leq \left(C_{0}^{(1)}\frac{J!}{J^{\a}}\frac{e^{(J-2)(\de+\underline{\Ga}_0(\la))}}{\ro_{0,1}^J}\right)
+(c\varepsilon)\!\left(C^{*(1)}\frac{\ro}{e^{(\de+\underline{\Ga}_0(\la))}}\frac{J!}{J^{\a}}\frac{e^{(J-2)(\de_0+\underline{\Ga}_0(\la))}}{\ro_{0,1}^J}\right)\!\left[\frac{(J-1)}{J}\!\!\int_{\la_0}^{\la}d\la'\frac{1}{|\la'|^2}e^{(J-2)\frac{(\la'-\la)}{\la'\la}}\right]^{\!\frac{1}{2}}\nn
 \eea
Observe that denoting, 
\[x=\frac{(\la'-\la)}{\la'\la}\ \ ;\ \ \frac{dx}{d\la'}=\frac{1}{\la'^2}\ ,\]
therefore
\bea
&&\ML\int_{\la_0}^{\la}d\la'\frac{1}{\la'^2}e^{(J-2)\frac{(\la'-{\la})}{\la'{\la}}}
=-\int_0^{x_0} dxe^{(J-2)x}=\frac{c}{J-2}
\eea
as $x_0=\frac{(\la_0-{\la})}{\la_0{\la}}<0$\ and, finally,

\bea
&&\left({ \QQ_2}^{(J-2)}(\la,\nu))+\QQb_2^{(J-2)}(\la,\nu))\right)^{\frac{1}{2}}\leq \left[C_{0}^{(1)}+(c'\varepsilon)C^{*(1)}\right]\!\left(\frac{J!}{J^{\a}}\frac{e^{(J-2)(\de+\underline{\Ga}({\la}))}}{\ro_{0,1}^J}\right)\nn\\
&&<\left(C^{*(1)}\frac{J!}{J^{\a}}\frac{e^{(J-2)(\de+\underline{\Ga}(\la))}}{\ro_{0,1}^J}\right)\ ,
 \eea
choosing $\varepsilon$ sufficiently small, which implies that it can be extended to any $\nu$ up to $\nu_*$ and to the largest value of $\la=\la_1$. Therefore the result holds everywhere and the control of the more delicate term of the error is done.
 \smallskip

\NI Let us consider now the second term in curly brackets. First we estimate
 \bea
 &&\ML\ML \left(\int_{C(\la';[\nu_0,\nu])}\!\!\tau_+^4\big|(\Lie_O^l{\!\ ^{(O)}{\tilde q}})\c({{\tilde W}^{(J-1-l)}})\big|^2\right)^{\!\!\frac{1}{2}}\nn\\
 &&\ML\ML \leq \int_{\nu_0}^{\nu}d\nu'\tau_+^4\int_{S(\la',\nu')}d\nu|\Lie_O^l{\!\ ^{(O)}{\tilde q}})|^2|{{\tilde W}^{(J-1-l)}})|^2\nn\\
 &&\ML\ML \leq \int_{\nu_0}^{\nu}d\nu'\tau_+^4|\Lie_O^l{\!\ ^{(O)}{\tilde q}})|_{4,S(\la,\nu')}^2|{{\tilde W}^{(J-1-l)}})|_{4,S(\la',\nu')}^2\nn\\
 &&\ML\ML\leq c_1\frac{(l+1)!}{(l+1)^{\a}}\frac{e^{(l-1)(\de+\underline{\Ga}(\la))}}{\ro_{0,1}^{l+1}}\left(C_1\frac{(J-l)!}{(J-l)^{\a}}\frac{e^{(J-l-2)(\de+\Ga(\la',\nu))}}{\ro_{0,1}^{J-l}}\right)\frac{1}{|\la'|^3}\c\nn\\
 &&\ML\ML\int_{\nu_0}^{\nu}d\nu'\frac{\tau_+^4}{r^{3+3}}\nn\\
 &&\ML\ML\leq \left(C_1\frac{J!}{J^{\a}}\frac{e^{(J-2)(\de+\underline{\Ga}(\la'))}}{\ro_{0,1}^{J}}\right)\left(c_1\frac{e^{-\de}}{\ro_{0,1}}\right)
 \frac{1}{|\la'|^3}\left[\frac{J^{\a}}{J!}\frac{(l+1)!}{(l+1)^{\a}}\frac{(J-l)!}{(J-l)^{\a}}\right]\nn\\
 &&\ML\ML\int_{\nu_0}^{\nu}\frac{d\nu'}{\nu'^2}\nn\\
 &&\ML\ML\leq \left(C_1\frac{J!}{J^{\a}}\frac{e^{(J-2)(\de+\underline{\Ga}(\la))}}{\ro_{0,1}^{J}}\right)\left(c_1c_2\frac{e^{-\de}}{\ro_{0,1}}\right)
 \frac{1}{|\la'|^3}\left[\frac{1}{J-1}\frac{J^{\a}}{J!}\frac{(l+1)!}{(l+1)^{\a}}\frac{(J-l)!}{(J-l)^{\a}}\right]\ .\nn
 \eea
 Therefore
  \bea\label{16543}
 &&\ML\ML\left[\sum_{k=0}^{J-1}\sum_{l=0}^{k\wedge(J-2)}\cbin{k}{l}\int_{\la_0}^{{\la}}d\la'\left(\int_{C(\la';[\nu_0,\nu])}\!\!\tau_+^4\big|(\Lie_O^l{\!\ ^{(O)}{\tilde q}})\c({{\tilde W}^{(J-1-l)}})\big|^2\right)^{\!\!\frac{1}{2}}\right]\nn\\
 &&\ML\ML\leq\left(C_1\frac{J!}{J^{\a}}\frac{e^{(J-2)(\de+\Ga({{\la}},\nu))}}{\ro_{0,1}^{J}}\right)\left(c_1c_2\frac{e^{-\de}}{\ro_{0,1}}\right)\c\nn\\
 &&\ML\ML\left[
 \sum_{k=0}^{J-1}\sum_{l=0}^{k\wedge(J-2)}\cbin{k}{l}\left(\frac{1}{J-1}\frac{J^{\a}}{J!}\frac{(l+1)!}{(l+1)^{\a}}\frac{(J-l)!}{(J-l)^{\a}}\right)\int_{\la_0}^{{\la}}\frac{d\la'}{|\la'|^3}e^{-(J-2)(\Ga(\la)-\Ga(\la'))}
 \right]\nn\\
 &&\ML\ML\leq\left(C_1\frac{J!}{J^{\a}}\frac{e^{(J-2)(\de+\Ga({{\la}},\nu))}}{\ro_{0,1}^{J}}\right)\left(c_1c_2c_3\frac{e^{-\de}}{\ro_{0,1}}\right)\left[\frac{1}{(J-1)}
 \sum_{k=0}^{J-1}\sum_{l=0}^{k\wedge(J-2)}\frac{k!}{l!(k-l)!}\frac{J^{\a}}{J!}\frac{(l+1)!}{(l+1)^{\a}}\frac{(J-l)!}{(J-l)^{\a}}\right]\nn\\
 \eea

Let us estimate the  term: 
\bea
&& \sum_{k=0}^{J-1}\sum_{l=0}^{k\wedge(J-2)}\frac{k!}{J!}\frac{(J-l)!}{(k-l)!}\frac{J^{\a}}{(J-l)^{\a}}\frac{1}{(l+1)^{\a-1}}\leq\sum_{k=0}^{J-1}\sum_{l=0}^{k\wedge(J-2)}\frac{J^{\a}}{(J-l)^{\a}}\frac{1}{(l+1)^{\a-1}}\nn
 \eea
We estimate this term with $\a=3$,
\bea
&&\sum_{k=0}^{J-1}\sum_{l=0}^{k\wedge(J-2)}\frac{J^{\a}}{(J-l)^{\a}}\frac{1}{(l+1)^{\a-1}}\leq C\int_0^{J-1}dk\int_0^k \frac{J^3}{(J-l)^{3}}\frac{dl}{(l+1)^{2}}=\nn\\
&&C\int_0^{J-1}dk\frac{J^3(\frac{(J+1)^2}{(J-l)^2} +\frac{4(J+1)}{J-l} -\frac{2(J+1)}{l+1}-6\ln(J-l)+6\ln(l+1))}{2(J+1)^4}\Big{|}_0^{k}\leq\nn\\
&&C'\int_0^{J-1}dk\frac{J^3\frac{(J+1)^2}{(J-k)^2}}{2(J+1)^4}\leq C''\int_0^{J-1}\frac{J+1}{(J-k)^2}dk\leq C''' (J+1)\nn
\eea
\NI And exploiting equation \ref{16543}, we obtain 
\bea
&&\ML\ML\left.\sum_{k=0}^{k\wedge(J-2)}\left[\sum_{l=1}^k\cbin{k}{l}
 \int_{\la_0}^{\la}d\la'\left(\int_{C(\la';[\nu_0,\nu])}\!\!\tau_+^4|(\Lie_O^l{\!\ ^{(O)}{\tilde q}})\lie_O^{l}({\tilde W}^{(J-1-l)})|^2\right)^{\!\!\frac{1}{2}}\right]\right\}\nn\\ \eql{16655}
&&\ML\ML\leq\left(C^{*(1)}\frac{J!}{J^{\a}}\frac{e^{(J-2)(\de+\underline{\Ga}(\la))}}{\ro_{0,1}^J}\right)
\eea
It is easy to prove that all the other terms of the error can be treated in the same way, hence we can consider proved Theorem \ref{Qconserv1}
\subsection{Step 5, the estimate of the norms of $\nabb^J\Psi$ in the interior region.}\label{165}
We state first the following lemmas, 
\medskip

\NI {\bf Lemma \ref{new1}}
{\em Assuming the following estimate for the norms of $\Psi(\lie_O^{J-1}W)$ which follow immediately from the  ${\cal Q}$ norms estimates in Section \ref{Qxxx},
\bea
\big|\Psi(\lie_O^{J-1}W)\big|_{p,S}\leq \left(C_0\frac{J!}{J^{\a}}\frac{e^{(J-2)(\de+\underline{\Ga}(\la)}}{\ro_{0,1}^J}\right)\ ,
\eea
the following estimate holds for the norms of $\Lie_O^{J-1}\Psi(W)$,
\bea
\big|\Lie_O^{J-1}\Psi\big|_{p,S}\leq \left(C_1\frac{J!}{J^{\a}}\frac{e^{(J-2)(\de+\underline{\Ga}(\la)}}{\ro_{0,1}^J}\right)\nn
\eea}
\begin{Le}\label{new2}
from the result of the previous lemma the following estimates hold,
\bea
\big|\nabb_O^{J-1}\Psi\big|_{p,S}\leq \left(C_2\frac{J!}{J^{\a}}\frac{e^{(J-2)(\de+\underline{\Ga}(\la)}}{\ro_{0,1}^J}\right).\ \eql{new2xx}
\eea

\end{Le}
\NI {\bf Proof:} Their proof is an adapted repetition of the Proof of Lemma \ref{L4.2}  , see subsection \ref{subxx}.
\medskip

\NI Once we have proved these lemmas we can prove Lemma \ref{fromLieOtoNabb}
we recall here,
\smallskip

\NI{\bf Lemma \ref{fromLieOtoNabb}}
 
\NI 
Assuming the norm estimates for the null Riemann component of lemma \ref{new2} and the  estimate of corollary \ref{cor9.9bis}  for the auxiliary coefficient, $h$, the following estimates hold:
\begin{theorem}

\bea\label{911}
\big| |\la|^{\underline{\phi}(\Psi)}r^{\phi(\Psi)+(J-1)-\frac{2}{p}}\nabb^{J-1}\Psi\big|_{p,S}\leq 
\left(C_4\frac{J!}{J^{\a}}\frac{e^{(J-2)(\de+\underline{\Ga}(\la)}}{\ro^J}\right)
\eea
with $\ro<\ro_{0,1}$, $C_4$ of order $O(1)$ for $J\geq 5$, and $O(\varepsilon)$ with $\ep_0^2<\varepsilon<\ep_0\ $ for $J< 5$,\footnote{If $J\geq 1$, for $J=0$ $\ro$ becomes $\ro-{\underline\ro}$, see \cite{Kl-Ni:book}.}
\bea
&&\ML\phi(\a)=\phi(\b)=\frac{7}{2}\ ,\ \phi(\ro)=\phi(\si)=3\ , \phi(\bb)=2\ ,\ \phi(\aa)=1\nn\\
&&\ML\underline{\phi}(\a)=\underline{\phi}(\b)=0\ ,\ \underline{\phi}(\ro)=\underline{\phi}(\si)=\frac{1}{2}\ , \underline{\phi}(\bb)=\frac{3}{2}\ ,\ \underline{\phi}(\aa)=\frac{5}{2}\ .\ \ 
\eea
\medskip
\end{theorem}
\NI {\bf Proof:}

\NI Recall that 
\bea
&&\ML\ML\ML\ML |\nabb^J{\Psi}|^2=\sum_{A_1}\sum_{A_2}..\sum_{A_J}e^{\mu_1}_{A_1}e^{\nu_1}_{A_1}e^{\mu_2}_{A_2}e^{\nu_2}_{A_1}\c\c\c e^{\mu_J}_{A_J}e^{\nu_J}_{A_J}\nabb_{\mu_1}\nabb_{\mu_2}\c\c\nabb_{\mu_J}
{\Psi}_{\mu}\nabb_{\nu_1}\nabb_{\nu_2}\c\c\nabb_{\nu_J}{\Psi}_{\nu}g^{\mu\nu}\nn\\
&&\ML\ML\ML\ML =g^{\mu\nu}\sum_{A_1}\sum_{A_2}..\sum_{A_J}\left(e^{\mu_1}_{A_1}e^{\mu_2}_{A_2}\c\c e^{\mu_J}_{A_J}\nabb_{\mu_1}\nabb_{\mu_2}\c\c\nabb_{\mu_J}{\Psi}_{\mu}\right)\left(e^{\nu_1}_{A_1}e^{\nu_2}_{A_2}\c\c e^{\nu_J}_{A_J}\nabb_{\nu_1}\nabb_{\nu_2}\c\c\nabb_{\nu_J}{\Psi}_{\nu}\right)\nn\\
\eea
and
\bea
&&\ML\ML\ML\ML |\nabb^J{\Psi}|^4=g^{\mu\nu}\sum_{A_1}\sum_{A_2}..\sum_{A_J}\left(e^{\mu_1}_{A_1}e^{\mu_2}_{A_2}\c\c e^{\mu_J}_{A_J}\nabb_{\mu_1}\nabb_{\mu_2}\c\c\nabb_{\mu_J}{\Psi}_{\mu}\right)\left(e^{\nu_1}_{A_1}e^{\nu_2}_{A_2}\c\c e^{\nu_J}_{A_J}\nabb_{\nu_1}\nabb_{\nu_2}\c\c\nabb_{\nu_J}{\Psi}_{\nu}\right)\nn\\
&&\ML\ML\ML\ML g^{\tau\si}\sum_{B_1}\sum_{B_2}..\sum_{B_J}\left(e^{\mu_1}_{B_1}e^{\mu_2}_{B_2}\c\c e^{\mu_J}_{B_J}\nabb_{\mu_1}\nabb_{\mu_2}\c\c\nabb_{\mu_J}{\Psi}_{\tau}\right)\left(e^{\nu_1}_{B_1}e^{\nu_2}_{B_2}\c\c e^{\nu_J}_{B_J}\nabb_{\nu_1}\nabb_{\nu_2}\c\c\nabb_{\nu_J}{\Psi}_{\si}\right)\nn\\
&&\ML\ML\ML\ML=g^{\mu\nu}g^{\tau\si}\sum_{A_1}\sum_{A_2}..\sum_{A_J}\sum_{B_1}\sum_{B_2}..\sum_{B_J}\left(e^{\mu_1}_{A_1}e^{\mu_2}_{A_2}\c\c e^{\mu_J}_{A_J}\nabb_{\mu_1}\nabb_{\mu_2}\c\c\nabb_{\mu_J}{\Psi}_{\mu}\right)\left(e^{\mu_1}_{B_1}e^{\mu_2}_{B_2}\c\c e^{\mu_J}_{B_J}\nabb_{\mu_1}\nabb_{\mu_2}\c\c\nabb_{\mu_J}{\Psi}_{\tau}\right)\nn\\
&&\ML\ML\ML\ML\c\left[\left(e^{\nu_1}_{A_1}e^{\nu_2}_{A_2}\c\c e^{\nu_J}_{A_J}\nabb_{\nu_1}\nabb_{\nu_2}\c\c\nabb_{\nu_J}{\Psi}_{\nu}\right)\left(e^{\mu_1}_{B_1}e^{\mu_2}_{B_2}\c\c e^{\mu_J}_{B_J}\nabb_{\mu_1}\nabb_{\mu_2}\c\c\nabb_{\mu_J}{\Psi}_{\tau}\right)\right]\nn\\
&&\ML\ML\ML\ML\leq \big|g^{\mu\nu}g^{\tau\si}\sum_{A_1}\sum_{A_2}..\sum_{A_J}\sum_{B_1}\sum_{B_2}..\sum_{B_J}\left(e^{\mu_1}_{A_1}e^{\mu_2}_{A_2}\c\c e^{\mu_J}_{A_J}\nabb_{\mu_1}\nabb_{\mu_2}\c\c\nabb_{\mu_J}{\Psi}_{\mu}\right)\left(e^{\mu_1}_{B_1}e^{\mu_2}_{B_2}\c\c e^{\mu_J}_{B_J}\nabb_{\mu_1}\nabb_{\mu_2}\c\c\nabb_{\mu_J}{\Psi}_{\tau}\right)\big|\nn\\
&&\ML\ML\ML\ML\c\sup_{A_1,...,A_J,B_1,...,B_J}\bigg|\left[\left(e^{\nu_1}_{A_1}e^{\nu_2}_{A_2}\c\c e^{\nu_J}_{A_J}\nabb_{\nu_1}\nabb_{\nu_2}\c\c\nabb_{\nu_J}{\Psi}_{\nu}\right)\left(e^{\mu_1}_{B_1}e^{\mu_2}_{B_2}\c\c e^{\mu_J}_{B_J}\nabb_{\mu_1}\nabb_{\mu_2}\c\c\nabb_{\mu_J}{\Psi}_{\tau}\right)\right]\bigg|\nn\\
&&\ML\ML\ML\ML\leq \big|g^{\mu\nu}g^{\tau\si}\sum_{A_1}\sum_{A_2}..\sum_{A_J}\sum_{B_1}\sum_{B_2}..\sum_{B_J}\left(e^{\mu_1}_{A_1}e^{\mu_2}_{A_2}\c\c e^{\mu_J}_{A_J}\nabb_{\mu_1}\nabb_{\mu_2}\c\c\nabb_{\mu_J}{\Psi}_{\mu}\right)\left(e^{\mu_1}_{B_1}e^{\mu_2}_{B_2}\c\c e^{\mu_J}_{B_J}\nabb_{\mu_1}\nabb_{\mu_2}\c\c\nabb_{\mu_J}{\Psi}_{\tau}\right)\big|\nn\\
&&\ML\ML\ML\ML\c\big|\nabb^J{\Psi}\big|^2\ .
\eea
Therefore
\bea
&&\ML\ML\ML\ML\ML\ML |\nabb^J{\Psi}|^2\leq c\big|g^{\mu\nu}g^{\tau\si}\sum_{A_1}\sum_{A_2}..\sum_{A_J}\sum_{B_1}\sum_{B_2}..\sum_{B_J}\left(e^{\mu_1}_{A_1}e^{\mu_2}_{A_2}\c\c e^{\mu_J}_{A_J}\nabb_{\mu_1}\nabb_{\mu_2}\c\c\nabb_{\mu_J}{\Psi}_{\mu}\right)\left(e^{\mu_1}_{B_1}e^{\mu_2}_{B_2}\c\c e^{\mu_J}_{B_J}\nabb_{\mu_1}\nabb_{\mu_2}\c\c\nabb_{\mu_J}{\Psi}_{\tau}\right)\big|\nn\\
&&\ML\ML\ML\ML\ML\ML\ML\ML\ML \leq c_1^J\frac{1}{r^{2J}}\sup_{\mu\nu\tau\si}\big|g^{\mu\nu}g^{\tau\si}\big|\sup_{\mu}\left|\left(\sum_{i_1}\sum_{i_2}..\sum_{i_J}O^{\mu_1}_{i_1}O^{\mu_2}_{i_2}\c\c O^{\mu_J}_{i_J}\nabb_{\mu_1}\nabb_{\mu_2}\c\c\nabb_{\mu_J}
{\Psi}_{\mu}\right)\right|\sup_{\nu}\left|\left(\sum_{l_1}\sum_{l_2}..\sum_{l_J}
O^{\nu_1}_{l_1}O^{\nu_2}_{l_2}\c\c O^{\nu_J}_{l_J}\nabb_{\nu_1}\nabb_{\nu_2}\c\c\nabb_{\nu_J}{\Psi}_{\nu}\right)\right|\ .\nn
\eea
Denoting
\[O^{\mu} \equiv \sum_{l}O^{\mu}_{l}\]
we need a more explicit expression of
\[\sup_{\mu}\big|O^{\mu_1}O^{\mu_2}\c\c O^{\mu_J}\nabb_{\mu_1}\nabb_{\mu_2}\c\c\nabb_{\mu_J}{\Psi}_{\mu}\big|\ .\]
\bea
&&\ML\ML\ML\ML O^{\mu_1}O^{\mu_2}\c\c O^{\mu_J}\nabb_{\mu_1}\nabb_{\mu_2}\c\c\nabb_{\mu_J}{\Psi}_{\mu}\nn\\
&&\ML\ML\ML\ML=O^{\mu_1}O^{\mu_2}\c\c O^{\mu_J}\de^{\ro_1}_{\mu_1}\de^{\ro_2}_{\mu_2}\c\c\de^{\ro_J}_{\mu_j}\nabb_{\ro_1}\nabb_{\ro_2}\c\c\nabb_{\ro_J}{\Psi}_{\mu}=O^{\mu_1}O^{\mu_2}\c\c O^{\mu_J}\de^{\ro_1}_{\mu_1}\nabb_{\ro_1}\de^{\ro_2}_{\mu_2}\nabb_{\ro_2}\c\c\de^{\ro_J}_{\mu_j}\nabb_{\ro_J}{\Psi}_{\mu}\nn\\
&&\ML\ML\ML\ML=O^{\mu_1}O^{\mu_2}\c\c O^{\mu_J}\left[\sum_{A_1}\sum_{A_2}..\sum_{A_J}(e_{A_1})_{\mu_1}e_{A_1}^{\ro_1}\nabb_{\ro_1}(e_{A_2})_{\mu_1}e_{A_2}^{\ro_2}\nabb_{\ro_2}\c\c(e_{A_J})_{\mu_J}e_{A_J}^{\ro_J}\nabb_{\ro_J}{\Psi}_{\mu}\right]\nn\\
&&\ML\ML\ML\ML=O^{\mu_2}\c\c O^{\mu_J}\left[\sum_{A_1}\sum_{A_2}..\sum_{A_J}\gggg(O,e_{A_1})\nabb_{A_1}(e_{A_2})_{\mu_2}\nabb_{A_2}\c\c(e_{A_J})_{\mu_J}\nabb_{A_J}{\Psi}_{\mu}\right]\ .
\eea
We have,
\bea
&&\ML\ML\ML\ML \sup_{\mu}\big|O^{\mu_1}O^{\mu_2}\c\c O^{\mu_J}\nabb_{\mu_1}\nabb_{\mu_2}\c\c\nabb_{\mu_J}{\Psi}_{\mu}\big|\nn\\
&&\ML\ML\ML\ML \leq |O|^{J-1}\left(\sup_A|\gggg(O,e_A)|\right)2^J\sup_{\mu}\sup_{A_2,A_3,...,A_J}\left|\nabb_{A_1}(e_{A_2})_{\mu_2}\nabb_{A_2}\c\c(e_{A_J})_{\mu_J}\nabb_{A_J}{\Psi}_{\mu}\right|\nn\\
&&\ML\ML\ML\ML\leq C_1^J\sup_{\mu}\sup_{A_2,A_3,...,A_J;\mu_2,\mu_3,...,\mu_J}\left|\nabb_{A_1}(e_{A_2})_{\mu_2}\nabb_{A_2}\c\c(e_{A_J})_{\mu_J}\nabb_{A_J}{\Psi}_{\mu}\right|\ .\eql{1698xx}
\eea
We are left with estimating 
\bea\label{nabbea}
\left|\nabb_{A_1}(e_{A_2})_{\mu_2}\nabb_{A_2}\c\c(e_{A_J})_{\mu_J}\nabb_{A_J}{\Psi}_{\mu}\right|\ ,
\eea
This goes basically as Lemma \ref{L4.1}, the main difference being that now the derivatives are $\nabb_A{\cal O}$ which are estimated in term of the $\nabb_O{\cal O}$ and, therefore, with a different $\ro$ namely a $\ro_{0,3}$ such that $\ro_{0,3}>\ro$ 

\NI This would imply, finally, that the $C_1^J$ in front is controlled by the factor $\left(\frac{\ro}{\ro_{0,3}}\right)^{\!J}$.
\medskip

\NI Let us reproduce the Lemma \ref{L4.1} applied to this case.

\NI We write $\nabb_{A_1}(e_{A_2})_{\mu_2}\nabb_{A_2}\c\c(e_{A_J})_{\mu_J}\nabb_{A_J}{\Psi}$, omitting the $\mu$ indices, as 
\[ \nabb_{A_1}(e_{A_2})\nabb_{A_2}\c\c(e_{A_J})\nabb_{A_J}{\Psi}\ .\]
As the $e_A$ are endowed with a given name, therefore numbered, we think at each one as defining a ``slot" namely the position at its left, therefore there are $J-1$ slots associated to the $(J\!-\!1)$ $e_A$'s we are considering; we add an extra slot which corresponds to the position immediately at the left of $\Psi$. Therefore we have $J$ slots and we imagine to distribute in all the possible ways the $J$ $\nabb_{A_i}$ in these slots. Clearly in this way we are counting more than the possible terms we can really  obtain  as each $\nabb_A$  cannot operate on the $e_A$'s at its left. Assuming the described distribution of the $\nabb$'s we have (the equality is formal and it has to be interpreted as an upper bounds when we consider the norms),
\bea
&&\ML\ML\nabb_{A_1}(e_{A_2})\nabb_{A_2}\c\c(e_{A_J})\nabb_{A_J}{\Psi}\nn\\
&&\ML\ML``=" \left\{\sum_{\ga_1,\ga_2,...,\ga_{J-2}\ga_{J}}^{\sum_{s=1}^{J}\!\!\ga_s=J;\ga_{J}\geq1}
\frac{J!}{\ga_1!\ga_2!\c\c\c\ga_{J-1}!\ga_{J}!}(\nabb_{A}^{\ga_1}e_{A_2})(\nabb_{A}^{\ga_2}e_{A_3})\c\c\c(\nabb_{A}^{\ga_{J-1}}e_{A_J})(\nabb_{A}^{\ga_{J}}\Psi)\right\}\ .\ \ \ \nn
\eea
We denote $\ga_{J-1}=q$ and rewrite the previous expression as
\bea
&&\ML\ML\nabb_{A_1}(e_{A_2})\nabb_{A_2}\c\c(e_{A_J})\nabb_{A_J}{\Psi}\nn\\
&&\ML\ML``=" \left\{\sum_{q=1}^{J}\cbin{J}{q}\nabb_A^{q}\Psi\!\!\!\!\!\!\!\!\sum_{\ga_1,\ga_2,...,\ga_{J-2}}^{\ \ \ \ \ \sum_{s=1}^{J-1}\!\!\ga_s=(J-q)}
\!\!\!\!\!\!\!\!\frac{(J\!-\!q)!}{\ga_1!\ga_2!\c\c\c\ga_{J-1}!}(\nabb_{A}^{\ga_1}e_{A_2})(\nabb_{A}^{\ga_2}e_{A_3})\c\c\c(\nabb_{A}^{\ga_{J-1}}e_{A_J})\right\}\ .\ \ \ \nn
\eea
Denoting $k=J-q$ we rewrite the previous expression as
\bea
&&\ML\ML\nabb_{A_1}(e_{A_2})\nabb_{A_2}\c\c(e_{A_J})\nabb_{A_J}{\Psi}\nn\\
&&\ML\ML``=" \left\{\sum_{k=0}^{J}\cbin{J}{k}\nabb_A^{J-k}\Psi\!\!\!\!\!\!\!\!\sum_{\ga_1,\ga_2,...,\ga_{J-1}}^{\ \ \ \ \ \sum_{s=1}^{J-1}\!\!\ga_s=k}
\!\!\!\!\frac{k!}{\ga_1!\ga_2!\c\c\c\ga_{J-2}!}(\nabb_{A}^{\ga_1}e_{A_2})(\nabb_{A}^{\ga_2}e_{A_3})\c\c\c(\nabb_{A}^{\ga_{J-1}}e_{A_J})\right\}\nn\\
&&\ML\ML=\nabb_A^{J}\Psi+\left\{\sum_{k=1}^{J}\cbin{J}{k}\nabb_A^{J-k}\Psi\!\!\!\!\!\!\!\!\sum_{\ga_1,\ga_2,...,\ga_{J-1}}^{\ \ \ \ \ \sum_{s=1}^{J-1}\!\!\ga_s=k}
\!\!\!\!\frac{k!}{\ga_1!\ga_2!\c\c\c\ga_{J-1}!}(\nabb_{A}^{\ga_1}e_{A_2})(\nabb_{A}^{\ga_2}e_{A_3})\c\c\c(\nabb_{A}^{\ga_{J-1}}e_{A_J})\right\}\ .\nn
\eea
Going to the norms, the estimate becomes
\bea
&&\ML\ML|\nabb_{A_1}(e_{A_2})\nabb_{A_2}\c\c(e_{A_J})\nabb_{A_J}{\Psi}|_{p,S}\leq|\nabb_A^{J}\Psi|_{p,S}\\
&&\ML\ML+\left\{\sum_{k=1}^{J}\cbin{J}{k}|\nabb_A^{J-k}\Psi|_{p,S}\!\!\!\!\!\!\!\!\sum_{\ga_1,\ga_2,...,\ga_{J-1}}^{\ \ \ \ \ \sum_{s=1}^{J-1}\!\!\ga_s=k}
\!\!\!\!\frac{k!}{\ga_1!\ga_2!\c\c\c\ga_{J-1}!}\bigg[|(\nabb_{A}^{\ga_1}e_{A_2})|_{\infty,S}|(\nabb_{A}^{\ga_2}e_{A_3})|_{\infty,S}\c\c\c|(\nabb_{A}^{\ga_{J-1}}e_{A_J})|_{\infty,S}\bigg]\right\}\ .\nn
\eea
We use now the results of theorem \ref{angh2} and of corollary \ref{angcorr} for the norms with $\nabb^J_A$ of the connection coefficients, and of the the Riemann null components estimates and the norms $\nabb^J_A$ of the $e_A$ versors, \footnote{In fact the weight factors in the following estimates are for some components a little better, see Theorem \ref{Thinitialdata} of Section \ref{S.s initial data}, but this is irrelevant here.}
\footnote{The estimate in the third line follows as (assuming $r=1$) although the $O$ generators are at the metric level, nevertheless in the estimates of the angular derivatives of $\ga_{ab}$ and of $X^a$ there is a loss of derivatives, see subsection \ref{xsec}, which implies the same estimates as for the connection coefficients.}
\bea\label{nabbio}
&&|\nabb_A^{J}\Psi|_{p,S}\leq C\frac{(J+1)!}{(J+1)^\a}\frac{e^{(J-1)\de}e^{(J-1)\underline{\Ga}(\la)}}{\ro_{0,3}^{J+1}}\nn\\
&& |r^{J+\phi({\cal O})}\nabb_A^{J}e_{A_i}|_{\infty,S}\leq C
\frac{(J+1)!}{(J+1)^\a}\frac{e^{(J-1)\de}e^{(J-1)\underline{\Ga}(\la)}}{\ro_{0,3}^{J+1}}\ .
\eea 
\medskip


\NI From these estimates it follows that 
\bea
&&\de(\sum_{s=1}^{J-1}\!\!\ga_s=k)\bigg[|(\nabb_{A}^{\ga_1}e_{A_2})|_{\infty,S}|(\nabb_{A}^{\ga_2}e_{A_3})|_{\infty,S}\c\c\c|(\nabb_{A}^{\ga_{J-1}}e_{A_J})|_{\infty,S}\bigg]\nn\\
&&\leq
\frac{e^{k(\de+\Ga)}}{\ro_{0,3}^{k+J-1}}\frac{(\ga_1+1)!(\ga_2+1)!\c\c\c(\ga_{J-1}+1)!e^{-(\de+\Ga)(\ep{(\ga_1)}+\ep{(\ga_2)}+\c\c\c+\ep{(\ga_{J-1})})}}{(1+\ga_1)^\a(1+\ga_2)^\a\c\c\c(1+\ga_{J-1})^\a}\nn
\eea
and we can rewrite the previous estimate,  as
\bea
&&\ML\ML\ML|\nabb_{A_1}(e_{A_2})\nabb_{A_2}\c\c(e_{A_J})\nabb_{A_J}{\Psi}|_{p,S}\leq|\nabb_A^{J}\Psi|_{p,S}
+\left\{\sum_{k=1}^{J-2}\cbin{J-1}{k}|\nabb_A^{J-k}\Psi|_{p,S}\frac{e^{k(\de+\Ga)}}{\ro_{0,3}^k}k!\ \c\right.\nn\\
&&\ \ \ \ \left.\c\left[\sum_{\ga_1,\ga_2,...,\ga_{J-2}}^{\ \ \ \ \ \sum_{s=1}^{J-2}\!\!\ga_s=k}
\!\!\!\!\frac{C^{(\ep{(\ga_1)}+\ep{(\ga_2)}+\c\c\c+\ep{(\ga_{J-1}))}}e^{-(\de_0+\underline{\Ga}_0(\la))(\ep{(\ga_1)}+\ep{(\ga_2)}+\c\c\c+\ep{(\ga_{J-2})})}}{(1+\ga_1)^\a(1+\ga_2)^\a\c\c\c(1+\ga_{J-1})^\a}\right]\right\}\nn\ .
\eea

\NI Applying the previous estimates for the various terms we have
\bea
&&\ML\ML\ML|\nabb_{A_1}(e_{A_2})\nabb_{A_2}\c\c(e_{A_J})\nabb_{A_J}{\Psi}|_{p,S}\leq \left\{C\frac{(J+1)!}{(J+1)^\a}\frac{e^{(J-1)(\de+\underline{\Ga}(\la))}}{\ro_{0,3}^{(J+1)}}\right.\nn\\
&&\ML\ML\ML\ML\left.+\sum_{k=1}^{J-2}\cbin{J-1}{k}C\frac{(J+1-k)!}{(J+1-k)^\a}\frac{e^{((J-k)-1)(\de+\Ga)}}{\ro_{0,3}^{J+1-k}}\frac{e^{k(\de+\underline{\Ga}(\la))}}{\ro_{0,3}^k}k!\big|\left[{\cal F}\right]\big|\right\}\eql{4.18x}
\eea
where
\bea
\ML\ML\left[{\cal F}\right]\equiv\left[\sum_{\ga_1,\ga_2,...,\ga_{J-1}}^{\ \ \ \ \ \sum_{s=1}^{J-1}\!\!\ga_s=k}
\!\!\!\!\frac{C^{(\ep{(\ga_1)}+\ep{(\ga_2)}+\c\c\c+\ep{(\ga_{J-2}))}}e^{-(\de+\underline{\Ga}(\la))(\ep{(\ga_1)}+\ep{(\ga_2)}+\c\c\c+\ep{(\ga_{J-2})})}}{\ro_{0,3}^{J-1}(1+\ga_1)^\a(1+\ga_2)^\a\c\c\c(1+\ga_{J-1})^\a}\right]\ .\ \ \ \ 
\eea
Therefore we rewrite \ref{4.18x} as
\bea
&&\ML\ML\ML\ML\ML\ML|\nabb_{A_1}(e_{A_2})\nabb_{A_2}\c\c(e_{A_J})\nabb_{A_J}{\Psi}|_{p,S}\leq 
\left(C_3\frac{(J+1)!}{(J+1)^\a}\frac{e^{(J-1)(\de+\underline{\Ga}(\la))}}{\ro_{0,3}^{(J+1)}}\right)\!\!\left\{1+\sum_{k=1}^{J-1}\cbin{J-1}{k}\frac{(J+1-k)!k!}{(J+1)!}\frac{(J+1)^{\a}}{(J-k)^{\a}}
\big|\left[{\cal F}\right]\big|\right\}\nn\\
&&\ML\ML\ML\ML\ML\leq \left(C\frac{(J+1)!}{(J+1)^\a}\frac{e^{(J-1)(\de+\underline{\Ga}(\la))}}{\ro_{0,3}^{(J+1)}}\right)\!
\!\left\{1+\sum_{k=1}^{J-1}\frac{(J+1-k)(J-k)}{(J+1)J}\frac{(J+1)^{\a}}{(J-k)^{\a}}\big|\left[{\cal F}\right]\big|\right\}\nn\\
&&\ML\ML\ML\ML\ML\leq  \left(C\frac{(J+1)!}{(J+1)^\a}\frac{e^{(J-1)(\de+\underline{\Ga}(\la))}}{\ro_{0,3}^{(J+1)}}\right)\!
\!\left\{1+\sum_{k=1}^{J-1}\frac{(J+1)^{\a-2}}{(J-k)^{\a-2}}\big|\left[{\cal F}\right]\big|\right\}\ .\eql{4.16sss}
\eea
Provide $\a$ is sufficiently large we have 
\bea
\big|\left[{\cal F}\right]\big|\leq C^{J-1}
\eea
so that
\bea
&&\ML\ML\ML|\nabb_{A_1}(e_{A_2})\nabb_{A_2}\c\c(e_{A_J})\nabb_{A_J}{\Psi}|_{p,S}\leq 
C\!\left(\frac{(J+1)!}{(J+1)^\a}\frac{e^{(J-1)(\de+\underline{\Ga}(\la)}}{\ro_{0,4}^{(J+1)}}\right)\!\!
\left(\frac{\ro_{0,4}}{\ro_{0,3}}\right)^{\!\!J+1}\!\!C^{J+1}\nn\\
&&\ML\ML\ML\leq 
C\left(\left(\frac{\ro_{0,4}}{\ro_{0,3}}\right)^{\!\!J+1}\!\!C^{J+1}\right)\!\!\left(\frac{(J+1)!}{(J+1)^\a}\frac{e^{(J-1)(\de+\underline{\Ga}(\la))}}{\ro_{0,4}^{J+1}}\right)\leq C\!\left(\frac{(J+1)!}{(J+1)^\a}\frac{e^{(J-1)(\de+\underline{\Ga}(\la))}}{\ro_{0,4}^{J+1}}\right)\ .\ \ \ \ \ \ \ \ \ \ \ \ \ \ \ \ 
\eea
\NI Provided $\ro_{0,4}$ sufficiently small. Therefore from \ref{1698xx} we have
\bea
&&\ML\ML\ML\ML \sup_{\mu}\big|O^{\mu_1}O^{\mu_2}\c\c O^{\mu_J}\nabb_{\mu_1}\nabb_{\mu_2}\c\c\nabb_{\mu_J}{\Psi}_{\mu}\big|\nn\\
&&\ML\ML\ML\ML \leq |O|^{J-1}\left(\sup_A|\gggg(O,e_A)|\right)2^J\sup_{\mu}\sup_{A_2,A_3,...,A_J}\left|\nabb_{A_1}(e_{A_2})_{\mu_2}\nabb_{A_2}\c\c(e_{A_J})_{\mu_J}\nabb_{A_J}{\Psi}_{\mu}\right|\nn\\
&&\ML\ML\ML\ML\leq C^J\sup_{\mu}\sup_{A_2,A_3,...,A_J;\mu_2,\mu_3,...,\mu_J}\left|\nabb_{A_1}(e_{A_2})_{\mu_2}\nabb_{A_2}\c\c(e_{A_J})_{\mu_J}\nabb_{A_J}{\Psi}_{\mu}\right|\nn\\
&&\ML\ML\ML\ML\leq C^JC\!\left(\frac{(J+1)!}{(J+1)^\a}\frac{e^{(J-1)(\de+\underline{\Ga}(\la))}}{\ro_{0,4}^{J+1}}\right)\leq C\!\left(\frac{(J+1)!}{(J+1)^\a}\frac{e^{(J-1)(\de+\underline{\Ga}(\la))}}{\ro_{0,5}^{J+1}}\right)
\eea
\NI for suitable $\ro_{0,5}<\ro_{0,4}$ and
\bea
&&\ML\ML\ML\ML\ML\ML |\nabb^J{\Psi}|^2\leq c\big|g^{\mu\nu}g^{\tau\si}\sum_{A_1}\sum_{A_2}..\sum_{A_J}\sum_{B_1}\sum_{B_2}..\sum_{B_J}\left(e^{\mu_1}_{A_1}e^{\mu_2}_{A_2}\c\c e^{\mu_J}_{A_J}\nabb_{\mu_1}\nabb_{\mu_2}\c\c\nabb_{\mu_J}{\Psi}_{\mu}\right)\left(e^{\mu_1}_{B_1}e^{\mu_2}_{B_2}\c\c e^{\mu_J}_{B_J}\nabb_{\mu_1}\nabb_{\mu_2}\c\c\nabb_{\mu_J}{\Psi}_{\tau}\right)\big|\nn\\
&&\ML\ML\ML\ML\ML\ML\ML\ML\ML \leq c_1^J\frac{1}{r^{2J}}\sup_{\mu\nu\tau\si}\big|g^{\mu\nu}g^{\tau\si}\big|\sup_{\mu}\left|\left(\sum_{i_1}\sum_{i_2}..\sum_{i_J}O^{\mu_1}_{i_1}O^{\mu_2}_{i_2}\c\c O^{\mu_J}_{i_J}\nabb_{\mu_1}\nabb_{\mu_2}\c\c\nabb_{\mu_J}
{\Psi}_{\mu}\right)\right|\sup_{\nu}\left|\left(\sum_{l_1}\sum_{l_2}..\sum_{l_J}
O^{\nu_1}_{l_1}O^{\nu_2}_{l_2}\c\c O^{\nu_J}_{l_J}\nabb_{\nu_1}\nabb_{\nu_2}\c\c\nabb_{\nu_J}{\Psi}_{\nu}\right)\right|\nn\\
&&\ML\ML\ML\ML\ML\ML\ML\ML\ML \leq C\!\left(\frac{(J+1)!}{(J+1)^\a}\frac{e^{(J-1)(\de+\underline{\Ga}(\la))}}{\ro^{J+1}}\right)
\eea
with $\ro<\ro_{0,1}$, where we omit the weight factors $|\la|^{\a}r^{\b}$ as they come out in a obvious way. Therefore Lemma \ref{fromLieOtoNabb} has been proven.
\medskip

\NI To summarize, the previous lemma proves that, apart some technicalities, we have an estimate  of the following kind 
\bea
|r^J\nabb^J{\Psi}|\leq C^J|\lie_O^J\Psi|.
\eea
\NI To control this $C^J$ factor we have to lower the $\ro$ of the estimate of $|\lie_O^J\Psi|$. 

\subsection{Proof of Lemma \ref{from|LieTLieOtoDTNabb}}\label{LLL16.6}

\NI We repeat here the statement of Lemma \ref{from|LieTLieOtoDTNabb}:
\medskip

\NI{\bf Lemma \ref{from|LieTLieOtoDTNabb}}
{\em 
Let us assume 
 the estimates of theorem \ref{324567},
 then the null Riemann components  $\Psi$ satisfy the following estimates  for $p+q=J$ on ${\cal K}$   with $\ro_1<\ro<\ro_{0,1}$ depending only the first $s$ derivatives of the initial data, with $s\leq 7$  and with a suitable choice of the  $ {\tilde C}^{(1)}$ constant of order $O(\varepsilon)$ for $J \leq7$\footnote{If $J\geq 1$, for $J=0$ $\ro$ becomes $\ro-{\underline\ro}$, see \cite{Kl-Ni:book}.} and $O(1)$ otherwise., \footnote{Here $\ro_1$ plays the same role of $\ro$ for the angular derivatives estimates.}
\bea
\big| |\la|^{\underline{\phi}(\Psi)}r^{\phi(\Psi)+k-\frac{2}{p}}D_S^p\nabb^{k}\Psi\big|_{L^p(S)}\leq {\tilde C}^{(1)}e^{(J-2)\de}e^{(J-2)\underline{\Ga}(\la)}\frac{J!}{J^\a}\frac{1}{\ro_1^J}\ ,
\eea

\bea
&&\ML\phi(\a)=\phi(\b)=\frac{7}{2}\ ,\ \phi(\ro)=\phi(\si)=3\ , \phi(\bb)=2\ ,\ \phi(\aa)=1\nn\\
&&\ML\underline{\phi}(\a)=\underline{\phi}(\b)=0\ ,\ \underline{\phi}(\ro)=\underline{\phi}(\si)=\frac{1}{2}\ , \underline{\phi}(\bb)=\frac{3}{2}\ ,\ \underline{\phi}(\aa)=\frac{5}{2}\ .\ \ 
\eea}

\smallskip

\NI {\bf Proof:} Preliminary we recall Theorem \ref{324567}, whose Proofs are again an adapted repetition of the Proof of Lemma \ref{L4.2}.

\NI {\bf Theorem \ref{324567}}
\NI{\em Under the assumptions  \ref{inizz1} and \ref{inizz2} on the mixed derivatives of the initial data connection coefficients on $C_0\cup\Cb_0$, the following estimates hold, with $\tilde{C}^{(1)}$ of order $O(\varepsilon)$ for any $J \leq 7$ and order $O(1)$ otherwise.
\bea
| |\la|^{\underline{\phi}(\Psi)}r^{\phi(\Psi)-\frac{2}{p}}\Psi(\lie_S^p\lie_O^kW)|_{p,S}(\la,\nu)\leq {\tilde C}^{(1)}\frac{J!}{J^{\a}}\frac{e^{(J-2)(\de+\underline{\Ga}(\la))}}{\ro_{0,1}^J}\  .
\eea
\bea
\big||\la|^{\underline{\phi}(\Psi)}r^{\phi(\Psi)-\frac{2}{p}}\Lie_S^p\Lie_O^{k}\Psi\big|_{p,S}\leq {\tilde C}^{(1)}\frac{J!}{J^{\a}}\frac{e^{(J-2)(\de+\underline{\Ga}(\la))}}{\ro_{0,1}^J}\ .
\eea
\smallskip}


\NI Let us now prove lemma \ref{from|LieTLieOtoDTNabb}, for the sake of simplicity we omit the $r^{\phi(\c)}$ term.  From the result of the previous lemma the following estimates hold,

\NI The proof is a repetition of lemma \ref{fromLieOtoNabb}, recall, in fact that 
\bea
&&\ML\ML\ML\ML |D_S^p\nabb^k{\Psi}|^2=\sum_{A_1}\sum_{A_2}..\sum_{A_k}e^{\mu_1}_{A_1}e^{\nu_1}_{A_1}e^{\mu_2}_{A_2}e^{\nu_2}_{A_1}\c\c\c e^{\mu_k}_{A_k}e^{\nu_k}_{A_k}D_S^p\nabb_{\mu_1}\nabb_{\mu_2}\c\c\nabb_{\mu_k} 
{\Psi}_{\mu}D_S^p\nabb_{\nu_1}\nabb_{\nu_2}\c\c\nabb_{\nu_J}{\Psi}_{\nu}g^{\mu\nu}\nn\\
&&\ML\ML\ML\ML =g^{\mu\nu}\sum_{A_1}\sum_{A_2}..\sum_{A_k}\left(e^{\mu_1}_{A_1}e^{\mu_2}_{A_2}\c\c e^{\mu_k}_{A_k}D_S^p\nabb_{\mu_1}\nabb_{\mu_2}\c\c\nabb_{\mu_k}{\Psi}_{\mu}\right)\left(e^{\nu_1}_{A_1}e^{\nu_2}_{A_2}\c\c e^{\nu_k}_{A_k}D_S^p\nabb_{\nu_1}\nabb_{\nu_2}\c\c\nabb_{\nu_k}{\Psi}_{\nu}\right)\nn\\
\eea
\NI and
\bea
&&\ML\ML\ML\ML |D_S^p\nabb^J{\Psi}|^4=g^{\mu\nu}\sum_{A_1}\sum_{A_2}..\sum_{A_k}\left(e^{\mu_1}_{A_1}e^{\mu_2}_{A_2}\c\c e^{\mu_k}_{A_k}D_S^p\nabb_{\mu_1}\nabb_{\mu_2}\c\c\nabb_{\mu_k}{\Psi}_{\mu}\right)\left(e^{\nu_1}_{A_1}e^{\nu_2}_{A_2}\c\c e^{\nu_k}_{A_k}D_S^p\nabb_{\nu_1}\nabb_{\nu_2}\c\c\nabb_{\nu_k}{\Psi}_{\nu}\right)\nn\\
&&\ML\ML\ML\ML g^{\tau\si}\sum_{B_1}\sum_{B_2}..\sum_{B_k}\left(e^{\mu_1}_{B_1}e^{\mu_2}_{B_2}\c\c e^{\mu_k}_{B_k}D_S^p\nabb_{\mu_1}\nabb_{\mu_2}\c\c\nabb_{\mu_k}{\Psi}_{\tau}\right)\left(e^{\nu_1}_{B_1}e^{\nu_2}_{B_2}\c\c e^{\nu_k}_{B_k}D_S^p\nabb_{\nu_1}\nabb_{\nu_2}\c\c\nabb_{\nu_k}{\Psi}_{\si}\right)\nn\\
&&\ML\ML\ML\ML=g^{\mu\nu}g^{\tau\si}\sum_{A_1}\sum_{A_2}..\sum_{A_k}\sum_{B_1}\sum_{B_2}..\sum_{B_k}\left(e^{\mu_1}_{A_1}e^{\mu_2}_{A_2}\c\c e^{\mu_k}_{A_k}D_S^p\nabb_{\mu_1}\nabb_{\mu_2}\c\c\nabb_{\mu_k}{\Psi}_{\mu}\right)\left(e^{\mu_1}_{B_1}e^{\mu_2}_{B_2}\c\c e^{\mu_k}_{B_k}D_S^p\nabb_{\mu_1}\nabb_{\mu_2}\c\c\nabb_{\mu_k}{\Psi}_{\tau}\right)\nn\\
&&\ML\ML\ML\ML\c\left[\left(e^{\nu_1}_{A_1}e^{\nu_2}_{A_2}\c\c e^{\nu_k}_{A_k}D_S^p\nabb_{\nu_1}\nabb_{\nu_2}\c\c\nabb_{\nu_k}{\Psi}_{\nu}\right)\left(e^{\mu_1}_{B_1}e^{\mu_2}_{B_2}\c\c e^{\mu_k}_{B_k}D_S^p\nabb_{\mu_1}\nabb_{\mu_2}\c\c\nabb_{\mu_k}{\Psi}_{\tau}\right)\right]\nn\\
&&\ML\ML\ML\ML\leq \big|g^{\mu\nu}g^{\tau\si}\sum_{A_1}\sum_{A_2}..\sum_{A_k}\sum_{B_1}\sum_{B_2}..\sum_{B_k}\left(e^{\mu_1}_{A_1}e^{\mu_2}_{A_2}\c\c e^{\mu_k}_{A_k}D_S^p\nabb_{\mu_1}\nabb_{\mu_2}\c\c\nabb_{\mu_k}{\Psi}_{\mu}\right)\left(e^{\mu_1}_{B_1}e^{\mu_2}_{B_2}\c\c e^{\mu_k}_{B_k}D_S^p\nabb_{\mu_1}\nabb_{\mu_2}\c\c\nabb_{\mu_k}{\Psi}_{\tau}\right)\big|\nn\\
&&\ML\ML\ML\ML\c\sup_{A_1,...,A_k,B_1,...,B_k}\bigg|\left[\left(e^{\nu_1}_{A_1}e^{\nu_2}_{A_2}\c\c e^{\nu_k}_{A_k}D_S^p\nabb_{\nu_1}\nabb_{\nu_2}\c\c\nabb_{\nu_k}{\Psi}_{\nu}\right)\left(e^{\mu_1}_{B_1}e^{\mu_2}_{B_2}\c\c e^{\mu_k}_{B_k}D_S^p\nabb_{\mu_1}\nabb_{\mu_2}\c\c\nabb_{\mu_k}{\Psi}_{\tau}\right)\right]\bigg|\nn\\
&&\ML\ML\ML\ML\leq \big|g^{\mu\nu}g^{\tau\si}\sum_{A_1}\sum_{A_2}..\sum_{A_k}\sum_{B_1}\sum_{B_2}..\sum_{B_k}\left(e^{\mu_1}_{A_1}e^{\mu_2}_{A_2}\c\c e^{\mu_k}_{A_k}D_S^p\nabb_{\mu_1}\nabb_{\mu_2}\c\c\nabb_{\mu_k}{\Psi}_{\mu}\right)\left(e^{\mu_1}_{B_1}e^{\mu_2}_{B_2}\c\c e^{\mu_k}_{B_k}D_S^p\nabb_{\mu_1}\nabb_{\mu_2}\c\c\nabb_{\mu_k}{\Psi}_{\tau}\right)\big|\nn\\
&&\ML\ML\ML\ML\c\big|D_S^p\nabb^k{\Psi}\big|^2\ .
\eea
\NI Therefore
\bea
&&\ML\ML\ML\ML\ML\ML |D_S^p\nabb^k{\Psi}|^2\leq c\big|g^{\mu\nu}g^{\tau\si}\sum_{A_1}\sum_{A_2}..\sum_{A_k}\sum_{B_1}\sum_{B_2}..\sum_{B_k}\left(e^{\mu_1}_{A_1}e^{\mu_2}_{A_2}\c\c e^{\mu_k}_{A_k}D_S^p\nabb_{\mu_1}\nabb_{\mu_2}\c\c\nabb_{\mu_k}{\Psi}_{\mu}\right)\left(e^{\mu_1}_{B_1}e^{\mu_2}_{B_2}\c\c e^{\mu_k}_{B_k}D_S^p\nabb_{\mu_1}\nabb_{\mu_2}\c\c\nabb_{\mu_k}{\Psi}_{\tau}\right)\big|\nn\\
&&\ML\ML\ML\ML\ML\ML\ML\ML\ML \leq c_1^k\frac{1}{r^{2k}}\sup_{\mu\nu\tau\si}\big|g^{\mu\nu}g^{\tau\si}\big|\sup_{\mu}\left|\left(\sum_{i_1}\sum_{i_2}..\sum_{i_k}O^{\mu_1}_{i_1}O^{\mu_2}_{i_2}\c\c O^{\mu_k}_{i_k}D_S^p\nabb_{\mu_1}\nabb_{\mu_2}\c\c\nabb_{\mu_k}
{\Psi}_{\mu}\right)\right|\sup_{\nu}\left|\left(\sum_{l_1}\sum_{l_2}..\sum_{l_k}
O^{\nu_1}_{l_1}O^{\nu_2}_{l_2}\c\c O^{\nu_k}_{l_k}D_S^p\nabb_{\nu_1}\nabb_{\nu_2}\c\c\nabb_{\nu_k}{\Psi}_{\nu}\right)\right|\ .\nn
\eea
\NI Denoting
\[O^{\mu} \equiv \sum_{l}O^{\mu}_{l},\]
\NI we need a more explicit expression of
\[\sup_{\mu}\big|O^{\mu_1}O^{\mu_2}\c\c O^{\mu_k}D_S^p\nabb_{\mu_1}\nabb_{\mu_2}\c\c\nabb_{\mu_k}{\Psi}_{\mu}\big|\ .\]
\bea
&&\ML\ML\ML\ML O^{\mu_1}O^{\mu_2}\c\c O^{\mu_k}D_S^p\nabb_{\mu_1}\nabb_{\mu_2}\c\c\nabb_{\mu_k}{\Psi}_{\mu}\nn\\
&&\ML\ML\ML\ML=O^{\mu_1}O^{\mu_2}\c\c O^{\mu_k}D_S^p\de^{\ro_1}_{\mu_1}\de^{\ro_2}_{\mu_2}\c\c\de^{\ro_k}_{\mu_k}\nabb_{\ro_1}\nabb_{\ro_2}\c\c\nabb_{\ro_k}{\Psi}_{\mu}=O^{\mu_1}O^{\mu_2}\c\c O^{\mu_k}D_S^p\de^{\ro_1}_{\mu_1}\nabb_{\ro_1}\de^{\ro_2}_{\mu_2}\nabb_{\ro_2}\c\c\de^{\ro_k}_{\mu_k}\nabb_{\ro_k}{\Psi}_{\mu}\nn\\
&&\ML\ML\ML\ML=O^{\mu_1}O^{\mu_2}\c\c O^{\mu_k}\left[\sum_{A_1}\sum_{A_2}..\sum_{A_k}D_S^p(e_{A_1})_{\mu_1}e_{A_1}^{\ro_1}\nabb_{\ro_1}(e_{A_2})_{\mu_1}e_{A_2}^{\ro_2}\nabb_{\ro_2}\c\c(e_{A_k})_{\mu_k}e_{A_k}^{\ro_k}\nabb_{\ro_k}{\Psi}_{\mu}\right]\nn\\
&&\ML\ML\ML\ML=O^{\mu_2}\c\c O^{\mu_k}\left[\sum_{A_1}\sum_{A_2}..\sum_{A_k}\gggg(O,D_S^p e_{A_1})\nabb_{A_1}(e_{A_2})_{\mu_2}\nabb_{A_2}\c\c(e_{A_k})_{\mu_k}\nabb_{A_k}{\Psi}_{\mu}\right]\nn\\
&&\ML\ML\ML\ML+O^{\mu_2}\c\c O^{\mu_k}\left[\sum_{A_1}\sum_{A_2}..\sum_{A_k}\gggg(O,e_{A_1})\gggg(D_S^pe_{A_1},e_D)\nabb_{D}(e_{A_2})_{\mu_2}\nabb_{A_2}\c\c(e_{A_k})_{\mu_k}\nabb_{A_k}{\Psi}_{\mu}\right]\nn\\
&&\ML\ML\ML\ML+O^{\mu_2}\c\c O^{\mu_k}\left[\sum_{A_1}\sum_{A_2}..\sum_{A_k}\gggg(O,e_{A_1})D_S^p\nabb_{A_1}(e_{A_2})_{\mu_2}\nabb_{A_2}\c\c(e_{A_k})_{\mu_k}\nabb_{A_k}{\Psi}_{\mu}\right]
\eea
\NI We have,
\bea
&&\ML\ML\ML\ML \sup_{\mu}\big|O^{\mu_1}O^{\mu_2}\c\c O^{\mu_k}\nabb_{\mu_1}D_S^p\nabb_{\mu_2}\c\c\nabb_{\mu_k}{\Psi}_{\mu}\big|\nn\\
&&\ML\ML\ML\ML\ML\ML \leq |O|^{k-1}\left(\sup_A|\gggg(O,D_S^pe_A)|+\sup_{A,D}|\gggg(O,e_A)||\gggg(D_S^pe_A,e_D)|\right)2^k\sup_{\mu}\sup_{A_2,A_3,...,A_k}\left|\nabb_{A_1}(e_{A_2})_{\mu_2}\nabb_{A_2}\c\c(e_{A_k})_{\mu_k}\nabb_{A_k}{\Psi}_{\mu}\right|\nn\\
&&\ML\ML\ML\ML\ML\ML +|O|^{k-1}\left(\sup_A|\gggg(O,e_A)|\right)2^k\sup_{\mu}\sup_{A_2,A_3,...,A_k}\left|D_S^p\nabb_{A_1}(e_{A_2})_{\mu_2}\nabb_{A_2}\c\c(e_{A_k})_{\mu_k}\nabb_{A_k}{\Psi}_{\mu}\right|\nn\\
&&\ML\ML\ML\ML\ML\ML\leq C_1^k\sup_{\mu}\sup_{A_2,A_3,...,A_k;\mu_2,\mu_3,...,\mu_k}\left[\left|\nabb_{A_1}(e_{A_2})_{\mu_2}\nabb_{A_2}\c\c(e_{A_k})_{\mu_k}\nabb_{A_k}{\Psi}_{\mu}\right|
+\left|D_S^p\nabb_{A_1}(e_{A_2})_{\mu_2}\nabb_{A_2}\c\c(e_{A_k})_{\mu_k}\nabb_{A_k}{\Psi}_{\mu}\right|\right]\ .\ \ \ \ \ \ \ \ \ \ \ \eql{1698xxxk}
\eea
\NI We are left with estimating \footnote{notice that in order to estimate $D_S^p\nabb_{A_1}(e_{A_2})_{\mu_2}\nabb_{A_2}\c\c(e_{A_k})_{\mu_k}\nabb_{A_k}{\Psi}_{\mu}$ we need to exploit the estimates of $D_S^P\nabb^Q {\cal O}$ which we have already obtained in Lemma \ref{cor10.3xx}.}
\[\left|D_S^p\nabb_{A_1}(e_{A_2})_{\mu_2}\nabb_{A_2}\c\c(e_{A_k})_{\mu_k}\nabb_{A_k}{\Psi}_{\mu}\right|\ ,\]
\NI This is a simple repetition with obvious modifications of what already done in the proof of Lemma \ref{fromLieOtoNabb} and we do not repeat here. The remaining part of the proof mimics the one of this lemma and we do not repeat it here.

\subsection{Proof of lemma  \ref{cor10.3xxx}}\label{158}

\NI We recall Lemma \ref{cor10.3xxx}:
\medskip

\NI {\bf Corollary \ref{cor10.3xxx}}
{\em Under the assumptions of Lemma \ref{from|LieTLieOtoDTNabb} the following estimates hold, with $p+k=J$: }

\bea
&&\big| |\la|^{\underline{\phi}(\Psi)}r^{\phi(\Psi)+(J-1)-\frac{2}{p}}D_{\nu}^p\nabb^{k}\Psi\big|_{L^p(S)}\leq \tilde{C}^{(1)}e^{(J-2)\de}e^{(J-2)\underline{\Ga}(\la)}\frac{J!}{J^\a}\frac{1}{\ro_2^p\ro_1^{k+1}}\nn\\ 
&&\big| |\la|^{\underline{\phi}(\Psi)+p}r^{\phi(\Psi)+k-\frac{2}{p}}D_{\la}^p\nabb^{k}\Psi\big|_{L^p(S)}\leq \tilde{C}^{(1)}e^{(J-2)\de}e^{(J-2)\underline{\Ga}(\la)}\frac{J!}{J^\a}\frac{1}{\ro_2^p\ro_1^{k+1}}\ .\ \ \ \ \ \ \nn
\eea
With suitable $\ro_2<\ro_1$.}

\medskip

\NI {\bf Remark:}  

\NI{\em In the following proof the term $r^{\phi(\Psi)}$ has been neglected but in fact should be reinstated.}
\medskip

\NI{\bf Proof:} The proof is a consequence of the following lemma,
\begin{Le}\label{cor10.3}
Defining $Q\geq (2|\oom|_{\infty,S}+1+\si|X|_{\infty,S}+O(\ep))$, then the following estimates hold,
\bea
\big||\ub|^k|u|^{l}r^{-\frac{2}{p}}
D_S^{p-l}D_{\la}^{l}\nabb^k\Psi\big|_{p,S}\leq  C^{(1)}e^{(J-2)\de}e^{(J-2)\underline{\Ga}(\la)}Q^{l}\frac{1}{\ro_1^{p+k+1}}.\nn\\
\eea
\end{Le}
{\bf Proof:} we have to prove the estimates by induction satrting from $l=0$;
recall that from Lemma 10.7, omitting the factor $r^{-\frac{2}{p}}$ always present, we have, with $p+k=J-1$
\bea
&&\ML\ML\big|\ub^k D_S^p\nabb^{k}\Psi\big|_{L^p(S)}\leq C^{(1)}e^{(J-2)\de}e^{(J-2)\underline{\Ga}(\la)}\frac{1}{\ro_{1}^{J}}\ \ \ \ \
\eea
We show that the estimates hold for $l=1$, in fact , if $\Psi=\aa$, omitting lower order terms, it follows,
recalling that
\[D_{\la}=\oom D_3+X\nabb\ ;\ \oom D_S=\frac{1}{2}(\ub D_{\nu}+u D_{\la})-\frac{u}{2}X\nabb\]
\bea
&&\ML\ML\ML\ML\big|\ub^k|u| D_S^{p-1}D_{\la}\nabb^{k}\aa\big|_{L^p(S)}\leq \big|\ub^k D_S^{p-1}(2\oom D_S+\ub D_{\nu}+uX\nabb)\nabb^{k}\aa\big|_{L^p(S)}\nn\\
&&\ML\ML\ML\ML\leq 2\big|\ub^k D_S^{p-1}\oom D_S\nabb^{k}\aa\big|_{L^p(S)}+\big|\ub^{k} D_S^{p-1} \ub D_{\nu}\nabb^{k}\aa\big|_{L^p(S)}+\big|\ub^{k} D_S^{p-1}u X\nabb^{k+1}\aa\big|_{L^p(S)}\nn\\
&&\ML\ML\ML\ML\leq 
\sum_{l=0}^{p-1}\cbin{p-1}{l}
\big|\ub^{k}(D_S^l\oom)D_S^{p-l}\nabb^k\aa\big|_{L^p(S)}+\sum_{l=0}^{p-1}\cbin{p-1}{l}
\big|\ub^{k}(D_S^l\ub)D_S^{p-1-l}D_{\nu}\nabb^k\aa\big|_{L^p(S)}\nn\\
&&\ML\ML\ML\ML+\sum_{l=0}^{p-1}\cbin{p-1}{l}
\big|\ub^{k}(D_S^luX)D_S^{p-1-l}\nabb^{k+1}\aa\big|_{L^p(S)}\nn\\
&&\ML\ML\ML\ML\leq 2\big|\oom\ub^k D_S^{p}\nabb^{k}\aa\big|_{L^p(S)}+\big|\ub^{k+1}D_S^{p-1} D_{\nu}\nabb^{k}\aa\big|_{L^p(S)}+\big|\ub^{k}uXD_S^{p-1}\nabb^{k+1}\aa\big|_{L^p(S)}\nn\\
&&\ML\ML\ML\ML+ \left\{\sum_{l=1}^{p-1}\cbin{p-1}{l}
\big|\ub^{k}(D_S^l\oom)D_S^{p-l}\nabb^k\aa\big|_{L^p(S)}+\sum_{l=1}^{p-1}\cbin{p-1}{l}
\big|\ub^{k}(D_S^l\ub)D_S^{p-1-l}D_{\nu}\nabb^k\aa\big|_{L^p(S)}\right.\nn\\
&&\ML\ML\ML\ML\left.+\sum_{l=1}^{p-1}\cbin{p-1}{l}
\big|\ub^{k}(D_S^luX)D_S^{p-1-l}\nabb^{k+1}\aa\big|_{L^p(S)}\right\}\ .
\eea
We write in a symbolic way,
\bea
D_{\nu}\nabb^k\a=\nabb^{k+1}\a +\{l.o.t.\}
\eea
and we will omit for the moment the contribution due to the $\{l.o.t.\}$ part. Moreover we write approximately,
\bea
D_S\ub=\frac{1}{2}(u\pr_3+\ub\pr_4)u=\frac{1}{2\oom}(u\pr_{u}+\ub\pr_{\ub})\ub+O(\nabb \ub)=
\frac{1}{2\oom}(u\pr_{u}+\ub\pr_{\ub})\ub=\frac{\ub}{2\oom}=\frac{\ub}{4}\nn
\eea
and
\bea
D_S^l\ub=\frac{\ub}{2^{2l}}\ .
\eea
Moreover 
\bea
D_S\oom=\frac{1}{2}(u\pr_3+\ub\pr_4)\oom=\frac{\oom}{2}(u\pr_3+\ub\pr_4)\log\oom
=-\oom(\ub\om+u\omb)\equiv{\cal O}
\eea
and we recall that we have the estimates for  $D_S^j{\cal O}$ as we control $D_{\nu}^j\om$, $D_{\nu}^j\omb$, $D_{\la}^j\om$, $D_{\la}^j\omb$. Therefore we have
\bea
&&\ML\ML\sum_{l=1}^{p-1}\cbin{p-1}{l}
\big|\ub^{k}(D_s^l\ub)D_S^{p-1-l}D_{\nu}\nabb^k\aa\big|_{L^p(S)} \leq
\sum_{l=1}^{p-1}\cbin{p-1}{l}\frac{1}{2^{2l}}
\big|\ub^{k+1}D_S^{p-1-l}\nabb^{k+1}\aa\big|_{L^p(S)}\nn\\
&&\ML\ML\leq  C_4e^{(J-2)\de}e^{(J-2)\underline{\Ga}(\la)}\sum_{l=1}^{p-1}\cbin{p-1}{l}\frac{1}{2^{2l}}\frac{1}{\ro_1^{p-1-l}\ro_1^{k+2}}
\leq O(\ro_1) C_4e^{(J-2)\de}e^{(J-2)\underline{\Ga}(\la)}\frac{1}{\ro_1^p\ro_1^{k+1}}\nn\\
\eea
\bea
&&\ML\ML \sum_{l=1}^{p-1}\cbin{p-1}{l}
\big|\ub^{k}(D_S^l\oom)D_S^{p-l}\nabb^k\aa\big|_{L^p(S)}\leq \sum_{l=1}^{p-1}\cbin{p-1}{l}
|(D_S^l\oom)|_{\infty,S}\big|\ub^{k}D_S^{p-l}\nabb^k\aa\big|_{L^p(S)}\nn\\
&&\ML\ML  \sum_{l=1}^{p-1}\cbin{p-1}{l}
|(D_S^l\oom)|_{\infty,S}C_4e^{(p-l+k-2)(\de+\underline{\Ga}(\la))}\frac{(p-l+k+1)!}{(p-l+k+1)^{\a}}\frac{1}{\ro_1^{p-l+k+1}}\ \ \
\eea
Moreover \footnote{The factor $(l+1)$ instead of $l$ is due to the delicate estimates of $D_{\nu}^j\om$ and of $D_{\la}^j\omb$.}
\bea
|D_S^l\oom|_{\infty,S}=|D_S^{l-1}{\cal O}|_{\infty,S}\leq ce^{(l-1)(\de+\underline{\Ga}(\la))}\frac{(l+1)!}{(l+1)^{\a}}\frac{1}{\ro_1^{l+1}}\eql{estimp}
\eea
{\bf Remark:}  

\NI {\em Notice that in the estimate \ref{estimp} turns out that in the denominator there is $\ro_1$ and not $\ro$; this follows as we already know that the norms of $\D_S\Psi$ have $\ro_1$, see Lemma \ref{cor10.3}, this implies the the estimates of the $\D_S$ of the connection coefficients  $\D_S^J{\cal O}$ have the same $\ro_{1}$ .}

\NI Therefore,
\bea
&&\ML\ML\ML\ML \sum_{l=1}^{p-1}\cbin{p-1}{l}
\big|\ub^{k}(D_S^l\oom)D_S^{p-l}\nabb^k\aa\big|_{L^p(S)}\nn\\
&&\ML\ML\ML\ML \leq \sum_{l=1}^{p-1}\cbin{p-1}{l}
ce^{(l-1)(\de+\underline{\Ga}(\la))}\frac{(l+1)!}{(l+1)^{\a}}\frac{1}{\ro^{l+1}}C_4e^{(p-l+k-2)(\de+\underline{\Ga}(\la))}\frac{(p-l+k+1)!}{(p-l+k+1)^{\a}}\frac{1}{\ro_1^{p-l+k+1}}\\
&&\ML\ML\ML\ML\ML\ML \leq C_4e^{(p+k+1-2)(\de+\underline{\Ga}(\la))}\frac{(p+k+1)!}{(p+k+1)^{\a}}\frac{1}{\ro_1^l\ro_1^{p-l+k+1}}\left[\frac{(p+k+1)^{\a}}{(p+k+1)!}\frac{ce^{-\de}}{\ro_1}\sum_{l=1}^{p-1}\cbin{p-1}{l}\left(\frac{\ro_1}{\ro}\right)^{l+1}\frac{(l+1)!}{(l+1)^{\a}}\frac{(p-l+k+1)!}{(p-l+k+1)^{\a}}\right]\nn\\
&&\ML\ML\ML\ML\ML\ML \leq C_4e^{(p+k+1-2)(\de+\underline{\Ga}(\la))}\frac{(p+k+1)!}{(p+k+1)^{\a}}\frac{1}{\ro_1^{p+k+1}}\left[\frac{(p+k+1)^{\a}}{(p+k+1)!}\frac{ce^{-\de}}{\ro_1}\sum_{l=1}^{p-1}\left(\frac{\ro_1}{\ro}\right)^{l+1}\frac{(p-1)!}{(p-1-l)!l!}\frac{(l+1)!}{(l+1)^{\a}}\frac{(p-l+k+1)!}{(p-l+k+1)^{\a}}\right]\nn\\
&&\ML\ML\ML\ML\ML\ML\ML \leq \left(C_4e^{(p+k+1-2)(\de+\underline{\Ga}(\la))}\frac{(p+k+1)!}{(p+k+1)^{\a}}\frac{1}{\ro_1^{p+k+1}}\right)\left[\frac{ce^{-\de}}{\ro_1}\sum_{l=1}^{p-1}\left(\frac{\ro_1}{\ro_{0,10}}\right)^{l+1}\frac{1}{(l+1)^{\a-1}}\frac{(p+k+1)^{\a}}{(p-l+k+1)^{\a}}\frac{(p-1)!}{(p-1-l)!}\frac{(p-l+k+1)!}{(p+k+1)!}\right]\nn
\eea
{\bf Remark:} 

\NI {\em 
The result is that the sum over $l$ is bounded and the factor in front $\frac{ce^{-\de}}{\ro_1}$ guarantees it is a small correction choosing $\de$ sufficiently large.}

\NI More precisely
\bea
&&\ML\ML\ML\ML\sum_{l=1}^{\frac{p}{2}}\left(\frac{\ro_1}{\ro}\right)^{l+1}\frac{1}{(l+1)^{\a-1}}\frac{(p+k+1)^{\a}}{(p-l+k+1)^{\a}}\frac{(p-1)!}{(p-1-l)!}\frac{(p-l+k+1)!}{(p+k+1)!}\nn\\
&&\ML\ML\ML\ML\leq \left(\frac{\ro_1}{\ro}\right)2^{\a}\sum_{l=1}^{\frac{p}{2}}\frac{1}{(l+1)^{\a-1}}\left[\frac{(p-1)!}{(p+k+1)!}\frac{(p-l+k+1)!}{(p-1-l)!}\right]\leq c\left(\frac{\ro_1}{\ro}\right)2^{\a}\sum_{l=1}^{\frac{p}{2}}\frac{1}{(l+1)^{\a-1}}\nn\\
&&\ML\ML\ML\ML\leq\left(\frac{\ro_1}{\ro}\right)2^{\a}c_1\leq C_1\ .
\eea
\bea
&&\ML\ML\ML\ML\sum_{l=\frac{p}{2}+1}^{p}\left(\frac{\ro_1}{\ro}\right)^{l+1}\frac{1}{(l+1)^{\a-1}}\frac{(p+k+1)^{\a}}{(p-l+k+1)^{\a}}\frac{(p-1)!}{(p-1-l)!}\frac{(p-l+k+1)!}{(p+k+1)!}\nn\\
&&\ML\ML\ML\ML\leq \left(\frac{\ro_1}{\ro}\right)^{\frac{p}{2}}p^{\a}\sum_{l=\frac{p}{2}+1}^{p}
\frac{1}{(l+1)^{\a-1}}\leq C_2 \ .
\eea
\medskip
The estimate of the third sum is done exactly in the same way and we do not repeat it here. The final result is that 
\bea
&&\ML\ML\ML\ML\left\{\sum_{l=1}^{p-1}\cbin{p-1}{l}
\big|\ub^{k}(D_S^l\oom)D_S^{p-l}\nabb^k\aa\big|_{L^p(S)}+\sum_{l=1}^{p-1}\cbin{p-1}{l}
\big|\ub^{k}(D_S^l\ub)D_S^{p-1-l}D_{\nu}\nabb^k\aa\big|_{L^p(S)}\right.\nn\\
&&\ML\ML\ML\ML\left.+\sum_{l=1}^{p-1}\cbin{p-1}{l}
\big|\ub^{k}(D_S^luX)D_S^{p-1-l}D_{\nu}\nabb^{k+1}\aa\big|_{L^p(S)}\right\}\leq O(\ep) C_4e^{(J-2)\de}e^{(J-2)\underline{\Ga}(\la)}\frac{1}{\ro_1^p\ro_1^{k+1}}\nn\\ .
\eea
Therefore we have
\bea
&&\ML\ML\big|\ub^k|u| D_S^{p-1}D_{\la}\nabb^{k}\aa\big|_{L^p(S)}\nn\\
&&\ML\ML \leq 2\big|\oom\ub^k D_S^{p}\nabb^{k}\aa\big|_{L^p(S)}+\big|\ub^{k+1}D_S^{p-1}
\nabb^{k+1}\aa\big|_{L^p(S)}+\nn\\
&&\big|\ub^{k}uXD_S^{p-1}\nabb^{k+1}\aa\big|_{L^p(S)}+O(\ep) C_4e^{(J-2)\de}e^{(J-2)\underline{\Ga}(\la)}\frac{1}{\ro_1^p\ro_1^{k+1}}\nn\\
&&+\ML\ML \leq 2|\oom|_{\infty,S} C_4e^{(J-2)\de}e^{(J-2)\Ga(\la,\nu)}{\ro_1^{p}\ro_1^{k+1}}+  C_4e^{(J-2)\de}e^{(J-2)\underline{\Ga}(\la)}\frac{1}{\ro_1^{p-1}\ro_1^{k+2}}\nn\\
&&+|X|_{\infty,S}\frac{|u|}{|\ub|} C_4e^{(J-2)\de}e^{(J-2)\underline{\Ga}(\la)}\frac{1}{\ro_1^{p-1}\ro_1^{k+2}}+
O(\ep) C_4e^{(J-2)\de}e^{(J-2)\underline{\Ga}(\la)}\frac{1}{\ro_1^p\ro_1^{k+1}}\nn\\
&&\ML\ML\leq  C_4e^{(J-2)\de}e^{(J-2)\underline{\Ga}(\la)}\frac{ (2|\oom|_{\infty,S}+1+\si|X|_{\infty,S}+O(\ep))}{\ro_1^{p}\ro_1^{k+1}} \leq \nn\\
&& C^{(1)}e^{(J-2)\de}e^{(J-2)\underline{\Ga}(\la)}\left(\frac{Q\ro_3}{\ro_1}\right)\frac{1}{\ro_1^{p-1}\ro_3\ro_1^{k+1}}\ .
\eea
Where $\ro_3$ plays no substantial role and could be omitted, we introduced it only for the sake of simplicity in the calculation
\bea
Q\geq (2|\oom|_{\infty,S}+1+\si|X|_{\infty,S}+O(\ep))\ .
\eea

\NI If $\Psi\neq\aa$, omitting lower order terms, where ${\tilde\Psi}$ can be $\aa$,
\bea
&&\ML\ML\ML\ML\big|\ub^k |u| D_S^{p-1}D_{\la}\nabb^{k}\Psi\big|_{L^p(S)}
=\frac{|u|}{|\ub|}\big|\ub^{k+1}D_S^{p-1}\nabb^{k+1}{\tilde\Psi}\big|_{L^p(S)}\nn\\
&&\ML\ML\ML\ML\leq \frac{|u|}{|\ub|} C_4e^{(J-2)\de}e^{(J-2)\underline{\Ga}(\la)}\frac{1}{\ro_1^{p-1}\ro_1^{k+2}}\leq 
\si\frac{\ro_3}{\ro_1} C_4e^{(J-2)\de}e^{(J-2)\underline{\Ga}(\la)}\frac{1}{\ro_1^{p-1}\ro_3\ro_1^{k+1}}\nn\\ .
\eea
Therefore we assume,
\bea
&&\ML\ML\ML\ML\big|\ub^k|u|^{l-1} D_S^{p-(l-1)}D_{\la}^{l-1}\nabb^{k}\Psi\big|_{L^p(S)}\leq
\left(\frac{Q\ro_3}{\ro_1}\right)^{l-1} C_4e^{(J-2)\de}e^{(J-2)\underline{\Ga}(\la)}\frac{1}{\ro_1^{p-(l-1)}\ro_3^{l-1}\ro_1^{k+1}}\nn\\
\eql{toassume}
\eea
and we want to prove the estimate for $\big|\ub^k|u|^{l} D_S^{p-l}D_{\la}^{l}\nabb^{k}\Psi\big|_{L^p(S)}$ namely
\bea
&&\ML\ML\ML\ML\big|\ub^k|u|^{l} D_S^{p-l}D_{\la}^{l}\nabb^{k}\Psi\big|_{L^p(S)}\leq
\left(\frac{Q\ro_3}{\ro_1}\right)^{l} C_4e^{(J-2)\de}e^{(J-2)\underline{\Ga}(\la)}\frac{1}{\ro_1^{p-l}\ro_3^{l}\ro_1^{k+1}}\nn\\
\eea
\medskip

{\bf Proof:} Let $\Psi=\aa$ then
\bea\label{lotuno}
&&\ML\ML\ML\ML\ML\big|\ub^k|u|^l D_S^{p-l}D_{\la}^l\nabb^{k}\aa\big|_{L^p(S)}\leq \big|\ub^k|u|^{l-1} D_S^{p-l}(2\oom D_S+\ub D_{\nu}+uX\nabb)D_{\la}^{l-1}\nabb^{k}\aa\big|_{L^p(S)}\nn\\
&&\ML\ML\ML\ML\ML\leq 2\big|\ub^k|u|^{l-1} D_S^{p-l}\oom D_SD_{\la}^{l-1}\nabb^{k}\aa\big|_{L^p(S)}+\big|\ub^{k}|u|^{l-1} D_S^{p-l}\ub D_{\nu}D_{\la}^{l-1}\nabb^{k}\aa\big|_{L^p(S)}
+\big|\ub^{k} D_S^{p-l}u X\nabb D_{\la}^{l-1}\nabb^{k}\aa\big|_{L^p(S)}\nn\\
&&\ML\ML\ML\ML\ML\leq 
2\sum_{s=0}^{p-l}\cbin{p-l}{s}\big|\ub^{k}|u|^{l-1}(D_S^s\oom)D_S^{p-(l-1)-s}D_{\la}^{l-1}\nabb^k\aa\big|_{L^p(S)}
+\sum_{s=0}^{p-l}\cbin{p-l}{s}\big|\ub^{k}|u|^{l-1}(D_S^s\ub)D_S^{p-l-s}D_{\nu}D_{\la}^{l-1}\nabb^k\aa\big|_{L^p(S)}\nn\\
&&\ML\ML\ML\ML\ML+\sum_{s=0}^{p-l}\cbin{p-l}{s}
\big|\ub^{k}|u|^{l-1}(D_S^s uX)D_S^{p-l-s}\nabb D_{\la}^{l-1}\nabb^{k}\aa\big|_{L^p(S)}\nn\\
&&\ML\ML\ML\ML\ML\ML\leq 2\big|\oom\ub^k|u|^{l-1} D_S^{p-(l-1)}D_{\la}^{l-1}\nabb^{k}\aa\big|_{L^p(S)}+\big|\ub^{k+1}|u|^{l-1} D_S^{p-l}D_{\nu}D_{\la}^{l-1}\nabb^{k}\aa\big|_{L^p(S)}
+\big|\ub^{k}|u|^l XD_S^{p-l}\nabb D_{\la}^{l-1}\nabb^{k}\aa\big|_{L^p(S)}\nn\\
&&\ML\ML\ML\ML\ML\ML+
\left\{2\sum_{s=1}^{p-l}\cbin{p-l}{s}\big|\ub^{k}|u|^{l-1}(D_S^s\oom)D_S^{p-(l-1)-s}D_{\la}^{l-1}\nabb^k\aa\big|_{L^p(S)}
+\sum_{s=1}^{p-l}\cbin{p-l}{s}\big|\ub^{k}|u|^{l-1}(D_S^s\ub)D_S^{p-l-s}D_{\nu}D_{\la}^{l-1}\nabb^k\aa\big|_{L^p(S)}\right.\nn\\
&&\ML\ML\ML\ML\ML\ML\left.+\sum_{s=1}^{p-l}\cbin{p-l}{s}
\big|\ub^{k}|u|^{l-1}(D_S^s uX)D_S^{p-l-s}\nabb D_{\la}^{l-1}\nabb^{k}\aa\big|_{L^p(S)}\right\}\nn\\
&&\ML\ML\ML\ML\ML\ML\leq 2\big|\oom\ub^k|u|^{l-1} D_S^{p-(l-1)}D_{\la}^{l-1}\nabb^{k}\aa\big|_{L^p(S)}+\big|\ub^{k+1}|u|^{l-1} D_S^{p-l}D_{\nu}D_{\la}^{l-1}\nabb^{k}\aa\big|_{L^p(S)}
+\big|\ub^{k}|u|^l XD_S^{p-l}\nabb D_{\la}^{l-1}\nabb^{k}\aa\big|_{L^p(S)}\nn\\
&&\ML\ML\ML\ML\ML\ML+\left\{l.o.t.  (I)\right\}\nn\\
&&\ML\ML\ML\ML\ML\ML\leq 2\big|\oom\ub^k|u|^{l-1} D_S^{p-(l-1)}D_{\la}^{l-1}\nabb^{k}\aa\big|_{L^p(S)}+\big|\ub^{k+1}|u|^{l-1} D_S^{p-l}D_{\la}^{l-1}D_{\nu}\nabb^{k}\aa\big|_{L^p(S)}
+\big|\ub^{k}|u|^l XD_S^{p-l}D_{\la}^{l-1}\nabb^{k+1}\aa\big|_{L^p(S)}\nn\\
&&\ML\ML\ML\ML\ML\ML+\left\{\big|\ub^{k+1}|u|^{l-1} D_S^{p-l}[D_{\nu},D_{\la}^{l-1}]\nabb^{k}\aa\big|_{L^p(S)}+\big|\ub^{k}|u|^l XD_S^{p-l}[\nabb,D_{\la}^{l-1}]\nabb^{k}\aa\big|_{L^p(S)}\right\}+\bigg\{l.o.t.  (I)\bigg\}\\
&&\ML\ML\ML\ML\ML\ML\leq 2\big|\oom\ub^k|u|^{l-1} D_S^{p-(l-1)}D_{\la}^{l-1}\nabb^{k}\aa\big|_{L^p(S)}+\big|\ub^{k+1}|u|^{l-1} D_S^{p-l}D_{\la}^{l-1}D_{\nu}\nabb^{k}\aa\big|_{L^p(S)}
+\big|\ub^{k}|u|^l XD_S^{p-l}D_{\la}^{l-1}\nabb^{k+1}\aa\big|_{L^p(S)}\nn\\
&&\ML\ML\ML\ML\ML+\bigg\{l.o.t. (II)\bigg\}+\bigg\{l.o.t.  (I)\bigg\}\nn
\eea
where
\bea\label{casino}
&&\ML\ML\ML\ML \bigg\{l.o.t. (II)\bigg\}=\left\{\big|\ub^{k+1}|u|^{l-1} D_S^{p-l}[D_{\nu},D_{\la}^{l-1}]\nabb^{k}\aa\big|_{L^p(S)}+\big|\ub^{k}|u|^l XD_S^{p-l}[\nabb,D_{\la}^{l-1}]\nabb^{k}\aa\big|_{L^p(S)}\right\}\ .\ \ \ \ \ \ \ \ \ \ \ \ \ \ 
\eea
Let us examine in more detail $\bigg\{l.o.t. (II)\bigg\}$, we have for the first term,
\bea
&&\ML\ML\ML\big|\ub^{k+1}|u|^{l-1} D_S^{p-l}[D_{\nu},D_{\la}^{l-1}]\nabb^{k}\aa\big|_{L^p(S)}\leq \sum_{s=0}^{l-2}\big|\ub^{k+1}|u|^{l-1} D_S^{p-l}D_{\la}^{l-2-s}[D_{\nu},D_{\la}]D_{\la}^s\nabb^{k}\aa\big|_{L^p(S)}\ ,\nn
\eea
$D_{\la}^s\nabb^{k}\aa$ is a $(s+k+2)$ tensor, therefore we have
\bea
&&\ML\ML [D_{\nu},D_{\la}]D_{\la}^s\nabb^{k}\aa_{\tau_1...\tau_{k+2}}=\sum_{j=1}^{s+k+2}R^{{\tilde\tau}_j}(e_{\la},e_{\nu})_{\tau_j}
(D^s\nabb^{k}\aa)_{\tau_1...{\tilde\tau}_j...\tau_{s+k+2}}\nn\\
&&\ML\ML ``\leq" (s+k+2)RD_{\la}^{s-1}D\nabb^{k}\aa\
\eea
where $D$ could be $D_{\la}, D_{\nu},\nabb$ and we can move, apart lower order terms on the r.h.s. near $\nabb^k$. Therefore
\bea
&&\ML\ML\ML\ML\big|\ub^{k+1}|u|^{l-1} D_S^{p-l}[D_{\nu},D_{\la}^{l-1}]\nabb^{k}\aa\big|_{L^p(S)}\leq 
\sum_{s=0}^{l-2}(s+k+2)\big|\ub^{k+1}|u|^{l-1} D_S^{p-l}D_{\la}^{l-2-s}RD_{\la}^{s-1}D\nabb^{k}\aa\big|_{L^p(S)}\nn\\
&&\ML\ML\ML\ML\leq \sum_{s=0}^{l-2}(s+k+2)\sum_{t=0}^{l-2-s}\cbin{l-2-s}{t}\big|\ub^{k+1}|u|^{l-1}D_S^{p-l}(D_{\la}^t R)D_{\la}^{l-2-s-t}D_{\la}^{s-1}D\nabb^{k}\aa\big|_{L^p(S)}\eql{toimprove}\\
&&\ML\ML\ML\ML\leq \sum_{s=0}^{l-2}(s+k+2)\sum_{t=0}^{l-2-s}\sum_{q=0}^{p-l}\cbin{l-2-s}{t}\cbin{p-l}{q}\big|\ub^{k+1}|u|^{l-1}(D_S^qD_{\la}^t R)D_S^{p-l-q}D_{\la}^{l-3-t}D\nabb^{k}\aa\big|_{L^p(S)}\nn
\eea
Observe that there are $p+k-2$ derivatives and as we have to introduce a $|\c|_{\infty,S}$ norm the number of derivatives turns out to be $p+k-1$. Let us try a rough estimate, assuming $D=D_{\la}$,
\bea
&&\ML\ML\ML\ML\big|\ub^{k+1}|u|^{l-1} D_S^{p-l}[D_{\nu},D_{\la}^{l-1}]\nabb^{k}\aa\big|_{L^p(S)}\nn\\ 
&&\ML\ML\ML\ML\leq \sum_{s=0}^{l-2}(s+k+2)\sum_{t=0}^{l-2-s}\sum_{q=0}^{p-l}\cbin{l-2-s}{t}\cbin{p-l}{q}||u|^t(D_S^qD_{\la}^t R)|_{\infty,S}\big|\ub^{k+1}|u|^{l-t-1}D_S^{p-l-q}D_{\la}^{l-2-t}\nabb^{k}\aa\big|_{L^p(S)}\nn\\
&&\ML\ML\ML\ML\leq \sum_{s=0}^{l-2}(s+k+2)\sum_{t=0}^{l-2-s}\sum_{q=0}^{p-l}\cbin{l-2-s}{t}\cbin{p-l}{q}\left(\frac{Q\ro_3}{\ro_1}\right)^t\frac{1}{\ro_1^{q+1}\ro_3^t}e^{(q+t)(\de+\underline{\Ga}(\la))}\frac{(q+t+2)!}{(q+t+2)^a}\c\nn\\
&&\ML\ML\ML\c\left(\frac{Q\ro_3}{\ro_1}\right)^{l-2-t}\frac{1}{\ro_1^{p-l-q}\ro_3^{l-2-t}\ro_1^{k+1}}e^{(p+k-q-t-3)(\de+\underline{\Ga}(\la))}\frac{(p+k-q-t-1)!}{(p+k-q-t-1)^{\a}}\nn\\
&&\ML\ML\ML\ML\ML\ML\ML\leq e^{(p+k-3)(\de+\Ga)}\left(\frac{Q\ro_3}{\ro_1}\right)^{l-2}\frac{1}{\ro_1^{p-l+1}\ro_3^{l-2}\ro_1^{k+1}}\left[\sum_{s=0}^{l-2}(s+k+2)\sum_{t=0}^{l-2-s}\sum_{q=0}^{p-l}\cbin{l-2-s}{t}\cbin{p-l}{q}\frac{(q+t+2)!}{(q+t+2)^a}\frac{(p+k-q-t-1)!}{(p+k-q-t-1)^{\a}}\right]\nn\\
&&\ML\ML\ML\ML\ML\ML\ML\leq e^{(p+k-1)(\de+\Ga)}\left(\frac{\ro_1e^{-2(\de+\Ga)}}{Q^2}\right)\left(\frac{Q\ro_3}{\ro_1}\right)^{l}\frac{1}{\ro_1^{p-l}\ro_3^{l}\ro_1^{k+1}}\nn\\
&&\left[\sum_{s=0}^{l-2}(s+k+2)\sum_{t=0}^{l-2-s}\sum_{q=0}^{p-l}\cbin{l-2-s}{t}\cbin{p-l}{q}\frac{(q+t+2)!}{(q+t+2)^a}\frac{(p+k-q-t-1)!}{(p+k-q-t-1)^{\a}}\right]\nn\\
&&\ML\ML\ML\ML\leq \left(\frac{\ro_1e^{-2(\de+\Ga)}}{Q^2}\right)\left(\left(\frac{Q\ro_3}{\ro_1}\right)^{l}\frac{1}{\ro_1^{p-l}\ro_3^{l}\ro_1^{k+1}}\right)\left(e^{(p+k-1)(\de+\underline{\Ga}(\la))}\frac{(p+k+1)!}{(p+k+1)^{\a}}\right)\c\nn\\
&&\ML\ML\ML\ML\c\frac{(p+k+1)^{\a}}{(p+k+1)!}\left[\sum_{s=0}^{l-2}(s+k+2)\sum_{t=0}^{l-2-s}\sum_{q=0}^{p-l}\cbin{l-2-s}{t}\cbin{p-l}{q}\frac{(q+t+2)!}{(q+t+2)^a}\frac{(p+k-q-t-1)!}{(p+k-q-t-1)^{\a}}\right]\nn\\
&&\ML\ML\ML\ML\leq \left(\frac{\ro_1e^{-2(\de+\Ga)}}{Q^2}\right)\left(C_4e^{(J-2)\de}e^{(J-2)\underline{\Ga}(\la)}\left(\frac{Q\ro_3}{\ro_1}\right)^{l}\frac{1}{\ro_1^{p-l}\ro_3^{l}\ro_1^{k+1}}\right)\c\nn\\
&&\ML\ML\ML\ML\c\frac{(p+k+1)^{\a}}{(p+k+1)!}\left[\sum_{s=0}^{l-2}(s+k+2)\sum_{t=0}^{l-2-s}\sum_{q=0}^{p-l}\cbin{l-2-s}{t}\cbin{p-l}{q}\frac{(q+t+2)!}{(q+t+2)^a}\frac{(p+k-q-t-1)!}{(p+k-q-t-1)^{\a}}\right]\nn\\
\eea
We have 
\bea
&&\ML\ML D^{p-2-s}AB=\sum_{h=0}^{p-2-s}\cbin{p-2-s}{h}(D^hA)D^{p-2-s-h}B\nn\\
&&\ML\ML=D^{p-l}D^{l-2-s}AB=D^{p-l}\sum_{t=0}^{l-2-s}\cbin{l-2-s}{t}(D^tA)D^{l-2-s-t}B\nn\\
&&\ML\ML=\sum_{q=0}^{p-l}\sum_{t=0}^{l-2-s}\cbin{p-l}{q}\cbin{l-2-s}{t}(D^{t+q}A)D^{p-2-s-(t+q)}B\nn\\
&&\ML\ML=\sum_{h=0}^{p-2-s}\left[\de(h-(t+q))\sum_{q=0}^{p-l}\sum_{t=0}^{l-2-s}\cbin{p-l}{q}\cbin{l-2-s}{t}\right](D^hA)D^{p-2-s-h}B\nn
\eea
Therefore
\bea
\left[\de(h-(t+q))\sum_{q=0}^{p-l}\sum_{t=0}^{l-2-s}\cbin{p-l}{q}\cbin{l-2-s}{t}\right]
=\cbin{p-2-s}{h}\ .
\eea
and
\bea
&&\ML\ML \left[\sum_{s=0}^{l-2}(s+k+2)\sum_{t=0}^{l-2-s}\sum_{q=0}^{p-l}\cbin{l-2-s}{t}\cbin{p-l}{q}\frac{(q+t+2)!}{(q+t+2)^a}\frac{(p+k-q-t-1)!}{(p+k-q-t-1)^{\a}}\right]\nn\\
&&\ML\ML=\sum_{s=0}^{l-2}(s+k+2)\sum_{h=0}^{p-2-s}\cbin{p-2-s}{h}\frac{(h+2)!}{(h+2)^a}\frac{(p+k-h-1)!}{(p+k-h-1)^{\a}}
\eea
and
\bea
&&\ML\ML\big|\ub^{k+1}|u|^{l-1} D_S^{p-l}[D_{\nu},D_{\la}^{l-1}]\nabb^{k}\aa\big|_{L^p(S)}\leq\nn\\
&&\ML\ML\leq \left(\frac{\ro_1e^{-2(\de+\Ga)}}{Q^2}\right)\left(C_4e^{(J-2)\de}e^{(J-2)\underline{\Ga}(\la)}\left(\frac{Q\ro_3}{\ro_1}\right)^{l}\frac{1}{\ro_1^{p-l}\ro_3^{l}\ro_1^{k+1}}\right)\c\nn\\ 
&&\ML\ML\c \frac{(p+k+1)^{\a}}{(p+k+1)!}\left[\sum_{s=0}^{l-2}(s+k+2)\sum_{h=0}^{p-2-s}\cbin{p-2-s}{h}\frac{(h+2)!}{(h+2)^a}\frac{(p+k-h-1)!}{(p+k-h-1)^{\a}}\right]\nn\\
&&\ML\ML\leq \left(\frac{\ro_1e^{-2(\de+\Ga)}}{Q^2}\right)\left(C_4e^{(J-2)\de}e^{(J-2)\underline{\Ga}(\la)}\left(\frac{Q\ro_3}{\ro_1}\right)^{l}\frac{1}{\ro_1^{p-l}\ro_3^{l}\ro_1^{k+1}}\right)\c\nn\\ 
&&\ML\ML\c \left[\sum_{s=0}^{l-2}(s+k+2)\sum_{h=0}^{p-2-s}\frac{(p+k+1)^{\a}}{(h+2)^{\a}(p+k-h-1)^{\a}}\frac{(p-2-s)!}{(p-2-s-h)!}\frac{(h+2)!}{h!}\frac{(p+k-h-1)!}{(p+k+1)!}\right]\nn
\eea
Let us look at the term in parenthesis considering the sum over $h$ only up to $\frac{p}{2}$,
\bea
&&\ML\ML\left[\sum_{s=0}^{l-2}(s+k+2)\sum_{h=0}^{\frac{p}{2}}\frac{(p+k+1)^{\a}}{(h+2)^{\a}(p+k-h-1)^{\a}}\frac{(p-2-s)!}{(p-2-s-h)!}\frac{(h+2)!}{h!}\frac{(p+k-h-1)!}{(p+k+1)!}\right]\nn\\
&&\ML\ML\leq 2^{\a}\frac{l(l+k)}{(p+k)(p+k+1)}\sum_{h=0}^{\frac{p}{2}}\frac{1}{(h+2)^{\a-2}}\left[\frac{(p-l)!}{(p-l-h)!}\frac{(p+k-h-1)!}{(p+k-1)!}\right]\nn\\
&&\ML\ML\leq 2^{\a}c\sum_{h=0}^{\frac{p}{2}}\frac{1}{(h+2)^{\a-2}}\leq C\ ,
\eea
recalling that $l\leq p$ and $\a>3$\ . Assume now $h$ between $\frac{p}{2}+1$ and $p$ then
\bea
&&\ML\ML\left[\sum_{s=0}^{l-2}(s+k+2)\sum_{h=\frac{p}{2}+1}^{p}\frac{(p+k+1)^{\a}}{(h+2)^{\a}(p+k-h-1)^{\a}}\frac{(p-2-s)!}{(p-2-s-h)!}\frac{(h+2)!}{h!}\frac{(p+k-h-1)!}{(p+k+1)!}\right]\nn\\
&&\ML\ML\leq l(l+k)\sum_{h=\frac{p}{2}+1}^{p}\frac{1}{(p+k-h-1)^{\a}}\frac{(p+k+1)^{\a}}{(h+2)^{\a}(p+k+1)(p+k)}\left[\frac{(p-2-s)!}{(p-2-s-h)!}\frac{(p+k-h-1)!}{(p+k-1)!}\right]\nn\\
&&\ML\ML\ML\ML\leq c\frac{l(l+k)}{(p+k+1)(p+k)}\sum_{h=\frac{p}{2}+1}^{p}\frac{1}{(p+k-h-1)^{\a}}\frac{(p+k+1)^{\a}}{(h+2)^{\a}}\leq c_1\sum_{h=\frac{p}{2}+1}^{p}\frac{1}{(p+k-h-1)^{\a}}\frac{(p+k+1)^{\a}}{(h+2)^{\a}}\leq C\ .\nn
\eea
In fact the last sum can easily be estimated in the following way, assume $p\geq k$,
\bea
&&\ML\ML \sum_{h=\frac{p}{2}+1}^{p}\frac{1}{(p+k-h-1)^{\a}}\frac{(p+k+1)^{\a}}{(h+2)^{\a}}
\leq \frac{(2p)^{\a}}{\left(\frac{p}{2}\right)^{\a}}\sum_{h=\frac{p}{2}+1}^{p}\frac{1}{(p+k-h-1)^{\a}}
\leq 2^{2\a}c\ .\nn
\eea
Assume $p<k$ then
\bea
&&\ML\ML \sum_{h=\frac{p}{2}+1}^{p}\frac{1}{(p+k-h-1)^{\a}}\frac{(p+k+1)^{\a}}{(h+2)^{\a}}
\leq \frac{(2k)^{\a}}{k^{\a}}\sum_{h=\frac{p}{2}+1}^{p}\frac{1}{(h+2)^{\a}}\leq 2^{\a}c_1\ .\ \ \ \ \ \ \ \ \ \ \ \ 
\eea
Therefore our rough estimate suggests  that
\bea
&&\ML\ML\big|\ub^{k+1}|u|^{l-1} D_S^{p-l}[D_{\nu},D_{\la}^{l-1}]\nabb^{k}\aa\big|_{L^p(S)}\leq \left(2^{2\a}\frac{\ro_1e^{-2(\de+\underline{\Ga}(\la))}}{Q^2}\right)\nn\\
&&\left( C_4e^{(J-2)\de}e^{(J-2)\Ga(\la,\nu)}\left(\frac{Q\ro_3}{\ro_1}\right)^{l}\frac{1}{\ro_1^{p-l}\ro_3^{l}\ro_1^{k+1}}\right)\nn\\
&&\ML\ML\leq \left(2^{2\a}\frac{\ro_1e^{-2(\de+\Ga)}}{Q}\right)\left( C_4e^{(J-2)\de}e^{(J-2)\Ga(\la,\nu)}\left(\frac{Q\ro_3}{\ro_1}\right)^{l-1}\frac{1}{\ro_1^{p-(l-1)}\ro_3^{l-1}\ro_1^{k+1}}\right)\nn\\
&&\ML\ML\leq O(\ep) C_4e^{(J-2)\de}e^{(J-2)\underline{\Ga}(\la)}\left(\frac{Q\ro_3}{\ro_1}\right)^{l-1}\frac{1}{\ro_1^{p-(l-1)}\ro_3^{l-1}\ro_1^{k+1}}\ \nn
\eea
and, more in general that,
\bea
\bigg\{l.o.t. (II)\bigg\}\leq O(\ep) C_4e^{(J-2)\de}e^{(J-2)\underline{\Ga}(\la)}\left(\frac{Q\ro_3}{\ro_1}\right)^{l-1}\frac{1}{\ro_1^{p-(l-1)}\ro_3^{l-1}\ro_1^{k+1}}\nn\\ 
\eea
The estimate of $\bigg\{l.o.t. (I)\bigg\}$ goes as the one done in the case $l=1$ and there is no need to repeat and again,
\bea
\bigg\{l.o.t. (I)\bigg\}\leq O(\ep) C_4e^{(J-2)\de}e^{(J-2)\underline{\Ga}(\la)}\left(\frac{Q\ro_3}{\ro_1}\right)^{l-1}\frac{1}{\ro_1^{p-(l-1)}\ro_3^{l-1}\ro_1^{k+1}}\nn\\ .
\eea
Going back we have, therefore, 
\bea
&&\ML\ML\ML\ML\ML\big|\ub^k|u|^l D_S^{p-l}D_{\la}^l\nabb^{k}\aa\big|_{L^p(S)}\nn\\
&&\ML\ML\ML\ML\ML\leq 2\big|\oom\ub^k|u|^{l-1} D_S^{p-(l-1)}D_{\la}^{l-1}\nabb^{k}\aa\big|_{L^p(S)}+\big|\ub^{k+1}|u|^{l-1} D_S^{p-l}D_{\la}^{l-1}D_{\nu}\nabb^{k}\aa\big|_{L^p(S)}
+\big|\ub^{k}|u|^l XD_S^{p-l}D_{\la}^{l-1}\nabb^{k+1}\aa\big|_{L^p(S)}\nn\\
&&\ML\ML\ML\ML\ML+O(\ep) C_4e^{(J-2)\de}e^{(J-2)\underline{\Ga}(\la)}\left(\frac{Q\ro_3}{\ro_1}\right)^{l-1}\frac{1}{\ro_1^{p-(l-1)}\ro_3^{l-1}\ro_1^{k+1}}\nn\\
&&\ML\ML\ML\ML\ML\ML\ML\ML\leq 2|\oom|_{\infty,S}\big|\ub^k|u|^{l-1} D_S^{p-(l-1)}D_{\la}^{l-1}\nabb^{k}\aa\big|_{L^p(S)}+\big|\ub^{k+1}|u|^{l-1} D_S^{p-l}D_{\la}^{l-1}\nabb^{k+1}\aa\big|_{L^p(S)}
+\frac{|u|}{|\ub|}|X|_{\infty,S}\big|\ub^{k+1}|u|^{l-1} D_S^{p-l}D_{\la}^{l-1}\nabb^{k+1}\aa\big|_{L^p(S)}\nn\\
&&\ML\ML\ML\ML\ML+O(\ep) C_4e^{(J-2)\de}e^{(J-2)\underline{\Ga}(\la)}\left(\frac{Q\ro_3}{\ro_1}\right)^{l-1}\frac{1}{\ro_1^{p-(l-1)}\ro_3^{l-1}\ro_1^{k+1}}\nn\\
&&\ML\ML\ML\ML\ML\ML\ML\leq \left(2|\oom|_{\infty,S}+1+\si|X|_{\infty,S}\right)\left(\frac{Q\ro_3}{\ro_1}\right)^{l-1} C_4e^{(J-2)\de}e^{(J-2)\underline{\Ga}(\la)}\frac{1}{\ro_1^{p-(l-1)}\ro_3^{l-1}\ro_1^{k+1}}+\nn\\
&&O(\ep) C_4e^{(J-2)\de}e^{(J-2)\Ga(\la,\nu)}\left(\frac{Q\ro_3}{\ro_1}\right)^{l-1}\frac{1}{\ro_1^{p-(l-1)}\ro_3^{l-1}\ro_1^{k+1}}\nn\\
&&\ML\ML\ML\ML\ML\ML\ML\leq \left(2|\oom|_{\infty,S}+1+\si|X|_{\infty,S}+O(\ep)\right)\frac{\ro_3}{\ro_1}\left(\frac{Q\ro_3}{\ro_1}\right)^{l-1} C_4e^{(J-2)\de}e^{(J-2)\underline{\Ga}(\la)}\frac{1}{\ro_1^{p-l}\ro_3^{l}\ro_1^{k+1}}\nn\\
&&\leq   C_4e^{(J-2)\de}e^{(J-2)\underline{\Ga}(\la)}\left(\frac{Q\ro_3}{\ro_1}\right)^{l}\frac{1}{\ro_1^{p-l}\ro_3^{l}\ro_1^{k+1}}\ .\ \ 
\eea
\medskip

Let $\Psi\neq \aa$ then
\bea
&&\ML\ML\ML\ML\big|\ub^k|u|^l D_S^{p-l}D_3^l\nabb^{k}\Psi\big|_{L^p(S)}\leq
\big|\ub^k|u|^l D_S^{p-l}D_3^{l-1}\nabb^{k}D_3\Psi\big|_{L^p(S)}+\left\{\big|\ub^k|u|^l D_S^{p-l}D_3^{l-1}[D_3,\nabb^{k}]\Psi\big|_{L^p(S)}\right\}\nn\\
&&\ML\ML\ML\ML\leq
\big|\ub^k|u|^l D_S^{p-l}D_3^{l-1}\nabb^{k+1}{\tilde\Psi}\big|_{L^p(S)}+\left\{\big|\ub^k|u|^l D_S^{p-l}D_3^{l-1}[D_3,\nabb^{k}]\Psi\big|_{L^p(S)}\right\}\nn\\
&&\ML\ML\ML\ML\leq
\frac{|u|}{|\ub|}\big|\ub^{k+1}|u|^{l-1} D_S^{p-l}D_3^{l-1}\nabb^{k+1}{\tilde\Psi}\big|_{L^p(S)}+\left\{\big|\ub^k|u|^l D_S^{p-l}D_3^{l-1}[D_3,\nabb^{k}]\Psi\big|_{L^p(S)}\right\}
\eea
The term $\left\{\big|\ub^k|u|^l D_S^{p-l}D_3^{l-1}[D_3,\nabb^{k}]\Psi\big|_{L^p(S)}\right\}$ has to be estimated as before, 
 therefore
\bea
&&\ML\ML\ML\ML\big|\ub^k|u|^l D_S^{p-l}D_3^l\nabb^{k}\Psi\big|_{L^p(S)}\nn\\
&&\ML\ML\ML\ML\leq  \si\left(\frac{3\ro_3}{\ro_1}\right)^{\!l-1}\frac{\ro_3}{\ro_1}\! C_4e^{(J-2)\de}e^{(J-2)\underline{\Ga}(\la)}\frac{1}{\ro_1^{p-l}\ro_3^{l}\ro_1^{k+1}}+O(\ep) C_4e^{(J-2)\de}e^{(J-2)\underline{\Ga}(\la)}\nn\\
&&\left(\frac{Q\ro_3}{\ro_1}\right)^{l-1}\frac{1}{\ro_1^{p-(l-1)}\ro_3^{l-1}\ro_1^{k+1}}\nn\\
&&\ML\ML\ML\ML\leq \left(\si+O(\ep)\right)\frac{\ro_3}{\ro_1}\left(\frac{Q\ro_3}{\ro_1}\right)^{l-1} C_4e^{(J-2)\de}e^{(J-2)\underline{\Ga}(\la)}\frac{1}{\ro_1^{p-l}\ro_3^{l}\ro_1^{k+1}}\nn\\
&&\leq  C_4e^{(J-2)\de}e^{(J-2)\underline{\Ga}(\la)}\left(\frac{Q\ro_3}{\ro_1}\right)^{l}\frac{1}{\ro_1^{p-l}\ro_3^{l}\ro_1^{k+1}}\ . \ \ \ \ \ \ \nn
\eea
Therefore, assuming the previous Proofs of lower order terms, we have proved for any $\Psi$ and any $l$ the estimate,
\bea
\big|\ub^k|u|^l D_S^{p-l}D_3^l\nabb^{k}\Psi\big|_{L^p(S)}\leq\left(\frac{Q\ro_3}{\ro_1}\right)^{l} C_4e^{(J-2)\de}e^{(J-2)\underline{\Ga}(\la)}\frac{1}{\ro_1^{p-l}\ro_3^{l}\ro_1^{k+1}}\nn\\
\eea
and, therefore for $l=p$ we have
\bea\label{107estimate}
&&\ML\ML\big|\ub^k|u|^pD_3^p\nabb^{k}\Psi\big|_{L^p(S)}\leq\left(\frac{Q\ro_3}{\ro_1}\right)^{p} C_4e^{(J-2)\de}e^{(J-2)\underline{\Ga}(\la)}\frac{1}{\ro_3^{p}\ro_1^{k+1}}\nn\\
&&\leq Q^p C_4e^{(J-2)\de}e^{(J-2)\Ga(\la,\nu)}\frac{1}{\ro_1^{p}\ro_1^{k+1}}\ ,\ \ \ \ \ \ \ \ \ \ \ \ \ 
\eea

\subsubsection{The estimate of the other terms with commutators}
To complete previous discussions we want to control the term, see equation \ref{casino} 
\[\big|\ub^{k}|u|^l XD_S^{p-l}[\nabb,D_{\la}^{l-1}]\nabb^{k}\aa\big|_{L^p(S)}\ .\]
\bea
&&\ML\ML\ML\ML \big|\ub^{k}|u|^l XD_S^{p-l}[\nabb,D_{\la}^{l-1}]\nabb^{k}\aa\big|_{L^p(S)}
\leq |X|_{\infty,S} \big|\ub^{k}|u|^l D_S^{p-l}[\nabb,D_{\la}^{l-1}]\nabb^{k}\aa\big|_{L^p(S)}\nn\\
&&\ML\ML\ML\ML\leq |X|_{\infty,S}\sum_{s=0}^{l-2}\big|\ub^{k}|u|^{l} D_S^{p-l}D_{\la}^{l-2-s}
[\nabb,D_{\la}]D_{\la}^s\nabb^{k}\aa\big|_{L^p(S)}
\eea
$D_{\la}^s\nabb^{k}\aa$ is a $(s+k+2)$ tensor, therefore we have
\bea
&&\ML\ML [\nabb_{\mu},D_{\la}]D_{\la}^s\nabb^{k}\aa_{\tau_1...\tau_{k+2}}=
\left[\sum_{j=1}^{s+k+2}{\underline C}^{\si_j}_{\mu\la_j}(D^s\nabb^{k}\aa)_{\la_1\c\c\si_j\c\c\la_s\tau_1...\tau_{k+2}}+\oom\chib^{\ro}_{\mu}\nabb_{\ro}(D^s\nabb^{k}\aa)_{\la_1\c\c\c\c\la_s\tau_1...\tau_{k+2}}\right]\nn\\
&&\ML\ML ``\leq"(s+k+2){\underline C}(D^s\nabb^{k}\aa)+\oom\chib\c\nabb D_{\la}^s\nabb^{k}\aa\ .
\eea
Therefore
\bea
&&\ML\ML\ML\ML \big|\ub^{k}|u|^l XD_S^{p-l}[\nabb,D_{\la}^{l-1}]\nabb^{k}\aa\big|_{L^p(S)}\nn\\
&&\ML\ML\ML\ML\leq |X|_{\infty,S}\sum_{s=0}^{l-2}(s+k+2)\big|\ub^{k}|u|^{l} D_S^{p-l}D_{\la}^{l-2-s}{\underline C}D^s\nabb^{k}\aa\big|_{L^p(S)}\nn\\
&&\ML\ML\ML\ML+\leq |X|_{\infty,S}\sum_{s=0}^{l-2}\big|\ub^{k}|u|^{l} D_S^{p-l}D_{\la}^{l-2-s}
\oom\chib\c\nabb D_{\la}^s\nabb^{k}\aa\big|_{L^p(S)}\nn\\
&&\ML\ML\ML\ML\leq |X|_{\infty,S}\sum_{s=0}^{l-2}(s+k+2)\big|\ub^{k}|u|^{l} D_S^{p-l}D_{\la}^{l-2-s}{\underline C}D^s\nabb^{k}\aa\big|_{L^p(S)}\nn\\
&&\ML\ML\ML\ML+\leq |X|_{\infty,S}\sum_{s=0}^{l-2}\big|\ub^{k}|u|^{l} D_S^{p-l}D_{\la}^{l-2-s}
\oom\chib D_{\la}^s\nabb^{k+1}\aa\big|_{L^p(S)}+\bigg\{ \ l.o.t \bigg\}\nn\\
&&\ML\ML\ML\ML\leq \si |X|_{\infty,S}\sum_{s=0}^{l-2}(s+k+2)\big|\ub^{k+1}|u|^{l-1} D_S^{p-l}D_{\la}^{l-2-s}{\underline C}D^s\nabb^{k}\aa\big|_{L^p(S)}\nn\\
&&\ML\ML\ML\ML+\leq \si |X|_{\infty,S}\sum_{s=0}^{l-2}\big|\ub^{k+1}|u|^{l-1} D_S^{p-l}D_{\la}^{l-2-s}
\oom\chib D_{\la}^s\nabb^{k+1}\aa\big|_{L^p(S)}+\bigg\{\ l.o.t \bigg\}\nn\ .
\eea
The first sum, treating ${\underline C}$ as $R$ is identical to the previous one and can be estimated in the same way, therefore we are left with estimating 
\bea
\si |X|_{\infty,S}\sum_{s=0}^{l-2}\big|\ub^{k+1}|u|^{l-1} D_S^{p-l}D_{\la}^{l-2-s}
\oom\chib D_{\la}^s\nabb^{k+1}\aa\big|_{L^p(S)}
\eea 
as the $\bigg\{\ l.o.t  \bigg\}$ is easy to control and has already been neglected. The second sum has the same structure as the first sum but simpler as it has  $\oom\chib$ instead of $\underline C$ and without the factor $(s+k+2)$ therefore it is easier to estimate.

\NI We are left to estimate $ \bigg\{l.o.t.  (I)\bigg\} $, see \ref{lotuno}, this term can be easily estimated following the same steps performed in calculation of $\bigg\{l.o.t.  (II)\bigg\}$

\NI Corollary  \ref{cor10.3xxx} follows choosing 
\bea\label{rotre!}
\ro_2\leq \ro_1/Q.
\eea 
\medskip

\NI Therefore in conclusion we have proved , omitting all the lower order terms,
\bea
&&\ML\ML\big|\ub^k|u|^pD_3^p\nabb^{k}\Psi\big|_{L^p(S)}\leq H(\c\c\c)\frac{1}{\ro_2^{p}\ro_1^{k+1}}\nn\\
&&\ML\ML\big|\ub^k|\ub|^pD_4^p\nabb^{k}\Psi\big|_{L^p(S)}\leq H(\c\c\c)\frac{1}{\ro_2^p\ro_1^{k+1}}\ ,
\eea

\medskip

\NI {\bf Remark:} 

\NI {\em In fact the second part of this result can be obtained proceeding as for the first part   posing $D_{\nu}$ instead of $D_{\la}$ and $|\ub|$ instead of $|u|$.}
\medskip


\section{Proof of Lemmas \ref{l3l} \label{L17.3} and  \ref{mixedderivatives3}}
\NI The proof is a generalisation of the proof of theorem \ref{T11.1} and goes by induction, we assume these estimates hold for for $(J,P)$ and we prove they hold also for for $(J+1,P-1)$, with $J+P=N$.

\NI Let us start from the quantity we want to estimate:
\[\nabb^{p_0}\ddb_{\nu}\nabb^{p_1}\ddb_{\nu}\nabb^{p_2}\c\c\c\ddb_{\nu}\nabb^{p_{J-1}}\ddb_{\nu}\nabb^{p_{J}}U\]
where
\[U=\oom^{-1}\tr\chi\ \ \ ,\ \ \ \sum_{s=0}^Jp_s=P\ \ \ , \ \ \ J+P=N\ .\]

\NI We write the previous term in the following way, recalling \ref{109bisabcd},
\bea
&&\ML\ML\ML\nabb^{p_0}\ddb_{\nu}\nabb^{p_1}\ddb_{\nu}\nabb^{p_2}\c\c\c\ddb_{\nu}\nabb^{p_{J-1}}\ddb_{\nu}\nabb^{p_{J}}U\eql{5.109}\\
&&\ML\ML\ML\ML=(\nabb^{p_0}\ddb_{\nu}\nabb^{p_1}\ddb_{\nu}\nabb^{p_2}\c\c\c\ddb_{\nu}\nabb^{p_{J-1}+p_J})(\ddb_{\nu}U)
+(\nabb^{p_0}\ddb_{\nu}\nabb^{p_1}\ddb_{\nu}\nabb^{p_2}\c\c\c\ddb_{\nu}\nabb^{p_{J-1}})([\nabb^{p_{J}},\ddb_{\nu}]U)\nn\\
&&\ML\ML\ML\ML=(\nabb^{p_0}\ddb_{\nu}\nabb^{p_1}\ddb_{\nu}\nabb^{p_2}\c\c\c\ddb_{\nu}\nabb^{p_{J-1}+p_J})\!\left(-\frac{\tr\chi}{2}{U}-\oom^{-1}|\hat{{\chi}}|^2\right)
+(\nabb^{p_0}\ddb_{\nu}\nabb^{p_1}\ddb_{\nu}\nabb^{p_2}\c\c\c\ddb_{\nu}\nabb^{p_{J-1}})\left([\ddb_{\nu},\nabb^{p_{J}}]U\right)\ .\nn
\eea
We use this last expression to prove inductively the desired estimate
Recall that, see Appendix \ref{AS.3},
\bea
[\nabb_{\mu},\ddb_{\nu}]U_{\nu_1...\nu_k}=-\sum_{j=1}^k{C}^{\si_j}_{\mu\nu_j}U_{\nu_1..{\si}_j..\nu_k}+\chi_{\mu}^{\ro}(\nabb_{\ro}U)_{\nu_1..\nu_k}
\eea
where
\bea
{C}^{\si_j}_{\mu\nu_j}=\oom\!\left[(\chi_{\mu\nu_j}\etab^{\si_j}\!-\!\chi_{\mu}^{\si_j}\etab_{\nu_j})
+\oom\theta^C_{\mu}\theta^D_{\nu_j}R^{{\si}_j}(\c,e_C,e_4,e_D)\right]\ .
\eea
Therefore
\bea
&&\ML\ML\ML\ML[\nabb^{p_{J}},\ddb_{\nu}]U=\sum_{q=0}^{{p_{J}}-1}\nabb^q[\nabb,\ddb_{\nu}]\nabb^{{p_{J}}-q-1}U
=\sum_{q=0}^{{p_{J}}-1}\nabb^q\left(-({p_{J}}-q-1)C\nabb^{{p_{J}}-q-1}U+\oom\chi\nabb^{{p_{J}}-q}U\right)\nn\\
&&\ML\ML\ML\ML=-\sum_{q=0}^{{p_{J}}-1}({p_{J}}-q-1)\sum_{h=0}^q\cbin{q}{h}(\nabb^hC)\nabb^{{p_{J}}-1-h}U+\sum_{q=0}^{{p_{J}}-1}\sum_{j=0}^q\cbin{q}{j}(\nabb^j\chi)\nabb^{{p_{J}}-h}U\ .
\eea
Using the relation,
\bea
&&\ML\ML\ML[\nabb^{p_{J}},\ddb_{\nu}]U
=-\sum_{q=0}^{p_{J}-1}(p_{J}-q-1)\sum_{h=0}^q\cbin{q}{h}(\nabb^hC)\nabb^{p_{J}-1-h}U+\sum_{q=0}^{p_{J}-1}\sum_{h=0}^q\cbin{q}{h}(\nabb^h\chi)\nabb^{p_{J}-h}U\ \nn
\eea
we estimate the first term of the r.h.s. of equation \ref{5.109} to check the inductive assumptions; defining 
\[\sum_{s=0}^Jp_s=P\ \ ,\ \ \sum_{s=1}^{J-1}l_s=L\ \  ,\ \ \sum_{s=0}^{J-1}h_s=H\ ,\]
we have
\bea
&&\ML\ML|r^{1+J+P+2\si(P)-\frac{2}{p}}\nabb^{p_0}\ddb_{\nu}\nabb^{p_1}\ddb_{\nu}\nabb^{p_2}\c\c\c\ddb_{\nu}\nabb^{p_{J-1}+p_J}\left(\tr\chi{U}\right)|_{p,S}\nn\\
&&\ML\ML\leq 
\sum_{l_1=0}^1\c\c\c\sum_{l_{J-2}=0}^1\sum_{l_{J-1}=0}^1\sum_{h_0=0}^{p_0}\ \ \ \sum_{h_1=0}^{p_1}\c\c\c\sum_{h_{J-2}=0}^{p_{J-2}}\sum_{h_{J-1}=0}^{p_{J-1}+p_J}\cbin{p_0}{h_0}\cbin{p_1}{h_1}\c\c\c\cbin{p_{J-2}}{h_{J-2}}\cbin{p_{J-1}+p_J}{h_{J-1}}\nn\\
&&\ML\ML\big|r^{1+J+P+2\si(P)-\frac{2}{p}}(\nabb^{h_0}\ddb_{\nu}^{l_1}\nabb^{h_1}\c\c\c\ddb_{\nu}^{l_{J-2}}\nabb^{h_{J-2}}\ddb_{\nu}^{l_{J-1}}\nabb^{h_{J-1}}\tr\chi)\c\nn\\
&&\ML\ML\c(\nabb^{p_0-h_0}\ddb_{\nu}^{1-l_1}\nabb^{p_1-h_1}\c\c\c\ddb_{\nu}^{1-l_{J-2}}\nabb^{p_{J-2}-h_{J-2}}\ddb_{\nu}^{1-l_{J-1}}\nabb^{p_{J-1}+p_J-h_{J-1}}U)\big|_{p,S}\\
&&\ML\ML\ML\ML\leq \sum_{l_1=0}^1\c\c\c\sum_{l_{J-2}=0}^1\sum_{l_{J-1}=0}^1\sum_{h_0=0}^{p_0}\ \ \ \sum_{h_1=0}^{p_1}\c\c\c\sum_{h_{J-2}=0}^{p_{J-2}}\sum_{h_{J-1}=0}^{p_{J-1}+p_J}\chi\left(L+H\leq \left[\frac{J+P}{2}\right]\right)\cbin{p_0}{h_0}\cbin{p_1}{h_1}\c\c\c\cbin{p_{J-2}}{h_{J-2}}\cbin{p_{J-1}+p_J}{h_{J-1}}\nn\\
&&\ML\ML\ML\ML\frac{1}{r^{1+2\si(H)+2\si(P-H)-2\si(P)}}\big|r^{1+L+H+2\si(H)}(\nabb^{h_0}\ddb_{\nu}^{l_1}\nabb^{h_1}\c\c\c\ddb_{\nu}^{l_{J-2}}\nabb^{h_{J-2}}\ddb_{\nu}^{l_{J-1}}\nabb^{h_{J-1}}\tr\chi)\big|_{\infty,S}\nn\\
&&\ML\ML\ML\ML\big|r^{1+J+P-(L+H)+2\si(P-H)-\frac{2}{p}}(\nabb^{p_0-h_0}\ddb_{\nu}^{1-l_1}\nabb^{p_1-h_1}\c\c\c\ddb_{\nu}^{1-l_{J-2}}\nabb^{p_{J-2}-h_{J-2}}\ddb_{\nu}^{1-l_{J-1}}\nabb^{p_{J-1}+p_J-h_{J-1}}U)\big|_{p,S}\nn\\
&&\ML\ML\ML\ML+\sum_{l_1=0}^1\c\c\c\sum_{l_{J-2}=0}^1\sum_{l_{J-1}=0}^1\sum_{h_0=0}^{p_0}\ \ \ \sum_{h_1=0}^{p_1}\c\c\c\sum_{h_{J-2}=0}^{p_{J-2}}\sum_{h_{J-1}=0}^{p_{J-1}+p_J}\chi\left(L+H\geq \left[\frac{J+P}{2}\right]+1\right)\cbin{p_0}{h_0}\cbin{p_1}{h_1}\c\c\c\cbin{p_{J-2}}{h_{J-2}}\cbin{p_{J-1}+p_J}{h_{J-1}}\nn\\
&&\ML\ML\ML\ML\frac{1}{r^{1+2\si(H)+2\si(P-H)-2\si(P)}}\big|r^{1+L+H+2\si(H)-\frac{2}{p}}(\nabb^{h_0}\ddb_{\nu}^{l_1}\nabb^{h_1}\c\c\c\ddb_{\nu}^{l_{J-2}}\nabb^{h_{J-2}}\ddb_{\nu}^{l_{J-1}}\nabb^{h_{J-1}}\tr\chi)\big|_{p,S}\nn\\
&&\ML\ML\ML\ML\big|r^{1+J+P-(L+H)+2\si(P-H)}(\nabb^{p_0-h_0}\ddb_{\nu}^{1-l_1}\nabb^{p_1-h_1}\c\c\c\ddb_{\nu}^{1-l_{J-2}}\nabb^{p_{J-2}-h_{J-2}}\ddb_{\nu}^{1-l_{J-1}}\nabb^{p_{J-1}+p_J-h_{J-1}}U)\big|_{\infty,S}\ .\nn
\eea
Let us consider the first sum, $\big[(I)\big]$, and, denoting $K=H+L$, we have, using as inductive assumptions the estimates of Lemma \ref{mixedderivatives2}, with $J-1$ instead of $J$, 
\bea
&&\ML\ML\ML\ML\ML\ML\ML\big[(I)\big]\leq \sum_{k=0}^{J+P}\left[\sum_{l_1=0}^1\c\c\c\sum_{l_{J-2}=0}^1\sum_{l_{J-1}=0}^1\sum_{h_0=0}^{p_0}\ \ \ \sum_{h_1=0}^{p_1}\c\c\c\sum_{h_{J-2}=0}^{p_{J-2}}\sum_{h_{J-1}=0}^{p_{J-1}+p_J}\cbin{p_0}{h_0}\cbin{p_1}{h_1}\c\c\c\cbin{p_{J-2}}{h_{J-2}}\cbin{p_{J-1}+p_J}{h_{J-1}}\de(H+L-k)\right]\\
&&\ML\ML\ML\ML\ML\ML\ML\chi\left(L+H\leq \left[\frac{J+P}{2}\right]\right)\left[\frac{1}{r^{1+\si(H)+\si(P-H)-\si(P)}}\frac{(L+H+1)!}{(L+H+1)^{\a}}\frac{e^{(L+H-1)(\de+\Ga)}}{\ro_2^{L+H+1}}\right]
\left[\frac{(J+P-1-(L+H))!}{(J+P-1-(L+H))^{\a}}\frac{e^{(J+P-(L+H)-3)(\de+\Ga)}}{\ro_2^{(J+P-(L+H)-1)}}\right]\nn
\eea
Observe that, denoting with $\nab$ a generic derivative operator we have, recalling that
$\sum_{s=0}^jp_s=P\ \ ,\ \ \sum_{s=1}^{J-1}l_s=L\ \  ,\ \ \sum_{s=0}^{J-1}h_s=H$,
\bea
&&\ML\ML\ML\ML\nab^{(J-1)+P}AB=\sum_{k=0}^{J+P-1}\cbin{J+P-1}{k}\nab^kA\nab^{J+P-1-k}B=\nab^{p_0}\nab\nab^{p_1}\nab\c\c\c\nab^{p_{j-1}+p_j}AB\nn\\
&&\ML\ML\ML\ML\ML=\sum_{l_1=0}^1\c\c\c\sum_{l_{J-2}=0}^1\sum_{l_{J-1}=0}^1\sum_{h_0=0}^{p_0}\ \ \ \sum_{h_1=0}^{p_1}\c\c\c\sum_{h_{J-2}=0}^{p_{J-2}}\sum_{h_{J-1}=0}^{p_{J-1}+p_J}\cbin{p_0}{h_0}\cbin{p_1}{h_1}\c\c\c\cbin{p_{J-2}}{h_{J-2}}\cbin{p_{J-1}+p_J}{h_{J-1}}\nn\\
&&\ML\ML\ML\ML\ML\c\left(\nab^{h_0}\nab^{l_1}\nab^{h_1}\c\c\c\nab^{l_{J-2}}\nab^{h_{J-2}}\nab^{l_{J-1}}\nab^{h_{J-1}}A\right)
\left(\nab^{p_0-h_0}\nab^{1-l_1}\nab^{p_1-h_1}\c\c\c\nab^{1-l_{J-2}}\nab^{p_{J-2}-h_{J-2}}\nab^{1-l_{J-1}}\nab^{p_{J-1}+p_J-h_{J-1}}B\right)\nn\\
&&\ML\ML\ML\ML\ML= \sum_{k=0}^{J+P-1}\left[\sum_{l_1=0}^1\c\c\c\sum_{l_{J-2}=0}^1\sum_{l_{J-1}=0}^1\sum_{h_0=0}^{p_0}\ \ \ \sum_{h_1=0}^{p_1}\c\c\c\sum_{h_{J-2}=0}^{p_{J-2}}\sum_{h_{J-1}=0}^{p_{J-1}+p_J}\cbin{p_0}{h_0}\cbin{p_1}{h_1}\c\c\c\cbin{p_{J-2}}{h_{J-2}}\cbin{p_{J-1}+p_J}{h_{J-1}}\de(H+L-k)\right]\nn\\
&&\ML\ML\ML\c\nab^{H+L}A\nab^{J+P-1-(H+L)}B\\
&&\ML\ML\ML\ML\ML= \sum_{k=0}^{J+P-1}\left[\sum_{l_1=0}^1\c\c\c\sum_{l_{J-2}=0}^1\sum_{l_{J-1}=0}^1\sum_{h_0=0}^{p_0}\ \ \ \sum_{h_1=0}^{p_1}\c\c\c\sum_{h_{J-2}=0}^{p_{J-2}}\sum_{h_{J-1}=0}^{p_{J-1}+p_J}\cbin{p_0}{h_0}\cbin{p_1}{h_1}\c\c\c\cbin{p_{J-2}}{h_{J-2}}\cbin{p_{J-1}+p_J}{h_{J-1}}\de(H+L-k)\right]\nn\\
&&\ML\ML\ML\c\nab^{k}A\nab^{J+P-1-k}B\ ,
\eea
which implies
\bea
\ML\ML\ML\ML\ML\ML\left[\sum_{l_1=0}^1\c\c\c\sum_{l_{J-2}=0}^1\sum_{l_{J-1}=0}^1\sum_{h_0=0}^{p_0}\ \ \ \sum_{h_1=0}^{p_1}\c\c\c\sum_{h_{J-2}=0}^{p_{J-2}}\sum_{h_{J-1}=0}^{p_{J-1}+p_J}\cbin{p_0}{h_0}\cbin{p_1}{h_1}\c\c\c\cbin{p_{J-2}}{h_{J-2}}\cbin{p_{J-1}+p_J}{h_{J-1}}\de(H+L-k)\right]=\cbin{J+P-1}{k}\ .\ \ \ \eql{cbinrel}
\eea
Therefore
\bea
&&\ML\ML\ML\ML \big[(I)\big]\leq \sum_{k=0}^{\left[\frac{J+P-1}{2}\right]}\cbin{J+P-1}{k}
\frac{{\hat F}}{r}\left[\frac{(k+1)!}{(k+1)^{\a}}\frac{e^{(k-1)(\de+\Ga)}}{\ro_2^{k+1}}\right]
\left[\frac{(J+P-1-k)!}{(J+P-1-k)^{\a}}\frac{e^{(J+P-k-3)(\de+\Ga)}}{\ro_2^{(J+P-k-1)}}\right]\\
&&\ML\ML\ML\ML + \sum_{k=\left[\frac{J+P-1}{2}\right]+1}^{J+P-1}\cbin{J+P-1}{k}
\frac{{\hat F}}{r}\left[\frac{(k+1)!}{(k+1)^{\a}}\frac{e^{(k-1)(\de+\Ga)}}{\ro_2^{k+1}}\right]\nn
\left[\frac{(J+P-1-k)!}{(J+P-1-k)^{\a}}\frac{e^{(J+P-k-3)(\de+\Ga)}}{\ro_2^{(J+P-k-1)}}\right].\\
\eea
Let us estimate the first sum:
\bea
&&\ML\ML\ML\ML\ML\ML\leq \left({\hat F}\frac{(J+P)!}{(J+P)^{\a}}\frac{e^{(J+P-2)(\de+\Ga)}}{\ro_2^{J+P}}\right)\left(\frac{e^{-2(\de+\Ga)}}{r}\right)
\left[\frac{(J+P)^{\a}}{(J+P)!}\sum_{k=0}^{\left[\frac{J+P-1}{2}\right]}\frac{(J+P-1)!}{(J+P-1-k)!k!}\frac{(k+1)!}{(k+1)^{\a}}\frac{(J+P-1-k)!}{(J+P-1-k)^{\a}}\right]\ \nn
\eea
and
\bea
&&\ML\ML\ML\left[\frac{(J+P)^{\a}}{(J+P)!}\sum_{k=0}^{\left[\frac{J+P-1}{2}\right]}\frac{(J+P-1)!}{(J+P-1-k)!k!}\frac{(k+1)!}{(k+1)^{\a}}\frac{(J+P-1-k)!}{(J+P-1-k)^{\a}}\right]\\
&&\ML\ML\ML\ML\leq c\left[\frac{(J+P-1)!(J+P)^{\a}}{(J+P)!(J+P-1-\left[\frac{J+P-1}{2}\right])^{\a}}\sum_{k=0}^{\left[\frac{J+P-1}{2}\right]}\frac{1}{(k+1)^{\a-1}}\right]\leq C(\a)\nn
\eea
provided $\a>2$\ . The second sum of the same term can be treated in the same way. We estimate in the same way the second term of equation \ref{5.109}.
 Let us consider now the most delicate term, namely,
  \[(\nabb^{p_0}\ddb_{\nu}\nabb^{p_1}\ddb_{\nu}\nabb^{p_2}\c\c\c\ddb_{\nu}\nabb^{p_{J-1}})\left([\ddb_{\nu},\nabb^{p_{J}}]U\right)\]
The estimate proceeds in the following way,
\bea
&&\ML\big|[(II)]\big|\equiv \left|r^{1+J+P+2\si(P)-\frac{2}{p}}\nabb^{p_0}\ddb_{\nu}\nabb^{p_1}\ddb_{\nu}\nabb^{p_2}\c\c\c\ddb_{\nu}\nabb^{p_{J-1}}\left([\ddb_{\nu},\nabb^{p_{J}}]U\right)\right|_{p,S}\nn\\
&&\ML\ML\ \ \leq \left|r^{1+J+P+2\si(P)-\frac{2}{p}}\nabb^{p_0}\ddb_{\nu}\nabb^{p_1}\ddb_{\nu}\nabb^{p_2}\c\c\c\ddb_{\nu}\nabb^{p_{J-1}}\bigg(\sum_{h=0}^{p_{J}-1}\nabb^h[\ddb_{\nu},\nabb]\nabb^{p_{J}-1-h}U\bigg)\right|_{p,S}\nn\\
&&\ML\ML\ \ \leq \sum_{h=0}^{p_{J}-1}\left|r^{1+J+P+2\si(P)-\frac{2}{p}}\nabb^{p_0}\ddb_{\nu}\nabb^{p_1}\ddb_{\nu}\nabb^{p_2}\c\c\c\ddb_{\nu}\nabb^{p_{J-1}+h}\bigg([\ddb_{\nu},\nabb]\nabb^{p_{J}-1-h}U\bigg)\right|_{p,S}\nn
\eea
As the following holds
\bea
[\ddb_{\nu},\nabb]\nabb^{p_{J}-1-h}U=\left(-({p_{J}}-1-h)C\nabb^{{p_{J}}-1-h}U+\oom\chi\nabb^{{p_{J}}-h}U\right)\ ,
\eea
\bea
&&\ML\ML\ML\ML\big|[(II)]\leq\nn\\
&&\ML\ML\ML\ML\leq \sum_{h=0}^{p_{J}-1}\left|r^{1+J+P+2\si(P)-\frac{2}{p}}\nabb^{p_0}\ddb_{\nu}\nabb^{p_1}\ddb_{\nu}\nabb^{p_2}\c\c\c\ddb_{\nu}\nabb^{p_{J-1}+h}\bigg(-({p_{J}}-1-h)C\nabb^{{p_{J}}-1-h}U+\oom\chi\nabb^{{p_{J}}-h}U\bigg)\right|_{p,S}\nn\\
&&\ML\ML\ML\ML\leq \sum_{h=0}^{p_{J}-1}({p_{J}}-1-h)\left|r^{1+J+P+2\si(P)-\frac{2}{p}}\nabb^{p_0}\ddb_{\nu}\nabb^{p_1}\ddb_{\nu}\nabb^{p_2}\c\c\c\ddb_{\nu}\nabb^{p_{J-1}+h}\big(C\nabb^{{p_{J}}-1-h}U\big)\right|_{p,S}\nn\\
&&\ML\ML\ML\ML+\sum_{h=0}^{p_{J}-1}\left|r^{1+J+P+2\si(P)-\frac{2}{p}}\nabb^{p_0}\ddb_{\nu}\nabb^{p_1}\ddb_{\nu}\nabb^{p_2}\c\c\c\ddb_{\nu}\nabb^{p_{J-1}+h}\big(\oom\chi\nabb^{{p_{J}}-h-1}\nabb U\big)\right|_{p,S}\nn\\
&&\ML\ML\ML\ML\leq \sum_{h=0}^{p_{J}-1}({p_{J}}-1-h)\sum_{l_1=0}^1\c\c\c\sum_{l_{J-2}=0}^1\sum_{l_{J-1}=0}^1
\sum_{h_0=0}^{p_0}\ \ \ \sum_{h_1=0}^{p_1}\c\c\c\sum_{h_{J-2}=0}^{p_{J-2}}\sum_{h_{J-1}=0}^{p_{J-1}+h}\cbin{p_0}{h_0}\cbin{p_1}{h_1}\c\c\c\cbin{p_{J-2}}{h_{J-2}}\cbin{p_{J-1}+h}{h_{J-1}}\nn\\
&&\ML\ML\ML\ML\ML\ML\ML\ML\ML\ML\left|r^{1+J+P+2\si(P)-\frac{2}{p}}(\nabb^{h_0}\ddb_{\nu}^{l_1}\nabb^{h_1}\c\c\c\ddb_{\nu}^{l_{J-2}}\nabb^{h_{J-2}}\ddb_{\nu}^{l_{J-1}}\nabb^{h_{J-1}} C)\big(\nabb^{p_0-h_0}\ddb_{\nu}^{1-l_1}\nabb^{p_1-h_1}\c\c\c\ddb_{\nu}^{1-l_{J-2}}\nabb^{p_{J-2}-h_{J-2}}\ddb_{\nu}^{1-l_{J-1}}\nabb^{p_{J-1}+p_J-h_{J-1}-1}U\big)\right|_{p,S}\nn\\
&&\ML\ML\ML\ML\ML\ML\ML+\sum_{h=0}^{p_{J}-1}\sum_{l_1=0}^1\c\c\c\sum_{l_{J-2}=0}^1\sum_{l_{J-1}=0}^1
\sum_{h_0=0}^{p_0}\ \ \ \sum_{h_1=0}^{p_1}\c\c\c\sum_{h_{J-2}=0}^{p_{J-2}}\sum_{h_{J-1}=0}^{p_{J-1}+h}\cbin{p_0}{h_0}\cbin{p_1}{h_1}\c\c\c\cbin{p_{J-2}}{h_{J-2}}\cbin{p_{J-1}+h}{h_{J-1}}\nn\\
&&\ML\ML\ML\ML\ML\ML\ML\ML\ML\ML\left|r^{1+J+P+2\si(P)-\frac{2}{p}}(\nabb^{h_0}\ddb_{\nu}^{l_1}\nabb^{h_1}\c\c\c\ddb_{\nu}^{l_{J-2}}\nabb^{h_{J-2}}\ddb_{\nu}^{l_{J-1}}\nabb^{h_{J-1}}\oom\chi)\big(\nabb^{p_0-h_0}\ddb_{\nu}^{1-l_1}\nabb^{p_1-h_1}\c\c\c\ddb_{\nu}^{1-l_{J-2}}\nabb^{p_{J-2}-h_{J-2}}\ddb_{\nu}^{1-l_{J-1}}\nabb^{p_{J-1}+p_J-h_{J-1}-1}\nabb U\big)\right|_{p,S}\nn
\eea
Let us consider the part of the sum with $L+{H}\leq \left[\frac{J+P}{2}\right]$, 
\bea
&&\ML\ML\ML\ML\big|[(II)]\leq \\
&&\ML\ML\ML\ML\ML\ML\ML\ML\ML\sum_{l_1=0}^1\c\c\c\sum_{l_{J-2}=0}^1\sum_{l_{J-1}=0}^1
\sum_{h_0=0}^{p_0}\ \ \ \sum_{h_1=0}^{p_1}\c\c\c\sum_{h_{J-2}=0}^{p_{J-2}}\cbin{p_0}{h_0}\cbin{p_1}{h_1}\c\c\c\cbin{p_{J-2}}{h_{J-2}}\left(\sum_{h=0}^{p_{J}-1}({p_{J}}-1-h)\sum_{h_{J-1}=0}^{p_{J-1}+h}\cbin{p_{J-1}+h}{h_{J-1}}\right)
\chi\left(L+H\leq \left[\frac{J+P}{2}\right]\right)\nn\\
&&\ML\ML\ML\ML\ML\left|r^{1+L+H}(\nabb^{h_0}\ddb_{\nu}^{l_1}\nabb^{h_1}\c\c\c\ddb_{\nu}^{l_{J-2}}\nabb^{h_{J-2}}\ddb_{\nu}^{l_{J-1}}
\nabb^{h_{J-1}} C)\right|_{\infty,S}\c\nn\\
&&\ML\ML\ML\ML\ML\c\left|r^{J+P-(L+H)+2\si(P)-\frac{2}{p}}\big(\nabb^{p_0-h_0}\ddb_{\nu}^{1-l_1}\nabb^{p_1-h_1}\c\c\c\ddb_{\nu}^{1-l_{J-2}}\nabb^{p_{J-2}-h_{J-2}}\ddb_{\nu}^{1-l_{J-1}}\nabb^{p_{J-1}+p_J-h_{J-1}-1}U\big)\right|_{p,S}\nn\\
&&\ML\ML\ML\ML\ML\ML\ML\ML+\sum_{h=0}^{p_{J}-1}\sum_{l_1=0}^1\c\c\c\sum_{l_{J-2}=0}^1\sum_{l_{J-1}=0}^1
\sum_{h_0=0}^{p_0}\ \ \ \sum_{h_1=0}^{p_1}\c\c\c\sum_{h_{J-2}=0}^{p_{J-2}}\sum_{h_{J-1}=0}^{p_{J-1}+h}\cbin{p_0}{h_0}\cbin{p_1}{h_1}\c\c\c\cbin{p_{J-2}}{h_{J-2}}\cbin{p_{J-1}+h}{h_{J-1}}
\chi\left(L+H\leq \left[\frac{J+P}{2}\right]\right)\nn\\
&&\ML\ML\ML\ML\ML\left|r^{1+L+H}(\nabb^{h_0}\ddb_{\nu}^{l_1}\nabb^{h_1}\c\c\c\ddb_{\nu}^{l_{J-2}}\nabb^{h_{J-2}}\ddb_{\nu}^{l_{J-1}}
\nabb^{h_{J-1}}\oom\chi)\right|_{\infty,S}\c\nn\\
&&\ML\ML\ML\ML\ML\c\left|r^{J+P-(L+H)+2\si(P)-\frac{2}{p}}\big(\nabb^{p_0-h_0}\ddb_{\nu}^{1-l_1}\nabb^{p_1-h_1}\c\c\c\ddb_{\nu}^{1-l_{J-2}}\nabb^{p_{J-2}-h_{J-2}}\ddb_{\nu}^{1-l_{J-1}}\nabb^{p_{J-1}+p_J-h_{J-1}-1}\nabb U\big)\right|_{p,S}\nn\\
&&\ML\ML\ML\ML\ML\ML\ML\ML\ML\ML\leq\sum_{l_1=0}^1\c\c\c\sum_{l_{J-2}=0}^1\sum_{l_{J-1}=0}^1
\sum_{h_0=0}^{p_0}\ \ \ \sum_{h_1=0}^{p_1}\c\c\c\sum_{h_{J-2}=0}^{p_{J-2}}\cbin{p_0}{h_0}\cbin{p_1}{h_1}\c\c\c\cbin{p_{J-2}}{h_{J-2}}\left(\sum_{h=0}^{p_{J}-1}({p_{J}}-1-h)\sum_{h_{J-1}=0}^{p_{J-1}+h}\cbin{p_{J-1}+h}{h_{J-1}}\right)
\chi\left(L+H\leq \left[\frac{J+P}{2}\right]\right)\nn\\
&&\ML\ML\ML\ML\ML\left|r^{1+L+H}(\nabb^{h_0}\ddb_{\nu}^{l_1}\nabb^{h_1}\c\c\c\ddb_{\nu}^{l_{J-2}}\nabb^{h_{J-2}}\ddb_{\nu}^{l_{J-1}}
\nabb^{h_{J-1}} C)\right|_{\infty,S}\c\nn\\
&&\ML\ML\ML\ML\ML\c\left|r^{J+P-(L+H)+2\si(P)-\frac{2}{p}}\big(\nabb^{p_0-h_0}\ddb_{\nu}^{1-l_1}\nabb^{p_1-h_1}\c\c\c\ddb_{\nu}^{1-l_{J-2}}\nabb^{p_{J-2}-h_{J-2}}\ddb_{\nu}^{1-l_{J-1}}\nabb^{p_{J-1}+p_J-h_{J-1}-1}U\big)\right|_{p,S}\nn\\
&&\ML\ML\ML\ML\ML\ML\ML\ML+\sum_{h=0}^{p_{J}-1}\sum_{l_1=0}^1\c\c\c\sum_{l_{J-2}=0}^1\sum_{l_{J-1}=0}^1\sum_{h_0=0}^{p_0}\ \ \ \sum_{h_1=0}^{p_1}\c\c\c\sum_{h_{J-2}=0}^{p_{J-2}}\sum_{h_{J-1}=0}^{p_{J-1}+h}\cbin{p_0}{h_0}\cbin{p_1}{h_1}\c\c\c\cbin{p_{J-2}}{h_{J-2}}\cbin{p_{J-1}+h}{h_{J-1}}
\chi\left(L+H\leq \left[\frac{J+P}{2}\right]\right)\nn\\
&&\ML\ML\ML\ML\ML\left|r^{1+L+H}(\nabb^{h_0}\ddb_{\nu}^{l_1}\nabb^{h_1}\c\c\c\ddb_{\nu}^{l_{J-2}}\nabb^{h_{J-2}}\ddb_{\nu}^{l_{J-1}}
\nabb^{h_{J-1}}\oom\chi)\right|_{\infty,S}\c\nn\\
&&\ML\ML\ML\ML\ML\c\left|r^{J+P-(L+H)+2\si(P)-\frac{2}{p}}\big(\nabb^{p_0-h_0}\ddb_{\nu}^{1-l_1}\nabb^{p_1-h_1}\c\c\c\ddb_{\nu}^{1-l_{J-2}}\nabb^{p_{J-2}-h_{J-2}}\ddb_{\nu}^{1-l_{J-1}}\nabb^{p_{J-1}+p_J-h_{J-1}-1}\nabb U\big)\right|_{p,S}\nn\\
&&\ML\ML\ML\ML\ML\ML\ML\ML\ML\ML\leq\sum_{l_1=0}^1\c\c\c\sum_{l_{J-2}=0}^1\sum_{l_{J-1}=0}^1\sum_{h_0=0}^{p_0}\ \ \ \sum_{h_1=0}^{p_1}\c\c\c\sum_{h_{J-2}=0}^{p_{J-2}}\cbin{p_0}{h_0}\cbin{p_1}{h_1}\c\c\c\cbin{p_{J-2}}{h_{J-2}}\left(\sum_{h=0}^{p_{J}-1}({p_{J}}-1-h)\sum_{h_{J-1}=0}^{p_{J-1}+h}\cbin{p_{J-1}+h}{h_{J-1}}\right)
\chi\left(L+H\leq \left[\frac{J+P}{2}\right]\right)\nn\\
&&\ML\ML\ML\ML\ML\frac{1}{r^{2\si(P-H-1)}}\left|r^{3+L+H}(\nabb^{h_0}\ddb_{\nu}^{l_1}\nabb^{h_1}\c\c\c\ddb_{\nu}^{l_{J-2}}\nabb^{h_{J-2}}\ddb_{\nu}^{l_{J-1}}
\nabb^{h_{J-1}} C)\right|_{\infty,S}\c\nn\\
&&\ML\ML\ML\ML\ML\c\left|r^{J+P-(L+H)+2\si(P-H-1)-\frac{2}{p}}\big(\nabb^{p_0-h_0}\ddb_{\nu}^{1-l_1}\nabb^{p_1-h_1}\c\c\c\ddb_{\nu}^{1-l_{J-2}}\nabb^{p_{J-2}-h_{J-2}}\ddb_{\nu}^{1-l_{J-1}}\nabb^{p_{J-1}+p_J-h_{J-1}-1}U\big)\right|_{p,S}\nn\\
&&\ML\ML\ML\ML\ML\ML\ML\ML+\sum_{h=0}^{p_{J}-1}\sum_{l_1=0}^1\c\c\c\sum_{l_{J-2}=0}^1\sum_{l_{J-1}=0}^1\sum_{h_0=0}^{p_0}\ \ \ \sum_{h_1=0}^{p_1}\c\c\c\sum_{h_{J-2}=0}^{p_{J-2}}\sum_{h_{J-1}=0}^{p_{J-1}+h}\cbin{p_0}{h_0}\cbin{p_1}{h_1}\c\c\c\cbin{p_{J-2}}{h_{J-2}}\cbin{p_{J-1}+h}{h_{J-1}}
\chi\left(L+H\leq \left[\frac{J+P}{2}\right]\right)\nn\\
&&\ML\ML\ML\ML\ML\ML\ML\left|r^{1+L+H+\si(H)}(\nabb^{h_0}\ddb_{\nu}^{l_1}\nabb^{h_1}\c\c\c\ddb_{\nu}^{l_{J-2}}\nabb^{h_{J-2}}\ddb_{\nu}^{l_{J-1}}
\nabb^{h_{J-1}}\oom\chi)\right|_{\infty,S}\c\nn\\
&&\ML\ML\ML\ML\ML\ML\ML\c\frac{1}{r^{1+\si(H)}}\left|r^{3+(J-1)+P-(L+H)-\frac{2}{p}}\big(\nabb^{p_0-h_0}\ddb_{\nu}^{1-l_1}\nabb^{p_1-h_1}\c\c\c\ddb_{\nu}^{1-l_{J-2}}\nabb^{p_{J-2}-h_{J-2}}\ddb_{\nu}^{1-l_{J-1}}\nabb^{p_{J-1}+p_J-h_{J-1}-1}\nabb U\big)\right|_{p,S}\nn
\eea

\medskip



\NI{\bf Remark:} 

\NI {\em Observe that in the term product of the two previous factors there is one derivative $\ddb_{\nu}$ less and the same number of $\nabb$ derivatives, therefore the binomial coefficient of the second factor is $((J-1)+P-(L+H)+1)!=(J+P-(L+H))!$\ . Viceversa in the previous term of the sum where $C$ appears there is one derivative $\ddb_{\nu}$ less and also one derivative $\nabb$ less.}
\NI \beaa\ML\ML\ML\ML\mbox{Therefore}\ \ \ \ \ \ \ \ \ \ \ \ \ \ \ \ \ \ \ \ \ \ \ \ \ \ \ \ \ \ \ \ \ \ \ \ \ \ \ \ \ \ \ \ \ \ \ \ \ \ \ \ \ \ \ \ \ \ \ \ \ \ \ \ \ \ \ \ \ \ \ \ \ \ \ \ \ \ \ \ \ \ \ \ \ \ \ \eeaa
\bea
&&\ML\ML\ML\ML\big|[(II)]\leq \\
&&\ML\ML\ML\ML\ML\ML\ML\ML\leq\sum_{l_1=0}^1\c\c\c\sum_{l_{J-2}=0}^1\sum_{l_{J-1}=0}^1\sum_{h_0=0}^{p_0}\ \ \ \sum_{h_1=0}^{p_1}\c\c\c\sum_{h_{J-2}=0}^{p_{J-2}}\cbin{p_0}{h_0}\cbin{p_1}{h_1}\c\c\c\cbin{p_{J-2}}{h_{J-2}}\left(\sum_{h=0}^{p_{J}-1}({p_{J}}-1-h)\sum_{h_{J-1}=0}^{p_{J-1}+h}\cbin{p_{J-1}+h}{h_{J-1}}\right)
\chi\left(L+H\leq \left[\frac{J+P}{2}\right]\right)\nn\\
&&\ML\ML\ML\ML\ML\frac{1}{r^{2\si(P-H-1)}}\left[\frac{(L+H+2)!}{(L+H+2)^{\a}}\frac{e^{(L+H)(\de+\underline{\Ga}(\la))}}{\ro_2^{L+H+2}}
\frac{(J+P-(L+H)-2)!}{(J+P-(L+H)-2)^{\a}}\frac{e^{(J+P-(L+H)-4)(\de+\underline{\Ga}(\la))}}{\ro_2^{J+P-(L+H)-2}}\right]\nn\\
&&\ML\ML\ML\ML\ML\ML\ML\ML+\sum_{h=0}^{p_{J}-1}\sum_{l_1=0}^1\c\c\c\sum_{l_{J-2}=0}^1\sum_{l_{J-1}=0}^1\sum_{h_0=0}^{p_0}\ \ \ \sum_{h_1=0}^{p_1}\c\c\c\sum_{h_{J-2}=0}^{p_{J-2}}\sum_{h_{J-1}=0}^{p_{J-1}+h}\cbin{p_0}{h_0}\cbin{p_1}{h_1}\c\c\c\cbin{p_{J-2}}{h_{J-2}}\cbin{p_{J-1}+h}{h_{J-1}}
\chi\left(L+H\leq \left[\frac{J+P}{2}\right]\right)\nn\\
&&\ML\ML\ML\ML\ML\frac{1}{r^{1+\si(H)}}\left[\frac{(L+H+1)!}{(L+H+1)^{\a}}\frac{e^{((L+H)-1)(\de+\underline{\Ga}(\la))}}{\ro_2^{L+H+1}}\frac{(J+P-1-(L+H))!}{(J+P-1-(L+H))^{\a}}\frac{e^{(J+P-(L+H)-3)(\de+\underline{\Ga}(\la))}}{\ro_2^{J+P-(L+H)-1}}\right]\nn
\eea
which we rewrite, denoting $k=L+H$,
\bea
&&\ML\ML\ML\ML\big|[(II)]\leq \\
&&\ML\ML\ML\ML\ML\ML\ML\ML\leq\sum_{k=0}^{\left[\frac{J+P}{2}\right]}\de\left(L+H-k\right)\left\{\sum_{l_1=0}^1\c\c\c\sum_{l_{J-2}=0}^1\sum_{l_{J-1}=0}^1\sum_{h_0=0}^{p_0}\ \ \ \sum_{h_1=0}^{p_1}\c\c\c\sum_{h_{J-2}=0}^{p_{J-2}}\cbin{p_0}{h_0}\cbin{p_1}{h_1}\c\c\c\cbin{p_{J-2}}{h_{J-2}}\left(\sum_{h=0}^{p_{J}-1}({p_{J}}-1-h)\sum_{h_{J-1}=0}^{p_{J-1}+h}\cbin{p_{J-1}+h}{h_{J-1}}\right)\right\}\nn\\
&&\ML\ML\ML\ML\ML\frac{1}{r^{2\si(P-H-1)}}\left[\frac{(L+H+2)!}{(L+H+2)^{\a}}\frac{e^{(L+H)(\de+\underline{\Ga}(\la))}}{\ro_2^{L+H+2}}
\frac{(J+P-(L+H)-2)!}{(J+P-(L+H)-2)^{\a}}\frac{e^{(J+P-(L+H)-4)(\de+\underline{\Ga}(\la))}}{\ro_2^{J+P-(L+H)-2}}\right]\nn\\
&&\ML\ML\ML\ML\ML\ML\ML\ML+\sum_{k=0}^{\left[\frac{J+P}{2}\right]}\de\left(L+H-k\right)\left\{\sum_{h=0}^{p_{J}-1}\sum_{l_1=0}^1\c\c\c\sum_{l_{J-2}=0}^1\sum_{l_{J-1}=0}^1\sum_{h_0=0}^{p_0}\ \ \ \sum_{h_1=0}^{p_1}\c\c\c\sum_{h_{J-2}=0}^{p_{J-2}}\sum_{h_{J-1}=0}^{p_{J-1}+h}\cbin{p_0}{h_0}\cbin{p_1}{h_1}\c\c\c\cbin{p_{J-2}}{h_{J-2}}\cbin{p_{J-1}+h}{h_{J-1}}\right\}\nn\\
&&\ML\ML\ML\ML\ML\frac{1}{r^{1+\si(H)}}\left[\frac{(L+H+1)!}{(L+H+1)^{\a}}\frac{e^{((L+H)-1)(\de+\underline{\Ga}(\la))}}{\ro_2^{L+H+1}}\frac{(J+P-1-(L+H))!}{(J+P-1-(L+H))^{\a}}\frac{e^{(J+P-(L+H)-3)(\de+\underline{\Ga}(\la))}}{\ro_2^{J+P-(L+H)-1}}\right]
\eea
The following holds, see before eqs. (15.11),
\bea
\sum_{k=J}^{N-2}(N-k-1)\cbin{k}{J}=\cbin{N}{N-2-J}
\eea
Observe that, denoting
\[h_{J-1}\equiv J\ \ ,\ \ p_{J-1}+h\equiv k\ \ ,\ \ p_{J-1}+p_J\equiv N \]
\bea
&&\ML\ML\ML\ML\ML\sum_{h=0}^{p_{J}-1}({p_{J}}-1-h)\sum_{h_{J-1}=0}^{p_{J-1}+h}\cbin{p_{J-1}+h}{h_{J-1}}
=\sum_{k=p_{J-1}}^{N-1}(N-k-1)\sum_{J=0}^k\cbin{k}{J}
=\sum_{J=0}^{N-1}\sum_{k=J}^{N-1}(N-k-1)\cbin{k}{J}\nn\\
&&\ML\ML\ML\ML\ML=\sum_{J=0}^{N-1}\sum_{k=J}^{N-2}(N-k-1)\cbin{k}{J}=\sum_{J=0}^{N-2}\sum_{k=J}^{N-2}(N-k-1)\cbin{k}{J}=\sum_{J=0}^{N-2}\cbin{N}{N-2-J}\nn\\
&&\ML\ML\ML\ML\ML=\sum_{J=0}^{N-2}\cbin{N}{N-2-J}=\sum_{h_{J-1}=0}^{ p_{J-1}+p_J-2}\cbin{ p_{J-1}+p_J}{h_{J-1}+2}=\sum_{{\hat h}_{J-1}=2}^{ p_{J-1}+p_J}\cbin{ p_{J-1}+p_J}{{\hat h}_{J-1}}\nn\\
&&\ML\ML\ML\ML\ML\leq \sum_{{\hat h}_{J-1}=0}^{ p_{J-1}+p_J}\cbin{ p_{J-1}+p_J}{{\hat h}_{J-1}}\ ,
\eea
The following holds, see before 
\bea
\sum_{H=J}^{N-1}\cbin{H}{k}=\cbin{N}{k+1}=\cbin{N}{N-k-1}
\eea
Therefore denoting $J+P-1-(p_j-h)=H$
\bea
&&\ML\ML\ML\ML\ML\sum_{h=0}^{p_{J}-1}\cbin{J+P-1-(p_J-h)}{k}=\sum_{H=J+P-1-p_J}^{J+P-2}\cbin{H}{k}
\leq \sum_{H=k}^{J+P-2}\cbin{H}{k}=\cbin{J+P-1}{k+1}\ ,\ \ \ \ \ \ \ 
\eea
where the inequality has a meaning if we sum only non negative terms, which is the present case; therefore the previous estimate can be rewritten as 
\bea
&&\ML\ML\ML\ML\big|[(II)]\leq \\
&&\ML\ML\ML\ML\ML\ML\ML\ML\leq\sum_{k=0}^{\left[\frac{J+P}{2}\right]}\de\left(L+H-k\right)\left\{\sum_{l_1=0}^1\c\c\c\sum_{l_{J-2}=0}^1\sum_{l_{J-1}=0}^1\sum_{h_0=0}^{p_0}\ \ \ \sum_{h_1=0}^{p_1}\c\c\c\sum_{h_{J-2}=0}^{p_{J-2}}\sum_{{\hat h}_{J-1}=2}^{ p_{J-1}+p_J}\cbin{p_0}{h_0}\cbin{p_1}{h_1}\c\c\c\cbin{p_{J-2}}{h_{J-2}}\cbin{ p_{J-1}+p_J}{{\hat h}_{J-1}}\right\}\nn\\
&&\ML\ML\ML\ML\ML\frac{1}{r^{2\si(P-H-1)}}\left[\frac{(L+H+2)!}{(L+H+2)^{\a}}\frac{e^{((L+H))(\de+\underline{\Ga}(\la))}}{\ro_2^{L+H+2}}
\frac{(J+P-(L+H)-2)!}{(J+P-(L+H)-2)^{\a}}\frac{e^{(J+P-(L+H)-4)(\de+\underline{\Ga}(\la))}}{\ro_2^{J+P-(L+H)-2}}\right]\nn\\
&&\ML\ML\ML\ML\ML\ML\ML\ML+\sum_{k=0}^{\left[\frac{J+P}{2}\right]}\de\left(L+H-k\right)\left\{\sum_{h=0}^{p_{J}-1}\sum_{l_1=0}^1\c\c\c\sum_{l_{J-2}=0}^1\sum_{l_{J-1}=0}^1\sum_{h_0=0}^{p_0}\ \ \ \sum_{h_1=0}^{p_1}\c\c\c\sum_{h_{J-2}=0}^{p_{J-2}}\sum_{h_{J-1}=0}^{p_{J-1}+h}\cbin{p_0}{h_0}\cbin{p_1}{h_1}\c\c\c\cbin{p_{J-2}}{h_{J-2}}\cbin{p_{J-1}+h}{h_{J-1}}\right\}\nn\\
&&\ML\ML\ML\ML\ML\frac{1}{r^{1+\si(H)}}\left[\frac{(L+H+1)!}{(L+H+1)^{\a}}\frac{e^{((L+H)-1)(\de+\underline{\Ga}(\la))}}{\ro_2^{L+H+1}}\frac{(J+P-1-(L+H))!}{(J+P-1-(L+H))^{\a}}\frac{e^{(J+P-(L+H)-3)(\de+\underline{\Ga}(\la))}}{\ro_2^{J+P-(L+H)-1}}\right]\ .\nn
\eea
Recall that, see \ref{cbinrel},
\bea
&&\ML\ML\ML\ML\ML\ML\de\left(L+H-k\right)\left\{\sum_{l_1=0}^1\c\c\c\sum_{l_{J-2}=0}^1\sum_{l_{J-1}=0}^1\sum_{h_0=0}^{p_0}\ \ \ \sum_{h_1=0}^{p_1}\c\c\c\sum_{h_{J-2}=0}^{p_{J-2}}\sum_{{\hat h}_{J-1}=2}^{ p_{J-1}+p_J}\cbin{p_0}{h_0}\cbin{p_1}{h_1}\c\c\c\cbin{p_{J-2}}{h_{J-2}}\cbin{ p_{J-1}+p_J}{{\hat h}_{J-1}}\right\}\nn\\
&&\ML\ML\ML\ML\ML\ML\ML``\leq"\left(L+H-k\right)\left\{\sum_{l_1=0}^1\c\c\c\sum_{l_{J-2}=0}^1\sum_{l_{J-1}=0}^1\sum_{h_0=0}^{p_0}\ \ \ \sum_{h_1=0}^{p_1}\c\c\c\sum_{h_{J-2}=0}^{p_{J-2}}\sum_{{\hat h}_{J-1}=0}^{ p_{J-1}+p_J}\cbin{p_0}{h_0}\cbin{p_1}{h_1}\c\c\c\cbin{p_{J-2}}{h_{J-2}}\cbin{ p_{J-1}+p_J}{{\hat h}_{J-1}}\right\}=\cbin{J+P-1}{k}\nn
\eea
and
\bea
&&\ML\ML\ML\ML\de\left(L+H-k\right)\left\{\sum_{h=0}^{p_{J}-1}\sum_{l_1=0}^1\c\c\c\sum_{l_{J-2}=0}^1\sum_{l_{J-1}=0}^1\sum_{h_0=0}^{p_0}\ \ \ \sum_{h_1=0}^{p_1}\c\c\c\sum_{h_{J-2}=0}^{p_{J-2}}\sum_{h_{J-1}=0}^{p_{J-1}+h}\cbin{p_0}{h_0}\cbin{p_1}{h_1}\c\c\c\cbin{p_{J-2}}{h_{J-2}}\cbin{p_{J-1}+h}{h_{J-1}}\right\}\nn\\
&&\ML\ML\ML\ML=\sum_{h=0}^{p_{J}-1}\cbin{J+P-1-(p_J-h)}{k}
\eea
and substituting in the previous expression, we obtain
\bea
&&\ML\ML\ML\ML\ML\ML\ML\big|[(II)]\big|\leq\sum_{k=0}^{\left[\frac{J+P}{2}\right]}\cbin{J+P-1}{k}\!
\left[\frac{(k+2)!}{(k+2)^{\a}}\frac{e^{k(\de+\underline{\Ga}(\la))}}{\ro^{k+2}}
\frac{(J+P-k-2)!}{(J+P-k-2)^{\a}}\frac{e^{(J+P-k-4)(\de+\underline{\Ga}(\la))}}{\ro_2^{J+P-k-2}}\right]\nn\\
&&\ML\ML\ML\ML\ML\ML\ML+\frac{1}{r}\sum_{k=0}^{\left[\frac{J+P}{2}\right]}\sum_{h=0}^{p_{J}-1}\chi\bigg(k\leq(J+P+h-1-p_J)\bigg)\cbin{J+P-1-(p_J-h)}{k}\!
\left[\frac{(k+1)!}{(k+1)^{\a}}\frac{e^{(k-1)(\de+\underline{\Ga}(\la))}}{\ro_2^{k+1}}\frac{(J+P-k-1)!}{(J+P-k-1)^{\a}}\frac{e^{(J+P-k-3)(\de+\underline{\Ga}(\la))}}{\ro_2^{J+P-k-1}}\right]\ .\nn
\eea
Let us estimate the two sums separately
\bea
&&\ML\ML\ML\ML\ML\ML\ML\sum_{k=0}^{\left[\frac{J+P}{2}\right]}\cbin{J+P-1}{k}\!
\left[\frac{(k+2)!}{(k+2)^{\a}}\frac{e^{k(\de+\underline{\Ga}(\la))}}{\ro_2^{k+2}}
\frac{(J+P-k-2)!}{(J+P-k-2)^{\a}}\frac{e^{(J+P-k-4)(\de+\underline{\Ga}(\la))}}{\ro_2^{J+P-k-2}}\right]\nn\\
&&\ML\ML\ML\ML\ML\ML\ML\leq c\left(\frac{(J+P)!}{(J+P)^{\a}}\frac{e^{(J+P-2)(\de+\underline{\Ga}(\la))}}{\ro_2^{J+P}}\right)\!\left(e^{-2(\de+\underline{\Ga}(\la))}\right)
\c
\left\{\frac{(J+P)^{\a}}{(J+P)!}\sum_{k=0}^{\left[\frac{J+P}{2}\right]}\left[\cbin{J+P-1}{k}\!\frac{(k+2)!}{(k+2)^{\a}}
\frac{(J+P-k-2)!}{(J+P-k-2)^{\a}}\right]\right\}\nn
\eea
and
\bea
&&\ML\ML\ML\ML\ML\ML\ML\bigg\{\c\c\c\bigg\}\leq \frac{(J+P)^{\a}}{{(J+P-\left[\frac{J+P}{2}\right]-2)^{\a}}}\frac{(J+P-1)!}{(J+P)!}
\sum_{k=0}^{\left[\frac{J+P}{2}\right]}\frac{(k+2)(k+1)}{(J+P-k-1)(k+2)^{\a}}\leq C(\a)\ \ ,\ \ \ \ \ 
\eea
provided $\a\geq 1$. 
The estimate of the second term is slightly more delicate
\bea
&&\ML\ML\ML\ML\ML\ML\ML\sum_{k=0}^{\left[\frac{J+P}{2}\right]}\sum_{h=0}^{p_{J}-1}\chi\bigg(k\leq(J+P+h-1-p_J)\bigg)\cbin{J+P-1-(p_J-h)}{k}\!
\left[\frac{(k+1)!}{(k+1)^{\a}}\frac{e^{(k-1)(\de+\Ga)}}{\ro_2^{k+1}}\frac{(J+P-k-1)!}{(J+P-k-1)^{\a}}\frac{e^{(J+P-k-3)(\de+\underline{\Ga}(\la))}}{\ro_2^{J+P-k-1}}\right]\nn\\
&&\ML\ML\ML\ML\ML\leq  c\left(\frac{(J+P)!}{(J+P)^{\a}}\frac{e^{(J+P-2)(\de+\Ga)}}{\ro_2^{J+P}}\right)\!\left({e^{-2(\de+\underline{\Ga}(\la))}}\right)
\left\{\sum_{k=0}^{\left[\frac{J+P}{2}\right]}\sum_{h=0}^{p_{J}-1}\chi\bigg(k\leq(J+P+h-1-p_J)\bigg)\c\right.\nn\\
&&\ \ \ \ \ \ \c\left.\frac{(J+P)^{\a}}{(J+P)!}\cbin{J+P-1-(p_J-h)}{k}\!\left[\frac{(k+1)!}{(k+1)^{\a}}\frac{(J+P-k-1)!}{(J+P-k-1)^{\a}}\right]\right\}
\eea
Observe that
\bea
&&\ML\ML\ML\ML\bigg\{\c\c\c\bigg\}\leq \left\{\sum_{k=0}^{\left[\frac{J+P}{2}\right]}
\frac{(J+P)^{\a}}{(J+P)!}\cbin{J+P-1}{k+1}\!\left[\frac{(k+1)!}{(k+1)^{\a}}\frac{(J+P-k-1)!}{(J+P-k-1)^{\a}}\right]\right\}\nn\\
&&\ML\ML\ML\ML\leq\frac{(J+P)^{\a}}{(J+P)!}\sum_{k=0}^{\left[\frac{J+P}{2}\right]}\frac{(J+P-1)!}{(J+P-k-2)!(k+1)!}\!\left[\frac{(k+1)!}{(k+1)^{\a}}\frac{(J+P-k-1)!}{(J+P-k-1)^{\a}}\right]\nn\\
&&\ML\ML\ML\ML\leq\sum_{k=0}^{\left[\frac{J+P}{2}\right]}\frac{1}{(k+1)^{\a}}\left[\frac{(J+P)^{\a}(J+P-1)!(J+P-k-1)!}{(J+P-k-1)^{\a}(J+P)!(J+P-k-2)!}\right]\leq c(\a)\sum_{k=0}^{\left[\frac{J+P}{2}\right]}\frac{1}{(k+1)^{\a}}\leq C(\a)\ .\nn\\
\eea
provided $\a>1$\ \ \ . The estimate of the sum with $k>\left[\frac{J+P}{2}\right]$ is done in the same way and has already been done in previous cases and we do not repeat here.

\subsection{ Proof of Lemma \ref{mixedderivatives3}}

We write, with $\sum_{s=0}^{J}p_s=P$,
\bea
&&\ML\nabb^{p_0}\ddb_{\nu}\nabb^{p_1}\ddb_{\nu}\nabb^{p_2}\c\c\c\ddb_{\nu}\nabb^{p_{J-1}}\ddb_{\nu}\nabb^{p_{J}}\Psi\nn\\
&&\ML\ML=\ddb_{\nu}\nabb^{p_0+p_1}\ddb_{\nu}\nabb^{p_2}\c\c\c\ddb_{\nu}\nabb^{p_{J-1}}\ddb_{\nu}\nabb^{p_{J}}\Psi
+[\nabb^{p_0},\ddb_{\nu}]\nabb^{p_1}\ddb_{\nu}\nabb^{p_2}\c\c\c\ddb_{\nu}\nabb^{p_{J-1}}\ddb_{\nu}\nabb^{p_{J}}\Psi\nn\\
&&\nn\\
&&\ML\ML\ML\ML=\ddb_{\nu}^2\nabb^{p_0+p_1+p_2}\ddb_{\nu}\nabb^{p_3}\ddb_{\nu}\c\c\c\ddb_{\nu}\nabb^{p_{J-1}}\ddb_{\nu}\nabb^{p_{J}}\Psi
+\ddb_{\nu}[\nabb^{p_0+p_1},\ddb_{\nu}]\nabb^{p_2}\c\c\c\ddb_{\nu}\nabb^{p_{J-1}}\ddb_{\nu}\nabb^{p_{J}}\Psi\nn\\
&&\ML\ML\ML+[\nabb^{p_0},\ddb_{\nu}]\nabb^{p_1}\ddb_{\nu}\nabb^{p_2}\c\c\c\ddb_{\nu}\nabb^{p_{J-1}}\ddb_{\nu}\nabb^{p_{J}}\Psi\nn\\
&&\nn\\
&&\ML\ML\ML\ML=\ddb_{\nu}^3\nabb^{p_0+p_1+p_2+p_3}\ddb_{\nu}\nabb^{p_4}\ddb_{\nu}\c\c\c\ddb_{\nu}\nabb^{p_{J-1}}\ddb_{\nu}\nabb^{p_{J}}\Psi
+\ddb_{\nu}^2[\nabb^{p_0+p_1+p_2},\ddb_{\nu}]\nabb^{p_3}\ddb_{\nu}\c\c\c\ddb_{\nu}\nabb^{p_{J-1}}\ddb_{\nu}\nabb^{p_{J}}\Psi\nn\\
&&\ML\ML\ML+\ddb_{\nu}[\nabb^{p_0+p_1},\ddb_{\nu}]\nabb^{p_2}\c\c\c\ddb_{\nu}\nabb^{p_{J-1}}\ddb_{\nu}\nabb^{p_{J}}\Psi
+[\nabb^{p_0},\ddb_{\nu}]\nabb^{p_1}\ddb_{\nu}\nabb^{p_2}\c\c\c\ddb_{\nu}\nabb^{p_{J-1}}\ddb_{\nu}\nabb^{p_{J}}\Psi\nn\\
&&\nn\\
&&\ML\ML\ML\ML=\ddb_{\nu}^4\nabb^{p_0+p_1+p_2+p_3+p_4}\ddb_{\nu}\nabb^{p_5}\c\c\c\ddb_{\nu}\nabb^{p_{J-1}}\ddb_{\nu}\nabb^{p_{J}}\Psi
+\ddb_{\nu}^3[\nabb^{p_0+p_1+p_2+p_3},\ddb_{\nu}]\nabb^{p_4}\ddb_{\nu}\c\c\c\ddb_{\nu}\nabb^{p_{J-1}}\ddb_{\nu}\nabb^{p_{J}}\Psi\nn\\
&&\ML\ML\ML+\ddb_{\nu}^2[\nabb^{p_0+p_1+p_2},\ddb_{\nu}]\nabb^{p_3}\ddb_{\nu}\c\c\c\ddb_{\nu}\nabb^{p_{J-1}}\ddb_{\nu}\nabb^{p_{J}}\Psi
+\ddb_{\nu}[\nabb^{p_0+p_1},\ddb_{\nu}]\nabb^{p_2}\c\c\c\ddb_{\nu}\nabb^{p_{J-1}}\ddb_{\nu}\nabb^{p_{J}}\Psi\nn\\
&&\ML\ML\ML+[\nabb^{p_0},\ddb_{\nu}]\nabb^{p_1}\ddb_{\nu}\nabb^{p_2}\c\c\c\ddb_{\nu}\nabb^{p_{J-1}}\ddb_{\nu}\nabb^{p_{J}}\Psi\nn\\
&&......\nn\\
&&\ML\ML\ML\ML\ML\ML\ML=\ddb_{\nu}^J\nabb^{p_0+p_1+p_2+\c\c+p_{J-1}+p_J}\Psi+\ddb_{\nu}^{J-1}[\nabb^{p_0+p_1+p_2+\c\c p_{J-1}},\ddb_{\nu}]\nabb^{p_{J}}\Psi
+\ddb_{\nu}^{J-2}[\nabb^{p_0+p_1+p_2+\c\c p_{J-2}},\ddb_{\nu}]\nabb^{p_{J-1}}\ddb_{\nu}\nabb^{p_{J}}\Psi\nn\\
&&\ML\ML\ML\ML\ML\ML+\ddb_{\nu}^{J-3}[\nabb^{p_0+p_1+p_2+\c\c p_{J-3}},\ddb_{\nu}]\nabb^{p_{J-2}}\ddb_{\nu}\nabb^{p_{J-1}}\ddb_{\nu}\nabb^{p_{J}}\Psi
+\ddb_{\nu}^{J-4}[\nabb^{p_0+p_1+p_2+\c\c p_{J-4}},\ddb_{\nu}]\nabb^{p_{J-3}}\ddb_{\nu}\nabb^{p_{J-2}}\ddb_{\nu}\nabb^{p_{J-1}}\ddb_{\nu}\nabb^{p_{J}}\Psi\nn\\
&&\ML\ML\ML\ML\ML\ML+\c\c\c\c\nn\\
&&\ML\ML\ML\ML\ML+\ddb_{\nu}^3[\nabb^{p_0+p_1+p_2+p_3},\ddb_{\nu}]\nabb^{p_4}\ddb_{\nu}\c\c\c\ddb_{\nu}\nabb^{p_{J-1}}\ddb_{\nu}\nabb^{p_{J}}\Psi
+\ddb_{\nu}^2[\nabb^{p_0+p_1+p_2},\ddb_{\nu}]\nabb^{p_3}\ddb_{\nu}\c\c\c\ddb_{\nu}\nabb^{p_{J-1}}\ddb_{\nu}\nabb^{p_{J}}\Psi\nn\\
&&\ML\ML\ML\ML\ML\ML+\ddb_{\nu}[\nabb^{p_0+p_1},\ddb_{\nu}]\nabb^{p_2}\c\c\c\ddb_{\nu}\nabb^{p_{J-1}}\ddb_{\nu}\nabb^{p_{J}}\Psi
+[\nabb^{p_0},\ddb_{\nu}]\nabb^{p_1}\ddb_{\nu}\nabb^{p_2}\c\c\c\ddb_{\nu}\nabb^{p_{J-1}}\ddb_{\nu}\nabb^{p_{J}}\Psi\nn\\
&&......\nn\\
&&\ML\ML\ML\ML=\ddb_{\nu}^J\nabb^P\Psi+\sum_{t=1}^J\ddb_{\nu}^{J-t}[\nabb^{\sum_{s=0}^{J-t}p_s},\ddb_{\nu}]
\nabb^{p_{J-t+1}}\ddb_{\nu}\nabb^{p_{J-t+2}}\ddb_{\nu}\c\c\nabb^{p_{J-1}}\ddb_{\nu}\nabb^{p_J}\Psi
\eea
Therefore the final result is 
\bea
&&\ML\ML\ML\ML\nabb^{p_0}\ddb_{\nu}\nabb^{p_1}\ddb_{\nu}\nabb^{p_2}\c\c\c\ddb_{\nu}\nabb^{p_{J-1}}\ddb_{\nu}\nabb^{p_{J}}\Psi\\
&&\ML\ML\ML\ML=\ddb_{\nu}^J\nabb^P\Psi+\sum_{t=1}^J\ddb_{\nu}^{J-t}[\nabb^{\sum_{s=0}^{J-t}p_s},\ddb_{\nu}]
\nabb^{p_{J-t+1}}\ddb_{\nu}\nabb^{p_{J-t+2}}\ddb_{\nu}\c\c\nabb^{p_{J-1}}\ddb_{\nu}\nabb^{p_J}\Psi\ .\nn
\eea
and we have the following estimate,
\bea
&&\ML\ML\ML\ML\big||\la|^{{\underline\phi}(\Psi)}r^{1+J+P+\phi(\Psi)-\frac{2}{p}}\nabb^{p_0}\ddb_{\nu}\nabb^{p_1}\ddb_{\nu}\nabb^{p_2}\c\c\c\ddb_{\nu}\nabb^{p_{J-1}}\ddb_{\nu}\nabb^{p_{J}}\Psi\big|_{p,S}\leq\big||\la|^{{\underline\phi}(\Psi)}r^{1+J+P+\phi(\Psi)-\frac{2}{p}}\ddb_{\nu}^J\nabb^P\Psi\big|_{p,S}\nn\\
&&\ML\ML\ML\ML\ +\sum_{t=1}^J\big||\la|^{{\underline\phi}(\Psi)}r^{1+J+P+\phi(\Psi)-\frac{2}{p}}\ddb_{\nu}^{J-t}[\nabb^{\sum_{s=0}^{J-t}p_s},\ddb_{\nu}]
\nabb^{p_{J-t+1}}\ddb_{\nu}\nabb^{p_{J-t+2}}\ddb_{\nu}\c\c\nabb^{p_{J-1}}\ddb_{\nu}\nabb^{p_J}U\big|_{p,S}\ .\nn
\eea
We make the following inductive assumptions, for $J+P<N-1$,
\bea
\ML\ML\ML\ML||\la|^{{\underline\phi}(\Psi)}r^{1+J+P+\phi(\Psi)-\frac{2}{p}}\nabb^{p_0}\ddb_{\nu}\nabb^{p_1}\ddb_{\nu}\nabb^{p_2}\c\c\c\ddb_{\nu}\nabb^{p_{J-1}}\ddb_{\nu}\nabb^{p_{J}}\Psi|_{p,S}\leq {F}\!\left(\frac{(J+P+1)!}{(J+P+1)^{\a}}\frac{e^{(J+P-1)(\de+\underline{\Ga}(\la))}}{\ro_2^{J+P+1}}\right)\ .\ \ \ \ 
\eea
and use the following expression for the commutator
\bea
\ML\ML\ML\ML[\nabb^{p_{J}},\ddb_{\nu}]\Psi
\!&=&\!-\sum_{q=0}^{p_{J}-1}(p_{J}-q-1)\sum_{h=0}^q\cbin{q}{h}(\nabb^hC)\nabb^{p_{J}-1-h}U+\sum_{q=0}^{p_{J}-1}\sum_{h=0}^q\cbin{q}{h}(\nabb^h\chi)\nabb^{p_{J}-h}\Psi\nn\\
\!&=&\!-\sum_{h=0}^{p_{J}-2}\cbin{p_J}{p_J-2-h}(\nabb^hC)\nabb^{p_{J}-1-h}U+\sum_{q=0}^{p_{J}-1}\sum_{h=0}^q\cbin{q}{h}(\nabb^h\chi)\nabb^{p_{J}-h}\Psi\ \ \ \ \ \ \ \ \ \ \ \ \ \ \ \ \nn\\
\!&=&\!-\sum_{h=0}^{p_{J}-2}\cbin{p_J}{p_J-2-h}(\nabb^hC)\nabb^{p_{J}-1-h}\Psi+\sum_{h=0}^{p_{J}-1}\cbin{p_{J}}{p_{J}-1-h}(\nabb^h\chi)\nabb^{p_{J}-h}\Psi\ \ \ \ \ \ \ \ \ \ \ \ \ \ \ \ \nn\\
\!&=&\!-\sum_{q=0}^{p_{J}-2}\cbin{p_J}{q}(\nabb^{p_j-2-q}C)\nabb^{q+1}\Psi+\sum_{q=0}^{p_{J}-1}\cbin{p_{J}}{q}(\nabb^{p_J-1-q}\chi)\nabb^{q+1}\Psi\ \ \ \ \ \ \ \ \ \ \ \ \ \ \ \ \
\eea
where
\bea
C\equiv C^{\si}_{\mu\nu}=(\chi_{\mu\nu}\etab^{\si}\!-\!\chi_{\mu}^{\si}\etab_{\nu})+\theta^C_{\mu}\theta^D_{\nu}R^{{\si}}(\c,e_C,e_4,e_D)\ .
\eea
We start the estimate,
\bea
&&\ML\ML\ML\ML\ML\ML\sum_{t=1}^J\big||\la|^{{\underline\phi}(\Psi)}r^{1+J+P+\phi(\Psi)-\frac{2}{p}}\ddb_{\nu}^{J-t}[\nabb^{\left(\sum_{s=0}^{J-t}p_s\right)},\ddb_{\nu}]
\nabb^{p_{J-t+1}}\ddb_{\nu}\nabb^{p_{J-t+2}}\ddb_{\nu}\c\c\nabb^{p_{J-1}}\ddb_{\nu}\nabb^{p_J}\Psi\big|_{p,S}\nn\\
&&\ML\ML\ML\ML\ML\ML\leq \sum_{t=1}^J\sum_{h=0}^{\left(\sum_{s=0}^{J-t}p_s\right)-2}\cbin{\left(\sum_{s=0}^{J-t}p_s\right)}{\left(\sum_{s=0}^{J-t}p_s\right)-2-h}\c\nn\\
&&\ML\ML\ML\ML\c\big||\la|^{{\underline\phi}(\Psi)}r^{1+J+P+\phi(\Psi)-\frac{2}{p}}\ddb_{\nu}^{J-t}(\nabb^{h}C)(\nabb^{\left(\sum_{s=0}^{J-t}p_s-1-h\right)}\nabb^{p_{J-t+1}}\ddb_{\nu}\nabb^{p_{J-t+2}}\ddb_{\nu}\c\c\nabb^{p_{J-1}}\ddb_{\nu}\nabb^{p_J}\Psi)\big|_{p,S}\nn\\
&&\ML\ML\ML\ML\ML\ML+ \sum_{t=1}^J\sum_{h=0}^{\left(\sum_{s=0}^{J-t}p_s\right)-1}\cbin{\left(\sum_{s=0}^{J-t}p_s\right)}{\left(\sum_{s=0}^{J-t}p_s\right)-1-h}\c\nn\\
&&\ML\ML\ML\ML\big||\la|^{{\underline\phi}(\Psi)}r^{1+J+P+\phi(\Psi)-\frac{2}{p}}\ddb_{\nu}^{J-t}(\nabb^{h}\chi)(\nabb^{\left(\sum_{s=0}^{J-t}p_s-h\right)}\nabb^{p_{J-t+1}}\ddb_{\nu}\nabb^{p_{J-t+2}}\ddb_{\nu}\c\c\nabb^{p_{J-1}}\ddb_{\nu}\nabb^{p_J}\Psi)\big|_{p,S}\ .\nn
\eea
Recalling previous definitions we define
\[\sum_{s=0}^Jp_s=P\ \ \ ,\ \ \ \sum_{s=0}^{J-t}p_s=P_{(J-t)}, \ \ \ J+P=N\ ,\]
and we estimate  the second sum as
\bea
&&\ML\ML\ML\ML\ML\big|(II)\big|_{p,S}\leq\\
&&\ML\ML\ML\ML\leq\sum_{t=1}^J\sum_{h=0}^{P_{(J-t)}-1}\sum_{l=0}^{J-t}\cbin{P_{(J-t)}}{P_{(J-t)}-1-h}
\cbin{J-t}{l}\c\nn\\
&&\ML\ML\c\big||\la|^{{\underline\phi}(\Psi)}r^{1+J+P+\phi(\Psi)-\frac{2}{p}}(\ddb_{\nu}^{l}\nabb^{h}\chi)(\ddb_{\nu}^{J-t-l}\nabb^{(P_{(J-t+1)}-h)}\ddb_{\nu}\nabb^{p_{J-t+2}}\ddb_{\nu}\c\c\nabb^{p_{J-1}}\ddb_{\nu}\nabb^{p_J}\Psi)\big|_{p,S}\nn\\
&&\ML\ML\ML\ML\leq\sum_{t=1}^J\sum_{h=0}^{P_{(J-t)}-1}\sum_{q=0}^{J-t}\cbin{P_{(J-t)}}{h+1}
\cbin{J-t}{q}\chi\bigg((q+h)\leq\left[\frac{J+P}{2}\right]\bigg)\c\nn\\
&&\ML\ML\big||\la|^{{\underline\phi}(\Psi)}r^{1+J+P+\phi(\Psi)-\frac{2}{p}}(\ddb_{\nu}^{q}\nabb^{h}\chi)(\ddb_{\nu}^{J-t-q}\nabb^{(P_{(J-t+1)}-h)}\ddb_{\nu}\nabb^{p_{J-t+2}}\ddb_{\nu}\c\c\nabb^{p_{J-1}}\ddb_{\nu}\nabb^{p_J}\Psi)\big|_{p,S}\nn\\
&&\ML\ML\ML\ML+\sum_{t=1}^J\sum_{h=0}^{P_{(J-t)}-1}\sum_{q=0}^{J-t}\cbin{P_{(J-t)}}{h+1}
\cbin{J-t}{q}\chi\bigg((q+h)>\left[\frac{J+P}{2}\right]\bigg)\c\nn\\
&&\ML\ML\big||\la|^{{\underline\phi}(\Psi)}r^{1+J+P+\phi(\Psi)-\frac{2}{p}}(\ddb_{\nu}^{q}\nabb^{h}\chi)(\ddb_{\nu}^{J-t-q}\nabb^{(P_{(J-t+1)}-h)}\ddb_{\nu}\nabb^{p_{J-t+2}}\ddb_{\nu}\c\c\nabb^{p_{J-1}}\ddb_{\nu}\nabb^{p_J}\Psi)\big|_{p,S}\ .\nn
\eea
Let us consider the first sum 
\bea
&&\ML\ML\ML\ML\ML\big|(II)_1\big|_{p,S}\leq\sum_{t=1}^J\sum_{h=0}^{P_{(J-t)}-1}\sum_{q=0}^{J-t}\cbin{P_{(J-t)}}{h+1}
\cbin{J-t}{q}\chi\bigg((q+h)\leq\left[\frac{J+P}{2}\right]\bigg)\c\nn\\
&&\ML\ML\ML\ML\ML\big|r^{1+q+h}(\ddb_{\nu}^{q}\nabb^{h}\chi)|_{\infty,S}
\big||\la|^{{\underline\phi}(\Psi)}r^{1+J+P-(q+h)-t+\phi(\Psi)-\frac{2}{p}}(\ddb_{\nu}^{J-t-q}\nabb^{(P_{(J-t+1)}-h)}\ddb_{\nu}\nabb^{p_{J-t+2}}\ddb_{\nu}\c\c\nabb^{p_{J-1}}\ddb_{\nu}\nabb^{p_J}\Psi)\big|_{p,S}\nn\\
&&\ML\ML\ML\ML\ML\leq\sum_{t=1}^J\sum_{h=0}^{P_{(J-t)}-1}\sum_{q=0}^{J-t}\cbin{P_{(J-t)}}{h+1}
\cbin{J-t}{q}\chi\bigg((q+h)\leq\left[\frac{J+P}{2}\right]\bigg)\nn\\
&&\ML\ML\ML\left(\frac{(q+h+1)!}{(q+h+1)^{\a}}\frac{e^{(q+h-1)(\de+\underline{\Ga}(\la))}}{\ro_2^{q+h+1}}\right)
\c{F}\!\left(\frac{(J+P+1-(q+h)-t)!}{(J+P+1-(q+h)-t)^{\a}}\frac{e^{(J+P-(q+h)-t-1)(\de+\underline{\Ga}(\la))}}{\ro_2^{J+P-(q+h)-t+1}}\right)\nn\\
&&\ML\ML\ML\ML\ML\leq {f}\!\left(\frac{(J+P+1)!}{(J+P+1)^{\a}}\frac{e^{(J+P-1)(\de+\underline{\Ga}(\la))}}{\ro_2^{J+P+1}}\right)\sum_{t=1}^J\left(\frac{e^{-(t+1)(\de+\underline{\Ga}(\la))}}{\ro_2^{-t+1}}\right)\left\{\frac{(J+P+1)^{\a}}{(J+P+1)!}\right.\nn\\
&&\ML\ML\ML\ML\ML\left.\left[\sum_{h=0}^{P_{(J-t)}-1}\sum_{q=0}^{J-t}\cbin{P_{(J-t)}}{h+1}
\cbin{J-t}{q}\chi\bigg((q+h)\leq\left[\frac{J+P}{2}\right]\bigg)\left(\frac{(q+h+1)!}{(q+h+1)^{\a}}\right)
\c\!\left(\frac{(J+P+1-(q+h)-t)!}{(J+P+1-(q+h)-t)^{\a}}\right)\right]\right\}\ .\nn
\eea
Let us denote $(q+h)=k$ then, observing that, denoting ${\hat h}=h+1\ \ ,\ \ {\hat k}=k+1$
\bea
&&\ML\ML\ML\ML\sum_{h=0}^{P_{(J-t)}-1}\sum_{q=0}^{J-t}\cbin{P_{(J-t)}}{h+1}
=\sum_{{\hat h}=1}^{P_{(J-t)}}\sum_{q=0}^{J-t}\cbin{P_{(J-t)}}{{\hat h}}
\cbin{J-t}{q}``\leq"\sum_{{\hat h}=0}^{P_{(J-t)}}\sum_{q=0}^{J-t}\cbin{P_{(J-t)}}{{\hat h}}\cbin{J-t}{q}\nn\\
&&\ML\ML\ML\ML\leq \sum_{{\hat k}=0}^{P_{(J-t)}+J-t}\cbin{P_{(J-t)}+J-t}{\hat k}
\eea
\bea
&&\ML\ML\ML\ML\ML\bigg[\c\bigg]
\leq\left[\sum_{{\hat k}=0}^{\left[\frac{J+P}{2}\right]}\cbin{P_{(J-t)}+J-t}{\hat k}
\frac{{\hat k}!}{{\hat k}^{\a}}\frac{(J+P+2-{\hat k}-t)!}{(J+P+2-{\hat k}-t)^{\a}}\right]\nn
\eea
and
\bea
&&\ML\ML\ML\ML\ML\left\{\frac{(J+P+1)^{\a}}{(J+P+1)!}\c\right.\nn\\
&&\ML\ML\ML\ML\ML\left.\left[\sum_{h=0}^{P_{(J-t)}-1}\sum_{q=0}^{J-t}\cbin{P_{(J-t)}}{h+1}
\cbin{J-t}{q}\chi\bigg((q+h)\leq\left[\frac{J+P}{2}\right]\bigg)\left(\frac{(q+h+1)!}{(q+h+1)^{\a}}\right)
\c\!\left(\frac{(J+P+1-(q+h)-t)!}{(J+P+1-(q+h)-t)^{\a}}\right)\right]\right\}\nn\\
&&\ML\ML\ML\ML\ML\leq \frac{(J+P+1)^{\a}}{(J+P+1)!}\left[\sum_{{\hat k}=0}^{\left[\frac{J+P}{2}\right]\wedge({J+P_{(J-t)}})}\cbin{P_{(J-t)}+J-t}{\hat k}
\frac{{\hat k}!}{{\hat k}^{\a}}\frac{(J+P+2-{\hat k}-t)!}{(J+P+2-{\hat k}-t)^{\a}}\right]\nn\\
&&\ML\ML\ML\ML\ML\leq {(J+P+1)^{\a}}\left[\sum_{{\hat k}=0}^{\left[\frac{J+P}{2}\right]\wedge({J+P_{(J-t)}})}\frac{1}{{\hat k}^{\a}}\frac{(J+P_{(J-t)}-t)!(J+P+2-{\hat k}-t)!}{(J+P_{(J-t)}-{\hat k}-t)!(J+P+1)!}
\frac{1}{(J+P+2-{\hat k}-t)^{\a}}\right]\nn\\
&&\ML\ML\ML\ML\ML\leq C(\a)\frac{(J+P+1)^{\a}}{(\frac{J+P}{2}-t)^{\a}}\ ,
\eea
provided,  $\a>1$. Therefore
\bea
&&\ML\ML\ML\ML\sum_{t=1}^J\left(\frac{e^{-(t+1)(\de+\Ga)}}{\ro^{-t}}\right)\left\{\frac{(J+P+2)^{\a}}{(J+P+2)!}\right.\\
&&\ML\ML\ML\ML\ML\left.\left[\sum_{h=0}^{P_{(J-t)}-1}\sum_{q=0}^{J-t}\cbin{P_{(J-t)}}{h+1}
\cbin{J-t}{q}\chi\bigg((q+h)\leq\left[\frac{J+P}{2}\right]\bigg)\left(\frac{(q+h+2)!}{(q+h+2)^{\a}}\right)
\c\!\left(\frac{(J+P+2-(q+h)-t)!}{(J+P+2-(q+h)-t)^{\a}}\right)\right]\right\}\nn\\
&&\ML\ML\ML\ML\leq e^{-\de}C(\a)\sum_{t=1}^J\left(\ro^t{e^{-t(\de+\Ga)}}\right)\frac{(J+P+1)^{\a}}{(\frac{J+P}{2}-t)^{\a}}\leq e^{-\de}C_1(\a)\ ,\nn
\eea
completing the proof of the estimate.\footnote{There are, in fact, other sums to estimate, but it is easy to realize that all the estimates go in the same way.}
Therefore we have proved that the following estimate holds with $J+P\leq N-1$,
\bea
\ML\ML\ML||\la|^{{\underline\phi}(\Psi)}r^{1+J+P+\phi(\Psi)-\frac{2}{p}}\nabb^{p_0}\ddb_{\nu}\nabb^{p_1}\ddb_{\nu}\nabb^{p_2}\c\c\c\ddb_{\nu}\nabb^{p_{J-1}}\ddb_{\nu}\nabb^{p_{J}}\Psi|_{p,S}\leq {F_1}\!\left(\frac{(J+P+1)!}{(J+P+1)^{\a}}\frac{e^{(J+P-1)(\de+\underline{\Ga}(\la))}}{\ro_2^{J+P+1}}\right) \ ,\ \ \ \ \ \eql{finesta1}
\eea
provided that the following estimate holds with $F_2<F_1$,
\bea
\ML||\la|^{{\underline\phi}(\Psi)}r^{1+J+P+\phi(\Psi)-\frac{2}{p}}\ddb_{\nu}^J\nabb^P\Psi|_{p,S}\leq {F_2}\!\left(\frac{(J+P+1)!}{(J+P+1)^{\a}}\frac{e^{(J+P-1)(\de+\underline{\Ga}(\la))}}{\ro_2^{J+P+1}}\right)\ .\ \ \ \ \ \ \eql{finesta2}
\eea

\section{Appendix to Section \ref{S.12}}\label{S.17}
\subsection{Proof of the estimate \ref{oomest1}} 
We have to control the right hand side of the following expression,
\bea
&&\ML\ML|\nabb^N\oom|_{p,S}\leq 
N!\sum_{k=1}^{\infty}\frac{1}{k!}\left[\sum_{\ga_1=0}^{\frac{N}{2}}\sum_{\ga_2,\c,\c,\c,\ga_k}^{\sum_{s=2}^k\ga_s=N-\ga_1; \ga_s\in[0,N]}\frac{1}{\ga_1!\ga_2!\c\c\ga_k!}\big|(\nabb^{\ga_1}\log\oom)(\nabb^{\ga_2}\log\oom)\c\c\c(\nabb^{\ga_k}\log\oom)\big|_{p,S}\right.\nn\\
&&\ML\ML\ \ \ \ \ \ \ \ \ \ \ \ \ \ \ \ \ \ \ \ \ \ \ \ \left.\sum_{\ga_1=\frac{N}{2}+1}^{N}\sum_{\ga_2,\c,\c,\c,\ga_k}^{\sum_{s=2}^k\ga_s=N-\ga_1; \ga_s\in[0,N]}\frac{1}{\ga_1!\ga_2!\c\c\ga_k!}\big|(\nabb^{\ga_1}\log\oom)(\nabb^{\ga_2}\log\oom)\c\c\c(\nabb^{\ga_k}\log\oom)\big|_{p,S}\right]\ .\nn
\eea
Let us consider the second sum
\bea
&&\ML\ML\ML\ML\ML \ \ N!\sum_{k=1}^{\infty}\frac{1}{k!}\sum_{\ga_1=\frac{N}{2}+1}^{N}\sum_{\ga_2,\c,\c,\c,\ga_k}^{\sum_{s=2}^k\ga_s=N-\ga_1; \ga_s\in[0,N]}\frac{1}{\ga_1!\ga_2!\c\c\ga_k!}\big|(\nabb^{\ga_1}\log\oom)(\nabb^{\ga_2}\log\oom)\c\c\c(\nabb^{\ga_k}\log\oom)\big|_{p,S}\nn\\
&&\ML\ML\ML\ML\ML\ML\ML\leq N!\sum_{\ga_1=\frac{N}{2}+1}^{N}\frac{\big|(\nabb^{\ga_1}\log\oom)\big|_{p,S}}{\ga_1!}\sum_{k=1}^{\infty}\frac{1}{k!}\sum_{\ga_2,\c,\c,\c,\ga_k}^{\sum_{s=2}^k\ga_s=N-\ga_1; \ga_s\in[0,N]}\frac{1}{\ga_2!\c\c\ga_k!}|(\nabb^{\ga_2}\log\oom)|_{\infty,S}\c\c\c|(\nabb^{\ga_k}\log\oom)|_{\infty,S}\nn\\
&&\ML\ML\ML\ML\ML\ML\ML\leq N!\sum_{\ga_1=\frac{N}{2}+1}^{N}{\tilde C}_3\frac{(\ga_1-1)!}{(\ga_1-1)^{\a}\ga_1!}\frac{e^{(\ga_1-3)(\de+\underline{\Ga}(\la))}}{\ro^{\ga_1-1}}\sum_{k=1}^{\infty}\frac{1}{k!}\sum_{\ga_2,\c,\c,\c,\ga_k}^{\sum_{s=2}^k\ga_s=N-\ga_1; \ga_s\in[0,N]}\frac{{\tilde C}_3^{\si(\ga_2)+\c\c+\si(\ga_k)}}{\ga_2^{\a}\c\c\ga_k^{\a}}\frac{e^{\si(\ga_2-2)(\ga_2-2)(\de+\underline{\Ga}(\la))}}{\ro^{\ga_2}}\c\c\c\frac{e^{\si(\ga_k-2)(\ga_k-2)(\de+\underline{\Ga}(\la))}}{\ro^{\ga_k}}\nn\\
&&\ML\ML\ML\ML\ML\ML\leq\left({\tilde C}_3\frac{(N-1)!}{(N-1)^{\a}}\frac{e^{(N-3)(\de+\underline{\Ga}(\la))}}{\ro^{N-1}}\right)
N\sum_{\ga_1=\frac{N}{2}+1}^{N}\frac{(N-1)^{\a}}{(\ga_1-1)^{\a}\ga_1}\frac{e^{-(N-\ga_1)(\de+\underline{\Ga}(\la))}}{\ro^{-(N-\ga_1)}}\sum_{k=1}^{\infty}\frac{1}{k!}\sum_{\ga_2,\c,\c,\c,\ga_k}^{\sum_{s=2}^k\ga_s=N-\ga_1; \ga_s\in[0,N]}
\frac{1}{\ga_2^{\a}\c\c\ga_k^{\a}}\frac{1}{\ro^{(N-\ga_1)}}\nn\\
&&\ML\ML\ML\ML\ML\ML\leq\left({\tilde C}_3\frac{(N-1)!}{(N-1)^{\a}}\frac{e^{(N-3)(\de+\underline{\Ga}(\la))}}{\ro^{N-1}}\right)
\sum_{\ga_1=\frac{N}{2}+1}^{N}\frac{N^{\a+1}}{\ga_1^{\a+1}}e^{-(N-\ga_1)(\de+\underline{\Ga}(\la))}\sum_{k=1}^{\infty}\frac{1}{k!}\sum_{\ga_2,\c,\c,\c,\ga_k}^{\sum_{s=2}^k\ga_s=N-\ga_1; \ga_s\in[0,N]}
\frac{1}{\ga_2^{\a}\c\c\ga_k^{\a}}\nn\\
&&\ML\ML\ML\ML\leq\left({\tilde C}_3\frac{(N-1)!}{(N-1)^{\a}}\frac{e^{(N-3)(\de+\underline{\Ga}(\la))}}{\ro^{N-1}}\right)
\left[\sum_{\ga_1=\frac{N}{2}+1}^{N}\frac{N^{\a+1}}{\ga_1^{\a+1}}e^{-(N-\ga_1)(\de+\underline{\Ga}(\la))}\sum_{k=1}^{\infty}\frac{1}{k!}\sum_{\ga_2,\c,\c,\c,\ga_k}^{\sum_{s=2}^k\ga_s=N-\ga_1; \ga_s\in[0,N]}
\frac{1}{\ga_2^{\a}\c\c\ga_k^{\a}}\right]\ .
\eea
Let us consider the first sum which we rewrite as
\bea
&&\ML\ML N!\sum_{k=1}^{\infty}\frac{1}{k!}\left[\sum_{\ga_2=0}^{\frac{N}{2}}\sum_{\ga_1=0}^{\frac{N}{2}}\sum_{\ga_3,\c,\c,\c,\ga_k}^{\sum_{s=2}^k\ga_s=N-\ga_1; \ga_s\in[0,N]}\frac{1}{\ga_1!\ga_2!\c\c\ga_k!}\big|(\nabb^{\ga_1}\log\oom)(\nabb^{\ga_2}\log\oom)\c\c\c(\nabb^{\ga_k}\log\oom)\big|_{p,S}\right.\nn\\
&&\ML\ML\left.\sum_{\ga_2=\frac{N}{2}+1}^{N}\sum_{\ga_1=0}^{\frac{N}{2}}\sum_{\ga_3,\c,\c,\c,\ga_k}^{\sum_{s=2}^k\ga_s=N-\ga_1; \ga_s\in[0,N]}\frac{1}{\ga_1!\ga_2!\c\c\ga_k!}\big|(\nabb^{\ga_1}\log\oom)(\nabb^{\ga_2}\log\oom)\c\c\c(\nabb^{\ga_k}\log\oom)\big|_{p,S}\right]
\eea
Let us consider the second sum of the first sum which we estimate as
\bea
&&\ML\ML\ML\ML\ML\ML\ML\ML\leq N!\sum_{k=1}^{\infty}\frac{1}{k!}\sum_{\ga_2=\frac{N}{2}+1}^{N}\frac{|(\nabb^{\ga_2}\log\oom)|_{p,S}}{\ga_2!}\sum_{\ga_1=0}^{\frac{N}{2}}\sum_{\ga_3,\c,\c,\c,\ga_k}^{\sum_{s=2}^k\ga_s=N-\ga_1-\ga_2; \ga_s\in[0,N]}\frac{1}{\ga_1!\ga_3!\c\c\ga_k!}\big|(\nabb^{\ga_1}\log\oom)|_{\infty,S}|(\nabb^{\ga_3}\log\oom)|_{\infty,S}\c\c\c|(\nabb^{\ga_k}\log\oom)|_{\infty,S}\nn\\
&&\ML\ML\ML\ML\ML\ML\ML\ML\leq N!\sum_{k=1}^{\infty}\frac{1}{k!}\sum_{\ga_2=\frac{N}{2}+1}^{N}\frac{|(\nabb^{\ga_2}\log\oom)|_{p,S}}{\ga_2!}\sum_{\ga_1\ga_3,\c,\c,\c,\ga_k}^{\sum_{s=1}^{k;s\neq 2}\ga_s=N-\ga_2; \ga_s\in[0,N]}\frac{1}{\ga_1!\ga_3!\c\c\ga_k!}\big|(\nabb^{\ga_1}\log\oom)|_{\infty,S}|(\nabb^{\ga_3}\log\oom)|_{\infty,S}\c\c\c|(\nabb^{\ga_k}\log\oom)|_{\infty,S}\ .\ \ \ \ \ \ 
\eea
The last sum is identically to the previous ``second sum" interchanging $\ga_1$ and $\ga_2$ and therefore can be bounded by
\bea
\leq\left({\tilde C}_3\frac{(N-1)!}{(N-1)^{\a}}\frac{e^{(N-3)(\de+\underline{\Ga}(\la))}}{\ro^{N-1}}\right)
\left[\sum_{\ga_1=\frac{N}{2}+1}^{N}\frac{N^{\a+1}}{\ga_1^{\a+1}}e^{-(N-\ga_1)(\de+\underline{\Ga}(\la))}\sum_{k=1}^{\infty}\frac{1}{k!}\sum_{\ga_2,\c,\c,\c,\ga_k}^{\sum_{s=2}^k\ga_s=N-\ga_1; \ga_s\in[0,N]}
\frac{1}{\ga_2^{\a}\c\c\ga_k^{\a}}\right]
\eea
as this has to be repeated $k$ times obtaining
\bea
&&\ML\ML\ML\leq\left({\tilde C}_3\frac{(N-1)!}{(N-1)^{\a}}\frac{e^{(N-3)(\de+\underline{\Ga}(\la))}}{\ro^{N-1}}\right)
\left[\sum_{\ga_1=\frac{N}{2}+1}^{N}\frac{N^{\a+1}}{\ga_1^{\a+1}}e^{-(N-\ga_1)(\de+\underline{\Ga}(\la))}\sum_{k=1}^{\infty}\frac{k}{k!}\sum_{\ga_2,\c,\c,\c,\ga_k}^{\sum_{s=2}^k\ga_s=N-\ga_1; \ga_s\in[0,N]}
\frac{1}{\ga_2^{\a}\c\c\ga_k^{\a}}\right]\nn\\
&&\ML\ML\ML\leq\left({\tilde C}_3\frac{(N-1)!}{(N-1)^{\a}}\frac{e^{(N-3)(\de+\underline{\Ga}(\la))}}{\ro^{N-1}}\right)
\left[\sum_{\ga_1=\frac{N}{2}+1}^{N}\frac{N^{\a+1}}{\ga_1^{\a+1}}e^{-(N-\ga_1)(\de+\underline{\Ga}(\la))}\sum_{k=1}^{\infty}\frac{1}{(k-1)!}\sum_{\ga_2,\c,\c,\c,\ga_k}^{\sum_{s=2}^k\ga_s=N-\ga_1; \ga_s\in[0,N]}
\frac{1}{\ga_2^{\a}\c\c\ga_k^{\a}}\right]\nn\\
&&\ML\ML\ML\leq\left({\tilde C}_3\frac{(N-1)!}{(N-1)^{\a}}\frac{e^{(N-3)(\de+\underline{\Ga}(\la))}}{\ro^{N-1}}\right)
\left(1+ce^{-\de}\right)\ .
\eea
We are left with a term where all the $\ga_i$ are $\leq \left[\frac{N}{2}\right]$ which is easier to estimate 
so that  the final result is
\bea
|\nabb^N\oom|_{p,S}\leq \left({\tilde C}_3\frac{(N-1)!}{(N-1)^{\a}}\frac{e^{(N-3)(\de+\underline{\Ga}(\la))}}{\ro^{N-1}}\right)
\left(1+ce^{-\de}\right)\ .
\eea
As $\pr_{\nu}\oom=\oom\ddb_4\oom=-2\oom\om$, to control  $\pr^N_{\nu}\oom$ we write
\bea
&&\ML\ddb^N_{\nu}\oom=\ddb^N_{\nu}e^{\log\oom}=\sum_{k=0}^{\infty}\frac{1}{k!}\ddb^N_{\nu}(\log\oom)^k\nn\\
&&\ML\ML=N!\sum_{k=1}^{\infty}\frac{1}{k!}\sum_{\ga_1,\c,\c,\c,\ga_k}^{\sum_{s=1}^k\ga_s=N; \ga_s\in[0,N]}\frac{1}{\ga_1!\ga_2!\c\c\ga_k!}(\ddb_{\nu}^{\ga_1}\log\oom)(\ddb_{\nu}^{\ga_2}\log\oom)\c\c\c(\ddb_{\nu}^{\ga_k}\log\oom)\nn\\
&&=\nn\\
&&\ML\ML=N!\sum_{k=1}^{\infty}\frac{(-2)^k}{k!}\sum_{\ga_1,\c,\c,\c,\ga_k}^{\sum_{s=1}^k\ga_s=N; \ga_s\in[0,N]}\frac{1}{\ga_1!\ga_2!\c\c\ga_k!}(\ddb_{\nu}^{\ga_1-1}\om)(\ddb_{\nu}^{\ga_2-1}\om)\c\c\c(\ddb_{\nu}^{\ga_k-1}\om)\ .\nn
\eea
As the following estimates hold, see Theorem \ref{T11.1},
\bea
\big|r^{2+J-\frac{2}{p}}\ddb_4^J\om\big|_{p,S}(\la,\nu)\leq {\underline F}_5\!\left(\frac{(J+1)!}{(J+1)^{\a}}\frac{e^{((J+1)-2)(\de+\underline{\Ga}(\la))}}{\ro^{J+1}}\right)\ ,
\eea
proceeding as before it is easy to prove the following estimate
\bea
|\pr^N_{\nu}\oom|_{p,S}\leq c\!\left(\frac{N!}{N^{\a}}\frac{e^{(N-2)(\de+\underline{\Ga}(\la))}}{\ro^{N}}\right)\ .
\eea

\subsection{Proof of the estimate \ref{12.14az}} From the transport equation
\bea
\ML\ML\ML\frac{\Dbb}{\pr\nu}(\nabb^J|X|^2)+(J+2)\ \!\frac{\oom\tr\chi}{2}\c\nabb^J|X|^2+J(\oom\chih)\c\nabb^J|X|^2
-2(\nabb^JX)\c\oom\chih\c X+2(\nabb^JX)\c\ze=\bigg\{good\bigg\}_{\!X}\ \ 
\eea
where
\bea
&&\ML\ML\bigg\{good\bigg\}_{\!X}
=-\left\{\left[\sum_{k=0}^{J-2}\cbin{J}{k}(\nabb^{J-1-k}\oom\chi)\c\nabb^{k+1}|X|^2-\sum_{k=0}^{J-1}\cbin{J+1}{k}(\nabb^{J-1-k}C)\nabb^k|X|^2\right]\right.\nn\\
&&\ML\ML\left.-\left[\sum_{k=1}^J\cbin{J}{k}(\nabb^k\oom\tr\chi)\nabb^{J-k}|X|^2+\sum_{\ga_1\ga_2\ga_3}^{\sum_{s=1}^4\!\ga_s=J; \ga_1,\ga_3\neq J}\frac{J!}{\ga_1!\ga_2!\ga_3!}(\nabb^{\ga_1}X)\c(\nabb^{\ga_2}\oom\chih)\c(\nabb^{\ga_3}X)\right.\right.\nn\\
&&\ML\ML\ \ \left.\left.+2\sum_{k=0}^{J-1}\sum_{l=0}^{J-k}\frac{J!}{k!(J-k)!}(\nabb^kX)\nabb^{J-k}\ze\right]\right\}.\nn
\eea
 it follows, as done many times before,
\bea
&&\ML\ML\frac{\pr|\nabb^J|X|^2|^p}{\pr\nu}+p(J+2)\frac{\oom\tr\chi}{2}|\nabb^J|X|^2|^p\nn\\
&&\ML\ML=-p|\nabb^J|X|^2|^{p-2}\left[J(\nabb^J|X|^2)\c\oom\chih\c(\nabb^J|X|^2)
+(\nabb^J|X|^2)\big(2(\nabb^JX)\c\oom\chih\c X-2(\nabb^JX)\c\ze\big)\right.\nn\\
&&\ML\ML\left.+(\nabb^J|X|^2)\c\bigg\{good\bigg\}_{\!X}\right]\nn\\
&&\ML\ML\leq p|\nabb^J|X|^2|^{p-1}J|\oom\chih\c(\nabb^J|X|^2)|
+2p|\nabb^J|X|^2|^{p-1}|\nabb^JX\c\oom\chih\c X|+2p|\nabb^J|X|^2|^{p-1}|\nabb^JX\c\ze|\nn\\
&&\ML\ML+p|\nabb^J|X|^2|^{p-1}\big|\big\{good\big\}_{\!X}\big|
\eea
which we rewrite as
\bea
&&\ML\ML\frac{\pr|\nabb^J|X|^2|^p}{\pr\nu}+p(J+2)\frac{{\oom\tr\chi}}{2}|\nabb^J|X|^2|^p\\
&&\ML\ML\leq p|\nabb^J|X|^2|^{p-1}\left\{\left[J|\oom\chih\c(\nabb^J|X|^2)\right]
+\left[2\!\left(|\nabb^JX\c\oom\chih\c X|+|\nabb^JX\c\ze|\right)
+\big|\big\{good\big\}_{\!X}\big|\right]\right\}\ .\nn
\eea
and in a more compact way as
\bea
&&\ML\ML\ML\ML\frac{\pr|\nabb^J|X|^2|^p}{\pr\nu}+p(J+2)\frac{{\oom\tr\chi}}{2}|\nabb^J|X|^2|^p\leq p|\nabb^J|X|^2|^{p-1}
|L_X|\ .\ \ \ \ \ \ \ \ \ \ 
\eea
where
\bea
\ML\ML|L_X|=J|\oom\chih\c(\nabb^J|X|^2)|+\left[2\!\left(|\nabb^JX\c\oom\chih\c X|+|\nabb^JX\c\ze|\right)
+p|\nabb^J|X|^2|^{p-1}\big|\big\{good\big\}_{\!X}\big|\right]\ \ \ 
\eea
From it,
\bea
&&\ML\ML\ML\frac{\partial|r^{(J+2)-\frac{2}{p})}\nabb^{J}|X|^2|^p_{p,S}}{\partial\nu}=
\int_S\left(\frac{\pr|r^{(J+2-\frac{2}{p})}\nabb^{J}{|X|^2}|^p}{\pr\nu}+\oom\tr\chi|r^{(J+2-\frac{2}{p})}\nabb^{J}|X|^2|^p\right)d\mu_{\ga}\nn\\
&&\ML\ML\ML\ML=\int_S\left(r^{(J+2-\frac{2}{p})p}\!\!\left[\frac{\pr|\nabb^{J}|X|^2|^p}{\pr\nu}
+p(J+2)\frac{\overline{\oom\tr\chi}}{2}|\nabb^{J}{|X|^2}|^p\right]+(\oom\tr\chi-\overline{\oom\tr\chi})|r^{(J+2-\frac{2}{p})}\nabb^{J}{|X|^2}|^p\right)d\mu_{\ga}\nn\\
&&\ML\ML\ML\leq p\int_S r^{(J+2-\frac{2}{p})p}|\nabb^J|X|^2|^{p-1}\!\left[|L_X|+\bigg(\frac{J+2-\frac{2}{p}}{2}|\oom\tr\chi-{\overline{{\oom}\tr\chi}}|\bigg)|\nabb^{J}{|X|^2}|\right]\ ,\nn
\eea
which we rewrite,
\bea
\frac{\partial|r^{(J+2-\frac{2}{p})}\nabb^{J}|X|^2|^p_{p,S}}{\partial\nu}\leq p\!\int_S r^{(J+2-\frac{2}{p})p} |\nabb^{J}|X|^2|^{p-1}|\Ls_X|d\mu_{\ga}\ ,
\eea
where
\bea
|\Ls_X|=\left[J\big(|\oom\tr\chi-{\overline{{\oom}\tr\chi}}|+|\oom\chih|\big)|\nabb^{J}{|X|^2}|+|\tilde{L}_X|\right]\ 
\eea
and
\bea
\tilde{L}_X=\left[2\!\left(|\nabb^JX\c\oom\chih\c X|+|\nabb^JX\c\ze|\right)+\big|\big\{good\big\}_{\!X}\big|\right]\ .
\eea
Proceeding as before we have, denoting ${\cal H}=|\oom\chih|+|\overline{{\oom}\tr\chi}|+|\oom\chih|$,
\bea
&&\ML\ML\frac{\partial|r^{(J+2-\frac{2}{p})}\nabb^{J}|X|^2|_{p,S}}{\partial\nu}\leq |r^{(J+2-\frac{2}{p})}\nabb^{J}\Ls_X|_{p,S}\leq J|{\cal H}|_{\infty,S}|r^{(J+2-\frac{2}{p})}\nabb^{J}|X|^2|_{p,S}+|r^{(J+2-\frac{2}{p})}{\tilde L}_X|_{p,S}\nn
\eea
and integrating
\bea
\ML\ML\ML|r^{(J+2-\frac{2}{p})}\nabb^{J}|X|^2|_{p,S}(\la,\nu)\leq \int_{\nu_0}^{\nu}d\nu' e^{J\left(\int_{\nu_0}^{\nu}{\cal H}-\int_{\nu_0}^{\nu'}{\cal H}\right)}|r^{(J+2-\frac{2}{p})}{\tilde L}_X|_{p,S}
\leq \int_{\nu_0}^{\nu}d\nu' e^{J{\tilde C}_0\frac{\nu-\nu'}{\nu\nu'}}|r^{(J+2-\frac{2}{p})}{\tilde L}_X|_{p,S}\ .\ \ \ \ \ 
\eea
Observing that 
\bea
\ML\nabb^J|X|^2=\sum_{k=0}^J\cbin{J}{k}\nabb^k|X|\nabb^{J-k}|X|=2(\nabb^J|X|)|X|+\sum_{k=1}^{J-1}\cbin{J}{k}\nabb^k|X|\nabb^{J-k}|X|\ \ ,  \nn
\eea
we write
\bea
&&\ML\ML|r^{(J+2-\frac{2}{p})}\left(2(\nabb^J|X|)|X|+\sum_{k=1}^{J-1}\cbin{J}{k}\nabb^k|X|\nabb^{J-k}|X|\right)|_{p,S}(\la,\nu)\nn\\
&&\ML\ML \leq \int_{\nu_0}^{\nu}d\nu' e^{J\left(\int_{\nu_0}^{\nu}{\cal H}-\int_{\nu_0}^{\nu'}{\cal H}\right)}|r^{(J+2-\frac{2}{p})}{\tilde L}_X|_{p,S}
\leq \int_{\nu_0}^{\nu}d\nu' e^{J{\tilde C}_0\frac{\nu-\nu'}{\nu\nu'}}|r^{(J+2-\frac{2}{p})}{\tilde L}_X|_{p,S}\ \ \ \ \ \ \ \ \ \ \ \  
\eea
and
\bea
\ML\ML\ML\ML|r^{(J+2-\frac{2}{p})}\nabb^J|X|)|X||_{p,S}(\nu,\la)\leq \frac{1}{2}\sum_{k=1}^{J-1}\cbin{J}{k}\big|r^{(J+2-\frac{2}{p})}\nabb^k|X|\nabb^{J-k}|X|\big|_{p,S}(\nu,\la)
+\frac{1}{2}\int_{\nu_0}^{\nu}d\nu' e^{J{\tilde C}_0\frac{\nu-\nu'}{\nu\nu'}}|r^{(J+2-\frac{2}{p})}{\tilde L}_X|_{p,S}(\nu',\la)\ .\nn
\eea
We consider the  $\sup$  of this inequality with respect to 
\[\frac{r|X|}{\log r}\in [0, M\equiv sup_{\nu}|\frac{r}{\log r}X|_{\infty,S}]\] 
considering $|X|$ as an independent variable and observing that 
$M=O(\varepsilon)$,
 obtaining,
\bea
&&\ML\ML\sup|{r^{(J+1-\frac{2}{p})}}{\log r}\nabb^J|X|\!\left(\frac{r|X|}{\log r}\right)|_{p,S}(\nu,\la)\\
&&\ML\ML\leq \frac{1}{2}\sum_{k=1}^{J-1}\cbin{J}{k}\big|{r^{(J+2-\frac{2}{p})}}\nabb^k|X|\nabb^{J-k}|X|\big|_{p,S}(\nu,\la)
+\frac{1}{2}\sup\int_{\nu_0}^{\nu}d\nu' e^{J{\tilde C}_0\frac{\nu-\nu'}{\nu\nu'}}|{r^{(J+2-\frac{2}{p})}}{\tilde L}_X|_{p,S}(\nu',\la)\ .\nn
\eea
obtaining
\bea
&&\ML\ML\ML{\log r}|{r^{(J+1-\frac{2}{p})}}\nabb^J|X||_{p,S}\leq M^{-1}\frac{1}{2}\sum_{k=1}^{J-1}\cbin{J}{k}\big|{r^{(J+2-\frac{2}{p})}}\nabb^k|X|\nabb^{J-k}|X|\big|_{p,S}\nn\\
&&\ML\ML\ML+M^{-1}\frac{1}{2}\int_{\nu_0}^{\nu}d\nu' e^{J{\tilde C}_0\frac{\nu-\nu'}{\nu\nu'}}
\left[M(\log r)|{r^{(J+1-\frac{2}{p})}}\nabb^JX\c\oom\chih|_{p,S}+|{r^{(J+2-\frac{2}{p})}}\nabb^JX\c\ze|_{p,S}+\sup_{|X|}\big|{r^{(J+2-\frac{2}{p})}}\big\{good\big\}_{\!X}\big|_{p,S}\right]\ .\nn
\eea
We make the iterative assumption if ,
 \bea
|\frac{r^{(J+1-\frac{2}{p})}}{\log r}\nabb^{J}|X||_{p,S}(\la,\nu)\leq c\!\left(\frac{J!}{J^{\a}}\frac{e^{(J-2)(\Ga_1+\de)}}{\ro^J}\right)\ , \eql{Itass}
\eea
with $c$ of order $O(\epsilon)$ if $J\leq 7$ and of order $O(1)$ otherwise, where 
\bea
\Ga_1=\Ga_1(\nu)=C_1\frac{\nu-\nu_0}{\nu\nu_0}\ .
\eea
To show the consistency of this assumption we proceed in the following way: first we consider the non integrated term
\bea
&&\ML\ML \frac{1}{2M}(\log r)^2\sum_{k=1}^{J-1}\cbin{J}{k}\big|\frac{r^{(J+2-\frac{2}{p})}}{(\log r)^2}\nabb^k|X|\nabb^{J-k}|X|\big|_{p,S}\nn\\
&&\ML\ML\leq \frac{1}{2M}(\log r)^2\sum_{k=1}^{\left[\frac{J-1}{2}\right]}\frac{J!}{(J-k)!k!}\big|\frac{r^{k+1}}{\log r}\nabb^k|X||_{\infty,S}|\frac{r^{J-k+1-\frac{2}{p}}}{\log r}\nabb^{J-k}|X|\big|_{p,S}\nn\\
&&\ML\ML+\frac{1}{2M}(\log r)^2\sum_{k=\left[\frac{J-1}{2}\right]+1}^{J-1}\frac{J!}{(J-k)!k!}\big|\frac{r^{k+1-\frac{2}{p}}}{\log r}\nabb^k|X||_{p,S}|\frac{r^{J-k+1}}{\log r}\nabb^{J-k}|X|\big|_{\infty,S}\nn\\
&&\ML\ML\leq \frac{1}{M}(\log r)^2\sum_{k=1}^{\left[\frac{J-1}{2}\right]}\frac{J!}{(J-k)!k!}\big|\frac{r^{k+1}}{\log r}\nabb^k|X||_{\infty,S}|\frac{r^{J-k+1-\frac{2}{p}}}{\log r}\nabb^{J-k}|X|\big|_{p,S}\nn\\
&&\ML\ML\leq  M^{-1}c(\log r)^2\sum_{k=1}^{\left[\frac{J-1}{2}\right]}\frac{J!}{(J-k)!k!}\frac{(k+1)!}{(k+1)^{\a}}\frac{e^{(k-1)(\Ga_1+\de)}}{\ro^{k+1}}\frac{(J-k)!}{(J-k)^{\a}}\frac{e^{(J-k-2)(\Ga_1+\de)}}{\ro^{J-k}}\nn\\
&&\ML\ML\leq c(\log r)^2\frac{J!}{J^{\a}}\frac{e^{(J-2)(\Ga_1+\de_1)}}{\ro^{J}}\left(M^{-1}\frac{e^{-(\Ga_1+\de)}}{\ro}\right)\sum_{k=1}^{\left[\frac{J-1}{2}\right]}\frac{1!}{(k+1)^{\a-1}}\frac{J^{\a}}{(J-k)^{\a}}\nn\\
&&\ML\ML\leq (\log r)^2\frac{1}{100}\left(\frac{J!}{J^{\a}}\frac{e^{(J-2)(\Ga_1+\de)}}{\ro^{J}}\right),
\eea
choosing $\de$ sufficiently large and $\a\!>\!2$. Let us control the delicate integral parts,
\bea
&&\ML\ML \int_{\nu_0}^{\nu}d\nu' e^{J{\tilde C}_0\frac{\nu-\nu'}{\nu\nu'}}|(\log r){r^{(J+1-\frac{2}{p})}}\nabb^JX\c\oom\chih|_{p,S}
\leq \int_{\nu_0}^{\nu}d\nu' e^{J{\tilde C}_0\frac{\nu-\nu'}{\nu\nu'}}(\log r)^2|\frac{r^{(J+1-\frac{2}{p})}}{\log r}\nabb^JX|_{p,S}|\oom\chih|_{\infty,S}\nn\\
&&\ML\ML\leq c\varepsilon_0\int_{\nu_0}^{\nu}\frac{d\nu'(\log\nu')^2}{\nu'^2}e^{J{\tilde C}_0\frac{\nu-\nu'}{\nu\nu'}}|\frac{r^{(J+1-\frac{2}{p})}}{\log r}\nabb^JX|_{p,S}\\
&&\ML\ML\leq
c\varepsilon_0\left(\frac{J!}{J^{\a}}\frac{e^{(J-2)(\Ga_1(\la,\nu)+\de)}}{\ro^J}\right)\int_{\nu_0}^{\nu}\frac{d\nu'(\log\nu')^2}{\nu'^2}e^{\left[J{\tilde C}_0\frac{\nu-\nu'}{\nu\nu'}-(J-2)(\Ga_1(\nu)-\Ga_1(\nu'))\right]}\nn\\
&&\ML\ML\leq c\varepsilon_0\left(\frac{J!}{J^{\a}}\frac{e^{(J-2)(\Ga_1+\de_1)}}{\ro^J}\right)\ ,\nn
\eea
provided $C_1>{\tilde C}_0$. The second integral term is more delicate 
\bea
&&\ML\ML\int_{\nu_0}^{\nu}d\nu' e^{J{\tilde C}_0\frac{\nu-\nu'}{\nu\nu'}}|{r^{(J+2-\frac{2}{p})}}\nabb^JX\c\ze|_{p,S}
\leq\int_{\nu_0}^{\nu}d\nu' e^{J{\tilde C}_0\frac{\nu-\nu'}{\nu\nu'}}(\log r)\big|\frac{r^{(J+1-\frac{2}{p})}}{\log r}\nabb^JX\big|_{p,s}\big|r\ze\big|_{\infty,S}\nn\\
&&\ML\ML\leq c\varepsilon_0\left(\frac{J!}{J^{\a}}\frac{e^{(J-2)(\Ga_1(\la,\nu)+\de)}}{\ro^J}\right)(\log r)(\la,\nu)\!\int_{\nu_0}^{\nu}\frac{d\nu'}{\nu'}e^{\left[J{\tilde C}_0\frac{\nu-\nu'}{\nu\nu'}-(J-2)(\Ga_1(\nu)-\Ga_1(\nu'))\right]}\nn\\
&&\ML\ML\leq c(\log r)^2\varepsilon_0\left(\frac{J!}{J^{\a}}\frac{e^{(J-2)(\Ga_1(\nu)+\de_1)}}{\ro^J}\right)\ .
\eea
Therefore
\bea
\ML M^{-1}\frac{1}{2}\int_{\nu_0}^{\nu}d\nu' e^{J{\tilde C}_0\frac{\nu-\nu'}{\nu\nu'}}
|\frac{r^{(J+2-\frac{2}{p})}}{\log r}\nabb^JX\c\ze|_{p,S}\leq \frac{c}{2}(\log r)^2\left(\frac{J!}{J^{\a}}\frac{e^{(J-2)(\Ga_1(\nu)+\de_1)}}{\ro^J}\right).\ \ \ 
\eea
The correction terms can be treated as before, the only request they imply is that $\de_1\geq \de$. They contribute as 
\[\frac{1}{100}c(\log r)^2\left(\frac{J!}{J^{\a}}\frac{e^{(J-2)(\Ga_1(\la,\nu)+\de_1)}}{\ro^J}\right)\ .\]
Collecting all the estimates together we conclude that
\bea
&&\ML\ML{\log r}|{r^{(J+1-\frac{2}{p})}}\nabb^J|X||_{p,S}\leq (\log r)^2\frac{1}{100}\left(\frac{J!}{J^{\a}}\frac{e^{(J-2)(\Ga_1+\de)}}{\ro^{J}}\right)+\\
&&\ML\ML\ML+\frac{c}{2}(\log r)^2\left(\frac{J!}{J^{\a}}\frac{e^{(J-2)(\Ga_1(\nu)+\de_1)}}{\ro^J}\right)
+\frac{1}{100}c(\log r)^2\left(\frac{J!}{J^{\a}}\frac{e^{(J-2)(\Ga_1(\nu)+\de_1)}}{\ro^J}\right)\nn\\
&&\ML\ML\ML \leq c(\log r)^2\left(\frac{J!}{J^{\a}}\frac{e^{(J-2)(\Ga_1(\nu)+\de_1)}}{\ro^J}\right)
\eea
which implies the inequality \ref{12.14az} for all $J$,\footnote{With some extra work it is possible to prove that
$$|\frac{r^{J-\frac{2}{p}}}{\log r}\pr^JX_a|_{p,S}\leq c\varepsilon_0\left(\frac{J!}{J^{\a}}\frac{e^{(J-2)(\Ga_1+\de)}}{\ro^J}\right)\ .$$} 
\beaa
|\frac{r^{(J+1-\frac{2}{p})}}{\log r}\nabb^{J}|X||_{p,S}(\la,\nu)\leq c\left(\frac{J!}{J^{\a}}\frac{e^{(J-2)(\Ga_1+\de)}}{\ro^J}\right)\ . 
\eeaa
\section{Appendix to Section \ref{S.13}}\label{S.18}


\NI\subsection{ Proof of Lemma \ref{Lemma14.1}}
 To prove this lemma we derive first the corresponding equations for the norms $\big|r^{2-\frac{2}{p}}\ze\big|_{p,S}$ and $\big||\la|^2r^{\frac{3}{2}+\ep-\frac{2}{p}}\bb\big|_{p,S}$; they are obtained in the following way, denoting $L_1=[\ddb_{\la}\nabb\log\oom]$,
\bea
&&\ML\ML\frac{\pr}{\pr\la}\big|r^{2-\frac{2}{p}}\ze\big|^p=p|r^{2-\frac{2}{p}}\ze|^{p-2}
(r^{2-\frac{2}{p}}\ze)\c\frac{\pr}{\pr\la}(r^{2-\frac{2}{p}}\ze)\nn\\
&&\ML\ML\ML\ML=p|r^{2-\frac{2}{p}}\ze|^{p-2}(r^{2-\frac{2}{p}}\ze)\c\left\{-r^{2-\frac{2}{p}}\big(\oom\tr\chib\zeta+2\oom\chibh\c\ze+\oom\bb+L_1\big)+(2-\frac{2}{p})r^{1-\frac{2}{p}}\frac{\pr r}{\pr\la}\ze\right\}\nn\\
&&\ML\ML\ML\ML=p|r^{2-\frac{2}{p}}\ze|^{p-2}(r^{2-\frac{2}{p}}\ze)\c\left\{-r^{2-\frac{2}{p}}\big(\oom\tr\chib\zeta+2\oom\chibh\c\ze+\oom\bb+L_1\big)+(2-\frac{2}{p})r^{2-\frac{2}{p}}\frac{1}{2}\overline{\oom\tr\chib}\ze\right\}\nn\\
&&\ML\ML\ML\ML\ML=p|r^{2-\frac{2}{p}}\ze|^{p-2}\left\{\left[-(\oom\tr\chib)|r^{2-\frac{2}{p}}\ze|^2-2\oom(r^{2-\frac{2}{p}}\ze)\c\chibh\c(r^{2-\frac{2}{p}}\ze)
-r^{2(2-\frac{2}{p})}\oom\bb\c\ze-r^{2(2-\frac{2}{p})}L_1\c\ze\right]
+(1-\frac{1}{p})\overline{\oom\tr\chib}|r^{2-\frac{2}{p}}\ze|^2\right\}\nn\\
&&\ML\ML\ML\ML\ML=\left(-p\oom\tr\chib+p\overline{\oom\tr\chib}-\overline{\oom\tr\chib}\right)|r^{2-\frac{2}{p}}\ze|^p
-p2\oom(r^{2-\frac{2}{p}}\ze)\c\chibh\c(r^{2-\frac{2}{p}}\ze)|r^{2-\frac{2}{p}}\ze|^{p-2}-p|r^{2-\frac{2}{p}}\ze|^{p-2}r^{2(2-\frac{2}{p})}\oom\bb\c\ze\nn\\
&&\ML\ML\ML\ML-p|r^{2-\frac{2}{p}}\ze|^{p-2}r^{2(2-\frac{2}{p})}L_1\c\ze\nn\\
&&\ML\ML\ML\ML\ML=\left((p-1)(\overline{\oom\tr\chib}-\oom\tr\chib)-(\oom\tr\chib)\right)|r^{2-\frac{2}{p}}\ze|^p-2p\oom(r^{2-\frac{2}{p}}\ze)\c\chibh\c(r^{2-\frac{2}{p}}\ze)|r^{2-\frac{2}{p}}\ze|^{p-2}
-p|r^{2-\frac{2}{p}}\ze|^{p-2}r^{2(2-\frac{2}{p})}\oom\bb\c\ze\nn\\
&&\ML\ML\ML\ML-p|r^{2-\frac{2}{p}}\ze|^{p-2}r^{2(2-\frac{2}{p})}L_1\c\ze\ .
\eea
Therefore
\bea
&&\ML\ML\ML\frac{\pr}{\pr\la}|r^{2-\frac{2}{p}}\ze|_{p,S}^p=\frac{\pr}{\pr\la}\int_S|r^{2-\frac{2}{p}}\ze|^p
=\int_S\left(\frac{\pr}{\pr\la}|r^{2-\frac{2}{p}}\ze|^p+\oom\tr\chib|r^{2-\frac{2}{p}}\ze|^p\right)\\
&&\ML\ML\ML=-(p-1)\int_S(\oom\tr\chib-\overline{\oom\tr\chib})|r^{2-\frac{2}{p}}\ze|^p
-p\int_S2\oom(r^{2-\frac{2}{p}}\ze)\c\chibh\c(r^{2-\frac{2}{p}}\ze)|r^{2-\frac{2}{p}}\ze|^{p-2}+|r^{2-\frac{2}{p}}\ze|^{p-2}r^{2(2-\frac{2}{p})}\oom\bb\c\ze\nn\\
&&\ML\ML\ML\leq p\int_S\big|(|\oom\tr\chib-\overline{\oom\tr\chib}|+|2\oom\chibh|)|r^{2-\frac{2}{p}}\ze|\big||r^{2-\frac{2}{p}}\ze|^{p-1}
+p\int_S\left(|r^{(2-\frac{2}{p})}\oom\bb||r^{2-\frac{2}{p}}\ze|^{p-1}+|r^{(2-\frac{2}{p})}L_1||r^{2-\frac{2}{p}}\ze|^{p-1}\right)\ .\nn
\eea
Applying Hoder inequality to the r.h.s. we obtain
\bea
&&\ML\ML\ML\frac{\pr}{\pr\la}|r^{2-\frac{2}{p}}\ze|_{p,S}^p\leq \big|(|\oom\tr\chib-\overline{\oom\tr\chib}|+|2\oom\chibh|)|r^{2-\frac{2}{p}}\ze|\big|_{p,S}\ \!p\big|r^{2-\frac{2}{p}}\ze\big|_{p,S}^{p-1}\nn\\
&&+\big|r^{2-\frac{2}{p}}\oom\bb\big|_{p,S}\ \!p\big|r^{2-\frac{2}{p}}\ze\big|_{p,S}^{p-1}+\big|r^{2-\frac{2}{p}}L_1\big|_{p,S}\ \!p\big|r^{2-\frac{2}{p}}\ze\big|_{p,S}^{p-1}
\eea
and from it
\bea
&&\ML\ML\ML\frac{\pr}{\pr\la}|r^{2-\frac{2}{p}}\ze|_{p,S}\leq (|\oom\tr\chib-\overline{\oom\tr\chib}|_{\infty,S}+|2\oom\chibh|_{\infty,S})\big|r^{2-\frac{2}{p}}\ze|\big|_{p,S}+\big|r^{2-\frac{2}{p}}\oom\bb\big|_{p,S}+\big|r^{2-\frac{2}{p}}L_1\big|_{p,S}\ .\ \ \ \ \ \ \ \ \ \ \ 
\eea
\NI Let us obtain the differential inequality for $\bb$, with $L_2=-[\oom(\divv\aa+\nabb\log\oom\c\aa)]$,

\[\ddb_{\la}\bb+2\oom\tr\chib\bb+2\oom\omb\bb-\oom\aa\c\ze-[\oom(\divv\aa+\nabb\log\oom\c\aa)]=0\]
\bea
&&\ML\ML\frac{\pr}{\pr\la}\big|r^{4-\frac{2}{p}}\bb\big|^p=p|r^{4-\frac{2}{p}}\bb|^{p-2}
(r^{4-\frac{2}{p}}\bb)\c\frac{\pr}{\pr\la}(r^{4-\frac{2}{p}}\bb)\nn\\
&&\ML\ML=p|r^{4-\frac{2}{p}}\bb|^{p-2}(r^{4-\frac{2}{p}}\bb)\c\left\{-r^{4-\frac{2}{p}}\big(2\oom\tr\chib\ \!\bb+2\oom\omb\bb-\oom\ze\c\aa+L_2\big)
+(4-\frac{2}{p})r^{3-\frac{2}{p}}\frac{\pr r}{\pr\la}\bb\right\}\nn\\
&&\ML\ML=p|r^{4-\frac{2}{p}}\bb|^{p-2}(r^{4-\frac{2}{p}}\bb)\c\left\{-r^{4-\frac{2}{p}}\big(2\oom\tr\chib\ \!\bb+2\oom\omb\bb-\oom\ze\c\aa+L_2\big)
+(4-\frac{2}{p})r^{4-\frac{2}{p}}\frac{1}{2}\overline{\oom\tr\chib}\ \!\bb\right\}\nn\\
&&\ML\ML=p|r^{4-\frac{2}{p}}\bb|^{p-2}\left\{\left[-(2\oom\tr\chib)|r^{4-\frac{2}{p}}\bb|^2-2\oom\omb|r^{4-\frac{2}{p}}\bb|^2
+r^{2(4-\frac{2}{p})}\oom\ \!\ze\c\aa\c\bb-r^{2(4-\frac{2}{p})}L_2\c\bb\right]\right.\nn\\
&&\ML\left.+(2-\frac{1}{p})\overline{\oom\tr\chib}|r^{4-\frac{2}{p}}\bb|^2\right\}\nn\\
&&\ML\ML=\left((2p-1)(\overline{\oom\tr\chib}-\oom\tr\chib)-(\oom\tr\chib)\right)|r^{4-\frac{2}{p}}\bb|^p-2p\oom\omb|r^{4-\frac{2}{p}}\bb|^p
+p|r^{4-\frac{2}{p}}\bb|^{p-2}r^{2(4-\frac{2}{p})}\oom\ \!\ze\c\aa\c\bb\nn\\
&&\ML\ML-p|r^{4-\frac{2}{p}}\bb|^{p-2}r^{2(4-\frac{2}{p})}L_2\c\bb\ .
\eea
Therefore
\bea
&&\ML\ML\ML\ML\ML\frac{\pr}{\pr\la}|r^{4-\frac{2}{p}}\bb|_{p,S}^p=\frac{\pr}{\pr\la}\int_S|r^{4-\frac{2}{p}}\bb|^p
=\int_S\left(\frac{\pr}{\pr\la}|r^{4-\frac{2}{p}}\bb|^p+\oom\tr\chib|r^{4-\frac{2}{p}}\bb|^p\right)\nn\\
&&\ML\ML\ML\ML\ML\ML\ML=-(2p-1)\!\int_S(\oom\tr\chib-\overline{\oom\tr\chib})|r^{4-\frac{2}{p}}\bb|^p
-p\int_S2\oom\omb|r^{4-\frac{2}{p}}\bb|^p-|r^{4-\frac{2}{p}}\bb|^{p-2}r^{2(4-\frac{2}{p})}\oom\ \!\ze\c\aa\c\bb
+|r^{4-\frac{2}{p}}\bb|^{p-2}r^{2(4-\frac{2}{p})}L_2\c\bb\nn\\
&&\ML\ML\ML\ML\ML\ML\leq 2p\!\int_S\big|(|\oom\tr\chib-\overline{\oom\tr\chib}|+|2\oom\omb|)|r^{4-\frac{2}{p}}\bb|\big||r^{4-\frac{2}{p}}\bb|^{p-1}+p\int_S\left(|r^{(4-\frac{2}{p})}\oom\ \!\ze\c\aa||r^{4-\frac{2}{p}}\bb|^{p-1}+|r^{(4-\frac{2}{p})}L_2||r^{4-\frac{2}{p}}\bb|^{p-1}\right)\ .
\nn\eea
Applying Hoder inequality to the r.h.s. we obtain
\bea
&&\ML\ML\ML\frac{\pr}{\pr\la}|r^{4-\frac{2}{p}}\bb|_{p,S}^p\leq \big|(|\oom\tr\chib-\overline{\oom\tr\chib}|+|2\oom\omb|)|r^{4-\frac{2}{p}}\bb|\big|_{p,S}\ \!2p\big|r^{4-\frac{2}{p}}\bb\big|_{p,S}^{p-1}\nn\\
&&+\big|r^{4-\frac{2}{p}}\oom\ze\c\aa\big|_{p,S}\ \!p\big|r^{4-\frac{2}{p}}\bb\big|_{p,S}^{p-1}+\big|r^{4-\frac{2}{p}}L_2\big|_{p,S}\ \!p\big|r^{4-\frac{2}{p}}\bb\big|_{p,S}^{p-1}
\eea
and from it
\bea
\ML\ML\ML\ML\frac{\pr}{\pr\la}|r^{4-\frac{2}{p}}\bb|_{p,S}\leq (2|\oom\tr\chib-\overline{\oom\tr\chib}|_{\infty,S}+|4\oom\omb|_{\infty,S})\big|r^{4-\frac{2}{p}}\bb|\big|_{p,S}+\big|r^{4-\frac{2}{p}}\oom\ \!\ze\c\aa\big|_{p,S}+\big|r^{4-\frac{2}{p}}L_2\big|_{p,S}\ .\ \ \ \ \ \ \ \ \ \ \ 
\eea
Collecting the results we have the following p.d.e. inequalities,
\bea
&&\ML\ML\ML\ML\frac{\pr}{\pr\la}|r^{2-\frac{2}{p}}\ze|_{p,S}\leq (|\oom\tr\chib-\overline{\oom\tr\chib}|_{\infty,S}+|2\oom\chibh|_{\infty,S})\big|r^{2-\frac{2}{p}}\ze|\big|_{p,S}+\big|r^{2-\frac{2}{p}}\oom\bb\big|_{p,S}+\big|r^{2-\frac{2}{p}}L_1\big|_{p,S}\ \ \ \ \ \ \ \ \ \ \ \ \\
&&\ML\ML\ML\ML\frac{\pr}{\pr\la}|r^{4-\frac{2}{p}}\bb|_{p,S}\leq (2|\oom\tr\chib-\overline{\oom\tr\chib}|_{\infty,S}+|4\oom\omb|_{\infty,S})\big|r^{4-\frac{2}{p}}\bb|\big|_{p,S}+\big|r^{4-\frac{2}{p}}\oom\ \!\ze\c\aa\big|_{p,S}+\big|r^{4-\frac{2}{p}}L_2\big|_{p,S}\ .\nn
\eea
Integrating along $\Cb_0$ from $S_0=S(\la_0,\nu_0)$ to $S(\la,\nu_0)$ we obtain
\bea
&&\ML\ML\ML\ML|r^{2-\frac{2}{p}}\ze|_{p,S}(\la,\nu_0)\leq c\left(|r^{2-\frac{2}{p}}\ze|_{p,S}(\la_0,\nu_0)
+\int_{\la_0}^{\la}d\la'\left(\big|r^{2-\frac{2}{p}}\oom\bb\big|_{p,S}+\big|r^{2-\frac{2}{p}}L_1\big|_{p,S}\right)\!(\la',\nu_0)\right)\ \ \ \ \ \ \ \ \ \ \\
&&\ML\ML\ML\ML|r^{4-\frac{2}{p}}\bb|_{p,S}(\la,\nu_0)\leq c\left(|r^{4-\frac{2}{p}}\bb|_{p,S}(\la_0,\nu_0)
+\int_{\la_0}^{\la}d\la'\left(\big|r^{4-\frac{2}{p}}\oom\ \!\ze\c\aa\big|_{p,S}+\big|r^{4-\frac{2}{p}}L_2\big|_{p,S}\right)\!(\la',\nu_0)\right)\ .\ \ \ \ \ \ \ \ \nn
\eea
Before getting the estimates for $\ze$ and $\bb$ we modify 
  the second one multiplying both sides with 
$|\la|^{\frac{3}{2}+\de}/r^{2-\ep}$, with $\ep>0,\de\geq 0$ which we are allowed to do as this function is increasing in $r$ if $\nu_0\leq 2r$. In fact, assume $|\la|=2r-\nu_0$ on $\Cb_0$ if $r\geq \nu_0/2$ which implies on $\Cb_0$, $r\geq t$ , therefore points outside the null cone with vertex in the origin (in Minkowski spacetime). It follows
\bea
&&\ML\ML\frac{d}{dr}\frac{|\la|^{\frac{3}{2}+\de}}{r^{2-\ep}}=2({3}/{2}+\de)\frac{(2r-\nu_0)^{\frac{1}{2}+\de}}{r^{2-\ep}}
-(2-\ep)\frac{(2r-\nu_0)^{\frac{3}{2}+\de}}{r^{3-\ep}}\\
&&\ML\ML=(2-\ep)\frac{(2r-\nu_0)^{\frac{1}{2}+\de}}{r^{2-\ep}}\left[\frac{2({3}/{2}+\de)}{2-\ep}-\frac{(2r-\nu_0)}{r}\right]\nn\\
&&\ML\ML=(2-\ep)\frac{(2r-\nu_0)^{\frac{1}{2}+\de}}{r^{2-\ep}}\left[\frac{2({3}/{2}+\de)}{2-\ep}-\frac{r-t}{r}\right]
\geq (2-\ep)\frac{(2r-\nu_0)^{\frac{1}{2}+\de}}{r^{2-\ep}}\left[({3}/{2}+\de)-1\right]\geq 0\ .\nn
\eea
 Therefore 
\bea
&&\ML\ML\ML\ML\ML\ML|r^{2-\frac{2}{p}}\ze|_{p,S}(\la,\nu_0)\leq c\left(|r^{2-\frac{2}{p}}\ze|_{p,S}(\la_0,\nu_0)
+\int_{\la_0}^{\la}d\la'\left(\big|r^{2-\frac{2}{p}}\oom\bb\big|_{p,S}+\big|r^{2-\frac{2}{p}}L_1\big|_{p,S}\right)\!(\la',\nu_0)\right)\ \ \ \ \ \ \ \ \ \ 
\eql{indataest}\\
&&\ML\ML\ML\ML\ML\ML||\la|^{\frac{3}{2}+\de}r^{2+\ep-\frac{2}{p}}\bb|_{p,S}(\la,\nu_0)\leq c\!\left(||\la|^{\frac{3}{2}+\de}r^{2+\ep-\frac{2}{p}}\bb|_{p,S}(\la_0,\nu_0)\!
+\!\int_{\la_0}^{\la}d\la'\left(\big||\la|^{\frac{3}{2}+\de}r^{2+\ep-\frac{2}{p}}\oom\ \!\ze\c\aa\big|_{p,S}+\big||\la|^{\frac{3}{2}+\de}r^{2+\ep-\frac{2}{p}}L_2\big|_{p,S}\right)\!\!\right)\!
 .\nn
\eea
the integral estimates \ref{indataest} to be meaningful require that the integrals are bounded, which happens if we require that on $\Cb_0$, with ${\tilde\de}>\de\geq 0$, 
\bea
\ML\ML\ML \ddb_{\la}\nabb\oom=O\left(r^{-(2+\ep)}|\la|^{-(1+{\tilde\de})}\right)\ \ ,\ \ 
\aa=O\left(r^{-(1+\ep)}|\la|^{-(\frac{5}{2}+{\tilde\de})}\right) .\ \ \ \ \ \ \eql{Cb0cond}
\eea
and prove that $\bb=O\left(r^{-(2+\ep)}|\la|^{-\frac{3}{2}}\right)\ \ ,\ \ \ze=O\left(r^{-2}\right)$.
We can choose $\de=0$ and the two integral inequalities become
\bea
&&\ML\ML\ML\ML\ML|r^{2+\ep-\frac{2}{p}}\ze|_{p,S}(\la,\nu_0)\leq c\!\left(|r^{2+\ep-\frac{2}{p}}\ze|_{p,S}(\la_0,\nu_0)
+\int_{\la_0}^{\la}d\la'\left(\big|r^{2+\ep-\frac{2}{p}}\oom\bb\big|_{p,S}+\big|r^{2+\ep-\frac{2}{p}}L_1\big|_{p,S}\right)\!\right)\ \ \ \ \ \ \ \ \ \eql{indataest2}\\
&&\ML\ML\ML\ML\ML\ML||\la|^{\frac{3}{2}}r^{2+\ep-\frac{2}{p}}\bb|_{p,S}(\la,\nu_0)\leq c\!\left(||\la|^{\frac{3}{2}}r^{2+\ep-\frac{2}{p}}\bb|_{p,S}(\la_0,\nu_0)\!
+\!\int_{\la_0}^{\la}d\la'\left(\big||\la|^{\frac{3}{2}}r^{2+\ep-\frac{2}{p}}\oom\ \!\ze\c\aa\big|_{p,S}+\big||\la|^{\frac{3}{2}}r^{2+\ep-\frac{2}{p}}L_2\big|_{p,S}\right)\!\!\right)\! .\nn
\eea
It is easy to prove that the expected result holds, namely
\[\bb=O\left(r^{-(2+\ep)}|\la|^{-\frac{3}{2}}\right)\ \ ,\ \ \ze=O\left(r^{-2}\right)\ ,\]
assuming \ref{Cb0cond}  and moreover that on $S_0$ we have
\bea
|r^{2-\frac{2}{p}}\ze|_{p,S}(\la_0,\nu_0)\leq c\varepsilon\ \ ,\ \ ||\la|^{\frac{3}{2}}r^{2+\ep-\frac{2}{p}}\bb|_{p,S}(\la_0,\nu_0)\leq c\varepsilon\ .
\eea
{\bf Remarks:} {\em 

\NI i) It is clear that this estimate has to be generalized to all the $\nabb$ derivatives of $\ze$ and $\bb$ with the standard mechanism we used to prove the estimates for the $\nabb$ derivatives in the interior. The way to proceed should be again an inductive one, we assume the desired estimates for $\nabb^J\ze$ and $\nabb^J\bb$, with $J<N$ and prove the same estimates for $J=N$. This implies a correct choice of ${\underline{\Ga}}_0$.
\smallskip

\NI ii) Observe that the way to estimate $\ze$ on $\Cb_0$ is different from the one used on $C_0$. Here the only loss of derivatives is due to the presence of the term $\bb$. }
\subsubsection{The initial data on $S_0=S(\la_0,\nu_0)$}

From the previous considerations concerning the estimates of the quantities on  $\Cb_0$ it follows that the following behaviours have to be required on $S_0$, to satisfy conditions \ref{rieminidecay2} 
\smallskip

\NI i) From the estimates on $\Cb_0$, $J>0$:
\bea
&&|r^{2+J-\frac{2}{p}}\nabb^J\ze|_{p,S}(\la_0,\nu_0)\leq c_1\varepsilon\left(\frac{(J+1)!}{(J+1)^{\a}}\frac{e^{(J-2)(\de_0+\Ga_{0})}}{\ro_{0,0}^J}\right)\nn\\
&&||\la|^{\frac{3}{2}}r^{2+\ep-\frac{2}{p}}\nabb^J\bb|_{p,S}(\la_0,\nu_0)\leq c_2\varepsilon\left(\frac{(J+1)!}{(J+1)^{\a}}\frac{e^{(J-2)(\de_0+\Ga_{0})}}{\ro_{0,0}^J}\right)\nn\\
&&|r^{2+J-\frac{2}{p}}\nabb^J\tr\chib|_{p,S}(\la_0,\nu_0)\leq c_3\varepsilon\left(\frac{J!}{J^{\a}}\frac{e^{(J-2)(\de_0+\Ga_{0,0})}}{\ro_{0}^J}\right)\nn\\
&&||\la|^{\frac{3}{2}}r^{1+J+\ep-\frac{2}{p}}\nabb^J\chibh|_{p,S}(\la_0,\nu_0)\leq c_4\varepsilon\left(\frac{J!}{J^{\a}}\frac{e^{(J-2)(\de_0+\Ga_{0})}}{\ro_{0,0}^J}\right)\ ,\ \ \ \ \ \ \ \ \ \ \eql{S0cond1}
\eea
together with the requirements that on $S_0$
\bea
\nabb\tr\chib-\ze\tr\chib=O\left(\varepsilon|\la_0|^{\frac{3}{2}+\ep}r_0^{2}\right)\ . \eql{S0const}
\eea
{\bf Remarks:} {\em 
\smallskip

\NI i) The first two conditions of \ref{S0cond1} are required to obtain the estimates for $\nabb^J\ze$ on $\Cb_0$. The last  condition, \ref{S0const}, is needed to have $\bb$ with the right estimates on $\Cb_0$ , this is, basically a repetition of what has been done for $\b$ on $C_0$. 
To clarify this point let us recall the expression of $\bb$ 
\[\bb=-(\nabb\tr\chib-\divv\chib-\zeta\tr\chib+\ze\c\chib)\ ,\]
and from it the following one
\bea
\nabb^J\bb=-\frac{1}{2}\nabb^{J+1}\tr\chib+\nabb^{J}\divv\chibh+\frac{1}{2}\nabb^J(\zeta\tr\chib)-\nabb^J(\ze\c\chibh)\ .
\eea
From the requirement that the l.h.s. decays as $O\left(|\la|^{\frac{3}{2}}r^{2+\ep}\right)$ it follows that this is automatically satisfied by $\nabb^J\divv\chibh$ and by $\nabb^J(\ze\c\chibh)$ while it is not from the terms, considered separately, $-\frac{1}{2}\nabb^{J+1}\tr\chib$ and $\frac{1}{2}\nabb^J(\zeta\tr\chib)$; to fulfill the condition we have, therefore, to require condition \ref{S0const}.
\smallskip

\NI ii) Notationally we recall that the various $\ro$'s used in the initial data estimates satisfy the following inequalities
\bea
\ro_0<\ro_{0,0,1}<\ro_{0,0}\ ,\eql{roineq}
\eea
where $\ro_0$ will be used when we obtain the internal estimates.}

\medskip

\NI{\bf The condition for $\underline{\omega}$ on $S_0$:}
\medskip

\NI We start from 
\bea
\frac{d}{d\nu}\omb-2\oom\om\omb=F
\eea
where
\bea
F=\ze\c\nabb\log\oom\!+\!\frac{3}{2}|\ze|^2\!-\!\frac{1}{2}|\nabb\log\oom|^2\!-\frac{1}{2}\!\big({\bf
K}\!+\!\frac{1}{4}\tr\chi\tr\chib\!-\!\frac{1}{2}\chih\c\chibh\big)
\eea
Let us observe that
\bea
\frac{d}{d\nu}\left(e^{\int_{\nu_0}^{\nu}(-2\oom\om)d\nu'}\omb\right)=e^{\int_{\nu_0}^{\nu}(-2\oom\om)d\nu'}\left[-2\oom\om\omb+\frac{d}{d\nu}\omb\right]=e^{\int_{\nu_0}^{\nu}(-2\oom\om)d\nu'}F(\nu)\ .\ \ \ \ 
\eea
Integrating
\bea
e^{\int_{\nu_0}^{\nu}(-2\oom\om)d\nu'}\omb(\nu)-\omb(\nu_0)=\int_{\nu_0}^{\nu}d\nu'e^{\int_{\nu_0}^{\nu'}(-2\oom\om)d\nu''}F(\nu')
\eea
Therefore
\bea
\omb(\nu)=e^{\int_{\nu_0}^{\nu}(2\oom\om)d\nu'}\omb(\nu_0)+e^{\int_{\nu_0}^{\nu}(2\oom\om)d\nu'}\int_{\nu_0}^{\nu}d\nu'e^{\int_{\nu_0}^{\nu'}(-2\oom\om)d\nu''}F(\nu')
\eea
and to require the $\lim_{\nu\rightarrow\infty}\omb(\nu)=0$ we need that
\bea
e^{\int_{\nu_0}^{\infty}(2\oom\om)d\nu'}\left[\omb(\nu_0)+\int_{\nu_0}^{\infty}d\nu'e^{\int_{\nu_0}^{\nu'}(-2\oom\om)d\nu''}F(\nu')
\right]=0
\eea
which implies that
\bea
&&\omb(\nu_0)=-\int_{\nu_0}^{\infty}d\nu'e^{\int_{\nu_0}^{\nu'}(-2\oom\om)d\nu''}F(\nu')=\nn\\
&&=-\int_{\nu_0}^{\infty}d\nu'e^{\int_{\nu_0}^{\nu'}(-2\oom\om)d\nu''}\left[\ze\c\nabb\log\oom\!+\!\frac{3}{2}|\ze|^2\!-\!\frac{1}{2}|\nabb\log\oom|^2\!-\frac{1}{2}\!\big({\bf K}\!+\!\frac{1}{4}\tr\chi\tr\chib\!-\!\frac{1}{2}\chih\c\chibh\big)\ .\nn
\right]\!(\nu').
\eea
\NI In conclusion what we proved in the previous subsections complete the proof of Theorem \ref{Thinitialdata}.
\newpage
\bibliographystyle{math}


\end{document}